# Millisecond Pulsars in Close Binaries

Habilitationsschrift, Universität Bonn

vorgelegt von

Dr. Thomas M. Tauris

© Juni 2014

zur Erlangung der Venia Legendi
der Hohen Mathematisch-Naturwissenschaftlichen Fakultät
der *Rheinischen Friedrich-Wilhelms-Universität Bonn*

# Foreword

In this *Habilitationsschrift* I present my research carried out over the last four years at the Argelander-Insitut für Astronomie (AIfA) and the Max-Planck-Institut für Radioastronomie (MPIfR). The thesis summarizes my main findings and has been written to fulfill the requirements for the Habilitation qualification at Universität Bonn. Although my work is mainly focused on the topic of millisecond pulsars, there is a fairly broad spread of research areas ranging from the formation of neutron stars in various supernova events, to their evolution, for example, via accretion processes in binary and triple systems, and finally to their possible destruction in merger events.

The thesis is organized in the following manner: A general introduction to neutron stars and millisecond pulsars is given in Chapter 1. A selection of key papers published in 2011–2014 are presented in Chapters 2–10, ordered within five main research areas. A summary of the main results of the thesis is given in Chapter 11, together with a brief description of ongoing projects and a future outlook.

Disclaimer: Although I contributed with the major part of all the work presented here, I did not take part in the observations, data analysis and gravity test calculations presented in Chapter 4. Nor did I take part in the observations or data analysis in Chapter 7. Slight differences (typos or minor language editing) may in some cases appear in comparison with the actual journal versions of the published papers. Hence, for citing these papers please refer to the journals.

I am profoundly grateful to Norbert Langer (AIfA) for many inspiring discussions on stellar and binary evolution and for generous financial support. I also thank him for his patience with my occasional rusty memory of certain fundamental processes in stars. The idea of getting me back into astrophysics, and to come to Bonn, was fostered by Michael Kramer (MPIfR) in 2010. I am deeply thankful for this splendid initiative. The main purpose of inviting me to Bonn was to link pulsar observations with stellar evolution and thus create tighter relations between these two large research groups in Bonn. I hope that I succeeded in this task. In addition, I cordially thank *all* my pulsar colleagues for many stimulating discussions, and in particular Paulo Freire for his eternal enthusiasm, Norbert Wex for always offering his expertise and help when needed, and John Antoniadis for many discussions. I thank Ed van den Heuvel (my mentor in this field for more than 20 years) and all my other collaborators, and my PhD students Alina Istrate and Matthias Kruckow for their fine work. Last, but not least, I thank my family (and in particular Birgitte) for always supporting me, being indulgent and accepting the many evenings and weekends spend on research. *Af hjertet tak!*

Thomas M. Tauris
Bonn, June 2014

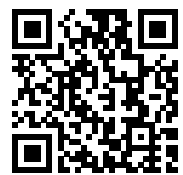

# List of publications included in this Habilitationsschrift

- **Tauris** & van den Heuvel (2014), ApJ Letters 781, L13
  *Formation of the Galactic Millisecond Pulsar Triple System PSR J0337+1715
  – A Neutron Star with Two Orbiting White Dwarfs*

- Freire & **Tauris** (2014), MNRAS Letters 438, 86
  *Direct formation of millisecond pulsars from rotationally delayed accretion-induced
  collapse of massive white dwarfs*

- Lazarus, **Tauris**, Knispel, Freire, et al. (2014), MNRAS 437, 1485
  *Timing of a young mildly recycled pulsar with a massive white dwarf companion*

- **Tauris**, Langer, Moriya, Podsiadlowski, Yoon & Blinnikov (2013), ApJ Letters 778, L23
  *Ultra-stripped Type Ic Supernovae from Close Binary Evolution*

- **Tauris**, Sanyal, Yoon & Langer (2013), A&A 558, 39
  *Evolution towards and beyond accretion-induced collapse of massive white dwarfs and
  formation of millisecond pulsars*

- Antoniadis, Freire, Wex, **Tauris**, et al. (2013), Science 340, 448
  *A Massive Pulsar in a Compact Relativistic Binary*

- **Tauris** (2012), Science 335, 561
  *Spin-Down of Radio Millisecond Pulsars at Genesis*

- **Tauris**, Langer & Kramer (2012), MNRAS 425, 1601
  *Formation of millisecond pulsars with CO white dwarf companions – II.
  Accretion, spin-up, true ages and comparison to MSPs with He white dwarf companions*

- **Tauris**, Langer & Kramer (2011), MNRAS 416, 2130
  *Formation of millisecond pulsars with CO white dwarf companions – I.
  PSR J1614-2230: evidence for a neutron star born massive*

# Additional publications in 2011–2014

- **Tauris**, Kaspi, Breton, Deller, et al. (2014), SKA Science Book, Chapter
  *Understanding the Neutron Star Population*

- Papitto, Torres, Rea & **Tauris** (2014), A&A, in press
  *Spin frequency distributions of binary millisecond pulsars*

- Guillemot & **Tauris** (2014), MNRAS 439, 2033
  *On the non-detection of γ-rays from energetic millisecond pulsars*
  *– dependence on viewing geometry*

- Ng **et al.** (2014), MNRAS 439, 1865
  *The High Time Resolution Universe pulsar survey*
  *– X. Discovery of four millisecond pulsars and updated timing solutions of a further 12*

- Chen, Chen, **Tauris** & Han (2013), ApJ 775, 27
  *Formation of Black Widows and Redbacks*
  *– Two Distinct Populations of Eclipsing Binary Millisecond Pulsars*

- Ivanova **et al.** (2013), A&ARv 21, 59
  *Common envelope evolution: where we stand and how we can move forward*

- Antoniadis **et al.** (2012), MNRAS 423, 3316
  *The relativistic pulsar-white dwarf binary PSR J1738+0333*
  *– I. Mass determination and evolutionary history*

- Mason **et al.** (2012), MNRAS 422, 199
  *The evolution and masses of the neutron star and donor star in the*
  *high mass X-ray binary OAO 1657−415*

- Freire **et al.** (2011), Science 334, 1107
  *Fermi Detection of a Luminous γ-Ray Pulsar in a Globular Cluster*

- Lazaridis, Verbiest, **Tauris**, Stappers, et al. (2011), MNRAS 414, 3134
  *Evidence for gravitational quadrupole moment variations in the companion of PSR J2051−0827*

# Manuscripts currently in preparation – to be submitted shortly

- **Tauris**, Langer & Podsiadlowski (2014), MNRAS, in prep.
  *Ultra-stripped Type Ic SNe: electron capture vs iron core collapse*

- Istrate, **Tauris**, & Langer (2014), A&A, in prep.
  *Evolution of low-mass X-ray binaries near the bifurcation period*
  *– Formation of millisecond pulsars with helium white dwarfs in tight binaries*

- Istrate, **Tauris**, Langer & Antoniadis (2014), A&A Letters(?), in prep.
  *Formation of low-mass helium white dwarfs in close binaries*

- Istrate, **Tauris**, Antoniadis & Langer (2014), A&A, in prep.
  *Formation of the 2 $M_\odot$ pulsar J0348+0432*

- **Tauris** et al. (2014), ApJ Letters(?), in prep.
  *Hypervelocity Stars Originating from Disrupted Binaries*

# Contents

# INTRODUCTION

Neutron stars (NSs) are unique creations in nature. They are formed violently as the remnants of massive stars which undergo a supernova (SN) explosion once their nuclear fusion has been exhausted. NSs allow for fundamental studies in all disciplines of physics in the most extreme regimes – under conditions which are impossible to imitate in any laboratory on Earth. They host the densest matter in the observable Universe and possess relativistic magnetospheres with outflowing energetic plasma winds. Therefore, being a rapidly spinning and magnetized NS, only about 10 km in radius and emitting energy at a rate $\sim 10^5$ times that of our Sun, radio pulsars constitute an interesting challenge for astrophysicists. Millisecond pulsars (MSPs) are of special interest since they are old NSs which have been spun up to very high rotation frequencies (in some cases $> 700$ Hz) via accretion of mass and angular momentum from a companion star in a binary system. It is these MSPs which constitute the main topic of this Habilitation thesis.

The evolution and interactions of binary stars play a central role in many areas of modern astrophysics: from the progenitors of different classes of SN explosions, accretion processes in X-ray binaries and formation of MSPs, to understanding gravitational wave sources such as colliding NSs/black holes, and gamma-ray bursts (GRBs, the most violent and energetic events in the known Universe), as well as aspects of nucleosynthesis and chemical enrichment of the interstellar medium in galaxies. MSPs represent the end point of binary stellar evolution. Their observed orbital and stellar properties are fossil records of their evolutionary history and hence binary pulsar systems are key probes of stellar astrophysics with the related forces and interactions of matter at work. Furthermore, MSPs are ultra stable clocks which allow for unprecedented tests of gravitational theories in the strong-field regime and they help us to investigate alternative gravity theories that try to explain dark matter and dark energy.

The first direct detection of gravitational waves from merging NSs in binaries is expected from the LIGO/VIRGO observatories within the next 2–3 years. This will open an entirely new window to the Universe and serve as a unique investigation tool for fundamental astrophysics. An analogy can be made to the 1960's when new technologies broadened the horizon beyond optical astronomy and made it possible to detect radio, infrared and X-rays from the Universe. The different frequencies of these electromagnetic waves led to the discovery of extreme objects like quasars and pulsars which have shown to be excellent test grounds for investigating physics. Gravitational waves will lead to a similar revolution in astrophysics, and in Bonn we are preparing for this upcoming challenge. In this thesis work, I will concentrate on the formation and evolution of binary NSs before they merge; for ongoing and future projects related to the detection rates of LIGO/VIRGO sources, see the final section (Chapter 11).

Since their discovery in the late 1960's the total population of known NSs has grown to more than 2600 sources. The last five decades of observations have yielded many surprises and demonstrated that the observational properties of NSs are remarkably diverse. The recent era of multi-wavelength observations has revealed a greater variety of possibly distinct observational classes of NSs than ever before. In addition to isolated NSs, these compact objects are also found in binaries and even triple stellar systems which stimulates theoretical research on their origin and evolution, besides enabling precise mass measurements via their companion stars. With emission spanning the entire electromagnetic spectrum, and some NSs showing strange transient behaviour and even dramatic high-energy outbursts, such incredible range and diversity is not only unpredicted, but in many ways astonishing given the perhaps naively simple nature of the NS – the last stellar bastion before the total collapse to a black hole. But why do NSs exhibit so much 'hair'? Even within the field there is confusion about the sheer number of different NS class nomenclature. For a recent review on the members of the NS zoo and the possible unification of the various flavours, I refer to Kaspi (2010) and references therein. An overview of the formation and evolution of NSs in binaries is given, for example, in Bhattacharya & van den Heuvel (1991) and Tauris & van den Heuvel (2006).

In the rest of this chapter, I give a broad introduction to NSs: their discovery and their various manifestations as compact objects with an emphasis on their basic observational properties as radio pulsars, their formation and evolution, their structures and masses, as well as a brief description of the spin-up process (*recycling*) of MSPs.
My recent research is presented in the subsequent chapters covering five areas:

i) Ultra-stripped SNe in close binaries (Chapter 2),

ii) Massive NSs in close binaries (Chapters 3 and 4),

iii) Spin-up of MSPs (Chapters 5–7),

iv) Formation of a triple MSP (Chapter 8), and

v) Accretion-induced collapse of white dwarfs and formation of MSPs (Chapters 9 and 10).

In Chapter 11, I give a summary and highlight ongoing projects and further outlook.

## 1.1 Discovery of neutron stars

NSs were introduced as a purely theoretical concept, apparently even *before* Chadwick's discovery of the neutron in 1932. According to a well-known recollection of Rosenfeld (1974), he met in Copenhagen with Bohr and Landau in the spring of 1932 to discuss possible implications of Chadwick's discovery of the neutron in Cambridge a few weeks earlier. However, in a recent historical investigation by Yakovlev et al. (2013) it is argued that, in fact, Landau, Bohr and Rosenfeld met in Copenhagen already in 1931 and that they were discussing a paper submitted by Landau earlier that year (i.e. before the discovery of the neutron). In that paper, Landau suggested the existence of dense stars that look like one giant nucleus.
A well-documented and an even more prophetic statement was made a few years later in a famous paper by Baade & Zwicky (1934) who also explained the formation of NSs:

> *"With all reserve we advance the view that supernovae represent the transitions from ordinary stars into neutron stars, which in their final stages consist of extremely closely packed neutrons."*

This incredible prediction was not confirmed until the late 1960's. The first radio pulsar was observed on November 28, 1967, by Jocelyn Bell Burnell and Antony Hewish (Hewish et al. 1968). As a spurious coincidence, it is worth mentioning that just two weeks before this ground-braking discovery, Pacini (1967) had pointed out that if NSs were spinning and had large magnetic fields then electromagnetic waves would be emitted. Actually, earlier that very same year Shklovsky (1967) examined X-ray and optical observations of Scorpius X-1 and concluded correctly that the radiation comes from an accreting NS system. The final proof of Baade and Zwicky's prediction on the formation of NSs via SNe came with the discovery of the Crab and Vela pulsars in 1968 – both pulsars are located inside gaseous SN remnants.

## 1.2 The kaleidoscopic neutron star population

Fig. 1 shows a plot of all currently known radio pulsars with measured values of spin period ($P$) and their time derivative ($\dot{P}$). The classic *radio pulsars* (red dots) are concentrated in the region with $P \simeq 0.2 - 2$ sec and $\dot{P} \simeq 10^{-16} - 10^{-13}$. They have magnetic fields of the order $B \simeq 10^{10} - 10^{13}$ G and lifetimes as radio sources of a few $10^7$ yr. The population plotted in blue circles are binary pulsars and they clearly indicate a connection to the rapidly spinning MSPs. The pulsars marked with stars indicate young pulsars observed inside, or near, their gaseous SN ejecta remnants. Then there are the "drama queens" of the NS population: the *magnetars* which undergo bursts and are powered by their huge magnetic energy reservoirs. Also plotted are the mysterious *rotating radio transients* (RRATs), the *isolated X-ray dim NSs* (XDINS) and the *central compact objects* (CCOs). In addition, there are other exotic radio pulsars such as the *intermittent pulsars*, the *black widows* and *redbacks*, and pulsars in triple systems. A main challenge of the past decade was – and continues to be – to find a way to unify this variety into a coherent physical picture. The NS zoo raises essential questions like: What determines whether a NS will be born with, for example, magnetar-like properties or as a Crab-like pulsar? What are the branching ratios for the various varieties, and, given estimates of their lifetimes, how many of each are there in the Galaxy? Can individual NSs evolve from one species to another (and why)? How does a NS interact with a companion star during and near the end of mass transfer (recycling)? And what limits its final spin period? Ultimately such questions are fundamental to understanding the fate of massive stars, close binary evolution and the nature of core-collapse SNe, while simultaneously relating to a wider variety of interesting fundamental physics and astrophysics questions ranging from the nature of matter in extremely high magnetic fields, to the equation-of-state of ultra-dense matter and details of accretion processes.

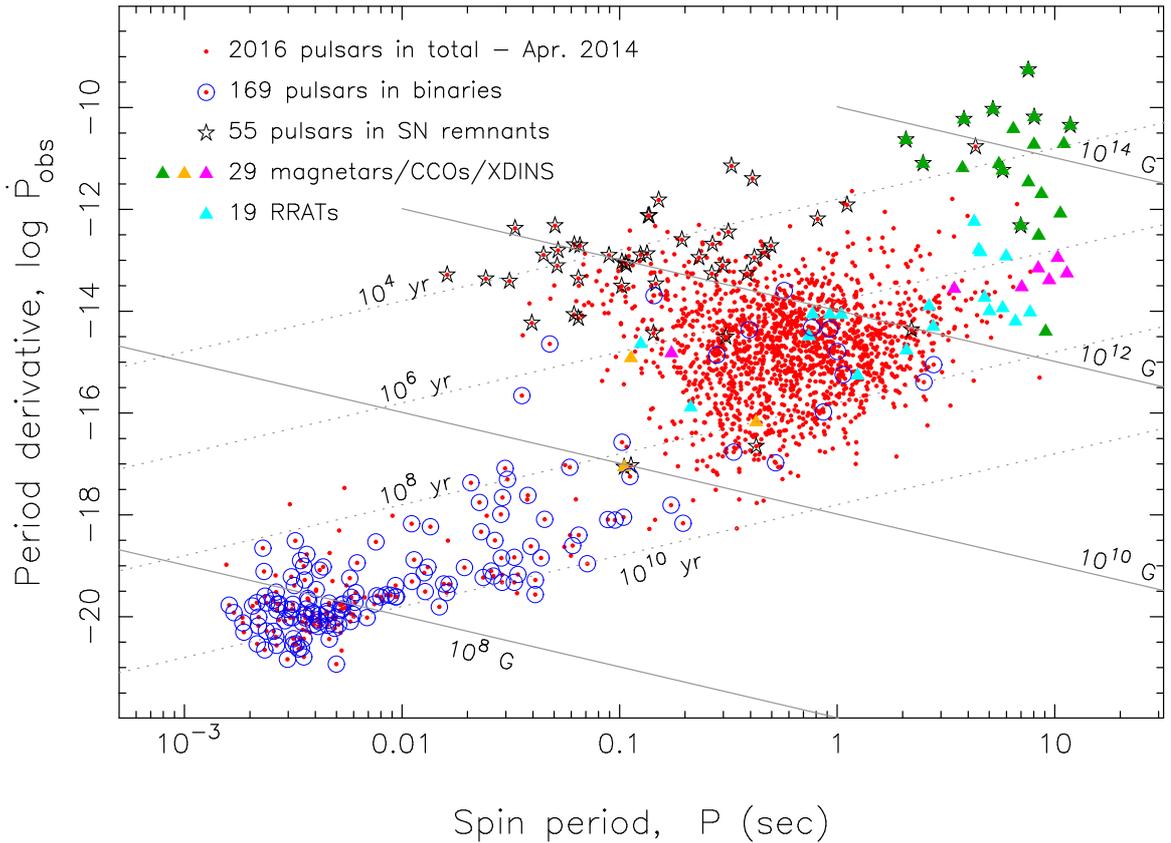

Figure 1: All currently known pulsars with measured values of $P$ and $\dot{P}$. Data from the ATNF Pulsar Catalogue (Manchester et al. 2005) – version 1.49, April 2014. For a more complete listing of known magnetars, see the McGill Online Magnetar catalog (Olausen & Kaspi 2014). Lines of constant characteristic age and constant B-field are marked (see Sections 1.4 and 1.5 for explanations).

## 1.3 Basic observational properties of radio pulsars

The "*pulsar lighthouse model*" is illustrated in Fig. 2 and explained in more detail in the following subsections. For further descriptions of the properties of radio pulsars I refer to Lorimer & Kramer (2004), and references therein.

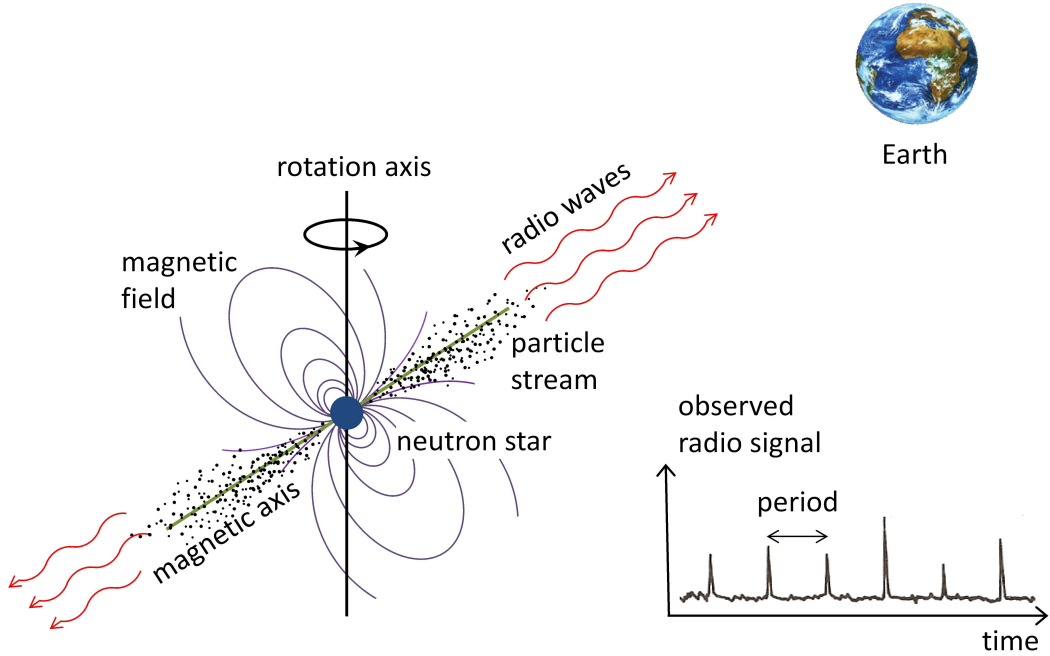

Figure 2: The pulsar lighthouse model. The rotating NS emits a beam of radio waves from above the polar regions of its magnetic axis, which is inclined with respect to its rotation axis. At the observatory on Earth periodic radio signals are recorded, revealing the spin period of the pulsar.

### 1.3.1 Spin periods, energy loss and lifetimes

Radio pulsars have observed spin periods between 1.4 ms and 8.5 sec. Some pulsar periods are known with 16 significant digits, i.e. down to the attosecond level (e.g. PSR J0437−4715 which has a spin period, $P = 0.005\,757\,451\,924\,362\,137(2)$ sec, Verbiest et al. 2008). Hence, pulsars represent excellent clocks and it is even possible to construct a pulsar-based timescale that has a precision comparable to the best modern atomic clocks (Hobbs et al. 2012). Of course, pulsars continuously lose rotational energy (see Section 1.4) which steadily increases their spin period, but this can be corrected for by measuring their (stable) spin period derivative, $\dot{P}$. With an ensemble of high-precision-timing pulsars, a so-called pulsar timing array (PTA, Hobbs et al. 2010b) can be constructed to detect low-frequency (nano-Hz) gravitational waves passing through the Milky Way, originating from merging supermassive black holes in distant galaxies.

The interval of observed spin period derivatives spans more than 10 orders of magnitudes (see Fig. 1) with a mean value of about $\dot{P} \simeq 10^{-15}$ (always positive and thus a slow down of the spin rate). The loss of rotational energy is mainly caused by magnetodipole radiation (see Section 1.4) which therefore enables an estimate of the dipole component of the pulsar's B-field. Using this method radio pulsars have estimated B-fields between $10^7 - 10^{14}$ G (see Fig. 1).

Only a tiny fraction of the rotational energy loss, $|\dot{E}_{\rm rot}|$ of a spinning radio pulsar is emitted as radio waves, from which it is detected. The energy loss is dominated by magnetodipole radiation with a frequency equal to the rotational frequency of the pulsar (see Section 1.4). For example, for the Crab pulsar the radio flux density of the detected signal at 436 MHz is $\sim 0.48$ Jy[1] corresponding to a luminosity of $L_{\rm radio} \sim 10^{31}$ erg s$^{-1}$, at a distance of 2.0 kpc, which is $10^7$ times smaller than $|\dot{E}_{\rm rot}| \approx 5 \times 10^{38}$ erg s$^{-1}$. It should be noted that a large fraction of $|\dot{E}_{\rm rot}|$ also goes to light-up the Crab nebula via injection of relativistic particles. Furthermore, a small fraction is needed for the observed magnetospheric emission of optical-, X- and $\gamma$-rays (via synchrotron radiation generated from gyration of charged particles).

---

[1] 1 Jansky $\equiv 10^{-23}$ erg s$^{-1}$ cm$^{-2}$ Hz$^{-1}$ St$^{-1}$.

The average lifetime of a normal (i.e. *non-recycled*) pulsar is a few 10 Myr. The radio emission process terminates once the electrostatic potential drop across the polar cap, $\Delta\phi \propto B/P^2$ decreases below a critical value for maintaining the required electron-positron pair production: $\gamma + B \to e^- + e^+$ (i.e. when the pulsar crosses the so-called "death line" in the $P\dot{P}$–diagram, cf. Beskin, Gurevich & Istomin 1988; Chen & Ruderman 1993). Recycled pulsars, which constitute the MSPs in focus of this thesis, have small B-fields and rapid spin. Hence, their ratio of $E_{\rm rot}/|\dot{E}_{\rm rot}|$ is very large and thus they remain active radio sources on a Hubble time.

### 1.3.2 Pulsar spectra, duty cycles and beaming fractions

Pulsars generally have rather steep radio-frequency spectra, $S_\nu \propto \nu^\alpha$ where $S_\nu$ is the flux density and the spectral index, $\alpha$ is typically about $-1.5$, and even steeper at high frequencies ($> 1$ GHz). Given the radio luminosity, one can calculate the surface intensity of the radio emission, $I_\nu$ and use a Planck function to demonstrate that if the radio emission was caused by thermal black body radiation one would obtain an extremely high brightness temperature ($T \approx 10^{28}$ K, leading to absurdly large particle energies, $E = kT \sim 10^{24}$ eV). Therefore the radiation mechanism of a radio pulsar must be *coherent*. (Most models invoke curvature radiation or a maser mechanism.)

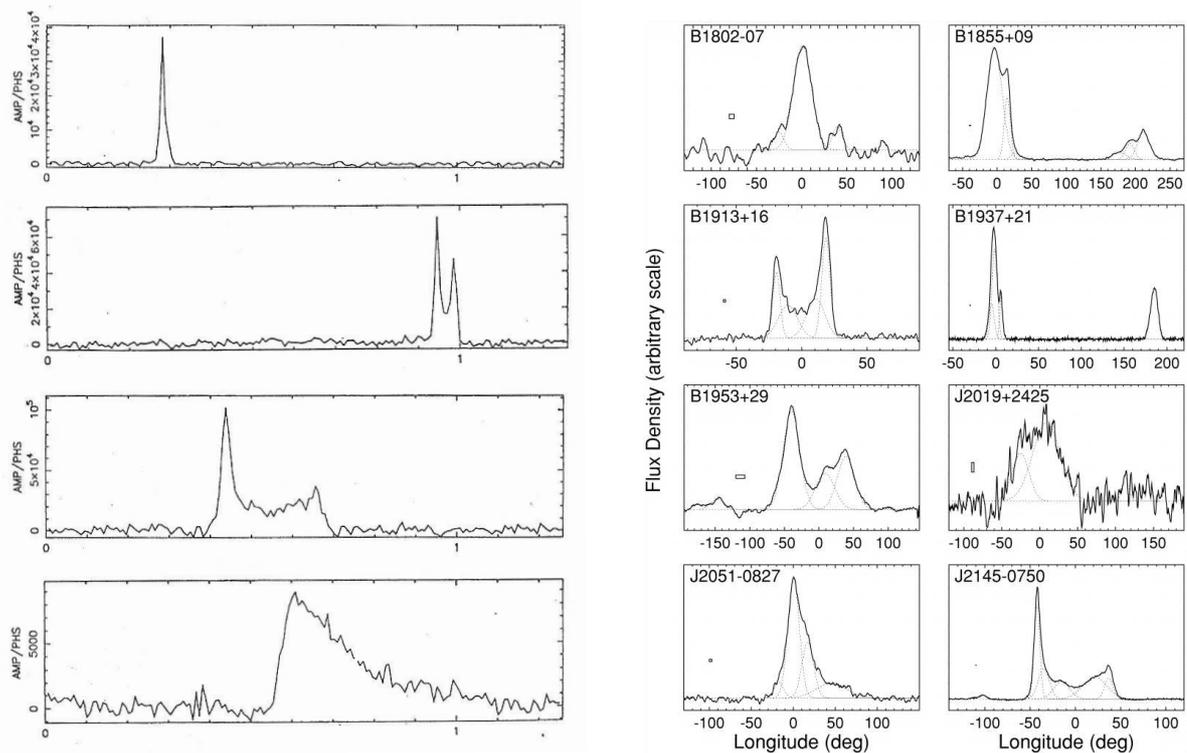

Figure 3: Radio flux density in arbitrary units plotted as a function of rotational phase for 12 pulsars. Left: Pulsar profiles of PSR J0206−4028 (top), J2346−0609, J2145−0750 and J1801−2304 (bottom) obtained from the Parkes Radio Telescope at a frequency of 436 MHz in the mid-1990's (Tauris 1997). The upper two pulsars are slow ($P \sim 1$ sec). The third one (PSR J2145−0750) is an MSP with $P = 16$ ms. Note the exponential tail due to interstellar scattering in the profile of J1801−2304 ($DM \sim 1000$ cm$^{-3}$ pc, see Section 1.3.4). Right: Pulse profiles of MSPs observed at a frequency of 1.4 GHz by the Effelsberg Radio Telescope (Kramer et al. 1998). Here, one full rotation = 360° in longitude of rotational phase. The pulse profiles can be complex with multiple components. PSR B1855+09 and B1937+21 even exhibit an interpulse, i.e. a component of emission from the opposite magnetic pole (at a longitude $\sim 180°$) compared to that of the main pulse.

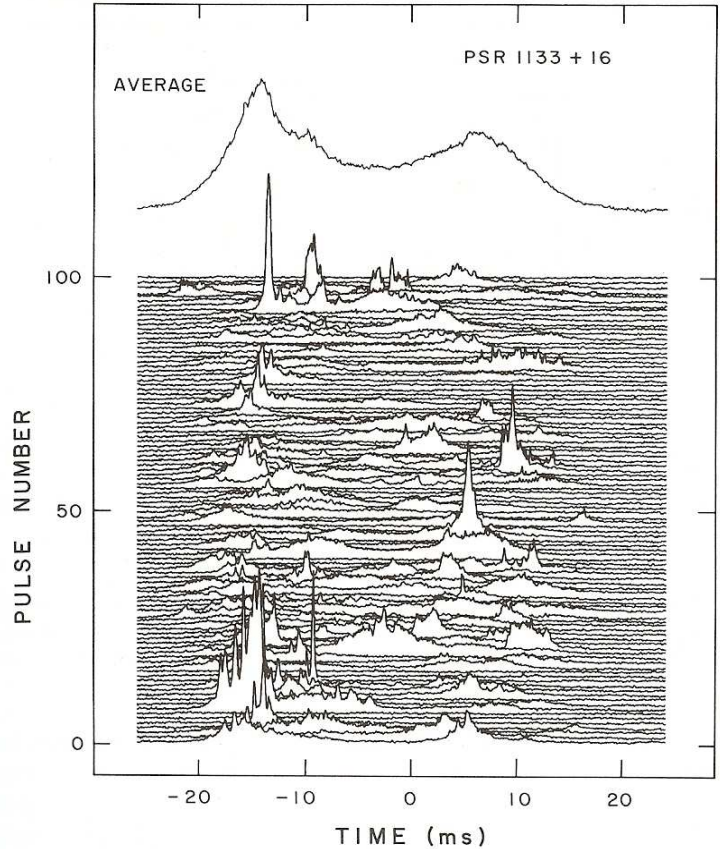

Figure 4: A sequence of 100 pulses from PSR 1133+16 recorded at 600 MHz by Cordes (1979). Consecutive individual pulses are plotted vertically to show their large variations. The *average* pulse behaviour, however, is extremely stable. An average of 500 pulses is shown at the top.

The shape of the observed radio pulse profile depends on the intersection of the line-of-sight across the emission region and the geometric structure of the pulsar beam (Lyne & Manchester 1988; Manchester 1995; Rankin 1983; 1990). Many pulsars have linear polarized profiles – up to 100%. Circular polarization is also seen but not as frequent nor as strong as the linear polarization. Polarization measurements of pulsars make it possible to determine the inclination angle between the magnetic and the rotational axes – an angle which affects the braking torque acting on the pulsar (e.g. Tauris & Manchester 1998; and references therein).

The duty cycle of radio pulsars (fraction of rotational phase with measurable emission) is typically 1–5%, although substantially larger for MSPs (Kramer et al. 1998), see Fig. 3. Similarly, the beaming fraction, the portion of the sky illuminated by a given pulsar, is increasing with spin period (Tauris & Manchester 1998). The observed pulse shapes are quite different in nature and often include two or more subpulses (Fig. 3). The micro-structure of each pulse (or subpulse) can be very complex, as shown in Fig. 4. However, the *average* pulse profile is remarkably stable – a feature which is essential for precise pulsar timing.

### 1.3.3  Timing pulsars

The concept of pulsar timing is straight forward in principle: one measures pulse time-of-arrivals (TOAs) at the observatory and compares them with time kept by a stable reference. During the data analysis the recorded TOAs must be transformed to the corresponding proper time of emission, $T$ in the pulsar reference frame (Taylor & Weisberg 1989; Lorimer & Kramer 2004). Such a transformation includes a dispersive delay from the interstellar medium, a transformation to the Solar System barycenter, relativistic and general relativistic time delays (Rømer, Einstein and Shapiro delays) for both the Solar System, and for the emitting pulsar system if the pulsar orbits a companion star. For some pulsars it is the post-Keplerian parameters of the binary system that allows for testing gravitational physics (Kramer et al. 2006b; Kramer & Wex 2009; Freire, Kramer & Wex 2012; Antoniadis et al. 2013; Wex 2014), see also Chapter 4.

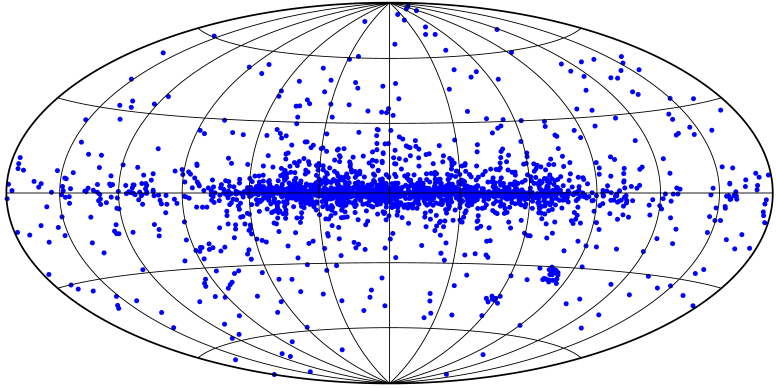

Figure 5: The Galactic distribution of all 2328 currently known pulsars. Data from the ATNF Pulsar Catalogue (Manchester et al. 2005) – version 1.50, June 2014. Figure provided by Norbert Wex.

Under the assumption of a deterministic spin-down law, the rotational phase of a pulsar can be written as:

$$\phi(T) = \phi_0 + \Omega T + \frac{1}{2}\dot{\Omega}T^2 + \frac{1}{6}\ddot{\Omega}T^3 + ....$$ (1)

In order to obtain a timing solution it is necessary to assign a pulse number to each recorded TOA. Some of the observations can be separated by weeks, months, or even years. Hence, between two consecutive observations the pulsar may have rotated as many as $10^7 - 10^{10}$ turns, and to extract the maximum information content from the data, these integer numbers of turns must be recovered *exactly*. The spin period derivative, $\dot{P}$ is an essential parameter for pulsars. All their physical parameters that can be extracted basically depend on this parameter – e.g. B-field, energy loss, braking torque, age, electrostatic potential gap across the polar cap region of the magnetosphere, etc. To measure $\dot{P}$ one must assure that the time interval between two neighbouring observations, multiplied by the error in the measured rotational frequency of the pulsar, is much less than one, i.e. $(t_2 - t_1)\,\Delta\Omega/2\pi \ll 1$.

### 1.3.4 Distance measurements and the distribution of Galactic pulsars

Fig. 5 shows the distribution of pulsars in Galactic coordinates. The clear clustering of sources in the Galactic plane shows that pulsars are indeed of Galactic origin, and that they are the likely remnants of massive OB stars. However, the broad scatter in their distribution is very significant. While their OB progenitor stars are found close to the Galactic plane (with a velocity dispersion of only $\sim 10 - 20$ km s$^{-1}$), pulsars have a velocity dispersion exceeding $100$ km s$^{-1}$. The explanation for this is that (most) NSs apparently receive a momentum kick at birth, resulting in a mean birth velocity of $\sim 400$ km s$^{-1}$ (Lyne & Lorimer 1994), although a fraction of these runaway velocities can also be explained by the disruption of binary systems (Tauris & Takens 1998).

For some nearby pulsars with good timing properties it is possible to obtain a parallax measurement. However, for the majority of pulsars one has to rely on distance estimates obtained from the amount of dispersion of the radio waves as they propagate through the ionized interstellar medium along the line-of-sight, i.e. the delay in the TOA of the different frequency components of the pulse, see Fig. 6. Since this method requires a model for the distribution of free electrons in the Milky Way (e.g. Cordes & Lazio 2002) the resulting distance estimates are, in general, only accurate to within 20–50%.

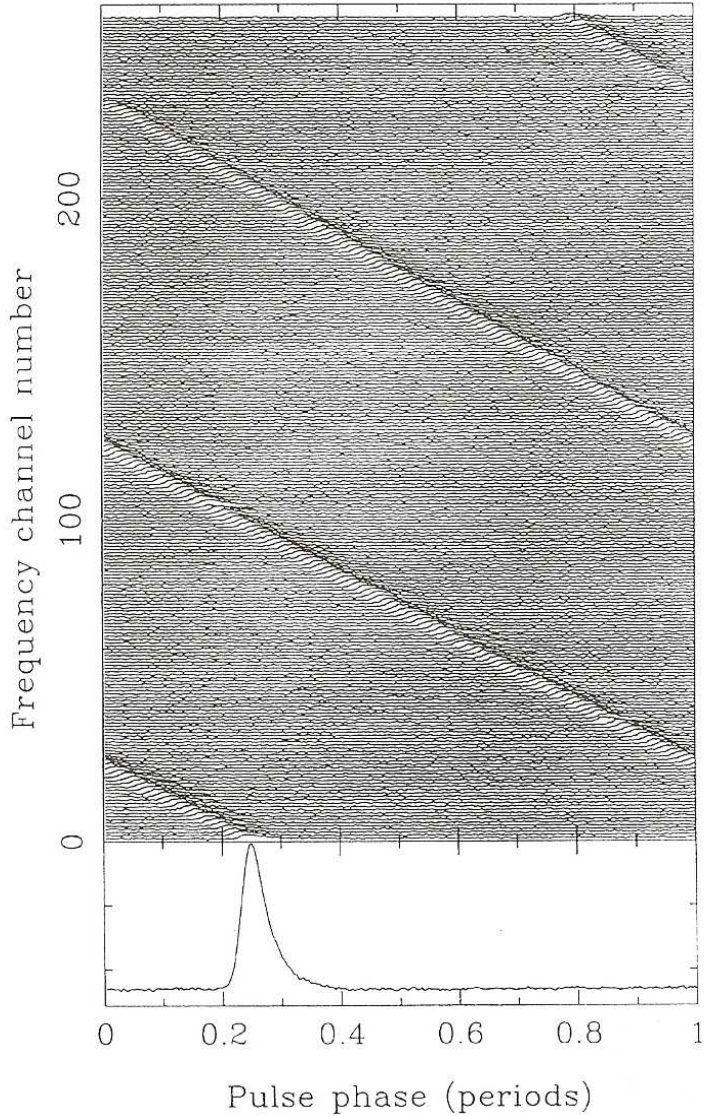

Figure 6: Dispersion of the radio pulse due to its propagation through the interstellar medium. Shown is the signal at each frequency channel at the observatory (channel no.1 @ 420 MHz → channel no.256 @ 452 MHz) as a function of pulse phase for the Vela pulsar, PSR B088−45. After Lorimer (1994).

## 1.4 The magnetized rotating neutron star

Given that radio pulsars are rapidly rotating, strongly magnetized NSs which have an inclined magnetic field axis with respect to their rotation axis these objects have a time-varying magnetic dipole moment. Therefore, pulsars radiate significant amounts of energy in the form of dipole waves (electromagnetic waves with a frequency equal to the spin frequency of the pulsar). The energy-loss rate due to magnetic dipole radiation is given by:

$$\dot{E}_{\text{dipole}} = -\frac{2}{3c^3}|\ddot{\vec{m}}|^2 \qquad \wedge \qquad |\ddot{\vec{m}}| = BR^3\Omega^2\sin\alpha \qquad (2)$$

where $\vec{m}$ is the magnetic moment of the NS, $B$ is the magnetic flux density at its surface (equator), $R$ is the radius, $\Omega = 2\pi/P$ is the angular velocity with $P$ being the pulsar spin period, $c$ is the speed of light in vacuum, and the magnetic inclination angle is $0 < \alpha \leq 90°$. The total loss of rotational energy of a radio pulsar is caused by a combination of magnetic dipole radiation (Pacini 1967), the presence of plasma currents in the magnetosphere (the Goldreich-Julian term, Goldreich & Julian 1969; Spitkovsky 2006) and gravitational wave radiation (e.g. Wade et al. 2012):

$$\dot{E}_{\rm rot} = \dot{E}_{\rm dipole} + \dot{E}_{\rm GJ} + \dot{E}_{\rm gw} \qquad (3)$$

Theoretically, the value of $\dot{E}_{\rm GJ}$ is found by considering the outward Poynting energy flux $S \sim c \cdot B^2/4\pi$ crossing the light cylinder, $r_{\rm lc} = c/\Omega$ and giving rise to the observed high-frequency radiation as well as emission of relativistic particles. Observations have shown that $\dot{E}_{\rm GJ}$ is of the same order as $\dot{E}_{\rm dipole}$ within a factor of a few (Kramer et al. 2006a; Lorimer et al. 2012; Camilo et al. 2012), and it is usually a very good approximation that $\dot{E}_{\rm gw} \ll \dot{E}_{\rm rot}$ (Abadie et al. 2010). Hence, the dipole component of pulsar B-fields is traditionally found simply by equating the loss rate of rotational energy ($\dot{E}_{\rm rot} = I\Omega\dot{\Omega} = -4\pi^2 I\dot{P}/P^3$) to the energy loss caused by magnetic dipole radiation (Eq. 2) yielding:

$$B_{\rm dipole} = \sqrt{\frac{3\,c^3\,I}{8\pi^2\,R^6}\,P\dot{P}} \quad \simeq \quad 3.2 \times 10^{19}\,\sqrt{P\dot{P}}\ \text{Gauss} \qquad (4)$$

where the numerical constant is calculated for the equatorial B-field of an orthogonal rotator ($\alpha = 90°$), and assuming $R = 10$ km and a NS moment of inertia of $I = 10^{45}$ g cm$^2$. In Tauris, Langer & Kramer (2012; see Chapter 6) we discuss this equation further and derive an alternative expression. For a discussion of Eq. (4) in the context of efficient $\gamma$-ray emitting MSPs, see Guillemot & Tauris (2014).

## 1.5 Pulsar evolutionary tracks and true ages

The slow-down of pulsar spin is characterized by the observable braking index, $n \equiv \Omega\ddot{\Omega}/\dot{\Omega}^2$ which relates the braking torque ($N = dJ_{\rm spin}/dt = I\dot{\Omega}$) to the rotational angular velocity via $\dot{\Omega} \propto -\Omega^n$. By integrating this pulsar spin-deceleration equation one can obtain evolutionary tracks and isochrones in the $P\dot{P}$–diagram (e.g. Tauris & Konar 2001; Lazarus et al. 2014; see Chapter 7). For a pulsar evolving with a constant value of $n$ the kinematic solution at time $t$ (positive in the future, negative in the past) is given by:

$$P = P_0 \left[ 1 + (n-1)\frac{\dot{P}_0}{P_0}\,t \right]^{1/(n-1)} \qquad (5)$$

and

$$\dot{P} = \dot{P}_0 \left( \frac{P}{P_0} \right)^{2-n} \qquad (6)$$

where $P_0$ and $\dot{P}_0$ represent values at $t = 0$. Similarly, the true age can be written as:

$$t = \frac{P}{(n-1)\dot{P}} \left[ 1 - \left( \frac{P_0}{P} \right)^{n-1} \right] \qquad (7)$$

The so-called *characteristic age* (Manchester & Taylor 1977; Shapiro & Teukolsky 1983) is defined as: $\tau \equiv P/(2\dot{P})$. However, this expression is only a good age estimate for pulsars which have evolved with a constant $n = 3$ and for which $P_0 \ll P$. For many pulsars, and in particular for magnetars and recycled pulsars (MSPs), $\tau$ is not a good true age estimator. For example, some young NSs associated with SN remnants have $\tau$ values of several Myr, although SN remnants are believed to be dissolved into the interstellar medium after less than 50 kyr (which is therefore an upper limit on the true age of these NSs). Another extreme example is the MSPs which in some cases have $\tau > 30$ Gyr. Ironically, in many cases $\tau$ can better be thought of as an estimate of the *remaining* lifetime of a pulsar given that $\tau = E_{\rm rot}/|\dot{E}_{\rm rot}|$. In Tauris (2012); Tauris, Langer & Kramer (2012) we discuss in much more detail the issue of $\tau$ being a poor true age estimator for MSPs, see Chapters 5 and 6.

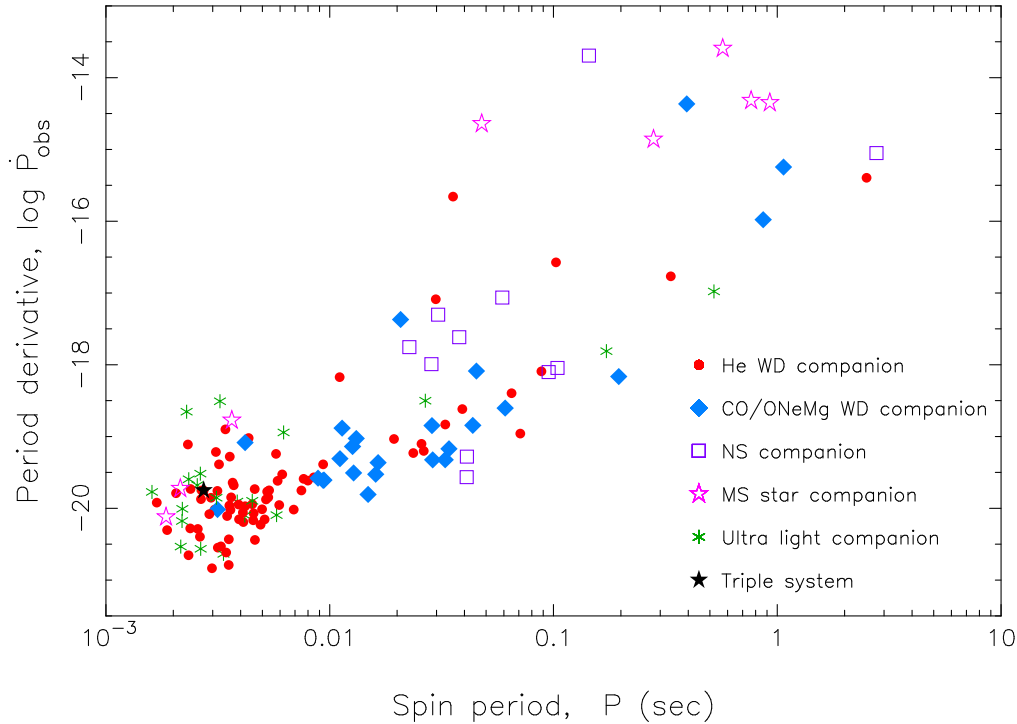

Figure 7: The various types of companions to known binary pulsars in the $P\dot{P}$–diagram. Adapted from Tauris, Langer & Kramer (2012). New data taken from the ATNF Pulsar Catalogue (Manchester et al. 2005) – version 1.49, April 2014.

## 1.6   Millisecond pulsars, their companions and the recycling process

The first ideas of a so-called recycling process of old NSs date back to the mid-1970's following the discovery of the Hulse-Taylor pulsar (Bisnovatyi-Kogan & Komberg 1974; Smarr & Blandford 1976). This concept of pulsar recycling was given a boost by the discovery of the first MSP (Backer et al. 1982; Alpar et al. 1982; Radhakrishnan & Srinivasan 1982). The idea is that the MSP obtains its rapid spin (and weak B-field) via a long phase of accretion of matter from a companion star in a low-mass X-ray binary (LMXB). As a result of the high incidence of binaries found in the following years among these fast spinning pulsars (see Fig. 1), this formation scenario has now become generally accepted (Bhattacharya & van den Heuvel 1991). Furthermore, the model was beautifully confirmed with the discovery of the first millisecond X-ray pulsar in the LMXB system SAX 1808.4–3658 (Wijnands & van der Klis 1998), and more recently by the detection of the so-called transitional MSPs which undergo changes between accretion and rotational powered states (Archibald et al. 2009; Papitto et al. 2013).

Roughly 15% of all known radio pulsars are MSPs (or at least mildly recycled pulsars) and the majority ($\sim 2/3$) of these have a companion star (Here we define an MSP as a pulsar with $P < 30$ ms and $\dot{P} < 10^{-16}$). Radio pulsars in general have been discovered in binary systems with a variety of companions: white dwarfs (WDs), NSs, main sequence stars, and even planets; see Fig. 7 for the distribution of pulsars with the various companion types in the $P\dot{P}$–diagram. The vast majority of the binary pulsar systems contain an MSP with a helium WD companion. However, there is a growing number of MSPs with a non- or semi-degenerate companion star which is being ablated by the pulsar wind, the so-called *black widows* and *redbacks* (Roberts 2013). This is evidenced by the radio signal from the pulsar being eclipsed for some fraction of the orbit (Fruchter, Stinebring & Taylor 1988; Stappers et al. 1996; Archibald et al. 2009). These companions are all low-mass stars with a mass between $0.02 - 0.3$ $M_\odot$ (Roberts 2013; Breton et al. 2013). In Chen et al. (2013) we have argued that that black widows and redbacks are two distinct populations and that they are not linked by an evolutionary path[2]. An important long standing question to answer is whether black widows are the progenitors of the isolated MSPs. Can ablation by the energetic pulsar flux lead to a complete evaporation of a low-mass companion (Kluzniak et al. 1988; Ruderman, Shaham & Tavani 1989)? Or will the companion star in an ultra low-mass binary pulsar system eventually disrupt via an internal instability when its mass becomes too small (Deloye & Bildsten 2003; Possenti 2013)?

Based on stellar evolution theory it is expected that pulsars can also be found with a helium star or a black hole companion. These systems have not yet been discovered but it is likely that the SKA (see Chapter 11) will reveal pulsars with such companions within the next decade.

In recent years, a few binary pulsars with peculiar properties have been discovered and which indicate a hierarchical triple system origin – e.g. PSR J1903+0327 (Champion et al. 2008; Freire et al. 2011b; Portegies Zwart et al. 2011). In 2013, two puzzling MSPs were discovered in eccentric binaries: PSR J2234+06 (Deneva et al. 2013) and PSR J1946+3417 (Barr et al. 2013). These two systems might also have a triple origin. However, their eccentricities and orbital periods have led us to suggest an alternative hypothesis of *direct* MSP formation via a rotationally delayed accretion-induced collapse of a massive WD (Freire & Tauris 2014; see Chapter 10). Besides from these intriguing systems, an exotic triple system MSP with two WD companions (PSR J0337+1715) was announced earlier this year by Ransom et al. (2014). This amazing system must have survived three phases of mass transfer and one SN explosion and challenges current knowledge of multiple stellar system evolution (Tauris & van den Heuvel 2014; , see Chapter 8).

### 1.6.1   Pulsar recycling

The progenitor systems of recycled pulsars are the X-ray binaries. The total population of known Galactic LMXBs and HMXBs is exceeding 300 sources (Liu, van Paradijs & van den Heuvel 2006; 2007; Patruno & Watts 2012; Chaty 2013). Many details of the recycling process, however, remain unclear. Some of the most important issues are discussed in Chapter 6 (Tauris, Langer & Kramer 2012; and references therein):

- The (accretion-induced?) decay of the surface B-field of a NS

- The maximum possible spin rate of an MSP

- The Roche-lobe decoupling phase

- The spin-up line and accretion torque reversals

- The progenitors of the isolated MSPs

There is empirical evidence that the surface B-field strength of a NS decays as a consequence of accretion. Nevertheless, the exact reason for this process is not well understood (Bhattacharya 2002). Nor is it known what dictates the fastest possible spin rate of a radio MSP. In Fig. 8 we have plotted the spin period distribution of radio MSPs and also included a comparison of the spin frequency distributions of different classes of their accretion-powered and nuclear-powered progenitors. Clearly, the radio MSPs seem to be slower spinning compared to their accreting progenitors – see Papitto et al. (2014) for a recent statistical analysis. In Tauris (2012) (Chapter 5) I have demonstrated that during the final stage of mass transfer MSPs may lose up to 50% of their rotational energy. The reason is that low-mass donor stars decouple from their Roche lobe on a timescale ($\sim 100$ Myr) which is comparable to the MSP spin-relaxation timescale. The resulting braking torque can explain the difference in the spin distributions of radio MSPs and their progenitors (although selection effects must be considered too).

Does the equation-of-state (EoS) of nuclear matter allow for the existence of sub-ms pulsars? Or is the current spin frequency limit, slightly above 700 Hz (Hessels et al. 2006), set by the onset of gravitational wave emission during accretion (Chakrabarty et al. 2003), or subsequent r-mode instabilities? Alternatively, the spin-up torque might saturate due to magnetosphere–disk conditions, thus preventing sub-ms MSPs to form (Lamb & Yu 2005).

---

[2]See, however, Benvenuto, De Vito & Horvath (2014) for a different point of view.

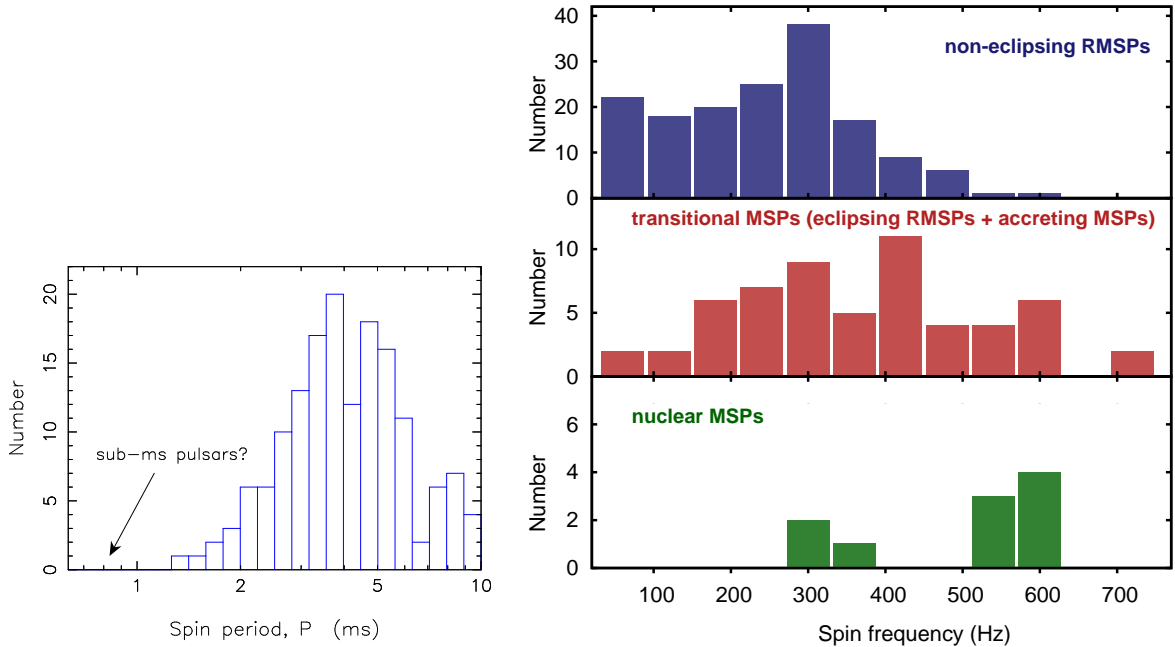

Figure 8: Left: The observed spin period distribution of radio MSPs. Data taken from the ATNF Pulsar Catalogue (Manchester et al. 2005) – version 1.49, April 2014. The discovery of sub-ms MSPs would have a huge scientific impact. Right: The spin frequency distribution of non-eclipsing rotation-powered radio MSPs (top), the so-called transitional MSPs (middle), and nuclear-powered MSPs (bottom); after Papitto et al. (2014).

### 1.6.2 The accretion torque

The mass transfered from the donor star carries with it angular momentum which eventually spins up the rotating NS once its surface B-field is low enough to allow for efficient accretion, i.e. following initial phases where accretion is prevented due to either the magnetodipole radiation pressure or propeller effects (Illarionov & Sunyaev 1975). The accretion torque acting on the spinning NS has a contribution from both material stress (dominant term), magnetic stress and viscous stress (Ghosh & Lamb 1992; Frank, King & Raine 2002; Shapiro & Teukolsky 1983; Dubus et al. 1999). The exchange of angular momentum ($\vec{J} = \vec{r} \times \vec{p}$) at the magnetospheric boundary eventually leads to a gain of NS spin angular momentum which can approximately be expressed via the acting torque:

$$N \approx \sqrt{GM r_{\mathrm{A}}} \, \dot{M} \, \xi \tag{8}$$

where $\xi \simeq 1$ is a numerical factor which depends on the flow pattern (Ghosh & Lamb 1979b; 1992), and

$$
\begin{aligned}
r_{\mathrm{A}} &\simeq \left( \frac{B^2 \, R^6}{\dot{M} \sqrt{2GM}} \right)^{2/7} \\
&\simeq 22 \text{ km} \; \cdot \; B_8^{4/7} \left( \frac{\dot{M}}{0.1 \, \dot{M}_{\mathrm{Edd}}} \right)^{-2/7} \left( \frac{M}{1.4 \, M_\odot} \right)^{-5/7}
\end{aligned}
\tag{9}
$$

is the Alfvén radius defined as the location where the magnetic energy density will begin to control the flow of matter (i.e. where the incoming material couples to the magnetic field lines and co-rotate with the NS magnetosphere). A typical value for the Alfvén radius in accreting X-ray MSPs, assuming $B \sim 10^8$ G and $\dot{M} \sim 0.01\,\dot{M}_{\mathrm{Edd}}$, is $\sim 40$ km corresponding to about $3\,R$. The expression above is found by equating the magnetic energy density ($B^2/8\pi$) to the ram pressure of the incoming matter and using the continuity equation (e.g. Pringle & Rees 1972). The sign of the accretion torque determining spin-up/spin-down depends on the location of the magnetospheric boundary (i.e. the inner edge of the accretion disk, roughly equal to $r_{\mathrm{A}}$) relative to the co-rotation radius and the light-cylinder radius, as well as the critical fastness parameter – see Tauris (2012); Tauris, Langer & Kramer (2012) (Chapters 5 and 6, and references therein) for a more detailed description.

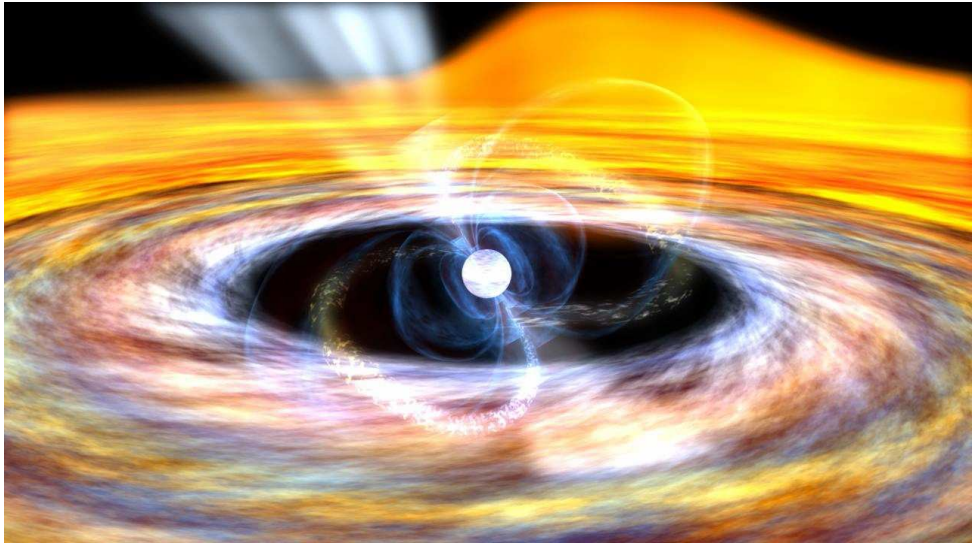

Figure 9: An artist's impression of an accreting X-ray MSP and its magnetosphere which truncates the inner part of the accretion disk. © NASA/Dana Berry.

## 1.7    The structure of neutron stars

The interior structure of NSs is subject to many research projects and much debate. The challenge is to understand the behaviour of nuclear matter at ultra-high densities inaccessible to laboratories on Earth. An illustration of the cross section is shown in Fig. 10. Whereas the crustal layers of a NS are fairly well understood the interior (core region) structure remains highly uncertain due to the unknown physics at work in regimes where the mass density exceeds that of normal nuclear matter, $\rho > \rho_{\mathrm{nuc}} \approx 2.8 \times 10^{14}$ g cm$^{-3}$. The main reasons for this uncertainty are the nucleon–nucleon interactions, the many-body problem and the presence of exotic particles at such high densities. The superfluid core is thought to include a soup mixture of hyperons (nucleon-like strange baryons), $\Delta$–resonance particles, and a pion/kaon condensate (e.g. Camenzind 2007; and references therein). It is even possible that so-called hybrid NSs exist (a modified version of pure quark stars), where the interior is subject to quark deconfinement (Glendenning 2000; Weber 2005). The transition from ordinary nuclear matter NSs into quark stars (or hybrid stars, in case only the core region undergoes a phase change) has been speculated to be accompanied with a transient 'quark-nova' which might be observable (Ouyed, Dey & Dey 2002).

Given the uncertainties about the physics of their interior, the NS equation-of-state (EoS) remains unknown (e.g. Steiner, Lattimer & Brown 2013; and references therein). However, observations of pulsars can constrain the EoS and thus help illuminating their interior composition. The EoS of NSs is often represented in mass–radius diagrams. Measurements of NS masses, radii and spin rates can help to constrain the region of possible solutions to the EoS – a key research goal in the MPIfR Pulsar Group. An example of such a plot is shown in Fig. 11.

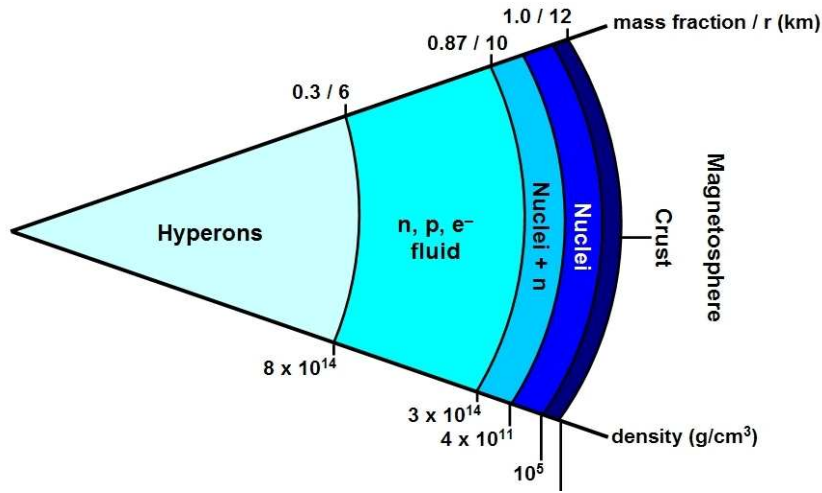

Figure 10: A simplified illustration of the cross section of a NS. The surface layer consists of $^{56}$Fe nuclei and $e^-$. In the outer crust the atomic nuclei become more and more neutron rich due to inverse $\beta$–decay and a filled Fermi sea of relativistic $e^-$ (Pauli-blocking). At a mass density of $\rho = 4 \times 10^{11}$ g cm$^{-3}$ it becomes energetically favorable for the neutrons to start popping out of the nuclei (the transition to the inner crust). The core region is determined by $\rho > \rho_{\rm nuc} \approx 2.8 \times 10^{14}$ g cm$^{-3}$. The inner core structure of NSs is unknown – largely due to uncertainties in the nucleon–nucleon interactions at high densities. The core is likely to host an exotic mixture of hyperons, and possibly even a central region with quark deconfinement, see text. After Bowers & Deeming (1984).

The current record high-mass NS is PSR J0348+0432 ($2.01 \pm 0.04$ $M_\odot$, Antoniadis et al. 2013; see Chapter 4) and the fastest known spinning NS is PSR J1748−2446ad (1.4 ms, Hessels et al. 2006). The NS radii can be constrained from observations by fitting black body spectra to the thermal radiation of young isolated NSs (Trümper 2005) or LMXBs in quiescent mode (Rutledge et al. 1999). However, these radii measurements are still suffering from uncertainties in distances estimates and chemical composition of the NS atmospheres (Ho & Heinke 2009). If LOFT (Large Observatory For X-ray Timing) is selected for a future ESA space mission there is realistic hope to nail down the NS EoS accurately[3].

## 1.8 Neutron star masses

The recent measurements of massive NSs ($\geq 2.0$ $M_\odot$) raise the question about their origin. Were these massive binary NSs born with a canonical mass of $1.3 - 1.4$ $M_\odot$ and then later on accreted up to $\sim 0.7$ $M_\odot$ from their companion star? Or were these NSs born relatively massive, such that they only needed to accrete little mass in order to obtain their measured mass values? This question is addressed in a couple of papers in Chapters 3 and 4 (Tauris, Langer & Kramer 2011; Antoniadis et al. 2013) (and for PSR J0348+0432, also in Istrate et al., in prep.). The distribution of possible remnant masses of compact objects left behind SN explosions is a longstanding question. From late stages of stellar evolution it is clear that the final pre-SN core mass (e.g. the region within the oxygen burning shells) is varying erratically as a function of initial ZAMS mass (Woosley, Heger & Weaver 2002; Langer 2012). A main reason for this is related to the varying number of convective oxygen burning shells (and the treatment of convection) which is sensitive to the input physics and the numerics of a given stellar evolution code. Another reason is the yet unknown details of the explosion physics. There seems to be some agreement in the literature that there are 3 peaks in the spectrum of NS birth masses. One narrow peak near 1.25 $M_\odot$ expected from electron capture SNe (the critical core mass of 1.37 $M_\odot$ minus the gravitational binding energy, see Section 1.9), and two broader peaks from iron core-collapse SNe at $1.10 - 1.45$ $M_\odot$ and $\sim 1.70$ $M_\odot$ from stars with ZAMS masses roughly in the intervals $10 - 20$ $M_\odot$ and $20 - 25$ $M_\odot$, respectively.

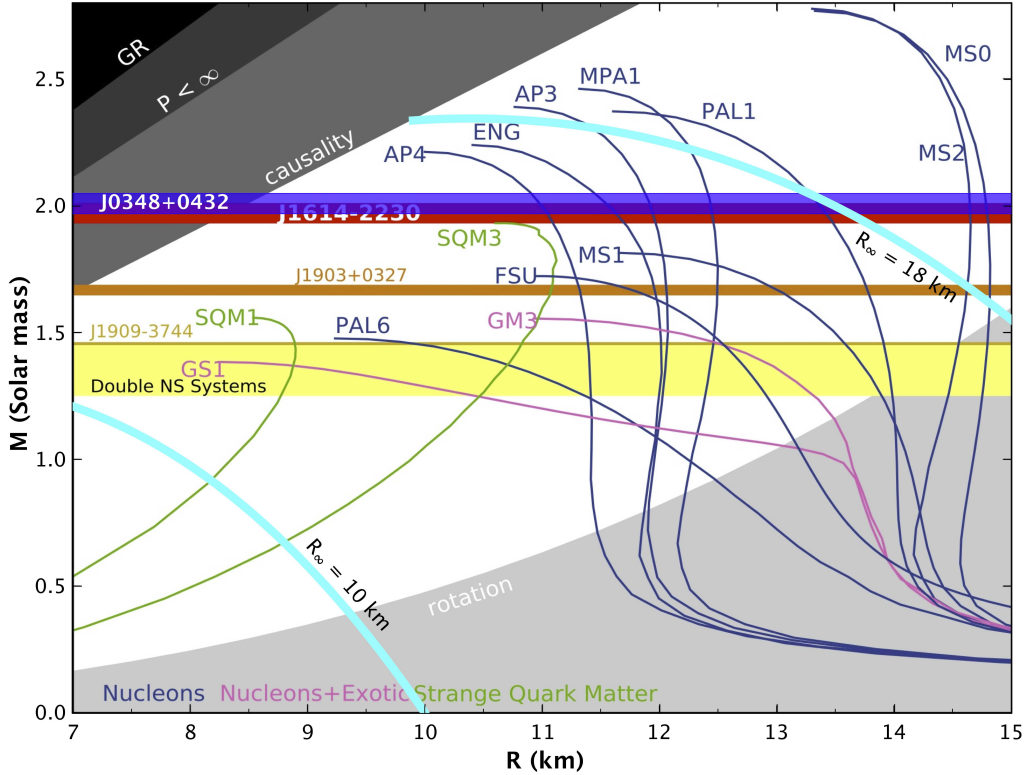

Figure 11: The NS EoS can be constrained from measurements of NS masses, radii and spins – see text. The two light-blue curves correspond to observed radii of $R_\infty = 18$ km (top) and 10 km (bottom). After original by Demorest et al. (2010) and modified by Norbert Wex.

## 1.9 Formation of neutron stars

NSs can be formed in different flavours of core-collapse SNe. The main categories are iron core-collapse SNe (Fe CCSNe, of Type II or Type Ib/c) and electron capture SNe (EC SNe), see e.g. Podsiadlowski et al. (2004); Langer (2012); Janka (2012). Whereas Fe CCSNe are expected for stars initially more massive than roughly 10 $M_\odot$ (depending on metallicity, binarity and treatment of core convective overshooting) which terminate their nuclear burning with an iron core above the Chandrasekhar mass limit of $\sim 1.4$ $M_\odot$, EC SNe originate from slightly less massive stars which produce ONeMg cores with a mass $> 1.37$ $M_\odot$, leading to loss of pressure support via electron captures on Mg and Ne when the central density reaches $\sim 4 \times 10^9$ g cm$^{-3}$ (Nomoto 1987). These two different core-collapse processes lead to different SN explosions and thus possibly distinct properties of the NS remnants. Whereas the Fe CCSNe produce NSs with masses between $1.10-1.70$ $M_\odot$ (Tauris, Langer & Kramer 2011; Chapter 3, and references therein) and, in general, accompanied with a large momentum kick at birth, the EC SNe all produce NSs with a relatively small mass near 1.25 $M_\odot$ (simply by conversion of the critical baryonic mass of 1.37 $M_\odot$ into a gravitational mass of 1.25 $M_\odot$ via release of the gravitational binding energy of the NS in the form of neutrinos) and possibly always with a small momentum kick. Theres is some evidence for this hypothesis in the observed population of NSs (Pfahl et al. 2002; van den Heuvel 2004; Schwab, Podsiadlowski & Rappaport 2010; Knigge, Coe & Podsiadlowski 2011). However, as we shall see in Chapter 2 (Tauris et al. 2013a) the picture is more complex and we suggest that Fe CCSNe might also produce NSs with small kicks, if they are formed in a tight binary system from an ultra-stripped progenitor star.

The physics of the EC SNe is related to an additional formation scenario in which NSs may be produced via the accretion-induced collapse (AIC) of a massive ONeMg WD, see e.g. Nomoto, Nariai & Sugimoto (1979); Taam & van den Heuvel (1986); Michel (1987); Canal, Isern & Labay (1990); Nomoto & Kondo (1991). In Bonn, we have investigated the possibility of forming MSPs, either indirectly or directly, via this NS formation channel. The AIC has the advantage of begin able to explain the possible existence of young NSs in an old stellar population like a globular cluster, cf. Chapter 9 (Tauris et al. 2013b). This can happen if these NSs were formed relatively recently via significant accretion onto a massive WD from a low-mass red giant star. In addition, we have demonstrated that rotationally delayed AIC – leading to the direct formation of an MSP – might explain a new class of puzzling MSPs in eccentric binaries, cf. Chapter 10 (Freire & Tauris 2014). Further observational tests of this scenario is in progress.

It is therefore of uttermost importance to explore the NS formation channels and the distribution of NS birth masses in order to gain more knowledge about the final stages of massive stellar evolution and explosion physics.

- **Chapter 2**
  Tauris, Langer, Moriya, Podsiadlowski, Yoon & Blinnikov (2013), ApJ Letters 778, L23
  *Ultra-stripped Type Ic Supernovae from Close Binary Evolution*

- Tauris, Langer & Podsiadlowski (2014), MNRAS, in prep.   [*not included here*]
  *Ultra-stripped Type Ic SNe: electron capture vs iron core collapse*

## 2. Ultra-stripped Type Ic Supernovae from Close Binary Evolution

**Tauris, Langer, Moriya, Podsiadlowski, Yoon & Blinnikov (2013)**
**ApJ Letters 778, L23**

### Abstract

Recent discoveries of weak and fast optical transients raise the question of their origin. We investigate the minimum ejecta mass associated with core-collapse supernovae (SNe) of Type Ic. We show that mass transfer from a helium star to a compact companion can produce an ultra-stripped core which undergoes iron core collapse and leads to an extremely fast and faint SN Ic. In this *Letter*, a detailed example is presented in which the pre-SN stellar mass is barely above the Chandrasekhar limit, resulting in the ejection of only $\sim 0.05 - 0.20$ $M_\odot$ of material and the formation of a low-mass neutron star. We compute synthetic light curves of this case and demonstrate that SN 2005ek could be explained by our model. We estimate that the fraction of such ultra-stripped to all SNe could be as high as $10^{-3} - 10^{-2}$. Finally, we argue that the second explosion in some double neutron star systems (for example, the double pulsar PSR J0737−3039B) was likely associated with an ultra-stripped SN Ic.

### 2.1 Introduction

In recent years, high-cadence surveys and dedicated supernova (SN) searches have increased the discovery rate of unusual optical transients. Examples of events with low peak luminosities ($\lesssim 10^{42}$ $\mathrm{erg\,s^{-1}}$) and rapidly decaying light curves comprise SN 2005ek (Drout et al. 2013), SN 2010X (Kasliwal et al. 2010) and SN 2005E (Perets et al. 2010). These SNe show diverse spectroscopic signatures (cf. Kleiser & Kasen 2014) and thus their explanation may require a diversity of models. Drout et al. (2013) concluded that SN 2005ek represents the most extreme small ratio of ejecta to remnant mass observed for a core-collapse SN to date.

Two classes of models have been suggested to explain these fast and faint events. The first one proposes a partial explosion (Bildsten et al. 2007) or the collapse of a white dwarf (Dessart et al. 2006), both involving a very low ejecta mass. The second relates to hydrogen-free core-collapse SNe with large amounts of fall back (Moriya et al. 2010) or with essentially no radioactive nickel (Dessart et al. 2011; Kleiser & Kasen 2014).

Also the afterglows of short $\gamma$−ray bursts (sGRBs, Fox et al. 2005; Berger 2010), which are suggested to originate from mergers of neutron stars (e.g. Blinnikov et al. 1984; Paczynski 1986; Eichler et al. 1989), are predicted theoretically to produce fast and faint "kilonovae" at optical/IR wavelengths powered by the ejection of $\sim 0.01$ $M_\odot$ of radioactive r-process material (Li & Paczyński 1998; Metzger et al. 2010; Barnes & Kasen 2013) – see Berger, Fong & Chornock (2013); Tanvir et al. (2013) for detections of a possible candidate. These mergers constitute the prime candidate sources of high frequency gravitational waves to be detected by the LIGO/VIRGO network within the next 5 years (Aasi et al. 2013), which fosters the search for their electromagnetic counterparts (e.g. Kasliwal & Nissanke 2013). A good understanding of faint and fast transients is thus urgently needed.

It is well-known that the outcome of stellar evolution in close binaries differs significantly from that of single stars. The main effects of mass loss/gain, and tidal forces at work, are changes in the stellar rotation rate, the nuclear burning scheme and the wind mass-loss rate (Langer 2012). As a result, the binary interactions affect the final core mass prior to collapse (Brown et al. 2001; Podsiadlowski et al. 2004), and therefore the type of compact remnant left behind and the amount of envelope mass ejected.

Whereas most Type Ib/c SNe are expected to originate in binary systems, from the initially more massive star which has been stripped of its hydrogen envelope by mass-transfer to its companion (Eldridge, Izzard & Tout 2008), these pre-SN stars typically have an envelope mass of 1 $M_\odot$ or more (Yoon, Woosley & Langer 2010). In a close X-ray binary, however, a second mass-transfer stage from a helium star to a neutron star (NS) can strip the helium star further prior to the SN (Dewi et al. 2002; Dewi & Pols 2003; Ivanova et al. 2003; and references therein). The fast decay of the Type Ic SN 1994I with an absolute magnitude of $M_V \approx -18$ was explained by Nomoto et al. (1994) from such a model, and for which they suggested a resulting pre-SN carbon-oxygen star of $\sim 2\ M_\odot$ and a corresponding ejecta mass of $\sim 0.9\ M_\odot$. The starting point of all these calculations (tight systems containing a naked helium star and a NS with orbital period, $P_{orb} < 2^d$) is a continuation of the expected outcome of common envelope evolution in high-mass X-ray binaries. Of particular interest is mass transfer by Roche-lobe overflow (RLO) initiated by the helium star expansion after core helium exhaustion (so-called Case BB RLO). Here, we investigate SN progenitors originating from the evolution of such helium star–NS binaries, and find them to be the most promising candidates to achieve maximally stripped pre-SN cores. We provide a detailed example and demonstrate that this scenario can produce an iron core collapse of a small, bare core of $\sim 1.5\ M_\odot$, leading to an ejecta mass of only $\sim 0.1\ M_\odot$. In Section 2.2 we present our model with computations extending beyond oxygen ignition, and study the final core structure prior to iron core collapse. We model light curves for the resulting fast and faint SN explosion and compare our results with SN 2005ek in Section 2.3. Finally, we estimate the rate of of these events, discuss double NS binaries and summarize our conclusions in Section 2.4.

## 2.2 Binary evolution and formation of a nearly naked pre-SN metal core

For our detailed calculations of the evolution of helium star–NS binaries, we applied the BEC stellar evolution code (Yoon, Woosley & Langer 2010; and references therein); for recent applications to X-ray binaries and further description of Case BB RLO, see e.g. Tauris, Langer & Kramer (2011; 2012; Lazarus et al. (2014). We assumed an initial helium star donor mass of $M_{He} = 2.9\ M_\odot$, a NS mass of $M_{NS} = 1.35\ M_\odot$ and orbital period, $P_{orb} = 0.10^d$. Fig. 12 shows the complete evolution of the helium star donor in the HR diagram. For the same model, we present the calculated mass-transfer rate, the Kippenhahn diagram and the final chemical structure in Figs. 13–15.

The points marked by letters along the evolutionary track in Fig. 12 correspond to: A) helium star zero-age main sequence ($t = 0$); B) core helium exhaustion at $t = 1.75$ Myr, defining the bottom of the giant branch with shell helium burning; C) onset of Case BB RLO at $t = 1.78$ Myr; $D_1$) core carbon burning during $t = 1.836 - 1.849$ Myr, leading to radial contraction and Roche-lobe detachment (as a result of the mirror principle when the core expands); $D_2$) consecutive ignitions of carbon burning shells during $t = 1.850 - 1.854$ Myr and Roche-lobe detachment again; E) maximum luminosity at $t = 1.854\,304$ Myr; F) off-center ($m/M_\odot \simeq 0.5$) ignition of oxygen burning at $t = 1.854\,337$ Myr ($T_c = 9.1 \times 10^8$ K; $\rho_c = 7.1 \times 10^7$ g cm$^{-3}$), marked by a bullet. Shortly thereafter (point G) the binary orbit becomes dynamically unstable when $\dot{M}_2 > 10^{-2}\ M_\odot$ yr$^{-1}$ and we took our model star out of the binary to continue its final evolution as an isolated star ($t \geq 1.854\,337$ Myr). At this stage $P_{orb} = 0.070^d$. Off-center oxygen burning ignites at $t = 1.854\,553$ Myr and 3–4 yr thereafter our computer code breaks down (see below). The evolution of our model is similar to those of Dewi et al. (2002) (cf. their fig. 6), who calculated their binary stellar models until carbon burning. Our model resembles also very well the cores of $9.5 - 11\ M_\odot$ single stars which were found to lead to iron core collapse (Umeda, Yoshida & Takahashi 2012; Jones et al. 2013). From fig. 4e of the latter work, we infer that our core is expected to undergo iron core collapse $\sim 10$ years after our calculations were terminated. Fig. 13 shows the mass-transfer rate, $|\dot{M}_2|$ as a function of time. The total duration of the mass-transfer phases is seen to last for about $\Delta t = 60\,000$ yr (excluding a couple of detached epochs), which causes the NS to accrete an amount $\Delta M_{NS} = (0.7 - 2.1) \times 10^{-3}\ M_\odot$, depending on the assumed accretion efficiency and the exact value of the Eddington accretion limit, $\dot{M}_{Edd}$. Here we assumed $\dot{M}_{Edd} = 3.9 \times 10^{-8}\ M_\odot$ yr$^{-1}$ (a typical value for accretion of helium rich matter) and allowed for the actual accretion rate to be somewhere in the interval 30%–100% of this value (see recent discussion by Lazarus et al. 2014). We note that the structure of the donor star is hardly affected by uncertainties in the NS accretion efficiency.

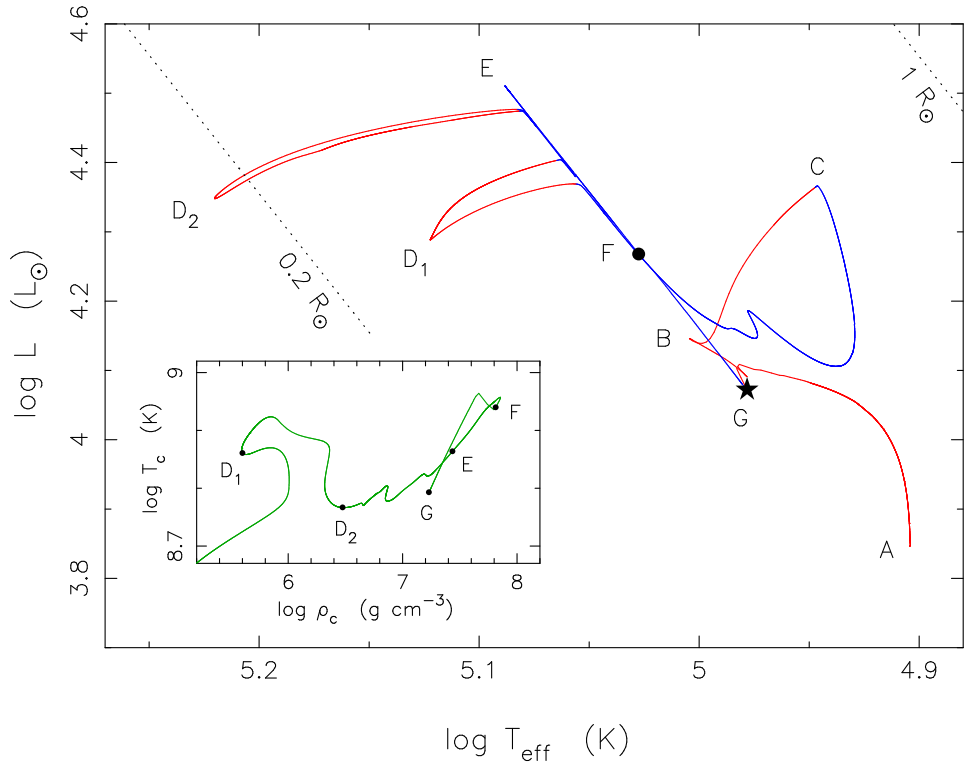

Figure 12: HR–diagram of the evolution of a 2.9 $M_\odot$ helium star which loses mass to a NS companion star prior to the core-collapse SN. The red part of the track corresponds to a detached system and the blue part is marking RLO. The evolutionary sequence A, B, C, $D_1$, $D_2$, E, F, G is explained in the text. The inner panel shows the final evolution in the central temperature-central density plane.

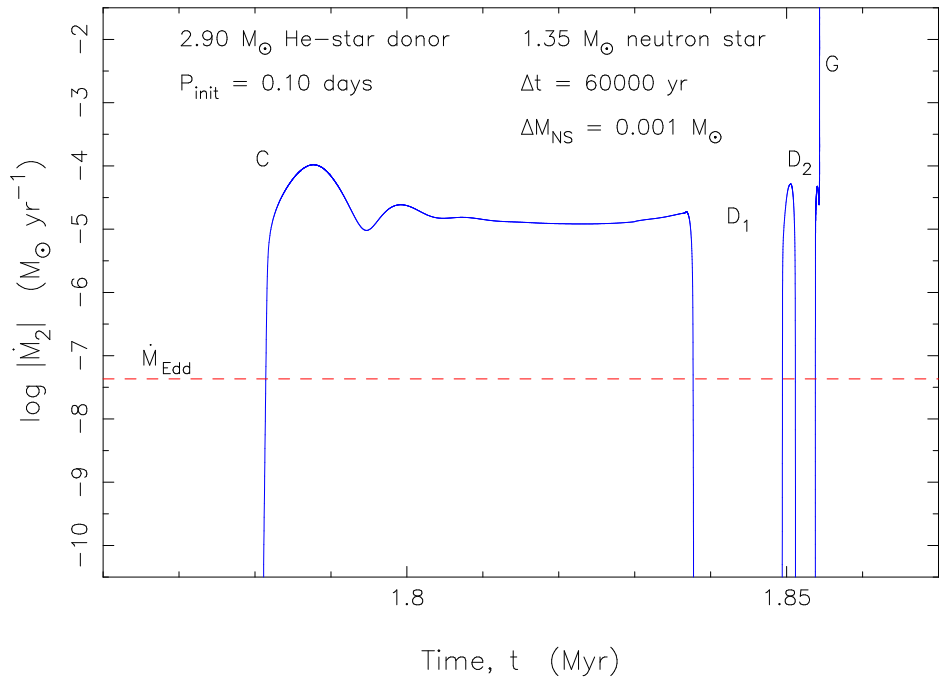

Figure 13: Mass-transfer rate as a function of stellar age for the helium star evolution plotted in Figs. 12 and 14. The major phase of mass transfer is the Case BB RLO lasting for about 56 000 yr. The mass-transfer rate is seen to be highly super-Eddington in this phase (the horizontal dashed line marks the Eddington accretion rate, $\dot{M}_{\rm Edd}$). The Case BB RLO is followed by a detached phase for about 12 000 yr because of core carbon burning ($D_1$). Following a second detached phase ($D_2$), a rigorous helium shell flash leads to the spike in point G – see text.

The tiny helium burning layer causes the star to expand vigorously, approximately at the same time the convective shell penetrates to the surface (cf. point G), resulting in numerical problems for our code. We therefore end our calculations without resolving this flash which is, in any case, not important for the core structure and the NS accretion due to the little amount of mass in the inflated envelope. As discussed in detail by Dewi & Pols (2003); Ivanova et al. (2003), the runaway mass transfer may lead to the onset of a common envelope (CE) evolution, where the final outcome is determined by the competition between the timescale for the onset of spiral-in and the remaining lifetime before gravitational collapse. In general, the duration of the CE and spiral-in phase is found to be $< 10^3$ yr (e.g. Podsiadlowski 2001), which is long compared to the estimated final lifetime of our model of about 10 yr (Jones et al. 2013). Therefore, while our model may still lose even more helium, the naked star is expected to undergo iron core collapse and produce a SN Ic (see Section 2.3.1 for a discussion of the SN type).

### 2.2.1 Final core structure

In Figs. 14–15 we present the final chemical structure of our stellar model, whose final mass is 1.50 $M_\odot$. It consists of an almost naked metal core of 1.45 $M_\odot$ which is covered by a helium-rich envelope of only 0.05 $M_\odot$. The Kippenhahn diagram in Fig. 14 shows the interior structure and evolution (energy production and convection zones) of the 2.9 $M_\odot$ helium star which undergoes Case BB RLO, and later Case BBB RLO (RLO re-initiated following shell carbon burning), and leaves behind a silicon-rich core. Seven carbon burning shells follow the core carbon burning phase. Oxygen ignites off-center (near a mass coordinate of $m/M_\odot \simeq 0.5$) due to an inversion of the temperature profile produced by neutrino cooling in the inner core. The convection zone on top of the initial oxygen burning shell (at $\log(t_* - t) \simeq 1.3$) reaches out to $m/M_\odot \simeq 1.2$. Hence, the final chemical structure of our model (Fig. 15) consists of a thick OMgSi outer core of $\sim 0.8\ M_\odot$, which is sandwiched by ONeMg-layers in the inner core ($\sim 0.4\ M_\odot$) and towards the envelope (0.14 $M_\odot$). This is surrounded by a CO-layer of 0.1 $M_\odot$ and a helium-rich envelope of $\sim 0.05\ M_\odot$. The total amount of helium in this envelope is 0.033 $M_\odot$.
The ultra-stripped nature of our model is achieved because the helium star is forced to lose (almost) its entire envelope in a very tight orbit where a NS can fit in.

## 2.3 Observational consequences

### 2.3.1 SN light curves and spectral type

We modelled the SN light curve evolution based on the SN progenitor star presented in Section 2.2, with SN kinetic energy, nickel mass and mass cut as free parameters. The light curve calculations were performed by a one-dimensional multi-group radiation hydrodynamics code `STELLA` (e.g. Blinnikov et al. 2006). Fig. 16 shows the obtained B-band light curves. We applied two mass cuts ($M_{cut} = 1.3\ M_\odot$ and 1.4 $M_\odot$) to the progenitor star with a corresponding SN ejecta mass ($M_{ej} = M_* - M_{cut}$) of 0.2 $M_\odot$ and 0.1 $M_\odot$, respectively.
As seen in Fig. 16, the calculation assuming a mass cut of $M_{cut} = 1.3\ M_\odot$, a nickel mass of $M_{Ni} = 0.05\ M_\odot$ and a SN explosion energy of $E_{ej} = 5 \times 10^{50}$ erg agrees well with the extremely rapidly declining Type Ic SN 2005ek (Drout et al. 2013). The multi-color light curve evolution of our model also matches that of SN 2005ek. These parameters are fairly consistent with those estimated by Drout et al. (2013) ($M_{ej} \sim 0.3\ M_\odot$, $E_{ej} \sim 2.5 \times 10^{50}$ erg, and $M_{Ni} \sim 0.03\ M_\odot$). As the rise time ($t_{rise} \propto M_{ej}^{3/4}\, E_{ej}^{-1/4}$, Arnett 1982) of SN 2005ek is not well-constrained by the observations, $M_{ej}$ and $E_{ej}$ remain uncertain and we expect some degeneracy in the light curve models. Therefore, the set of parameters shown in Fig. 16 for SN 2005ek may not be unique.
Recent modelling of synthetic SN spectra by Hachinger et al. (2012) suggests that more than 0.06 $M_\odot$ of helium is needed for helium lines to become visible in optical/IR spectra. Our final stellar model contains only 0.033 $M_\odot$ of helium. This amount could be further reduced by mass transfer and winds, as well as by explosive helium burning during the SN explosion. Therefore, the SN corresponding to our model is expected to be observed as a SN Ic, rather than a SN Ib.

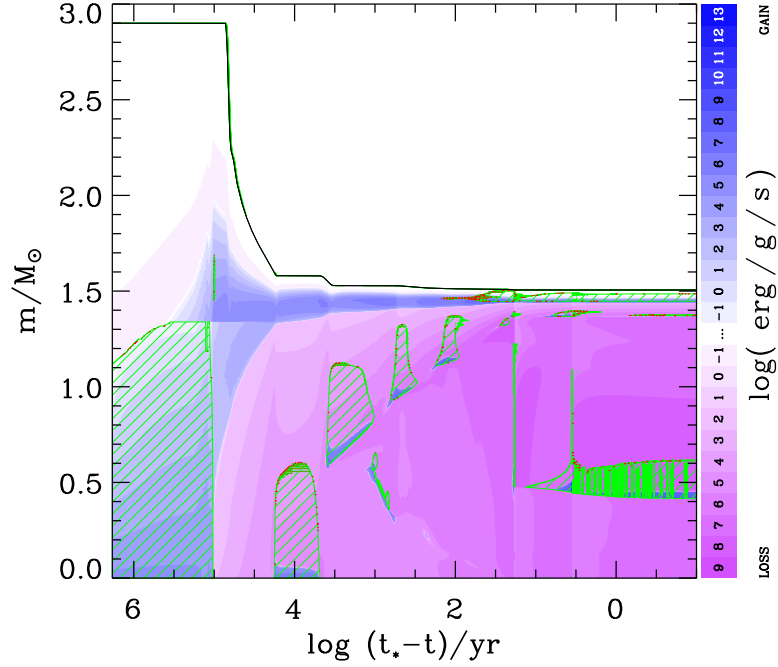

Figure 14: The Kippenhahn diagram of the $2.9\,M_\odot$ helium star undergoing Case BB/BBB RLO in Figs. 12–13. The plot shows cross-sections of the helium star in mass-coordinates from the centre to the surface of the star, along the y-axis, as a function of stellar age on the x-axis. The value $(t_* - t)$/yr is the remaining time of our calculations, spanning a total time of $t_* = 1.854356$ Myr. The green hatched areas denote zones with convection; red color indicates semi-convection. The intensity of the blue/purple color indicates the net energy-production rate. Shortly after off-centered oxygen ignition (at $m/M_\odot \simeq 0.5$, when $\log(t_* - t) = 1.3$) we evolved the star further as an isolated star for $\sim 20$ yr until our code crashed, about 10 yr prior to core collapse.

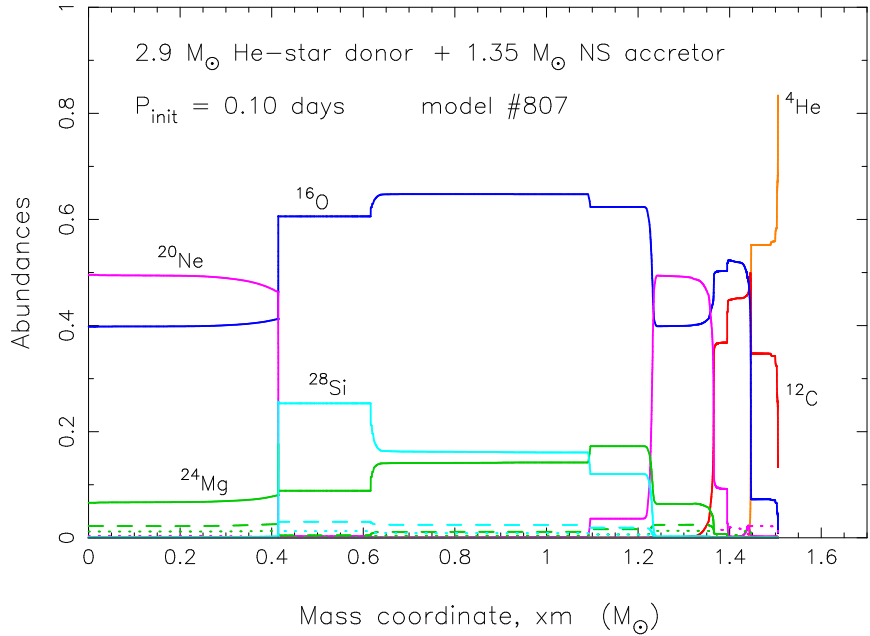

Figure 15: The chemical abundance structure of an ultra-stripped SN Ic progenitor star (from our last calculated model #807 at $t = 1.854356$ Myr), This naked $1.50\,M_\odot$ pre-collapsing star has a hybrid structure with an ONeMg inner core enclothed by a thick OMgSi outer core, which again is enclothed by shells of ONeMg and CO, and outermost a tiny envelope with $0.033\,M_\odot$ of helium.

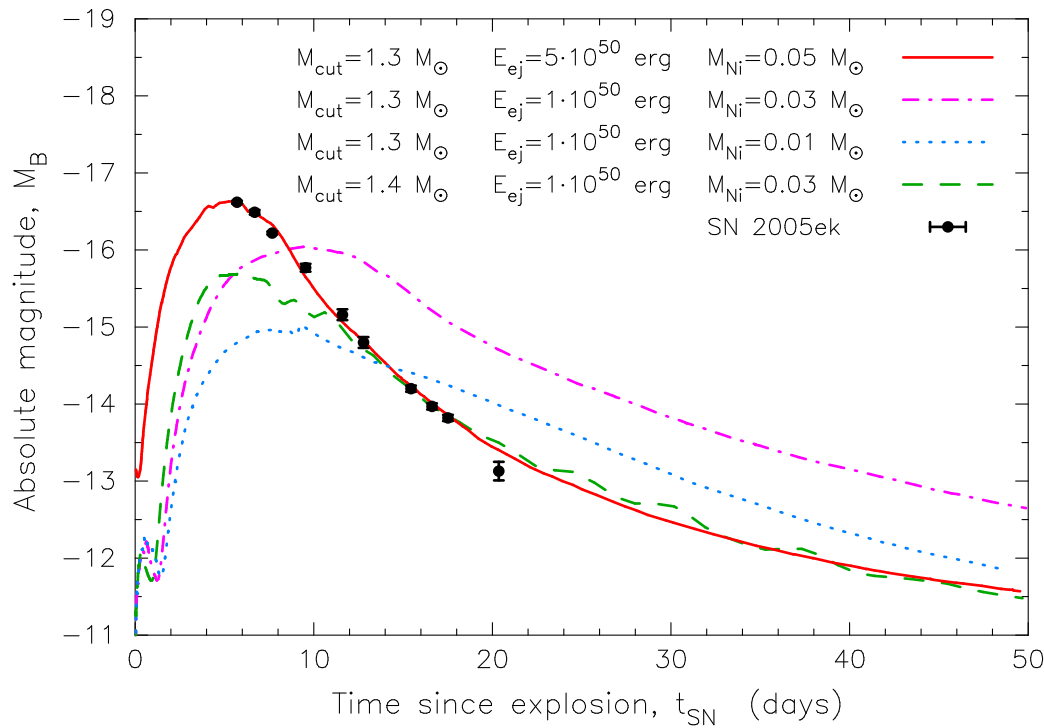

Figure 16: B-band SN light curves of ultra-stripped SNe Ic from the progenitor calculation obtained by binary evolution. The different curves correspond to various combinations of mass cut ($M_{cut}$), explosion energy ($E_{ej}$) and amount of $^{56}$Ni synthesized ($M_{Ni}$). The total bolometric luminosities, $L_{bol}$ of the four curves peak in the range $0.4 - 2.3 \times 10^{42}$ erg s$^{-1}$. The data for SN 2005ek is taken from Drout et al. (2013). The explosion date is arbitrarily chosen to match the light curve.

### 2.3.2   Comparison to regular SNe Ib/c

Yoon, Woosley & Langer (2010) showed that the stellar radii of regular SNe Ib/c progenitors in binary systems are generally larger for lower final masses. This is because the helium envelope expands farther for a more compact carbon-oxygen core during the evolutionary stages beyond core helium exhaustion. Relatively low-mass progenitors of SNe Ib/c with such extended envelopes will appear systematically more luminous than more massive SNe Ib/c progenitors (Yoon et al. 2012).

However, the ultra-stripped SN Ic progenitors considered in the present study have a very compact nature at the pre-SN stage despite their low final masses, due to the close proximity of the compact companion and the resultant tiny amount of helium left in the envelope. The final radius of about 0.4 $R_\odot$, bolometric luminosity of about 12 000 $L_\odot$ and a corresponding high surface temperature of about 95 000 K, as predicted by our model (Fig. 12), imply that they are visually faint ($M_V \simeq +1.9$), and it will be extremely difficult to identify them in pre-SN optical/IR images. Depending on the column density of the accretion disk corona and the circumbinary material from the ejected envelope, along the line-of-sight to Earth, these ultra-stripped systems may be bright X-ray sources prior to the SN.

If the mass transfer ends before the core collapse, our stripped progenitor is not expected to develop an optically thick wind. Therefore, the photosphere of the shock breakout will form near the hydrostatic stellar surface. The corresponding X-ray flash will have a harder energy spectrum, $kT \approx 10$ keV, and a shorter duration, $\delta t_s \approx 30$ sec (e.g. according to recent simulations and analytical estimates by Tolstov, Blinnikov & Nadyozhin 2013 and Sapir, Katz & Waxman 2013), than in the case of regular SN Ib/c (such as SN 2008D) for which $kT \approx 0.1$ keV and $\delta t_s \approx 200$ sec (e.g. Soderberg et al. 2008; Modjaz et al. 2009). This is because the progenitors of regular SNe Ib/c have either a larger radius or an optically thick wind, making the shock breakouts occur at relatively low temperatures.

### 2.3.3 Rates of ultra-stripped SNe Ic

The expected rate of ultra-stripped SNe Ic as discussed above can be estimated as follows. The evolution of a binary system must avoid a merger during the CE phase which produces the helium star–NS binary, but the resulting orbit must be close enough such that the compact star is able to peel-off the helium shell during the mass transfer initiated by the helium star expansion.

Immediate descendants of such systems are close double NS binaries which eventually merge due to gravitational wave radiation. The merger rate of those in a Milky Way-like galaxy is estimated to be between $\sim$ a few Myr$^{-1}$ and up to $\sim 100$ Myr$^{-1}$ (e.g. Abadie et al. 2010; and references therein). Compared to a total Galactic SN rate of $\sim 1 - 2$ per century, this would mean that only one in $10^2 - 10^4$ SNe is of this kind. On the other hand, we have to include similar systems producing black hole–NS binaries and also add the number of systems (a factor of a few) which were disrupted as a consequence of the dynamical effects of the second SN explosion, and are therefore not contributing to the estimated merger rates quoted above.

In addition, there are several other channels including a helium star and a relatively compact companion star that could also produce ultra-stripped SNe Ic: (1) helium star–white dwarf binaries and (2) the progenitors of intermediate-mass/low-mass X-ray binaries (LMXBs) and binary millisecond pulsars, where the companion is a relatively compact main-sequence star. The binary pulsar PSR J1141$-$6545 provides an example for the first of these evolutionary channels (Case BB RLO would occur between stages 7 and 8 in fig. 1, in Tauris & Sennels 2000). The existence of a non-recycled (and thereby short-lived) pulsar in this system with $P_{\rm orb} = 0.20^{\rm d}$ suggests that the rate for this channel could be at least comparable to that of the double NS merger channel. The progenitors of LMXBs with $P_{\rm orb} < 1$ day are also potential systems producing, at least partly, stripped SNe Ic. Their birthrate is probably $\lesssim 10^{-5}$ yr$^{-1}$ (Pfahl, Rappaport & Podsiadlowski 2003), but since the vast majority of these systems become disrupted in the SN (i.e. before the observed LMXB phase), the rate for stripped SNe Ic from this channel could be up to a factor of 10 higher. We note that the stripping effect is less certain for a giant helium star experiencing mass transfer to a low-mass main-sequence star and more research is needed to investigate this.

While it is difficult to estimate the rate of ultra-stripped SNe Ic, these arguments taken together suggest that a realistic estimate of the ratio of ultra-stripped SNe Ic to the total SN rate is likely to be in the range of $10^{-2} - 10^{-3}$.

### 2.3.4 Resulting neutron star birth properties

The extreme stripping producing the bare pre-SN core investigated here leads to an iron core-collapse SN with little ejecta mass, $M_{\rm ej}$. It is expected that the NS left behind will have a gravitational mass in the range $1.18 - 1.31$ $M_{\odot}$, depending on $M_{\rm cut}$ and the yet unknown equation-of-state (EoS) of NS matter and its associated release of gravitational binding energy (Lattimer & Yahil 1989). These values are in accordance with the NS masses measured in some double NS systems (e.g. Kramer et al. 2006b). Including the uncertainties in $M_{\rm cut}$ and the EoS, the total range of NS birth masses produced in iron core-collapse SNe may thus range from $1.10 - 1.70$ $M_{\odot}$ (see Tauris, Langer & Kramer 2011; for a discussion of the upper limit). The magnitude of any kick imparted to the newborn NS from an ultra-stripped iron core-collapse SN is uncertain. It may depend on, for example, $M_{\rm cut}$ (and $E_{\rm ej}$) and thus on the the timescale of the explosion compared to those of the non-radial hydrodynamic instabilities producing large kicks (e.g. Podsiadlowski et al. 2004; 2005; Janka 2012). For a symmetric explosion, the post-SN eccentricity of our system will be $e \simeq 0.07 - 0.13$, also in agreement with constraints obtained from some of the known double NS systems. In case our modelled pre-SN core had a slightly smaller mass, the outcome would have been an electron capture SN (Podsiadlowski et al. 2004; 2005). In general, we conclude that an ultra-stripped core, following an evolution similar to the one presented here, is an evident SN Ic progenitor candidate of any second iron core-collapse or electron capture SN forming a close-orbit double NS system. The young radio pulsar PSR J0737−3039B, in the double pulsar system (Kramer et al. 2006b), is an example of a NS which is most likely to have formed this way.

## 2.4 Conclusions

We have shown that post-CE mass stripping in a helium star–NS binary can produce a SN progenitor with a total mass of only $\sim 1.50\ M_\odot$ and an envelope mass of barely $0.05\ M_\odot$. The resulting iron core collapse of our example sequence leads to a Type Ic SN with an ejecta mass in the range $0.05 - 0.20\ M_\odot$. We estimate that one in every 100–1000 SNe may be of this ultra-stripped type. Through synthetic light curve calculations, we have demonstrated that SN 2005ek is a viable candidate for such an event. Given the current ambitious observational efforts to search for peculiar and weak optical transients, it seems probable that more such ultra-stripped SN Ic with diminutive ejecta will be detected within the coming years. Finally, we conclude that ultra-stripped cores are evident SN Ic progenitors of both iron core-collapse and electron capture SNe producing the second NS in close-orbit double NS systems.

- **Chapter 3**
  Tauris, Langer & Kramer (2011), MNRAS 416, 2130
  *Formation of Millisecond Pulsars with CO White Dwarf Companions – I.*
  *PSR J1614-2230: Evidence for a Neutron Star Born Massive*

- **Chapter 4**
  Antoniadis, Freire, Wex, Tauris, et al. (2013), Science 340, 448
  *A Massive Pulsar in a Compact Relativistic Binary*

- Istrate, Tauris, Antoniadis & Langer (2014), A&A, in prep.    [*not included here*]
  *Formation of the 2 $M_\odot$ pulsar J0348+0432*

# 3. Formation of Millisecond Pulsars with CO White Dwarf Companions – I. PSR J1614-2230: Evidence for a Neutron Star Born Massive

**Tauris, Langer & Kramer (2011)**

**MNRAS 416, 2130**

## Abstract

The recent discovery of a $2\,M_\odot$ binary millisecond pulsar (Demorest et al. 2010) has not only important consequences for the equation-of-state of nuclear matter at high densities but also raises the interesting question if the neutron star PSR J1614−2230 was born massive. The answer is vital for understanding neutron star formation in core collapse supernovae. Furthermore, this system raises interesting issues about the nature of the progenitor binary and how it evolved during its mass exchanging X-ray phase. In this paper we discuss the progenitor evolution of PSR J1614−2230. We have performed detailed stellar evolution modelling of intermediate-mass X-ray binaries undergoing Case A Roche-lobe overflow (RLO) and applied an analytic parameterization for calculating the outcome of either a common envelope evolution or the highly super-Eddington isotropic re-emission mode. We find two viable possibilities for the formation of the PSR J1614−2230 system: either it contained a $2.2 - 2.6\,M_\odot$ giant donor star and evolved through a common envelope and spiral-in phase or, more likely, it descended from a close binary system with a $4.0 - 5.0\,M_\odot$ main sequence donor star via Case A RLO. We conclude that the neutron star must have been born with a mass of $\sim 1.95\,M_\odot$ or $1.7 \pm 0.15\,M_\odot$, respectively – which significantly exceeds neutron star birth masses in previously discovered radio pulsar systems. Based on the expected neutron star birth masses from considerations of stellar evolution and explosion models, we find that the progenitor star of PSR J1614−2230 is likely to have been more massive than $20\,M_\odot$.

## 3.1 Introduction

Neutron stars are formed as compact remnants of massive stars ($10 - 30\,M_\odot$) which explode in supernovae at the end of their stellar life (Woosley, Heger & Weaver 2002; Heger et al. 2003). In order to better understand the mechanisms of the electron capture and core collapse supernovae knowledge of the distribution of birth masses of neutron stars is vital. However, in order to weigh a neutron star it must be a member of a binary system. This introduces an uncertainty in determining the original birth mass of the neutron star since these neutron stars are often observed in X-ray binaries or, at a later stage, as recycled pulsars and hence *after* they have undergone a phase of mass accretion from their companion star. The most precisely measured masses of neutron stars are obtained in double neutron star systems via general relativistic effects. The related post-Keplerian parameters include periastron advance, redshift/time dilation, orbital period derivative and Shapiro delay (e.g. Will 2009). Shapiro delays of radio signals from pulsars (Stairs et al. 1998) have the advantage of being measurable also in low eccentricity systems if the orbital inclination is such that the pulses passes in the vicinity of its companion. This method yields the opportunity to weigh both neutron stars accurately – and hence also determine the mass of the last formed neutron star which has not accreted any material. So far, such measurements have revealed that even the most massive of these neutron stars (the non-recycled pulsars) do not exceed a mass of $1.39\,M_\odot$ (Thorsett & Chakrabarty 1999; Schwab, Podsiadlowski & Rappaport 2010). There is, however, some evidence from neutron stars in X-ray binaries, e.g. Vela X-1, that suggests neutron stars can be born more massive than this value.

Binary millisecond pulsars are known to be key sources of research in fundamental physics. They host the densest matter in the observable Universe and possess very rapid spins as well as relativistic magnetospheres with outflowing plasma winds. Being ultra stable clocks they also allow for unprecedented tests of gravitational theories in the strong-field regime (Kramer

Table 1: Physical parameters of the binary millisecond pulsar PSR J1614−2230 (the data was taken from Demorest et al. 2010).

| Parameter | value |
|---|---|
| Pulsar mass | $1.97 \pm 0.04 \, M_\odot$ |
| White dwarf mass | $0.500 \pm 0.006 \, M_\odot$ |
| Orbital period | 8.6866194196(2) days |
| Projected pulsar semimajor axis | 11.2911975 light sec |
| Orbital eccentricity | $1.30 \pm 0.04 \times 10^{-6}$ |
| Inclination angle | $89.17 \pm 0.02$ deg. |
| Dispersion-derived distance | 1.2 kpc |
| Pulsar spin period | 3.1508076534271 ms |
| Period derivative | $9.6216 \times 10^{-21}$ |

& Wex 2009). Equally important, however, binary millisecond pulsars represent the end point of stellar evolution, and their observed orbital and stellar properties are fossil records of their evolutionary history. Thus one can use binary pulsar systems as key probes of stellar astrophysics.

Recent Shapiro delay measurements of PSR J1614−2230 (Demorest et al. 2010) allowed a precise mass determination of this record high-mass pulsar (neutron star) and its white dwarf companion. Characteristic parameters of the system are shown in Table 1. It is well established that the neutron star in binary millisecond pulsar systems forms first, descending from the initially more massive of the two binary stellar components. The neutron star is subsequently spun-up to a high spin frequency via accretion of mass and angular momentum once the secondary star evolves (Alpar et al. 1982; Radhakrishnan & Srinivasan 1982; Bhattacharya & van den Heuvel 1991). In this recycling phase the system is observable as a low-mass X-ray binary (e.g. Nagase 1989) and towards the end of this phase as an X-ray millisecond pulsar (Wijnands & van der Klis 1998; Archibald et al. 2009). Although this formation scenario is now commonly accepted many aspects of the mass-transfer process and the accretion physics (e.g. the accretion efficiency and the details of non-conservative evolution) are still not well understood (Lewin & van der Klis 2006).

In this paper we investigate the progenitor evolution of PSR J1614−2230. We are mainly focusing on the important X-ray binary phase starting from the point where the neutron star has already formed. However, we shall also briefly outline the previous evolution from the zero-age main sequence (ZAMS) binary until this stage since this evolution is important for the birth mass of the neutron star. In Section 3.2 we discuss the three different possibilities for mass transfer toward a neutron star from an intermediate-mass star of $2.2 - 5.0 \, M_\odot$, for the Roche-lobe overflow (RLO) Cases A, B and C. The evolution of the original ZAMS binary until the X-ray phase is briefly discussed in Section 3.3. In Section 3.4 we compare our results with the outcome of the independent work by Lin et al. (2011) and also discuss our results in a broader context in relation to neutron star birth masses predicted by stellar evolution and supernova explosions. Our conclusions are given in Section 3.5.

In Paper II (Tauris, Langer & Kramer 2012) we continue the discussion of PSR J1614−2230 in view of general aspects of accretion onto neutron stars during the recycling process of millisecond pulsars.

## 3.2   Mass transfer in X-ray binaries

Consider a close interacting binary system which consists of a non-degenerate donor star and a compact object, in our case a neutron star. If the orbital separation is small enough the (evolved) non-degenerate star fills its inner common equipotential surface (Roche-lobe) and becomes a donor star for a subsequent epoch of mass transfer toward the, now, accreting neutron star. In this phase the system is observed as an X-ray binary. When the donor star fills its Roche-lobe it is perturbed by removal of mass and it falls out of hydrostatic and thermal equilibrium. In the process of re-establishing equilibrium the star will either grow or shrink – depending on the properties of its envelope layers as discussed below – first on a dynamical (adiabatic) timescale and subsequently on a slower thermal timescale. However, any exchange and loss of mass in such an X-ray binary system will also lead to alterations of the orbital dynamics, via modifications in the orbital angular momentum, and hence changes in the size of the critical Roche-lobe radius of the donor star. The stability of the mass-transfer process therefore depends on how these two radii evolve (i.e. the radius of the star and the Roche-lobe radius). The various possible modes of mass exchange and loss include, for example, direct fast wind mass loss, Roche-lobe overflow, with or without isotropic re-emission, and common envelope evolution (e.g. van den Heuvel 1994a; Soberman, Phinney & van den Heuvel 1997; and references therein). The RLO mass transfer can be initiated while the donor star is still on the main sequence (Case A RLO), during hydrogen shell burning (Case B RLO) or during helium shell burning (Case C RLO). The corresponding evolutionary timescales for these different cases will in general proceed on a nuclear, thermal or dynamical timescale, respectively, or a combination thereof. We now investigate each of these three cases with the aim of reproducing the parameters of PSR J1614−2230.

### 3.2.1 Case C RLO - dynamical unstable mass transfer

Donor stars in systems with wide orbits ($P_{\rm orb} \simeq 10^2 - 10^3$ days) prior to the mass-transfer phase develop a deep convective envelope as they become giant stars before filling their Roche-lobe. The response to mass loss for these stars with outer layers of constant low entropy and negative adiabatic mass-radius exponents ($\xi = \partial \ln R / \partial \ln M < 0$) is therefore expansion which causes the stars to overfill their Roche-lobes even more. To exacerbate this problem, binaries also shrink in size if mass transfer occurs from a donor star somewhat more massive than the accreting neutron star. This causes further overfilling of the donor star Roche-lobe resulting in enhanced mass loss etc. This situation is clearly a vicious circle that leads to a runaway mass transfer and the formation of a contact binary with a common envelope (CE) followed by a spiral-in phase, e.g. Paczyński (1976), Iben & Livio (1993).

A simple estimate of the reduction of the orbit can be found by equating the binding energy of the envelope of the AGB giant donor to the difference in orbital energy before and after the CE-phase. The idea is that the motion of the neutron star, once captured in the CE, results in friction and thus dissipation of orbital energy which can be used to expel the CE. Following the formalism of Webbink (1984) and de Kool (1990), the binding energy of the envelope at the onset of RLO mass transfer can be written as: $-GM_2M_{\rm env}/(\lambda\,R_2)$, where $M_2$ is the mass of the donor star, with envelope mass $M_{\rm env}$, and $R_2 = R_{\rm L}$ is the Roche-lobe radius of the donor star at the onset of the CE-phase. This radius is often calculated in terms of its dimensionless Roche-lobe radius, $r_{\rm L}$ (Eggleton 1983) such that $R_2 \simeq R_{\rm L} = a_0 \cdot r_{\rm L}$, where $a_0$ is the initial orbital separation.

The total binding energy of the envelope includes both the negative gravitational binding energy and the positive thermal energy. Besides from the thermal energy of a simple perfect gas, the latter term also includes the energy of radiation, terms due to ionization of H and He and dissociation of $H_2$, as well as the contribution from the Fermi energy of the degenerate electrons (Han, Podsiadlowski & Eggleton 1994; 1995). The value of the $\lambda$-parameter can thus be calculated from stellar structure models (Dewi & Tauris 2000; 2001; Tauris & Dewi 2001; Xu & Li 2010b;a; Loveridge, van der Sluys & Kalogera 2011; Ivanova 2011). Given the radius of the donor star and the $\lambda$-parameter enables one to estimate the change in orbital separation as a result of the neutron star spiral-in and ejection of the envelope. Let $\eta_{\rm ce}$ describe the efficiency of ejecting the envelope via drag forces, i.e. of converting orbital energy ($E_{\rm orb} = -GM_2M_{\rm NS}/2\,a$) into the kinetic energy that provides the outward motion of the envelope: $E_{\rm env} \equiv \eta_{\rm ce}\,\Delta E_{\rm orb}$ and one finds the well-known expression for the ratio of the change in orbital separation:

$$\frac{a}{a_0} = \frac{M_{\rm core}M_{\rm NS}}{M_2}\frac{1}{M_{\rm NS} + 2M_{\rm env}/(\eta_{\rm ce}\lambda r_{\rm L})} \tag{10}$$

where $M_{core} = M_2 - M_{env}$ is the core mass of the evolved donor star (essentially the mass of the white dwarf to be formed); $M_{NS}$ is the mass of the neutron star and $a$ is the final orbital separation after the CE-phase. Strictly speaking, when considering the energy budget the "effective efficiency parameter" should also include the excess energy of the ejected matter at infinity – although this effect is probably small. Recent work (Zorotovic et al. 2010; de Marco et al. 2011) suggests that the efficiency parameter is of the order $30\%$, i.e. $\eta_{ce} \simeq 0.3$, although its uncertainty is large. The value may not be universal and could, for example, depend on the stellar mass ratio in a given binary.

During the very short spiral-in and ejection phase of a common envelope evolution ($\sim 10^3$ yr) it is a good approximation to assume that the neutron star does not accrete any significant amount of matter given that its accretion is limited by the Eddington luminosity corresponding to a maximum accretion rate of $\sim 10^{-8}\, M_\odot\, yr^{-1}$, depending on the exact chemical composition of the accreted material and the geometry of its flow.

The possibility of hypercritical accretion onto the neutron star during the spiral-in phase has been suggested to lead to significant mass increase and possible collapse of the neutron star into a black hole (Chevalier 1993; Brown 1995; Fryer, Benz & Herant 1996). However, there is observational evidence that, at least in some cases, this is not the case. The recent determination of the low pulsar mass in PSR J1802−2124 of $1.24 \pm 0.11\, M_\odot$ (Ferdman et al. 2010) clearly demonstrates that this 12.6 millisecond recycled pulsar did not accrete any significant amount of matter. This pulsar is in a tight binary with an orbital period of only 16.8 hours and it has a carbon-oxygen white dwarf (CO WD) companion of mass $0.78 \pm 0.04\, M_\odot$. With such a small orbital period combined with a massive white dwarf there seems to be no doubt that this system evolved through a common envelope and spiral-in phase and the low pulsar mass reveals that very little mass has been accumulated by the neutron star during this phase. However, the recycling of this pulsar does require some $10^{-2}\, M_\odot$ of accreted material (see further discussion in Paper II). While in principle, PSR J1802−2124 could have formed via accretion-induced collapse of an ONeMg WD, so far there are no models which suggest this.

Before we proceed to discuss the case of PSR J1614−2230 let us introduce a new parameterization for calculating the outcome of a common envelope evolution. We define a mass ratio parameter $k \equiv q_0/q$ where $q_0$ and $q$ represent the initial and final ratio, respectively, of the donor star mass to the neutron star mass. Assuming the neutron star mass to be constant during the CE-phase we can also write $k = M_2/M_{core}$. The value for $k$ is thus the mass of the donor star, in units of its core mass, at the onset of the RLO. This allows for a convenient rewriting of Eq. (10):

$$\frac{a}{a_0} = \frac{k^{-1}}{1 + 2q(k-1)/\eta_{ce}\lambda r_L} = [k + 2q_0(k-1)/\eta_{ce}\lambda r_L]^{-1} \qquad (11)$$

The post-CE value for the mass ratio $q = M_{WD}/M_{NS} \simeq 0.25$ is the present value in J1614−2230, which is directly determined from measurements (see Table 1). Hence we have $k = M_2/M_{WD}$. Taking $M_{WD} = 0.50\, M_\odot$ as the core mass we must first determine the value of $M_2$ (and thus $k$) from stellar evolution calculations. To this purpose we used a detailed one-dimensional hydrodynamic stellar evolution code. This code has been described in detail e.g. in Heger, Langer & Woosley (2000). Using solar chemical abundances ($Z = 0.02$) and a mixing-length parameter of $\alpha = l/H_p = 1.5$ (Langer 1991) we find $2.4 \leq M_2/M_\odot \leq 2.6$, see Fig. 17, if we disregard core convective overshooting. Including a core convective overshooting parameter of $\delta_{OV} = 0.20$ (Claret 2007) allows for donor masses as low as $2.2\, M_\odot$ to produce a final WD mass of $0.50\, M_\odot$. Hence, $4.4 \leq k \leq 5.2$ and we can now use Eq. (11) to find the pre-CE orbital separation, $a_0$ and hence the radius of the Roche-lobe filling donor star.

In Fig. 18 we demonstrate that indeed PSR J1614−2230 could have evolved from a CE. The shaded rectangular area shows the parameter space of solutions. The $k$-values are constrained by the initial donor mass, $M_2$ which in turn is constrained by the observed white dwarf mass, $M_{WD}$. The upper limit for the radius of the donor star at the onset of the RLO is simply its maximum possible radius on the AGB, $R_{max}$. We notice from the curves in the figure that only $\lambda$-values larger than about 2 are in agreement with this constraint (i.e. of having a donor radius less than $R_{max}$). The reason for this is the relatively wide orbit of PSR J1614$-$2230 with an orbital separation of 24.1 $R_{\odot}$. Hence, only a modest amount of orbital energy was released during spiral-in – almost independent of the pre-CE separation, $a_0$ since usually $a \ll a_0$ – and therefore the binding energy of the donor star envelope cannot have been too large for a successful envelope ejection ($E_{bind} \propto \lambda^{-1}$). The lower limit of the progenitor star radius at $\sim 300 \, R_{\odot}$ is therefore determined by exactly this requirement of having an envelope with small enough binding energy (in this case corresponding to $\lambda \geq 2$) such that it can be successfully ejected during the spiral-in phase. (For a graphical example of a slightly more massive donor star of 3 $M_{\odot}$, see Fig. 17 in Dewi & Tauris, 2000). If the donor radius is smaller at the time of the onset of the CE then its $\lambda$-value is too small (i.e. its envelope binding energy is too large, on an absolute scale, to allow ejection from the available orbital energy release). The outcome is a merger event – possibly leading to a Thorne-Żytkow object (Thorne & Żytkow 1977). A similar fate is expected for donor stars of late Case B RLO. These stars also possess a deep convective envelope, resulting in a CE evolution. However, they are less evolved than stars on the AGB and have much smaller $\lambda$-values and hence more tightly bound envelopes which strengthens the case for a merger.

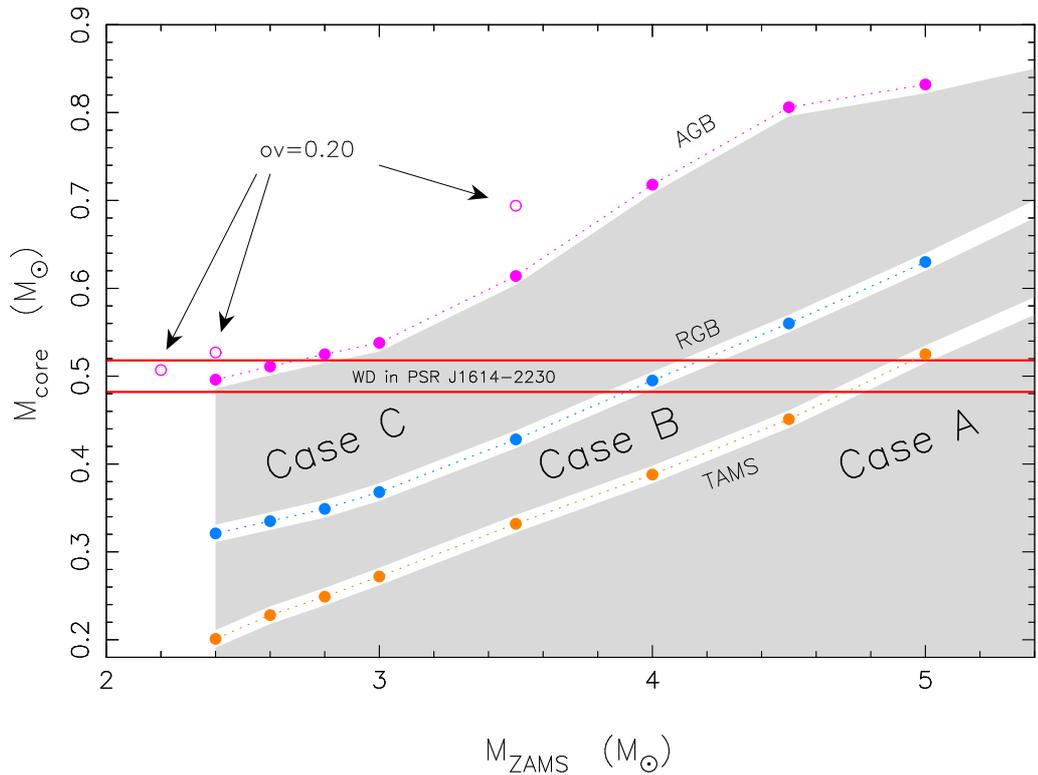

Figure 17: Stellar core mass as a function of the initial ZAMS mass, calculated at different evolutionary epochs. Mass transfer initiated either before the TAMS, the tip of the RGB or the AGB, corresponds to RLO Case A, Case B and Case C, respectively. Filled circles represent models without convective core-overshooting. The open circles show a few examples of core masses at the tip of the AGB assuming a convective core-overshooting parameter of $\delta_{OV} = 0.20$. The horizontal red lines indicate the measured mass interval (within 3-$\sigma$ error bars) of the white dwarf in PSR J1614$-$2230. This white dwarf descends from the core of the donor star in the X-ray binary.

To summarize, based on the orbital dynamics and the masses of the two stellar components, Case C RLO (leading to a CE and spiral-in) is possible to have occurred in PSR J1614$-$2230. This would have the implication that the neutron star was born massive with a mass close to its presently observed mass of 1.97 $M_{\odot}$. However, see further discussion in Sections 3.3 and 3.4, and also in paper II.

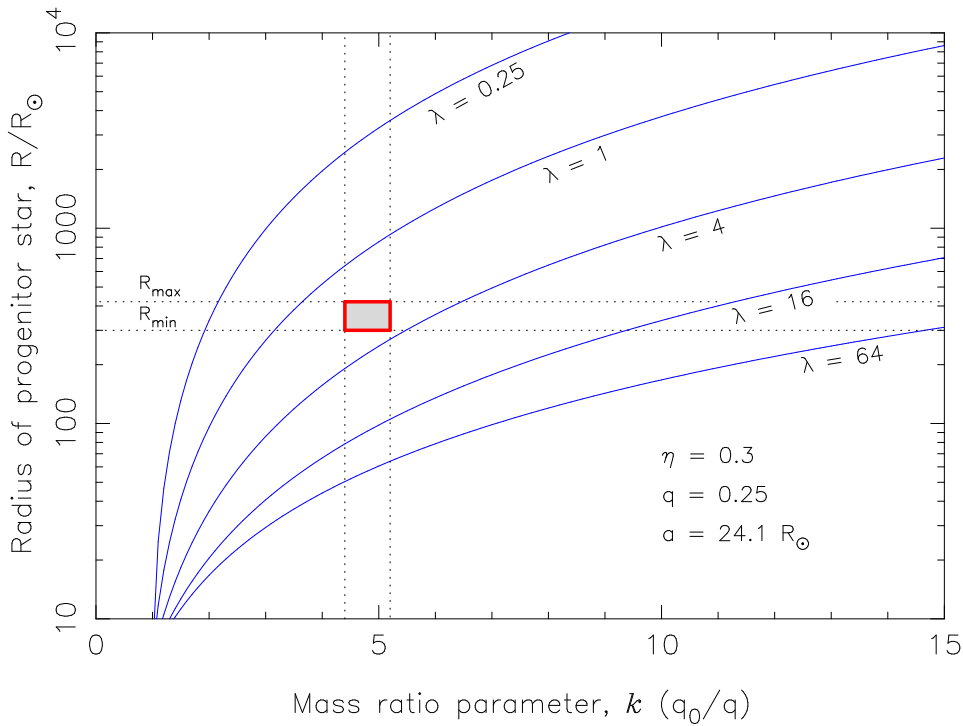

Figure 18: Constraints on stellar parameters assuming common envelope evolution in the PSR J1614−2230 progenitor system. The shaded area in the red box indicates the allowed parameter space for the radius of the progenitor star at onset of RLO and the mass ratio parameter, $k$. The various curves correspond to different $\lambda$-values of the binding energy of the envelope (see text).

#### 3.2.1.1 Case BB RLO following a common envelope?

One might ask whether the $0.50\,M_\odot$ CO WD in PSR J1614−2230 could have formed via Case BB RLO from a $\sim 1\,M_\odot$ helium star which had previously lost its hydrogen envelope in a common envelope phase. In this case the neutron star could have accreted significantly and needed not be so massive at birth (only about $1.5\,M_\odot$ if it accreted $0.5\,M_\odot$ subsequently). We have tested this hypothesis and find that it is not a possible formation channel. The reason is that there is not enough orbital energy available (by an order of magnitude) to eject the envelope of the WD progenitor already on the red giant branch (RGB). Assume the progenitor of the naked $1\,M_\odot$ helium core was a $5-7\,M_\odot$ star (the exact mass depends e.g. on the assumed amount of core convective overshooting). For such a star on the RGB we find that the (absolute) binding energy of the envelope is $1.3 - 2.2 \times 10^{48}$ erg – according to the formula for the binding energy in Section 3.2.1 and using Fig. 1 in Dewi & Tauris (2000) for estimating the largest possible $\lambda$-values to be $0.4-0.6$. The Case BB RLO (from the less massive helium star to the more massive neutron star) would have widened the orbit to its currently observed orbital period of 8.69 days. Assuming a conservative mass transfer the initial orbital separation would have been about $10.7\,R_\odot$. This value is then equivalent to the post-CE orbital separation corresponding to an orbital energy of $-2.7 \times 10^{47}$ erg. Hence, even if all available orbital energy was released from an initial separation at infinity there would be far too little energy to eject the envelope of the RGB star ($\Delta E_{\rm orb} \ll E_{\rm env}$). Only on the AGB is the envelope loose enough to allow ejection from spiral-in of the neutron star. But in this case the core has already burned its helium into carbon and oxygen and then we are back where we began in Section 3.2.1.

#### 3.2.2 Early Case B RLO - thermal timescale mass transfer

If a $3-5\,M_\odot$ donor star fills its Roche-lobe shortly after leaving the main sequence (*early* case B) its envelope is still radiative and the binary may survive thermal timescale mass transfer. This was shown a decade ago in three independent papers: King & Ritter (1999) and Podsiadlowski &

Rappaport (2000) studied the formation and evolution of Cyg X-2 and Tauris, van den Heuvel & Savonije (2000) investigated the formation of binary millisecond pulsars with a CO WD companion. Although Tauris, van den Heuvel & Savonije (2000) and Podsiadlowski, Rappaport & Pfahl (2002) have demonstrated that one can form systems with a $0.50 \, M_\odot$ CO WD and $P_{\rm orb} = 8.7$ days (as observed in PSR J1614$-$2230) they both assumed in their calculations an initial canonical neutron star mass of $1.30 - 1.40 \, M_\odot$, which does not apply in this scenario since the neutron star only accretes a few $0.01 \, M_\odot$ during the Case B rapid mass-transfer phase, thus disqualifying the neutron star from reaching its present mass of $1.97 \, M_\odot$. If the initial mass ratio between the donor star and the neutron star is of the order $q_0 \simeq 2 - 3$ the orbit shrinks significantly in response to mass loss, as mentioned previously. This leads to highly super-Eddington mass-transfer rates and hence we can apply the isotropic re-emission mode of mass transfer (Bhattacharya & van den Heuvel 1991). In this model matter flows over from the donor star ($M_2$) to the accreting neutron star ($M_{\rm NS}$) in a conservative manner and thereafter a certain fraction, $\beta$ of this matter is ejected from the vicinity of the neutron star with the specific orbital angular momentum of the neutron star (for example, in a jet as observed in SS433, see also King & Begelman 1999). Integrating the orbital angular momentum balance equation one can find the change in orbital separation during the isotropic re-emission RLO (e.g. Tauris 1996; King et al. 2001):

$$\frac{a}{a_0} = \left( \frac{q_0(1-\beta)+1}{q(1-\beta)+1} \right)^{\frac{3\beta-5}{1-\beta}} \left( \frac{q_0+1}{q+1} \right) \left( \frac{q_0}{q} \right)^2 \qquad (12)$$

where it is assumed that $\beta$ remains constant during the mass-transfer phase. Indeed Tauris, van den Heuvel & Savonije (2000) showed in their Fig. 1 that in intermediate-mass X-ray binary (IMXB) systems very little mass is accreted onto the neutron star since the timescale for the mass-transfer phase is very short ($\sim 1$ Myr) leading to highly super-Eddington mass-transfer rates by 3–4 orders of magnitude and hence $\beta > 0.999$. It is therefore interesting to consider Eq. (12) in the limit where $\beta \to 1$ and we find (see also King et al. 2001) for the change in orbital period:

$$\lim_{\beta \to 1} \left( \frac{P}{P_0} \right) = \left( \frac{k\,q+1}{q+1} \right)^2 k^3 \, e^{3q(1-k)} \qquad (13)$$

under the above mentioned assumptions and by applying Kepler's third law.

In Fig. 19 we demonstrate that early Case B mass transfer is not possible to have occurred in the progenitor binary of PSR J1614$-$2230. The constraints on $k$ for Case B mass transfer can be found from Fig.17: For Case B mass transfer (between evolutionary epochs TAMS and the RGB) a progenitor star of $4.0 - 4.5 \, M_\odot$ is needed to yield a core mass of $0.50 \, M_\odot$ (the observed mass of the white dwarf in PSR J1614$-$2230). Hence, we find $8 \le k \le 9$ for this scenario. Recalling $q = M_{\rm WD}/M_{\rm NS} = 0.25$ and given the orbital period of $P = 8.69$ days then, according to Eq. (13), this would require an initial orbital period, $P_0 \simeq 0.7$ days which is not possible for Case B mass transfer – the minimum initial period for (early) Case B RLO is shown as the red line in Fig. 19. In fact with such a short initial orbital period the donor star would fill its Roche-lobe radius instantly on the ZAMS. Even if we expand the interval of donor star masses to the entire range $2.5 < M_2/M_\odot < 6.0$ early Case B RLO would still not be possible to explain the parameters of PSR J1614$-$2230. We can therefore safely rule out all variations of Case B mass transfer (early, late and Case BB).

### 3.2.3 Case A RLO – mass transfer from a main sequence star

In order to reproduce PSR J1614$-$2230 via Case A RLO we notice from Fig.17 that we must at first glance require an initial donor mass of almost $M_2 \simeq 5 \, M_\odot$ in order to end with a final white dwarf mass of $0.50 \, M_\odot$. However, the evolution of Case A RLO is somewhat complex and not straight forward to analyse analytically (see Tauris & Langer, in prep., for further details). The estimated TAMS core masses from Fig.17 are not necessarily good indicators for the final mass of the white dwarf remnants evolving from Case A donors in X-ray binaries for two reasons: 1) forced mass loss from the Roche-lobe filling donor star results in a lower core mass as the donor now evolves less massive, and 2) the formation of an outgoing hydrogen shell source during the final phase (phase AB, see below) of the mass transfer causes the core mass to grow with the helium ashes left behind. Therefore, to obtain the final mass of the white dwarf requires detailed numerical stellar models. The overall effect is that the core mass will have grown somewhat by the time the system detaches from the RLO. Hence, the white dwarf remnant left behind is expected to be slightly more massive than the donor core mass at the TAMS. For this reason the ZAMS mass interval found from Fig. 17 for a Case A donor star of PSR J1614−2230 should be considered as an upper limit and in the following we explore donor masses down to $4.0\,M_\odot$. (Below this limit the WD remnant becomes too light, see Section 3.2.3.1). Stars more massive than $5.0\,M_\odot$ could leave behind a core mass even less than $0.50\,M_\odot$ if the mass transfer is initiated well before reaching the TAMS. However, these binaries would not be dynamically stable with a neutron star accretor, see Section 3.2.3.1.

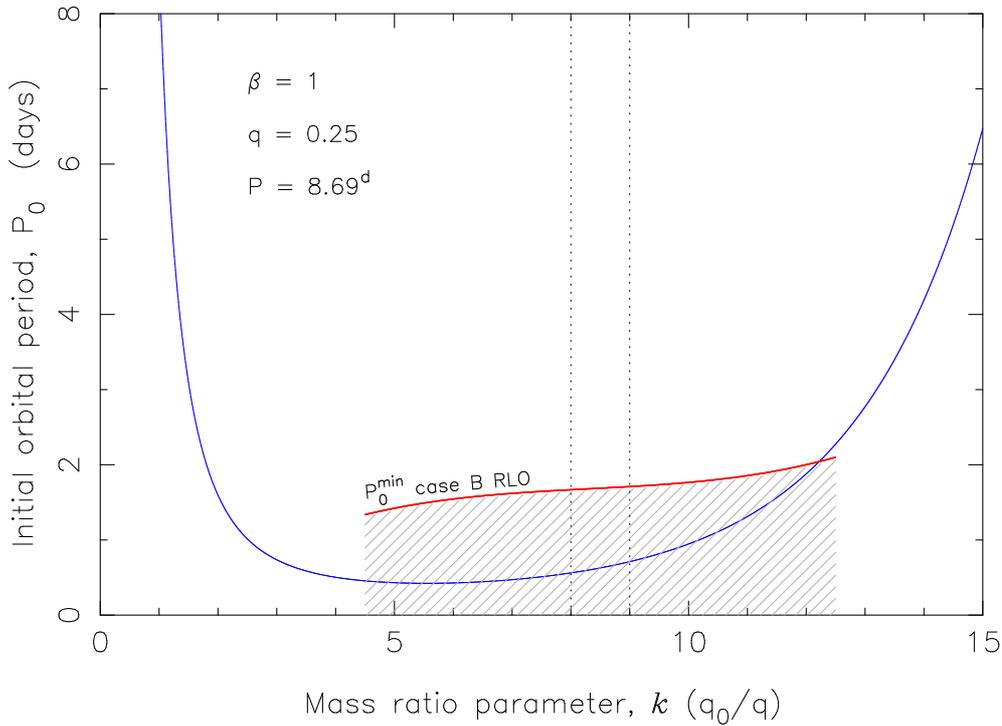

Figure 19: The blue line represents the initial orbital period of the progenitor X-ray binary as a function of the mass ratio parameter $k$ using the isotropic re-emission model for early Case B RLO. The two vertical dotted lines indicate the interval of possible values of $k$ for PSR J1614−2230. The hatched region is excluded for Case B RLO since the donor star would have filled its Roche-lobe before reaching the stage of shell hydrogen burning. The original donor star mass is $M_2 = k \times 0.50\,M_\odot$. Note, the $k$-values for Case B RLO are larger than the $k$-values expected for Case C RLO in Fig. 18. The reason for this is that the core of a Case B donor star has not had time to evolve to the large core masses found in Case C RLO (see Fig. 17), resulting in higher required donor masses, $M_2$ for Case B compared to a Case C scenario.

Our analysis reveals the parameter space of Case A binaries which produce the characteristic parameters of PSR J1614−2230. Figs. 20–23 show an example of a calculation of a possible progenitor X-ray binary. This IMXB started out with an initial donor star of mass $4.50\,M_\odot$ and a $1.68\,M_\odot$ neutron star accretor having an initial orbital period of 2.20 days. We assumed here a convective core-overshooting of $\delta_{\rm OV} = 0.20$. In Fig. 20 and Fig. 21 we demonstrate that the system experiences three phases of mass transfer (hereafter denoted phases A1, A2 and AB, respectively). In phase A1 the mass transfer proceeds on the thermal timescale (see Langer et al. 2000; for further discussions on thermally unstable mass transfer). The reason for this is the initially large mass ratio between the heavier donor star and the lighter neutron star which causes the orbit to shrink in response to mass transfer. As mentioned earlier, the outcome in this situation is that the donor star overfills its Roche-lobe even more – leading to further mass loss – and within 1 Myr it loses more than $3\,M_\odot$ at a rate exceeding $10^{-5}\,M_\odot\,\mathrm{yr}^{-1}$. This rate is still less than the estimated limit for which photons are trapped, leading possibly to rapid neutrino cooling and hypercritical accretion. It has been demonstrated, for example by Fryer, Benz & Herant (1996) and King & Begelman (1999), that it takes an accretion rate of at least a few times $10^{-4}\,M_\odot\,\mathrm{yr}^{-1}$ before the outward diffusion of photons cannot overcome the inward advection of photons in the accreted matter. (The exact limit is uncertain and depends, for example, on the amount and the geometry of outflows). Although the donor star is driven out of thermal equilibrium during phase A1 it manages to retain hydrostatic equilibrium and the system can in this case avoid a so-called delayed dynamical instability (Hjellming & Webbink 1987; Kalogera & Webbink 1996) which would have resulted in a common envelope and most likely a merger event.

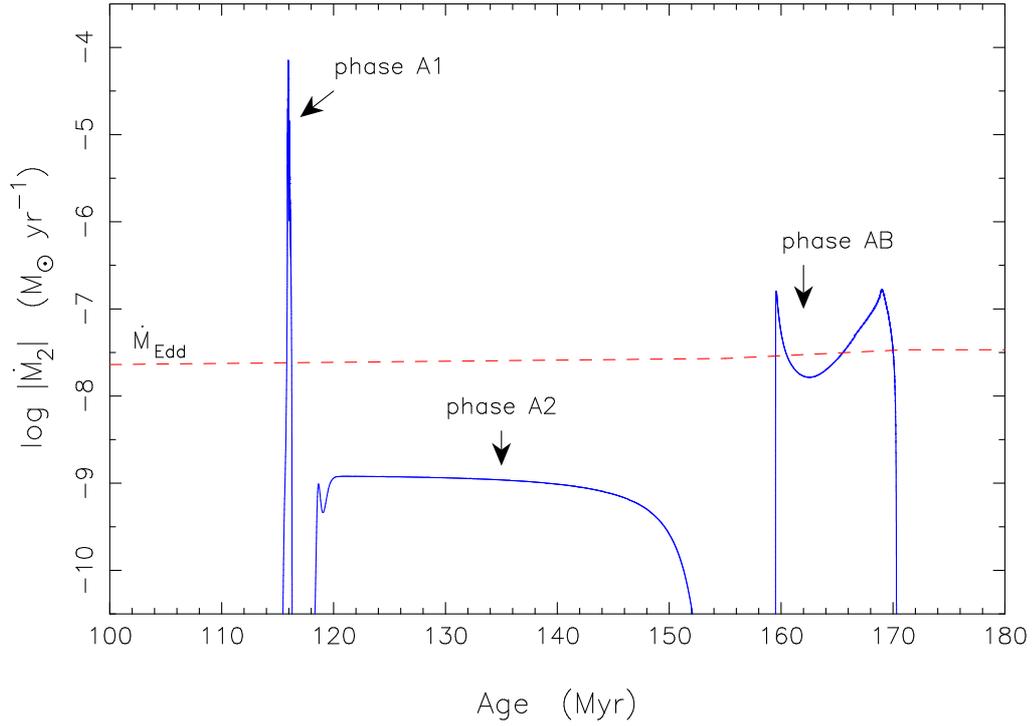

Figure 20: Case A RLO in a binary initially made of a $4.50\,M_\odot$ donor star with a $1.68\,M_\odot$ neutron star and an initial orbital period of $P_{\rm orb} = 2.20$ days. The graph shows the mass-transfer rate from the donor star as a function of its age. Three phases of mass transfer (A1, A2 and AB – see text) are identified.

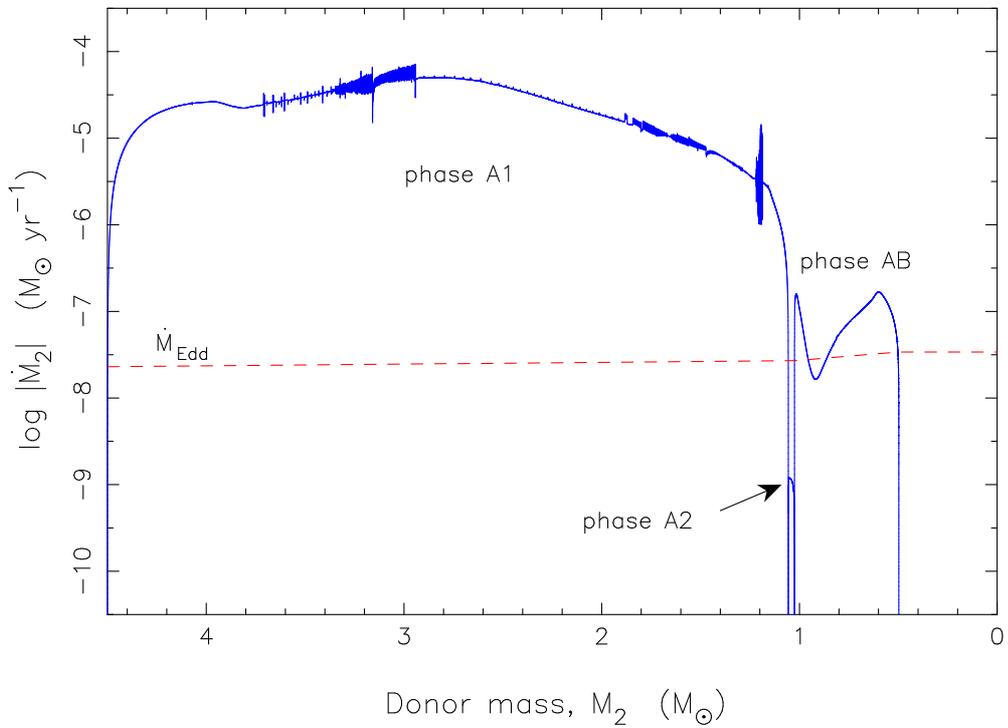

Figure 21: The mass-transfer rate as a function of the decreasing donor star mass for the stellar evolution calculation shown in Fig. 20. Very little mass ($\sim 0.01\,M_\odot$) is accreted by the neutron star during phase A1 which proceeds on a thermal timescale.

The final mass of the neutron star in our example is $1.99\,M_\odot$. The neutron star has thus accreted a total of $0.31\,M_\odot$. The amount accreted in each phase is found from Fig. 20 by integrating the area under the blue line which falls below the Eddington accretion limit (red dashed line). Hardly any accretion takes place during the very short (thermal timescale), ultra super-Eddington phase A1. Phases A2 and AB proceed on nuclear timescales dictated by core burning of the remaining hydrogen and, later on, hydrogen shell burning, respectively.

Fig. 22 shows the track of this IMXB donor star in the HR-diagram on its path to forming a carbon-oxygen white dwarf (CO WD) orbiting a millisecond pulsar. The Case A RLO mass transfer is initiated at an orbital period of 2.20 days. At this stage the core of the donor star still has a central hydrogen mass abundance of $X_c = 0.09$. The error of putting the donor star on the ZAMS is small given that the progenitor star of the neutron star (i.e. the primary star of the ZAMS binary) most likely had a mass of at least $20\,M_\odot$ (see arguments in Sections 3.3 and 3.4) and hence a lifetime of less than 10 Myr, which is short compared to the main sequence lifetime of a $4.50\,M_\odot$ star. Binaries with shorter initial periods will have less evolved donor stars when entering the mass-exchange phase. This leads to lower helium cores masses which are then often below the threshold for igniting the triple-$\alpha$ process. Hence, these systems will leave behind pulsars with a low-mass helium WD companion, as first pointed out by Podsiadlowski, Rappaport & Pfahl (2002) – for further discussions, see Tauris & Langer, in prep.

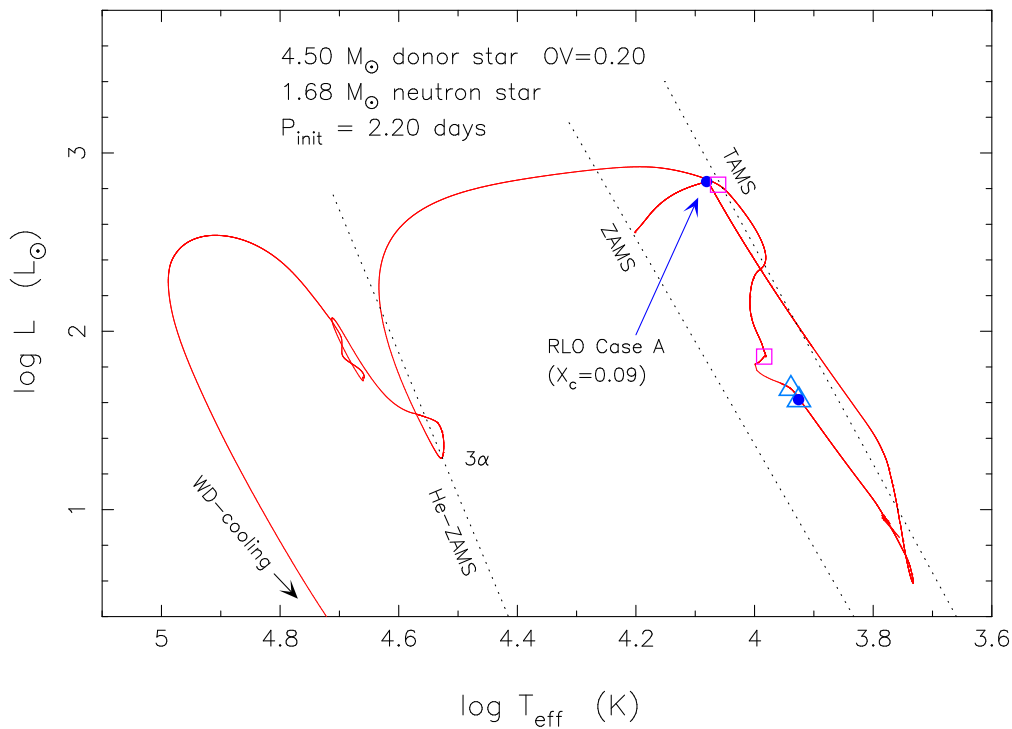

Figure 22: Evolution of the mass-losing donor star in the HR-diagram. Starting and termination points of the three phases of mass transfer are shown by filled circles, open triangles and open squares, corresponding to phases A1, A2 and AB, respectively (cf. Fig. 20). The core helium burning is ignited after detachment from the Roche lobe, see text.

When the donor star settles on the He-ZAMS its luminosity is entirely generated by the triple-$\alpha$ process in the core – marked in Fig. 22 with the symbol "$3\alpha$". At this stage the donor star has an age of 180 Myr. The ignition of the triple-$\alpha$ process actually occurs gradually already from $(\log T_{\rm eff}, \log(L/L_\odot)) = (4.2, 2.9)$ when $T_{\rm core} > 10^8$ K, shortly after the detachment from the Roche-lobe, and overlaps with the ceasing stages of the shell hydrogen burning. The curly loop at $\log(L/L_\odot) \simeq 2$ indicates the beginning of the shell helium burning phase at an age $t = 300$ Myr.

The orbital evolution is shown in Fig. 23 where the orbital period is plotted as a function of decreasing donor star mass. The final orbital period of our system is 8.67 days. The widening of the orbit is quite significant in phase AB where the mass ratio, $q$, is small. It is also during phase AB that the neutron star (NS) gains the majority of its accreted mass – see green line in Fig. 23. The donor star of the X-ray binary Cyg X-2 is an example of a hydrogen shell burning star near the end of phase AB (Podsiadlowski & Rappaport 2000). Interestingly enough, a massive neutron star ($\sim 1.8\,M_\odot$) has been inferred for this source as well, see for example Casares, Charles & Kuulkers (1998).

The chemical abundance profile of the resulting CO WD is shown in Fig. 24. We notice that the inner core ($\sim 0.28\,M_\odot$) contains almost 90% oxygen (mass fraction). The CO WD is seen to have a $\sim 0.04\,M_\odot$ helium envelope and a tiny content of up to 8% hydrogen in the outermost $10^{-4}\,M_\odot$ (amounting to a total of $1.7 \times 10^{-5}\,M_\odot$). The oxygen content of the WD is larger than found by Lin et al. (2011) using the MESA code (see their Fig. 6, panel d). To further test the chemical composition of our WD we compared a run of a similar binary system (4.50 $M_\odot$ donor star, 1.68 $M_\odot$ NS, $P_{\rm orb} = 2.20$ days, $\delta_{\rm OV} = 0.20$) using the Eggleton stellar evolution code. This computation yielded an oxygen content in agreement with our result (Evert Glebbeek, private communication). The reason for the different oxygen yield in Lin et al. (2011) could very well be caused by the use of different $^{12}$C$(\alpha, \gamma)^{16}$O reaction rates.

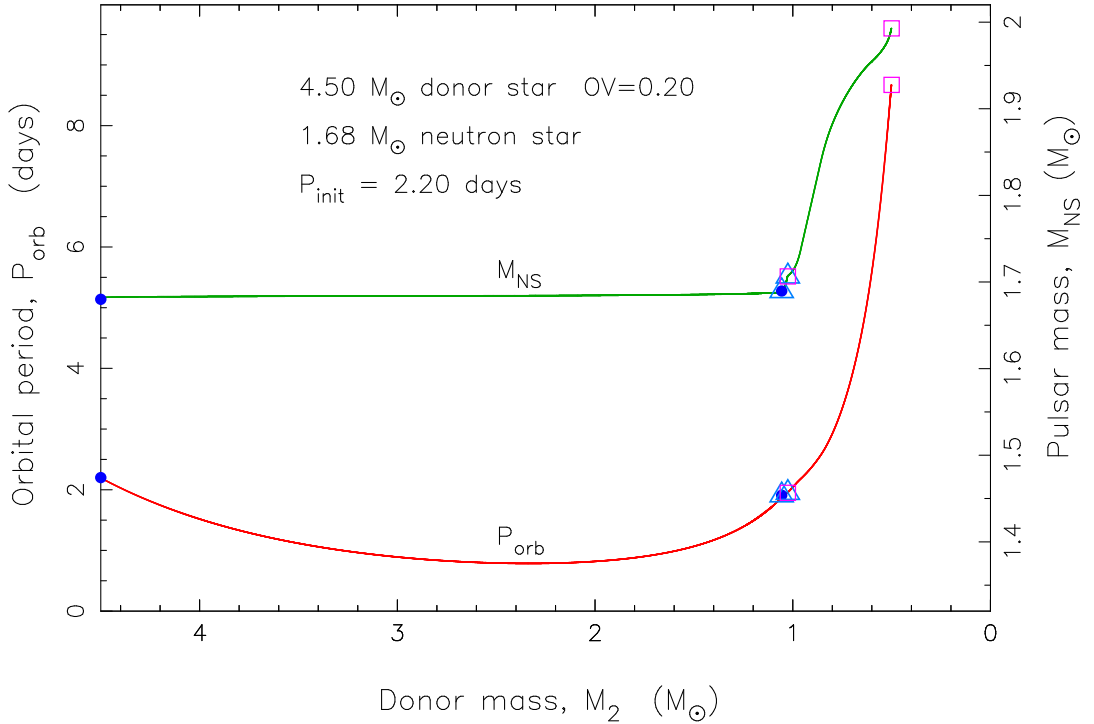

Figure 23: Orbital period (red line) and pulsar mass (green line) as a function of decreasing donor star mass. The symbols are equivalent to those defined in Fig. 22.

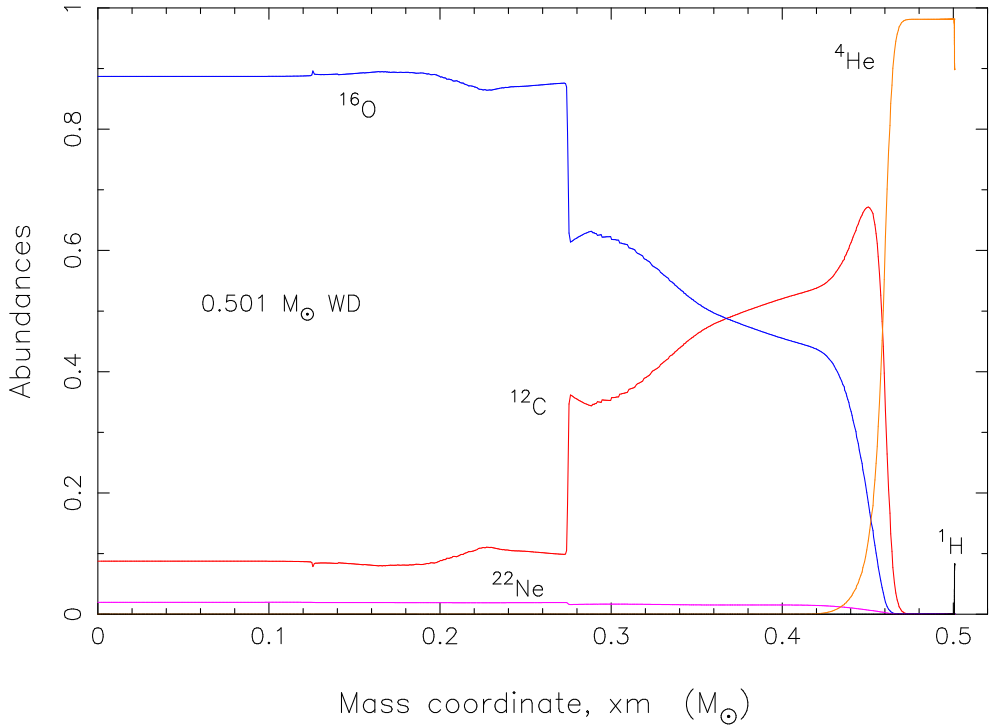

Figure 24: The chemical abundance structure of the CO WD formed in the Case A RLO shown in Figs. 20-23. This profile could well resemble the structure of the CO WD companion to PSR J1614−2230 – possibly applicable to a WD cooling model. The radius of the WD, once it settles on the cooling track, is 9500 km yielding a surface gravity of $\log(g) = 7.9$.

To summarize, the final outcome of our example shown for Case A evolution is a $0.501\,M_\odot$ CO WD orbiting a $1.99\,M_\odot$ (millisecond) pulsar with an orbital period of 8.67 days – almost exactly in agreement with the observed parameters of PSR J1614−2230, see Table 1.

### 3.2.3.1 Case A RLO parameter space leading to the formation of PSR J1614−2230

We have demonstrated above that both Case C and Case A RLO during the X-ray phase can reproduce the observed parameters of PSR J1614−2230 for suitable initial masses of the two components and their orbital period. In order to search the entire parameter space of Case A systems we explored a range of binaries by altering the stellar masses, the orbital period and the accretion efficiency. In Fig. 25 we show the grid of resulting NS+WD systems in the final orbital period versus final neutron star mass plane. This plot was obtained by varying the initial mass of the neutron star as well as the accretion efficiency for a fixed value of the donor star mass, $M_2 = 4.50\,M_\odot$ at the onset of the X-ray phase. It is obvious that the final mass of the neutron star is a growing function of its initial mass as well as the efficiency of accretion.

In order to be able to compare with previous work we have in this plot defined the accretion efficiency as a value in percent of the Eddington mass accretion limit ($\dot{M}_{\rm Edd}$) for pure hydrogen on a neutron star with a radius of 10 km, such that a value of 100% corresponds to the canonical accretion rate of $1.5 \times 10^{-8}\,M_\odot\,{\rm yr}^{-1}$. A value larger than 100% corresponds to either accretion at slight super-Eddington rates or accretion of matter with a larger mean molecular weight per electron (e.g. an accretion efficiency value of 200% corresponds to accreting pure helium at the Eddington limit).

Many of the grid points in Fig. 25 are not obtained from actual stellar evolution calculations. For example, the evolution leading to grid points based on an initial neutron star mass of $1.4\,M_\odot$ were dynamically unstable in our models, leading to runaway mass transfer (see below, and also Podsiadlowski, Rappaport & Pfahl 2002). Nevertheless, one can still compare with the calculations in Podsiadlowski, Rappaport & Pfahl (2002). Using a neutron star with an initial mass of $1.4\,M_\odot$ orbiting a donor star of mass $4.5\,M_\odot$ with an initial orbital period of 2.38 days, and assuming an accretion efficiency of 50%, these authors end up with a NS+WD binary with a final neutron star mass of $1.507\,M_\odot$, a white dwarf mass of $0.471\,M_\odot$ and an orbital period of 3.43 days. (Also in their work they find that such an X-ray binary is barely on the edge of stability and note that this system may be dynamically unstable). The result of their calculation is shown in our figure with an open black circle and the agreement with our result is indeed quite good (cf. the orange neighbour point in our grid just below their point). Our result is based on one of our calculations with a $4.50\,M_\odot$ donor star and a neutron star mass high enough to avoid a dynamical instability, for example of mass $1.7\,M_\odot$. The effect of changing the neutron star mass and/or the accretion efficiency can easily be found analytically from an extrapolation of our calculated model using Eq. (12) for each of the three phases of mass transfer by adapting the new values of $\beta$, $q$ and $q_0$. The underlying assumption that the amount of mass transferred from the donor star remains roughly constant (i.e. independent on the neutron star mass) has been tested by us and shown to be correct. This was done by directly comparing the result of an extrapolated model with a calculated model. See Tauris & Langer (in prep.) for further discussion.

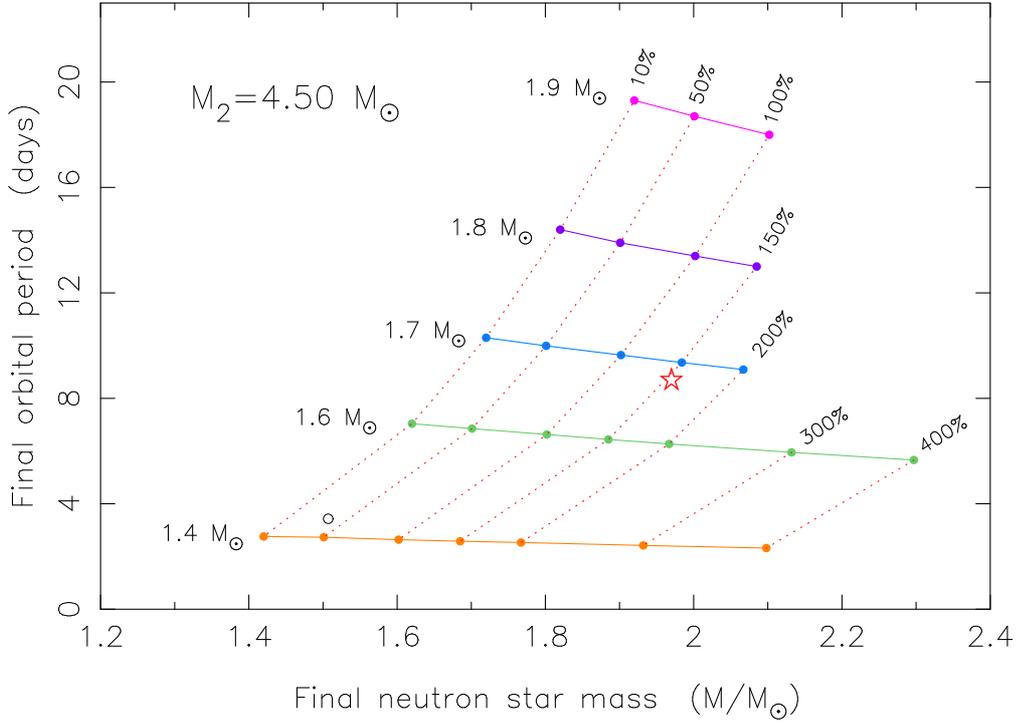

Figure 25: The final orbital period as a function of final neutron star mass for a grid of X-ray binaries evolving from a $4.50\,M_\odot$ donor star through Case A RLO. In all models the CO WD is formed with a mass of about $0.51 \pm 0.01\,M_\odot$. The initial orbital period was in all cases about 2.2 days, corresponding to a core hydrogen content of $\sim 10\%$ at the time of RLO. The variables are the *initial* neutron star mass (solid lines) and the accretion efficiency (dotted lines). The observed values of PSR J1614−2230 are shown with a red star. Our calculations show that indeed PSR J1614−2230 could have evolved from a $4.5\,M_\odot$ donor star and a neutron star born with a mass of $\sim 1.7\,M_\odot$, accreting at an efficiency of 150% – see text.

In our stellar evolution code the Eddington accretion limit (i.e. the accretion efficiency) depends on the chemical composition of the accreted matter as well as the radius of the neutron star, both of which are time dependent – see description in Paper II, and the red dashed lines in Figs. 20–21.

It is important to notice from Fig. 25 how the final orbital period is correlated with the initial mass of the neutron star (increasing in value upwards in the grid diagram, see solid lines). The grid clearly shows that, in the frame of Case A RLO, the neutron star in PSR J1614−2230 cannot have been born with the canonical birth mass of about $1.3\,M_\odot$. This conclusion was also found by Lin et al. (2011).

Using our stellar evolution code we find that our IMXBs are only stable against dynamical mass transfer for initial mass ratios up to $q_0 \simeq 2.7 - 3$, e.g. corresponding to initial donor masses at most $3.5 - 4.0\,M_\odot$ for a $1.3\,M_\odot$ neutron star and (what is important for PSR J1614−2230) donor stars up to $5.0\,M_\odot$ for a $1.7\,M_\odot$ neutron star. Therefore we adapt $5.0\,M_\odot$ as the upper limit for the initial mass of the donor star. The lower limit for the mass of the donor star is constrained by the mass of the CO WD in PSR J1614−2230. We find a lower limit of about $4.0\,M_\odot$ (a $3.5\,M_\odot$ donor star in the region of relevant initial orbital periods leaves behind a WD mass of only $0.39\,M_\odot$ which is 20% smaller than needed for PSR J1614−2230).

The effect on the final orbital period and neutron star mass, imposed by changing only the donor star mass and keeping all other parameters fixed, can be visualized by moving the entire grid in Fig. 25 up or down for a less massive and a more massive donor star, respectively. We find that a $4.0\,M_\odot$ donor star would need to be in a binary with a neutron star of initial mass of $1.55\,M_\odot$ in order to reproduce PSR J1614−2230. The $5.0\,M_\odot$ donor star would need a neutron star of initial mass of $1.77\,M_\odot$ in order to reproduce PSR J1614−2230. In both cases the required accretion efficiency value is about 160-170%. However, the precise limits depend on, for example, the uncertain strength and the assumed underlying physics of the tidal torques and resulting spin-orbit couplings (which may help to stabilize the orbital evolution in X-ray binaries with even higher mass ratios, Tauris & Savonije 2001).

There is evidence from comparison of numerical binary stellar evolution calculations and observations of recycled pulsars that the accretion efficiency in some cases is rather low (see Fig. 4b, §5.7 and §6.4 in Tauris & Savonije 1999), possibly as a result of the propeller effect and/or accretion disk instabilities. A low accretion efficiency would have the implication for PSR J1614−2230 that it was born with an even higher mass. For a $4.5\,M_\odot$ donor star and assuming an accretion efficiency of 50% (according to the definition given earlier) this would result in a neutron star birth mass of about $1.9\,M_\odot$. However, in this case one can only reproduce the observed orbital period of 8.69 days by increasing the specific orbital angular momentum of the lost material beyond the value expected for matter in the vicinity of the accreting neutron star (i.e. by matter ejected from a location elsewhere within the binary system). To illustrate this one can see from Fig. 25 that an initial neutron star mass of $1.9\,M_\odot$ and an accretion efficiency of 50% would yield a final orbital period of about 18 days (much larger than observed in PSR J1614−2230). This problem can be solved if one instead assumes that the material lost from the system has the *average* specific orbital angular momentum of the binary. The reason is that the orbit widens mainly during the mass-transfer phase AB, when the accreting neutron star is more massive than the donor star, and in our model the ejected material has the low specific orbital angular momentum of the neutron star during this phase.

### 3.3 Evolution of progenitor binaries from the ZAMS to the X-ray phase

In the previous section we presented evidence for two different formation scenarios (hereafter simply called Case A and Case C) for the formation of PSR J1614−2230. Hence, we know the required parameters at the onset of the RLO, for the X-ray binary containing a neutron star and a non-degenerate star, and one can then try to calculate backwards and estimate the initial configuration of ZAMS binaries which may eventually form a system like PSR J1614−2230. Our brief description presented here is only qualitative. An analysis including, for example, dynamical effects of asymmetric supernovae (SNe) is rather cumbersome. To obtain a set of more detailed and finetuned parameters of the progenitor binaries one would need to perform a population synthesis (which is beyond the scope of this paper). Nevertheless, below we present the main ideas. The results of our simple analysis are shown in Table 2 and illustrated in Fig. 26.

A mass reversal between the two interacting stars during the evolution from the ZAMS to the X-ray phase can be ruled out in both scenarios as a consequence of the large difference in ZAMS mass between the two stars. The secondary star, the donor star of the X-ray binary, had a mass of $2.2 - 2.6\,M_\odot$ or $4.50\,M_\odot$, respectively. The primary star had at least a mass above the threshold of $\sim 10\,M_\odot$ for producing a NS in a close binary. In Section 3.4.2 we argue that the mass of the progenitor of the NS in PSR J1614−2230 was most likely more than $20\,M_\odot$ since it left behind a massive NS. Hence, there is no doubt that the NS is the remnant of the original primary star, $M_1$ (i.e. the initially more massive) in the ZAMS binary and the donor star in the X-ray phase thus descends from the secondary ZAMS star, $M_2$. Based on this argument, due to the small mass ratio $M_2/M_1 \simeq 0.1 - 0.25$ we would expect the progenitor binary to have evolved through a CE and spiral-in phase in both Case A and Case C on their path from the ZAMS to the SN stage. The initial ZAMS orbital period was then probably quite large ($> 10^3$ days) in order to let the primary star evolve to a late evolutionary stage before the onset of the CE. This would both facilitate the ejection of the envelope – which is more loosely bound at late evolutionary stages – and ensure that the helium core of the neutron star progenitor evolved "clothed" – see Section 3.4.2. We will now briefly discuss each of the two scenarios: Case C and Case A.

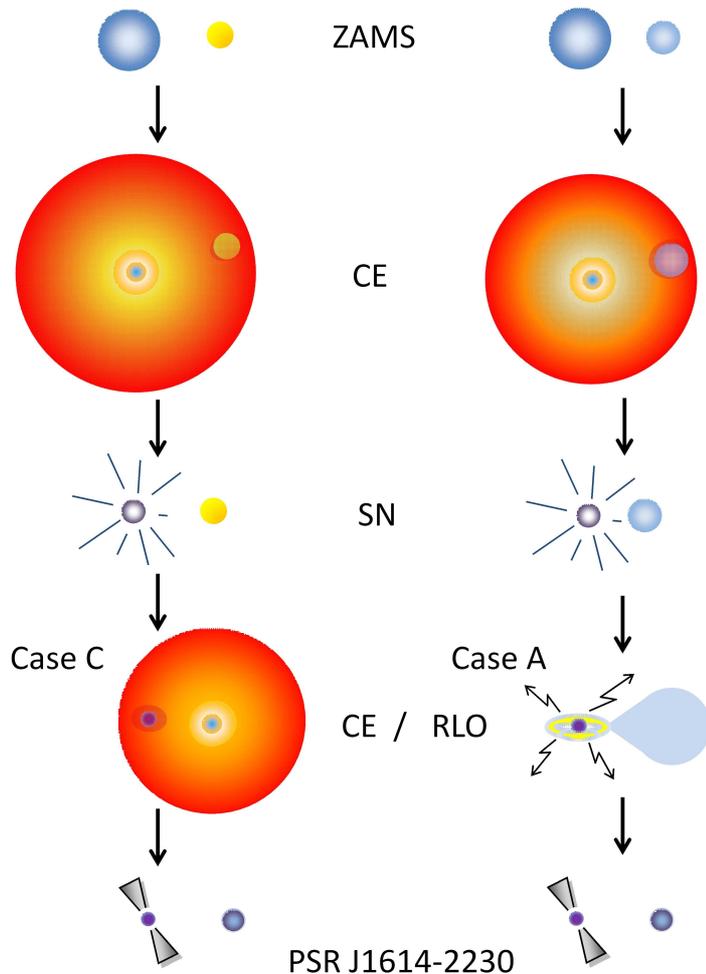

Figure 26: An illustration of the progenitor evolution leading to the formation of PSR J1614−2230 for both Case A and Case C. Only a few evolutionary epochs are shown for simplicity.

### 3.3.1 Case C progenitor binary

The aim here is to obtain a $2.2 - 2.6\,M_\odot$ non-degenerate AGB star orbiting a NS (see Section 3.2.1). Hence, the binary must have been very wide ($\sim 10^3$ days) following the SN in order to allow the donor star to ascend the AGB before initiating mass transfer. A wide post-SN orbit could have been the result of a large kick imparted to the newborn NS at birth (Lyne & Lorimer 1994). However, tidal interactions in a highly eccentric orbit may significantly reduce the orbital separation (Sutantyo 1974) thus preventing the companion star from ascending the AGB before it fills its Roche-lobe. The other alternative, namely that the wide post-SN orbit simply reflects that the pre-SN orbit was wide too is also possible, but not very likely either. Such a wide system would rarely survive any kick imparted to the newborn neutron star – and a small kick originating from an electron capture supernova would be in contradiction with the high neutron star mass (Podsiadlowski et al. 2004; van den Heuvel 2004). Furthermore, the expected outcome of the first CE phase (between the ZAMS and SN stages) is not a wide system. The reason is that the binding energy of the massive primary star's envelope is too large, on an absolute scale, to allow for an early ejection. Hence, there could probably not have been a "mild in-spiral" as a result of an easy ejection of the envelope resulting in only a modest conversion of orbital energy and allowing the orbit to remain fairly wide. An alternative possibility is that the initial orbit was so wide that the two stars did not exchange mass during the giant phase of the primary star. The subsequent kick in the SN then shot the newborn NS into a closer orbit around the secondary star (Kalogera 1998). However, this scenario would require a very fortunate finetuning of both the kick magnitude and the direction, making it unlikely too.

### 3.3.2 Case A progenitor binary

To produce an X-ray binary with an orbital period of only 2.20 days and a $4.50\,M_\odot$ donor star seems more likely compared to the case described above. The short orbital period, both before and after the SN, is a simple consequence of the in-spiral during the first CE-phase when the primary star was a giant. We therefore conclude at this stage, that based on binary evolution considerations PSR J1614−2230 seems more likely to have evolved from a Case A RLO X-ray binary and that the initial ZAMS system was composed of a $\geq 20\,M_\odot$ primary with a $4-5\,M_\odot$ secondary in a wide orbit. In the next section we discuss the stellar evolution of the NS progenitor in much more detail and in Paper II we discuss progenitor systems based on the spin-up of the neutron star.

### 3.4 Discussion

The precise measurement of the high neutron star mass in PSR J1614−2230 leads to interesting implications for the nuclear physics behind the equation-of-state (e.g. Lattimer & Prakash 2010). Equally important, the result has caused renewed interest in modelling close binary evolution – in particular the mass-transfer phase. In a recent paper Kiziltan, Kottas & Thorsett (2011) discuss the possibility of "*alternative* evolution" in order to explain the observed mass of PSR J1614−2230. However, in Section 3.2 of this paper we have demonstrated that this is not required and PSR J1614−2230 may have followed standard evolution paths expected from stellar astrophysics.

The challenge in reproducing massive binary millisecond pulsars with a CO white dwarf companion (like PSR J1614−2230) is to get all three fundamental observable parameters correct: the masses of the two compact objects and the orbital period. In this paper we have demon-strated a methodical approach to do this involving all three mass-transfer scenarios (RLO Cases A, B and C).

Table 2: Evolution and characteristics of the possible progenitor binaries of PSR J1614−2230. The different columns correspond to the different cases of RLO in the X-ray binary phase. In this table the evolution with time goes from top to bottom (see also Fig. 26).

| Initial ZAMS | ↓ | ↓ | ↓ | ↓ | Comments |
|---|---|---|---|---|---|
| Primary mass $(M_\odot)$ | 20–25 | – | 20–25 | 20–25 | In all cases the evolution from the |
| Secondary mass $(M_\odot)$ | 2.2–2.6 | – | 4.0–5.0 | 4.50 | ZAMS to the X-ray phase goes |
| Orbital period (days) | $10^3$ | – | $10^3$ | $10^3$ | through a (first) CE-phase |
| Initial X-ray binary | **Case C** | **Case B** | **Case A** | **Case A***  | *Our example shown in this paper |
| Neutron star mass $(M_\odot)$ | 1.97 | – | 1.55–1.77 | 1.68 | The results of this paper |
| Donor mass $(M_\odot)$ | 2.2–2.6 | – | 4.0–5.0 | 4.50 | |
| Orbital period (days) | $> 10^{2**}$ | – | 2.0–2.3 | 2.20 | ** Depending on details of the first CE |
| Binary millisecond pulsar | ↓ | ↓ | ↓ | ↓ | **PSR J1614−2230** |
| Pulsar mass $(M_\odot)$ | 1.97 | – | 1.95–2.05 | 1.99 | 1.97 |
| White dwarf mass $(M_\odot)$ | 0.50 | – | 0.47–0.53 | 0.501 | 0.500 |
| Orbital period (days) | 0.1–20 | – | 3–16 | 8.67 | 8.69 |

In a recent paper Lin et al. (2011) systematically computed the evolution of a large number of Case A and early Case B IMXB binaries and applied their results to understand the formation of PSR J1614−2230. They conclude that a system like PSR J1614−2230 requires a minimum initial neutron star mass of at least $1.6 \pm 0.1 \, M_\odot$, as well as an initial donor mass of $4.25 \pm 0.10 \, M_\odot$ and orbital period of $\sim 49 \pm 2 \, \mathrm{hr}$ ($2.05 \pm 0.1 \, \mathrm{days}$). In general their Case A results are in fine agreement with our Case A results. The main difference is their rather narrow range of required donor star masses ($4.25 \pm 0.10 \, M_\odot$) compared to our wider interval of $4.0$–$5.0 \, M_\odot$. This minor discrepancy could arise from using different values of the convective core-overshooting parameter, the mixing length parameter and/or the chemical composition of the donor star[4]. However, it is interesting to notice the broad agreement in the final results given that the stellar evolution codes are different.

It is important to emphasize that other observed recycled pulsars with a CO WD companion and $P_{\mathrm{orb}} \simeq 5$–$15$ days (such as PSR J0621+1002 and PSR J2145−0750) are, in general, not expected to have a massive neutron star. Tauris, van den Heuvel & Savonije (2000) demonstrated that these systems can be reproduced by early Case B RLO with a typical $1.3 \, M_\odot$ neutron star and van den Heuvel (1994b) argued for a CE evolution scenario using a neutron star mass of $1.4 \, M_\odot$. Systems with a massive neutron star, like PSR J1614−2230 investigated here, require the neutron star to be born massive – possibly followed by an extended phase of mass transfer allowing for significant further accretion of matter.

### 3.4.1 Neutron star birth masses in pulsar binaries

The interval of known radio pulsar masses ranges from $1.17 \, M_\odot$ in the double neutron star binary PSR J1518+4909 (3-$\sigma$ upper limit, Janssen et al. 2008) to $1.97 \, M_\odot$ in PSR J1614−2230, discussed in this paper. The most massive of the non-recycled companions in double neutron star systems is the unseen companion in PSR 1913+16 which has a mass of $1.389 \, M_\odot$ (Weisberg, Nice & Taylor 2010). Interestingly, the observed pulsar in this binary is the most massive of the (mildly) recycled pulsars detected in any of the ten double neutron star systems. It has a mass of $1.440 \, M_\odot$. However, the relatively slow spin period of this pulsar (59 ms) hints that only about $10^{-3} \, M_\odot$ was needed in the recycling process (see Paper II) and thus $1.44 \, M_\odot$ is the previously known upper limit derived for the *birth* mass of any neutron star detected in a binary pulsar system. Only a few of the $\sim 120$ binary pulsars with WD companions have measured masses – see Paper II – and just a handful of these are more massive than $1.44 \, M_\odot$. But even in those cases the mass determinations are often very inaccurate and also include the mass accreted from the progenitor of their WD companion. However, in this paper we have demonstrated that the birth mass of the neutron star in PSR J1614−2230 is at least $1.7 \pm 0.15 \, M_\odot$. This result is important for understanding the physics of core collapse supernova and neutron star formation.

### 3.4.2 Predictions from stellar evolution and SN explosion models

The difficulty of predicting an upper limit for the birth mass of a neutron star is mainly caused by unknown details of stellar evolution and explosion physics, as well as the neutron star equation-of-state. For example, internal mixing processes, wind mass loss and key nuclear reaction rates are still not known accurately. Additionally, even models with the same input physics predict a rather erratic behaviour of the final iron core mass as a function of ZAMS mass – cf. Fig. 17 in Woosley, Heger & Weaver (2002) and Fig. 4 in Timmes, Woosley & Weaver (1996). Moreover, rotation and metallicity (and perhaps magnetic fields) affect the final remnant mass too. Rotation induced chemical homogeneous evolution will cause rapidly rotating stars to develop larger cores (e.g. Yoon, Langer & Norman 2006) and a low metallicity content also causes stars to develop somewhat larger cores as a result of reduced wind mass loss (Vink, de Koter & Lamers 2001) – see also Linden et al. (2010) and Mirabel et al. (2011) who discuss the relatively high number of black hole HMXBs at low metallicities (compared to HMXBs with a neutron star companion).

---

[4]These parameters are not stated in their present publication.

As for the explosion physics there are, for example, uncertainties in deriving the explosion energy as function of the pre-supernova structure, and in the determination of the mass cut separating the initial compact remnant from the ejected matter. Also the amount of fall back material is uncertain.

Despite the above-mentioned uncertainties we can identify two main factors which determine the remnant mass of a given early-type star: 1) its ZAMS mass, and 2) whether or not it loses its hydrogen-rich envelope (e.g. as a result of mass transfer in a binary) before or during core helium burning. Both of these factors influence the carbon abundance at core helium exhaustion which plays a crucial role for the subsequent carbon, neon and oxygen burning stages which again determines the size of the silicon and iron cores (and thus the mass of the newborn neutron star).

### 3.4.2.1   To burn carbon convectively or not – the role of the $^{12}$C/$^{16}$O-ratio at central helium depletion

Following Brown et al. (2001), the $^{12}$C/$^{16}$O-ratio at central helium depletion is determined by the competition between the formation of carbon via the triple-$\alpha$ process ($3\,\alpha \rightarrow {}^{12}$C) and the destruction of carbon, mainly via $\alpha$-capture: $^{12}$C$(\alpha,\gamma)^{16}$O. More massive stars perform core helium burning at higher temperatures and lower densities, which decrease the net carbon yield. In stars initially more massive than about $19-20\,M_\odot$ the resulting core carbon abundance is too small (in mass fraction $C_c \leq 0.15$) to provide a long lasting convective core carbon burning phase (Woosley & Weaver 1995). The short lasting (radiative) core carbon burning phase results in less energy carried away by neutrinos. As a consequence, the entropy in the core remains fairly high which leads to more massive pre-SN cores. A low carbon abundance also causes the carbon burning shells to be located further out which increases the size of the carbon-free core and thus increases the mass of the iron core to form. Single stars with ZAMS masses less than $20\,M_\odot$, on the other hand, deplete core helium burning with a relatively high $^{12}$C/$^{16}$O-ratio. These stars undergo significant convective carbon burning which subsequently leads to relatively small iron cores and hence low-mass neutron stars.

### 3.4.2.2   Single stars / wide orbit binaries (Case C RLO)

Based on the core carbon burning dichotomy discussed above, Timmes, Woosley & Weaver (1996) found a bimodal distribution of neutron star birth masses with narrow peaks at $1.28 \pm 0.06\,M_\odot$ and $1.73 \pm 0.08\,M_\odot$ (gravitational masses), for single star progenitors (type II SNe) below and above the critical ZAMS mass of $\sim 20\,M_\odot$. Based on assumptions of rather soft equations-of-state for nuclear matter at high densities Brown et al. (2001) did not predict high-mass neutron stars ($\sim 1.7\,M_\odot$) to be formed and therefore concluded that all single ZAMS stars in the interval $20-25\,M_\odot$ end their lives as black holes – see also Fryer (1999; 2006). However, given the discovery of the massive pulsar PSR J1614$-$2230 this picture has to change. Following our work in this paper the birth mass of this neutron star is at least $1.7 \pm 0.15\,M_\odot$ and given the expected wide orbit of its progenitor star (see Section 3.3) we conclude that it formed from an original ZAMS star of mass $20-25\,M_\odot$ which underwent Case C RLO (leading to the CE-phase prior to the SN – see Fig. 26). Hence, in general we expect $20-25\,M_\odot$ single (or wide orbit stars) to form high-mass neutron stars. Beyond $25\,M_\odot$, the large binding energies of the mantle is expected to result in large fall back and the production of a black hole.

For the sake of completeness, at the other end of the scale where stars with an initial mass of about $10\,M_\odot$ end their lives the minimum gravitational mass expected for a newborn neutron star is about $1.25\,M_\odot$ for an electron capture SN (e.g. Podsiadlowski et al. 2004). However, for single stars, the initial mass range for stars to undergo electron capture SN is very small, such that only a few percent of all core collapse supernovae are expected to go through this channel (Poelarends et al. 2008). It is interesting to notice that small iron core-collapse SNe of type II also seem to allow for the formation of $\sim 1.15\,M_\odot$ neutron stars (Timmes, Woosley & Weaver

1996).

To summarize, one would expect three peaks in the distribution of initial neutron star masses from the evolution of single stars or stars in wide orbit binaries (as pointed out by van den Heuvel 2004): one (small) peak at $\sim 1.25\,M_\odot$ from electron capture SNe of $8-10\,M_\odot$ stars (leading to a small kick), one peak at $1.25-1.4\,M_\odot$ from iron core-collapse SNe of $10-20\,M_\odot$ stars and one high-mass peak at $\geq 1.7\,M_\odot$ from iron core-collapse SNe of $20-25\,M_\odot$ stars. Above a ZAMS mass of $25\,M_\odot$ single stars are expected to form black holes – unless the metallicity is very high (above solar). Rapid rotation, on the other hand, may produce black holes from stars less massive than $25\,M_\odot$.

### 3.4.2.3    Close binary stars (Case A and Case B RLO)

The progenitors of neutron stars in close binaries lose their hydrogen envelope as a result of mass transfer (cf. Section 3.2) and thus end up as type Ib/c SNe. As shown by Brown et al. (2001), see also Wellstein & Langer (1999), stars which lose their envelope before or early during core helium burning (as in Case A/B RLO in a close binary) evolve as "naked" helium stars. There are four reasons why "naked" helium stars, at least at solar metallicity, develop small cores. Firstly, the lack of a hydrogen burning shell on top of the helium core prevents the core mass from growing during core helium burning. Secondly, the lack of this shell also prevents the convective part of the core from growing during core helium burning. In single stars, the growing convective core brings in fresh helium, which during late core helium burning is mostly used to convert carbon to oxygen. "Naked" helium stars therefore end core helium burning with a much higher carbon abundance (compared to single or wide orbit stars which evolve helium cores as "clothed"), and this will lead to lower mass iron cores as discussed above. Thirdly, helium stars emit strong winds as so called Wolf-Rayet stars, which reduces the final CO-core mass. And finally, not only are the core masses of "naked" helium stars reduced, compared to the situation of single or wide orbit stars, also the masses and the binding energies of the surrounding envelope are reduced, which leads to a corresponding reduction in the amount of material that falls back after the SN explosion (Woosley & Weaver 1995). All these effects lead to smaller neutron star masses in close binaries where the progenitor of the neutron star evolved through Case A/B RLO, compared to formation from a single star or wide binary (Case C RLO) evolution.

Timmes, Woosley & Weaver (1996) calculated remnant masses from explosions of the pre-SN helium star models of Woosley, Langer & Weaver (1995). They estimate these models correspond to ZAMS masses up to $35\,M_\odot$ for stars in close binaries which lose their hydrogen envelope before or early during core helium burning. Timmes, Woosley & Weaver (1996) predict an upper limit for the gravitational mass of these neutron stars to be about $1.4\,M_\odot$ (see their Fig. 6c for type Ib SN). Depending on the details of the fall back of material these values could be as high as $\leq 1.6\,M_\odot$. In fact, according to Brown et al. (2001) even stars initially as massive as $40-60\,M_\odot$ may end their lives as a neutron star if they lose their envelope early in their evolution (Case A or Case B RLO).

At the low end of progenitor masses, Podsiadlowski et al. (2004) showed that electron capture supernovae from stars in close binary systems may originate from a much broader mass range than in the single star (or wide binary) case due to the avoidance of the second dredge-up phase.

In a recent paper Yoon, Woosley & Langer (2010) investigated the evolution of stars from binaries leading to type Ib/c SNe using the most recent theoretical models of (reduced) Wolf-Rayet winds. The result is larger final helium core masses which may lead to somewhat larger iron core masses and thus larger post-SN remnant masses. An investigation of this question is currently in progress.

All in all, we see that the expected neutron star mass distribution from close binary stars is very different from that expected from single stars. In general, close binaries are thought to produce lower mass neutron stars. In this context, our conclusion that that the first mass transfer in the progenitor evolution of PSR J1614−2230 was of Case C (a wide binary) is fully consistent with the high initial neutron star mass of $1.7\,M_\odot$ we derived for this system above.

### 3.4.3 Ramifications from PSR J1614−2230, Vela X-1 and the black widow pulsar

Mass determinations of Vela X-1 (Barziv et al. 2001; Rawls et al. 2011) suggest that this neutron star has an observed mass of $1.77 \pm 0.08\,M_\odot$. The companion star to Vela X-1 is a B0.5 Ib supergiant (HD 77581) with a mass of about $23\,M_\odot$ which implies that the present mass of the neutron star is very close to its birth mass. (Even a hypothetical strong wind accretion at the Eddington limit would not have resulted in accretion of more than about $10^{-2}\,M_\odot$ given the short lifetime of its massive companion). We therefore conclude that not only was the neutron star in PSR J1614−2230 born massive ($1.7 \pm 0.15\,M_\odot$) also the neutron star Vela X-1 was born with a mass of at least $1.7\,M_\odot$. Furthermore, both of these neutron stars were produced from progenitors with a ZAMS mass above $20\,M_\odot$ and these progenitors did not lose their envelope before core helium burning exhaustion (i.e. they evolved via Case C RLO prior to the SN – not to be confused with the later X-ray phase discussed in Section 3.2). Unlike PSR J1614−2230, which had a relatively low- or intermediate-mass companion, Vela X-1 has a massive companion and this system is therefore not expected to have evolved through a CE prior to the SN.

A recent analysis by van Kerkwijk, Breton & Kulkarni (2011) of the so-called "black widow pulsar" yielded a mass of $2.4\,M_\odot$. Although the uncertainties of this result are rather large, such a mass would be difficult to explain if the neutron star was born with the canonical mass of about $1.4\,M_\odot$.

### 3.4.4 Future observational constraints

Future observations could push the empirical upper limit of the possible neutron star birth mass to even higher values. This could, for example, be achieved by measurements of a binary radio pulsar which reveal a massive neutron star in a double neutron star binary or, perhaps more likely, in a very tight binary with an ONeMg WD. In both cases the evolutionary timescale of the progenitor of the last formed compact object would be so short that no substantial amount of mass could be accreted by the mildly recycled pulsar. Hence, in these cases the observed mass of the pulsar would be almost equal to its birth mass. Also detection of an even more massive neutron star than Vela X-1 in a HMXB system would push the limit upwards.

We note that it is much more difficult and uncertain to estimate the birth mass of a pulsar in a close binary with a low-mass helium WD companion. Even in binaries with $P_{\rm orb} < 1$ day where the mass transfer rate is expected to have been sub-Eddington (e.g. Tauris & Savonije 1999) it is difficult posterior to estimate the effects on the accretion rate, and thus infer the birth mass of the neutron star, due to the poorly known occurrence of accretion disk instabilities and strength of the propeller effects – cf. further discussion in Paper II.

## 3.5 Conclusions

We have investigated the formation of PSR J1614−2230 by detailed modelling of the mass exchanging X-ray phase of the progenitor system. We have introduced a new analytic parameterization for calculating the outcome of either a CE evolution or the highly super-Eddington isotropic re-emission mode, which depends only on the present observable mass ratio, $q$ and the ratio between the initial and final donor star mass, $k$. Using a detailed stellar evolution code we calculated the outcome of a number of IMXBs undergoing Case A RLO. Based on the orbital dynamics and observational constraints on the stellar masses we find that PSR J1614−2230 could have evolved from a neutron star with either a $2.2 - 2.6\,M_\odot$ asymptotic giant donor star through a CE evolution (initiated by Case C RLO), or from a system with a $4.0 - 5.0\,M_\odot$ donor star via Case A RLO. Simple qualitative arguments on the evolution from the ZAMS to the X-ray phase suggest that Case A is the most likely of the two scenarios. The methods used in this paper, for RLO Cases A, B and C, could serve as a recipe for investigations of the progenitor system of other massive binary millisecond pulsars with heavy white dwarf companions to be discovered in the future.

We conclude that the neutron star in PSR J1614−2230 was born significantly more massive

($1.7 \pm 0.15\,M_\odot$) than neutron stars found in previously known radio pulsar binaries – a fact which is important for understanding stellar evolution of massive stars in binaries and the explosion physics of core-collapse SNe. Finally, based on this high value for the neutron star birth mass we argue that the progenitor star of PSR J1614$-$2230 had a ZAMS mass of $20-25\,M_\odot$ and did not lose its envelope before core helium exhaustion.

In Paper II we continue the discussion of the formation of PSR J1614$-$2230 in view of the spin-up process and include general aspects of accretion onto a neutron star during the recycling process and apply our results to other observed millisecond pulsars.

## 4. A Massive Pulsar in a Compact Relativistic Binary

### Antoniadis, Freire, Wex, Tauris, Lynch, van Kerkwijk, Kramer et al. (2013) Science 340, 448

### Abstract

Many physically motivated extensions to general relativity (GR) predict significant deviations in the properties of spacetime surrounding massive neutron stars. We report the measurement of a 2.01±0.04 solar mass ($M_\odot$) pulsar in a 2.46-hr orbit with a 0.172±0.003 $M_\odot$ white dwarf. The high pulsar mass and the compact orbit make this system a sensitive laboratory of a previously untested strong-field gravity regime. Thus far, the observed orbital decay agrees with GR, supporting its validity even for the extreme conditions present in the system. The resulting constraints on deviations support the use of GR-based templates for ground-based gravitational wave detectors. Additionally, the system strengthens recent constraints on the properties of dense matter and provides insight to binary stellar astrophysics and pulsar recycling.

### 4.1 Introduction

Neutron stars (NSs) with masses above $1.8\,M_\odot$ manifested as radio pulsars are valuable probes of fundamental physics in extreme conditions unique in the observable Universe and inaccessible to terrestrial experiments. Their high masses are directly linked to the equation-of-state (EOS) of matter at supra-nuclear densities (Lattimer & Prakash 2004; Demorest et al. 2010) and constrain the lower mass limit for production of astrophysical black holes (BHs). Furthermore, they possess extreme internal gravitational fields which result in gravitational binding energies substantially higher than those found in more common, $1.4\,M_\odot$ NSs. Modifications to GR, often motivated by the desire for a unified model of the four fundamental forces, can generally imprint measurable signatures in gravitational waves (GWs) radiated by systems containing such objects, even if deviations from GR vanish in the Solar System and in less massive NSs (Damour & Esposito-Farèse 1993; Damour & Esposito-Farèse 1996; Will 1993).

However, the most massive NSs known today reside in long-period binaries or other systems unsuitable for GW radiation tests. Identifying a massive NS in a compact, relativistic binary is thus of key importance for understanding gravity-matter coupling under extreme conditions. Furthermore, the existence of a massive NS in a relativistic orbit can also be used to test current knowledge of close binary evolution.

### 4.2 Results

#### 4.2.1 PSR J0348+0432 & optical observations of its companion

PSR J0348+0432, a pulsar spinning at 39 ms in a 2.46-hr orbit with a low-mass companion, was detected by a recent survey (Boyles et al. 2013; Lynch et al. 2013) conducted with the Robert C. Byrd Green Bank Telescope (GBT). Initial timing observations of the binary yielded an accurate astrometric position, which allowed us to identify its optical counterpart in the Sloan Digital Sky Survey (SDSS) archive (Section 4.5, SOM). The colors and flux of the counterpart are consistent with a low-mass white dwarf (WD) with a helium core at a distance of $d \sim 2.1$ kpc. Its relatively high apparent brightness ($g' = 20.71 \pm 0.03$ mag) allowed us to resolve its spectrum using the Apache Point Optical Telescope. These observations revealed deep Hydrogen lines, typical of low-mass WDs, confirming our preliminary identification. The radial velocities of the WD mirrored that of PSR J0348+0432, also verifying that the two stars are gravitationally bound.

In December 2011 we obtained phase-resolved spectra of the optical counterpart using the FORS2 spectrograph of the Very Large Telescope (VLT). For each spectrum, we measured the radial velocity which we then folded modulo the system's orbital period. Our orbital fit to the velocities constrains the semi-amplitude of their modulation to be $K_{\rm WD} = 351 \pm 4\,{\rm km\,s^{-1}}$ (Fig. 27; see also Materials & Methods). Similarly, the orbital solution from radio-pulsar timing yields $K_{\rm PSR} = 30.008235 \pm 0.000016\,{\rm km\,s^{-1}}$ for the pulsar. Combined, these constraints imply a mass ratio, $q = M_{\rm PSR}/M_{\rm WD} = K_{\rm WD}/K_{\rm PSR} = 11.70 \pm 0.13$.

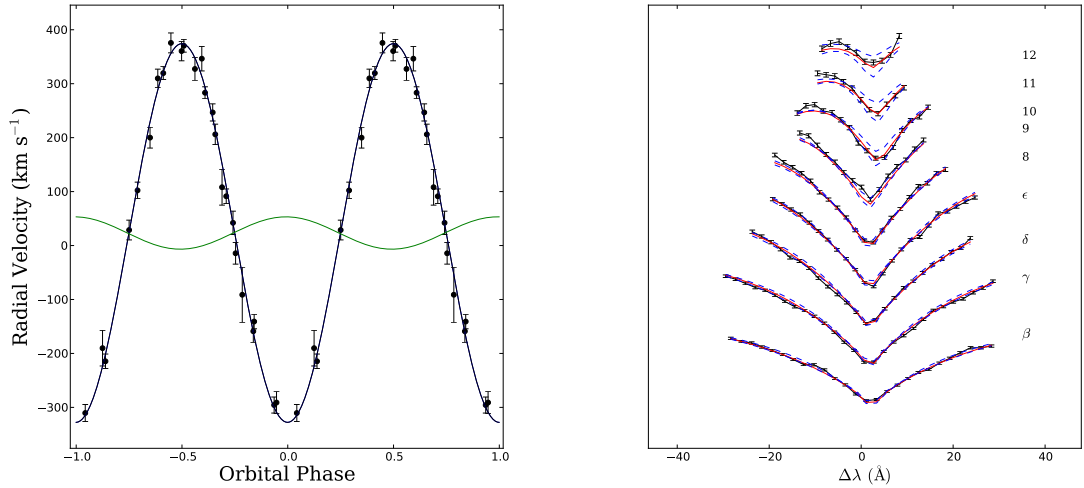

Figure 27: Radial velocities and spectrum of the white dwarf companion to PSR J0348+0432.
**Left:** Radial velocities of the WD companion to PSR J0348+0432 plotted against the orbital phase (shown twice for clarity). Over-plotted is the best-fit orbit of the WD (blue line) and the mirror orbit of the pulsar (green). **Right:** Details of the fit to the Balmer lines (H$\beta$ to H12) in the average spectrum of the WD companion to PSR J0348+0432 created by the coherent addition of 26 individual spectra shifted to zero velocity. Lines from H$\beta$ (bottom) to H12 are shown. The red solid lines are the best-fit atmospheric model (see text). Two models with $(T_{\rm eff}, \log_{10} g) = (9900\,{\rm K}, 5.70)$ and $(T_{\rm eff}, \log_{10} g)$ =(10200 K, 6.30), each $\sim 3$-$\sigma$ off from the best-fit central value (including systematics) are shown for comparison (dashed blue lines).

Modeling of the Balmer-series lines in a high signal-to-noise average spectrum formed by the coherent addition of individual spectra (Fig. 27, right panel) shows that the WD has an effective temperature of $T_{\rm eff} = (10120 \pm 47_{\rm stat} \pm 90_{\rm sys})\,{\rm K}$ and a surface gravity of $\log_{10}(g\,[{\rm cm\,s^{-2}}]) = (6.035 \pm 0.032_{\rm stat} \pm 0.060_{\rm sys})$ dex. Here the systematic error is an overall estimate of uncertainties due to our fitting technique and flux calibration (Section 4.5, SOM). We found no correlation of this measurement with orbital phase and no signs of rotationally-induced broadening in the spectral lines (Section 4.5, SOM). Furthermore, we searched for variability using the ULTRA-CAM instrument (Dhillon et al. 2007) on the 4.2-m William-Herschel Telescope at La Palma, Spain. The lightcurves, spanning 3 hours in total, have a root-mean-square scatter of $\sim 0.53$, 0.07 and 0.08 mag in $u'$, $g'$ and $r'$ respectively and show no evidence for variability over the course of the observations. The phase-folded light-curve shows no variability either. Additionally, our calibrated magnitudes are consistent with the SDSS catalogue magnitudes, implying that the WD shone at a constant flux over this $\sim 5\,{\rm yr}$ timescale (Section 4.5, SOM).

### 4.2.2 Mass of the white dwarf

The surface gravity of the WD scales with its mass and the inverse square of its radius ($g \equiv GM_{\rm WD}/R_{\rm WD}^2$). Thus, the observational constraints combined with a theoretical finite-temperature mass-radius relation for low-mass WDs yield a unique solution for the mass of the companion (van Kerkwijk et al. 2005). Numerous such models exist in the literature, the most detailed of which are in good agreement for very low mass WDs ($< 0.17 - 0.18\,{\rm M_\odot}$), but differ substantially for higher masses [e.g. Serenelli et al. (2001); Panei et al. (2007); Kilic et al. (2010)]. The main reason for this is the difference in the predicted size of the hydrogen envelope, which determines whether the main energy source of the star is residual hydrogen burning (for "thick" envelopes) or the latent heat of the core (for "thin" envelopes).

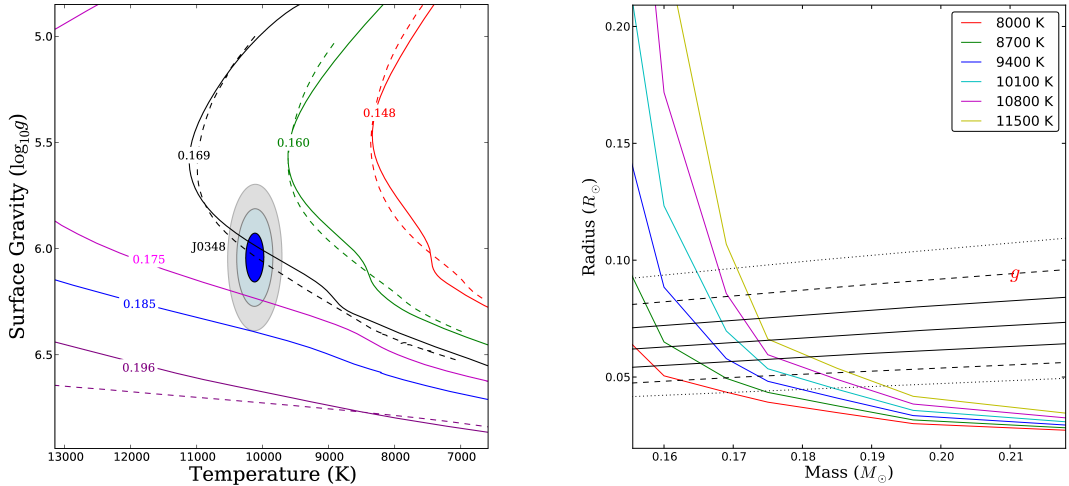

**Figure 28:** Mass measurement of the white dwarf companion to PSR J0348+0432.
**Left:** Constraints on effective temperature, $T_{eff}$, and surface gravity, $g$, for the WD companion to PSR J0348+0432 compared with theoretical WD models. The shaded areas depict the $\chi^2 - \chi^2_{min} = 2.3$, 6.2 and 11.8 intervals (equivalent to 1, 2 and 3-$\sigma$) of our fit to the average spectrum. Dashed lines show the detailed theoretical cooling models of Serenelli et al. (2001). Continuous lines depict tracks with thick envelopes for masses up to $\sim 0.2\,M_\odot$ that yield the most conservative constraints for the mass of the WD. **Right:** Finite-temperature mass-radius relations for our models together with the constraints imposed from modeling of the spectrum (see text). Low mass – high temperature points are an extrapolation from lower temperatures.

In the most widely accepted scenario, WDs lose their thick hydrogen envelope only if their mass exceeds a threshold. The exact location of the latter is still uncertain but estimated to be around $\sim 0.17 - 0.22\,M_\odot$ [e.g. Serenelli et al. (2001); Panei et al. (2007); Kilic et al. (2010)]. Two other pulsars with WD companions, studied in the literature, strongly suggest that this transition threshold is indeed most likely close to $0.2\,M_\odot$ (van Kerkwijk et al. 2005; Antoniadis et al. 2012). In particular, the WD companion of PSR J1909−3744 has a well-determined mass of $0.20\,M_\odot$ (Jacoby et al. 2005), a large characteristic age of a several Gyr and a WD companion that appears to be hot (van Kerkwijk et al. 2005), suggesting that its envelope is thick. For this reason we base the WD mass estimate on cooling tracks with thick hydrogen atmospheres for masses up to $0.2\,M_\odot$, which we constructed using the "MESA" stellar evolution code (Paxton et al. 2011; , Section 4.5). Initial models were built for masses identical to the ones in Serenelli et al. (2001) — for which previous comparisons have yielded good agreement with observations (Antoniadis et al. 2012) — with the addition of tracks with 0.175 and 0.185 $M_\odot$ for finer coverage (Fig. 28). For masses up to 0.169 $M_\odot$ our models show excellent agreement with Serenelli et al. (2001); our 0.196 $M_\odot$ model though is quite different, because it has a thick envelope instead of a think one. Being closer to the constraints for the WD companion to PSR J0348+0432, it yields a more conservative mass constraint: $M_{WD} = 0.165 - 0.185$ at 99.73% confidence (Fig. 29 & Table 3), which we adopt. The corresponding radius is $R_{WD} = 0.046 - 0.092\,R_\odot$ at 99.73% confidence. Our models yield a cooling age of $\tau_{cool} \sim 2\,Gyr$.

### 4.2.3 Pulsar mass

The derived WD mass and the observed mass ratio $q$ imply a NS mass in the range 1.97 – 2.05 $M_\odot$ at 68.27% or 1.90 – 2.18 $M_\odot$ at 99.73% confidence. Hence, PSR J0348+0432 is only the second NS with a precisely determined mass around $2\,M_\odot$, after PSR J1614−2230 (Demorest et al. 2010). It has a 3-$\sigma$ lower mass limit 0.05 $M_\odot$ higher than the latter, and therefore provides a verification, using a different method, of the constraints on the EOS of

Table 3: Observed and derived parameters for the PSR J0348+0432 system.

Timing parameters for the PSR J0348+0432 system, indicated with their 1-$\sigma$ uncertainties as derived by TEMPO2 where appropriate (numbers in parentheses refer to errors on the last digits). The timing parameters are calculated for the reference epoch MJD 56000, and are derived from TOAs in the range MJD 54872 − 56208.

| Optical Parameters | |
|---|---|
| Effective temperature, $T_{\rm eff}$ (K) ........................... | $10120 \pm 47_{\rm stat} \pm 90_{\rm sys}$ |
| Surface gravity, $\log_{10}(g[\rm cm\,s^{-1}])$ ...................... | $6.035 \pm 0.032_{\rm stat} \pm 0.060_{\rm sys}$ |
| Semi-amplitude of orbital radial velocity, $K_{\rm WD}$ (km s$^{-1}$) | $351 \pm 4$ |
| Systemic radial velocity relative to the Sun, $\gamma$ (km s$^{-1}$) . | $-1 \pm 20$ |
| **Timing Parameters** | |
| Right ascension, $\alpha$ (J2000) ............................ | $03^{\rm h}\,48^{\rm m}\,43^{\rm s}.639000(4)$ |
| Declination, $\delta$ (J2000) ................................. | $+04°\,32'\,11''.4580(2)$ |
| Proper motion in right ascension, $\mu_\alpha$ (mas yr$^{-1}$) ....... | $+4.04(16)$ |
| Proper motion in declination, $\mu_\delta$ (mas yr$^{-1}$) .......... | $+3.5(6)$ |
| Parallax, $\pi_d$ (mas) ..................................... | $0.47^*$ |
| Spin frequency, $\nu$ (Hz) ................................. | $25.5606361937675(4)$ |
| First derivative of $\nu$, $\dot\nu$ ($10^{-15}$ Hz s$^{-1}$) ................ | $-0.15729(3)$ |
| Dispersion measure, DM (cm$^{-3}$ pc) .................... | $40.46313(11)$ |
| First derivative of DM, DM1 (cm$^{-3}$ pc yr$^{-1}$) .......... | $-0.00069(14)$ |
| Orbital period, $P_{\rm b}$ (d) ................................ | $0.102424062722(7)$ |
| Time of ascending node, $T_{\rm asc}$ (MJD) ................... | $56000.084771047(11)$ |
| Projected semi-major axis of the pulsar orbit, $x$ (lt-s) .. | $0.14097938(7)$ |
| $\eta \equiv e\sin\omega$ .............................................. | $(+1.9 \pm 1.0) \times 10^{-6}$ |
| $\kappa \equiv e\cos\omega$ ............................................ | $(+1.4 \pm 1.0) \times 10^{-6}$ |
| First derivative of $P_{\rm b}$, $\dot P_{\rm b}$ ($10^{-12}$ s s$^{-1}$) ................. | $-0.273(45)$ |
| **Derived Parameters** | |
| Galactic longitude, $l$ .................................... | $183°.3368$ |
| Galactic latitude, $b$ .................................... | $-36°.7736$ |
| Distance, $d$ (kpc) ..................................... | $2.1(2)$ |
| Total proper motion, $\mu$ (mas yr$^{-1}$) .................... | $5.3(4)$ |
| Spin period, $P$ (ms) ................................... | $39.1226569017806(5)$ |
| First derivative of $P$, $\dot P$ ($10^{-18}$ s s$^{-1}$) ................. | $0.24073(4)$ |
| Characteristic age, $\tau_c$ (Gyr) ............................ | $2.6$ |
| Transverse magnetic field at the poles, $B_0$ ($10^9$ G) ..... | $\sim 2$ |
| Rate or rotational energy loss, $\dot E$ ($10^{32}$ erg s$^{-1}$) ........ | $\sim 1.6$ |
| Mass function, $f$ (M$_\odot$) ............................... | $0.000286778(4)$ |
| Mass ratio, $q \equiv M_{\rm PSR}/M_{\rm WD}$ ......................... | $11.70(13)$ |
| White dwarf mass, $M_{\rm WD}$ (M$_\odot$) ....................... | $0.172(3)$ |
| Pulsar mass, $M_{\rm PSR}$ (M$_\odot$) ............................ | $2.01(4)$ |
| "Range" parameter of Shapiro delay, $r$ ($\mu$s) ........... | $0.84718^*$ |
| "Shape" parameter of Shapiro delay, $s \equiv \sin i$ ......... | $0.64546^*$ |
| White dwarf radius, $R_{\rm WD}$ (R$_\odot$) ...................... | $0.065(5)$ |
| Orbital separation, $a$ ($10^9$ m) ......................... | $0.832$ |
| Orbital separation, $a$ (R$_\odot$) ............................ | $1.20$ |
| Orbital inclination, $i$ .................................... | $40°.2(6)$ |
| $\dot P_{\rm b}$ predicted by GR, $\dot P_{\rm b}^{\rm GR}$ ($10^{-12}$ s s$^{-1}$) ................ | $-0.258^{+0.008}_{-0.011}$ |
| $\dot P_{\rm b}/\dot P_{\rm b}^{\rm GR}$ ................................................ | $1.05 \pm 0.18$ |
| Time until coalescence, $\tau_m$ (Myr) ...................... | $\sim 400$ |

$^*$ For these timing parameters we have adopted the optically derived parameters (see text for details).

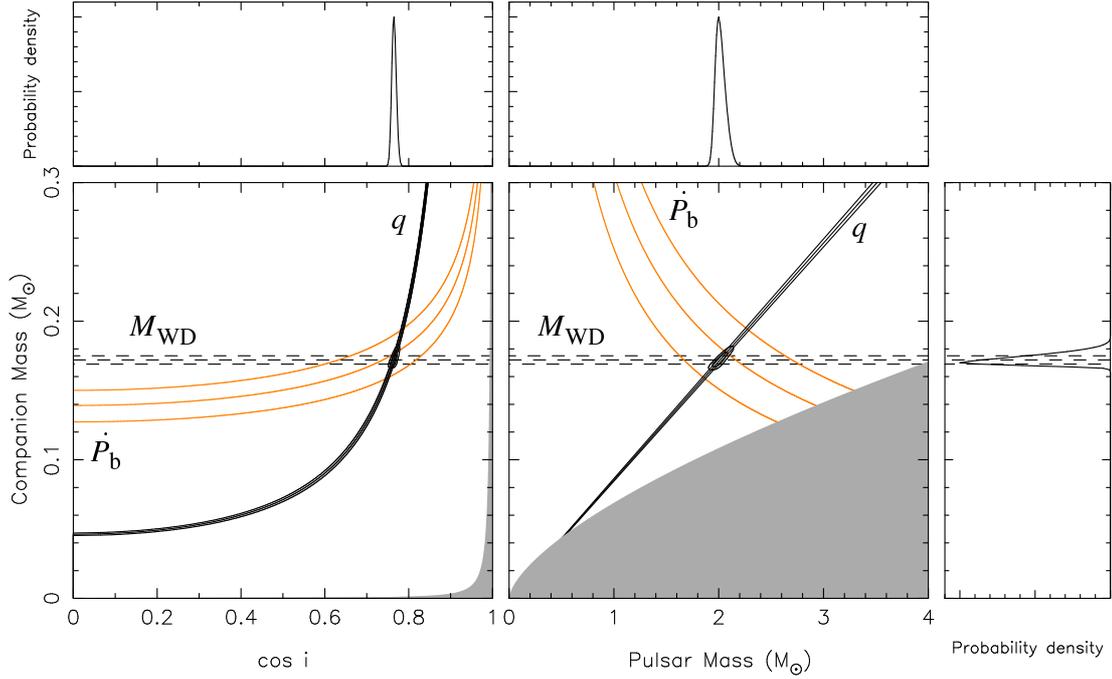

Figure 29: System masses and orbital-inclination constraints.
Constraints on system masses and orbital inclination from radio and optical measurements of PSR J0348+0432 and its WD companion. Each triplet of curves corresponds to the most likely value and standard deviations (68.27% confidence) of the respective parameters. Of these, two (the mass ratio $q$ and the companion mass $M_{\mathrm{WD}}$) are independent of specific gravity theories (in black). The contours contain the 68.27 and 95.45% of the two-dimensional probability distribution. The constraints from the measured intrinsic orbital decay ($\dot{P}_{\mathrm{b}}^{\mathrm{int}}$, in orange) are calculated *assuming* that GR is the correct theory of gravity. All curves intersect in the same region, meaning that GR passes this radiative test (Section 4.5, SOM). **Left**: $\cos i$–$M_{\mathrm{WD}}$ plane. The gray region is excluded by the condition $M_{\mathrm{PSR}} > 0$. **Right**: $M_{\mathrm{PSR}}$–$M_{\mathrm{WD}}$ plane. The gray region is excluded by the condition $\sin i \leq 1$. The lateral panels depict the one-dimensional probability-distribution function for the WD mass (right), pulsar mass (upper right) and inclination (upper left) based on the mass function, $M_{\mathrm{WD}}$ and $q$.

super-dense matter present in NS interiors (Özel et al. 2010; Demorest et al. 2010). For these masses and the known orbital period, GR predicts that the orbital period should decrease at the rate of $\dot{P}_{\mathrm{b}}^{\mathrm{GR}} = (-2.58^{+0.07}_{-0.11}) \times 10^{-13}\,\mathrm{s\,s^{-1}}$ (68.27% confidence) due to energy loss through GW emission.

### 4.2.4 Radio observations

Since April 2011 we have been observing PSR J0348+0432 with the 1.4 GHz receiver of the 305-m radio telescope at the Arecibo Observatory, using its four Wide-band Pulsar Processors (Dowd, Sisk & Hagen 2000). In order to verify the Arecibo data, we have been independently timing PSR J0348+0432 at 1.4 GHz using the 100-m radio telescope in Effelsberg, Germany. The two timing data sets produce consistent rotational models, providing added confidence in both. Combining the Arecibo and Effelsberg data with the initial GBT observations (Lynch et al. 2013), we derive the timing solution presented in Table 3. To match the arrival times, the solution requires a significant measurement of orbital decay, $\dot{P}_{\mathrm{b}} = (-2.73 \pm 0.45) \times 10^{-13}\,\mathrm{s\,s^{-1}}$ (68.27% confidence).

The total proper motion and distance estimate (Table 3) allows us to calculate the kinematic corrections to $\dot{P}_{\mathrm{b}}$ from its motion in the Galaxy, plus any contribution due to possible variations of Newton's gravitational constant $G$: $\delta\dot{P}_{\mathrm{b}} = 0.016 \pm 0.003 \times 10^{-13}\,\mathrm{s\,s^{-1}}$. This is negligible compared to the measurement uncertainty. Similarly, the small rate of rotational energy loss of the pulsar (Table 3) excludes any substantial contamination due to mass loss from the system; furthermore we can exclude substantial contributions to $\dot{P}_b$ from tidal effects (see SOM in Section 4.5 for details). Therefore, the observed $\dot{P}_b$ is caused by GW emission and its magnitude is entirely consistent with the one predicted by GR: $\dot{P}_b/\dot{P}_b^{GR} = 1.05 \pm 0.18$ (Fig. 29).

If we *assume* that GR is the correct theory of gravity, we can then derive the component masses from the intersection of the regions allowed by $q$ and $\dot{P}_b$ (Fig. 29): $M_{WD} = 0.177^{+0.017}_{-0.018}\,M_\odot$ and $M_{PSR} = 2.07^{+0.20}_{-0.21}\,M_\odot$ (68.27% confidence). These values are not too constraining yet. However, the uncertainty of the measurement of $\dot{P}_b$ decreases with $T^{-5/2}$ (where $T$ is the timing baseline); therefore this method will yield very precise mass measurements within a couple of years.

## 4.3 Discussion

### 4.3.1 PSR J0348+0432 as a testbed for gravity

There are strong arguments for GR not to be valid beyond a (yet unknown) critical point, like its incompatibility with quantum theory and its prediction of the formation of spacetime singularities under generic conditions. Therefore, it remains an open question if GR is the final description of macroscopic gravity. This strongly motivates testing gravity regimes that have not been tested before, in particular regimes where gravity is strong and highly non-linear. Presently, binary pulsars provide the best high-precision experiments to probe strong-field deviations from GR and the best tests of the radiative properties of gravity (Damour & Taylor 1992; Stairs 2003; Kramer et al. 2006b; Damour 2009; Freire et al. 2012). Among these systems PSR J0348+0432 has a special role: it is the first massive ($\sim 2\,M_\odot$) NS in a relativistic binary orbit. The orbital period of PSR J0348+0432 is only 15 seconds longer than that of the double pulsar system, but it has $\sim 2$ times more fractional gravitational binding energy than each of the double pulsar NSs. This places it far outside the presently tested binding energy range [see Fig. 30, left panel, and Table 5 in SOM, Section 4.5]. Because the magnitude of strong-field effects generally depends non-linearly on the binding energy, the measurement of orbital decay transforms the system into a gravitational laboratory for a previously untested regime, qualitatively very different from what was accessible in the past.

In physically consistent and extensively studied alternatives, gravity is generally mediated by extra fields (e.g. scalar) in addition to the tensor field of GR (Will 1993). A dynamical coupling between matter and these extra fields can lead to prominent deviations from GR that only occur at the high gravitational binding energies of massive NSs. One of the prime examples is the strong-field scalarization discovered in Damour & Esposito-Farese (1993). If GR is not valid, in the PSR J0348+0432 system where such an object is closely orbited by a weakly self-gravitating body, one generally expects a violation of the strong equivalence principle that in turn leads to a modification in the emission of gravitational waves. While in GR the lowest source multipole that generates gravitational radiation is the quadrupole, alternative gravity theories generally predict the presence of monopole and dipole radiation, on top of a modification of the other multipoles (Will 1993). For a binary system, the leading change in the orbital period is then given by the dipole contribution, which for a (nearly) circular orbit reads (Section 4.5, SOM):

$$\dot{P}_b^{dipolar} \simeq -\frac{4\pi^2 G}{c^3 P_b}\,\frac{M_{PSR}M_{WD}}{M_{PSR}+M_{WD}}\,(\alpha_{PSR}-\alpha_{WD})^2\,, \tag{14}$$

where $\alpha_{PSR}$ is the effective coupling strength between the NS and the ambient fields responsible for the dipole moment [e.g. scalar fields in scalar-tensor gravity], and $\alpha_{WD}$ is the same parameter for the WD companion. The WD companion to PSR J0348+0432 has a fractional gravitational binding energy ($E_{grav}/M_{WD}c^2$) of just $-1.2 \times 10^{-5}$, and is therefore a weakly self-gravitating object. Consequently, $\alpha_{WD}$ is practically identical to the linear field-matter coupling $\alpha_0$, which is well constrained ($|\alpha_0| < 0.004$) in Solar System experiments (Will 2006; Damour 2009).

For $\alpha_{PSR}$, the situation is very different. Even if $\alpha_0$ is vanishingly small, $\alpha_{PSR}$ can have values close to unity, due to a non-linear behaviour of gravity in the interaction between matter and the gravitational fields in the strong-gravity regime inside NSs (Damour & Esposito-Farese 1993;

Damour & Esposito-Farèse 1996). A significant $\alpha_{\rm PSR}$ for NSs up to $1.47\,{\rm M}_\odot$ has been excluded by various binary pulsar experiments (Freire et al. 2012; and Section 4.5). The consistency of the observed GW damping ($\dot{P}_{\rm b}$) with the GR predictions for PSR J0348+0432 (Table 3) implies $|\alpha_{\rm PSR} - \alpha_0| < 0.005$ (95% confidence) and consequently excludes significant strong-field deviations, even for massive NSs of $\sim 2\,{\rm M}_\odot$.

To demonstrate in some detail the implications of our results for possible strong-field deviations of gravity from Einstein's theory, we confront our limits on dipolar radiation with a specific class of scalar-tensor theories, in which gravity is mediated by a symmetric second-rank tensor field $g^{*}_{\mu\nu}$ and by a long-range (massless) scalar field $\varphi$. Scalar-tensor theories are well motivated and consistent theories of gravity, extensively studied in the literature (e.g. Fujii & Maeda 2003; Goenner 2012). For this reason, they are the most natural framework for us to illustrate the gravitational phenomena that can be probed with PSR J0348+0432.

Concerning the EOS of NS matter, in our calculations we use the rather stiff EOS ".20"of Haensel, Proszynski & Kutschera (1981) that supports (in GR) NSs of up to $2.6\,{\rm M}_\odot$. We make this choice for two reasons: i) a stiffer EOS generally leads to more conservative limits when constraining alternative gravity theories, and ii) it is able to support even more massive NSs than PSR J0348+0432, which are likely to exist (Freire et al. 2008; van Kerkwijk, Breton & Kulkarni 2011; Romani et al. 2012). Furthermore, in most of our conclusions a specific EOS is used only for illustrative purposes, and the obtained generic results are EOS independent.

Fig. 30 (right panel) illustrates how PSR J0348+0432 probes a non-linear regime of gravity that has not been tested before. A change in EOS and gravity theory would lead to a modified functional shape for the effective coupling strength, $\alpha_{\rm PSR}$. However, this would not change the general picture: even in the strong gravitational field of a $2\,{\rm M}_\odot$ NS gravity seems to be well described by GR and there is little space for any deviations, at least in the form of long-range fields, which influence the binary dynamics. Short range interactions, like massive Brans-Dicke gravity (Alsing et al. 2012) with a sufficiently large scalar mass (heavier than $\sim 10^{-19}\,{\rm eV}/c^2$), cannot be excluded by PSR J0348+0432. Nevertheless, as we will argue below, in combination with the upcoming ground-based GW detectors, this could lead to particularly illuminating insights into the properties of gravitational interaction.

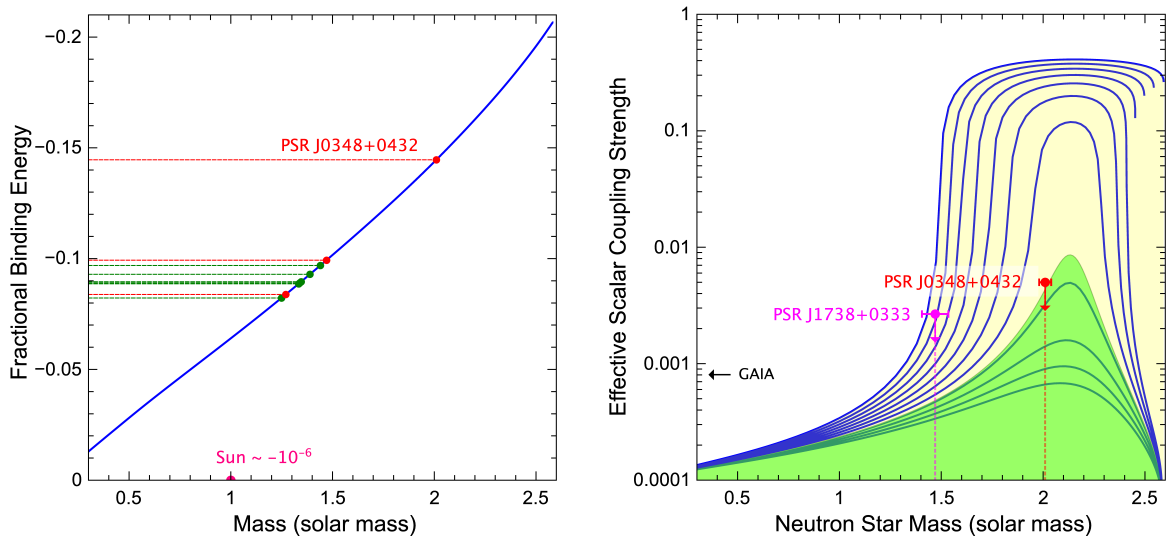

Figure 30: Probing strong field gravity with PSR J0348+0432.
**Left:** Fractional gravitational binding energy as a function of the inertial mass of a NS in GR (blue curve). The dots indicate the NSs of relativistic NS-NS (in green) and NS-WD (in red) binary-pulsar systems currently used for precision gravity tests (Section 4.5, SOM). **Right:** Effective scalar coupling as a function of the NS mass, in the "quadratic" scalar-tensor theory of Damour & Esposito-Farèse (1996). For the linear coupling of matter to the scalar field we have chosen $\alpha_0 = 10^{-4}$, a value well below the sensitivity of any near-future Solar System experiment (e.g. GAIA Hobbs et al. 2010a). The solid curves correspond to stable NS configurations for different values of the quadratic coupling $\beta_0$: $-5$ to $-4$ (top to bottom) in steps of 0.1. The yellow area indicates the parameter space allowed by the best current limit on $|\alpha_{\rm PSR} - \alpha_0|$ (Freire et al. 2012), while only the green area is in agreement with the limit presented here. PSR J0348+0432 probes deeper into the non-linear strong-field regime due to its high mass.

### 4.3.2 Constraints on the phase evolution of neutron-star mergers

The first likely direct GW detection from astrophysical sources by ground-based laser interferometers, like the LIGO (Laser Interferometer Gravitational Wave Observatory)[5] and the VIRGO[6] projects, will mark the beginning of a new era of GW astronomy (Sathyaprakash & Schutz 2009). One of the most promising sources for these detectors are in-spiralling compact binaries, consisting of NSs and BHs, whose orbits are decaying towards a final coalescence due to GW damping. While the signal sweeps in frequency through the detectors' typical sensitive bandwidth between about 20 Hz and a few kHz, the GW signal will be deeply buried in the broadband noise of the detectors (Sathyaprakash & Schutz 2009). To detect it, one will have to apply a matched filtering technique, i.e. correlate the output of the detector with a template wave form. Consequently, it is crucial to know the binary's orbital phase with high accuracy for searching and analyzing the signals from in-spiraling compact binaries. Typically, one aims to lose less than one GW cycle in a signal with $\sim 10^4$ cycles. For this reason, within GR such calculations have been conducted with great effort by various groups up to the 3.5 post-Newtonian order, i.e. all (non-vanishing) terms up to order $(v/c)^7$, providing sufficient accuracy for a detection (Blanchet 2006; Will 2011; Maggiore 2008).

If the gravitational interaction between two compact masses is different from GR, the phase evolution over the last few thousand cycles, which fall into the bandwidth of the detectors, might be too different from the (GR) template in order to extract the signal from the noise. In scalar-tensor gravity for instance, the evolution of the phase is driven by radiation reaction, which is modified because the system loses energy to scalar GWs (Will 1994; Damour & Esposito-Farèse 1998). Depending on the difference between the effective scalar couplings of the two bodies, $\alpha_A$ and $\alpha_B$, the 1.5 post-Newtonian dipolar contribution to the phase evolution could drive the GW signal many cycles away from the GR template. For this reason, it is desirable that potential deviations from GR in the interaction of two compact objects can be tested and constrained prior to the start of the advanced GW detectors. For "canonical" $1.4\,M_\odot$ NSs and long-range gravitational fields, this has already been achieved to a high degree in binary pulsar experiments, e.g. Damour & Esposito-Farèse (1998). So far, the best constraints on dipolar gravitational wave damping in compact binaries come from the observations of the millisecond pulsar PSR J1738+0333 (Freire et al. 2012). However, as discussed in detail above, these timing experiments are insensitive to strong-field deviations that might only become relevant in the strong gravitational fields associated with high-mass NSs. Consequently, the dynamics of a merger of a $2\,M_\odot$ NS with a "canonical" NS or a BH might have a significant contribution from dipolar GWs. With our constraints on dipolar radiation damping from the timing observations of PSR J0348+0432, given above, we can already exclude a deviation of more than $\sim 0.5$ cycles from the GR template during the observable in-spiral caused by additional long-range gravitational fields, for the whole range of NS masses observed in nature (see Fig. 31, and SOM in Section 4.5 for the details of the calculation). This compares to the precision of GR templates based on the 3.5 post-Newtonian approximation (Blanchet 2006; Maggiore 2008). Furthermore, in an extension of the arguments in Will (1994); Damour & Esposito-Farèse (1998) to massive NSs, our result implies that binary pulsar experiments are already more sensitive for testing such deviations than the upcoming advanced GW detectors.

Finally, as mentioned before, our results on PSR J0348+0432 cannot exclude dipolar radiation from short-range fields. Hence, if the range of the additional field in the gravitational interaction happens to lie between the wavelength of the GWs of PSR J0348+0432 and the wavelength of the merger signal ($\sim 10^9$ cm; $\sim 10^{-13}$ eV/$c^2$), then the considerations concerning the applicability of the GR template given here do not apply. On the other hand, in such a case the combination

---

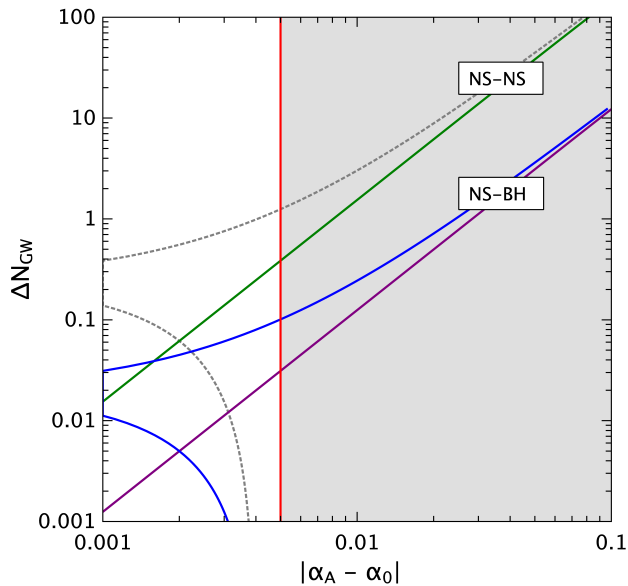

Figure 31: Constraints on the phase offset in gravitational wave cycles in the LIGO/VIRGO bands. Maximum offset in GW cycles in the LIGO/VIRGO band (20 Hz to a few kHz) between the GR template and the true phase evolution of the in-spiral in the presence of dipolar radiation, as a function of the effective coupling of the massive NS for two different system configurations: a $2\,M_\odot$ NS with a $1.25\,M_\odot$ NS (NS-NS), and a merger of a $2\,M_\odot$ NS with a $10\,M_\odot$ BH (NS-BH). In the NS-NS case, the green line is for $\alpha_B = \alpha_0$, and the gray dotted line represents the most conservative, rather unphysical, assumption $\alpha_0 = 0.004$ and $\alpha_B = 0$ (Section 4.5, SOM). In the NS-BH case, $\alpha_B$ is set to zero (from the assumption that the no-hair theorem holds). The blue line is for $\alpha_0 = 0.004$ (Solar System limit for scalar-tensor theories), and the purple line represents $\alpha_0 = 0$. The gray area to the right of the red line is excluded by PSR J0348+0432. In this plot there is no assumption concerning the EOS.

of binary pulsar and LIGO/VIRGO experiments can be used to constrain the mass of this extra field.

### 4.3.3 Formation, past and future evolution of the system

The measured spin period $P$ and spin-period derivative $\dot{P}$ of PSR J0348+0432, combined with the masses and orbital period of the system (Table 3), form a peculiar set of parameters that gives insight to binary stellar evolution. The short 2.46-hr orbital period is best understood from evolution via a common envelope where the NS is captured in the envelope of the WD progenitor, leading to efficient removal of orbital angular momentum on a short timescale of $\sim 10^3$ yr (Iben & Livio 1993). This implies that the NS was born with an initial mass close to its current mass of $2.01\,M_\odot$, because very little accretion was possible. Whereas the slow spin period of $\sim 39$ ms and the unusually strong magnetic field (Section 4.5, SOM) of a few $10^9$ G (Table 3) provide further support for this scenario, the low WD mass contradicts the standard common-envelope hypothesis by requiring a progenitor star mass smaller than $2.2\,M_\odot$, because more massive stars would leave behind more massive cores (Tauris, Langer & Kramer 2011; see SOM in Section 4.5). For such low donor star masses, however, the mass ratio of the binary components is close to unity, leading to dynamically stable mass transfer without forming a common envelope (Tauris & Savonije 1999; Podsiadlowski, Rappaport & Pfahl 2002). One potential solution to this mass discrepancy for common-envelope evolution is to assume that the original mass of the WD was $\geq 0.4\,M_\odot$ and that it was subsequently evaporated by the pulsar wind (Fruchter et al. 1988) when PSR J0348+0432 was young and energetic, right after its recycling phase (Section 4.5, SOM). Such an evolution could also help explain the formation of another puzzling system, PSR J1744−3922 (Breton et al. 2007). However, we find that this scenario is quite unlikely given that the observed spectrum of the WD in PSR J0348+0432 only displays hydrogen lines, which is not expected if the WD was indeed a stripped remnant of a much more massive helium or carbon-oxygen WD. Furthermore, it is unclear why this evaporation process should have come to a complete stop when the WD reached its current mass of $0.17\,M_\odot$. A speculative hypothesis to circumvent the above-mentioned problems would be a common-envelope evolution with hypercritical accretion, where $\sim 0.6\,M_\odot$ of material was efficiently transferred to a $1.4\,M_\odot$ NS (see Chevalier 1993; and SOM in Section 4.5).

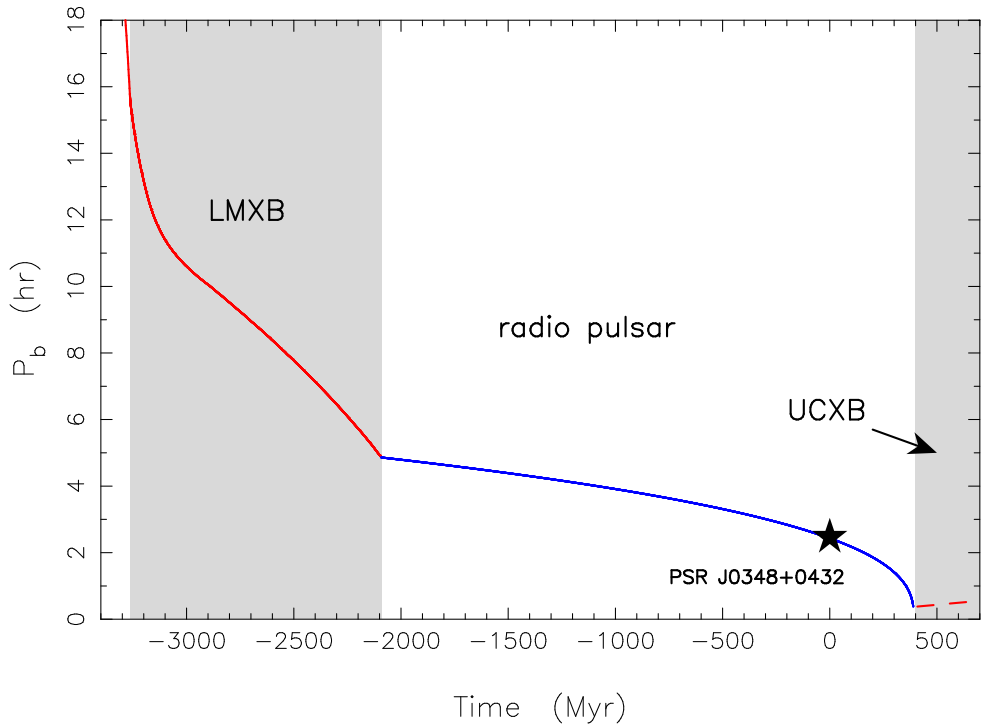

Figure 32: Past and future orbital evolution of PSR J0348+0432.
Formation of PSR J0348+0432 from our converging LMXB model calculation. The plot shows orbital period as a function of time (calibrated to present day). The progenitor detached from its Roche lobe about 2 Gyr ago (according to the estimated cooling age of the WD) when $P_b \simeq 5$ hr, and since then GW damping reduced the orbital period to its present value of 2.46 hr (marked with a star).

An alternative, and more promising, formation scenario is evolution via a close-orbit low-mass X-ray binary (LMXB) with a $1.0 - 1.6\,M_\odot$ donor star that suffered from loss of orbital angular momentum due to magnetic braking (Pylyser & Savonije 1989; Podsiadlowski, Rappaport & Pfahl 2002; van der Sluys, Verbunt & Pols 2005). This requires a finely tuned truncation of the mass-transfer process which is not yet understood in detail, but is also required for other known recycled pulsars (Section 4.5, SOM) with short orbital periods of $P_b \leq 8$ hr and low-mass helium WD companions with $M_{WD} \approx 0.14 - 0.18\,M_\odot$. The interplay between magnetic braking, angular momentum loss from stellar winds (possibly caused by irradiation) and mass ejected from the vicinity of the NS is poorly understood and current stellar evolution models have difficulties reproducing these binary pulsar systems. One issue is that the converging LMXBs most often do not detach but keep evolving with continuous mass transfer to more and more compact systems with $P_b \leq 1$ hr and ultra-light donor masses smaller than $0.08\,M_\odot$.

Using the Langer stellar evolution code (Section 4.5, SOM), we have attempted to model the formation of the PSR J0348+0432 system via LMXB evolution (Fig. 32). To achieve this, we forced the donor star to detach its Roche lobe at $P_b \sim 5$ hr, such that the system subsequently shrinks in size to its present value of $P_b \simeq 2.46$ hr due to GW radiation within 2 Gyr, the estimated cooling age of the WD. An illustration of the past and future evolution of PSR J0348+0432 from the two different formation channels is shown in Fig. 33.

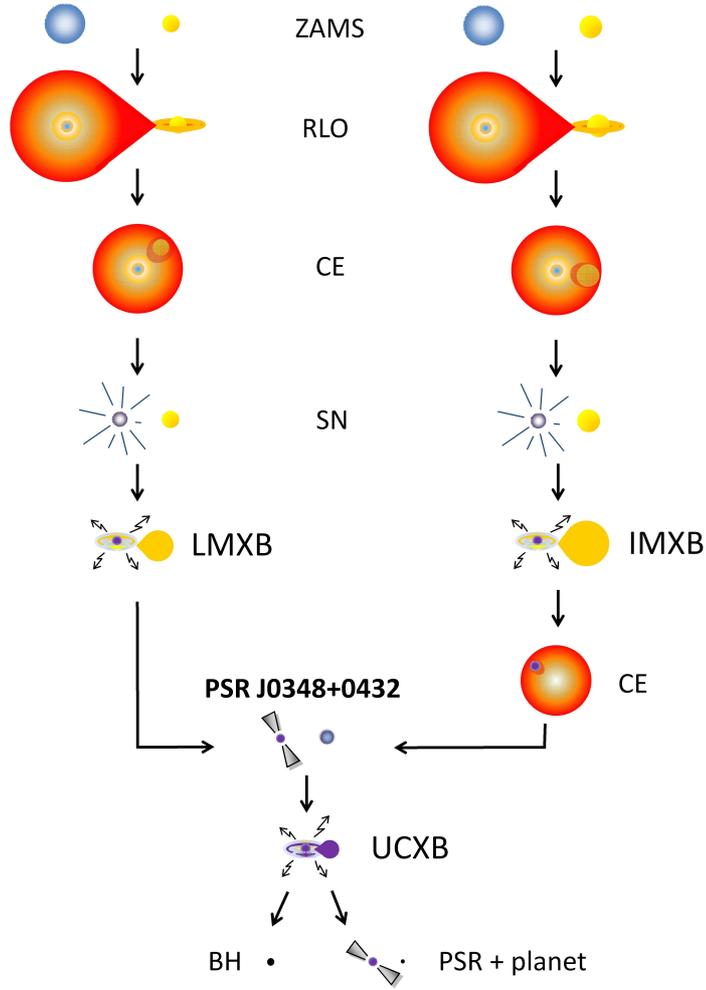

Figure 33: Possible formation channels and final fate of PSR J0348+0432.
An illustration of the formation and evolution of PSR J0348+0432. The zero-age main sequence (ZAMS) mass of the NS progenitor is likely to be $20 - 25 \, M_\odot$, whereas the WD progenitor had a mass of $1.0 - 1.6 \, M_\odot$ (LMXB) or $2.2 - 5 \, M_\odot$ (common envelope, CE), depending on its formation channel. In $\sim 400 \, \mathrm{Myr}$ (when $P_\mathrm{b} \simeq 23 \, \mathrm{min}$) the WD will fill its Roche lobe and the system becomes an ultra-compact X-ray binary (UCXB) leading to the formation of a BH or a pulsar with a planet.

An abnormality of PSR J0348+0432 in view of the LMXB model is its slow spin period of $P \sim 39 \, \mathrm{ms}$ and, in particular, the high value for the spin period derivative, $\dot{P} = 2.41 \times 10^{-19} \, \mathrm{s \, s^{-1}}$. These values correspond to an inferred surface magnetic flux density of $B \sim 2 \times 10^9 \, \mathrm{G}$, which is high compared to most other recycled pulsars (Tauris, Langer & Kramer 2012). However, a high $B$ value naturally explains the slow spin period of PSR J0348+0432 from a combination of spin-down during the Roche-lobe decoupling phase (Tauris 2012) and subsequent magnetic dipole radiation from this high-magnetic-field pulsar (Tauris, Langer & Kramer 2012; see SOM). Another intriguing question concerning this evolutionary channel is the spread in NS masses. In the five currently known NS-WD systems with $P_\mathrm{b} \leq 8 \, \mathrm{hr}$, the NS masses span a large range of values, ranging from $\sim 1.4$ up to $2.0 \, M_\odot$. The small masses imply that the mass transfer during the LMXB phase is extremely inefficient — only about 30% of the material leaving the donor is accreted by the NS (Jacoby et al. 2005; Antoniadis et al. 2012). If this is indeed the case, and one assumes that the physical processes that lead to the formation of these systems are similar, it is likely that PSR J0348+0432 was born with an initial mass of $1.7 \pm 0.1 \, M_\odot$, providing further support for a non-negligible fraction of NSs born massive (Tauris, Langer & Kramer 2011).

Emission of GWs will continue to shrink the orbit of PSR J0348+0432 and in $400 \, \mathrm{Myr}$ (when $P_\mathrm{b} \simeq 23 \, \mathrm{min}$) the WD will fill its Roche lobe and possibly leave behind a planet orbiting the pulsar (Bailes et al. 2011; van Haaften et al. 2012b). Alternatively, if PSR J0348+0432 is near the upper-mass limit for NSs then a BH might form via accretion-induced collapse of the massive NS in a cataclysmic, $\gamma$-ray burst-like event (Dermer & Atoyan 2006).

## 4.4 Materials & Methods

### 4.4.1 Radial velocities and atmospheric parameters

A detailed log of the VLT observations can be found in the SOM (Section 4.5, Figure 34 & Table 4). We extracted the spectra following closely the method used in Antoniadis et al. (2012) and compared them with template spectra to measure the radial velocities. Our best fits for the WD had reduced $\chi^2$ minimum values of $\chi^2_{\rm red,min} = 1.0 - 1.5$ (Section 4.5, SOM). Uncertainties were taken to be the difference in velocity over which $\chi^2$ increases by $\chi^2_{\rm red,min}$ to account for the fact that $\chi^2_{\rm red,min}$ is not equal to unity (Antoniadis et al. 2012). After transforming the measurements to the reference frame of the Solar System Barycenter (SSB), we folded them using the radio-timing ephemeris described below. We then fitted for the semi-amplitude of the radial velocity modulation, $K_{\rm WD}$, and the systemic radial velocity with respect to the SSB, $\gamma$, assuming a circular orbit and keeping the time of passage through the ascending node, $T_{\rm asc}$, fixed to the best-fit value of the radio-timing ephemeris. Our solution yields $K_{\rm WD} = 351 \pm 4\,{\rm km\,s^{-1}}$ and $\gamma = -1 \pm 20\,{\rm km\,s^{-1}}$ (Section 4.5, SOM).

Details of the Balmer lines in the average spectrum of PSR J0348+0432, created by the coherent addition of the individual spectra shifted to zero velocity, are shown in Fig. 27 (right panel). We modeled the spectrum using a grid of detailed hydrogen atmospheres (Koester 2010). These models incorporate the improved treatment of pressure broadening of the absorption lines presented in Tremblay & Bergeron (2009). As mentioned above, our fit yields $T_{\rm eff} = (10120 \pm 35_{\rm stat} \pm 90_{\rm sys})$ K for the effective temperature and $\log_{10} g = (6.042 \pm 0.032_{\rm stat} \pm 0.060_{\rm sys})$ for the surface gravity (Section 4.5, SOM). The $\chi^2$ map shown in Fig. 28 (left panel) is inflated to take into account systematic uncertainties. The average spectrum was also searched for rotational broadening. Using the analytic profile of Gray (2005) to convolve the model atmospheres, we scanned the grid of velocities $0 \leq v_{\rm r} \sin i \leq 2000\,{\rm km\,s^{-1}}$ with a step size of $100\,{\rm km\,s^{-1}}$. The result is consistent with no rotation and our $1$-$\sigma$ upper limit is $v_{\rm r} \sin i \leq 430\,{\rm km\,s^{-1}}$.

### 4.4.2 Modeling of the white-dwarf mass

Low-mass WDs are thought to form naturally within the age of the Universe via mass transfer in a binary, either through Roche-lobe overflow or common-envelope evolution. In both cases, the WD forms when the envelope mass drops below a critical limit, which depends primarily on the mass of the stellar core, forcing the star to contract and detach from its Roche lobe. After the contraction, the mass of the relic envelope is fixed for a given core mass, but further reduction of its size may occur shortly before the star enters the final cooling branch due to hydrogen shell flashes which force the star to re-expand to giant dimensions. Additional mass removal via Roche-lobe overflow as well as rapid shell hydrogen burning through the CNO cycle may then lead to a decrease of the envelope size and affect the cooling history and atmospheric parameters. To investigate the consequence of a reduced envelope size for the WD companion to PSR J0348+0432, we constructed WD models in which we treat the envelope mass as a free parameter (Section 4.5, SOM). For the WD companion to PSR J0348+0432, an envelope mass below the critical limit for hydrogen fusion is not likely for two main reasons:

First, for a pure helium composition, the observed surface gravity translates to a WD mass of $\sim 0.15\,{\rm M_\odot}$ and a cooling age of $\sim 20\,{\rm Myr}$, which is anomalously small. Such a small age would also imply a large increase in the birth and in-spiral rate of similar relativistic NS–WD systems (Kim et al. 2004). Furthermore, post-contraction flash episodes on the WD are not sufficient to remove the entire envelope. Therefore, creation of a pure helium WD requires large mass loss rates before the progenitor contracts, which is unlikely. For small progenitor masses ($\leq 1.5\,{\rm M_\odot}$) large mass loss prevents contraction and the star evolves to a semi-degenerate companion on a nuclear timescale that exceeds the age of the Universe. For more massive progenitors ($> 1.5\,M_\odot$) the core grows beyond $\sim 0.17\,M_\odot$ in a short timescale and ultimately leaves a too-massive WD. Second, even for envelope hydrogen fractions as low as $X_{\rm avg} = 10^{-6}$, the observed temperature and surface gravity cannot be explained simultaneously: The low surface gravity would again require a small mass of $\sim 0.15\,M_\odot$. However, in this case the surface hydrogen acts like an insulator, preventing the heat of the core from reaching the stellar surface. As a result, temperatures as high as $10000\,{\rm K}$ can only be reached for masses above $\sim 0.162\,M_\odot$.

Past a critical envelope mass, the pressure at the bottom of the envelope becomes high enough to initiate hydrogen-shell burning. The latter then becomes the dominant energy source and the evolutionary time-scale increases; the radius of the star grows by $\sim 50\%$ (depending on the mass), expanding further for larger envelopes. The dependence of the surface gravity on the radius implies that the observed value translates to a higher mass as the envelope mass increases. Therefore, the most conservative lower limit for the WD mass (and thus for PSR J0348+0432, given the fixed mass ratio) is obtained if one considers models with the absolute minimum envelope mass required for hydrogen burning. In this scenario, the mass of the WD is in the range $0.162 - 0.181\,M_\odot$ at 99.73% confidence (Section 4.5, SOM). Despite this constraint being marginally consistent with our observations (Section 4.5, SOM), it is not likely correct due to the high degree of fine-tuning.

For these reasons we have adopted the assumption that the WD companion to PSR J0348+0432 has a thick envelope as generally expected for WDs with such low surface gravity and high temperature.

### 4.4.3 Radio-timing analysis

The Arecibo observing setup (Section 4.5, SOM) and data reduction are similar to the well tested ones described in Freire et al. (2012). Special care is taken with saving raw search data, which allows for iterative improvement of the ephemeris and eliminates orbital-phase dependent smearing of the pulse profiles, which might contaminate the measurement of $\dot{P}_{\rm b}$ (Nice, Stairs & Kasian 2008). From this analysis we derive 7773 independent measurements of pulse times of arrival (TOAs) with a root-mean-square (rms) uncertainty smaller than $10\,\mu{\rm s}$. Similarly the Effelsberg observations yield a total of 179 TOAs with uncertainties smaller than $20\,\mu{\rm s}$.

We use the TEMPO2 timing package (Hobbs, Edwards & Manchester 2006) to derive the timing solution presented in Table 3, using 8121 available TOAs from GBT (Lynch et al. 2013), Arecibo and Effelsberg. The motion of the radiotelescopes relative to the barycenter of the Solar System was computed using the DE/LE 421 Solar System ephemeris (Folkner, Williams & Boggs 2009), published by the Jet Propulsion Laboratories. The orbit of PSR J0348+0432 has a very low eccentricity, therefore we use the "ELL1" orbital model (Lange et al. 2001) to describe the motion of the pulsar.

For the best fit, the reduced $\chi^2$ of the timing residuals (TOA minus model prediction) is 1.66, a result similar to what is obtained in timing observations of other millisecond pulsars. The overall weighted residual rms is $4.6\,\mu{\rm s}$. There are no unmodeled systematic trends in the residuals; either as a function of orbital phase or as a function of time. Therefore $\chi^2 > 1$ is most likely produced by under-estimated TOA uncertainties. We increased our estimated TOA uncertainties for each telescope and receiver to produce a reduced $\chi^2$ of unity on short timescales; for our dominant dataset (Arecibo) the errors were multiplied by a factor of 1.3.

This produces more conservative estimates of the uncertainties of the timing parameters; these have been verified using the Monte Carlo statistical method described in Freire et al. (2012): when all parameters are fitted, the Monte Carlo uncertainty ranges are very similar to those estimated by TEMPO2. As an example, TEMPO2 estimates $\dot{P}_{\rm b} = (-2.73 \pm 0.45) \times 10^{-13}\,{\rm s\,s}^{-1}$ (68.27% confidence) and the Monte Carlo method yields $\dot{P}_{\rm b} = (-2.72 \pm 0.45) \times 10^{-13}\,{\rm s\,s}^{-1}$ (68.27% confidence), in excellent agreement. The observed orbital decay appears to be stable; no higher derivatives of the orbital period are detected (Section 4.5, SOM).

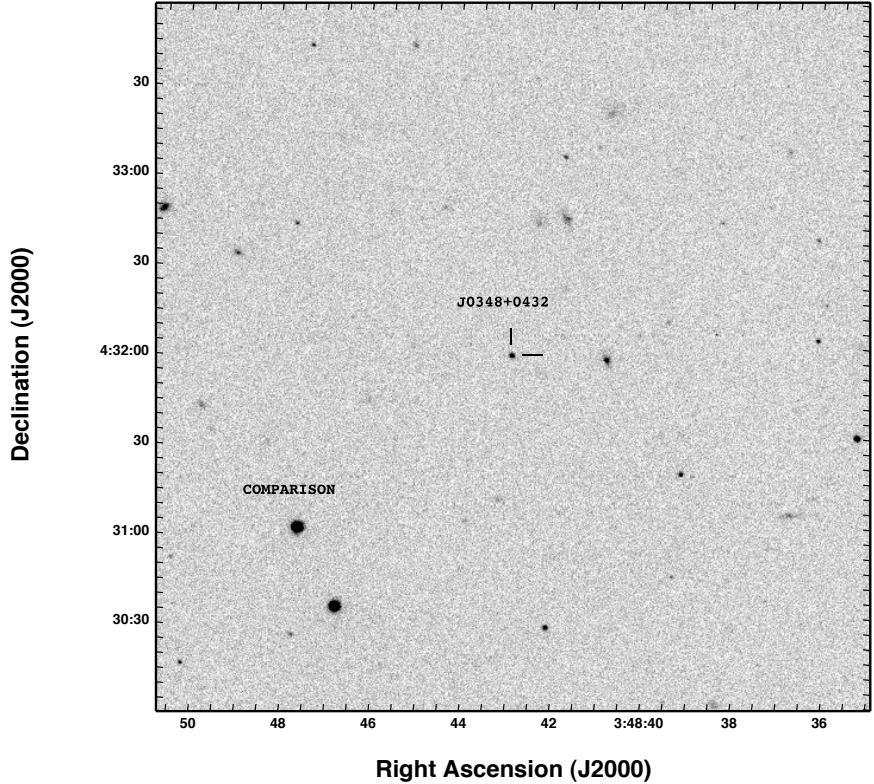

Figure 34: Finding chart for the PSR J0348+0432 system and the comparison star used in our analysis (see text), created from the archived SDSS $g'$ image.

## 4.5 Supporting Online Material (SOM)

### 4.5.1 VLT spectral observations and analysis

We observed the companion of PSR J0348+0432 during December 19 & 20, 2011 with the FORS2 (Appenzeller et al. 1998) instrument on Unite-Telescope 1 (Antú), using its blue sensitive E2V CCD detectors and the G1200B grism. This setup delivers a resolution of 0.36 Å per binned-by-two-pixel along dispersion and 0.″25 per binned-by-two-pixel along the spatial direction. Because of the short orbital period of the binary, we chose a relatively wide 1″ slit to avoid severe radial velocity smearing (by reducing the exposure time) and minimize possible dispersion losses not corrected by the dispersion corrector of the instrument. However, this choice may potentially result in systematic offsets in radial velocity measurements due to non-uniform illumination of the slit. To monitor these effects we rotated the slit by 134.°8 (north-through east) with respect to the parallactic angle to include a bright nearby star for local flux and velocity calibration (Fig. 34). Our setup covers the spectral range from ∼ 3700 to 5200 Å with a resolution ranging from ∼ 2 to 3 Å depending on the seeing. During the first night the conditions were good to photometric and the seeing varied between ∼ 0.″7 and 1.″2. The second night was sporadically plagued with thin cirrus and the seeing ranged from ∼ 0.″9 to 1.″7. Bias, flat and Mercury-Cadmium (HgCd) frames for wavelength calibration were collected during day-time after each run. We collected a total of 34 spectra of the white dwarf companion to PSR J0348+0432 and the nearby comparison star. Of these, 22 had 800-s exposures and were taken with the slit rotated by the angle mentioned above, 4 were taken with 850-s exposures during bad weather instances and 8 with the slit rotated by a slightly different angle during experimental stages. In addition, we collected 2 spectra of the comparison through a wider, 2.″5 slit and spectra of several flux standards at the beginning and the end of each run through both $1''$ and $2{.}''5$ slits.

We reduced the data using routines inside the Munich Image and Data Analysis System (MIDAS). Our analysis, from cosmetic corrections to extraction of spectra, is identical to that followed for the white dwarf companion to PSR J1738+0333 and is described in detail elsewhere (Antoniadis et al. 2012). The dispersion solution has root-mean-square (rms) residuals of $\sim 0.03$ Å for 18 lines. Flux calibration was performed separately for each night by comparing flux-standard observations (Table 4) with high S/N templates or appropriate white dwarf model spectra (Koester 2010). Overall, the response curves from each standard are consistent with each other, with the largest differences (up to 10%) observed at short wavelengths ($\lambda \leq 4000$ Å); we use their average for flux calibration. Prior to the latter, we corrected the narrow-slit spectra for wavelength dependent slit losses using the wide slit spectra of the comparison and accounted for atmospheric extinction using the average extinction curve for La Silla.

### 4.5.2 Radial velocities

We extracted the radial velocities of the white dwarf companion and the nearby comparison star following the procedure described in Antoniadis et al. (2012). First, we identified the nearby comparison star as being a type G1V star (with an uncertainty of about 2 subtypes) and used a high-resolution spectrum of the similar star HD 20807 (Bagnulo et al. 2003) as a template. For the PSR J0348+0432 white-dwarf companion we fitted a high S/N spectrum with a grid of DA model atmospheres (Koester 2010), used the best-fit template to measure radial velocities, averaged the zero-velocity spectra and finally re-fitted the average spectrum to determine the final template. We scanned a grid of velocities from $-800$ to $+800$ km s$^{-1}$ with a step-size of 5 km s$^{-1}$. The best fits had $\chi^2_{\rm red,min} = 1 - 1.5$ and $\chi^2_{\rm red,min} = 1.1 - 3.0$ for the comparison star. As described in the main paper, we scaled the errors to account for the fact that $\chi^2_{\rm red,min}$ was not equal to unity.

The velocities of the comparison star show a peak-to-peak variation of $\sim 40$ km s$^{-1}$, much higher than the typical 0.8 km s$^{-1}$ measurement error. While we find no evidence for binarity, the measurements form 5 distinct groups, each of which display a variability only marginally higher than the formal errors. These coincide with blocks of observations interrupted for target repositioning. The scatter of velocities is therefore clearly related to the instrument and most probably associated with positioning uncertainties. For this reason we chose to use velocities relative to the comparison.

The best-fit solution using all available (barycentred) data gave $K_{\rm WD} = 346 \pm 6$ km s$^{-1}$ and a systemic velocity of $\delta\gamma = +8 \pm 4$ km s$^{-1}$ relative to the comparison with $\chi^2_{\rm red,min} = 2.78$ for 32 degrees of freedom (dof). However, 8 of the observations used here were taken with the slit at a different angle and the white dwarf's velocity relative to the comparison is thus most likely contaminated with an extra systematic shift due to slit rotation. For this reason we neglect these data. Using the homogeneous set of observations only and further rejecting one outlier with spuriously shaped continuum (no. 15 in Table 4) we obtain $K_{\rm WD} = 345 \pm 4$ km s$^{-1}$ and $\delta\gamma = +23 \pm 5$ km s$^{-1}$ respectively with $\chi^2_{\rm red,min} = 0.99$ for 23 degrees of freedom (Fig. 27). After correcting for the small effect of orbital smearing ($\sin(\pi \langle t_{\rm exp}\rangle / P_{\rm b})/(\pi \langle t_{\rm exp}\rangle / P_{\rm b}) = 0.98636$) we find a semi-amplitude of $K_{\rm WD} = 351 \pm 4$ km s$^{-1}$. The best-fit systemic velocity of J0348+0432 using the raw white dwarf velocity measurements is $\gamma = -1 \pm 6$ km s$^{-1}$. Given the large scatter of the comparison's velocity we adopt $\gamma = -1 \pm 20$ km s$^{-1}$ with the uncertainty being a conservative estimate based on the scatter of the data.

### 4.5.3 Average spectrum and atmospheric parameters

We scanned a grid of models covering effective temperatures from $T_{\rm eff} = 8000$ to 25000 K with a step-size of 250 K and surface gravities ranging from $\log g = 5.00$ to $\log g = 8.00$ with a step-size of 0.25 dex. At each point of the grid we fitted for the normalization using a polynomial function of the wavelength to account for non-perfect flux calibration. Analysis of statistical errors is again identical to that followed for the white dwarf companion of PSR J1738+0333 (Antoniadis et al. 2012). We achieved the best fit to higher Balmer lines when excluding the continuum regions between $4000 - 4050$, $4180 - 4270$ and $4400 - 4790$ Å, which had small irregularities due to leftover detector imperfections: $T_{\mathrm{eff}} = 10120 \pm 35$ K and $\log g = 6.042 \pm 0.032$ ($1\sigma$) with $\chi^2_{\mathrm{red,min}} = 1.02$. To estimate the influence of systematics we varied the degree of the polynomial used for normalization (1st to 5th degree), the spectral regions used for the fit (lines-only to whole spectrum) and the assumed spectral resolution (by steps of $\sim 5\%$). We also searched for velocity smearing by checking the consistency of the solution in an average of spectra taken close to orbital conjunction and an average of spectra taken close to the nodes. Finally, we fitted each line (from H$\beta$ to H12) separately to verify the consistency of the fit over the spectrum and examined the influence of our flux calibration by fitting the average uncalibrated spectrum. Overall, all tests gave fits consistent within statistical errors with only few exceptions that had (higher) central values that differed by 120 K and 0.11 dex compared to the numbers above. The good agreement is probably due to the high S/N of the spectrum. The values adopted in the main paper are based on the solution using a third degree polynomial and the systematic error is a conservative estimate based on the scatter of the different fits mentioned above.

Table 4: Observations log and radial velocity measurements
Notes: (1) refers to the barycentric mid-exposure time. (2) is the orbital phase, $\phi$, using the ephemeris in Table 3. (3) is the comparison's velocity in respect to the Solar System Barycenter and (4) the raw barycentric velocities of the white dwarf companion to PSR J0348+0432.

| Target | No. | $MJD^1_{bar}$ | $\phi^2$ | slit | exposure s | rotation deg | pos. angle deg | $v_R^3$ (km s$^{-1}$) | $v_{WD}^4$ (km s$^{-1}$) |
|---|---|---|---|---|---|---|---|---|---|
| PSR J0348+0432 | 1 | 55915.070159 | 0.9742 | 1″ | 799.96 | −135.0 | −150.0 | −18.24 ± 0.41 | −348.25 ± 7.17 |
| | 2 | 55915.080011 | 0.0704 | 1″ | 799.96 | −135.0 | −155.1 | −22.46 ± 0.39 | −345.54 ± 7.55 |
| | 3 | 55915.099359 | 0.2593 | 1″ | 799.95 | −135.0 | −166.5 | −29.37 ± 0.40 | −12.36 ± 9.71 |
| | 4 | 55915.109036 | 0.3538 | 1″ | 799.96 | −135.0 | −172.8 | −25.21 ± 0.39 | +159.76 ± 9.54 |
| | 5 | 55915.120877 | 0.4694 | 1″ | 799.96 | −135.0 | +179.2 | −19.26 ± 0.38 | +332.63 ± 8.37 |
| | 6 | 55915.130692 | 0.5652 | 1″ | 799.98 | −135.0 | +172.7 | −20.79 ± 0.38 | +295.18 ± 8.08 |
| | 7 | 55915.140364 | 0.6596 | 1″ | 799.97 | −135.0 | +166.4 | −19.54 ± 0.38 | +149.72 ± 8.77 |
| | 8 | 55915.150194 | 0.7556 | 1″ | 799.98 | −135.0 | +160.4 | −19.41 ± 0.38 | −45.85 ± 8.76 |
| | 9 | 55915.217387 | 0.4116 | 1″ | 799.96 | −134.8 | +132.1 | −8.28 ± 0.44 | +311.34 ± 11.94 |
| | 10 | 55915.227349 | 0.5089 | 1″ | 849.96 | −134.8 | +129.7 | −9.74 ± 0.43 | +360.61 ± 11.87 |
| | 11 | 55915.237602 | 0.6090 | 1″ | 849.97 | −134.8 | +127.3 | −5.13 ± 0.44 | +278.11 ± 11.03 |
| | 12 | 55915.247859 | 0.7092 | 1″ | 849.95 | −134.8 | +125.4 | −4.85 ± 0.46 | +86.48 ± 13.56 |
| | 13 | 55915.261379 | 0.8412 | 1″ | 799.96 | −134.8 | +123.1 | −3.10 ± 0.42 | −143.98 ± 13.42 |
| | 14 | 55915.271051 | 0.9356 | 1″ | 799.97 | −134.8 | +121.7 | −2.20 ± 0.46 | −297.87 ± 14.87 |
| | 15 | 55915.280729 | 0.0301 | 1″ | 799.96 | −134.8 | +120.4 | −0.34 ± 0.52 | −272.12 ± 18.87 |
| | 16 | 55915.290404 | 0.1245 | 1″ | 799.97 | −134.8 | +119.5 | +1.56 ± 0.61 | −188.66 ± 33.02 |
| | 17 | 55916.060700 | 0.6452 | 1″ | 799.98 | −134.8 | −146.8 | −49.62 ± 0.44 | +197.14 ± 15.88 |
| | 18 | 55916.070535 | 0.7412 | 1″ | 799.97 | −134.8 | −151.6 | −52.40 ± 0.50 | −10.28 ± 21.58 |
| | 19 | 55916.080364 | 0.8372 | 1″ | 799.96 | −134.8 | −156.8 | −50.10 ± 0.51 | −209.03 ± 20.37 |
| | 20 | 55916.091598 | 0.9469 | 1″ | 799.97 | −134.8 | −163.4 | −48.75 ± 0.49 | −339.79 ± 20.05 |
| | 21 | 55916.101421 | 0.0428 | 1″ | 799.96 | −134.8 | −169.6 | −43.97 ± 0.44 | −354.42 ± 15.82 |
| | 22 | 55916.111221 | 0.1384 | 1″ | 799.97 | −134.8 | −176.1 | −41.06 ± 0.42 | −255.63 ± 13.45 |
| | 23 | 55916.122672 | 0.2502 | 1″ | 799.96 | −134.8 | +176.6 | −44.69 ± 0.46 | −15.94 ± 18.30 |
| | 24 | 55916.132782 | 0.3490 | 1″ | 849.96 | −134.8 | +169.6 | −46.23 ± 0.46 | +153.76 ± 19.31 |
| | 25 | 55916.142895 | 0.4477 | 1″ | 799.96 | −134.8 | +163.1 | −43.02 ± 0.46 | +332.62 ± 18.44 |
| | 26 | 55916.154536 | 0.5613 | 1″ | 799.96 | −134.8 | +156.3 | −41.91 ± 0.49 | +285.43 ± 20.94 |
| | 27 | 55916.164373 | 0.6574 | 1″ | 799.97 | −134.8 | +151.1 | −40.10 ± 0.48 | +165.98 ± 18.92 |
| | 28 | 55916.174210 | 0.7534 | 1″ | 799.97 | −134.8 | +146.4 | −39.74 ± 0.48 | −54.07 ± 20.13 |
| | 29 | 55916.229170 | 0.2900 | 1″ | 799.96 | −134.8 | +128.5 | −12.85 ± 0.44 | +89.37 ± 15.28 |
| | 30 | 55916.238967 | 0.3857 | 1″ | 799.97 | −134.8 | +126.5 | −9.24 ± 0.43 | +300.61 ± 17.14 |

Table 4 – Continued

| Target | No. | MJD$_{bar}$ | $\phi$ | slit | exposure | rotation | pos. angle | $v_R$ | $v_{WD}$ |
|--------|-----|-------------|--------|------|----------|----------|------------|-------|----------|
|        |     |             |        |      | s        | deg      | deg        | (km s$^{-1}$) | (km s$^{-1}$) |
|        | 31 | 55916.250488 | 0.4982 | $1''$ | 799.96 | $-134.8$ | $+124.4$ | $-9.52 \pm 0.43$ | $+350.95 \pm 18.00$ |
|        | 32 | 55916.260274 | 0.5937 | $1''$ | 799.97 | $-134.8$ | $+122.8$ | $-15.17 \pm 0.55$ | $+331.23 \pm 22.73$ |
|        | 33 | 55916.270071 | 0.6894 | $1''$ | 799.97 | $-134.8$ | $+121.4$ | $-18.58 \pm 0.62$ | $+89.53 \pm 33.20$ |
|        | 34 | 55916.279869 | 0.7850 | $1''$ | 799.97 | $-134.8$ | $+120.2$ | $-19.10 \pm 0.76$ | $-110.32 \pm 50.99$ |
|        | 35 | 55915.181590 | 0.0641 | $2''\!.5$ | 799.97 | $-134.8$ | $+144.0$ | | |
|        | 36 | 55916.190536 | 0.9148 | $2''\!.5$ | 799.97 | $-134.8$ | $+139.4$ | | |
| EG 21  | 37 | 55915.031970 | | $1''$ | 21.99 | 0.0 | $-25.8$ | | |
|        | 38 | 55915.034856 | | $2''\!.5$ | 21.99 | 0.0 | $-24.5$ | | |
| HD 49798 | 39 | 55915.350264 | | $1''$ | 2.00 | 0.0 | $+72.6$ | | |
|        | 40 | 55916.343776 | | $1''$ | 2.00 | 0.0 | $+71.3$ | | |
|        | 41 | 55916.346515 | | $2''\!.5$ | 2.01 | 0.0 | $+72.3$ | | |
| LTT 3218 | 42 | 55916.352238 | | $2''\!.5$ | 22.01 | 0.0 | $+58.8$ | | |
|        | 43 | 55916.357062 | | $2''\!.5$ | 22.01 | 0.0 | $+62.4$ | | |
|        | 44 | 55916.369131 | | $1''$ | 35.00 | 0.0 | $+64.5$ | | |
| GD 108 | 45 | 55916.360948 | | $2''\!.5$ | 22.00 | 0.0 | $-176.4$ | | |
|        | 46 | 55916.365961 | | $1''$ | 35.00 | 0.0 | $+179.9$ | | |

### 4.5.4 Spectroscopic modeling and the "high $\log g$" problem

Spectroscopic modeling of the Balmer lines in *higher* mass white dwarfs shows a spurious increase in surface gravity for stars with temperatures between $\sim 8000$ and $11000$ K. This well-known problem is linked to the incomplete treatment of convection in 1-D atmospheric models and disappears with the use of 3-D model atmospheres (Tremblay et al. 2011; 2013). However, our modeling below shows that for the parameter space relevant to the PSR J0348+0432 companion, the atmosphere is not yet convective (e.g. convection sets in at $T_{\mathrm{eff}} \leq 9300$ K for $M_{\mathrm{WD}} = 0.17\,\mathrm{M}_\odot$). Therefore this problem is very unlikely to be relevant for the mass determination presented here.

### 4.5.5 Initial white-dwarf models

To construct the white dwarf models presented in the main paper, we evolved solar composition stars (metal mass-fraction of $Z = 0.02$) with masses between 1.0 and $1.5\,\mathrm{M}_\odot$ and applied a large mass-loss wind at various points on the Red Giant Branch (RGB). To constrain the upper limit of the envelope mass expected from natural binary evolution, we removed the mass before the star enters the asymptotic RGB, letting the star evolve and contract naturally to become a white dwarf. Our upper limits agree well with the results of previous studies (Driebe et al. 1998; Serenelli et al. 2001; Panei et al. 2007). Finally, to fully control the envelope mass of the white dwarf at the final stages of evolution we neglected hydrogen fusion through the CNO bi-cycle that is responsible for the hydrogen shell flashes[7].

In Fig. 35 we show the post-contraction white dwarf cooling age when $T_{\mathrm{eff}} = 10000$ K, as a function of the total hydrogen mass (after cessation of the mass transfer), for masses ranging from 0.155 to $0.185\,\mathrm{M}_\odot$. For low envelope masses, hydrogen burning cannot be initiated and the white dwarf quickly radiates the latent thermal energy of the core and cools in a few Myr. The thick-envelope modes presented in the main text were constructed as above.

### 4.5.6 Metallicity

The metallicity of the white dwarf plays an important role in both regulating the CNO luminosity and changing the chemical profile of the stellar envelope. Qualitatively, our main models described above are in good agreement with the $Z = 0.001$ models of Serenelli et al. (2002) for the parameter space relevant to the white dwarf companion to PSR J0348+0432. Specifically, their $0.172\,\mathrm{M}_\odot$ track has a thick envelope and predicts a surface gravity of $\log g = 6.13$ for $T_{\mathrm{eff}} = 10000\,\mathrm{K}$ which is reached at a cooling age of $\tau_{\mathrm{cool}} = 2.85\,\mathrm{Gyr}$. This agreement is not surprising given that CNO burning is neglected in our analysis, convective mixing has not yet set in at $T = 10000\,\mathrm{K}$ and consequently metals are absent from surface layers due to gravitational settling. Therefore we consider that any uncertainties due to metallicity are small and anyway included in our adopted errors.

---

[7]For a white dwarf at the final cooling branch, CNO luminosity accounts for less than 5% of the total energy budget. Hence, it is safe to neglect it without influencing the macroscopic characteristics of the models (Steinfadt, Bildsten & Arras 2010).

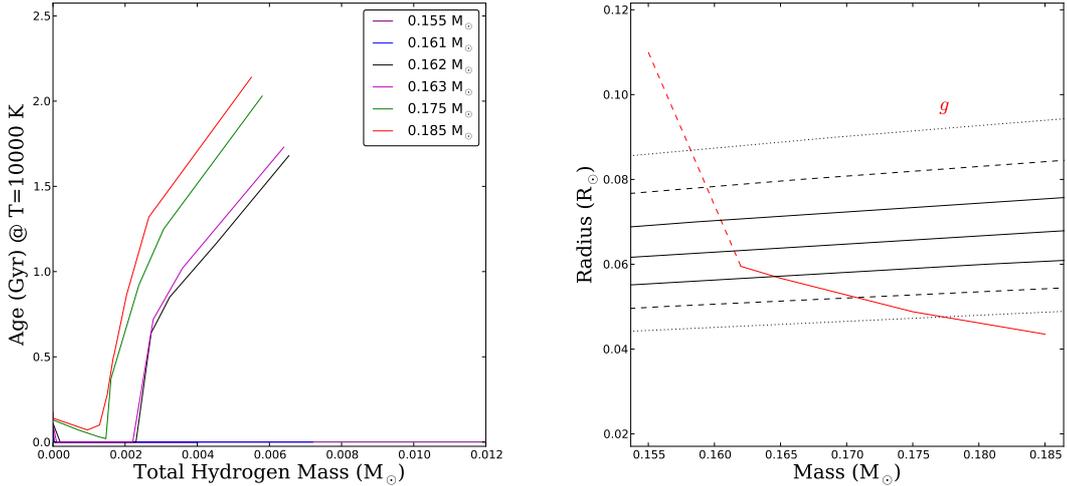

**Figure 35: Left:** White dwarf cooling age (measured from the onset of the core contraction) when the temperature reaches $T_{\mathrm{eff}} = 10000\,\mathrm{K}$ as a function of the total hydrogen mass of the star. Each line depicts a different total mass (from 0.155 to $0.185\,\mathrm{M_\odot}$). For each model, hydrogen burning through the pp-chain at the bottom of the stellar envelope cannot be initiated below a critical envelope mass limit. As a result the white dwarf cools in a few Myr. For models below $\sim 0.162\,\mathrm{M_\odot}$ a temperature of $10000\,\mathrm{K}$ cannot be reached regardless of the envelope size. **Right:** Finite-temperature mass-radius relation (for $10000\,\mathrm{K}$) for models that have the minimum envelope mass required for hydrogen burning (red line). Over-plotted are the most-likely value and 1, 2 and $3\sigma$ constraints on the surface gravity for PSR J0348+0432 (solid, dashed and dotted black lines respectively). For masses below $0.162\,\mathrm{M_\odot}$ the radius is an extrapolation from lower temperatures (in dashed red).

### 4.5.7 Input physics of the stellar evolution models

Stellar models used in our analysis were constructed using the 1-D stellar evolution code "`star`" provided with the Modules for Experiments in Stellar Astrophysics (MESA) (Paxton et al. 2011). `star` solves for the equations of hydrostatic equilibrium, nuclear energy generation, convection and time-dependent element diffusion using a self-adaptive non-Lagrangian mesh and analytic Jacobians. We used default options for the equation-of-state, radiative and neutrino opacities, thermonuclear and weak reaction rates described in Paxton et al. (2011) and references therein. We implemented the mixing length theory of convection from Henyey, Vardya & Bodenheimer (1965) that takes into account radiative losses near the outer layers of the star. Diffusion was taken into account using the method and coefficients from Thoul, Bahcall & Loeb (1994) and transport of material was calculated using the method described in Iben & MacDonald (1985) after grouping the elements in "classes" in terms of atomic mass ranges. Finally, boundary atmospheric conditions were calculated using the gray-atmosphere approach of Eddington (1926).

### 4.5.8 Photometry

A photometric campaign on the white dwarf companion to PSR J0348+0432 was carried out during February 1, 2012 using the ULTRACAM instrument (Dhillon et al. 2007) on the 4.2-m William-Herschel Telescope at La Palma, Spain. The data were reduced using the standard ULTRACAM pipeline (Fig. 36).

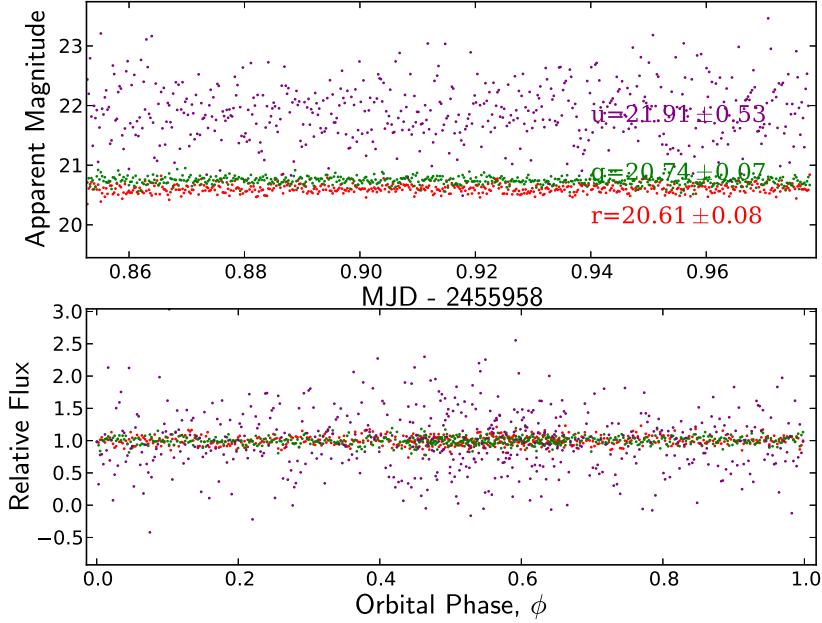

Figure 36: Photometric (upper) and phase-folded (lower) light-curve of the white dwarf companion to PSR J0348+0432 in $u'$, $g'$ and $r'$.

The lightcurves have an rms scatter of ~0.53, 0.07 and 0.08 mag in $u'$, $g'$ and $r'$ respectively and show no evidence for variability over the course of the observations. The phase-folded light-curve shows no variability either. Additionally, our calibrated magnitudes are consistent with the SDSS catalogue magnitudes implying no significant variability at the ~ 5 yr time-scale. Qualitatively, this result supports the use of the PSR J0348+0432 system as a gravitational laboratory (see timing analysis below). In what follows, we discuss the limits on various parameters in more detail. Three effects that cause phase-dependent variability are: deformation of the white dwarf by tides raised by the neutron star, irradiation by the pulsar wind, and Doppler boosting caused by the white dwarf's orbital motion. For a circular orbit, the combined modulation in photon rate $n_\gamma$ is given by

$$\Delta n_\gamma / n_{\gamma,0} \simeq f_{\rm ell} \left(\frac{R_{\rm WD}}{a}\right)^3 q \sin^2 i \cos(4\pi\phi) + f_{\rm db} \frac{K_{\rm WD}}{c} \sin i \cos(2\pi\phi) - f_{\rm irr} \frac{T_{\rm irr}^4}{32 T_{\rm eff}^4} \sin i \sin(2\pi\phi),$$
(15)

with $f_{\rm ell}$, $f_{\rm db}$, and $f_{\rm irr}$ factors of order unity describing the observability in a given filter, $0 \leq \phi \leq 1$ the orbital phase, and $T_{\rm irr} = L_{\rm psr0348}/4\pi a^2 \sigma \simeq 2400$ K the effective temperature corresponding to the pulsar flux incident on the white dwarf. We find that all terms should be small. For the tidal deformation, $f_{\rm ell} = -3(15 + u_1)(1 + \tau_1)/20(3 - u_1) = 1.75$, where we use linear approximations for limb and gravity darkening, with coefficients $u_1 = 0.36$ (Hermes et al. 2012a) and $\tau_1 = 1$ (appropriate for a radiative atmosphere). Thus, the expected modulation is $1.5 \times 10^{-3}$. For the Doppler boosting, approximating the white dwarf as a black-body emitter, $f_{\rm db} \simeq \alpha \exp \alpha / (\exp \alpha - 1) \simeq 2.6$, where $\alpha = hc/\lambda k T_{\rm eff} \simeq 2.8$ (van Kerkwijk et al. 2010), with $\lambda \simeq 550$ nm the typical observing wavelength. Hence, the expected amplitude is ~ $3 \times 10^{-3}$. Finally, for the irradiation, $f_{\rm irr} = (1 - A)f_{\rm db} \leq 2$, where the maximum is for albedo $A \simeq 0$. Thus, irradiation could cause a modulation of up to ~ $1.2 \times 10^{-4}$.

Fitting the observed lightcurves with a function of the form $\Delta n_\gamma / \langle n_\gamma \rangle = 1 + a_{\rm ell} \cos(4\pi\phi) + a_{\rm db} \cos(2\pi\phi) - a_{irr} \sin(2\pi\phi)$, we find good fits ($\chi^2_{\rm red} \simeq 1$) but no significant detections, with averaged amplitudes of the higher S/N $r$ and $g$ band lightcurves of $a_{\rm ell} = 0.003 \pm 0.003$, $a_{\rm db} = 0.003 \pm 0.003$ and $a_{\rm irr} = 0.006 \pm 0.004$. The marginal irradiation signal would correspond to a temperature difference between the irradiated and non-irradiated side of ~ 100 K, which is substantially larger than the expected difference of 2 K. Even if confirmed, however, this would not affect our inferred radial velocity amplitude or white dwarf parameters.

Finally, another possible source of variability is quadrupole moment variations of the white dwarf (Steinfadt, Bildsten & Arras 2010): these typically change the star's luminosity by a few per-cent (e.g. $\sim 20\%$ for the only three known cases of pulsating low-mass white dwarfs (Hermes et al. 2012b; 2013)) and result in changes of the orbital period $P_b$ through classical spin-orbit coupling (Applegate 1992). To our knowledge, all possible mechanisms for such variations would result in modulations much higher than the precision of our lightcurve. Therefore we can neglect this effect and assume that the star is in equilibrium. Our assumption is further supported by the lack of second or higher-order derivatives in the measured orbital period (see below) and recent theoretical findings (Córsico et al. 2012) that locate the instability strip for g-mode oscillations outside the parameter space relevant for the white dwarf companion to PSR J0348+0432.

The SDSS photometry places a constraint on the distance to the system. Adopting the model of Schlegel, Finkbeiner & Davis (1998) for the interstellar reddening and the $0.169\,M_\odot$ cooling track of Serenelli et al. (2001), we find that the luminosities (Fig. 36) are consistent with a distance of $d \simeq 2.1\,\mathrm{kpc}$ (and a reddening of $A_V \sim 0.7$). Given the uncertainties in the models the error is difficult to estimate but it should be better than $\sim 10\%$. Our estimate is also consistent with the distance of $d_{DM} \sim 2\,\mathrm{kpc}$ implied by the dispersion measure (DM) of the pulsar and the NE2001 model for the Galactic free electron density (Cordes & Lazio 2002).

### 4.5.9 Radio observations

The observing setup for the Arecibo telescope is identical to the well-tested setup described in Freire et al. (2012), with the exception of one WAPP now being centered at 1610 MHz instead of 1310 MHz; the former band is cleaner and its use improves the precision of our DM measurements. Also, as in the former case, data are taken in search mode and processed off-line. This allows for iterative improvement of the pulsar ephemeris which is important at the early stages when the timing parameters are not yet very precise. With each improved ephemeris, we de-disperse and re-fold the data, obtaining pulse profiles with higher S/N that yield more accurate pulse times-of-arrival (TOAs). This helps to avoid orbital-phase dependent smearing and timing artefacts, which may corrupt the determination of orbital parameters, particularly the orbital phase and orbital period variation (Nice, Stairs & Kasian 2008).

We de-disperse and fold the radio spectra following the procedure described in Freire et al. (2012). TOAs are derived every 4 minutes to preserve the orbital information in the signal. The pulse profile template, resulting from more than 1 hour of data, is displayed in Fig. 37. Although the pulse profile changes significantly from 350 to 2200 MHz (Lynch et al. 2013), the changes within the band of the L-wide receiver used for timing (1100-1660 MHz, also displayed in Fig. 37) are small enough for us to consider this single average profile taken at 1410 MHz as a good template for all the data. The latter is cross-correlated with every 4-minute/25 MHz-wide pulse profile in the Fourier domain (Lorimer & Kramer 2004; Taylor 1992) and the phase offset that yields the best match is used to derive the topocentric TOA of a reference sub-pulse (normally that closest to the start of each sub-integration). The results described below are obtained using 7773 TOAs with stated rms uncertainty smaller than $10\,\mu s$.

In order to verify the Arecibo data we have been timing PSR J0348+0432 with the 100-m radio telescope in Effelsberg, Germany, which has a very different observing system. The polarization characterization of the radio emission of PSR J0348+0432, displayed in the top part of Fig. 38, was made with this telescope. Overlaid on the polarization data is a theoretical Rotating Vector Model (RVM). It is generally difficult to fit a RVM model to polarization data from recycled pulsars, but for PSR J0348+0432 this model works surprisingly well. For instance, as explained in Lorimer & Kramer (2004), the covariance between the angle between the spin and magnetic axis, $\alpha$, and the angle between the spin axis and the line of sight $\zeta$, allows for a wide range of possible solutions (Fig. 38). However, if we assume that during the accretion episode that recycled the pulsar the spin axis of the pulsar was aligned with the orbital angular momentum (which has an angle $i = 40^\circ\!.2 \pm 0^\circ\!.6$ to the line of sight) then $\alpha \simeq 45^\circ$. The minimum angle

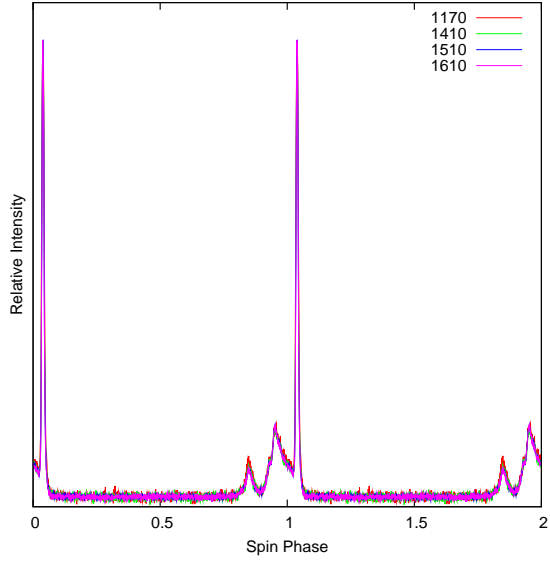

Figure 37: Pulse profiles for PSR J0348+0432 obtained with the WAPP spectrometers at frequencies of 1170, 1410, 1510 and 1610 MHz. Two full cycles are displayed for clarity. Their (almost) perfect overlap indicates that there is little pulse profile evolution between 1170 and 1610 MHz. The 1410 MHz pulse profile is the template used to derive all TOAs. The TOAs correspond to integer phases in this plot, which mark the maximum of the fundamental harmonic.

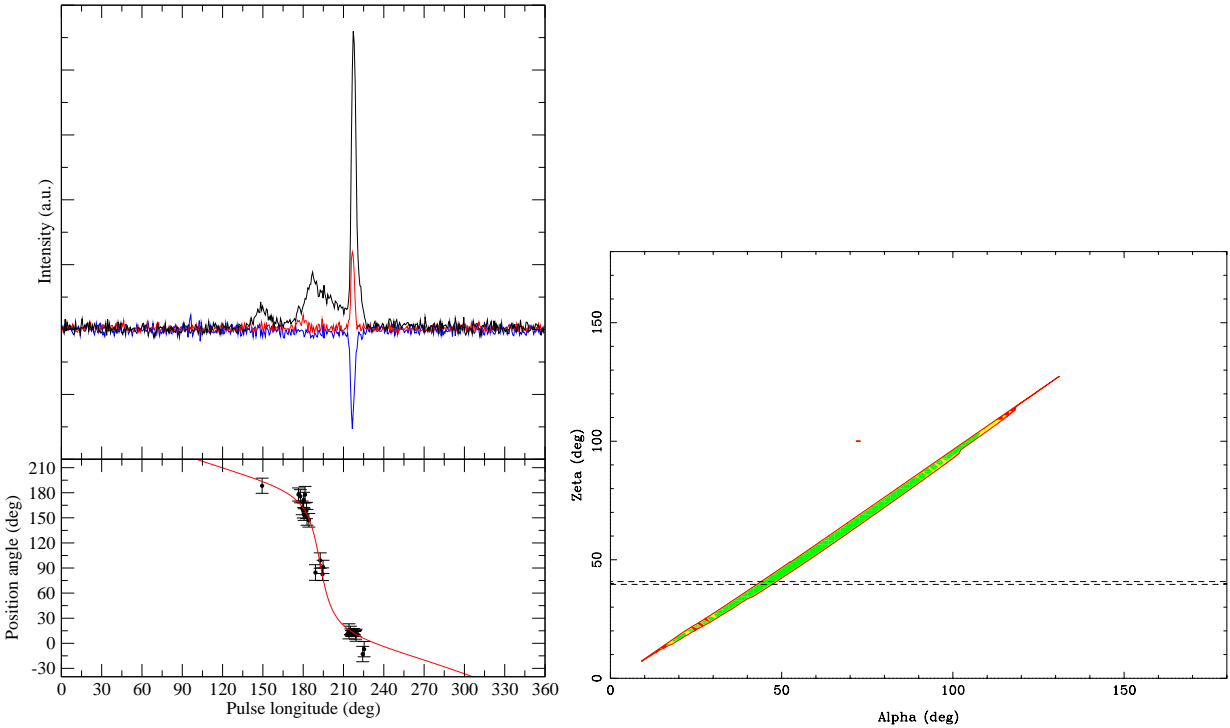

Figure 38: **Left:** Polarization profile of PSR J0348+0432 obtained with the Effelsberg Telescope. The upper panel shows total intensity ($I$, black), the linearly polarized intensity ($L$, red) and the circularly polarized intensity ($V$, blue). The lower panel shows the position angle of $L$ measured at pulse longitudes where $L$ exceeds $2\sigma$ measured from an off-pulse RMS. The red line shows the resulting fit of a Rotating Vector Model (RVM), which indicates an "outer line-of-sight" (see Lorimer & Kramer (2004) for details). **Right:** Map of the RVM parameters $\alpha$ (the angle between the spin axis and magnetic axis) and $\zeta$ (the angle between the line of sight and the spin axis). The green region corresponds to combinations of $\alpha, \zeta$ for which the RVM provides a good description of the polarimetry of PSR J0348+0432 . Based on the polarimetry alone we would have a large uncertainty regarding $\alpha$ and $\zeta$. However, if we assume that during the accretion episode that recycled the pulsar the spin axis of the pulsar was aligned with the orbital angular momentum (which has an angle $i = 40°\!.2 \pm 0°\!.6$ to the line of sight) then $\alpha \simeq 45°$. The minimum angle between the magnetic axis and the line of sight is then given by $\beta = \zeta - \alpha = -4°\!.8$.

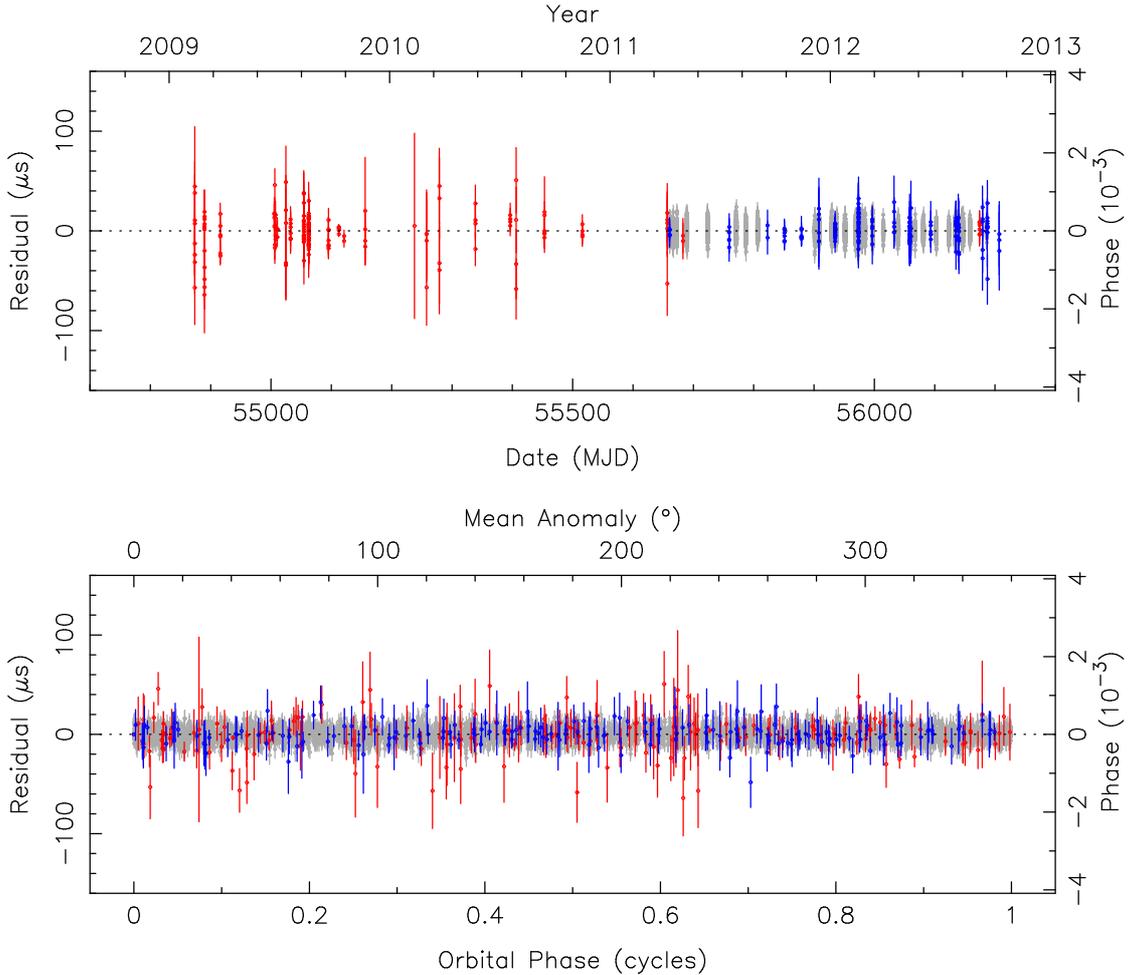

Figure 39: Post-fit residuals from the GBT (red), Arecibo (gray) and Effelsberg (blue) TOAs, obtained with the timing model presented in Table 3. **Top:** Residuals versus time. No significant un-modeled trends can be found in the TOA residuals. **Bottom:** Post-fit residuals versus orbital phase, which for this very low-eccentricity system is measured from the ascending node (i.e., the mean anomaly is equal to the orbital longitude). No significant trends can be identified in the residuals; indicating that the orbital model can describe the orbital modulation of the TOAs correctly. No dispersive delays or unaccounted Shapiro delay signatures are detectable near orbital phase 0.25 (superior conjunction), nor artifacts caused by incorrect de-dispersion or folding of the data.

between the magnetic axis and the line of sight is then given by $\beta = \zeta - \alpha = -5°$.

Apart from the polarimetry, the Effelsberg data yielded a total of 179 high-quality TOAs. As can be seen in Fig. 39, these follow the Arecibo timing very closely, providing added confidence in both.

### 4.5.10 Timing analysis

The combined timing dataset contains 8121 TOAs. The TOA residuals obtained with the best ephemeris (Table 3) are displayed as a function of time in the top panel and as a function of orbital phase in the bottom plot of Fig. 39. To derive the ephemeris in Table 3 (using TEMPO2) we increased the TOA uncertainties by factors of 1.3 for the GBT and Arecibo data and by 1.8 for the Effelsberg data. This results in the residuals of each dataset having a normalized $\chi^2$ of 1. Using these slightly increased (but more realistic) TOA uncertainties results in more conservative (i.e. larger) uncertainties for the fitted timing parameters. Globally, the residuals have a weighted rms of $4.6\,\mu s$ and the reduced $\chi^2$ is 1.019 for 8102 degrees of freedom. The TOA uncertainties presented in Fig. 39 are those used to derive the timing solution.

The orbit of PSR J0348+0432 has a very low eccentricity, therefore we use the "ELL1" orbital model (Lange et al. 2001) to parametrize it.[8] This parametrization yields Keplerian and post-Keplerian parameters very weakly correlated with each other. In order to estimate the intrinsic ("real") eccentricity of the binary (Table 4) we adopt $M_{\mathrm{WD}} = 0.172\,\mathrm{M_{\odot}}$ and $i = 40°.2$ obtained from the optical observations. This assumption is safe because GR is known to provide a sufficiently accurate description of spacetime around weakly self-gravitating objects (Bertotti, Iess & Tortora 2003). According to Freire & Wex (2010), the orthometric amplitude of the Shapiro delay (which quantifies the time amplitude of the *measurable* part of the Shapiro delay) is $h_3 = 42\,\mathrm{ns}$. Fitting for this quantity we obtain $h_3 = 69 \pm 53\,\mathrm{ns}$. This is 1-$\sigma$ consistent with the prediction but the low relative precision of this measurement implies that we cannot determine $M_{\mathrm{WD}}$ and $\sin i$ independently from the existing timing data. A precise measurement of the component masses of this system from Shapiro delay would require an improvement in timing precision that is much beyond our current capabilities.

### 4.5.11 Intrinsic orbital decay

As described in the main text, we detect an orbital decay consistent with the prediction of General Relativity. When we say that this decay is stable, we mean that we detect no higher-order variations of the orbital frequency $f_{\mathrm{b}} \equiv 1/P_{\mathrm{b}}$ nor large variations in $x \equiv a_p \sin i/c$:

$$\frac{d^2 f_{\mathrm{b}}}{dt^2} = -4.5 \pm 4.4 \times 10^{-23}\,\mathrm{Hz\,s^{-2}}, \tag{16}$$

$$\frac{d^3 f_{\mathrm{b}}}{dt^3} = +4.1 \pm 2.5 \times 10^{-36}\,\mathrm{Hz\,s^{-3}}, \tag{17}$$

$$\frac{dx}{dt} = +7.4 \pm 4.4 \times 10^{-15}\,\mathrm{s\,s^{-1}}, \tag{18}$$

where the values and 1-$\sigma$ uncertainties were obtained using the TEMPO implementation of the BTX orbital model. In systems where the quadrupole moment of the white dwarf changes, we should expect such timing effects (Lazaridis et al. 2011) plus significant photometric variations with orbital phase (discussed above). Since none are observed, the companion to PSR J0348+0432 is very likely to have a stable quadrupole moment.

The constraints on the total proper motion $\mu$ combined with the optically derived distance $d = 2.1 \pm 0.2\,\mathrm{kpc}$ allow us to calculate the two kinematic corrections to the observed $\dot{P}_{\mathrm{b}}$. The more important one is the Shklovskii effect (Shklovskii 1970):

$$\dot{P}_{\mathrm{b}}^{\mathrm{Shk}} = P_{\mathrm{b}}\frac{\mu^2 d}{c} = 0.0129^{+0.0025}_{-0.0021} \times 10^{-13}\,\mathrm{s\,s^{-1}}, \tag{19}$$

where we have adopted the 10% error-estimate on the distance. The second correction is caused by the difference of Galactic accelerations between the binary and the Solar System. Using the detailed procedure outlined in Freire et al. (2012), we obtain:

$$\dot{P}_{\mathrm{b}}^{\mathrm{Acc}} = P_{\mathrm{b}}\frac{a_c}{c} = 0.0037^{+0.0006}_{-0.0005} \times 10^{-13}\,\mathrm{s\,s^{-1}}. \tag{20}$$

A third correction could arise from a possible variation of the gravitational constant $\dot{G}$. Conservative limits are given by[9] Damour, Gibbons & Taylor (1988); Damour & Taylor (1991);

---

[8]The ELL1 timing model as implemented in the TEMPO2 software package is a modification of the DD timing model (Damour & Deruelle 1985; 1986) adapted to low-eccentricity binary pulsars. In terms of post-Keplerian observables, it contains all those which are numerically relevant for systems with $e \ll 1$. The "Einstein delay" term is not relevant for such systems and is therefore not taken into account.

[9]The effect of a variable gravitational constant on the orbital period decay $\dot{P}_{\mathrm{b}}$ of PSR J0348+0432 is actually mitigated by the change of the binding energy of the neutron star related to $\dot{G}$ (Nordtvedt 1990). Since this change is both EOS and gravity theory dependent, we keep the more conservative expression of Damour, Gibbons & Taylor (1988), which is anyway two orders of magnitude below our measurement precision of $\dot{P}_{\mathrm{b}}$.

Nordtvedt (1990):

$$\dot{P}_{\rm b}^{\dot{G}} = -2P_{\rm b}\frac{\dot{G}}{G} = (0.0003 \pm 0.0018) \times 10^{-13}\,{\rm s\,s^{-1}}, \qquad (21)$$

where we used the latest limit on $\dot{G}$ from Lunar Laser Ranging (Hofmann, Müller & Biskupek 2010).

Adding these corrections, we obtain a total of $\sim (+1.6 \pm 0.3) \times 10^{-15}\,{\rm s\,s^{-1}}$, or about 0.006 of the measured value. This is much smaller than the current measurement uncertainty and therefore we can conclude that, at the current precision limit, the observed value is intrinsic to the system. Its magnitude is entirely consistent with the GR prediction for the orbital decay caused by emission of gravitational waves: $\dot{P}_{\rm b}/\dot{P}_{\rm b}^{\rm GR} = 1.05 \pm 0.18$. This agreement is depicted graphically in a $\cos i - M_{\rm WD}$ and $M_{\rm PSR} - M_{\rm WD}$ diagram (Fig. 29); the consequences are discussed in the main paper and in detail further below.

### 4.5.12 Mass loss contribution to $\dot{P}_{\rm b}$

If the system is losing mass, that should cause a change in the orbital period (Damour & Taylor 1991):

$$\dot{P}_b^{\dot{M}} = 2\frac{\dot{M}_{\rm T}}{M_{\rm T}}P_{\rm b}, \qquad (22)$$

where $\dot{M}_{\rm T} = \dot{M}_{\rm PSR} + \dot{M}_{\rm WD}$ is the change of mass of both components.

We now estimate both mass loss terms. The pulsar is losing rotational energy at a rate given by $\dot{E} = 4\pi^2 I_{\rm PSR}\dot{P}P^{-3} = 1.6 \times 10^{32}\,{\rm erg\,s^{-1}}$, where $I_{\rm PSR}$ is the pulsar's moment of inertia, normally assumed to be $10^{45}\,{\rm g\,cm^2}$. This dominates the mass loss for the pulsar (Damour & Taylor 1991):

$$\frac{\dot{M}_{\rm PSR}}{M_{\rm T}} = \frac{\dot{E}}{M_{\rm T}c^2} = 4.1 \times 10^{-23}\,{\rm s^{-1}}. \qquad (23)$$

Most of this energy is emitted as a wind of relativistic particles, which we assume to be isotropic to first order. A fraction of this energy $F = R_{\rm WD}^2/4a^2 = 0.00074$ (where $a = xc(q+1)/\sin i = 8.32 \times 10^8\,{\rm m}$ is the separation between components) strikes the surface of the white dwarf. This is the energy available to power mass loss from the white dwarf. Conservation of energy requires that

$$\dot{E}F = \frac{1}{2}\dot{M}_{\rm WD}v^2, \qquad (24)$$

where $v$ is the velocity of the escaping particles. This equation shows that $\dot{M}$ increases as $v$ decreases, however $v$ must be at least equal to the escape velocity for the star to lose mass, i.e., $v^2/2 > GM_{\rm WD}/R_{\rm WD}$. Putting all the constraints together, we obtain:

$$\frac{\dot{M}_{\rm WD}}{M_{\rm T}} < 5.4 \times 10^{-21}\,{\rm s^{-1}}. \qquad (25)$$

Therefore, $\dot{M}_{\rm T} \simeq \dot{M}_{\rm WD}$. Evaluating equation (22), we obtain $\dot{P}_b^{\dot{M}} < 0.4 \times 10^{-16}$, which is $\sim 5 \times 10^2$ times smaller than the current uncertainty in the measurement of $\dot{P}_{\rm b}$.

### 4.5.13 Tidal contribution to $\dot{P}_{\rm b}$

We now calculate the orbital decay caused by tides. If these change the angular velocity of the white dwarf $\dot{\Omega}_{\rm WD}$, this will be compensated by a change in the orbital period of the system $\dot{P}_{\rm b}^{\rm T}$. We can relate the two because of conservation of angular momentum:

$$\dot{P}_{\rm b}^{\rm T} = \frac{3k\Omega_{\rm WD}}{2\pi q(q+1)}\left(\frac{R_{\rm WD}P_{\rm b}\sin i}{xc}\right)^2\frac{1}{\tau_s}, \qquad (26)$$

where $\tau_s = -\Omega_{\rm WD}/\dot{\Omega}_{\rm WD}$ is the synchronization timescale and $k \equiv I_{\rm WD}/(M_{\rm WD}R_{\rm WD}^2)$, where $I_{\rm WD}$ is the white dwarf moment of inertia. For idealized white dwarfs (particularly those with a mass much below the Chandrasekhar limit) sustained solely by degeneracy pressure of non-relativistic electrons, a polytropic sphere with $n = 1.5$ provides a good approximation. For such stars, we have $k = 0.2$ (Motz 1952). However, for this light white dwarf only the core is degenerate, and is surrounded by a deep non-degenerate layer that accounts for only about 5% of the mass of the star. Therefore, the mass distribution is much more centrally condensed than for an $n = 1.5$ polytrope and the moment of inertia is much smaller. We therefore use the output of our white dwarf model calculations (see Fig 28 and above) to estimate that factor. For the model closer to the mean of the white dwarf mass distribution, with $M_{\rm WD} = 0.169 \, {\rm M_\odot}$, $R_{\rm WD} = 0.069 \, {\rm R_\odot}$ and $T_{\rm eff} = 9950 \, {\rm K}$ we obtain $k = 0.0267$. We adopt this value in subsequent calculations.

The only unknown parameters in this expression are $\Omega_{\rm WD}$ and $\tau_s$. If $\tau_s$ were much smaller than the characteristic age of the pulsar $\tau_c = 2.6 \, {\rm Gyr}$ (which is similar to the cooling age of the white dwarf, i.e., this number is likely to be a good approximation to the true age of the system), then the white dwarf rotation would already be synchronized with the orbit ($\Omega_{\rm WD} = 2\pi/P_{\rm b}$). In this case the orbital decay would be slightly affected because, as the orbital period decreases, the white dwarf spin period would decrease at exactly the same rate in order to preserve tidal locking. The resulting exchange of angular momentum would change the orbital decay by a factor $\Delta \dot P_{\rm b}$ given by the ratio of the moment of inertia of the white dwarf and the binary:

$$\frac{\Delta \dot P_{\rm b}}{\dot P_{\rm b}^{\rm GR}} \simeq \frac{I_{\rm WD}}{I_{\rm b}} = \frac{k}{q(q+1)} \left( \frac{R_{\rm WD} \sin i}{xc} \right)^2 = 1.2 \times 10^{-4}. \tag{27}$$

This means that, were the system synchronized, $\Delta \dot P_{\rm b}$ would be an insignificant correction given our current measurement precision.

If the white dwarf is not yet synchronized, then $\tau_s > \tau_c$. In this case $\Omega_{\rm WD}$ can be much larger than $2\pi/P_{\rm b}$, but it must still be smaller than the break-up angular velocity, i.e., $\Omega_{\rm WD} < (GM_{\rm WD}/R_{\rm WD}^3)^{1/2} = 0.0142 \, {\rm rad \, s^{-1}}$. These conditions for $\Omega_{\rm WD}$ and $\tau_s$ yield $\dot P_{\rm b}^{\rm T} < 4.2 \times 10^{-16} \, {\rm s \, s^{-1}}$. Thus, even if the white dwarf were rotating near break-up velocity, $\dot P_{\rm b}^{\rm T}$ would still be two orders of magnitude smaller than the uncertainty in the measurement of $\dot P_{\rm b}$. We note, however, that the progenitor of the white dwarf was very likely synchronized with the orbit at formation, which had a period of $\sim 5$ hours (see below). When the white dwarf formed, fall-back of material within the Roche lobe into it would have spun it up, but not by more than 1 order of magnitude (e.g., Appendix B2.2 of Bassa et al. (2006)). Therefore, at formation $\Omega_{\rm WD}$ was of the order of $3.5 \times 10^{-3} \, {\rm rad \, s^{-1}}$; this would yield $\dot P_{\rm b}^{\rm T} < 1.0 \times 10^{-16} \, {\rm s \, s^{-1}}$.

### 4.5.14 Constraints on dipolar radiation and Scalar-Tensor gravity

In scalar-tensor gravity, like for most other alternatives to GR, the dominant contribution to the GW damping of the orbital motion of a binary system would come from the scalar dipolar waves, proportional to $(\alpha_A - \alpha_B)^2$, where $\alpha_A$ and $\alpha_B$ denote the effective scalar-coupling constants of the two masses $m_A$ and $m_B$, respectively, of the binary system. Such deviations should then become apparent as a modification in the orbital period decay observed in binary pulsars. In GR the emission of quadrupolar tensor waves enters the orbital dynamics at the 2.5 post-Newtonian (pN) level, which corresponds to corrections of order $(v/c)^5$ in the equations of motion, $v$ being a typical orbital velocity. A contribution from dipolar GWs enters already at the 1.5pN level, i.e. terms of order $(v/c)^3$. As an example, in scalar-tensor gravity the change in angular orbital frequency $n_{\rm b} \equiv 2\pi/P_{\rm b}$ for a circular orbit caused by gravitational wave damping up to 2.5pN order is given by Damour & Esposito-Farese (1992); Damour & Esposito-Farèse (1998)

$$\frac{\dot n_{\rm b}}{n_{\rm b}^2} = \frac{X_A X_B}{1 + \alpha_A \alpha_B} \left[ \frac{96}{5} \kappa \left( \frac{v}{c} \right)^5 + (\alpha_A - \alpha_B)^2 \left( \frac{v}{c} \right)^3 \right], \tag{28}$$

where

$$v \equiv [G_*(1 + \alpha_A \alpha_B)(m_A + m_B)n_{\rm b}]^{1/3}, \tag{29}$$

with $G_*$ denoting the bare gravitational constant, and $X_A \equiv m_A/(m_A + m_B)$ and $X_B \equiv m_B/(m_A + m_B)$. The quantity $\kappa$, where $\kappa = 1$ in GR, holds terms arising from the emission of scalar quadrupolar waves and higher order terms of the scalar dipolar emission (Damour & Esposito-Farèse 1998):

$$\kappa = 1 + \frac{1}{6}\left(\alpha_A X_B + \alpha_B X_A\right)^2 + d_1\left(\alpha_A - \alpha_B\right) + d_2\left(\alpha_A - \alpha_B\right)^2 \tag{30}$$

and[10]

$$d_1 = \frac{1}{6}(\alpha_A X_A + \alpha_B X_B)(X_A - X_B) + \frac{5}{48}\frac{\beta_B \alpha_A X_A - \beta_A \alpha_B X_B}{1 + \alpha_A \alpha_B}, \tag{31}$$

$$d_2 = \frac{5}{64} + \frac{253}{576}X_A X_B - \frac{39 + 49\,\alpha_A \alpha_B}{144(1 + \alpha_A \alpha_B)} - \frac{5(X_B \alpha_B^2 \beta_A + X_A \alpha_A^2 \beta_B)}{72(1 + \alpha_A \alpha_B)^2}. \tag{32}$$

GR is recovered for $G_* = G$ and $\alpha_A = \alpha_B = 0$. Equation (28) can directly be confronted with the results compiled in Table 4, in combination with the p.d.f. of the white dwarf mass in Fig. 29, where we use index $A$ for the pulsar and index $B$ for the white dwarf companion.[11] One finds from the mass ratio $q$ that $X_A = q/(q+1) = 0.9213 \pm 0.0008$ and $X_B = 1/(q+1) = 0.0787 \pm 0.0008$. Furthermore, since $G_*(m_A + m_B) \simeq Gm_B(q+1)$, one has $v/c = (0.001970 \pm 0.000016) \times (1 + \alpha_A \alpha_B)^{1/3}$. With the observed change in the orbital frequency $\dot{n}_b = -2\pi \dot{P}_b/P_b^2 = (2.23 \pm 0.36) \times 10^{-20}$, which agrees with GR, one can infer the following constraint on $(\alpha_A - \alpha_0)$ using equation (28):

$$|\alpha_A - \alpha_0| < 0.005 \quad (95\% \text{ C.L.}). \tag{33}$$

Our detailed calculations show that this limit is solely enforced by the dominant 1.5pN term of equation (28) (see also equation (1) in the main text), and is practically insensitive to the values assumed by $\beta_A$ and $\beta_B$. Consequently, as in Lazaridis et al. (2009); Freire et al. (2012) the limit (33) can be seen as a generic limit, that is independent of the EOS.

To illustrate how PSR J0348+0432 probes a new gravity regime, we present detailed calculations based on a specific EOS and a specific class of alternative gravity theories. As an EOS we use the rather stiff EOS ".20" of Haensel, Proszynski & Kutschera (1981), which supports (in GR) neutron stars of up to $2.6\,M_\odot$. Concerning the alternative gravity theories, we use the class of "quadratic" mono-scalar-tensor theories used in Damour & Esposito-Farese (1993); Damour & Esposito-Farèse (1996), where the (field-dependent) coupling strength $\alpha(\varphi)$ between the scalar field and matter contains two parameters: $\alpha(\varphi) = \alpha_0 + \beta_0 \varphi$. Every pair $(\alpha_0, \beta_0)$ represents a specific scalar-tensor theory of gravity. As discovered in Damour & Esposito-Farese (1993), for certain values of $\beta_0$, neutron stars can develop a significant scalarization, even for vanishingly small $\alpha_0$, if their mass exceeds a critical ($\beta_0$-dependent) value. For this reason, this class of gravity theories is particularly well suited to demonstrate how the limit (33) probes a new gravity regime that has not been tested before (see Fig. 30, right panel). The specific parameters and EOS in Fig. 30 have been chosen for demonstration purposes. A change in the EOS, for instance, would lead to a modification in the details of the functional shape of $\alpha_A$, but would not change the overall picture.

### 4.5.15 Constraints on the phase evolution of neutron-star mergers

So far, the best constraints on dipolar gravitational wave damping in compact binaries come from the observations of the millisecond pulsar PSR J1738+0333, a $1.47^{+0.07}_{-0.06}\,M_\odot$ neutron star in a tight orbit ($P_b \approx 8.5\,\text{h}$) with a spectroscopically resolved white-dwarf companion (Antoniadis et al. 2012; Freire et al. 2012). However, as discussed in detail above, such timing experiments

---

[10]To our knowledge, the quantity $d_2$ has been calculated here for the first time.

[11]Strictly speaking, when using the masses of Table 3 in equation (28) one has to keep in mind the difference between the bare gravitational constant $G_*$ and Newton's gravitational constant $G = G_*(1 + \alpha_0)^2$ as measured in a Cavendish-type experiment. However, since $\alpha_0^2 < 10^{-5}$, we can ignore this difference in our calculations.

Table 5: Fractional binding energies of neutron stars (Data for Fig 30, left panel).

| Neutron Star | Mass ($M_\odot$) | Reference | Fractional $E_{NS}^{bind}$ |
|---|---|---|---|
| | | pulsars with white dwarf companions | |
| PSR J0348+0432 | 2.01 | (this paper) | $-0.1446$ |
| PSR J1141−6545 | 1.27 | Bhat, Bailes & Verbiest (2008) | $-0.0838$ |
| PSR J1738+0333 | 1.47 | Antoniadis et al. (2012) | $-0.0993$ |
| | | pulsars with neutron star companions | |
| PSR J0737−3039A | 1.338 | Kramer et al. (2006b) | $-0.0890$ |
| PSR J0737−3039B | 1.249 | Kramer et al. (2006b) | $-0.0822$ |
| PSR B1534+12 | 1.333 | Stairs et al. (2002) | $-0.0887$ |
| . . . companion | 1.345 | Stairs et al. (2002) | $-0.0896$ |
| PSR B1913+16 | 1.440 | Weisberg, Nice & Taylor (2010) | $-0.0969$ |
| . . . companion | 1.389 | Weisberg, Nice & Taylor (2010) | $-0.0929$ |

Fractional binding energies $E_{NS}^{bind}$ of neutron stars in relativistic binaries, which are currently used in precision tests for gravity, and where the neutron star masses are determined with good ($<$ few %) precision. The masses are taken from the given references. The specific numbers for the fractional binding energy are based on the equation-of-state ".20" of Haensel, Proszynski & Kutschera (1981). A different equation-of-state gives different numbers, but does not change the fact that J0348+0432 significantly exceeds the tested binding energy range.

are insensitive to strong-field effects that might only become relevant in the strong gravitational fields of high-mass neutron stars. Consequently, the dynamics of a merger of a $2\,M_\odot$ neutron star with a "canonical" neutron star or a black hole (BH) might have a significant contribution from dipolar GW damping, leading to a modification of the orbital dynamics that is incompatible with the sophisticated GR templates used to search for GWs with ground-based GW detectors, like LIGO and VIRGO, (Sathyaprakash & Schutz 2009). With the results on PSR J0348+0432, in particular with the limit given in equation (33), this question can finally be addressed in some details. For this purpose, we decompose equation (28) into the 2.5pN contribution, that is matched by an appropriate GR template, and the 1.5pN contribution, that drives the phase evolution away from the 2.5pN dynamics. Following Will (1994); Damour & Esposito-Farèse (1998), we introduce the dimensionless orbital angular velocity

$$u \equiv \mathcal{M} n_b = \pi \mathcal{M} f_{GW} \, , \tag{34}$$

where $f_{GW}$ denotes the frequency of the GW and

$$\mathcal{M} \equiv \frac{G_* M}{c^3} \frac{(X_A X_B \kappa)^{3/5}}{(1 + \alpha_A \alpha_B)^{2/5}} \, . \tag{35}$$

To leading order, one then finds

$$\mathcal{M} \dot{u} = \frac{96}{5} (u^{11/3} + \mathcal{B} u^3) \, , \tag{36}$$

where

$$\mathcal{B} \equiv \frac{5}{96} \left( \frac{X_A X_B}{1 + \alpha_A \alpha_B} \right)^{2/5} \frac{(\alpha_A - \alpha_B)^2}{\kappa^{3/5}} \, . \tag{37}$$

The observed GW cycles in a frequency band $[f_{in}, f_{out}]$ can be computed as follows:

$$N_{GW} = \int_{t_{in}}^{t_{out}} f \, dt = \int_{f_{in}}^{f_{out}} (f/\dot{f}) \, df = \frac{1}{\pi} \int_{u_{in}}^{u_{out}} \frac{u}{\mathcal{M} \dot{u}} \, du \, . \tag{38}$$

Consequently, the difference between the 2.5pN dynamics and the 2.5pN + 1.5pN dynamics is given by

$$\Delta N_{\rm GW} = \frac{5}{32\pi} \left( \frac{1}{5u^{5/3}} - \frac{1}{3u\mathcal{B}} + \frac{1}{u^{1/3}\mathcal{B}^2} + \frac{\arctan\left(u^{1/3}/\mathcal{B}^{1/2}\right)}{\mathcal{B}^{5/2}} \right) \Bigg|_{u_{\rm in}}^{u_{\rm out}}, \qquad (39)$$

where we made no assumption about the size of the value for $\mathcal{B}$. For the LIGO/VIRGO band $u_{\rm in} \ll u_{\rm out}$.[12] Fig. 31 gives $\Delta N_{\rm GW}$ for the LIGO/VIRGO detectors, for which a typical bandwidth of 20 Hz to a few kHz was assumed, as a function of $|\alpha_A - \alpha_0|$ for two different systems, a $2/1.25\,{\rm M}_\odot$ NS-NS system and a $2/10\,{\rm M}_\odot$ NS-BH system. Concerning the NS-BH systems, we considered the class of alternative gravity theories where BHs are practically identical to GR, and consequently used $\alpha_B = 0$. For instance, this is the case in scalar-tensor gravity theories with negligible time dependence of the asymptotic scalar field (Damour & Esposito-Farèse 1998). For the NS-NS system an extreme case is represented by the assumption that only the massive neutron star has a significant scalar coupling strength $\alpha_A$, while the lighter companion behaves like a weakly self-gravitating body, meaning $\alpha_B = \alpha_0$. Besides this, for the NS-NS system we have also performed calculations using a hypothetical most conservative (maximal $\Delta N_{\rm GW}$) value for the effective coupling strength of the companion $B$, which is $\alpha_B = 0$. However, such an assumption seems unphysical for a non-zero $\alpha_0$, where $\alpha_B$ is expected to approach $\alpha_0$ (and not 0) for less massive stars. With the limit obtained from PSR J0348+0432 we find a conservative upper limit for the dipolar phase offset of $\sim 0.5$ (NS-NS) and 0.04 (NS-BH) cycles, an amount that would not jeopardize the detection of the gravitational wave signal in the LIGO/VIRGO band (Maggiore 2008).

### 4.5.16 Formation via a common envelope and spiral-in phase

Common-Envelope (CE) evolution (Paczyński 1976; Iben & Livio 1993) in X-ray binaries is initiated by dynamically unstable mass transfer, often as the result of a high mass-transfer rate and a large initial donor/accretor mass ratio, $q_{\rm i} \equiv M_2/M_{\rm PSR} > 1$. If the CE is initiated while the donor star is still early in its main sequence stage (i.e. if $P_{\rm b} < 1$ day), the outcome is expected to be a merger (Taam & Sandquist 2000). It is generally believed that a binary can only survive the CE evolution, and thereby successfully eject the envelope of the donor star, if the binding energy of the envelope, $E_{\rm bind}$, is less than the released orbital energy from the in-spiral process, $\Delta E_{\rm orb}$ (Webbink 1984). The orbital energy of PSR J0348+0432 is: $|E_{\rm orb}| = GM_{\rm PSR}M_{\rm WD}/2a \simeq 5.5 \times 10^{47}$ erg. Hence, even if assuming in-spiral from infinity to the current orbital separation, the amount of liberated orbital energy from the CE phase cannot exceed this value. From calculations of $E_{\rm bind}$ of intermediate-mass stars (Table 6), we find that $E_{\rm bind} \gg \Delta E_{\rm orb}$ during most of their evolutionary stages. Only if the donor star (i.e. the white dwarf progenitor) is an evolved giant is it possible to eject the envelope. However, in this case the core mass of such an evolved star, $M_{\rm core}$, is more massive than the observed white dwarf companion by at least a factor of $2-3$. (As argued in the main text, a reduction in white dwarf mass via evaporation from the pulsar wind seems to be ruled out for PSR J0348+0432 and therefore cannot help circumvent this discrepancy.)

From Fig. 40 we see that only a low-mass donor star with mass $M_2 \leq 2.2\,{\rm M}_\odot$, and not evolved beyond the terminal age main-sequence (TAMS), would leave behind $M_{\rm core} = M_{\rm WD} \simeq 0.17\,{\rm M}_\odot$. In this case, it is clear that energy sources other than $\Delta E_{\rm orb}$ must contribute to expel the envelope (since in this case $E_{\rm bind} \gg \Delta E_{\rm orb}$). Such an energy source could be the release of gravitational potential energy from material which accretes onto the neutron star during the CE. The amount of released energy per accreted unit mass is roughly $\Delta U/m \sim GM/R \sim 2 \times 10^{20}\,{\rm erg\,g^{-1}}$. Hence, assuming full absorption and 100% energy conversion of this released energy to eject the envelope, this would require accretion of $\sim 4 \times 10^{-5}\,{\rm M}_\odot$; a value which is not unrealistic given a timescale of the CE event of $\sim 10^3$ yr with Eddington limited accretion (a few $10^{-8} \, M_\odot \, \text{yr}^{-1}$).

---

[12]For a detector that is sensitive up to a few kHz, the frequency $f_{\rm GW}^{\rm out}$ is determined by the innermost circular orbit, which is $\sim 1350$ Hz for a $2/1.25\,{\rm M}_\odot$ system and $\sim 370$ Hz for a $2/10\,{\rm M}_\odot$ system (see Blanchet 2006).

As a consequence of this relatively short CE phase, the currently observed mass of $M_{\text{PSR}} = 2.01 \, M_\odot$ should be close to the original mass of the neutron star after its formation in a type Ib/c supernova. According to recent studies by Ugliano et al. (2012), neutron star birth masses of $2.0 \, M_\odot$ are indeed possible. As mentioned in the main text, however, having an initially massive neutron star would be a more serious problem for formation via a CE event with a $\leq 2.2 \, M_\odot$ donor star. Such a high value of $M_{\text{PSR}}$ would lead to a value of $q_i$ close to unity, in which case the Roche-lobe overflow (RLO) is expected to be dynamically *stable*, thereby avoiding the formation of a CE (Tauris & Savonije 1999; Podsiadlowski, Rappaport & Pfahl 2002). The only solution to this problem would be that the neutron star was originally born with a more typical mass of $\sim 1.4 \, M_\odot$, in which case $q_i$ would be sufficiently high to ensure formation of a CE. However, in that case one would have to accept the concept of hypercritical accretion (Chevalier 1993; Ivanova 2011), allowing the neutron star to accrete a large amount of mass $\sim 0.5 - 0.7 \, M_\odot$ on a timescale of $\sim 10^3$ yr. One could argue that PSR J0348+0432 would then be the best (and to our knowledge the only) candidate known in which hypercritical accretion might have been at work.

To summarize, given the many issues discussed above we find that a CE formation channel is less favorable to explain PSR J0348+0432 and we now proceed with investigating another solution, the LMXB formation channel.

### 4.5.17 Formation via a converging low-mass X-ray binary

As mentioned in the main text, a handful of binary pulsars exist with values of $P_b \leq 8$ hr and $M_{\text{WD}} \approx 0.14 - 0.18 \, M_\odot$, similar to those of PSR J0348+0432. These systems are tentatively thought to descend from low-mass X-ray binaries (LMXBs) in which the binary suffered from loss of orbital angular momentum caused by magnetic braking (Pylyser & Savonije 1989; Podsiadlowski, Rappaport & Pfahl 2002; van der Sluys, Verbunt & Pols 2005). However, there remains a general problem for reproducing these pulsar binaries using current stellar evolution codes. A main issue is that converging LMXBs most often do not detach but keep evolving with continuous mass transfer to more and more compact systems with $P_b \leq 1$ hr and ultra-light donor masses $M_2 < 0.08 \, M_\odot$. In a few instances, where fine-tuning may lead to detachment and the right values of $P_b$ and $M_2$, the donor star is typically too hydrogen rich to settle and cool as a compact He white dwarf [however, see sequence $d$ in fig. 16 of Podsiadlowski, Rappaport & Pfahl (2002) for an exception]. Our numerical studies are no exception from this general picture.

Table 6: Stellar envelope binding energies, $E_{\text{bind}}$, for given donors and evolutionary stages.

| Stage | $E_{\text{bind}}$ $(2.2 \, M_\odot)$ | $E_{\text{bind}}$ $(4.0 \, M_\odot)$ | CE outcome |
|-------|-----------|-----------|------------|
| $X_c = 0.40$ | $1.8 \times 10^{49}$ erg | $3.8 \times 10^{49}$ erg | merger* |
| $X_c = 0.20$ | $1.6 \times 10^{49}$ erg | $3.4 \times 10^{49}$ erg | merger* |
| $X_c = 0.02$ | $1.5 \times 10^{49}$ erg | $3.2 \times 10^{49}$ erg | merger* |
| TAMS | $1.6 \times 10^{49}$ erg | $3.6 \times 10^{49}$ erg | may survive if $L_{\text{acc}}$ can eject envelope |
| RGB | $9.8 \times 10^{47}$ erg | $2.1 \times 10^{48}$ erg | survives with $0.30 \leq M_{\text{WD}}/M_\odot \leq 0.50$ |
| AGB | $2.0 \times 10^{47}$ erg | $1.9 \times 10^{47}$ erg | survives with $0.44 \leq M_{\text{WD}}/M_\odot \leq 0.72$ |

Envelope binding energies of stars with a total mass of $2.2 \, M_\odot$ and $4.0 \, M_\odot$, respectively. In all cases $E_{\text{bind}}$ was calculated assuming $M_{\text{core}} = M_{\text{WD}} = 0.17 \, M_\odot$, except for the cases where the RLO was initiated at the tip of the RGB/AGB with resulting values of $M_{\text{WD}}$ as listed in the table.
* Note, that intermediate-mass donor stars on the main sequence ($X_c > 0$) with $P_b > 1$ day, or at the TAMS, may avoid the onset of a CE altogether and evolve as a stable intermediate mass X-ray binary, IMXB (Tauris, van den Heuvel & Savonije 2000).

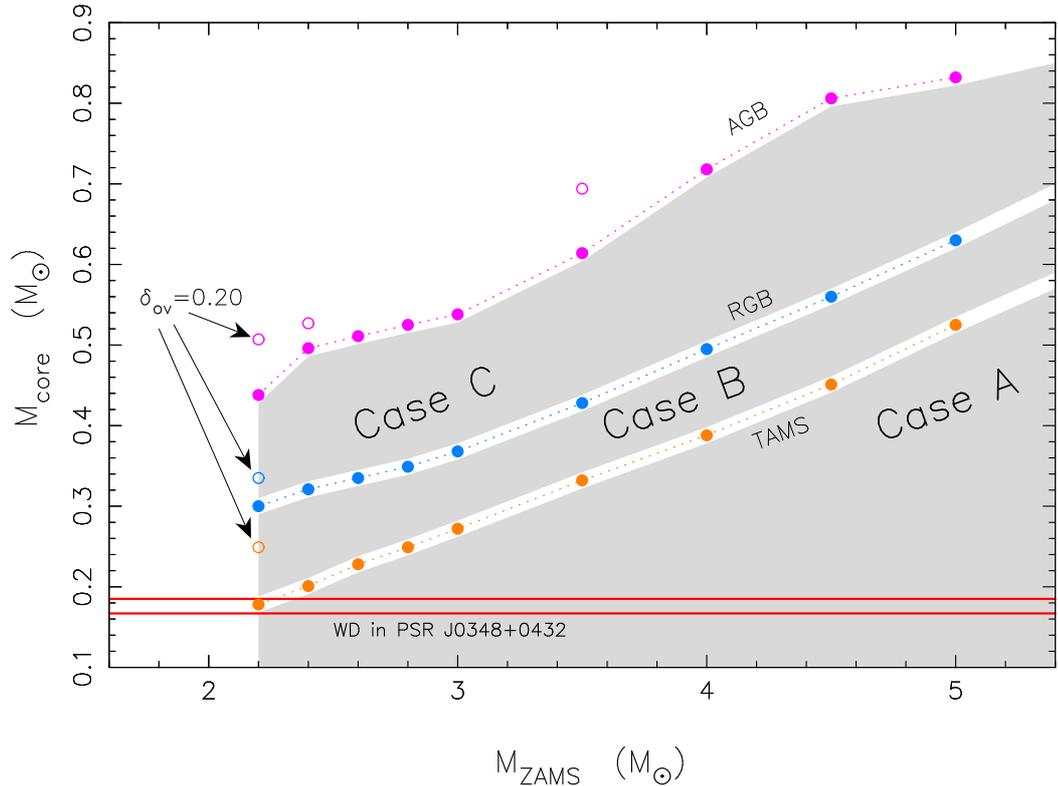

Figure 40: Stellar core mass at different evolutionary epochs as a function of zero-age main sequence (ZAMS) mass. Assuming $M_{core} = M_{WD} \simeq 0.17\,M_\odot$ (as observed in PSR J0348+0432) constrains the progenitor star ZAMS mass to be $\leq 2.2\,M_\odot$ and that its envelope was lost near the terminal-age main sequence (TAMS). All calculations were performed without convective core overshooting. Including this effect (for example, using $\delta_{OV} = 0.20$) would lower the required donor mass even more. [Figure adapted from Tauris, Langer & Kramer (2011)].

Using the Langer stellar evolution code [e.g. Wellstein & Langer (1999); Tauris, Langer & Kramer (2011)] we have attempted to model the formation and evolution of the PSR J0348+0432 system. Here we present a solution where we have forced the donor star to detach its Roche lobe at $P_b \sim 5$ hr, such that the system subsequently shrinks in size to its present value of $P_{orb} \simeq 2.46$ hr due to GW radiation within the estimated cooling age of the white dwarf ($t_{WD} \simeq 2$ Gyr, depending on cooling models and assumed metallicity). To be more precise, the estimated $t_{WD}$ is actually a lower limit on the timescale during which the detached system evolved via GW radiation since it takes $10^8 - 10^9$ yr for the detached pre-white dwarf to settle on the final cooling track. This can be compensated for by choosing a slightly larger $P_b$ at the ZAMS, which causes the system to detach from the LMXB in a somewhat wider orbit.

In Fig. 41 (and see also Fig. 32 in the main text) we show an example of our LMXB calculations. The model binary shown here consisted initially of a $1.75\,M_\odot$ neutron star and a $1.1\,M_\odot$ donor star with metallicity $Z = 0.02$, mixing length parameter, $\alpha = 2.0$ and ZAMS orbital period, $P_b = 2.55$ days. The initial $P_b$ depends on the modeling of magnetic braking. Here the value corresponds to onset of RLO at $P_b \simeq 0.65$ days, shortly after the donor star ceased central hydrogen burning. Our high value of the initial neutron star mass is motivated from studies which show that the accretion efficiency in LMXBs must be rather small — even for systems which are expected to have accreted at sub-Eddington levels (Antoniadis et al. 2012; Jacoby et al. 2005; Tauris & Savonije 1999). Hence, by adopting an accretion efficiency of only 30% we need an initial high-mass neutron star in order to reach the present mass of PSR J0348+0432. Note, that some neutron stars are indeed expected to have been born massive [for a discussion, see Tauris, Langer & Kramer (2011) and references therein]. The outcome of our calculations would possibly have been somewhat similar by assuming an accretion efficiency close to 100% and starting with $M_{PSR} = 1.3\,M_\odot$. To model the loss of orbital angular momentum due to mass loss from the system, we adopted the isotropic re-emission model (Bhattacharya & van den Heuvel 1991).

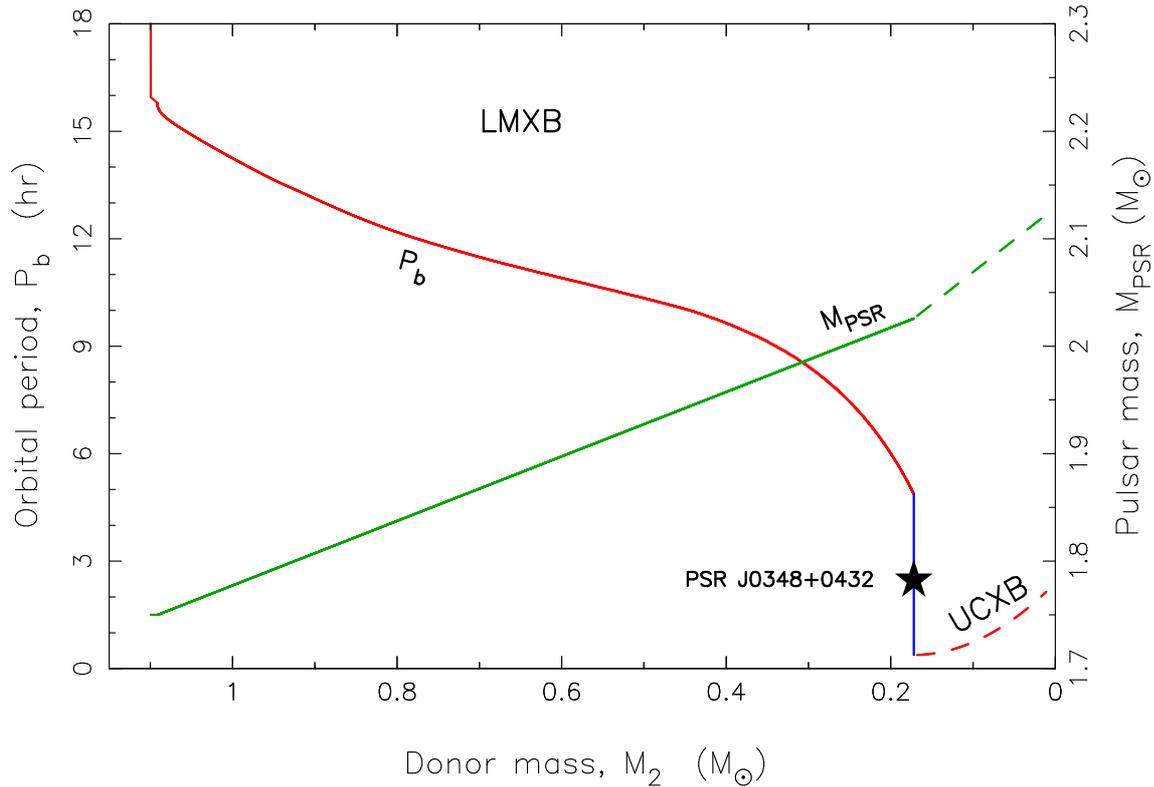

Figure 41: Formation of PSR J0348+0432 from a converging LMXB for the same model as shown in Fig. 32 in the main text. The plot shows how $P_b$ (red line) and the mass of the accreting neutron star (green line) evolved as a function of decreasing donor star mass (here assumed to be $1.1\,M_\odot$ on the ZAMS). The RLO was initiated when $P_b \simeq 16\,hr$ and detached when $P_b \simeq 4.9\,hr$. In this model the initial mass of the neutron star was assumed to be $1.75\,M_\odot$, although it may have been significantly lower if the neutron star accreted with an efficiency close to 100%. The present location of PSR J0348+0432 is marked with a star.

Based on its proper motion and radial velocity measurements, PSR J0348+0432 has an estimated 3D space velocity of $56 \pm 8\,km\,s^{-1}$ with respect to the Solar System. From Monte Carlo simulations of its past motion through our Galaxy [following the method described in Antoniadis et al. (2012); Freire et al. (2011b)], we find that this velocity corresponds to a peculiar velocity with respect to the local standard of rest at every transition of the Galactic plane of $75 \pm 6\,km\,s^{-1}$. This result is rather independent of the applied Galactic model. From subsequent simulations of the dynamical effects of the supernova explosion, we find that a relatively small kick magnitude of $w < 150\,km\,s^{-1}$ was imparted to the newborn neutron star, by probing a broad range of values of the pre-supernova orbital period and the masses of the collapsing naked He-core and its companion star (the white dwarf progenitor).

### 4.5.18 Spin evolution of PSR J0348+0432

A peculiarity of PSR J0348+0432, compared to other recycled pulsars with similar $P_b$ and $M_{WD}$, is its slow spin period, $P = 39\,ms$ and its high value of the spin period derivative, $\dot{P} = 2.41 \times 10^{-19}\,s\,s^{-1}$, cf. the unusual location of PSR J0348+0432 in both the $P\dot{P}$–diagram and the Corbet diagram (Figs. 42, 43). In particular, the Corbet diagram clearly displays the unique characteristics of PSR J0348+0432 with a small $P_b$ and a large value of $P$.

During the LMXB phase, a pulsar is generally expected to accrete much more mass and angular momentum than needed to be spun-up to a few milliseconds (Tauris, Langer & Kramer 2012). In the same process, its B-field should have decayed significantly — typically to values $\leq 10^8$ G. However, for some reason the B-field ($\sim 2 \times 10^9$ G) remained relatively high in PSR J0348+0432. In contrast, the other known binary radio pulsars with similar values of $P_{\rm b} \leq 8$ hr and $M_{\rm WD} \approx 0.14 - 0.18 \, M_\odot$ (e.g. PSRs J0751+1807 and J1738+0333), besides from the many black-widow-like systems, have low B-fields and spin periods of a few milliseconds, as expected from current theories of LMXB evolution.

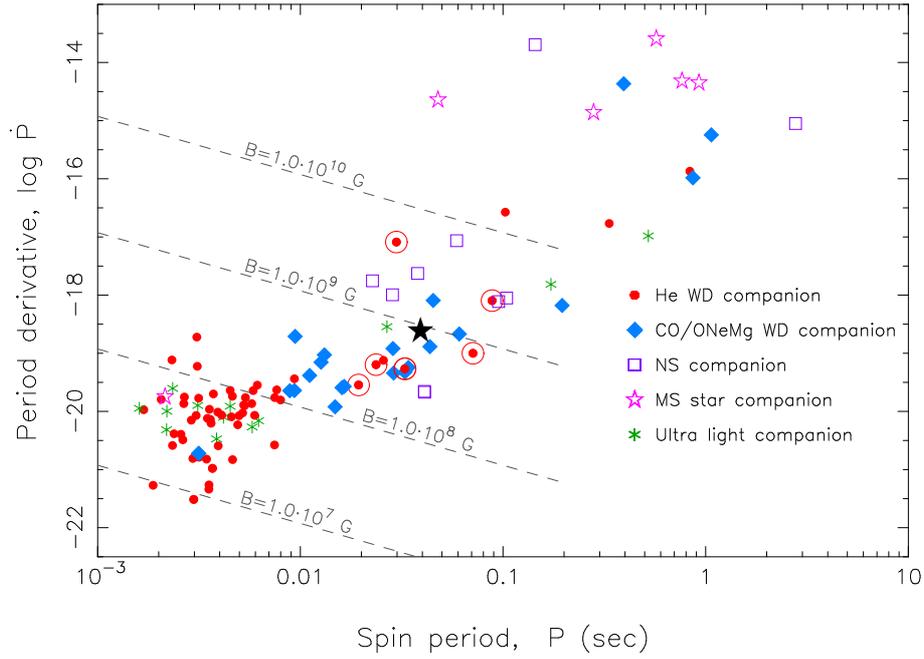

Figure 42: A $P\dot{P}$–diagram of the 111 known binary radio pulsars in the Galactic disk. The location of PSR J0348+0432 is marked with a black star in a region which is mainly dominated by slow spin and high B-field pulsars with massive white dwarf companions (marked with blue diamonds). The dashed lines of constant B-fields were calculated following Tauris, Langer & Kramer (2012) and assuming for simplicity $M_{PSR} = 1.4\ M_\odot$ and $\sin\alpha = \phi = \omega_c = 1$. All $\dot{P}$ values in this plot are intrinsic values obtained from kinematic corrections to the observed values. Data taken from the ATNF Pulsar Catalogue (Manchester et al. 2005), in Oct. 2012.

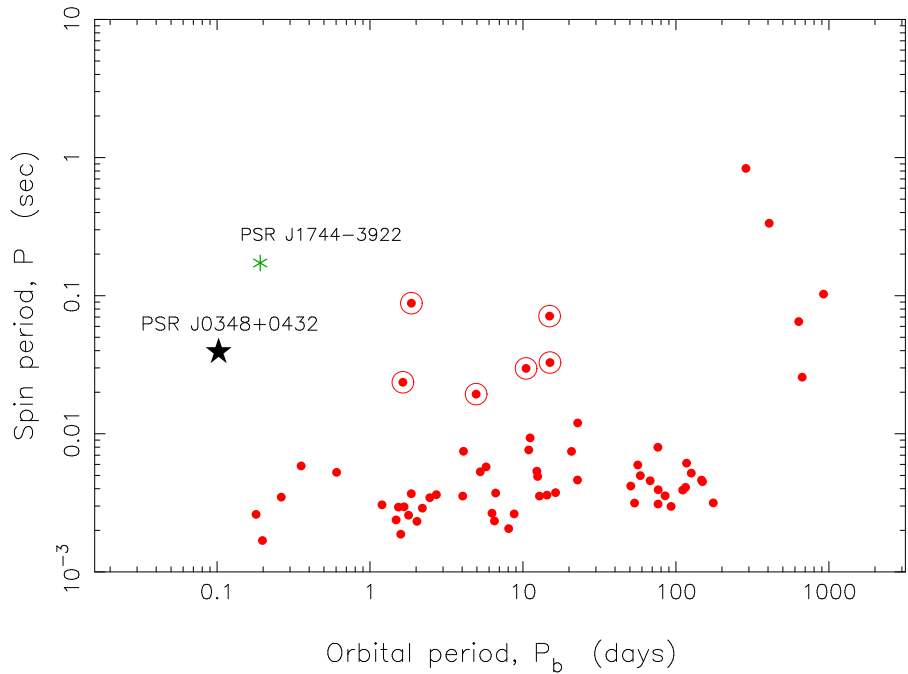

Figure 43: A $P_b - P$ (Corbet) diagram of the 63 known Galactic binary pulsars with a He white dwarf companion of mass $M_{WD} > 0.14\,M_\odot$. The unique location of PSR J0348+0432 is shown with a star. Another puzzling pulsar, PSR J1744−3922 (Breton et al. 2007) marked with a green asterisk, is included in this plot. These two pulsars seem to share high B-field properties with the 6 pulsars in circles. Data taken from the ATNF Pulsar Catalogue (Manchester et al. 2005), in Sep. 2012.

In Fig. 44 we have plotted the past and the future evolution of PSR J0348+0432. In the upper panel is seen the evolution of $P_{\rm b}$. In the lower panel is seen the spin evolution of the pulsar assuming different values of a constant braking index $2 \leq n \leq 5$. If the estimated cooling age of $\sim 2$ Gyr is correct (and adding to this value a pre-white dwarf contraction phase between RLO detachment and settling on the final cooling track, yielding an assumed total age of about $2 - 2.5$ Gyr) we can estimate that PSR J0348+0432 was recycled with an initial spin period of about $10 - 20$ ms. This relatively slow spin could be (partly) caused by enhanced braking of the spin rate, due to the high B-field of the pulsar, during the Roche-lobe decoupling phase when the progenitor of the white dwarf ceased its mass transfer (Tauris 2012). If the total post-LMXB age is $\sim 2.6$ Gyr then the pulsar could, at first sight, have been recycled with an initial spin period of 1 ms for $n \geq 3$. However, calculations of the pulsar spin-up line (Tauris, Langer & Kramer 2012) do not predict such a rapid spin for pulsars with high B-fields and which accreted with typical mass-accretion rates of $\dot{M} < 10^{-2} \, \dot{M}_{\rm Edd}$ (evident from both theoretical modeling of the LMXB RLO and observations of LMXB luminosities (Lewin & van der Klis 2006)).

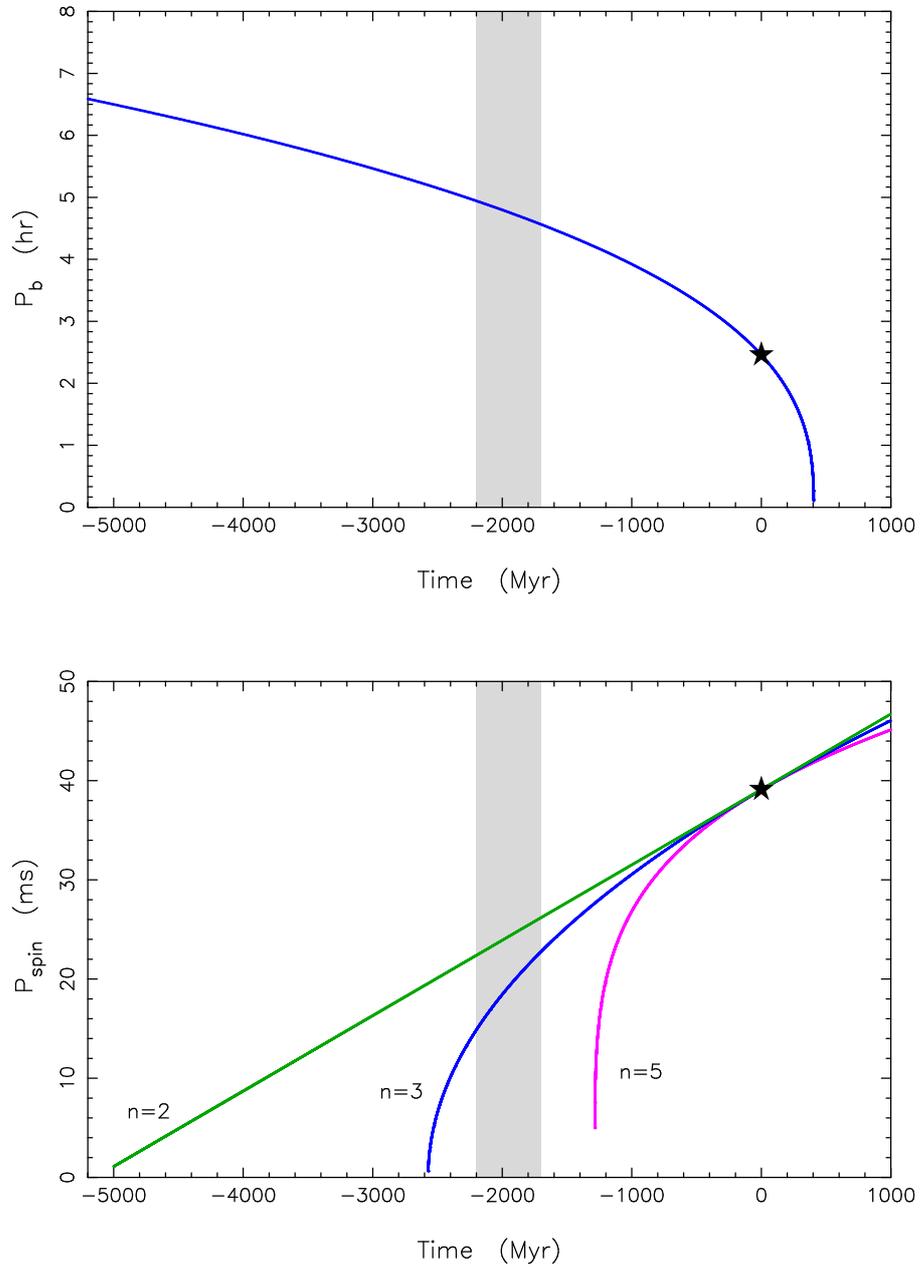

Figure 44: Orbital evolution (top) and spin evolution (bottom) of PSR J0348+0432 in the past and in the future. Different evolution tracks are plotted for different assumed values of a (constant) braking index, $n$. The grey shaded region marks the estimated white dwarf cooling age. The past is for a negative value of time, future is for a positive value of time. See text for a discussion.

# SPIN-UP OF MILLISECOND PULSARS

- **Chapter 5**
  Tauris (2012), Science 335, 561
  *Spin-Down of Radio Millisecond Pulsars at Genesis*

- **Chapter 6**
  Tauris, Langer & Kramer (2012), MNRAS 425, 1601
  *Formation of millisecond pulsars with CO white dwarf companions – II.*
  *Accretion, spin-up, true ages and comparison to MSPs with He white dwarf companions*

- **Chapter 7**
  Lazarus, Tauris, Knispel, Freire, et al. (2014), MNRAS 437, 1485
  *Timing of a young mildly recycled pulsar with a massive white dwarf companion*

## 5. Spin-Down of Radio Millisecond Pulsars at Genesis

**Tauris (2011)**

**Science 335, 561**

*Note, in the original paper there are no subsection headings.*

## Abstract

Millisecond pulsars are old neutron stars that have been spun up to high rotational frequencies via accretion of mass from a binary companion star. An important issue for understanding the physics of the early spin evolution of millisecond pulsars is the impact of the expanding magnetosphere during the terminal stages of the mass-transfer process. Here I report binary stellar evolution calculations that show that the braking torque acting on a neutron star, when the companion star decouples from its Roche-lobe, is able to dissipate > 50% of the rotational energy of the pulsar. This effect may explain the apparent difference in observed spin distributions between X-ray and radio millisecond pulsars and help account for the noticeable age discrepancy with their young white dwarf companions.

### 5.1 Introduction

Millisecond pulsars (MSPs) are rapidly spinning, strongly magnetized neutron stars. They form as the result of stellar cannibalism, where matter and angular momentum flow from a donor star to an accreting neutron star (Alpar et al. 1982; Bhattacharya & van den Heuvel 1991). During this process the system is detectable as an X-ray source (Bildsten et al. 1997). In some cases, X-ray pulsations reveal a fast spinning neutron star (Wijnands & van der Klis 1998); the 13 known accreting X-ray MSPs (AXMSPs) have an average spin period of $\langle P \rangle_{\rm AXMSP} = 3.3$ ms (Hessels 2008). These AXMSPs are thought to be the evolutionary progenitors of radio MSPs. When the donor star decouples from its Roche lobe[13] and the mass transfer generating the X-rays ceases, the radio emission is activated (Archibald et al. 2009). Until now more than 200 recycled radio MSPs have been detected in our galaxy, both in the field and in globular clusters, with spin periods between 1.4 and 20 ms. These MSPs, which are observed after the Roche-lobe decoupling phase (RLDP), have $\langle P \rangle_{\rm MSP} = 5.5$ ms (Hessels 2008) [see the supporting online material (SOM) in Section 5.6 for a critical discussion on selection effects and statistics]. It is unknown whether this spin difference is caused by the role of the RLDP or subsequent spin-down from magnetodipole radiation during the lifetime of the radio MSPs. The interplay between the magnetic field of a neutron star and the conducting plasma of the accreted material is known to provide the accretion torque necessary to spin up the pulsar (Lamb, Pethick & Pines 1973; Ghosh & Lamb 1992; Frank, King & Raine 2002). However, these interactions can also lead to a torque reversal under certain conditions (Illarionov & Sunyaev 1975), especially when the mass-transfer rate decreases (Ruderman, Shaham & Tavani 1989). Similarly, detailed studies of low-mass X-ray binaries (LMXBs) using stellar evolution codes have demonstrated (Webbink, Rappaport & Savonije 1983; Rappaport et al. 1995; Tauris & Savonije 1999; Podsiadlowski, Rappaport & Pfahl 2002) that these systems undergo a long, stable phase of mass transfer, which provides sufficient mass to spin up the accreting neutron star to spin periods of milliseconds. However, these previous studies did not combine such numerical stellar evolution calculations with computations of the resulting accretion torque at work. Here, I compute this torque during the termination of the mass-transfer phase by integrating these two methods, using the time-dependent mass-transfer rate to follow its effect on the pulsar spin rate.

---

[13]The Roche-lobe of a binary star is the innermost equipotential surface passing through the first Lagrangian point, L1. If a star fills its Roche-lobe, the unbalanced pressure at L1 will cause mass transfer to the other star (Bhattacharya & van den Heuvel 1991; and references therein).

*Thomas M. Tauris - Uni. Bonn*

## 5.2 Computations of the Roche-lobe decoupling phase (RLDP)

The computation of an LMXB donor star detaching from its Roche lobe is shown in Fig. 45 [see further details in the SOM (Section 5.6)]. The full length of the mass-transfer phase was about 1 billion years (Gyr); the RLDP happened in the last 200 million years (Myr) when the mass-transfer rate decreased rapidly. The original mass of the donor star was 1.0 $M_\odot$ and by the time it entered the final stage of the RLDP it had lost 99% of its envelope mass encapsulating a core of 0.24 $M_\odot$ (i.e. the mass of the hot white dwarf being formed after the RLDP). The orbital period at this stage was 5.1 days and the MSP had a mass of 1.53 $M_\odot$. By using the received mass-transfer rate, $\dot{M}(t)$ from my stellar models, the radius of the magnetospheric boundary of the pulsar, located at the inner edge of the accretion disk, can be written as $r_{mag} = \phi \cdot r_A$ [with $r_A$, the Alfvén radius[14], calculated (see SOM in Section 5.6) following standard prescriptions in the literature (Frank, King & Raine 2002; Shapiro & Teukolsky 1983)]. Knowledge of the relative location of $r_{mag}$, the corotation radius, $r_{co}$, and the light cylinder radius, $r_{lc}$, enabled a computation of the accretion torque acting on the pulsar (SOM, Fig. 47). After a long phase of mass transfer as an LMXB the pulsar was spinning at its equilibrium period when entering the RLDP. The rapid torque reversals (unresolved in the center panel of Fig. 45; see also SOM Fig. 47) originated from successive, small episodes of spin-up or spin-down depending on the relative location of $r_{co}$ and $r_{mag}$, which reflects the small fluctuations in $\dot{M}(t)$. However, this equilibrium was broken when $\dot{M}(t)$ decreased substantially on a short timescale. That resulted in $r_{mag}$ increasing on a timescale ($t_{RLDP}$) faster than the spin-relaxation timescale, $t_{torque}$, at which the torque would transmit the effect of deceleration to the neutron star and therefore $r_{mag} > r_{co}$. In this propeller phase (Illarionov & Sunyaev 1975), a centrifugal barrier arose and expelled material entering the magnetosphere whereby a braking torque acted to slow down the spin rate of the pulsar even further (see bottom panel in Fig. 45 and SOM Fig. 48). The spin-relaxation timescale is given by $t_{torque} = J/N$ which yields:

$$ t_{torque} \simeq 50 \text{ Myr} \cdot B_8^{-8/7} \left( \frac{\dot{M}}{0.1 \, \dot{M}_{Edd}} \right)^{-3/7} \left( \frac{M}{1.4 \, M_\odot} \right)^{17/7} \qquad (40) $$

where the spin angular momentum of the neutron star is $J = 2\pi I/P$ and the braking torque at the magnetospheric boundary is roughly given by $N \simeq \dot{M}\sqrt{GMr_{mag}}$ (SOM, Section 5.6). Here $G$ is the gravitational constant; $M$ and $I$ are the neutron star mass and moment of inertia, respectively; and $B_8$ is the surface magnetic flux density in units of $10^8$ G. In addition to this torque, the magnetic field drag on the disk (Rappaport, Fregeau & Spruit 2004) was included in the model, although this effect is usually less dominant. In these calculations, I assumed that the strength of the neutron star B-field had reached a constant, residual level before the propeller phase, a reasonable assumption given that less than 1% of the donor star envelope mass remained to be transferred. The propeller phase was terminated when $r_{mag} > r_{lc}$. At this point the MSP activated its radio emission and turned on a plasma wind which then inhibited any further accretion onto the neutron star (Ruderman, Shaham & Tavani 1989; Burderi et al. 2001). The duration of the RLDP in this example was $t_{RLDP} > 100$ Myr (including early stages before the propeller phase), which is a substantial fraction of the spin-relaxation time scale, $t_{torque} \approx 200$ Myr calculated by using Eq. 40 at the onset of the propeller phase. For this reason the RLDP has an important effect on the spin rate, $P_0$ and the characteristic age of the radio MSP at birth, $\tau_0$. In the case reported here, the radio MSP was born (recycled) with $P_0 \approx 5.4$ ms and an initial so-called characteristic age $\tau_0 \equiv P_0/2\dot{P}_0 \approx 15$ Gyr. The spin period before the RLDP was about 3.7 ms, implying that this MSP lost more than 50% of its rotational energy during the RLDP. If the pulsar had not broken its spin equilibrium during the RLDP, it would have been recycled with a slow spin period of $P_0 \approx 58$ ms and thus not become an MSP[15].

---

[14]The magnetospheric coupling parameter, $0.5 < \phi < 1.4$ is a numerical factor or order unity depending on the accretion flow, the disk model and the magnetic inclination angle of the pulsar (Ghosh & Lamb 1992; Frank, King & Raine 2002; Wang 1997; Lamb & Yu 2005; D'Angelo & Spruit 2010).

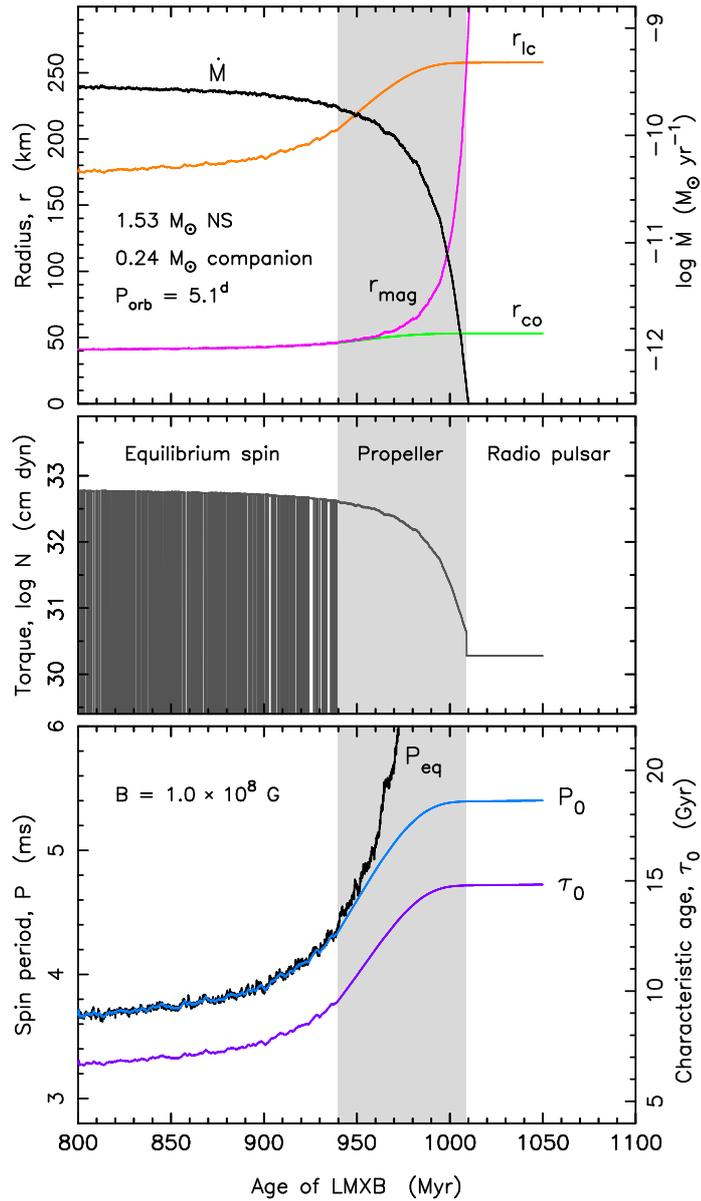

Figure 45: Final stages of mass transfer in an LMXB. The gradual decoupling of the donor star from its Roche lobe causes the received mass-transfer rate, $\dot{M}$ (i.e. the ram pressure of the inflowing material) to decrease whereby the magnetospheric boundary of the neutron star, $r_{mag}$, moves outward relative to the corotation radius, $r_{co}$, and the light cylinder radius, $r_{lc}$ (top). The alternating spin-up/spin-down torques during the equilibrium spin phase are replaced by a continuous spin-down torque in the propeller phase ($r_{co} < r_{mag} < r_{lc}$), until the pulsar activates its radio emission and the magnetodipole radiation remains as the sole braking mechanism, (center; see also SOM and Figs. 47 and 48). After an initial phase of spin-down, the spin equilibrium is broken which limits the loss of rotational energy of the pulsar and sets the value of the spin period, $P_0$, of the radio MSP at birth (bottom). Also shown in the bottom panel is the resulting characteristic age, $\tau_0$ of the recycled radio MSP, assuming a constant B-field of $10^8$ G during the RLDP. The gray shaded areas indicate the propeller phase.

---

[15]This "turn-off problem" has previously been debated elsewhere (Ruderman, Shaham & Tavani 1989; Lamb & Yu 2005; Ruderman et al. 1989).

### 5.3 Spin distributions of radio MSPs versus accreting MSPs

Radio pulsars emit magnetic dipole radiation as well as a plasma wind (Shapiro & Teukolsky 1983; Manchester & Taylor 1977). These effects cause rotational energy to be lost, and hence radio MSPs will slow down their spin rates with time after the RLDP. However, they cannot explain the apparent difference in spin distributions between AXMSPs and radio MSPs, because radio MSPs, which have weak surface magnetic field strengths, could not spin down by the required amount even in a Hubble time. The true age of a pulsar (Manchester & Taylor 1977) is given by $t = P/((n-1)\dot{P}) \left[1 - (P_0/P)^{n-1}\right]$. Assuming an evolution with a braking index $n = 3$ and $B = 1.0 \times 10^8$ G, the time scale $t$ is larger than 10 Gyr, using $P_0 = <P>_{\text{AXMSP}} = 3.3$ ms and $P(t) = <P>_{\text{MSP}} = 5.5$ ms. To make things worse, one has to add the main-sequence lifetime of the LMXB donor star, which is typically $3-12$ Gyr, thereby reaching unrealistic large total ages. Although the statistics of AXMSPs still has its basis on small numbers, and care must be taken for both detection biases (such as eclipsing effects of radio MSPs) and comparison between various subpopulations (SOM, Section 5.6), it is evident from both observations and theoretical work that the RLDP effect presented here plays an important role for the spin distribution of MSPs.

### 5.4 The characteristic age of recycled MSPs

The RLDP effect may also help explain a few other puzzles, for example, why characteristic (or spin-down) ages of radio MSPs often largely exceed cooling age determinations of their white dwarf companions (Lorimer et al. 1995). It has been suggested that standard cooling models of white dwarfs may not be correct (Alberts et al. 1996; Nelson, Dubeau & MacCannell 2004; van Kerkwijk et al. 2005), particularly for low-mass helium white dwarfs. These white dwarfs avoid hydrogen shell flashes at early stages and retain thick hydrogen envelopes at the bottom of which residual hydrogen burning can continue for several billion years after their formation, keeping the white dwarfs relatively hot ($\sim 10^4$ K) and thereby appearing much younger than they actually are. However, it is well known that the characteristic age is not a trustworthy measure of true age[16], and the RLDP effect exacerbates this discrepancy even further. In the model calculation presented in Fig. 45, it was assumed that $B = 1.0 \times 10^8$ G and $\phi = 1.0$. However, $P_0$ and $\tau_0$ depend strongly on both $B$ and $\phi$. This is shown in Fig. 46 where I have calculated the RLDP effect for different choices of $B$ and $\phi$ by using the same stellar donor model [i.e., same $\dot{M}(t)$ profile] as before. The use of other LMXB donor star masses, metallicities, and initial orbital periods would lead to other $\dot{M}(t)$ profiles (Tauris & Savonije 1999; Podsiadlowski, Rappaport & Pfahl 2002) and hence different evolutionary tracks. The conclusion is that recycled MSPs can basically be born with any characteristic age. Thus we are left with the cooling age of the white dwarf companion as the sole reliable, although still not accurate, measure as an age indicator.

### 5.5 Maximum spin rate of recycled MSPs

A final puzzle is why no sub-millisecond pulsars have been found among the 216 radio MSPs detected in total so far. Although modern observational techniques are sensitive enough to pick up sub-millisecond radio pulsations, the fastest spinning known radio MSP, J1748−2446ad (Hessels et al. 2006), has a spin frequency of only 716 Hz, corresponding to a spin period of 1.4 ms. This spin rate is far from the expected minimum equilibrium spin period (SOM, Section 5.6) and the physical mass shedding limit of about 1500 Hz. It has been suggested that gravitational wave radiation during the accretion phase halts the spin period above a certain level (Bildsten 1998; Chakrabarty et al. 2003). The RLDP effect presented here is a promising candidate for an alternative mechanism, in case a sub-ms AXMSP is detected (SOM, Section 5.6).

---

[16]This is the case e.g. if the pulsar spin period, $P$ is close to its initial spin period, $P_0$

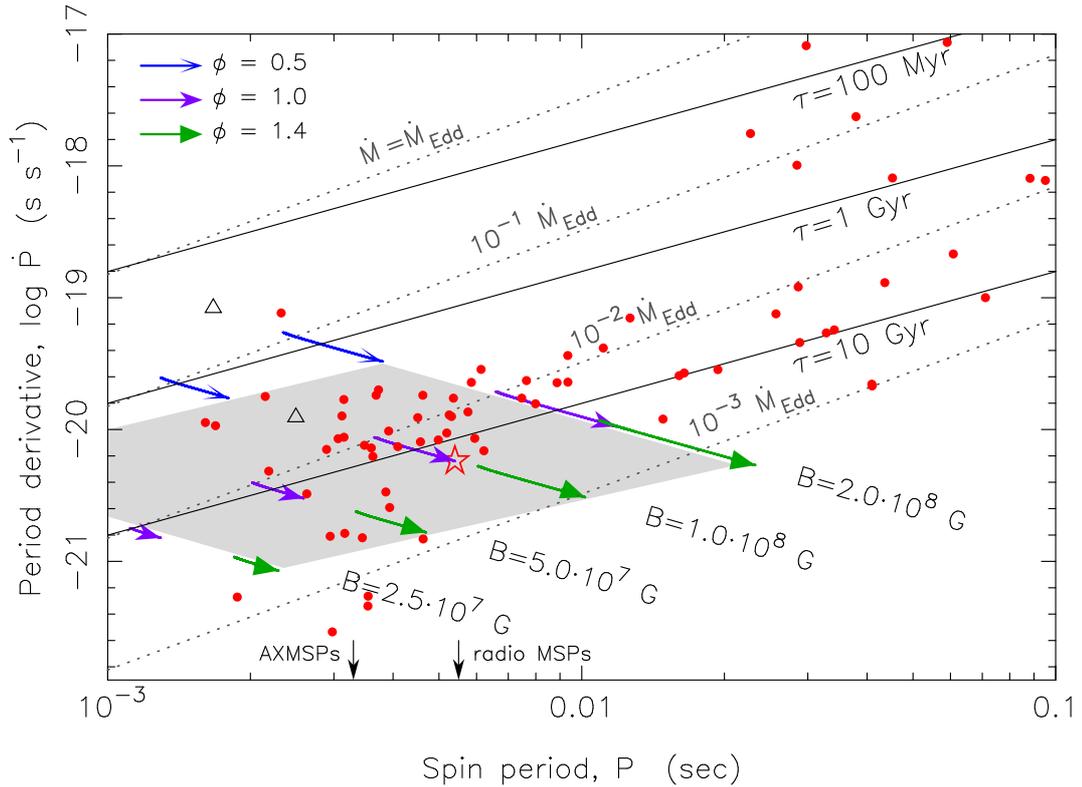

Figure 46: Evolutionary tracks during the Roche-lobe decoupling phase (RLDP). Computed tracks are shown as arrows in the $P\dot{P}$–diagram calculated by using different values of the neutron star B-field strength. The various types of arrows correspond to different values of the magnetospheric coupling parameter, $\phi$. The gray-shaded area indicates all possible birth locations of recycled MSPs calculated from one donor star model. The solid lines represent characteristic ages, $\tau$, and the dotted lines are spin-up lines calculated for a magnetic inclination angle, $\alpha = 90°$. The star indicates the radio MSP birth location for the case presented in Fig. 45. The two triangles indicate approximate locations (Patruno 2010b) of the AXMSPs SWIFT J1756.9−2508 (upper) and SAX 1808.4−3658 (lower). Observed MSPs in the Galactic field are shown as dots [data taken from the ATNF Pulsar Catalogue, December 2011]. All the measured $\dot{P}$ values are corrected for the Shklovskii effect, a kinematic projection effect that affects the apparent value of $\dot{P}$ for pulsars (Shklovskii 1970). If the transverse velocity of a given pulsar was unknown, I used a value of $72 \text{ km s}^{-1}$, the median value of the 44 measured MSP velocities. The average spin periods of AXMSPs and radio MSPs are indicated with arrows at the bottom of the diagram (see SOM, Section 5.6).

## 5.6 Supporting Online Material (SOM)

### 5.6.1 Numerical binary stellar evolution code

When modeling the evolution of an X-ray binary one should take into account a number of issues related to stellar evolution, for example, the stability of the mass-transfer process, possible spin-orbit couplings, accretion of material onto the neutron star and ejection of matter from the system. The major uncertainties are related to the specific orbital angular momentum of ejected matter – and for close systems also the treatment of spin-orbit couplings. I applied the Langer code (Braun 1997; Wellstein & Langer 1999) which is a binary stellar evolution code developed on the basis of a single star code (Langer 1998) and adapted to binary interactions. It is a 1-dimensional implicit Lagrangian code which solves the hydrodynamic form of the stellar structure and evolution equations (Kippenhahn & Weigert 1990). The evolution of two stars and, in case of mass transfer, the evolution of the mass-transfer rate and of the orbital separation are computed simultaneously through an implicit coupling scheme using standard prescriptions for the Roche-approximation (Eggleton 1983) and computation of the mass-transfer rate (Ritter 1988). The stellar models are computed using OPAL opacities and extended nuclear networks. An updated version of the code has been applied to low- and intermediate-mass X-ray binaries (Tauris, Langer & Kramer 2011). In close orbit LMXBs the mass-transfer rate is sub-Eddington at all times. To account for the relatively small masses of some radio pulsars descending from such systems (Tauris & Savonije 1999), I assumed that the received mass-transfer rate, $\dot{M} = 0.30\ \dot{M}_2$, where $\dot{M}_2$ is the RLO mass-transfer rate from the donor star. Assuming fully conservative RLO does not diminish the RLDP effect discussed here. In the presented model I assumed an initial neutron star mass of 1.30 $M_\odot$ and an orbital period of 0.82 days at the onset of the LMXB phase.

### 5.6.2 Magnetospheric boundary and resultant accretion torque

The light cylinder radius is defined by: $r_{\rm lc} \equiv c/\Omega$, where $\Omega = 2\pi/P$ and $c$ is the speed of light in vacuum. This radius marks the outer boundary of closed B-field lines (perpendicular to the spin axis) of the neutron star. The Alfvén radius is roughly a measure of the boundary of the pulsar magnetosphere, inside of which the flow pattern of accreted plasma is largely dictated by the B-field corotating with the neutron star. It is found by equating the magnetic energy density ($B^2/8\pi$) to the ram pressure of the spherically accreted matter and is approximately given by (Davidson & Ostriker 1973): $r_A = B^{4/7} R^{12/7} \left(\dot{M}\sqrt{2GM}\right)^{-2/7}$, where $R$ and $M$ are the radius and mass of the neutron star, respectively, and $G$ is the gravitational constant. The plasma flow onto the neutron star is not spherical, but instead forms an accretion disk where excess angular momentum is transported outwards by turbulent-enhanced viscous stresses (Ghosh & Lamb 1992; Frank, King & Raine 2002; Shapiro & Teukolsky 1983; Dubus et al. 1999). The location of the inner edge of the disk, i.e. the coupling radius of the magnetosphere, is then estimated as: $r_{\rm mag} = \phi \cdot r_A$, where (cf. Ghosh & Lamb 1992; Frank, King & Raine 2002; Wang 1997; Lamb & Yu 2005; D'Angelo & Spruit 2010). The corotation radius, $r_{\rm co} = \left(GM/\Omega^2\right)^{1/3}$ is defined as the radial distance at which the Keplerian angular frequency is equal to the angular spin frequency of the neutron star. Using binary stellar models a mass-transfer profile, $\dot{M}(t)$ is obtained, and which is needed, in combination with the relative locations of $r_{\rm co}$, $r_{\rm mag}$ and $r_{\rm lc}$, to calculate the resulting accretion torque as a function of time, given by:

$$N(t) = n(\omega) \left[\dot{M}(t)\sqrt{GM\,r_{\rm mag}(t)}\,\xi + \frac{\mu^2}{9\,r_{\rm mag}^3(t)}\right] - \frac{\dot{E}_{\rm dipole}(t)}{\Omega(t)} \qquad (41)$$

Here the second term in the parenthesis on the right-hand side is the drag from the accretion disk interaction with the B-field of the neutron star (Rappaport, Fregeau & Spruit 2004) and $\dot{E}_{\rm dipole} = (-2/3c^3)|\ddot{\mu}|^2$ is the loss of rotational energy due to emission of magnetodipole waves. Following previous work (Ghosh & Lamb 1992; Spruit & Taam 1993; D'Angelo & Spruit 2010) a dimensionless quantity, here defined by: $n(\omega) = \tanh\left(\frac{1-\omega}{\delta_\omega}\right)$, is introduced to model a gradual torque change in a transition zone near the magnetospheric boundary. It is typically a good assumption that the width of this zone is small (Spruit & Taam 1993), i.e. $\delta_\omega \ll 1$, corresponding to a step function $n(\omega) = \pm 1$ (Fig. 47). In case the X-ray pulsar is a fast rotator ($\Omega_{\rm PSR} > \Omega_{\rm K}(r_{\rm mag})$), yielding a fastness parameter $\omega > 1$, which corresponds to $r_{\rm mag} > r_{\rm co}$) a centrifugal barrier arises causing the magnetosphere to act like a propeller. The acceleration of the incoming gas to escape speed is accomplished by the "magnetic slingshot" mechanism (Blandford & Payne 1982), causing the rotational energy of the neutron star to be lost. It has been noted (Rappaport, Fregeau & Spruit 2004; D'Angelo & Spruit 2010; Spruit & Taam 1993), however, that for accretion at a fastness parameter near 1, the velocity excess of neutron star rotation over the Keplerian velocity may not be energetically sufficient to gravitationally unbind all accreting matter, and therefore the expelled gas will return to the disk and build up in mass until the magnetosphere is pushed inward and forces accretion. However, it is important to note that even if the accretion flow is not continuous in nature, in a relatively narrow interval of accretion near a fastness parameter of 1, this suggested cyclic behaviour disappears shortly after the onset of the propeller phase when the magnetosphere moves outward rapidly. Furthermore, recent work by D'Angelo & Spruit (2010) shows that cyclic accretion is supposed to confine a considerable amount of mass to the disk and even increase the braking torque, exacerbating the loss of rotational energy of the pulsar during the Roche-lobe decoupling phase (RLDP).

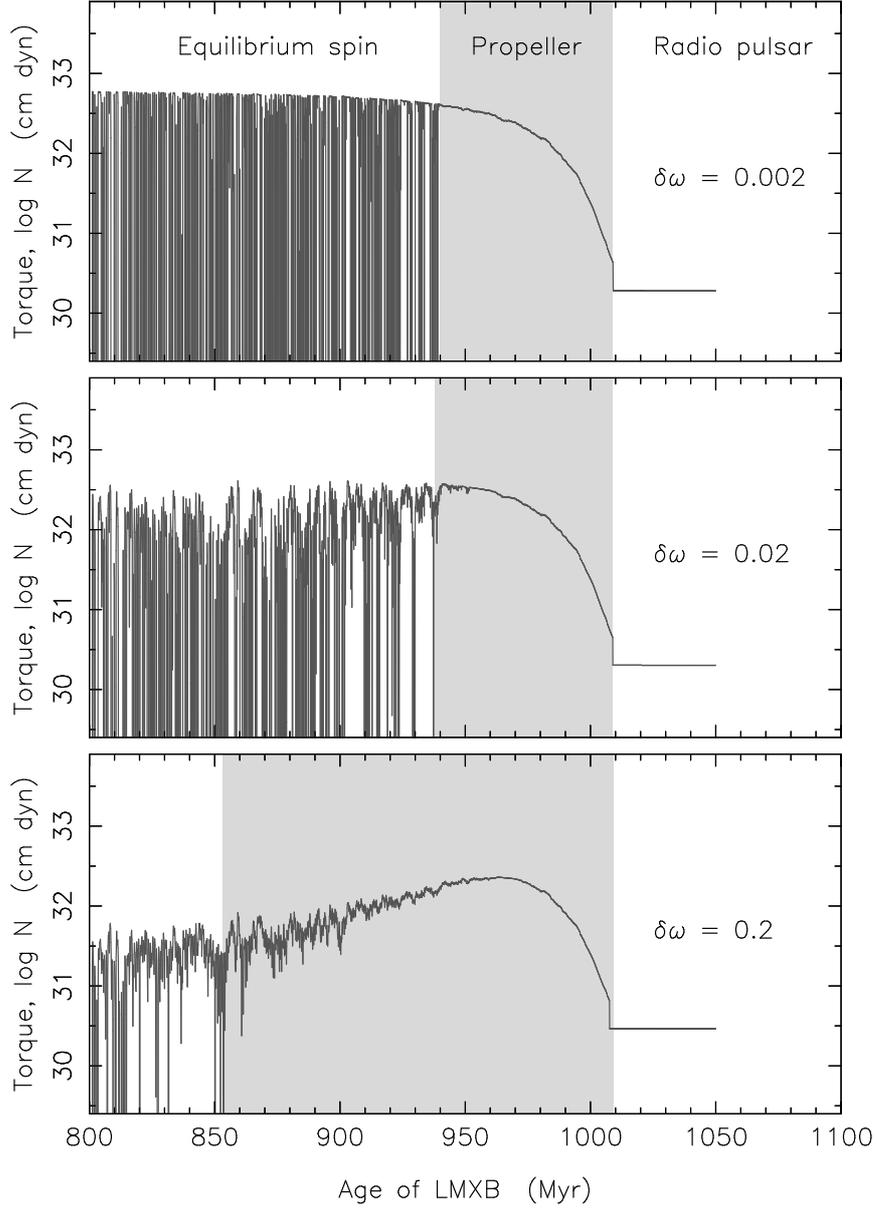

Figure 47: Braking torque at RLDP. The resultant braking torque acting on the pulsar during the Roche-lobe decoupling phase is calculated for different values of the transition zone parameter, $\delta_\omega = 0.002, 0.02$ and $0.2$ (top to bottom) and $\xi = 1$. The post-RLDP spin period, $P_0$ is computed to be: 5.4 ms, 5.3 ms and 4.7 ms, respectively. The pre-RLDP spin period was $P \approx 3.7$ ms. Hence, even at (unrealistic) wide transition zones the RLDP effect is still important. A small value $\delta_\omega \ll 1$ corresponds to the switch mode transition $n(\omega) = \pm 1$ used in Fig. 45.

### 5.6.3 Broken spin equilibrium

The small fluctuations in $\dot{M}(t)$ are reflected in the location of $r_{\text{mag}}$ which then causes rapid oscillations in the sign of the accretion torque, i.e. alternate episodes of spin-up or spin-down. This causes the pulsar spin period to fluctuate about the average value of $P_{\text{eq}}$ at any time as already predicted in earlier work (Elsner, Ghosh & Lamb 1980). This behaviour is demonstrated in Fig. 48 (which is a zoom-in of the bottom panel of Fig. 45 in the main paper). It is therefore expected that among the population of AXMSPs some systems will be detected with positive (spin-up) torques while others will have a negative (braking) torque acting. The torque reversal timescale is difficult to calculate accurately because of both numerical computational issues and unknown details of the accretion disk physics (Frank, King & Raine 2002; Dubus et al. 1999). The time steps during the calculated RLDP were less than $10^5$ yr and thus much smaller than $t_{\rm RLDP}$ (100 Myr). However, it should be noted that the torque reversals described here, which are related to fluctuations in $\dot{M}(t)$, appear on a much longer timescale compared to the frequent torque reversals observed in many X-ray binaries (Bildsten et al. 1997). These torque reversals occur on a timescale of a few years, or decades, and are possibly related to warped disks, disk-magnetosphere instabilities, X-ray irradiation effects of the donor star or transitions in the nature of the disk flow. From Fig. 48 it is clear that the spin of the neutron star is able to remain in equilibrium during the initial stages of the RLDP. However, the spin equilibrium is broken at some point when $\dot{M}(t)$ decreases too quickly for the torque to transmit the deceleration to the spinning neutron star (see arrow). From this point onwards, $r_{\rm co}$ cannot keep up with the rapidly increasing value of $r_{\rm mag}$. The result of the computation presented here is that more than 50% of the rotational energy ($E_{\rm rot} = 1/2 \, I\Omega^2$) of the pulsar is lost by the time the radio emission turns on. Other $\dot{M}(t)$ profiles would lead to different results, depending on the RLDP timescale relative to the spin-relaxation timescale in any given case.

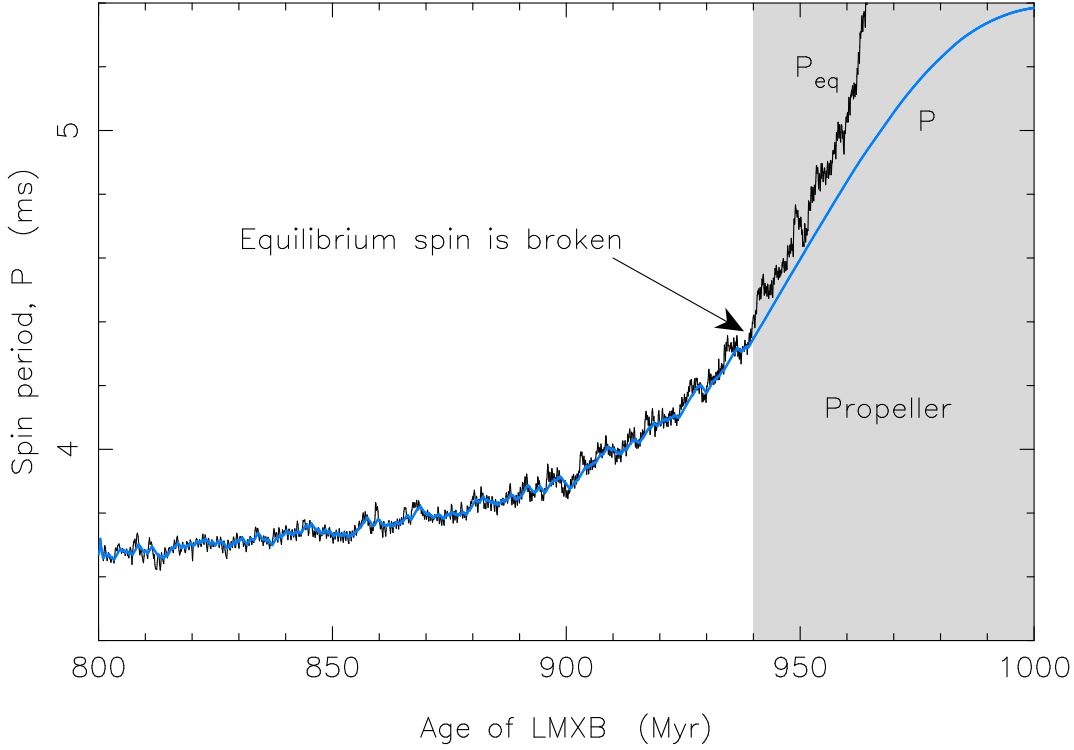

Figure 48: Transition from equilibrium spin to propeller phase. At early stages of the Roche-lobe decoupling phase (RLDP) the neutron star spin is able to remain in equilibrium despite the outward moving magnetospheric boundary caused by decreasing ram pressure. However, at a certain point (indicated by the arrow), when the mass-transfer rate decreases rapidly, the torque can no longer transmit the deceleration fast enough for the neutron star to remain in equilibrium. This point marks the onset of the propeller phase where $r_{\rm mag} > r_{\rm co}$ (and $P < P_{\rm eq}$) at all times. This plot is a zoom-in of the bottom panel of Fig. 45 in the main paper.

### 5.6.4 The minimum spin period

The fastest spin a recycled pulsar can obtain is the minimum equilibrium period (Bhattacharya & van den Heuvel 1991):

$$P_{\min} \simeq 0.71 \text{ ms} \cdot B_8^{6/7} \left( \frac{\dot{M}}{0.1 \, \dot{M}_{\text{Edd}}} \right)^{-3/7} \left( \frac{M}{1.4 \, M_\odot} \right)^{-5/7} R_{10}^{18/7} \tag{42}$$

where $R_{10}$ is the neutron star radius in units of 10 km and $\dot{M}_{\text{Edd}}$ is the Eddington accretion limit (of the order $3.0 \times 10^{-8} \, M_\odot \, \text{yr}^{-1}$ for spherical accretion, depending on the hydrogen mass fraction of the transferred material). Hence, in principle sub-millisecond pulsars should be able to form. There are, however, selection effects against detecting sub-ms AXMSPs. The reason is that these objects should preferentially be neutron stars with intrinsic weak B-fields (alternatively, their field could be temporarily buried in the crust at this stage (Cumming, Zweibel & Bildsten 2001)) and hence there may not be any observable modulation in the X-ray flux. The burst source XTE J1739−285 has been announced (Kaaret et al. 2007) to have a neutron star spin frequency of 1122 Hz, corresponding to a spin period of 0.89 ms. However, this result is still not confirmed. There are a number of suggestions – mainly related to gravitational wave radiation (Bildsten 1998; Chakrabarty et al. 2003) – why the spin rate of accreting neutron stars seems to be halted at a certain frequency cutoff near 730 Hz (cf. recent discussions in Chakrabarty 2008; Ho, Maccarone & Andersson 2011). If future observations reveal sub-ms AXMSPs, but no sub-ms radio MSPs, the RLDP effect suggested here could very well be responsible for such a division. Doppler smearing of radio pulsations in tight binary orbits which could cause a selection bias against detection of sub-ms radio pulsars is less serious in present day acceleration searches (at least for dispersion measures, $DM < 100$). The absence of AXMSPs with high spin frequencies is discussed by Lamb & Yu (2005) who make two statements: i) there is no clear evidence for pulsars to have sufficiently low B-fields and large enough mass quadrupoles that gravitational wave emission will limit their spin-up, and ii) the absence of AXMSPs with frequencies $> 750$ Hz may instead be caused by a combination of their B-field strength and mass-accretion rate, which leads to their magnetospheric boundary (at spin equilibrium) being located at a large distance, thus limiting $P_{\text{eq}}$. To settle the question one must perform detailed modeling of, for example, ultracompact X-ray binaries to probe the RLDP timescale relative to the spin-relaxation timescale. Such computations must include assumptions about the equation of state for (semi)degenerate donors (to calculate real time mass-transfer rates), irradiation effects (which are more important in such tight binaries with bloated donors), accretion disk instabilities and magnetic braking laws – a task beyond the scope of this work.

### 5.6.5 On the spin distributions and the (sub)populations of AXMSPs and radio MSPs

When comparing radio MSPs and AXMSPs one should be aware of differences in their binary properties and observational biases. Using data (ATNF Pulsar Catalogue) from the 61 binary Galactic disk pulsars with $P < 20$ ms (Hessels 2008), I find $\langle P \rangle_{\text{MSP}} = 5.42$ ms with a standard deviation, $\sigma = 3.88$ ms. For the now 14 known AXMSPs (Patruno 2010b; Papitto et al. 2011), I find $\langle P \rangle_{\text{AXMSP}} = 3.28$ ms with $\sigma = 1.43$ ms. The difference between $\langle P \rangle_{\text{MSP}} = 5.42$ ms and $\langle P \rangle_{\text{AXMSP}} = 3.28$ ms is statistically significant at the 99.9% confidence level using a t-test analysis (assuming normally distributed periods). However, when considering the spin distributions, it is important to keep in mind that these two populations are not trivial to compare, for two main reasons: i) observational biases, and ii) selection criteria for sources with different physical properties.

i) As an example, there is an observational bias against detecting the fastest rotating AXM-SPs since these pulsars, in general, have the lowest B-fields ($B < 10^8$ G) according to the spin-up theory (Bhattacharya & van den Heuvel 1991). As already mentioned, such weak B-field pulsars may not be able to channel the flow of accreted material – a necessary condition for detecting the modulations in the X-ray signal. Similarly, there is a bias against detecting fast spinning radio MSPs in tight binaries since these pulsars are likely to evaporate their companion star in the process and which may result in eclipses of their radio signal (for the two best studied cases PSR B1957+20 and PSR J2051−0827, however, this is only at the 10% level). One could also consider the removal from the AXMSP sample those sources with orbital periods shorter than a few hours since these systems often lead to the formation of isolated MSPs.

ii) Each population can be divided further into subpopulations depending on orbital period, B-field of the pulsar and nature of the companion star (e.g. ultra-light substellar (brown) dwarf, nuclear burning dwarf and, for radio MSPs, helium or carbon-oxygen white dwarf). Similarly, the AXMSPs can be divided into accretion powered and nuclear powered pulsars (persistent and transient burst sources, respectively). If the nuclear powered pulsars are added to the sample of accretion powered AXMSPs (Patruno 2010b) the result is $\langle P \rangle_{\mathrm{AXMSP}} = 2.86$ ms and $\sigma = 1.30$ ms for all 23 AXMSPs. Furthermore, knowledge of the AXMSP B-fields (or spin frequency derivatives) would reveal their location in the $P\dot{P}$–diagram and thus help disentangle the complex evolutionary links between the various subpopulations of AXMSPs and radio MSPs. Unfortunately, until now only two AXMSPs (Patruno 2010b) have their B-field estimated (see their location in Fig. 46). Although many radio MSPs are found with orbital periods up to a few hundred days, so far we have no samples of AXMSPs with orbital periods above one day (which is a puzzle in itself).

In light of the above discussion one could, for example, consider the subpopulation of the 38 Galactic disk binary radio MSPs which have orbital periods less than 30 days, $P < 15$ ms, and which do not have carbon-oxygen white dwarf companions, and compare these to the 7 accretion powered AXMSPs with orbital periods above 2 hours (most of which have minimum companion masses $> 0.08\ M_{\odot}$). In this case one finds $\langle P \rangle_{\mathrm{MSP}} = 4.11$ ms ($\sigma = 2.15$ ms) and $\langle P \rangle_{\mathrm{AXMSP}} = 2.57$ ms ($\sigma = 0.83$ ms). The difference between these average spin periods is significant at the 99.7% significance level. In any case, we are indeed still dealing with small number statistics and in particular when taking the subdivisions of the populations into account. Observationally, the RLDP-effect hypothesis presented here can be verified (to become an accepted theory), or falsified, when future surveys discover a significant number of AXMSPs and radio MSPs confined to an interval with similar orbital periods and companion star masses.

# 6. Formation of Millisecond Pulsars with CO White Dwarf Companions – II. Accretion, Spin-up, True Ages and Comparison to MSPs with He White Dwarf Companions

## Abstract

Millisecond pulsars are mainly characterised by their spin periods, B-fields and masses – quantities which are largely affected by previous interactions with a companion star in a binary system. In this paper, we investigate the formation mechanism of millisecond pulsars by considering the pulsar recycling process in both intermediate-mass X-ray binaries (IMXBs) and low-mass X-ray binaries (LMXBs). The IMXBs mainly lead to the formation of binary millisecond pulsars with a massive carbon-oxygen (CO) or an oxygen-neon-magnesium white dwarf (ONeMg WD) companion, whereas the LMXBs form recycled pulsars with a helium white dwarf (He WD) companion. We discuss the accretion physics leading to the spin-up line in the $P\dot{P}$–diagram and demonstrate that such a line cannot be uniquely defined. We derive a simple expression for the amount of accreted mass needed for any given pulsar to achieve its equilibrium spin and apply this to explain the observed differences of the spin distributions of recycled pulsars with different types of companions. From numerical calculations we present further evidence for significant loss of rotational energy in accreting X-ray millisecond pulsars in LMXBs during the Roche-lobe decoupling phase (Tauris 2012) and demonstrate that the same effect is negligible in IMXBs. We examine the recycling of pulsars with CO WD companions via Case BB Roche-lobe overflow (RLO) of naked helium stars in post common envelope binaries. We find that such pulsars typically accrete of the order $0.002 - 0.007\,M_\odot$ which is just about sufficient to explain their observed spin periods. We introduce isochrones of radio millisecond pulsars in the $P\dot{P}$–diagram to follow their spin evolution and discuss their true ages from comparison with observations. Finally, we apply our results of the spin-up process to complement our investigation of the massive pulsar PSR J1614−2230 from Paper I (Tauris, Langer & Kramer, 2011), confirming that this system formed via stable Case A RLO in an IMXB and enabling us to put new constraints on the birth masses of a number of recycled pulsars.

## 6.1 Introduction

Binary millisecond pulsars (BMSPs) represent the advanced phase of stellar evolution in close, interacting binaries. Their observed orbital and stellar properties are fossil records of their evolutionary history. Thus one can use binary pulsar systems as key probes of stellar astrophysics. It is well established that the neutron star component in binary millisecond pulsar systems forms first, descending from the initially more massive of the two binary zero-age main sequence (ZAMS) stars. The neutron star is subsequently spun up to a high spin frequency via accretion of mass and angular momentum once the secondary star evolves (Alpar et al. 1982; Radhakrishnan & Srinivasan 1982; Bhattacharya & van den Heuvel 1991). In this recycling phase the system is observable as a low-mass X-ray binary (e.g. Hayakawa 1985; Nagase 1989; Bildsten et al. 1997) and towards the end of this phase as an X-ray millisecond pulsar (Wijnands & van der Klis 1998; Archibald et al. 2009; Patruno & Watts 2012). Although this standard formation scenario is commonly accepted many aspects of the mass-transfer process and the accretion physics are still not understood in detail (Lewin & van der Klis 2006). Examples of such ambiguities include the accretion disk structure, the disk-magnetosphere transition zone, the accretion efficiency, the decay of the surface B-field of the neutron star and the outcome of common envelope evolution.

The current literature on the recycling process of pulsars is based on a somewhat simplified treatment of both the accretion process and the so-called equilibrium spin theory, as well as the application of a pure vacuum magnetic dipole model for estimating the surface B-field strength of a radio pulsar. These simplifications become a problem when trying to probe the formation and the evolution of observed recycled radio pulsars located near the classical spin-up line for Eddington accretion in the $P\dot{P}$–diagram (e.g. Freire et al. 2011a). In this paper we discuss the concept and the location of the spin-up line and investigate to which extent it depends on the assumed details of the accretion disk/magnetosphere interactions and the magnetic inclination angle of the pulsar. Furthermore, we include the plasma filled magnetosphere description by Spitkovsky (2006) for determining the surface B-field strengths of pulsars in application to spin-up theory and compare to the case of using the standard vacuum dipole model.

A key quantity to understand the formation of any given recycled radio pulsar is its spin period which is strongly related to the amount of mass accreted. The amount of accumulated mass is again determined by the mass-transfer timescale of the progenitor X-ray binary (and partly by the mode of the mass transfer and mass loss from the system), and hence it depends crucially on the nature of the donor star. At present the scenario is qualitatively well understood where low-mass X-ray binaries (LMXBs) and intermediate-mass X-ray binaries (IMXBs) in general lead to the formation of pulsars with a helium white dwarf (He WD) or a carbon-oxygen white dwarf (CO WD) companion, respectively (e.g. Tauris, van den Heuvel & Savonije 2000; Podsiadlowski, Rappaport & Pfahl 2002; Deloye 2008; Tauris 2011). However, here we aim to quantify this picture in more detail by re-analysing the spin-up process. As an example, we investigate in this paper the possibilities of spinning up a neutron star from Case BB RLO leading to a mildly recycled pulsar with a CO WD companion. We also discuss the increasing number of systems which apparently do not fit the characteristics of these two main populations of binary pulsars.

We follow the pulsar spin evolution during the final phase of the mass transfer by including calculations of the torque acting on an accreting pulsar during the Roche-lobe decoupling phase (Tauris 2012) for both an LMXB and an IMXB and compare the results. To complete the full history of pulsar spin evolution we subsequently consider the true ages of recycled radio pulsars (which are intimately related to their spin evolution) by calculating isochrones and discussing the distribution of pulsars in the $P\dot{P}$–diagram.

Although accreting X-ray millisecond pulsars (AXMSPs) are believed to be progenitors of BMSPs, all of the 14 observed AXMSPs have orbital periods less than one day whereas fully recycled radio BMSPs are observed with orbital periods all the way up to a few hundred days (see also Tauris 2012). This puzzle will be investigated in a future paper. For the evolution of ultra-compact X-ray binaries (UCXBs) leading to AXMSPs, and later to the formation of tight-orbit recycled radio BMSPs with ultra-light ($< 0.08\,M_\odot$) companions, we refer to Podsiadlowski, Rappaport & Pfahl (2002); Deloye & Bildsten (2003); van der Sluys, Verbunt & Pols (2005); van Haaften et al. (2012b). In this paper we focus on the formation and evolution of BMSPs with regular He WDs and CO/ONeMg WDs.

The discovery of PSR J1614−2230 (Hessels et al. 2005; Demorest et al. 2010) plays an important role for understanding BMSP formation. This pulsar system is interesting since it has a unique combination of a short pulsar spin period of 3.2 ms and a massive CO WD companion. A rapidly spinning pulsar is usually associated with a long phase of mass-transfer evolution in an LMXB, whereas a CO WD companion in a relatively close orbit (8.7 days) is evidently the outcome of an IMXB evolution. A possibly solution to this paradox is that PSR J1614−2230 evolved from Case A RLO of an IMXB which results in both a relatively long-lived X-ray phase ($> 10^7\,\text{yr}$), needed to spin up the pulsar effectively, and leaving behind a CO WD. In this case PSR J1614−2230 is the first BMSP known to have evolved via this path. Indeed, in Tauris, Langer & Kramer (2011), hereafter Paper I, we investigated the progenitor evolution of PSR J1614−2230 with emphasis on the X-ray phase where the binary contained a neutron star and a donor star. We found two viable possibilities for the progenitor of the PSR J1614−2230 system: either it contained a $2.2 − 2.6\,M_\odot$ asymptotic giant branch donor star and evolved through a common envelope and spiral-in phase initiated by Case C RLO, or it descended from a close binary system with a $4.0 − 5.0\,M_\odot$ main sequence donor star via Case A RLO as hinted above. The latter scenario was also found by Lin et al. (2011). The fact that PSR J1614−2230 was spun-up to (less than) 3.2 ms could indeed hint which one of the two formation scenarios is most likely. In order to test this idea and to further distinguish between Case A and Case C we turn our attention, here in Paper II, to the spin dynamics of this pulsar in the recycling process.

As discussed in Paper I, the distribution of neutron star birth masses is an important probe of both stellar evolution, binary interactions and explosion physics. For a number of BMSPs we are now able to put constraints on the birth mass of the pulsar given the derived amount of mass needed to spin up the observed recycled pulsar. In particular, it is of interest to see if we can identify further pulsars showing evidence of being born massive ($\sim 1.7 \, M_\odot$) like PSR J1614$-$2230.

In order to understand the many different observational properties of BMSPs we have combined a variety of subtopics here with the aim of presenting a full picture of the subject. Given the many facets included, our new findings and the resulting length of the manuscript, we have chosen throughout this paper to finalize individual subtopics with a direct comparison to observational data followed by discussions in view of our theoretical modelling.

Our paper is structured in the follow way: We begin with a brief, updated review of the formation channels of BMSPs with CO WDs (Section 6.2) and present a summary of the latest observational data in Section 6.3. In this section we also demonstrate an emerging unified picture of pulsar formation history which, however, is challenged by a number of interesting systems which share expected properties of both an LMXB and an IMXB origin. In Section 6.4 we investigate the recycling process in general with a focus on the location of the spin-up line in the $P\dot{P}$–diagram and also relate the initial spin of a rejuvenated pulsar to the amount of mass accreted. The theoretical modelling is continued in Section 6.5 where we highlight the effects of the Roche-lobe decoupling phase on the spin evolution of recycled pulsars. In Section 6.6 our results are compared to the observed spin period distributions and in Section 6.7 we investigate if BMSPs with CO WD companions obtained their fast spin periods *after* a common envelope evolution. In Section 6.8 we discuss our results in a broader context in relation to the spin evolution and the true ages of millisecond radio pulsars. In Section 6.9 we continue our discussion from Paper I on the formation and evolution of PSR J1614$-$2230 and in Section 6.10 we return to the question of neutron star birth masses. Our conclusions are summarized in Section 6.11. Finally, in the Appendix (Section 6.12) we present a new tool to identify the most likely nature of the companion star to any observed binary pulsar.

## 6.2 Formation of BMSPs with CO WD companions

According to stellar evolution theory one should expect radio pulsars to exist in binaries with a variety of different companions: white dwarfs (He WDs, CO WDs and ONeMg WDs), neutron stars, black holes, ultra light semi-degenerate dwarfs (i.e. substellar companions or even planets), helium stars, main sequence stars and, for very wide systems, sub-giant and giant stars (Bhattacharya & van den Heuvel 1991; Tauris & van den Heuvel 2006; and references therein). Of these possibilities, pulsars orbiting black holes, helium stars and giant stars still remain to be detected.

The majority of radio BMSPs have He WD companions. The formation of these systems is mainly channeled through LMXBs and have been well investigated in previous studies (e.g. Webbink, Rappaport & Savonije 1983; Pylyser & Savonije 1988; 1989; Rappaport et al. 1995; Ergma, Sarna & Antipova 1998; Tauris & Savonije 1999; Podsiadlowski, Rappaport & Pfahl 2002; Nelson, Dubeau & MacCannell 2004; van der Sluys, Verbunt & Pols 2005; Deloye 2008). These systems have orbital periods between less than 0.2 days and up to several hundred days. One of the most striking features of these systems is the relation between white dwarf masses and orbital periods (e.g. Tauris & Savonije 1999). However, BMSPs with He WD companions may also form in IMXBs if Case A RLO is not initiated too late during the main sequence evolution of the donor star (Tauris, van den Heuvel & Savonije 2000; Podsiadlowski, Rappaport & Pfahl 2002). In Section 6.3 we identify 6 BMSPs which may have formed via this channel.

BMSP systems with relatively heavy WDs (CO or ONeMg WDs) all have observed orbital periods $\leq 40$ days. For these systems there are a number of suggested formation channels which are briefly summarized below – see Paper I, and references therein, for a more rigorous discussion. In order to leave behind a CO WD, its progenitor must be more massive than about $3.5\,M_\odot$ if Roche-lobe overflow (RLO) is initiated while the donor star is still on the main sequence. If the donor star is an asymptotic giant branch (AGB) star when initiating mass transfer it can, for example, have a mass as low as $2.2\,M_\odot$ and still leave behind a $0.5\,M_\odot$ CO WD remnant – see fig. 1 in Paper I. CO WDs in close binaries have masses between $0.33 - 1.0\,M_\odot$, whereas the ONeMg WDs are slightly heavier with masses of $1.1 - 1.3\,M_\odot$. The upper limit for the initial mass of donor stars in X-ray binaries leaving an ONeMg WD is often assumed to be about $7 - 8\,M_\odot$ (Podsiadlowski et al. 2004), depending on the orbital period, metallicity, treatment of convection and the amount of convective overshooting. However, Wellstein & Langer (1999) demonstrated that even stars with an initial mass in excess of $13\,M_\odot$ may form an ONeMg WD if these progenitor stars are in a tight binary and thus lose their envelope at an early stage via Case A RLO. Stars exceeding the upper threshold mass for producing an ONeMg WD will leave behind a neutron star remnant via an electron capture supernova (Nomoto 1984; Podsiadlowski et al. 2004) or via a core-collapse supernova of type Ib/c for slightly more massive stars. For single stars, or very wide binaries, the critical ZAMS mass for neutron star production via type II supernovae is about $10\,M_\odot$ (Zhang, Woosley & Heger 2008).

In this paper we consider all systems with either a CO or an ONeMg WD as *one* population, denoted by CO WDs unless specified otherwise, and all scenarios listed below apply to BMSPs with both types of massive white dwarf companions. However, the required progenitor star masses are, in general, somewhat larger for the ONeMg WDs compared to the CO WDs. Despite these minor differences, the bottom line is that in all cases IMXBs are the progenitor systems of BMSPs with massive white dwarf companions.

### 6.2.1  IMXBs with Case A RLO

BMSPs with CO WDs, which formed via Case A RLO in IMXBs, evolve from donor stars with masses $3.5 - 5\,M_\odot$ and orbital periods of a few days (Tauris, van den Heuvel & Savonije (2000); Podsiadlowski, Rappaport & Pfahl (2002); Paper I). The final orbital periods of the BMSPs are $5 - 20$ days. The mass-transfer phase consists of three parts: A1, A2 and AB. Phase A1 is thermal time-scale mass transfer and lasts for about $1\,{\rm Myr}$. The majority of the donor envelope is transfered at a high rate exceeding $10^{-5}\,M_\odot\,{\rm yr}^{-1}$. However, the accretion onto the neutron star is limited by the Eddington luminosity, corresponding to an accretion rate of a few $10^{-8}\,M_\odot\,{\rm yr}^{-1}$, and thus the far majority of the transfered material (up to 99.9%) is ejected from the system. Phase A2 is driven by the nuclear burning of the residual hydrogen in the core. The resulting mass-transfer rate is very low ($< 10^{-9}\,M_\odot\,{\rm yr}^{-1}$) and only a few $10^{-2}\,M_\odot$ of material is transfered towards the accreting neutron star in a time-interval of $20 - 50\,{\rm Myr}$. Phase AB is caused by expansion of the donor star while undergoing hydrogen shell burning. The mass-transfer rate is $10^{-8} - 10^{-7}\,M_\odot\,{\rm yr}^{-1}$ and lasts for $\sim 10\,{\rm Myr}$. A few $0.1\,M_\odot$ is transfered towards the accreting neutron star – significantly more than in the previous two phases, A1 and A2. As we shall see later on, this phase is responsible for spinning up the neutron star to become a millisecond pulsar.

### 6.2.2  IMXBs with early Case B RLO

Wider orbit IMXB systems evolve from donor stars which are in the Hertzsprung gap at the onset of the RLO. The orbital periods at the onset of the Case B RLO are about $3 - 10$ days and donor masses of $2.5 - 5.0\,M_\odot$ are needed to yield a CO WD remnant in a BMSP system (Tauris, van den Heuvel & Savonije 2000; Podsiadlowski, Rappaport & Pfahl 2002). The final orbital periods of the BMSPs are about $3 - 50$ days. The mass-transfer rate in the IMXB phase is highly super-Eddington ($10^{-5}\,M_\odot\,{\rm yr}^{-1}$) and thus the far majority of the transfered material is ejected. However, a few $10^{-2}\,M_\odot$ are transfered to the neutron star – enough to create a mildly ($> 10$ ms) spun-up millisecond pulsar, see Section 6.4.

Binaries with either more massive donor stars ($> 5\,M_\odot$) or initial periods in excess of 10 days will undergo dynamically unstable RLO (see Paper I, and references therein) leading to a common envelope and spiral-in. The outcome of the in-spiral is probably a merger for Case B RLO, since for these stars the (absolute) binding energy of the envelope is too large to allow for its ejection – except if accretion luminosity of the neutron star can be efficiently converted into kinetic energy of the envelope.

### 6.2.3  IMXBs with Case C RLO and a common envelope

Donor stars in systems with very wide orbits ($P_{\rm orb} \simeq 10^2 - 10^3$ days) prior to the mass-transfer phase develop a deep convective envelope as they become giant stars before filling their Roche-lobe. The response to mass loss for these stars is therefore expansion which causes the stars to overfill their Roche-lobe even more. To exacerbate this problem, binaries also shrink in size if mass transfer occurs from a donor star somewhat more massive than the accreting neutron star. This causes further overfilling of the donor star Roche-lobe resulting in enhanced mass loss etc. This situation is a vicious circle that leads to a dynamically unstable, runaway mass transfer and the formation of a common envelope (CE) followed by a spiral-in phase, e.g. Paczyński (1976), Iben & Livio (1993), Ivanova et al. (2013). Whereas donor stars which initiate RLO during hydrogen shell burning still have relatively tightly bound envelopes, these stars here which initiate RLO on the AGB have only weakly bound envelopes (Han, Podsiadlowski & Eggleton 1994; Dewi & Tauris 2000). Hence, these donor stars can survive the spiral-in phase of the neutron star without merging. To leave a CO WD remnant from such wide-orbit systems the ZAMS masses of the progenitor stars (the donor stars of the IMXBs) must be $2.2 - 6\,M_\odot$ (Paper I), depending on the assumed amount of convective core-overshooting in the stellar models.

### 6.2.4  BMSPs with a CO WD from LMXBs

In addition to the formation channels from IMXBs, pulsars with a CO WD companion can, in rare cases, be produced from late Case B RLO in an LMXB system – e.g. see Table A1 in Tauris & Savonije (1999). The outcome is an extremely wide-orbit pulsar system ($P_{\rm orb} > 1000$ days) with a $0.47 - 0.67\,M_\odot$ CO WD and a slowly spinning pulsar. The pulsar B0820+02 ($P = 0.86$ sec, $P_{\rm orb} = 1232$ days) is an example of a system which followed this formation channel (cf. Section 6.3.2). It is an interesting fact that CO WDs produced from donor stars in IMXBs can be less massive than CO WDs produced in such LMXBs. In low-mass stars ($\leq 2.2\,M_\odot$) helium is ignited in a flash under degenerate conditions and CO WDs produced from these stars in LMXBs have minimum masses of $0.47\,M_\odot$, whereas in cores of intermediate-mass stars helium ignites non-degenerately and CO WDs from IMXBs can be made with masses down to $0.33\,M_\odot$ (Kippenhahn & Weigert 1990; Tauris, van den Heuvel & Savonije 2000; Podsiadlowski, Rappaport & Pfahl 2002).

### 6.2.5  BMSPs with a CO WD from AIC?

Under certain circumstances is may be possible for an accreting ONeMg WD to reach the Chandrasekhar mass limit and thereby implode to form a neutron star – the so-called accretion-induced collapse, AIC (Miyaji et al. 1980; Nomoto 1984; Canal, Isern & Labay 1990). A neutron star formed this way may leave behind a radio pulsar. The necessary conditions for a pulsar formed this way to further accrete from the companion star and leaving behind, for example a BMSP with a CO WD companion, is investigated in a separate paper (Tauris et al., in preparation). However, the recycling process, which is the main topic in this paper, remains the same.

## 6.3 Observational characteristics of BMSPs with CO WD companions

Before looking into the details of recycling a pulsar in an X-ray binary, we first summarize the observational properties of the end products, the BMSPs. This is important for our subsequent discussions and understanding of the theoretical modelling which constitutes the core of our work. Furthermore, we shall investigate if and how the BMSPs with CO WD companions can be distinguished from the large population of BMSPs with He WD companions. For this purpose we make use of the $P\dot{P}$–diagram, the Corbet diagram and information of orbital eccentricities.

### 6.3.1 The $P\dot{P}$–diagram

In order to understand the formation and evolution of recycled pulsars we have plotted in the $P\dot{P}$–diagram in Fig. 49 the distribution of 103 binary radio pulsars located in the Galactic disk. The various companion types are marked with different symbols (see also Table 7). The majority of the companion stars are helium white dwarfs (He WDs). However, there is also a significant population of the more massive CO and ONeMg WDs, as well as pulsars with another neutron star as companion. In some systems the companion is still a main sequence star. In other (very tight) systems the companion has evolved into a semi-degenerate substellar remnant ($< 0.08\,M_\odot$) – possibly aided by evaporation caused by the illumination from the pulsar following an ultra-compact LMXB phase (Podsiadlowski 1991; Deloye & Bildsten 2003; van der Sluys, Verbunt & Pols 2005; Bailes et al. 2011; van Haaften et al. 2012b).

Pulsar systems found in globular clusters are not useful as probes of stellar evolution and binary interactions since their location in a dense environment causes frequent exchange collisions whereby the orbital parameters are perturbed or the companion star is replaced with another star and thus information is lost regarding the mass transfer and spin-up of the pulsar. These pulsars could even have undergone accretion events from two or more different companions.

The location of PSR J1614−2230 in the $P\dot{P}$–diagram is quite unusual for a BMSP with a CO WD companion. It is located in an area which is otherwise only populated by BMSPs with He WD companions. This is important as it proves for the first time that efficient, full recycling (i.e. $P \leq 8\,\mathrm{ms}$) of a pulsar is also possible in X-ray systems leading to BMSPs with CO WD companions. In general, the BMSPs with CO WD companions are seen to have large values of both $P$ and $\dot{P}$ compared to BMSPs with He WD companions. This trend can be understood since the former systems evolve from IMXBs which often have short lasting RLO and thus inefficient spin-up – PSR J1614−2230 being the only known exception (this system evolved from an IMXB via Case A RLO, see Paper I).

In Table 8 we list all binary radio pulsars with CO WD (or ONeMg WD) companions – see also Hobbs et al. (2004). At the bottom of the Table we list an additional number of sources which we identify as CO WD candidate systems (alternatively, these candidates may be He WDs evolving from IMXBs). The three slowest rotating of the 20 pulsars with a CO WD companion are not recycled – either because they were formed *after* the WD (Tauris & Sennels 2000), or because they formed in a very wide orbit (cf. Section 6.2.4). Of the remaining 17 recycled pulsars with a CO WD companion, 16 have a measured value of $\dot{P}$. These are plotted in Fig. 50 together with those other systems which may also have evolved from IMXB systems and thus possibly host a CO WD companion too (cf. Section 6.3.2). With the exception of PSR J1841+0130 (which is young, see Section 6.8.3.4), these pulsars are distributed in a somewhat more linear manner in the $P\dot{P}$–diagram compared to BMSPs with a He WD companion.

### 6.3.2 The Corbet diagram - six unusual pulsars

Plotting BMSPs in the Corbet diagram (Corbet 1984) could lead to potential clues about their formation history. In the top panel of Fig. 51 we have shown the distribution of all binary pulsars. Little can be learnt from this plot when considering the entire ensemble of binary pulsars as one population. In the central plot, however, we display only those binary pulsars which have a He WD companion. The distribution of pulsars in this diagram shows some interesting features. The six pulsars marked with a circle are noticeable for having unusual

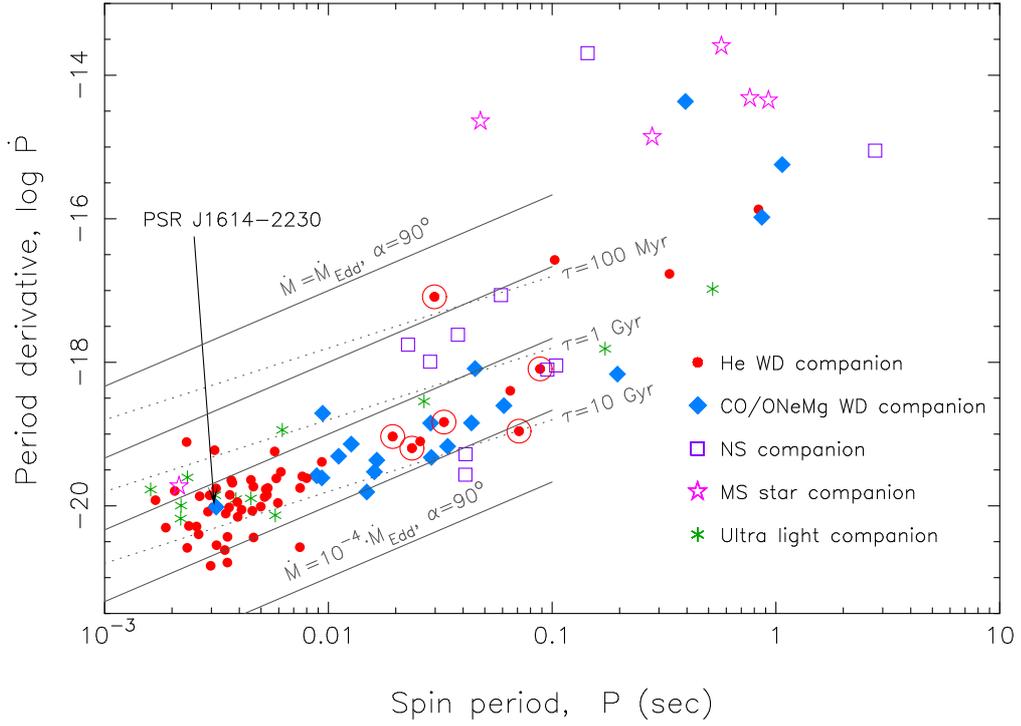

Figure 49: Distribution in the $P\dot{P}$–diagram of 103 binary radio pulsars in the Galactic disk. The five solid lines indicate theoretical spin-up lines using equation (52) for a $1.4\,M_\odot$ pulsar accreting with a rate of $\dot{M} = \dot{M}_{\mathrm{Edd}}$, $0.1\,\dot{M}_{\mathrm{Edd}}$, $0.01\,\dot{M}_{\mathrm{Edd}}$, $0.001\,\dot{M}$ and $\dot{M} = 10^{-4}\,\dot{M}_{\mathrm{Edd}}$, respectively (top to bottom). Also shown as dotted lines are characteristic ages of 100 Myr, 1 Gyr and 10 Gyr, respectively. The location of PSR J1614−2230 is peculiar in the sense that it has a massive CO WD companion (blue diamond) in a region where all other pulsars have a low-mass He WD companion (red filled circles). The six pulsars marked with circles may have evolved from IMXBs and some of these are potential candidates for having a CO WD companion – see Section 6.3.2. Data taken from the *ATNF Pulsar Catalogue*, http://www.atnf.csiro.au/research/pulsar/psrcat/ (Manchester et al. 2005), in May 2012.

Table 7: Binary radio pulsars and their companions. Data taken from the *ATNF Pulsar Catalogue* in May 2012.

| Population | Number |
|---|---|
| All known binary pulsars: | 186 |
|    Galactic disk | 110 |
|    Extra galactic (SMC) | 1 |
|    Globular clusters | 75 |
| Galactic disk* binary pulsars with measured $\dot{P}$: | 103 |
| Companion star: | |
|    He white dwarf | 56 |
|    CO/ONeMg white dwarf | 19 |
|    Neutron star | 10 |
|    Main sequence star* | 6 |
|    Ultra light (semi)deg. dwarf, or planet(s) | 12 |

* Including PSR J0045−7319 which is located in the Small Magellanic Cloud and has a *B*1V companion (Kaspi et al. 1994).

Table 8: Binary radio pulsars which are likely to have evolved from IMXB systems. Data taken from the *ATNF Pulsar Catalogue* in May 2012.

| PSR | $P_{\rm orb}$ | $P$ | $\dot{P}$ | ecc | $M_{\rm WD}$ |
|-----|------|-----|-----------|-----|------|
|     | days | ms | $10^{-19}$ | $10^{-4}$ | $M_\odot$ |
| J1952+2630 | 0.392 | 20.7 | n/a | n/a | 1.13 |
| J1757−5322 | 0.453 | 8.87 | 0.263 | 0.0402 | 0.67 |
| J1802−2124 | 0.699 | 12.6 | 0.726 | 0.0247 | 0.78* |
| B0655+64 | 1.03 | 196 | 6.85 | 0.0750 | 0.80 |
| J1435−6100 | 1.35 | 9.35 | 0.245 | 0.105 | 1.08 |
| J1439−5501 | 2.12 | 28.6 | 1.42 | 0.499 | 1.30* |
| J1528−3146 | 3.18 | 60.8 | 2.49 | 2.13 | 1.15 |
| J1157−5112 | 3.51 | 43.6 | 1.43 | 4.02 | 1.30* |
| J1337−6423 | 4.79 | 9.42 | 1.95 | 0.197 | 0.95 |
| J1603−7202 | 6.31 | 14.8 | 0.156 | 0.0928 | 0.34 |
| J2145−0750 | 6.84 | 16.1 | 0.298 | 0.193 | 0.50 |
| J1022+1001 | 7.81 | 16.5 | 0.433 | 0.970 | 0.85 |
| J0621+1002 | 8.32 | 28.9 | 0.473 | 24.6 | 0.67* |
| J1614−2230 | 8.69 | 3.15 | 0.0962 | 0.0130 | 0.50* |
| J1454−5846 | 12.4 | 45.2 | 8.17 | 19.0 | 1.05 |
| J0900−3144 | 18.7 | 11.1 | 0.491 | 0.103 | 0.42 |
| J1420−5625 | 40.3 | 34.1 | 0.675 | 35.0 | 0.44 |
| J1141−6545[a] | 0.198 | 394 | 4307 | 1719 | 1.02* |
| B2303+46[a] | 12.3 | 1066 | 569 | 6584 | 1.30* |
| B0820+02[b] | 1232 | 865 | 1050 | 118.7 | > 0.52 |
| J1622−6617[c] | 1.64 | 23.6 | 0.636 | n/a | 0.11 |
| J1232−6501[c] | 1.86 | 88.3 | 8.11 | 1.09 | 0.17 |
| J1745−0952[c] | 4.94 | 19.4 | 0.925 | 0.0985 | 0.13 |
| J1841+0130[c] | 10.5 | 29.8 | 81.7 | 0.819 | 0.11 |
| J1904+0412[c] | 14.9 | 71.1 | 1.10 | 2.20 | 0.26 |
| J1810−2005[c] | 15.0 | 32.8 | 1.47 | 0.192 | 0.33 |

* Mass obtained from timing. Other masses are median masses (i.e. calculated for $i = 60°$ and an assumed $M_{\rm NS} = 1.35\,M_\odot$).

[a] A non-recycled pulsar formed *after* its WD companion (Tauris & Sennels 2000).

[b] This system descends from a very wide-orbit LMXB (see text).

[c] CO WD candidate or He WD evolving from an IMXB.

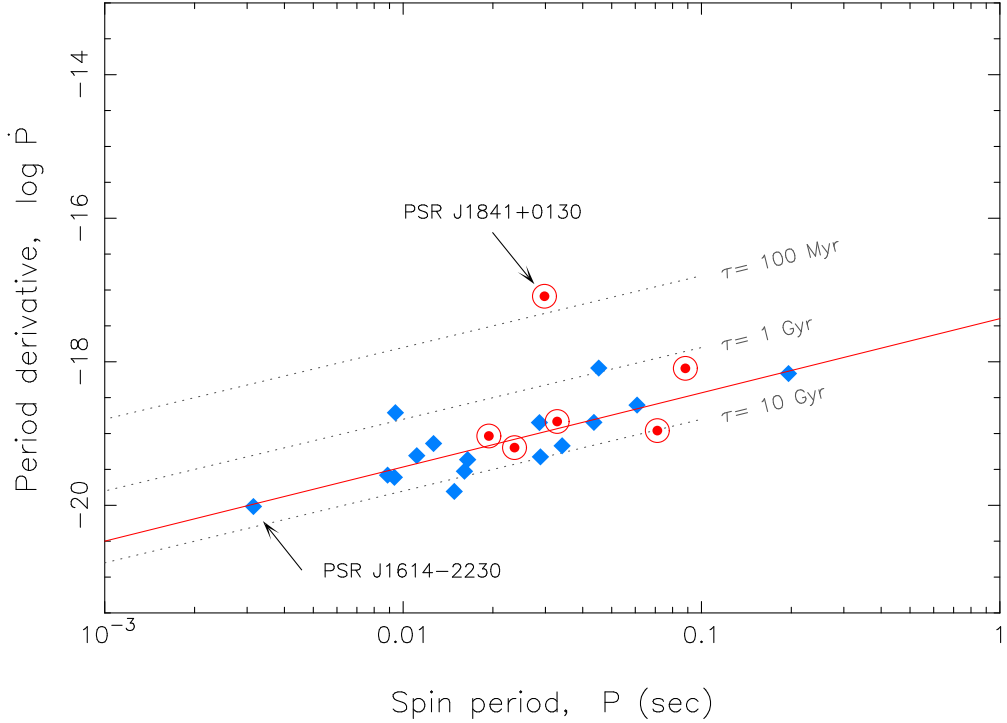

Figure 50: The observed recycled pulsars with CO WD companions (diamonds) seem to cluster somewhat along a straight line in the $P\dot{P}$–diagram. The full line is a linear regression fit to all these pulsars and the dotted lines represent constant characteristic ages. Other pulsars which are also candidates for having an IMXB origin are shown as dots in a circle (cf. Section 6.3.2). PSR J1841+0130 is a young example of such a candidate – see text.

slow spin periods compared to other pulsars with similar orbital periods, see Table 8 for further details. This fact hints an inefficient and shorter spin-up phase compared to the BMSPS with fast spin. Tauris (2011) suggested that these pulsars may perhaps have formed via the accretion induced collapse (AIC) of a WD. Here, we suggest instead that these pulsars (which all have $P_{\rm orb} \geq 1$ day) may have formed in IMXBs. In that case the companions are likely to be CO WDs, although He WD companions may also form from IMXBs (Tauris, van den Heuvel & Savonije 2000; Podsiadlowski, Rappaport & Pfahl 2002). Similar ideas of an IMXB origin have been proposed by Camilo et al. (2001) and Li (2002). The latter author argued that accretion disk instabilities in IMXBs may explain the slow spin periods and the high $\dot{P}$ values of these pulsars. The pulsars with $P_{\rm orb} > 200$ days all have slow spin periods – see Section 6.6.3 for a discussion on these systems. The bottom panel shows the binary pulsars with CO WD companions. The main thing to notice is that these systems have $P_{\rm orb} \leq 40^{\rm d}$ and slower pulsar spin periods compared to BMSPs with He WD companions, as first pointed out by Camilo (1996). The slower spin of pulsars in this population can be understood well which is discussed further in Section 6.6. A few pulsars marked with a circle or a triangle stand out from the rest of the population. These are PSR J1141−6545 and PSR B2303+46 where the neutron star formed after the white dwarf companion (Tauris & Sennels 2000) and therefore these pulsars remain non-recycled with slow spin periods, and PSR B0820+02 which formed from an extremely wide-orbit LMXB with short lasting and possibly inefficient RLO (see Sections 6.2.4 and 6.6.3).

### 6.3.2.1 PSR J1841+0130

PSR J1841+0130 (discovered by Lorimer et al. 2006a) is one of the six pulsars which we note to have an offset location in the Corbet diagram when comparing to BMSPs with He WD companions (Fig. 51, central panel). It has a slow spin period of 29 ms and an orbital period, $P_{\rm orb} = 10.5^{\rm d}$. The majority of other BMSPs with similar $P_{\rm orb}$ near 10 days have much faster spin periods between $2-10$ ms. The slow spin period of PSR J1841+0130 can be explained if it originated from an IMXB system rather than an LMXB system. The mass function of this pulsar, $f = 4.22 \times 10^{-4}\, M_\odot$ suggests that its companion star has a mass of $0.11\, M_\odot$ for an orbital inclination angle of $60°$. If the companion star is a CO WD formed in an IMXB it must have a mass $\geq 0.33\, M_\odot$ (cf. Section 6.2.4). This would require a small inclination angle, $i \leq 20°$ (depending on the neutron star mass). The probability for this to happen from a distribution of random orbital orientations is about 6 per cent. The situation is similar in PSR J1622−6617 which is another candidate systems to have evolved from an IMXB. However, IMXBs can also leave behind He WDs which have somewhat smaller masses. Alternatively, if PSR J1841+0130 originated from an LMXB system then according to the ($M_{\rm WD}$, $P_{\rm orb}$)-relation, e.g. Tauris & Savonije (1999), we would expect $i \approx 25°$ to obtain the predicted He WD mass of $0.26 \pm 0.01\, M_\odot$. (The probability for this is about 9 per cent.) Hence, even in this case the inclination angle would be rather low. As discussed in Sections 6.4.3 and 6.8.3.4, PSR J1841+0130 is also interesting for its ability to constrain spin-up physics and for its young age.

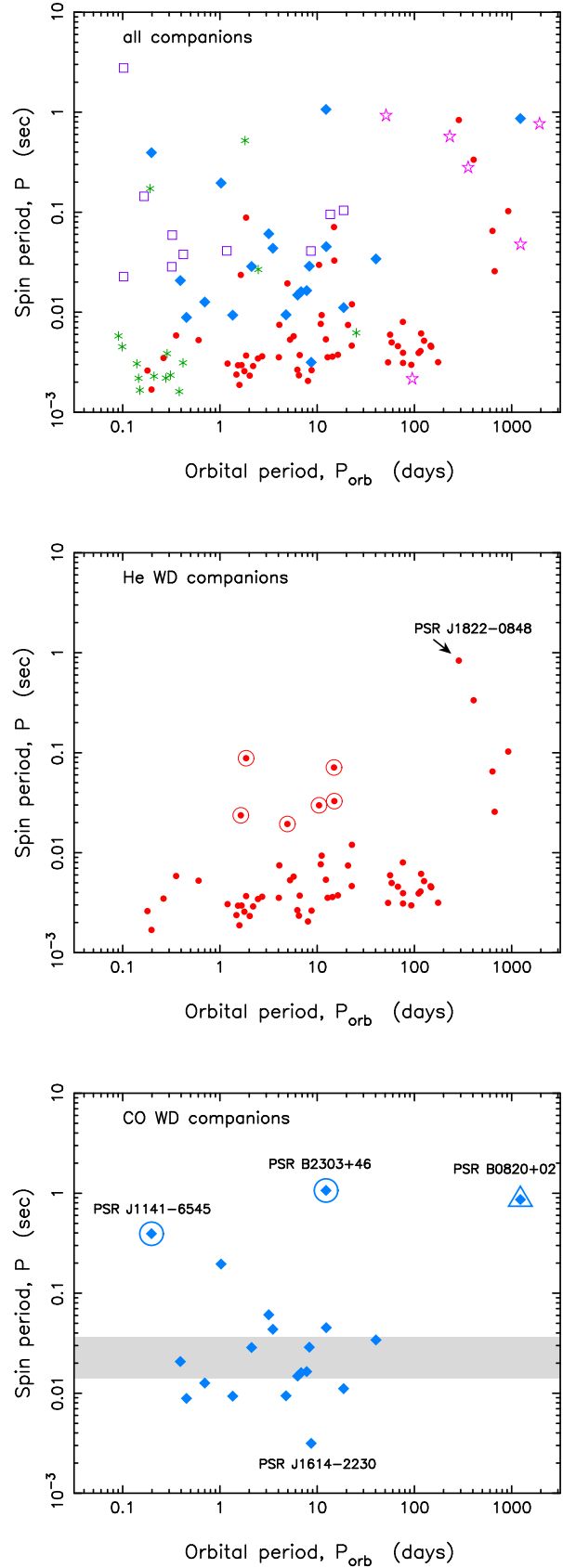

Figure 51: All binary radio pulsars in the Galactic disk plotted in a Corbet diagram (top), see Fig. 49 for an explanation of the symbols plotted. Due to their missing $\dot{P}$ measurements 8 of these pulsars were not shown in Fig. 49. The central and bottom panels show the distribution of pulsars with He WD (60 systems) and CO WD companions (20 systems), respectively. In the bottom panel the grey shaded region marks the range of spin periods obtained from one of our model calculations of Case BB RLO in a post-CE binary (Fig. 58). See text for further discussion.

### 6.3.3 The orbital eccentricity

Another fossil record of binary evolution is the orbital eccentricity (Phinney 1992; Phinney & Kulkarni 1994). In Fig. 52 we have plotted the eccentricity as a function of orbital period. The BMSPs with He WDs have in general somewhat lower eccentricities than systems with CO WDs. The spread is large for both of the two populations which also overlap. Among the systems with CO WD companions it is not surprising that PSR J1614−2230 has the lowest eccentricity since we have demonstrated (Paper I) that this system formed via a relatively long phase of stable RLO. It is interesting to notice that the four BMSPs with CO WDs and $P_{\rm orb} < 2$ days all have quite small eccentricities $\leq 10^{-5}$ even though they are believed to have formed via a CE.

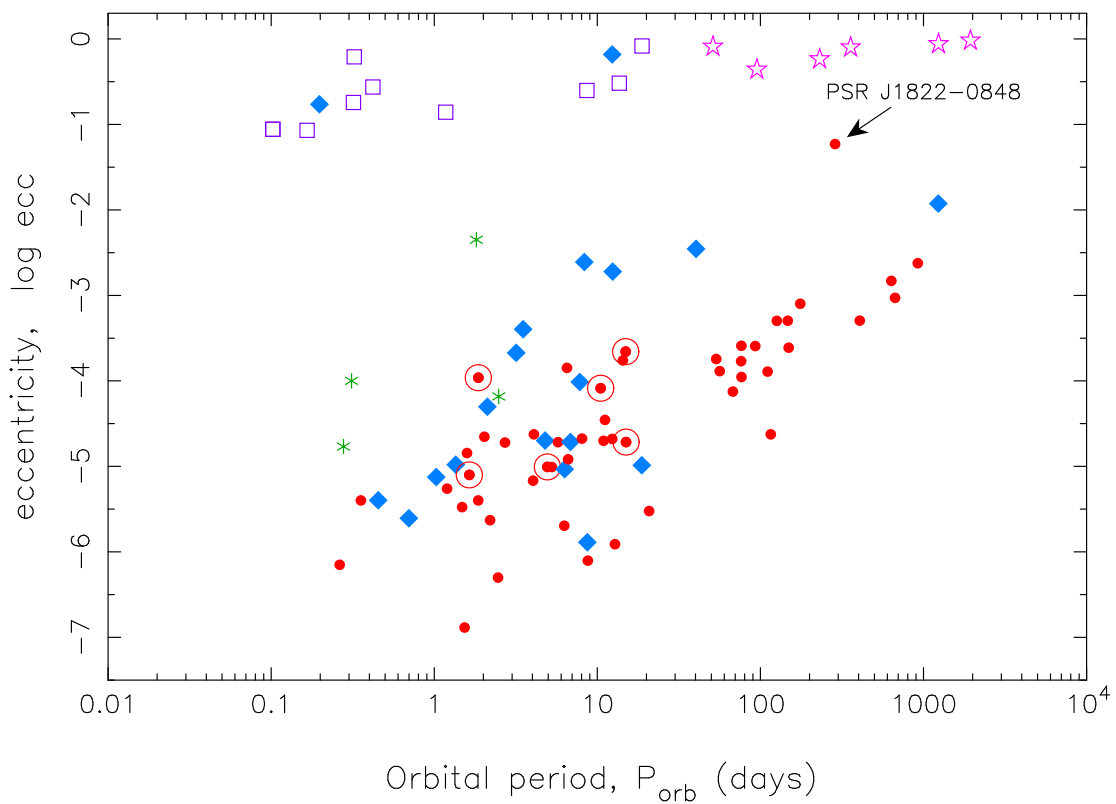

Figure 52: Eccentricities as a function of orbital period for binary radio pulsars in the Galactic disk. The various symbols are defined in Fig. 49. We notice PSR J1822−0848 which has an eccentricity two orders of magnitude larger than other wide-orbit binary pulsars with a He WD companion.

### 6.3.3.1 PSR J1822−0848

PSR J1822−0848 (Lorimer et al. 2006a) has an orbital period of 287 days and an eccentricity, $ecc = 0.059$. From Fig. 52 it is seen that this eccentricity is two orders of magnitude larger than that of other pulsars in similar wide-orbit systems. This fact, combined with its very slow spin period of 835 ms (see the central panel in Fig. 51) and its high value of $\dot{P} = 1.35 \times 10^{-16}$, hints that this pulsar may not have a He WD companion star. It is possible it belongs to the same class as PSR B0820+02 which is in a wide orbit with a CO WD (see Section 6.2.4). However, in that case it needs to be explained why some pulsars have He WD companions in even larger orbits compared to PSR J1822−0848. Alternatively, one could speculate that PSR J1822−0848 experienced weak spiral-in from an almost unbound common envelope of an AGB star. This could explain its non-recycled characteristics and why it has $P_{\rm orb} < 1000$ days.

Although, it has a relatively large eccentricity of 0.059 this value is still far too small to suggest that PSR J1822−0848 was born after the WD, like in the case of PSR 1141−6545 and PSR B2303+46 (Tauris & Sennels 2000). Even in the unrealistic case of a collapse of a naked core with no envelope, and without a kick, the present eccentricity of 0.059 for such a wide orbit would require a maximum instant mass loss of only $\sim 0.10\,M_\odot$ – a value which is even less than the release of the gravitational binding energy during the core collapse. Thus this scenario is not possible and we therefore conclude that PSR J1822−0848 was formed before its WD companion, like 99 per cent of all detected binary pulsars with WD companions.

## 6.4 Recycling the pulsar

We now proceed with a general examination of the recycling process of pulsars. We begin by analysing the concept of spin-up lines in the $P\dot{P}$–diagram and derive an analytic expression for the amount of accreted mass needed to spin-up a pulsar to a given equilibrium spin period. We use standard theory for deriving these expressions, but include it here to present a coherent and detailed derivation needed for our purposes. Furthermore, we apply the Spitkovsky (2006) solution to the pulsar spin-down torque which combines the effect of a plasma current in the magnetosphere with the magnetic dipole model. The physics of the disk–magnetosphere interaction is still not known in detail. The interplay between the neutron star magnetic field and the conducting plasma in the accretion disk is a rather complex process. For further details of the accretion physics we refer to Pringle & Rees (1972); Lamb, Pethick & Pines (1973); Davidson & Ostriker (1973); Ghosh & Lamb (1979a;b); Shapiro & Teukolsky (1983); Ghosh & Lamb (1992); Spruit & Taam (1993); Campana et al. (1998); Frank, King & Raine (2002); Rappaport, Fregeau & Spruit (2004); Bozzo et al. (2009); D'Angelo & Spruit (2010); Ikhsanov & Beskrovnaya (2012) and references therein.

### 6.4.1 The accretion disk

The accreted gas from a binary companion possess large specific angular momentum. For this reason the flow of gas onto the neutron star is not spherical, but leads to the formation of an accretion disk where excess angular momentum is transported outwards by (turbulent-enhanced) viscous stresses, e.g. Shapiro & Teukolsky (1983); Frank, King & Raine (2002). Depending on the mass-transfer rate, the opacity of the accreted material and the temperature of the disk, the geometric shape and flow of the material may take a variety of forms (thick disk, thin thick, slim disk, torus-like, ADAF). Popular models of the inner disk (Ghosh & Lamb 1992) include optically thin/thick disks which can be either gas (GPD) or radiation pressure dominated (RPD). The exact expression for the spin-up line in the $P\dot{P}$–diagram also depends on the assumed model for the inner disk – mainly as a result of the magnetosphere boundary which depends on the characteristics of the inner disk. Close to the neutron star surface the magnetic field is strong enough that the magnetic stresses truncate the Keplerian disk, and the plasma is channeled along field lines to accrete on to the surface of the neutron star. At the inner edge of the disk the magnetic field interacts directly with the disk material over some finite region. The physics of this transition zone from Keplerian disk to magnetospheric flow is important and determines the angular momentum exchange from the differential rotation between the disk and the neutron star.

Interestingly enough, it seems to be the case that the resultant accretion torque, acting on the neutron star, calculated using detailed models of the disk–magnetosphere interaction does not deviate much from simple expressions assuming idealized, spherical accretion and newtonian dynamics. For example, it has been pointed out by Ghosh & Lamb (1992) as a fortuitous coincidence that the equilibrium spin period calculated under simple assumptions of spherical flow resembles the more detailed models of an optically thick, gas pressure dominated inner accretion disk. Hence, as a starting point this allows us to expand on standard prescriptions in the literature with the aims of: 1) performing a more careful analysis of the concept of a spin-up line, 2) deriving an analytic expression for the mass needed to spin-up a given observed millisecond pulsar, and 3) understanding the effects on the spinning neutron star when the donor star decouples from its Roche-lobe. As we shall discuss further in Section 6.5, the latter issue depends on the location of the magnetospheric boundary (or the inner edge of the accretion disk) relative to the corotation radius and the light-cylinder radius of the pulsar. All stellar parameters listed below will refer to the neutron star unless explicitly stated otherwise.

### 6.4.2 The accretion torque – the basics

The mass transfered from the donor star carries with it angular momentum which eventually spins up the rotating neutron star once its surface magnetic flux density, $B$, is low enough to allow for efficient accretion, i.e. following initial phases where either the magneto-dipole radiation pressure dominates or propeller effects are at work.

The accretion torque acting on the spinning neutron star has a contribution from both material stress (dominant term), magnetic stress and viscous stress, and is given by: $N = \dot{J}_\star \equiv (d/dt)(I\Omega_\star)$, where $J_\star$ is the pulsar spin angular momentum, $\Omega_\star$ is its angular velocity and $I \approx 1\text{--}2 \times 10^{45}\ \mathrm{g\ cm^2}$ is its moment of inertia. The exchange of angular momentum ($\vec{J} = \vec{r} \times \vec{p}$) at the magnetospheric boundary eventually leads to a gain of neutron star spin angular momentum which can approximately be expressed as:

$$\Delta J_\star = \sqrt{GM r_{\mathrm{A}}}\,\Delta M\,\xi \tag{43}$$

where $\xi \simeq 1$ is a numerical factor which depends on the flow pattern (Ghosh & Lamb 1979b; 1992), $\Delta M = \dot{M} \times \Delta t$ is the amount of mass accreted in a time interval $\Delta t$ with average mass accretion rate $\dot{M}$ and

$$r_{\mathrm{A}} \simeq \left( \frac{B^2\,R^6}{\dot{M}\sqrt{2GM}} \right)^{2/7} \tag{44}$$

$$\simeq 22\,\mathrm{km}\quad B_8^{4/7} \left( \frac{\dot{M}}{0.1\,\dot{M}_{\mathrm{Edd}}} \right)^{-2/7} \left( \frac{M}{1.4\,M_\odot} \right)^{-5/7}$$

is the Alfvén radius defined as the location where the magnetic energy density will begin to control the flow of matter (i.e. where the incoming material couples to the magnetic field lines and co-rotate with the neutron star magnetosphere). Here $B$ is the surface magnetic flux density, $R$ is the neutron star radius, $M$ is the neutron star mass (see relation between $R$ and $M$ further below) and $B_8$ is $B$ in units of $10^8$ Gauss. A typical value for the Alfvén radius in accreting X-ray millisecond pulsars (AXMSPs), obtained from $B \sim 10^8$ G and $\dot{M} \sim 0.01\,\dot{M}_{\mathrm{Edd}}$, is $\sim 40\,\mathrm{km}$ corresponding to about $3\,R$. The expression above is found by equating the magnetic energy density ($B^2/8\pi$) to the ram pressure of the incoming matter and using the continuity equation (e.g. Pringle & Rees 1972). Furthermore, it assumes a scaling with distance, $r$ of the far-field strength of the dipole magnetic moment, $\mu$ as: $B(r) \propto \mu/r^3$ (i.e. disregarding poorly known effects such as magnetic screening (Vasyliunas 1979)). A more detailed estimation of the location of the inner edge of the disk, i.e. the coupling radius or magnetospheric boundary, is given by: $r_{\mathrm{mag}} = \phi \cdot r_A$, where $\phi$ is $0.5 - 1.4$ (Ghosh & Lamb 1992; Wang 1997; D'Angelo & Spruit 2010).

#### 6.4.2.1 The surface B-field strength of recycled radio pulsars

Before we proceed we need an expression for $B$. One can estimate the B-field of recycled MSPs based on their observed spin period, $P$ and its time derivative $\dot{P}$. The usual assumption is to apply the vacuum magnetic dipole model in which the rate of rotational energy loss ($\dot{E}_{\rm rot} = I\Omega\dot{\Omega}$) is equated to the energy-loss rate caused by emission of dipole waves (with a frequency equal to the spin frequency of the pulsar) due to an inclined axis of the magnetic dipole field with respect to the rotation axis of the pulsar:

$$\dot{E}_{\rm dipole} = (-2/3c^3)|\ddot{\mu}|^2 \tag{45}$$

The result is:

$$\begin{aligned} B_{\rm dipole} &= \sqrt{\frac{3c^3 I P\dot{P}}{8\pi^2 R^6}}\ \frac{1}{\sin\alpha} \\ &\simeq\ 1.6\times 10^{19}\,G\ \ \sqrt{P\dot{P}}\ \left(\frac{M}{1.4\,M_\odot}\right)^{3/2}\frac{1}{\sin\alpha} \end{aligned} \tag{46}$$

where the magnetic inclination angle is $0 < \alpha \leq 90°$. This is the standard equation for evaluating the B-field of a radio pulsar. Our numerical scaling constant differs by a factor of a few from the conventional one: $B = 3.2\times 10^{19}\,G\ \sqrt{P\dot{P}}$, which assumes $R = 10$ km and $I = 10^{45}\,{\rm g\,cm^2}$, both of which we believe are slightly underestimated values (our assumed relations between $I$, $M$ and $R$ are discussed in Section 6.4.2.2). Note also that some descriptions in the literature apply the polar B-field strength ($B_p = 2\,B$) rather than the equatorial B-field strength used here.

It is important to realize that the above expression does not include the rotational energy loss obtained when considering the spin-down torque caused by the $\vec{j}\times\vec{B}$ force exerted by the plasma current in the magnetosphere (e.g. Goldreich & Julian 1969; Spitkovsky 2008). For this reason the vacuum magnetic dipole model does not predict any spin-down torque for an aligned rotator ($\alpha = 0°$), which is not correct. The incompleteness of the vacuum magnetic dipole model was in particular evident after the discovery of intermittent pulsars by Kramer et al. (2006a) and demonstrated the need for including the plasma term in the spin-down torque. A combined model was derived by Spitkovsky (2006) and applying his relation relation between $B$ and $\alpha$ we can rewrite the above expression slightly:

$$\begin{aligned} B &= \sqrt{\frac{c^3 I P\dot{P}}{4\pi^2 R^6}}\ \frac{1}{1+\sin^2\alpha} \\ &\simeq\ 1.3\times 10^{19}\,G\ \ \sqrt{P\dot{P}}\ \left(\frac{M}{1.4\,M_\odot}\right)^{3/2}\sqrt{\frac{1}{1+\sin^2\alpha}} \end{aligned} \tag{47}$$

This new expression leads to smaller estimated values of $B$ by a factor of at least $\sqrt{3}$, or more precisely $\sqrt{\frac{3}{2}(2+\cot^2\alpha)}$, compared to the vacuum dipole model. As we shall shortly demonstrate, this difference in dependence on $\alpha$ is quite important for the location of the spin-up line in the $P\dot{P}$–diagram.

#### 6.4.2.2 The Eddington accretion limit

The Eddington accretion limit is given by:

$$\begin{aligned} \dot{M}_{\rm Edd} &= \frac{4\pi c\, m_p}{\sigma_{\rm T}}\,R\,\mu_{\rm e} \\ &\simeq\ 3.0\times 10^{-8}\,M_\odot\,{\rm yr}^{-1}\ \ R_{13}\left(\frac{1.3}{1+X}\right) \end{aligned} \tag{48}$$

where $c$ is the speed of light in vacuum, $m_p$ is the proton mass, $\sigma_{\rm T}$ is the Thomson scattering cross section, $\mu_{\rm e} = 2/(1+X)$ is the mean molecular weight per electron which depends on the hydrogen mass fraction, X of the accreted material, and $R_{13}$ is the neutron star radius in units of 13 km. The expression is found by equating the outward radiation pressure to the gravitational force per unit area acting on the nucleons of the accreted plasma. In general, luminosity is generated from both nuclear burning at the neutron star surface as well as from the release of gravitational binding energy of the accreted material, i.e. $L = (\epsilon_{\text{nuc}} + \epsilon_{\text{acc}})\,\dot{M}$. However, for accreting neutron stars $\epsilon_{\text{nuc}} \ll \epsilon_{\text{acc}}$ and thus we have neglected the contribution from nuclear processing. To estimate the neutron star radius we used a mass-radius exponent following a simple non-relativistic degenerate Fermi-gas polytrope ($R \propto M^{-1/3}$) with a scaling factor such that: $R = 15\,(M/M_\odot)^{-1/3}$ km, calibrated from PSR J1614−2230, cf. fig. 3 in Demorest et al. (2010). In our code the value of $\dot{M}_{\text{Edd}}$ is calculated according to the chemical composition of the transfered material at any given time. Note, the value of $\dot{M}_{\text{Edd}}$ is only a rough measure since the derivation assumes spherical symmetry, steady state accretion, Thompson scattering opacity and Newtonian gravity.

### 6.4.3 The spin-up line

The observed spin evolution of accreting neutron stars often shows rather stochastic variations on a short timescale (Bildsten et al. 1997). The reason for the involved dramatic torque reversals is not well known – see hypotheses listed at the beginning of Section 6.5. However, the long-term spin rate will eventually tend towards the *equilibrium* spin period, $P_{\text{eq}}$ – meaning that the pulsar spins at the same rate as the particles forced to corotate with the $B$-field at the magnetospheric boundary. The location of the associated so-called spin-up line for the rejuvenated pulsar in the $P\dot{P}$–diagram can be found by considering the equilibrium configuration when the angular velocity of the neutron star is equal to the Keplerian angular velocity of matter at the magnetospheric boundary where the accreted matter enters the magnetosphere, i.e. $\Omega_\star = \Omega_{\text{eq}} = \omega_c\,\Omega_{\text{K}}(r_{\text{mag}})$ or:

$$
\begin{aligned}
P_{\text{eq}} &= 2\pi \sqrt{\frac{r_{\text{mag}}^3}{GM}}\,\frac{1}{\omega_c} \\[2mm]
&\simeq 1.40\,\text{ms}\quad B_8^{6/7}\left(\frac{\dot{M}}{0.1\,\dot{M}_{\text{Edd}}}\right)^{-3/7}\left(\frac{M}{1.4\,M_\odot}\right)^{-5/7} R_{13}^{18/7}
\end{aligned}
\tag{49}
$$

where $0.25 < \omega_c \leq 1$ is the so-called critical fastness parameter which is a measure of when the accretion torque vanishes (depending on the dynamical importance of the pulsar spin rate and the magnetic pitch angle, Ghosh & Lamb 1979b). One must bear in mind that factors which may differ from unity were omitted in the numerical expression above. In all numerical expressions in this paper we assumed $\sin\alpha = \phi = \xi = \omega_c = 1$. Actually, the dependence on the neutron star radius, $R$ disappears in the full analytic formula obtained by inserting equations (44) and (47) into the top expression in equation (49), and using $r_{\text{mag}} = \phi \cdot r_{\text{A}}$, which yields:

$$
P_{\text{eq}} = \left(\frac{\pi c^9}{\sqrt{2}\,G^5}\,\frac{I^3 \dot{P}^3}{M^5 \dot{M}^3}\right)^{1/4}(1+\sin^2\alpha)^{-3/4}\,\phi^{21/8}\,\omega_c^{-7/4}
\tag{50}
$$

Notice, in the above step we needed to link the B-fields of accreting neutron stars to the B-fields (expressed by $P$ and $\dot{P}$) estimated for observed recycled radio pulsars. This connection can be approximated in the following manner: If the radio pulsar after the recycling phase is "born" with a spin period, $P_0$ which is somewhat close to $P_{\text{eq}}$ then we can estimate the location of its magnetosphere when the source was an AXMSP just prior to the accretion turn-off during the Roche-lobe decoupling phase (RLDP), *if* this event did not significantly affect the spin period of the pulsar, cf. discussion in Section 6.5. (Further details of our assumptions of the B-fields of accreting neutron stars are given in Section 6.4.4.1, and in Section 6.8 we discuss the subsequent spin evolution of recycled radio pulsars towards larger periods, $P > P_0$).

It is often useful to express the time derivative of the spin period as a function the equilibrium spin period, for example for the purpose of drawing the spin-up line in the $P\dot{P}$–diagram:

$$\dot{P} = \frac{2^{1/6}G^{5/3}}{\pi^{1/3}c^3} \frac{\dot{M}M^{5/3}P_{\mathrm{eq}}^{4/3}}{I} \; (1 + \sin^2\alpha) \; \phi^{-7/2}\,\omega_c^{7/3} \tag{51}$$

Given that $\dot{M}_{\mathrm{Edd}}$ is a function of the neutron star radius and using the relation between $M$ and $R$ stated in Section 6.4.2.2 we need a relation between the moment of inertia and the mass of the neutron star. According to the equations-of-state studied by Worley, Krastev & Li (2008) these quantities scale very close to linearly as: $I_{45} \simeq M/M_\odot$ (see their fig. 4) where $I_{45}$ is the moment of inertia in units of $10^{45}\,\mathrm{g\,cm^2}$. Towards the end of the mass-transfer phase the amount of hydrogen in the transfered matter is usually quite small ($X \leq 0.20$). The donor star left behind is basically a naked helium core (the proto WD). Hence, we can rewrite equation (51) and estimate the location of the spin-up line for a recycled pulsar in the $P\dot{P}$–diagram only as a function of its mass and the mass accretion rate:

$$\dot{P} = 3.7 \times 10^{-19} \; (M/M_\odot)^{2/3} \, P_{\mathrm{ms}}^{4/3} \left( \frac{\dot{M}}{\dot{M}_{\mathrm{Edd}}} \right) \tag{52}$$

assuming again $\sin\alpha = \phi = \omega_c = 1$, and where $P_{\mathrm{ms}}$ is the equilibrium spin period in units of milliseconds.

In the literature the spin-up line is almost always plotted without uncertainties. Furthermore, one should keep in mind the possible effects of the applied accretion disk model on the location of the spin-up line, cf. Section 6.4.1. In Fig. 53 we have plotted equation (51) for different values of $\alpha$, $\phi$ and $\omega_c$ to illustrate the uncertainties in the applied accretion physics to locate the spin-up line. The upper boundary of each band (or "line") is calculated for a neutron star mass $M = 2.0\,M_\odot$ and a magnetic inclination angle, $\alpha = 90°$. The lower boundary is calculated for $M = 1.0\,M_\odot$ and $\alpha = 0°$. The green hatched band corresponds to $\phi = 1$ and $\omega_c = 1$, which we used in our calculations throughout this paper. The blue and red hatched bands are upper and lower limits set by reasonable choices of the two parameters ($\phi$, $\omega_c$). In all cases we assumed a fixed accretion rate of $\dot{M} = \dot{M}_{\mathrm{Edd}}$. The location of the spin-up line is simply shifted one order of magnitude in $\dot{P}$ down (up) for every order of magnitude $\dot{M}$ is decreased (increased).

It is important to realize that there is no universal spin-up line in the $P\dot{P}$–diagram. Only an upper limit. Any individual pulsar has its own spin-up line/location which in particular depends on its unknown accretion history ($\dot{M}$). Also notice that the dependence on the magnetic inclination angle, $\alpha$ is much less pronounced when applying the Spitkovsky formalism for estimating the B-field of the MSP compared to applying the vacuum dipole model. The difference in the location of spin-up lines using $\alpha = 90°$ and $\alpha = 0°$ is only a factor of two in the Spitkovsky formalism. For a comparison, using the vacuum dipole model with a small magnetic inclination of $\alpha = 10°$ results in a spin-up line which is translated downwards by almost two orders of magnitude compared to its equivalent orthogonal rotator model, cf. the two red dashed lines in Fig. 53.

If we assume that accretion onto the neutron star is indeed Eddington limited then the three bands in Fig. 53 represent upper limits for the spin-up line for the given sets of ($\phi$, $\omega_c$). Thus we can in principle use this plot to constrain ($\phi$, $\omega_c$) and hence the physics of disk–magnetosphere interactions. The fully recycled pulsars B1937+21 and J0218+4232, and the mildly recycled pulsar PSR J1841+0130, are interesting since they are located somewhat in the vicinity of the green spin-up line. Any pulsar above the green line would imply that $\phi < 1$. The pulsar PSR J1823−3021A is close to this limit. Usually the derived value of $\dot{P}$ for globular cluster pulsars is influenced by the cluster potential. However, the $\dot{P}$ value for PSR J1823−3021A was recently constrained from Fermi LAT $\gamma$-ray observations (Freire et al. 2011a) and for this reason we have included it here. The high $\dot{P}$ values of the three Galactic field MSPs listed above: B1937+21, J0218+4232 and PSR J1841+0130, imply that these MSPs are quite young. We discuss their true ages in Sections 6.8.3.3 and 6.8.3.4.

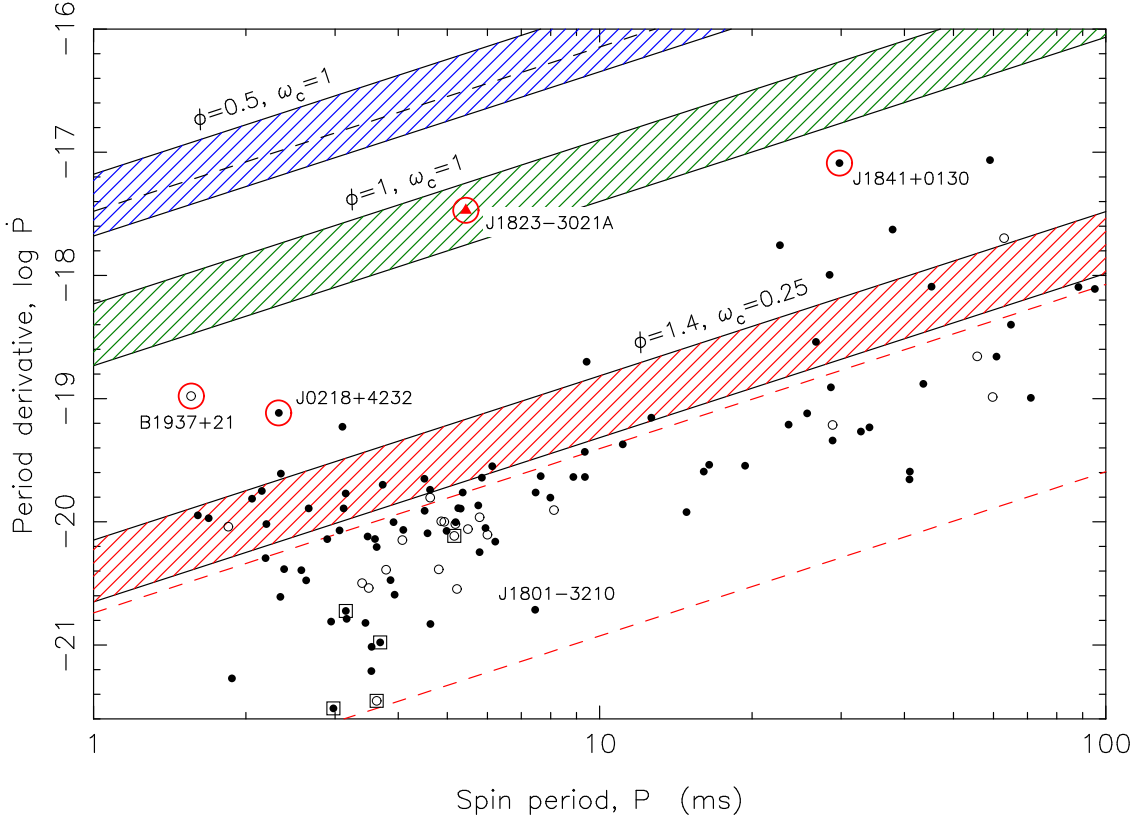

Figure 53: The spin-up line is shown as three coloured bands depending on the parameters ($\phi$, $\omega_c$). In all three cases the spin-up line is calculated assuming accretion at the Eddington limit, $\dot{M} = \dot{M}_{Edd}$ and applying the Spitkovsky torque formalism. Thus these "lines" represent upper limits for the given set of parameters. The width of each line (band) results from using a spread in neutron star mass and magnetic inclination angle from $(2.0\,M_\odot,\ \alpha = 90°)$ to $(1.0\,M_\odot,\ \alpha = 0°)$, upper and lower boundary, respectively. The dashed line within the upper blue band is calculated from $(2.0\,M_\odot,\ \alpha = 0°)$. Hence, the band width above this line reflects the dependence on $\alpha$, whereas the band width below this line shows the dependence on $M$. The two red dashed spin-up lines below the red band are calculated for a $1.4\,M_\odot$ neutron star using the vacuum magnetic dipole model for the radio pulsar torque and assuming $\phi = 1.4$, $\omega_c = 0.25$ and $\dot{M} = \dot{M}_{Edd}$ for $\alpha = 90°$ (upper) and $\alpha = 10°$ (lower). The observed distribution of binary and isolated radio pulsars in the Galactic disk (i.e. outside globular clusters) are plotted as filled and open circles, respectively – see Fig. 61 for further information. Also plotted is the pulsar J1823−3021A which is located in the globular cluster NGC 6624. This pulsar and the three other pulsars marked by a red circle are discussed in the text.

### 6.4.3.1  Formation of a submillisecond pulsar

The possible existence of submillisecond pulsars ($P < 1$ ms) is a long standing and intriguing question in the literature since it is important for constraining the equation-of-state of neutron stars (e.g. Haensel & Zdunik 1989). There has been intense, but so far unsuccessful, efforts to detect these objects in either radio or X-rays (D'Amico 2000; Keith et al. 2010; Patruno 2010a). We note from equation (49) that if the inner edge of the accretion disk (roughly given by $r_{mag}$) extends all the way down to the surface of the neutron star we obtain the smallest possible value for $P_{eq} \sim 0.6$ ms (depending on $M$ and assuming $\omega_c = 1$). It is possible to achieve $r_{mag} \simeq R$ from a combination of a large value of $\dot{M}$ and a small value of $B$, cf. equation (44).

### 6.4.4  Amount of mass needed to spin up a pulsar

### 6.4.4.1  Accretion-induced magnetic field decay

The widely accepted idea of accretion-induced magnetic field decay in neutron stars is based on observational evidence (e.g. Taam & van den Heuvel 1986; Shibazaki et al. 1989; van den

Heuvel & Bitzaraki 1994). During the recycling process the surface B-field of pulsars is reduced by several orders of magnitude, from values of $10^{11-12}\,G$ to $10^{7-9}\,G$. However, it is still not understood if this is caused by spin-down induced flux expulsion of the core proton fluxoids (Srinivasan et al. 1990), or if the B-field is confined to the crustal regions and decays due to diffusion and Ohmic dissipation, as a result of a decreased electrical conductivity when heating effects set in from nuclear burning of the accreted material (Geppert & Urpin 1994; Konar & Bhattacharya 1997), or if the field decay is simply caused by a diamagnetic screening by the accreted plasma – see review by Bhattacharya (2002) and references therein. Although there has been attempts to model or empirically fit the magnetic field evolution of accreting neutron stars (e.g. Shibazaki et al. 1989; Zhang & Kojima 2006; Wang et al. 2011) the results are quite uncertain. One reason is that it is difficult to estimate how much mass a given recycled pulsar has accreted. Furthermore, we cannot rule out the possibility that some of these neutron stars may originate from the accretion-induced collapse of a massive white dwarf, in which case they might be formed with a high B-field near the end of the mass-transfer phase.

To model the B-field evolution for our purpose, which is to relate the spin period of pulsars to the amount of mass accreted, we make the following assumptions:

(i) The B-field decays rapidly in the early phases of the accretion process via some unspecified process (see above).

(ii) Accreting pulsars accumulate the majority of mass while spinning at/near equilibrium.

(iii) The magnetospheric boundary, $r_{mag}$ is approximately kept at a fixed location for the majority of the spin-up phase, until the mass transfer ceases during the RLDP, cf. Section 6.5.

(iv) During the RLDP the B-field of an AXMSP can be considered to be constant since very little envelope material ($\sim 0.01\,M_\odot$) remains to be transfered from its donor at this stage.

### 6.4.4.2 Accreted mass vs final spin period relation

The amount of spin angular momentum added to an accreting pulsar is given by:

$$\Delta J_\star = \int n(\omega, t)\,\dot{M}(t)\,\sqrt{GM(t)r_{mag}(t)}\,\xi(t)\,dt \tag{53}$$

where $n(\omega)$ is a dimensionless torque, see Section 6.5 for a discussion. Assuming $n(\omega) = 1$, and $M(t)$, $r_{mag}(t)$ and $\xi(t)$ to be roughly constant during the major part of the spin-up phase we can rewrite the expression and obtain a simple formula (see also equation 43) for the amount of matter needed to spin up the pulsar:

$$\Delta M \simeq \frac{2\pi I}{P\sqrt{GMr_{mag}}\,\xi} \tag{54}$$

Note, that the initial spin angular momentum of the pulsar prior to accretion is negligible given that $\Omega_0 \ll \Omega_{eq}$. To include all numerical scaling factors properly we can insert equations (44), (47) and (51) into equation (54), recalling that $r_{mag} = \phi \cdot r_A$, and we find:

$$\Delta M_{eq} = I\left(\frac{\Omega_{eq}^4}{G^2 M^2}\right)^{1/3} f(\alpha, \xi, \phi, \omega_c) \tag{55}$$

where $f(\alpha, \xi, \phi, \omega_c)$ is some dimensionless number of order unity. Once again we can apply the relation between moment of inertia and mass of the neutron star (see Section 6.4.2.2) and we obtain a simple convenient expression to relate the amount of mass to be accreted in order to spin-up a pulsar to a given (equilibrium) rotational period:

$$\Delta M_{eq} = 0.22\,M_\odot\,\frac{(M/M_\odot)^{1/3}}{P_{ms}^{4/3}} \tag{56}$$

assuming that the numerical factor $f(\alpha, \xi, \phi, \omega_c) = 1$.

In the above derivation we have neglected minor effects related to release of gravitational binding energy of the accreted material – see e.g. equation (22) in Tauris & Savonije (1999), and in particular Bejger et al. (2011) and Bagchi (2011) for a more detailed, general discussion including various equations of state, general relativity and the critical mass shedding spin limit. However, since the exchange of angular momentum takes place near the magnetospheric boundary the expression in equation (56) refers to the baryonic mass accreted from the donor star. To calculate the increase in (gravitational) mass of the pulsar one must apply a reducing correction factor of $\sim 0.85 - 0.90$, depending on the neutron star equation-of-state (Lattimer & Yahil 1989). In Fig. 54 we show the amount of mass, $\Delta M_{\rm eq}$ needed to spin-up a pulsar to a given spin period. The value of $\Delta M_{\rm eq}$ is a strongly decreasing function of the pulsar spin period, $P_{\rm eq}$. For example, considering a pulsar with a final mass of $1.4\,M_\odot$ and a recycled spin period of either 2 ms, 5 ms, 10 ms or 50 ms requires accretion of $0.10\,M_\odot$, $0.03\,M_\odot$, $0.01\,M_\odot$ or $10^{-3}\,M_\odot$, respectively. Therefore, it is no surprise that observed recycled pulsars with massive companions (CO WD, ONeMg WD or NS) in general are much more slow rotators – compared to BMSPs with He WD companions – since the progenitor of their massive companions evolved on a relatively short timescale, only allowing for very little mass to be accreted by the pulsar, cf. Section 6.6.

### 6.4.4.3 Comparison with other work

The simple expression in equation (56) was also found by Alpar et al. (1982). Alternatively, one can integrate equation (54) directly which yields the maximum spin rate (minimum period) that can be attained by a neutron star accelerated from rest (Lipunov & Postnov 1984):

$$P_{\rm min} = \frac{3\pi\,I}{\sqrt{G\,r_{\rm mag}}\,\xi}\left(M^{3/2} - M_{\rm init}^{3/2}\right)^{-1} \tag{57}$$

where $M_{\rm init} = M - \Delta M_{\rm eq}$ is the mass of the neutron star prior to accretion. In this expression it is still assumed that $I$ and $r_{\rm mag}$ remain constant during the accretion phase. For comparison, the expression above is also plotted in Fig. 54 as a dashed line assuming $\Delta M_{\rm eq} \ll M$, $\xi = 1$ and $r_{\rm mag} = 22$ km (for example, if $B = 10^8$ G, $\dot{M} = 0.1\,\dot{M}_{\rm Edd}$, $M = 1.4\,M_\odot$, $\phi = 1$, see equation (44)). Although the latter expression is simplified with a chosen numerical value of $r_{\rm mag}$, and the two expressions differ by more than a factor of two if $P > 10$ ms, the overall match is fairly good. One the one hand, the value of $\Delta M_{\rm eq}$ should be regarded as a lower limit to the actual amount of material required to be transfered to the neutron star, even at sub-Eddington rates, since a non-negligible amount may be ejected from the pulsar magnetosphere due to magnetodipole wave pressure or the propeller effect (Illarionov & Sunyaev 1975). Furthermore, accretion disk instabilities (Pringle 1981; van Paradijs 1996) are also responsible for ejecting part of the transfered material. Recently it was demonstrated by Antoniadis et al. (2012) that the accretion efficiency in some cases is less than 50 per cent, even in short orbital period binaries accreting at sub-Eddington levels. On the other hand, we did not take into account the possibility of a more efficient angular momentum transfer early in the accretion phase where the value of $r_{\rm mag}$ (the lever arm of the torque) could have been larger if the B-field did not decay rapidly. In this respect it is somewhat surprising that Wang et al. (2011), using a model for accretion induced B-field decay, find that it requires more mass to be accreted to obtain a given spin period (see their fig. 6) compared to our work. This could perhaps (partly) be related to the fact that $\Omega$ is small during the initial accretion phases.

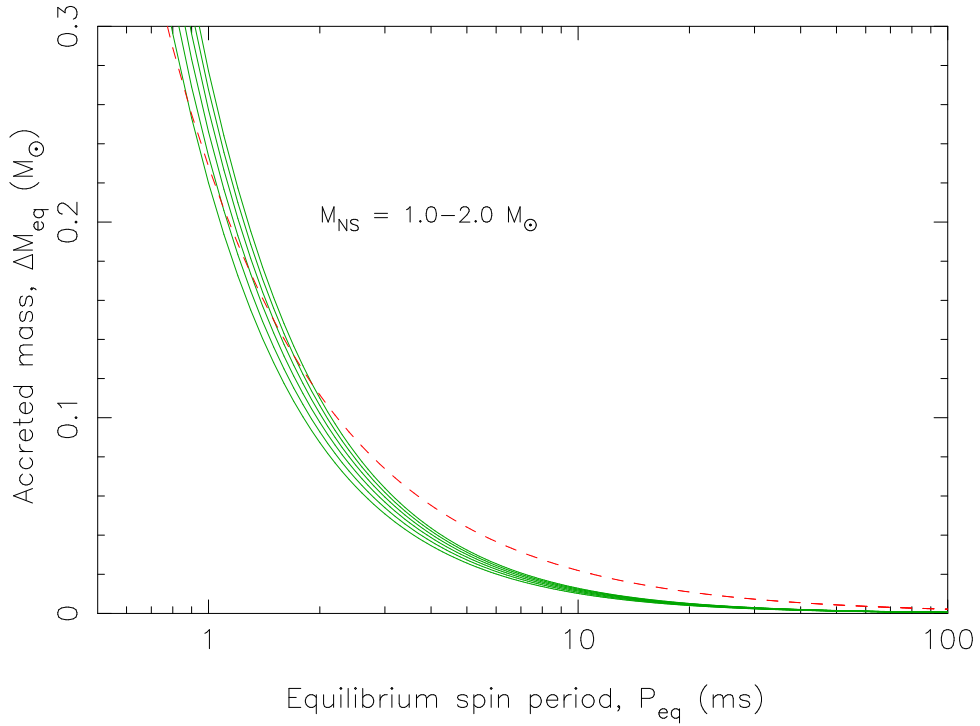

Figure 54: The amount of mass needed to spin up a pulsar as a function of its equilibrium spin period using equation (56). The different green curves correspond to various neutron star masses in steps of $0.2\,M_\odot$, increasing upwards. The dashed red curve is the expression from equation (57) – see text.

### 6.4.5 Spin-relaxation timescale

Above we have calculated the amount of mass needed to spin up a pulsar to a given equilibrium spin period. However, we must be sure that the accretion-torque can actually transmit this acceleration on a timescale shorter than the mass-transfer timescale. To estimate the spin-relaxation timescale (the time needed to spin up a slowly-rotating neutron star to spin equilibrium) one can simply consider: $t_{\rm torque} = J/N$ where $J = 2\pi I/P_{\rm eq}$ and $N = \dot{M}\sqrt{GMr_{\rm mag}}\,\xi$ which yields:

$$
\begin{aligned}
t_{\rm torque} &= I\left(\frac{4G^2M^2}{B^8R^{24}\dot{M}^3}\right)^{1/7}\frac{\omega_c}{\phi^2\,\xi} \\
&\simeq 50\,{\rm Myr}\quad B_8^{-8/7}\left(\frac{\dot{M}}{0.1\,\dot{M}_{\rm Edd}}\right)^{-3/7}\left(\frac{M}{1.4\,M_\odot}\right)^{17/7}
\end{aligned}
\tag{58}
$$

or equivalently, $t_{\rm torque} = \Delta M/\dot{M}$, see equation (54). If the duration of the mass-transfer phase, $t_X$ is shorter than $t_{\rm torque}$ then the pulsar will not be fully recycled. In the following section we examine the importance of $t_{\rm torque}$ relative to the RLO turn-off timescale at the end of the mass-transfer process and how their ratio may affect the pulsar spin period.

### 6.5 Roche-lobe decoupling phase – RLDP

An important effect on the final spin evolution of accreting pulsars, related to the Roche-lobe decoupling phase (RLDP), has recently been demonstrated (Tauris 2012). The rapidly decreasing mass-transfer rate during the RLDP results in an expanding magnetosphere which may cause a significant braking torque to act on the spinning pulsar. It was shown that this effect can explain why the recycled radio MSPs (observed *after* the RLDP) are significantly slower rotators compared to the rapidly spinning accreting X-ray MSPs (observed *before* the RLDP). This difference in spin periods was first noted by Hessels (2008). However, only for MSPs with $B > 10^8$ G can the difference in spin periods be partly understood from regular magneto-dipole and plasma current spin-down of the recycled radio MSPs. In this section we investigate the RLDP effect on IMXBs and LMXBs and demonstrate different outcomes for BMSPs with CO and He WD companions. The purpose of the computations is to follow the spin evolution of the accreting X-ray millisecond pulsar (AXMSP) when the donor star decouples from its Roche-lobe, and to calculate the initial spin period of the recycled radio pulsar once the mass-transfer ceases. We model the main effect (the RLDP which causes the magnetosphere to expand dramatically) on the general spin evolution and ignore other, less known or somewhat hypothetical, dynamical effects during this epoch, such as warped disks, disk-magnetosphere instabilities, variations in the mass transfer rate caused by X-ray irradiation effects of the donor star, or transitions between a Keplerian thin disk and a sub-Keplerian, advection-dominated accretion flow (ADAF) (e.g. Spruit & Taam 1993; Nelson et al. 1997; Yi, Wheeler & Vishniac 1997; van Kerkwijk et al. 1998; Li & Wickramasinghe 1998; Locsei & Melatos 2004; Dai & Li 2006; Camero-Arranz et al. 2012; and references therein). Although many of these effects may be less important during the RLDP they could perhaps explain some of the frequent torque reversals observed in X-ray pulsars (Bildsten et al. 1997).

### 6.5.1 Onset of a propeller phase

In the final stages of the X-ray phase, when the donor star is just about to detach from its Roche-lobe, the mass-transfer rate decreases. This causes the ram pressure of the incoming flow to decrease whereby the magnetospheric boundary, and thus the coupling radius $r_{\mathrm{mag}}$, moves outward. When this boundary moves further out than the corotation radius ($r_{\mathrm{mag}} > r_{\mathrm{co}}$) given by:

$$
\begin{aligned}
r_{\mathrm{co}} &= \left( \frac{GM}{\Omega_\star^2} \right)^{1/3} \\
&\simeq 17\,\mathrm{km} \cdot P_{\mathrm{ms}}^{2/3} \left( \frac{M}{1.4\,M_\odot} \right)^{1/3}
\end{aligned}
\tag{59}
$$

a centrifugal barrier arises since the plasma flowing towards the neutron star couples to the field lines in a super-Keplerian orbit. The material is therefore presumably ejected away from the neutron star in this propeller phase (Illarionov & Sunyaev 1975). This ejection of material causes exchange of angular momentum between the now relatively fast spinning neutron star and the slower spinning material at the edge of the inner disk. The result is a braking torque which acts to slow down the spin of the pulsar. This causes the corotation radius to move outwards too and the subsequent evolution is determined by the rate at which these two radii expand relative to one another. However, during the RLDP $\dot{M}$ decreases too rapidly for $r_{\mathrm{co}}$ to keep up with the rapidly expanding $r_{\mathrm{mag}}$ and the equilibrium spin phase comes to an end.

In our model we assumed, in effect, that the inner edge of the accretion disk, $r_{\mathrm{disk}}$ follows $r_{\mathrm{mag}}$ and that the centrifugally expelled material has sufficient kinetic energy to be gravitationally unbound during the propeller phase when $r_{\mathrm{mag}} > r_{\mathrm{co}}$. These assumptions may not be entirely correct (Spruit & Taam 1993; Rappaport, Fregeau & Spruit 2004). For example, D'Angelo & Spruit (2010; 2011; 2012) recently demonstrated that accretion disks may be trapped near the corotation radius, even at low accretion rates. In this state the accretion can either continue to be steady or become cyclic. Furthermore, based on energy considerations they point out that most of the gas will not be ejected if the Keplerian corotation speed is less than the local escape speed. This will be the case if $r_{\mathrm{disk}} < 1.26\,r_{\mathrm{co}}$. However, this trapped state only occurs in a narrow region around the corotation radius ($\Delta r \ll r_{\mathrm{co}}$) and given that $r_{\mathrm{mag}}/r_{\mathrm{co}}$ often reaches a factor of $3-5$ in our models at the end of the RLDP we find it convincing that the RLDP effect of braking the spin rates of the recycled pulsars is indeed present (see Tauris (2012) for further details).

### 6.5.1.1 The magnetosphere–disk interaction

In our numerical calculations of the propeller phase we included the effect of additional spin-down torques, acting on the neutron star, due to both magnetic field drag on the accretion disk (Rappaport, Fregeau & Spruit 2004) as well as magnetic dipole radiation (see equation (45)), although these effects are usually not dominant. It should be noted that the magnetic stress in the disk is also related to the critical fastness parameter, $\omega_c$ (Ghosh & Lamb 1979b). We follow Tauris (2012) and write the total spin torque as:

$$
N_{\mathrm{total}} = n(\omega) \left( \dot{M} \sqrt{GM r_{\mathrm{mag}}}\, \xi + \frac{\mu^2}{9 r_{\mathrm{mag}}^3} \right) - \frac{\dot{E}_{\mathrm{dipole}}}{\Omega}
\tag{60}
$$

where

$$n(\omega) = \tanh\left(\frac{1-\omega}{\delta_\omega}\right) \tag{61}$$

is a dimensionless function, depending on the fastness parameter, $\omega = \Omega_\star/\Omega_K(r_{mag}) = (r_{mag}/r_{co})^{3/2}$, which is introduced to model a gradual torque change in a transition zone near the magnetospheric boundary. The width of this zone has been shown to be small (Spruit & Taam 1993), corresponding to $\delta_\omega \ll 1$ and a step function behaviour $n(\omega) = \pm 1$. In our calculations presented here we used $\delta_\omega = 0.002$, $\xi = 1$, $r_{disk} = r_{mag}$ and also assumed the moment of inertia, $I$, to be constant during the RLDP. The latter is a good approximation since very little material is accreted during this termination stage of the RLO.

#### 6.5.1.2   Radio ejection phase

As the mass-transfer rate continues to decrease, the magnetospheric boundary eventually crosses the light-cylinder radius, $r_{lc}$ of the neutron star given by:

$$r_{lc} = c/\Omega_\star \simeq 48\,\mathrm{km} \ \cdot \ P_{ms} \tag{62}$$

When $r_{mag} > r_{lc}$ the plasma wind of the pulsar can stream out along the open field lines, providing the necessary condition for the radio emission mechanism to turn on (e.g. Michel 1991; Spitkovsky 2008). Once the radio millisecond pulsar is activated the presence of the plasma wind may prevent further accretion from the now weak flow of material at low $\dot{M}$ (Kluzniak et al. 1988). As discussed by Burderi et al. (2001); Burderi, D'Antona & Burgay (2002) this will be the case when the total spin-down pressure of the pulsar (from magneto-dipole radiation and the plasma wind) exceeds the inward pressure of the material from the donor star, i.e. if: $\dot{E}_{rot}/(4\pi r^2 c) > P_{disk}$. However, it is interesting to notice that the recycled pulsar PSR J1023+0038 (Archibald et al. 2009) recently (within a decade) turned on its radio emission and it would indeed be quite a coincidence if we have been lucky enough to catch that moment – unless recycled pulsars evolve through a final phase with recurrent changes between accretion and radio emission modes.

#### 6.5.2   Slow RLDP

In Fig. 55 we have shown a model calculation of the RLDP of an LMXB with an original donor star mass of $1.1\,M_\odot$. The outcome was the formation of a BMSP with $P = 5.2\,\mathrm{ms}$ (assuming $B = 1.0 \times 10^8\,\mathrm{G}$) and a He WD companion of mass $0.24\,M_\odot$, orbiting with a period of 5.0 days. It has been argued by Tauris & Savonije (1999); Antoniadis et al. (2012) that the majority of the transfered material in some LMXBs is lost from the system, even for accretion at sub-Eddington rates. Therefore, we assumed an effective accretion efficiency of 30 per cent in our LMXB model, i.e. $\dot{M} = 0.30\,|\dot{M}_2|$, where $|\dot{M}_2|$ is the RLO mass-transfer rate from the donor star. The accreted[17] mass-transfer rate $\dot{M}$ is shown in the upper left panel.

The three phases: *equilibrium spin* (corresponding to $r_{mag} \simeq r_{co}$), *propeller phase* ($r_{co} < r_{mag} < r_{lc}$) and *radio pulsar* ($r_{mag} > r_{lc}$) are clearly identified in this figure. During the equilibrium spin phase the rapidly alternating sign changes of the torque (partly unresolved on the graph in the lower left panel) reflect small oscillations around the semi-stable equilibrium, corresponding to successive small episodes of spin-up and spin-down. The reason is that the relative location of $r_{mag}$ and $r_{co}$ depends on the small fluctuations in $\dot{M}$. Despite applying an implicit coupling scheme in the code and ensuring that our time steps during the RLDP (about $8 \times 10^4\,\mathrm{yr}$) are much smaller than the duration of the RLDP ($\sim 10^8\,\mathrm{yr}$), our calculated mass-transfer rates are subject to minor numerical oscillations. However, these oscillations could in principle be physical within the frame of our simple model. Examples of physical perturbations that could cause real fluctuations in $\dot{M}$, but on a much shorter timescale, are accretion disk instabilities and clumps in the transfered material from the donor star (our donor stars have active, convective envelopes during the RLDP).

---

[17]Strictly speaking this is the estimated mass-transfer rate received from the inner edge of the accretion disk. Some of this material is ejected during the propeller phase.

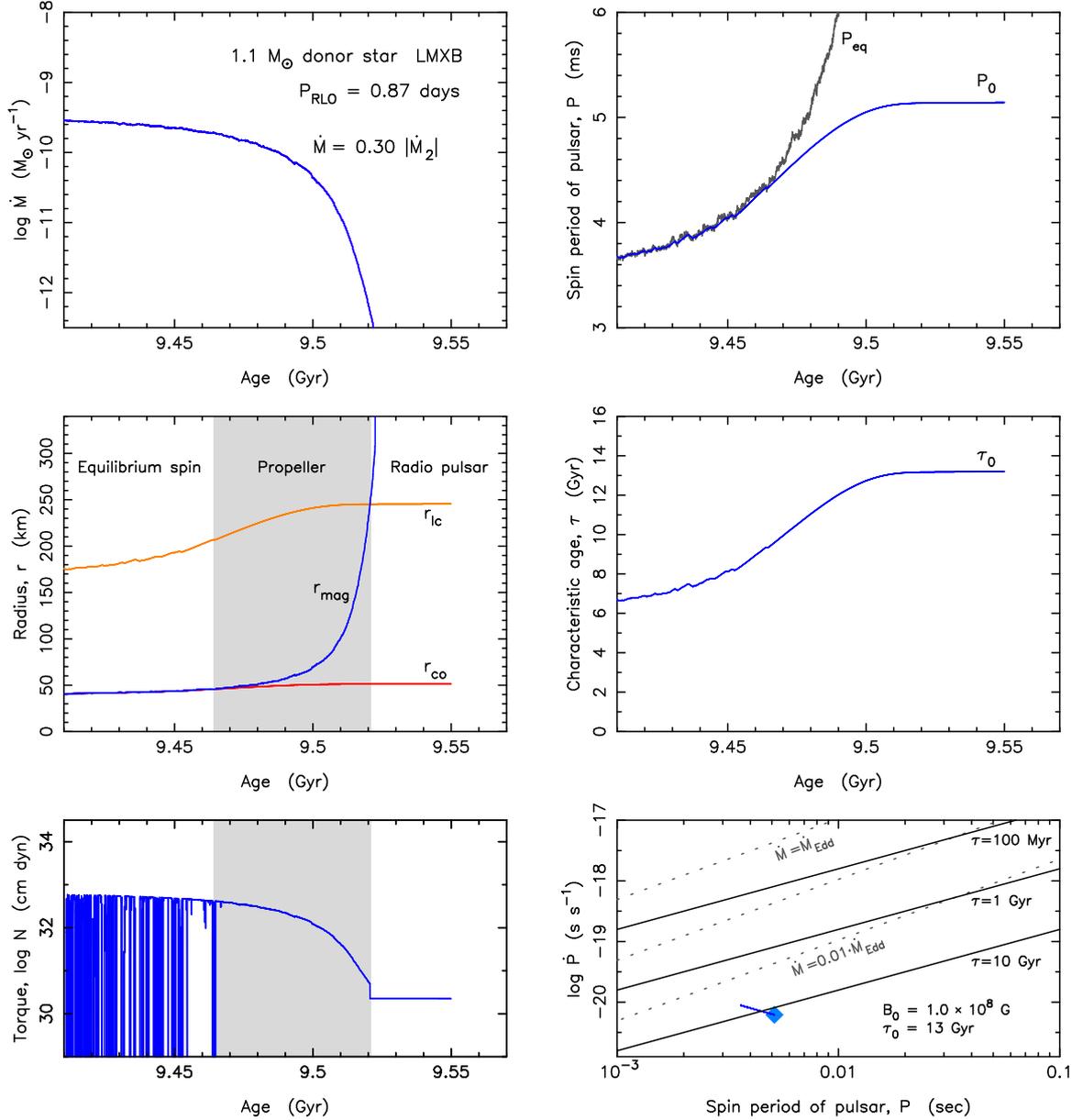

Figure 55: Example of a slow Roche-lobe decoupling phase (RLDP) and the final spin evolution of a pulsar formed with a He WD companion in an LMXB. The left side panels show: The accretion rate of the pulsar, $\dot{M}$ as a function of the total age of the donor star (top); the evolution of the location of the magnetospheric boundary, $r_{mag}$, the corotation radius, $r_{co}$ and the light-cylinder radius, $r_{lc}$ (centre); and the resulting braking torque (bottom). The right side panels show: The evolution of the pulsar spin period as a function of the total age of the donor star (top); the evolution of the pseudo characteristic age of the pulsar (centre); and the RLDP evolutionary track leading to the birth location in the $P\dot{P}$–diagram shown as a blue diamond (bottom). The three epochs of the RLDP: *equilibrium spin*, *propeller phase* and *radio pulsar* are easily identified in the left side panels. The propeller phase is grey shaded. In these model calculations we assumed $\sin\alpha = \phi = \xi = \omega_c = 1$ and $\delta_\omega = 0.002$. It is interesting to notice how the changes in the accretion rate, $\dot{M}$ affects the magnetospheric coupling radius, $r_{mag}$ which again affects the spin-down torque and the equilibrium spin period, cf. equations (44), (43) and (49). The reason why the spin of the pulsar (blue line, upper right panel) decouples from its equilibrium spin (grey line) is that the spin-down torque cannot transmit the deceleration on a timescale short enough for the pulsar to adapt to its new equilibrium. In this example, the duration of the RLDP is a significant fraction of the spin-relaxation timescale of the accreting neutron star and thus the RLDP has a momentous impact on the rotational evolution. The birth spin period, $P_0$ of this recycled pulsar is seen to deviate substantially from its original $P_{eq}$ calculated at the onset of the RLDP. Furthermore, the effect on the characteristic age for this BMSP is also quite significant. During the 56 Myr of the RLDP it doubles to become $\tau = 13$ Gyr at *birth*. Hence, many MSPs which appear to be old according to their $\tau$ may in fact be quite young. See text for further discussions.

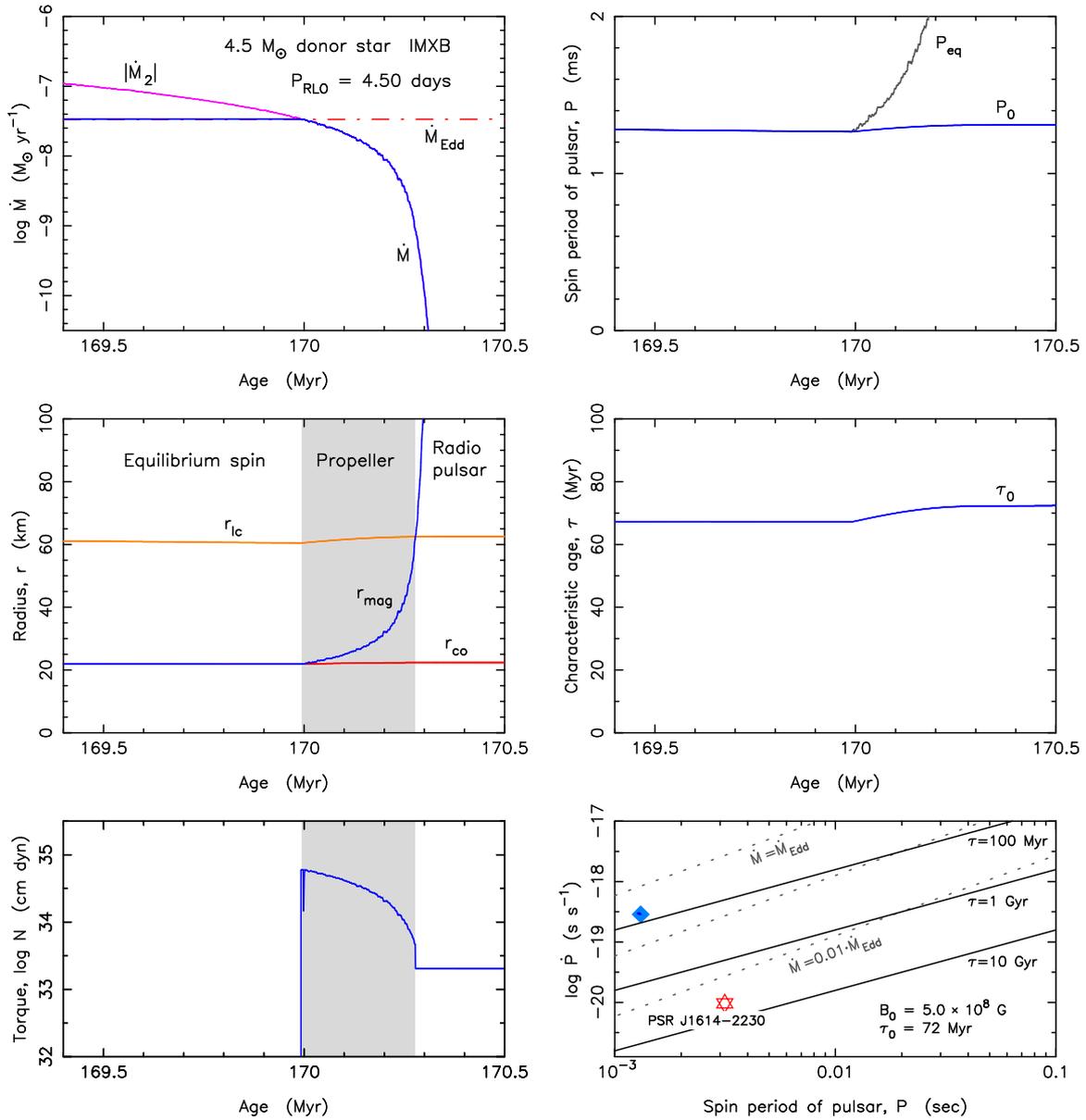

Figure 56: Example of a rapid Roche-lobe decoupling phase (RLDP) in an IMXB using our mass-transfer progenitor model of PSR J1614−2230. The spin evolution of the pulsar is shown during the RLDP when the IMXB donor star terminated its mass transfer. For a description of the various panels and their labels, see Fig. 55. In the $P\dot{P}$–diagram in the lower right panel the birth location, immediately following the RLDP, of the recycled radio pulsar is shown as a blue diamond for $\sin\alpha = \phi = \xi = \omega_c = 1$. The red star indicates the presently observed location of PSR J1614−2230. In Section 6.9.1 we argue that PSR J1614−2230 is more likely to have evolved with $\sin\alpha < 1$, $\phi > 1$ and $\omega_c < 1$ which results in a birth location much closer to its present location in the $P\dot{P}$–diagram. In IMXBs where the donor star evolves on a relatively short timescale, the RLDP is rapid. Therefore, the spin evolution of the accreting X-ray pulsar (see blue line in upper right panel) freezes up near the original value of $P_{eq}$, i.e. the recycled radio pulsar is born with a spin period, $P_0$ which is close to the equilibrium spin period of the accreting X-ray pulsar calculated at the time when the donor star begins to decouple from its Roche-lobe. In this IMXB model the mass-transfer rate, $\dot{M} > \dot{M}_{Edd}$ and the accretion onto the neutron star is therefore limited by the Eddington mass-accretion rate (red dot-dashed line in the upper left panel). This explains the why the spin of the pulsar, $P = P_{eq} \simeq const.$ (top right panel) until $\dot{M} < \dot{M}_{Edd}$, as opposed to the situation in Fig. 55.

At some point the equilibrium spin is broken. Initially, the spin can remain in equilibrium by adapting to the decreasing value of $\dot{M}$. However, further into the RLDP the result is that $r_{\rm mag}$ increases on a timescale faster than the spin-relaxation timescale, $t_{\rm torque}$ at which the torque can transmit the effect of deceleration to the neutron star, and therefore $r_{\rm mag} > r_{\rm co}$ (see Tauris 2012; and supporting online material therein). During the propeller phase the resultant accretion torque acting on the neutron star is negative, i.e. a spin-down torque. When the radio pulsar is activated and the accretion has come to an end, the spin-down torque is simply caused by the magneto-dipole radiation combined with the pulsar wind (see Section 6.4.2.1).

The aim here is to calculate the initial spin period of the recycled radio pulsar, $P_0$ once the mass-transfer ceases. In this LMXB model calculation the value of the spin period increased significantly during the RLDP. The reason for this is that the duration of the propeller phase (i.e. the RLDP) is a substantial fraction of the spin-relaxation timescale, $t_{\rm torque}$. Using equation (58) we find $t_{\rm torque} = 195\,{\rm Myr}$ whereas the RLDP lasts for $t_{\rm RLDP} \simeq 56\,{\rm Myr}$ which is a significant fraction of $t_{\rm torque}$ ($t_{\rm RLDP}/t_{\rm torque} = 0.29$). Therefore, the RLDP has quite a significant impact on the spin evolution of the neutron star. It is seen that the spin period increases from 3.7 ms to 5.2 ms during this short time interval, i.e. the pulsar loses 50 per cent of its rotational energy during the RLDP. As shown by Tauris (2012) this RLDP effect is important for understanding the apparent difference in spin period distributions between AXMSPs[18] and radio MSPs (Hessels 2008). The grey line labelled "$P_{\rm eq}$" in the upper right panel reflects the evolution of $P$ in case the neutron star was able to instantly re-adjust itself to the changing $P_{\rm eq}$ (equation 49) during the RLDP. However, the neutron star possesses a large amount of rotational inertia and such rapid changes in $P$ are not possible to transmit to the neutron star given the limited torque acting on it. Therefore the spin evolution terminates at the end of the blue line labelled "$P_0$". The small fluctuations of $P_{\rm eq}$ during the equilibrium spin phase reflect fluctuations in $\dot{M}(t)$.

The RLDP effect also causes a large impact on the characteristic age, $\tau_0$, of the radio pulsar at birth. In the central right panel we show how $\tau_0$ doubles in value from 6.7 Gyr to 13 Gyr during the RLDP (see also the solid line ending at the blue diamond in the $P\dot{P}$–diagram in the lower right panel). The characteristic age at birth is given by: $\tau_0 \equiv P_0/2\dot{P}_0$, where $\dot{P}_0$ can be estimated from the assumed B-field of the pulsar at this epoch. Actually, the "characteristic age" of the pulsar during the RLDP (which is plotted in the central right panel) is a pseudo age which is only relevant at the end point ($\tau_0$) when the accretion has stopped. In order to construct the track leading to $\tau_0$ we assumed $B^2 \propto P\dot{P} = const.$ from equation (46), leading to $\tau \propto P^2$. During the RLDP it is a good approximation to assume a (final) constant value of the neutron star surface magnetic flux density, $B$ since 99 per cent of the accretion onto the neutron star occurs before the final termination stage of the RLDP. In this case we assumed a final B-field strength of $B = 1 \times 10^8\,{\rm G}$ – a typical value for BMSPs with He WD companions. This example clearly reflects why characteristic ages of MSPs are untrustworthy as true age indicators since here the MSP is born with $\tau_0 \simeq 13\,{\rm Gyr}$ (see also Tauris (2012)). This fact is important since the characteristic ages of recycled pulsars are often compared to the cooling ages of their white dwarf companions (e.g. see discussion in van Kerkwijk et al. 2005).

### 6.5.3   Rapid RLDP ($P_{\rm eq}$ freeze-up)

In Fig. 56 we have plotted the RLDP during the final stages of the IMXB phase AB using our Case A model calculation of PSR J1614−2230 presented in Paper I. The initial conditions were a 4.5 $M_\odot$ donor star in an IMXB with an orbital period of 2.20 days. The final system was a BMSP with a 0.50 $M_\odot$ CO WD orbiting a 1.99 $M_\odot$ recycled pulsar with an orbital period of 8.7 days – resembling the characteristics of the PSR J1614−2230 system. The accretion rate onto the neutron star, $\dot{M}$ (see blue line in upper left panel) is Eddington limited (red dot-dashed line) and hence $\dot{M} = \min\left(|\dot{M}_2|, \dot{M}_{\mathrm{Edd}}\right)$ where $|\dot{M}_2|$ represents the mass-transfer rate from the donor star (purple line). When the age of the donor star is $\sim 170.0\,\mathrm{Myr}$ the mass-transfer rate becomes sub-Eddington and $\dot{M}$ decreases with $|\dot{M}_2|$. From this point onwards $r_{\mathrm{mag}}$ increases as expected, as a result of the decreasing ram pressure, and it increases at a rate where $r_{\mathrm{co}}$ cannot keep up with it (see central panel, left column) and therefore the pulsar enters the propeller phase. After less than 0.3 Myr the accretion phase, and thus the propeller phase, is terminated when $r_{\mathrm{mag}} > r_{\mathrm{lc}}$ and the radio ejection mechanism is activated.

---

[18]The majority of the AXMSPs have very small $P_{\rm orb}$ (typically 1–2 hours) and substellar companions ($<0.08\,M_\odot$) and hence most of these AXMSPs are not true progenitors of the general population of radio MSPs. However, the important thing is that the pulsars are spun up to rotational periods which seem to be almost independent of orbital period up to 200 days, as seen in the central panel of Fig. 51.

During the equilibrium spin phase the net accretion torque acting on the pulsar is close to zero. This is not only due to our transition zone approximation (equation (61)) where the torque vanishes when $r_{\mathrm{mag}} \simeq r_{\mathrm{co}}$. If the transition zone was arbitrary thin ($\delta_\omega \lll 1$), and $n(\omega)$ would be a step-like function going from $+1$ to $-1$ (i.e. switch mode), there would be rapidly alternating sign changes of the torque during the equilibrium spin phase. Whereas small fluctuations in $\dot{M}$ would lead to such rapid torque oscillations for sub-Eddington accretion (as seen in Fig. 55) the sharply defined Eddington limit applied here prevents such fluctuations during the equilibrium spin phase. However, the time-averaged torque would be close to zero in any case.

In this IMXB model calculation the value of the spin period only increased very little during the RLDP. The reason for this is that the duration of the RLDP ($t_{\mathrm{RLDP}} \simeq 0.28\,\mathrm{Myr}$) is relatively short compared to the spin-relaxation timescale ($t_{\mathrm{torque}} = 6.8\,\mathrm{Myr}$). The small ratio $t_{\mathrm{RLDP}}/t_{\mathrm{torque}} \simeq 0.04$ at the onset of the RLDP causes the spin period to "freeze" at the original value of $P_{\mathrm{eq}}$ (see also Ruderman, Shaham & Tavani 1989). Note, the values quoted above as well as rough values of $t_{\mathrm{torque}} = (2\pi/P)\,I/N$ and $r_{\mathrm{mag}} \simeq r_{\mathrm{co}} \simeq 22\,\mathrm{km}$ can be checked by reading numbers from the plots and using equations (44) and (59). The relevant numbers are: $B = 5 \times 10^8\,\mathrm{G}$ (a typical value for BMSPs with CO WD companions), $\dot{M} \simeq \dot{M}_{\mathrm{Edd}}$, $M \simeq 1.99\,M_\odot$, $P_{\mathrm{eq}} \simeq 1.28\,\mathrm{ms}$ and $\log N = 34.8$. For the spin-up process discussed here we assumed again $\sin\alpha = \phi = \xi = \omega_c = 1$.

### 6.5.3.1 Rapid RLDP with a helium star donor

As seen in Fig. 58 the helium star donor evolves on a short nuclear timescale compared to a normal hydrogen-rich donor. This leads to a rapid accretion turn-off lasting only a few $10^4\,\mathrm{yr}$ and therefore $t_{\mathrm{RLDP}}/t_{\mathrm{torque}} \ll 1$, which also in these systems causes $P_0$ to "freeze" at the original value of $P_{\mathrm{eq}}$.

### 6.5.4 RLDP in IMXB vs LMXB systems

The major difference between the RLDP in an IMXB (often leading to a BMSP with a CO WD) and the RLDP in an LMXB (leading to a BMSP with a He WD) is the time duration of the RLDP relative to the spin-relaxation timescale, i.e. the $t_{\mathrm{RLDP}}/t_{\mathrm{torque}}$-ratio. The mass-transfer in X-ray binaries proceeds much faster in IMXBs compared to LMXBs. Hence, also the termination of the RLDP is shorter in IMXBs, and therefore the $t_{\mathrm{RLDP}}/t_{\mathrm{torque}}$-ratio is smaller, compared to LMXBs. The conclusion is that, in general, we expect a significant RLDP effect only in LMXBs. (It may be possible, however, that the RLDP effect could play a more significant role in some IMXB systems if they leave behind a neutron star with a relatively high B-field, thereby decreasing $t_{\mathrm{torque}}$).

### 6.6 Observed spin-period distributions

We now investigate if the empirical pulsar spin data can be understood in view of the theoretical modelling presented in the last couple of sections.

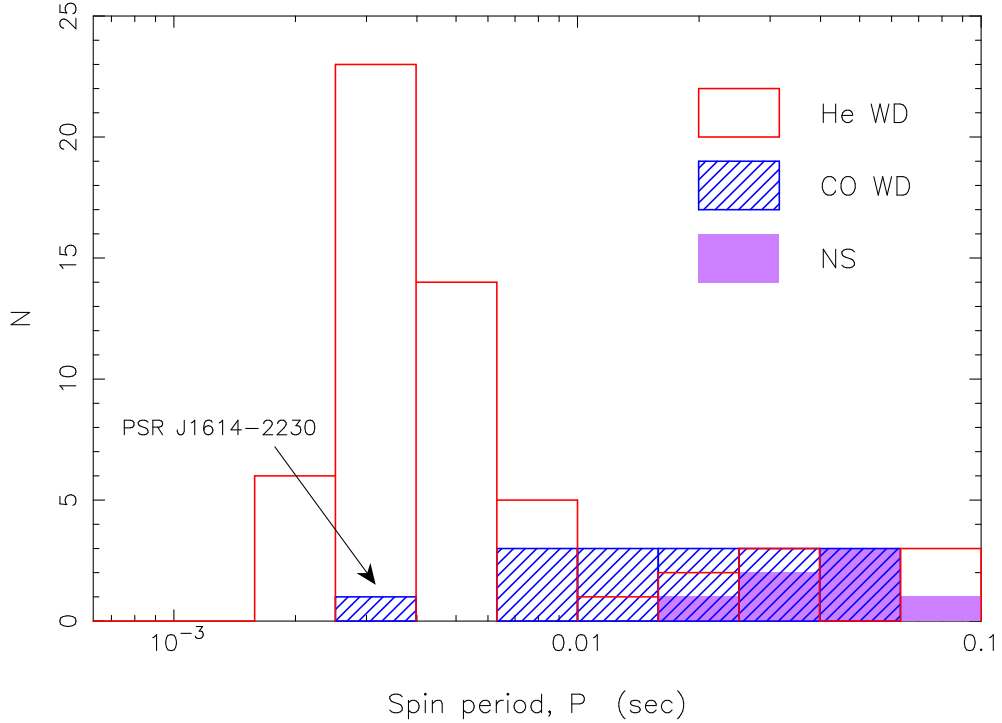

Figure 57: The observed spin period distribution of recycled pulsars in the Galactic disk with $P < 100$ ms for pulsars with He WD and CO WD companions. The median spin periods for these populations are 3.9 ms and 16 ms, respectively. For comparison we have also shown the spin distribution of those recycled pulsars with neutron star companions and $P < 100$ ms.

### 6.6.1 Recycled pulsars with WD companions

In Fig. 57 we have plotted histograms of the observed spin period distribution of recycled pulsars with He WD and CO WD companions. There are 57 known recycled pulsars in the Galactic disk with a He WD companion and a spin period $P < 100$ ms (with or without a measured $\dot{P}$) and these pulsars have a median spin period of 3.9 ms (the average spin period is 10 ms). The median spin period of the 16 known recycled pulsars with a CO WD companion and $P < 100$ ms, however, is 16 ms (the average value is 23 ms). All observed BMSPs have lost rotational energy due to magneto-dipole radiation and a pulsar wind since they appeared as recycled pulsars. Hence, their current spin period, $P$ is larger than their $P_{eq}$. Unfortunately, we have no empirical constraints on the spin-down torque acting on recycled radio MSPs and therefore their braking index remains unknown, cf. Section 6.8. Furthermore, in Section 6.5 we demonstrated that BMSPs formed in LMXBs lose rotational energy during the RLDP. Correcting for both of these spin-down effects we here simply assume $P = \sqrt{2}\,P_{eq}$. The resulting median values of $P_{eq}$ are thus found to be 2.8 ms and 11 ms, respectively. For a $1.4\,M_\odot$ neutron star these median equilibrium spin periods correspond to $\Delta M_{eq} = 0.06\,M_\odot$ and $\Delta M_{eq} = 0.01\,M_\odot$ for pulsars with a He WD or a CO WD companion, respectively. In other words, to explain the observed period distribution of these two classes of recycled pulsars we conclude that those pulsars with CO WD companions typically have accreted less mass by a factor of 6, compared to the pulsars with He WD companions. This conclusion remains valid for other scalings between the present spin period, $P$ and the initial spin period, $P_{eq}$ – even for assuming $P = P_{eq}$. We conclude from Fig. 57 that BMSPs with CO WDs usually have evolved via early Case B or Case C RLO (leading to CE evolution). The reason is that these two paths involve mass transfer on a much shorter timescale, and hence lead to smaller amounts of transfered mass, leading to less effective recycling and thereby longer spin periods, as seen among the observed pulsars in Fig. 57 (see also Section 6.7). Furthermore, the BMSPs with CO WDs also have relatively large values of $\dot{P}$. This reflects larger values of $B$, which is expected since less mass was accreted in these systems, cf. Section 6.4.4.1. The only exception known from the description outlined above is PSR J1614−2230 which is discussed in Section 6.9.

For BMSPs with He WD companions, we notice from the orbital period distribution in Fig. 51 (central panel) that the majority of these systems seem to have evolved through Case B RLO with an initial LMXB orbital period of $P_{\rm orb} > 1$ day (Tauris & Savonije 1999; Podsiadlowski, Rappaport & Pfahl 2002; Tauris 2011; and references therein).

### 6.6.2 Double neutron star systems

In Fig. 57 we have also plotted the spin distribution of recycled pulsars with neutron star companions and $P < 100$ ms. These systems originate from high-mass X-ray binaries (HMXBs), e.g. Tauris & van den Heuvel (2006). Their spin distribution is skewed to larger periods compared to the recycled pulsars with CO WD companions. This is expected since their massive stellar progenitors evolved even more rapidly than the donor stars of the IMXBs. It is generally believed that double neutron star systems evolved via a CE and spiral-in phase since the mass transfer in HMXBs is always dynamically unstable – see however Brown (1995); Dewi, Podsiadlowski & Sena (2006) for an alternative formation mechanism via a double-core scenario.

### 6.6.3 Formation of recycled pulsars with $P_{\rm orb} > 200$ days

As mentioned earlier (see also Fig. 51), pulsars with $P_{\rm orb} > 200$ days all have slow rotational spins. This fact could be explained if they had accreted little mass. If their progenitors were wide-orbit LMXBs they had experienced continuous super-Eddington mass transfer on a short timescale of the order 10 Myr (Tauris & Savonije 1999). However, even if their accretion rates had been restricted to the Eddington limit these accreting pulsars could still have received up to $0.3\,M_\odot$. This amount is 10 times larger than needed to spin-up a pulsar to 5 ms according to equation (56). Therefore, to explain the slow spin periods of these wide-orbit recycled pulsars one may speculate that the accretion process was highly inefficient due to, for example, enhanced accretion disk instabilities (van Paradijs 1996; Dubus et al. 1999). This could be related to the advanced evolutionary stage of their donor stars on the red giant branch (RGB). These stars had deep convective envelopes which may lead to enhanced clump formation in the transfered material.

Alternatively, one may ask if these systems descended from wide-orbit IMXBs which underwent a CE and a mild spiral-in. In this case their companion stars should be fairly massive CO WDs ($\geq 0.5\,M_\odot$, see fig. 1 in Paper I). However, quite a few of these wide-orbit recycled pulsars seem to have relatively low-mass He WDs according to their mass functions. In fact it was pointed out by Tauris (1996); Tauris & Savonije (1999); Stairs et al. (2005) that some of those WD masses may even be significantly less massive than expected from the $(M_{\rm WD}, P_{\rm orb})$-relation which applies to He WDs descending from LMXBs, although this analysis still has its basis in small numbers. It is therefore important to observationally determine the masses of the WDs in these wide-orbit pulsars systems and settle this evolutionary question.

### 6.7 Spinning up post common envelope pulsars

An interesting question is if our derived relation between pulsar spin and accreted mass (equation 56) can also explain the observed spin periods of the BMSPs with *massive* WD companions and short orbital periods, i.e. systems we expect to have evolved through a CE evolution. Let us consider the important system PSR J1802-2124 recently observed by Ferdman et al. (2010). The nature of its $0.78 \pm 0.04\,M_\odot$ CO WD companion, combined with an orbital period of 16.8 hr and a post-accretion pulsar mass of only $1.24 \pm 0.11\,M_\odot$ makes this one of our best cases for a system which evolved via a CE and spiral-in phase. The present spin period of the pulsar is 12.6 ms. Assuming an initial equilibrium spin period of 10 ms we find from equation (56) that $\Delta M_{\rm eq} \simeq 0.01\,M_\odot$ which is a much more acceptable value for a system undergoing CE-evolution (compared to the $0.10\,M_\odot$ needed for spinning up a BMSP with a He WD companion to a spin period of a few ms). However, one still has to account for the $0.01\,M_\odot$ accreted. It is quite likely that this accretion occurred *after* the CE-phase and we now consider this possibility.

### 6.7.1 Case BB RLO in post common envelope binaries

The total duration of the CE-phase itself is probably so short ($< 10^3$ yr, e.g. Podsiadlowski 2001; Passy et al. 2012; Ivanova et al. 2013) that no substantial accretion onto the neutron star is possible with respect to changing its rotational dynamics – even if the envelope of the donor star is not ejected from the system until long after the in-spiral of the secondary star which occurs on a dynamical timescale of a few $P_{orb}$. Thus we are left with two possibilities: 1) the recycling of the pulsar occurs during the subsequent stable RLO of a naked helium star following the CE (Case BB RLO), or 2) the recycling occurs due to wind accretion from the naked helium or carbon-oxygen core left behind by the donor star after it loses its envelope in the CE-phase.

In Fig. 58 we show one of our model calculations of Case BB RLO. In our example we assumed a $1.40\,M_\odot$ helium star ($Y = 0.98$) as the donor and a $1.35\,M_\odot$ neutron star as the accretor. The giant phase of a helium star is short lived and in this case the mass transfer lasts for about 170.000 yr. (This time interval, however, is much longer than the CE-event.) Assuming the accretion onto the neutron star to be Eddington limited will therefore lead to an accretion of only about $7 \times 10^{-3}\,M_\odot$. Although this is a small amount, it is still sufficient to spin up the recycled pulsar to a period of 14 ms, which is even faster than the observed median spin period of BMSPs with CO WD companions (see Fig. 57). Assuming the accretion efficiency to be only 30 per cent of $\dot{M}_{Edd}$ would still spin up the pulsar to 36 ms. A helium donor star with a lower mass can even spin up the pulsar below 14 ms. We found that for a $1.10\,M_\odot$ helium star donor the neutron star can accrete 50 per cent more (resulting in $P_{eq} = 11$ ms) due to the increased nuclear timescale of the shell burning of a lower mass helium star. Hence, based on rotational dynamics we conclude that Case BB RLO is a viable formation channel to explain most of the observed BMSPs with CO WD companions. A more systematic study of Case BB RLO for various system configurations is needed in order to find the minimum possible spin period of BMSPs with CO WD companions formed via this formation channel.

It should be noted that a pulsar may also be recycled via Case BA RLO, i.e. from RLO which is initiated while the companion star is still on the helium main sequence (see Dewi et al. 2002; for detailed calculations). In this case the mass-transfer phase will last up to ten times longer, compared to Case BB RLO, providing more than sufficient material to recycle the pulsar effectively. However, in order to initiate RLO while the helium star donor is still burning core helium on its main sequence requires a tight, fine-tuned interval of orbital separation following the common envelope. Typical orbital periods needed are 1-2 hours. Since Case BB RLO is initiated for a much wider interval of larger orbital periods (up to several tens of days) we expect many more systems to evolve via Case BB RLO compared to Case BA RLO.

### 6.7.2 Wind accretion prior to Case BB RLO

Finally, we tested if the neutron star could acquire any significant spin-up from wind accretion in the epoch between post-CE evolution and Case BB RLO. To produce CO or ONeMg WD remnants via Case BB RLO one would expect a typical helium star mass of $1.1 - 2.2\,M_\odot$ prior to the mass transfer. These stars have luminosities of $\log(L/L_\odot) = 2.5 - 3.3$ (Langer 1989) and spend less than 10 Myr on the He-ZAMS before they expand and initiate Case BB RLO. Their mass-loss rates are of the order $\dot{M}_{wind} \simeq 10^{-10 \pm 0.6}\,M_\odot\,\mathrm{yr}^{-1}$ (Jeffery & Hamann 2010). Therefore, even if as much as 10 per cent of this ejected wind mass was accreted onto the neutron star it would accrete a total of at most $\sim 10^{-4}\,M_\odot$ prior to the RLO. We therefore conclude that wind accretion prior to Case BB RLO is negligible for the recycling process of BMSPs with CO WD companions.

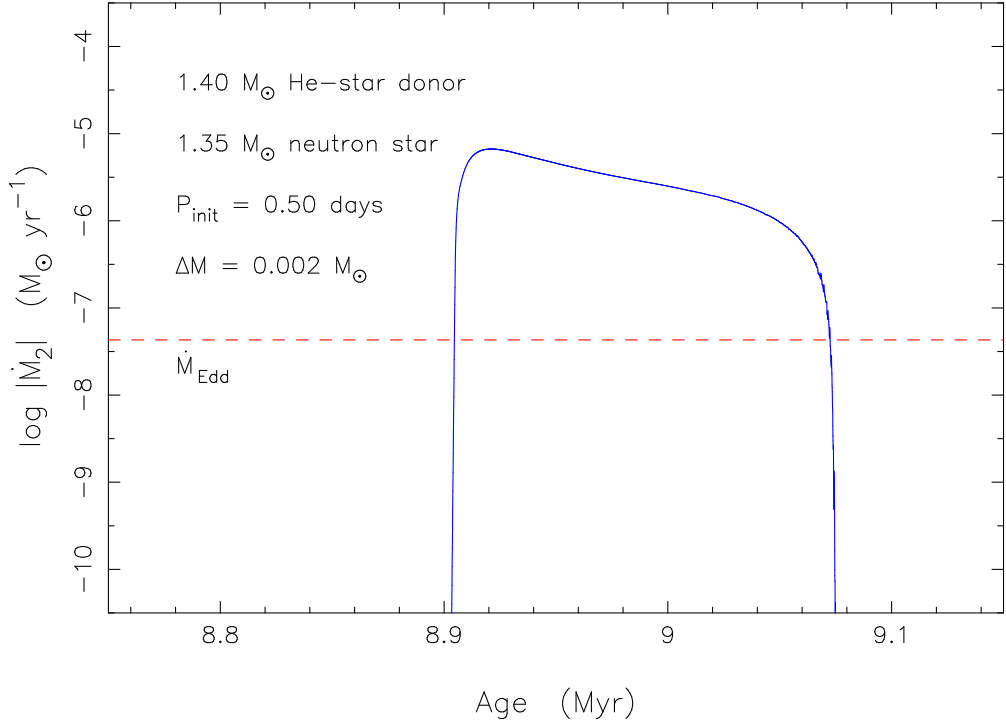

Figure 58: A model calculation of Case BB RLO. The plot shows mass-transfer rate as a function of age for a $1.4\,M_\odot$ helium star donor. The initial orbital period is 0.50 days and the neutron star mass is $1.35\,M_\odot$. After the RLO the donor star settles as a $0.89\,M_\odot$ CO WD orbiting the recycled pulsar with an orbital period of 0.95 days. The mass-transfer phase only lasts for about 170.000 yr. Even if the neutron star only accretes $\sim 0.002\,M_\odot$ (assuming an accretion efficiency of just 30%) it is still able to spin up the pulsar to a spin period of 36 ms. If the neutron star accretes all matter at the Eddington limit it can be spun up to 14 ms. A comparison can be made with the observed spin periods of BMSPs with CO WD companions in the bottom panel of Fig. 51.

### 6.7.3 Will the progenitor binaries make it to Case BB RLO?

Before drawing conclusions about the Case BB RLO we should verify that the progenitor binary survives to follow this path. In Paper I we argued Case BB RLO does not work for PSR J1614−2230 because the envelope of the CO WD progenitor star is too tightly bound to be ejected during the spiral-in phase[19]. Is that the case for all BMSPs? The answer is no. We find that if the post-CE orbital period (i.e. the observed $P_{orb}$ of a BMSP) is about 0.5 days or less then sufficient orbital energy will be released during spiral in to energetically allow for a successful ejection of the envelope of the original $5-7\,M_\odot$ red giant donor star. The total (absolute) binding energy of the envelope of such a star is $1.3-2.2 \times 10^{48}$ ergs which is equivalent to an orbital energy corresponding roughly to $P_{orb} \leq 0.5$ days. Since we must require $\Delta E_{orb} > E_{env}$ to avoid a merger we therefore conclude that Case BB RLO is in principle a viable formation channel for BMSPs with CO WD companions if these systems have short orbital periods. As argued earlier we consider PSR J1802−2124 (which has $P_{orb} = 16.8$ hours) as a strong case for a BMSP which evolved though a common envelope. Whether or not one can finetune the Case BB RLO evolution to account for the recycling of this pulsar, or if the recycling was due to wind accretion from the exposed post-CE CO core of an AGB star or, thirdly, if accretion-induced collapse of an ONeMg WD was at work, is an interesting question to address.

---

[19]We even considered a conservative Case BB RLO evolution which is far from valid given that the mass-transfer rate is highly super-Eddington, see Fig. 58 above.

## 6.8 True ages of millisecond pulsars

Knowledge of the true ages of recycled radio pulsars is important for comparing the observed population with the properties expected from the spin-up theory outlined in Section 6.4. All radio pulsars lose rotational energy with age and the braking index, $n$ is given by (Manchester & Taylor 1977):

$$\dot{\Omega} \propto -\Omega^n \tag{63}$$

which yields (for $n$ constant): $n = \Omega\ddot{\Omega}/\dot{\Omega}^2$. This deceleration law can also be expressed as: $\dot{P} \propto P^{2-n}$ and hence the slope of a pulsar evolutionary track in the $P\dot{P}$–diagram is simply given by: $2 - n$. Depending on the physical conditions under which the pulsar spins down $n$ can take different values as mentioned earlier. For example:

$$
\begin{array}{ll}
\text{gravitational wave radiation} & n = 5 \\
\text{B-decay or alignment or multipoles} & n > 3 \\
\text{perfect magnetic dipole} & n = 3 \\
\text{B-growth/distortion or counter-alignment} & n < 3
\end{array}
\tag{64}
$$

The combined magnetic dipole and plasma current spin-down torque may also result in $n \neq 3$ (Contopoulos & Spitkovsky 2006). A simple integration of equation (63) for a *constant* braking index ($n \neq 1$) yields the well-known expression:

$$t = \frac{P}{(n-1)\dot{P}} \left[ 1 - \left( \frac{P_0}{P} \right)^{n-1} \right] \tag{65}$$

where $t$ is the so-called true age of a pulsar, which had an initial spin period $P_0$ at time $t = 0$.

### 6.8.1 Isochrones in the $P\dot{P}$–diagram

Unfortunately, one cannot directly use the equation above to obtain evolutionary tracks in the $P\dot{P}$–diagram (even under the assumption of a constant $n$ and a known initial spin period, $P_0$). For chosen values of $t$, $n$ and $P_0$ one can find a whole family of solutions $(P, \dot{P})$ to be plotted as an isochrone, see the dot-dashed line in Fig. 59 (and see also Kiziltan & Thorsett 2010). However, there is only one point which is a valid solution for a given pulsar and the above equation does reveal which point is correct. The problem is that the variables in equation (65) are not independent and we do not know a priori the initial spin period derivative, $\dot{P}_0$. The evolution of the spin period is a function of both $t$, $n$, the initial period and its time derivative, i.e. $P(t, n, P_0, \dot{P}_0)$. Here, for simplicity, we assume $n$ to be constant. To determine $\dot{P}_0$ one must make assumptions about the initial surface magnetic flux density, $B_0$, the initial magnetic inclination angle, $\alpha_0$ and the initial ellipticity, $\varepsilon_0$ (the geometric distortion which is relevant for rotational deceleration due to gravitational wave radiation). Once $\dot{P}_0$ is known, one can combine equation (65) with the deceleration law, e.g. $\dot{P}P^{n-2} = const.$ (equation 63), and solve by integration for the evolution of a given pulsar. This way one can produce isochrones, for example as a function of the braking index for pulsars with the same given value of $\dot{P}_0$.

In Fig. 59 we have plotted pulsar isochrones for a pulsar with $P_0 = 2.5\,\text{ms}$ and $\dot{P}_0 = 6.26 \times 10^{-20}$ (e.g., corresponding to $B_0 = 1.15 \times 10^8\,\text{G}$ for $M = 1.4\,M_\odot$, $\sin\alpha_0 = 1$ and $\varepsilon_0 = 0$) by varying the value of the braking index such that $1 \leq n \leq 15$. Our numerical calculations were confirmed by analytic calculations for the special cases where $n = 2, 3$ or $5$ (see small circles at the $t = 12\,\text{Gyr}$ isochrone). The black dot indicates where the two 6 Gyr isochrones, calculated for $n = 3$ or $\dot{P}_0 = 6.26 \times 10^{-20}$, respectively, cross each other for a common solution and also in conjunction with the intersection of the evolutionary track calculated for the same values of $n$ and $\dot{P}_0$.

### 6.8.2 Characteristic versus true ages of MSPs

Introducing the characteristic age of a pulsar, $\tau \equiv P/(2\dot{P})$ one finds the relation between $\tau$ (the observable) and $t$:

$$\log \tau = \log t + \log \left( \frac{n-1}{2} \right) - \log \left( 1 - (\frac{P_0}{P})^{n-1} \right) \tag{66}$$

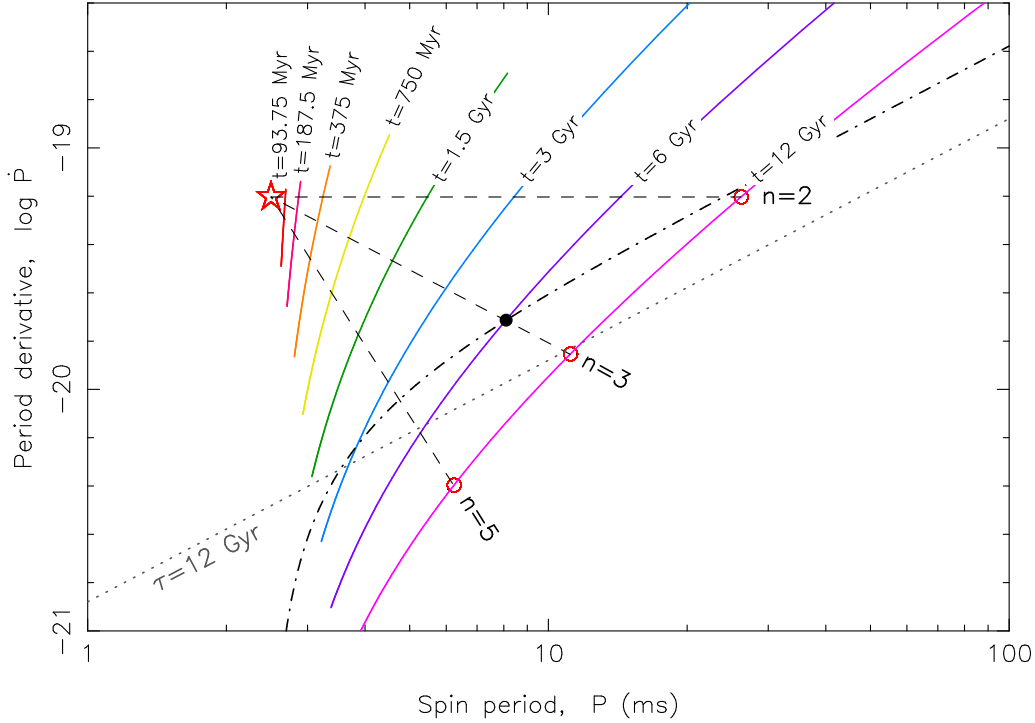

Figure 59: Isochrones of pulsar evolution for a pulsar with known initial $(P_0, \dot{P}_0)$ for different values of a constant braking index, $n$. Along any given isochrone $n$ increases continuously from $n = 1$ (top) to $n = 15$ (bottom). The initial position of the recycled test pulsar is shown by a red star and the dashed lines show evolutionary tracks for $n = 2$, 3 or 5, respectively. The dotted line yields a characteristic age, $\tau \equiv P/(2\dot{P})$ of 12 Gyr. This line would intersect with the 12 Gyr isochrone exactly at $n = 3$ if $P_0 \ll P$, which is not completely fulfilled. The dot-dashed isochrone line is a general solution to equation (65) for $n = 3$, $t = 6$ Gyr and $P_0 = 2.5$ ms, but unknown $\dot{P}_0$.

for evolution with a constant braking index, $n$. The asymptotic version of this relation (as $t \to \infty$ and $P \gg P_0$) is given by:

$$\log \tau = \begin{cases} \log t + \log 2 & \text{for } n = 5 \\ \log t & \text{for } n = 3 \\ \log t - \log 2 & \text{for } n = 2 \end{cases} \tag{67}$$

In Fig. 60 we have plotted $\tau$ as a function of $t$ on log-scales for braking indices of $n = 2$, 3 and 5, respectively. Initially, when $t$ is small and $P \simeq P_0$, $\tau$ is always greater than $t$. After a certain timescale $\tau$ can either remain larger or become smaller than $t$, depending on $n$. For $n = 5$ ($n > 3$) we always have $\tau > t$, for $n = 3$ we have $\tau \geq t$ and only for $n = 2$ ($n < 3$) we have the possibility that $\tau$ can be either greater or smaller than $t$. Having $n = 2$ corresponds to $\dot{P}$ being a constant, and solving for $\tau = t$ yields $P = 2P_0$ and $t = P_0/\dot{P}$. We notice that for a reasonable interval of braking index values $2 \leq n \leq 5$ the observed characteristic age will never deviate from the true age by more than a factor of two when $P \gg P_0$. However, to reach $P \gg P_0$ may take several Hubble timescales if $B_0$ is low.

The characteristic age at birth (after recycling) is given by: $\tau_0 = P_0/(2\,\dot{P}_0)$ and combining with equation (47) we get:

$$\tau_0 = \frac{P_0^2\,k^2}{2B_0^2} \tag{68}$$

where the constant $k = 9.2 \times 10^{18}\,\text{G s}^{-1/2}$ for $\sin\alpha_0 = 1$ and $M = 1.4\,M_\odot$. This expression verifies the well-known result that MSPs that are born relatively slowly spinning (large $P_0$ value) and/or have a relatively weak initial $B$-field (small $B_0$ value) are also those MSPs which are born with characteristic ages which differ the most from their true ages as shown in Fig. 60 (see also Kiziltan & Thorsett 2010). In other words, those pulsars with smallest values of $\dot{P}_0$ for a given $B_0$ or $P_0$ are those with $\tau_0 \gg t$ (which is hardly a surprise given that $\tau \equiv P/(2\dot{P})$). The extent to which they continue evolving with $\tau \gg t$ depends on their initial conditions ($P_0$, $B_0$) and well as $n$. Notice, for recycled pulsars $n$ is not measurable and for this reason its value remains unknown.

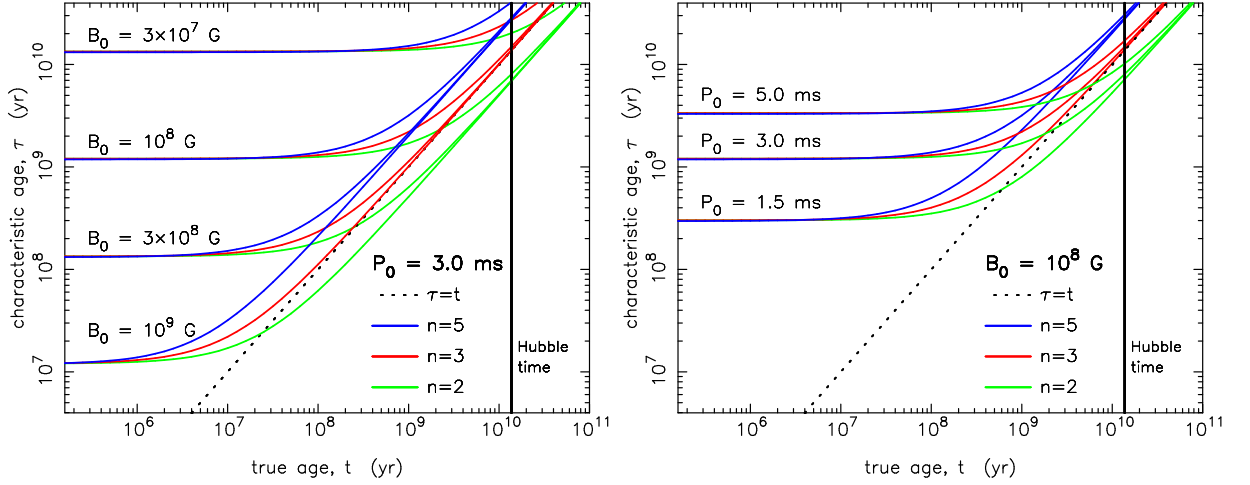

Figure 60: Evolutionary tracks of characteristic ages, $\tau$ calculated as a function of true ages, $t$ for recycled pulsars with a constant braking index of $n = 2$, $n = 3$, and $n = 5$. In all cases for $n$ we assumed a constant $M = 1.4\,M_\odot$, $\sin\alpha = 1$, a constant moment of inertia, $I$ and for $n = 5$ we also assumed a constant ellipticity, $\varepsilon \neq 0$. In the left panel we assumed in all cases an initial spin period of $P_0 = 3.0\,\mathrm{ms}$ and varied the value of the initial surface magnetic flux density, $B_0$. In the right panel we assumed in all cases $B_0 = 10^8\,\mathrm{G}$ and varied $P_0$. The dotted line shows a graph for $\tau = t$ and thus only pulsars located on (near) this line have characteristic ages as reliable age indicators. We notice that recycled pulsars with small values of $\dot{P}_0$, resulting from either small values of $B_0$ (left panel) and/or large values of $P_0$ (right panel), tend to have $\tau \gg t$, even at times exceeding the age of the Universe. In all degeneracy splittings of the curves the upper curve (blue) corresponds to $n = 5$, the central curve (red) corresponds to $n = 3$ and the lower curve (green) corresponds to $n = 2$. The asymptotic behaviour of the curves can easily be understood from equation (67) – see text for explanations.

### 6.8.3 Comparison with observations

#### 6.8.3.1 Kinematic corrections to $\dot{P}$

In a discussion of the ages and the evolution of millisecond pulsars in the $P\dot{P}$–diagram kinematic corrections must be included when considering the observed values of $\dot{P}$ ($\dot{P}_{\mathrm{obs}}$). The kinematic corrections to the intrinsic $\dot{P}$ ($\dot{P}_{\mathrm{int}}$) are caused by acceleration due to proper motion of nearby pulsars (Shklovskii 1970) and to vertical ($a_Z$) and differential rotational acceleration in our Galaxy ($a_{\mathrm{GDR}}$). The total corrections are given by:

$$\left(\frac{\dot{P}_{\mathrm{obs}}}{P}\right) = \left(\frac{\dot{P}_{\mathrm{int}}}{P}\right) + \left(\frac{\dot{P}_{\mathrm{shk}}}{P}\right) + \frac{a_Z}{c} + \frac{a_{\mathrm{GDR}}}{c} \qquad (69)$$

and following Damour & Taylor (1991) and Wolszczan et al. (2000) we can express these corrections as:

$$\begin{aligned}
\left(\frac{\dot{P}_{\mathrm{obs}}}{P}\right) = & \left(\frac{\dot{P}_{\mathrm{int}}}{P}\right) + \frac{\mu^2\,d}{c} - \frac{a_Z\,\sin b}{c} \\
& - \frac{v_0^2}{cR_0}\left[\cos l + \frac{(d/R_0) - \cos l}{1 + (d/R_0)^2 - 2(d/R_0)\cos l}\right]
\end{aligned} \qquad (70)$$

where $d$ is the distance to the pulsar, $\mu$ is the proper motion (related to the transverse velocity, $v_\perp = \mu d$), $a_Z$ is the vertical component of Galactic acceleration (e.g., from the Galactic potential model of Kuijken & Gilmore (1989)), $l$ and $b$ are the Galactic coordinates of the pulsar, $R_0 = 8.0\,\text{kpc}$ and $v_0 = 220\,\text{km s}^{-1}$ represent the distance to the Galactic centre and the orbital velocity of the Sun, and $c$ is the speed of light in vacuum. The corrections to $\dot{P}$ due to Galactic vertical and differential rotational accelerations are typically quite small (a few $10^{-22} \ll \dot{P}_{\text{int}}$) and can be ignored, except in a few cases (see below).

### 6.8.3.2 Evolutionary tracks and true age isochrones

In order to investigate if we can understand the distribution of MSPs in the $P\dot{P}$–diagram we have traced the evolution of eight hypothetical, recycled MSPs with different birth locations. In each case we traced the evolution for $2 \leq n \leq 5$ and plotted isochrones similar to those introduced in Fig. 59. The results are shown in Fig. 61 together with observed data. All the measured $\dot{P}$ values have been corrected for kinematic effects, as described above. If the transverse velocity of a given pulsar is unknown we used a value of $67\,\text{km s}^{-1}$ which we found to be the median value of the 49 measured velocities of binary pulsars[20]. In five cases (PSRs: J1231−1411, J1614−2230, J2229+2643, J1024−0719, J1801−1417, marked in squares in Fig. 61) we obtain $\dot{P}_{\text{int}} < 0$ which is not physically possible. The reason is probably an overestimate of the pulsar distance. In those cases we have recalculated $\dot{P}_{\text{int}}$ assuming only half the distance and included the corrections due to Galactic vertical and differential rotational acceleration.

Three main conclusions can be drawn from this diagram:

1) The overall distribution of observed pulsars follows nicely the banana-like shape of an isochrone with multiple choices for $\dot{P}_0$ (or $B_0$), see fat purple line. The chosen values of $P_0 = 3.0\,\text{ms}$, $n = 3$ and $t = 6\,\text{Gyr}$ are just for illustrative purposes only and not an attempt for a best fit to the observations. Fitting to one curve would not be a good idea given that MSPs are born with different initial spin periods, which depend on their accretion history, and given that the pulsars have different ages. The spread in the observed population is hinting that recycled pulsars are born at many different locations in the $P\dot{P}$–diagram.

2) The far majority of the recycled pulsars seem to have true ages between 3 and 12 Gyr, as expected since the population accumulates and the pulsars keep emitting radio waves for a Hubble time.

3) Pulsars with small values of the period derivative $\dot{P} \simeq 10^{-21}$ hardly evolve at all in the diagram over a Hubble time. This is a trivial fact, but nevertheless important since it tells us that these pulsars were basically born with their currently observed values of $P$ and $\dot{P}$ (first pointed out by Camilo, Thorsett & Kulkarni 1994). In this respect, it is interesting to notice PSR J1801−3210 (recently discovered by Bates et al. 2011) which must have been recycled with a relatively slow birth period, $P_0 \sim 7\,\text{ms}$ despite its low B-field $< 10^8\,\text{G}$ – see Fig. 53 for its location in the $P\dot{P}$–diagram.

One implication of the third conclusion listed above is that some radio MSPs must have been born with very small values of $B_0 \simeq 1 \times 10^7\,\text{G}$, if the inferred $B$-fields using equation (47) are correct. Given their weak B-fields such sources would most likely not be able to channel the accreted matter sufficiently to become observable as AXMSPs (Wijnands & van der Klis 1998). One case is the 1.88 ms radio pulsar J0034−0534 (Bailes et al. 1994) which has an observed $\dot{P} = 4.97 \times 10^{-21}$. However, this pulsar has a transverse velocity of $146\,\text{km s}^{-1}$ and correcting for the Shklovskii effect yields an intrinsic period derivative of only $\dot{P}_{\text{int}} = 5.36 \times 10^{-22}$. These values result in $B = 9 \times 10^6\,\text{G}$ (for $M = 1.4\,M_\odot$ and $\alpha = 90°$) and a characteristic age, $\tau = 55\,\text{Gyr}$. Once again we see that $\tau$ is not a good measure of the true age of a pulsar. If this pulsar has any gravitational wave emission ($\varepsilon \neq 0$) its derived surface $B$-field would be even smaller. We also notice a few pulsars which seem to have been recycled with $\dot{M} < 10^{-3}\,\dot{M}_{\text{Edd}}$ (approaching $10^{-4}\,\dot{M}_{Edd}$). Their progenitors could be the so-called weak LMXBs which are low-luminosity LMXBs (van der Klis 2006). It is also quite possible that their position in the $P\dot{P}$–diagram was strongly shifted during the RLDP (see fig. 2 in Tauris 2012) so that their average mass-accretion rate could have been larger prior to the RLDP. Alternatively, these MSPs demonstrate that $\phi > 1$ and $\omega_c < 1$ which could move the spin-up line substantially downwards, as demonstrated in Fig. 53.

---

[20]If we add to this sample the measured velocities of 13 isolated pulsars which show strong signatures of being recycled ($P < 100\,\text{ms}$ and $\dot{P} < 10^{-16}$) the median velocity becomes $69\,\text{km s}^{-1}$, which would make the Shklovskii corrections slightly larger and thus $\dot{P}_{\text{int}}$ somewhat smaller.

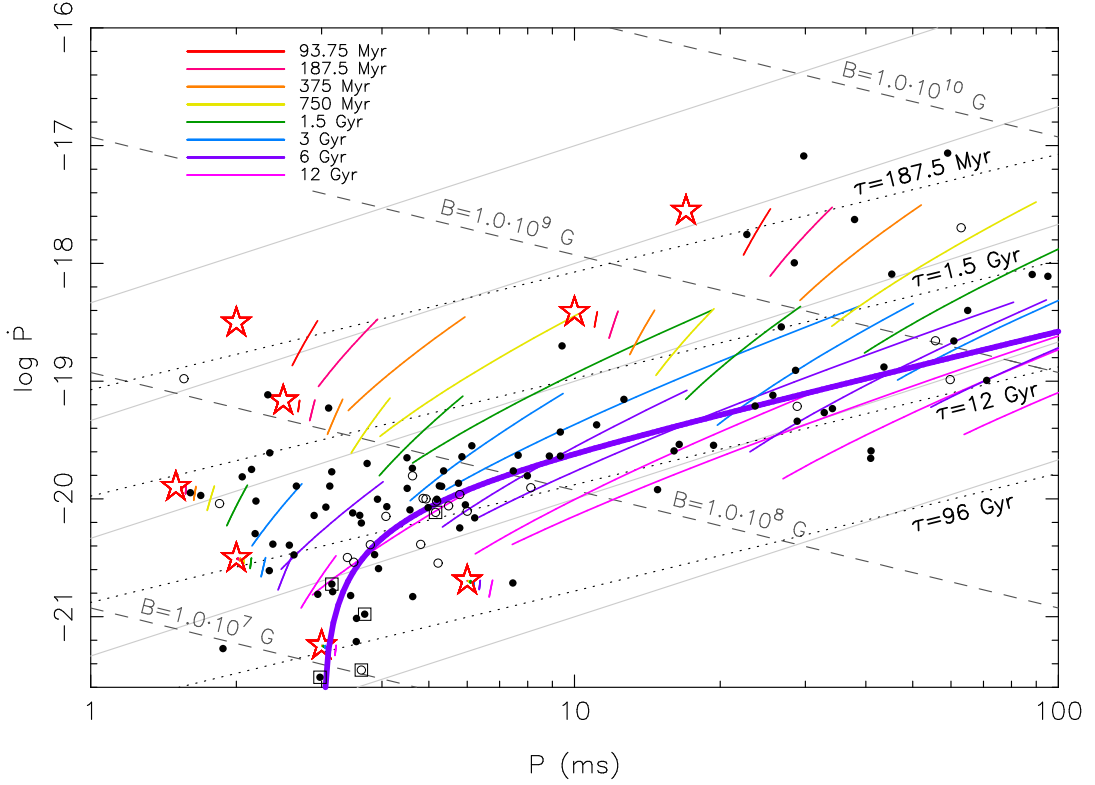

Figure 61: Isochrones of eight hypothetical recycled pulsars born at the locations of the red stars. The isochrones were calculated for different values of the braking index, $2 \leq n \leq 5$ (see Fig. 59 for details). Also plotted are inferred $B$-field values (dashed lines) and characteristic ages, $\tau$ (dotted lines). The thin grey lines are spin-up lines with $\dot{M}/\dot{M}_{Edd} = 1$, $10^{-1}$, $10^{-2}$, $10^{-3}$ and $10^{-4}$ (top to bottom, and assuming $\sin\alpha = \phi = \omega_c = 1$). In all calculations we assumed a pulsar mass of $1.4\,M_\odot$. It is noted that the majority of the observed population is found near the isochrones for $t = 3 - 12$ Gyr, as expected. The fat, solid purple line indicates an example of a $t = 6$ Gyr isochrone for pulsars with any value of $\dot{P}_0$, but assuming $P_0 = 3.0$ ms and $n = 3$. It is seen how the banana shape of such a type of an isochrone fits very well with the overall distribution of observed pulsars in the Galactic disk. Binary pulsars are marked with solid circles and isolated pulsars are marked with open circles. All values of $\dot{P}$ have been corrected for the Shklovskii effect using data from the *ATNF Pulsar Catalogue*. Pulsars in squares have been corrected for both a reduced distance and acceleration effects in the Galaxy. Pulsars with very small values of $\dot{P}$ are basically born on location in this diagram since their evolutionary timescale exceeds the Hubble time. Pulsars with relative small values of $P$ and large values of $\dot{P}$ have the potential to constrain white dwarf cooling models – see text for discussion.

It is obvious that a different sample of recycled pulsar birth locations in Fig. 61 would lead to somewhat different isochrones. However, we believe the qualitative interpretation is trustworthy and encourage further detailed statistical population analysis to be carried out (see Kiziltan & Thorsett 2010).

It is worth pointing out the difference in evolutionary timescales for the progenitor systems of BMSPs with He WDs or CO WDs, respectively. The IMXBs, leading mainly to BMSPs with CO WD companions, evolve on nuclear timescales of typically $100-300$ Myr. Hence, BMSPs with CO WDs could have true ages from zero to up to about 12 Gyr (assuming the first progenitor stars in our Galaxy formed $\sim 1\,\mathrm{Gyr}$ after Big Bang). The LMXBs, on the other hand, could easily have ages of $10\,\mathrm{Gyr}$ *before* they form, which is the typical timescale for a $1\,M_\odot$ star to evolve into a sub-giant which fills its Roche-lobe in an orbit with $P_{\mathrm{orb}}$ larger than a few days. One may then think that BMSPs with He WD companions must be much younger than BMSPs with CO WD companions (the BMSP age being calculated from after the RLDP). However, some LMXBs may have donor stars up to $2\,M_\odot$ and these stars evolve on a timescale of only $\sim 1.5\,\mathrm{Gyr}$. Hence, BMSPs with He WD companions can have almost similar true (post recycling) ages as BMSPs with CO WD companions.

### 6.8.3.3   PSR B1937+21 and J0218+4232: two recently fully recycled pulsars?

Although recycled pulsars with high values of $\dot{P}$ evolve fast across the $P\dot{P}$–diagram, we should expect to observe a few fully recycled pulsars with $\dot{P} \simeq 10^{-19}$. The only two cases in the Galactic disk known so far are PSR B1937+21 (Backer et al. 1982) and PSR 0218+4232 (Navarro et al. 1995), see Fig. 53. The latter is a 2.32 ms pulsar ($\dot{P}_{\mathrm{int}} = 7.66 \times 10^{-20}$) which not only has a small characteristic age ($\tau = 480\,\mathrm{Myr}$) but also must have a fairly young *true* age. Even if we assume it evolved with a constant value of $\dot{P}$ (and disregard the possibility of $n > 2$ which would have resulted in a larger initial value of $\dot{P}_0$ and hence a younger age) we find $t = 132 - 546\,\mathrm{Myr}$ when assuming $P_0$ is in the interval $1\text{–}2\,\mathrm{ms}$. As expected, this system hosts a white dwarf companion star which is still hot enough to be detected. Interestingly enough, all the cooling models applied by Bassa, van Kerkwijk & Kulkarni (2003) to this white dwarf seem to indicate a somewhat larger age (see their fig. 4). This discrepancy has the intriguing consequence that either their applied cooling models are overestimating the age of the white dwarf, or this pulsar evolved with $n < 2$ and evolved *upward* in the $P\dot{P}$–diagram.

PSR B1937+21 is the first MSP discovered and is not only young, like PSR J0218+4232, but also isolated. This fast spinning pulsar (1.56 ms) has $\dot{P}_{\mathrm{int}} = 1.05 \times 10^{-19}$ which yields a characteristic age of $\tau = 235\,\mathrm{Myr}$. We find upper limits ($n = 2$) for the true age to be in the interval $t = 78 - 229\,\mathrm{Myr}$ when assuming $P_0$ is in the interval $0.8\text{–}1.3\,\mathrm{ms}$. MSPs (or any pulsar with a low $\dot{P}$) are not believed to be formed directly from supernova explosions. They must interact with a companion star to spin-up and decrease their B-field by a large amount. Thus it is interesting that this pulsar was able to evaporate its companion star within a timescale of (a few) $10^8\,\mathrm{yr}$ – unless it evolved from an ultra-compact X-ray binary possibly leaving a planet around it (van Haaften et al. 2012a).

### 6.8.3.4   PSR J1841+0130: a young mildly recycled pulsar

PSR J1841+0130 is the 29 ms binary pulsar discussed earlier in Section 6.3.2.1. It has an unusual high value of $\dot{P}$, close to $10^{-17}$, which reveals that it is young and places it close to the spin-up line for $\dot{M} = \dot{M}_{\mathrm{Edd}}$ (assuming $\sin\alpha = \phi = \omega_c = 1$), see Fig. 53. The characteristic age of PSR J1841+0130 is 58 Myr. An upper limit to its true age is 77 Myr, which is found by assuming $n = 2$ and $P_0 = 10\,\mathrm{ms}$ ($P_0$ cannot have been much smaller than 10 ms under the above mentioned assumptions regarding the spin-up line). If we assume $n = 3$, the upper limit for the true age is 51 Myr. The WD companion is therefore hot enough to yield spectroscopic information that will reveal its true nature – i.e. whether this is a CO WD or a He WD, as discussed in Section 6.3.2.1. According to the dispersion measure, $DM = 125\,\mathrm{cm}^{-3}\,\mathrm{pc}$, of this pulsar its distance is about 3 kpc which might make such observations difficult.

## 6.9   Spinning up PSR J1614−2230

Recent Shapiro delay measurements of the radio millisecond pulsar J1614−2230 (Demorest et al. 2010) allowed a precise mass determination of this record high-mass neutron star and its white dwarf companion. A few key characteristic parameters of the system are shown in Table 9. In Paper I we discussed the binary evolution which led to the formation of this system (see also Lin et al. 2011). For estimating the amount of necessary mass accreted in order to recycle PSR J1614−2230 we can apply equation (56) and assuming, for example, an initial spin period of 2.0 ms following the RLO. We find that $\Delta M_{eq} = 0.11 \, M_\odot$. This result is in fine accordance with our Case A calculation in Paper I where a total of $0.31 \, M_\odot$ is accreted by the neutron star. (Recall that $\Delta M_{eq}$ is a minimum value for ideal, efficient spin-up.) For the Case C scenario it is difficult to reconcile the required accretion of $0.11 \, M_\odot$ with the very small amount which is expected to be accreted by the neutron star evolving through a CE-phase (usually assumed to be $< 10^{-2} \, M_\odot$). Furthermore, in Paper I we argued that Case BB RLO (stable RLO from a naked helium star following a CE) was not an option for this system.

Table 9: Selected physical parameters of the binary millisecond pulsar J1614−2230 (data taken from Demorest et al. 2010).

| Parameter | value |
|---|---|
| Pulsar mass | $1.97 \pm 0.04 \, M_\odot$ |
| White dwarf mass | $0.500 \pm 0.006 \, M_\odot$ |
| Orbital period | $8.6866194196(2)$ days |
| Orbital eccentricity | $1.30 \pm 0.04 \times 10^{-6}$ |
| Pulsar spin period | $3.1508076534271$ ms |
| Period derivative | $9.6216 \times 10^{-21}$ |

The next check is to see if the spin-relaxation time scale was shorter than the mass-transfer timescale for PSR J1614−2230. In order to estimate $t_{torque}$ we must have an idea of the B-field strength during the RLO. Since we have shown in Paper I that the mass-transfer rate in the final phase AB was near (slightly above) the Eddington limit, the value of $B$ prior to phase AB must have been somewhat larger than its current estimated value of $\sim 8.4 \times 10^7 \, G$, according to equation (47). Using $B = 4 - 10 \times 10^8 \, G$ we find $t_{torque} \simeq 2 - 9 \, Myr$. Since this timescale is shorter than the duration of RLO ($\sim 10 \, Myr$ for phase AB, see Paper I) we find that indeed it was possible for PSR J1614−2230 to spin up to its equilibrium period.

Based on both binary evolution considerations (Paper I) and the spin dynamics (this paper) we conclude that the Case A scenario is required to explain the existence of PSR J1614−2230. From Fig. 57 we notice that PSR J1614−2230 is an anomaly among the population of BMSPs with CO WD companions – it has an unusual rapid spin. This is explained by its formation via a stable and relatively long lasting Case A RLO (Paper I) which is, apparently, not a normal formation scenario for these systems.

### 6.9.1 Evolution of PSR J1614−2230 in the $P\dot{P}$–diagram

We have demonstrated in Paper I that PSR J1614−2230 mainly accreted its mass during the final phase (AB) of mass transfer. As mentioned above, in this phase the mass-transfer rate was high enough that the accretion onto the neutron star was limited by the Eddington limit, i.e. $\dot{M} = \dot{M}_{Edd}$. A natural question to address (see also Bhalerao & Kulkarni (2011)) is then how PSR J1614−2230 evolved to the present location in the $P\dot{P}$–diagram which, at first sight, is even below the spin-up line expected for $\dot{M} = 10^{-2} \, \dot{M}_{Edd}$. However, one must bear in mind the dependency of the parameters $\alpha$, $\phi$ and $\omega_c$ when discussing the location of the spin-up line (cf. Section 6.4.3).

If PSR J1614−2230 has a magnetic inclination angle, $\alpha < 90°$, and in case $\phi \approx 1.4$ and $\omega_c \approx 0.25$, then the problem is solved since the $\dot{M} = \dot{M}_{Edd}$ spin-up line (see orange line in Fig. 62) moves close to the current position of PSR J1614−2230. Therefore we do not see any reason to question the recycling model of BMSPs, based on the observations of the PSR J1614−2230 system, as proposed by Bhalerao & Kulkarni (2011). However, if $\phi \approx \omega_c \approx 1$ and $\alpha \approx 90°$ then we have to consider an alternative explanation which we shall now investigate.

It is well-known that the rotational evolution of millisecond pulsars is rather poorly understood since the braking index only has been measured accurately for a few young, non-recycled pulsars. As briefly mentioned in Section 6.8 there is a variety of mechanisms which can influence the braking index of a pulsar. For example, the geometry of the B-field, the decay of the B-field, a changing magnetic inclination angle (i.e. between the spin axis and the B-field axis of the neutron star), the dynamics the current flow in the pulsar magnetosphere and gravitational wave emission, e.g. Manchester & Taylor (1977) and Contopoulos & Spitkovsky (2006). We have made a simple toy model for two of these mechanisms to show that PSR J1614−2230 could in principle have accreted with $\dot{M} = \dot{M}_{\rm Edd}$ for $\sin \alpha \approx \phi \approx \omega_c \approx 1$ and subsequently evolve to its presently observed values of $P$ and $\dot{P}$ (near the corresponding $\dot{M} = 10^{-2}\,\dot{M}_{\rm Edd}$ spin-up line) after the radio pulsar turned on. In the first model we assumed pulsar spin-down as a result of gravitational wave radiation, which corresponds to $n = 5$. The emission of gravitational waves is caused by a time-varying quadrupole moment which arises, for example, if the accretion onto the neutron star created non-axisymmetric moments of inertia relative to the spin axis (Shapiro & Teukolsky 1983). In the second model we assumed an enhanced torque decay ($n > 3$) for a limited amount of time following the recycling phase. Such enhanced torque decay could be caused by either alignment between the B-field axis and the spin axis of the pulsar (but not complete alignment, in which case no pulsar would be seen) or further decay of the surface B-field through some unspecified mechanism. In our calculation for the second model we assumed that effectively $B$ or $\sin \alpha$ decreased by a factor of about five on a decay timescale of 200 Myr. In Fig. 62 we show that the present position in $P\dot{P}$–diagram can be reached from such toy models within 2 Gyr, which is in accordance with the cooling age of the CO WD inferred by Bhalerao & Kulkarni (2011). Notice, if PSR J1614−2230 accreted $0.31\,M_\odot$, as suggested in Paper I, it would in principle be possible for it to spin-up to a period of 0.92 ms (disregarding possible braking torque effects due to gravitational wave radiation, Bildsten (1998)) and thus the initial spin periods in our toy models are justified. However, we emphasize again that it is much less cumbersome to explain the current location of PSR J1614−2230 in the $P\dot{P}$–diagram simply by finetuning, in particular, $\phi$ and $\omega_c$. Therefore, future observations of systems like PSR J1614−2230 may help to constrain the disk-magnetosphere interactions.

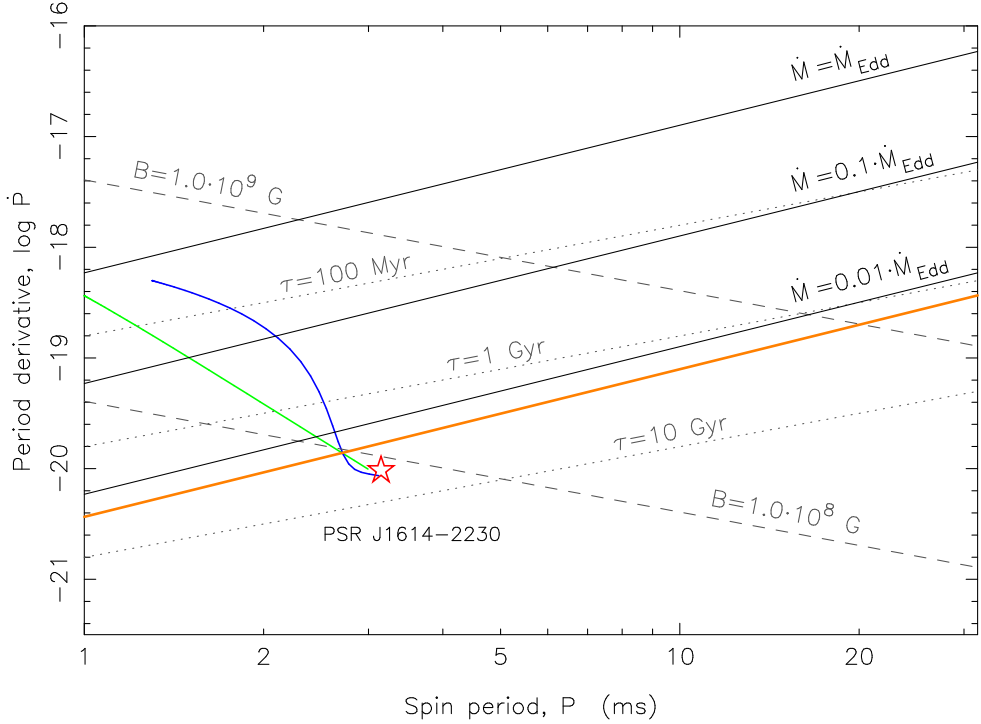

Figure 62: Simple models of how PSR J1614−2230 could have evolved from a birth near a spin-up line with $\dot{M} = \dot{M}_{\rm Edd}$ to its currently observed location, shown by the red star, within 2 Gyr. The orange line indicates the location of the $\dot{M} = \dot{M}_{\rm Edd}$ spin-up line if $\alpha = 10°$, $\phi = 1.4$ and $\omega_c = 0.25$. PSR J1614−2230 is located close to this line which suggests it was born (recycled) close to it. The three black solid lines were calculated by using $\alpha = 90°$, $\phi = 1$ and $\omega_c = 1$. If PSR J1614−2230 was born near such a spin-up line with $\dot{M} = \dot{M}_{\rm Edd}$ the need for subsequent spin-down after its birth would be severe and we have evolved a couple of toy models for this alternative: The green model represents gravitational wave radiation and the blue model represents a phase of enhanced torque decay (see text).

## 6.10 Neutron star birth masses

In Paper I, we concluded that the neutron star in PSR J1614−2230 must have been born in a supernova explosion (SN) with a mass of $1.7 \pm 0.15\,M_\odot$ if it subsequently evolved through an X-ray phase with Case A RLO. Based on evolutionary considerations we also found an alternative scenario where the post-SN system could have evolved via Case C RLO leading to a CE. However, we also argued that the CE scenario was less probable given the difficulties in the required evolution from the ZAMS to the X-ray phase. In this paper we rule out the CE scenario completely in view of our analysis of the rotational dynamics. The in-spiral of the neutron star in a CE occurs on such a short timescale that no significant spin-up is possible, and wind accretion from the proto WD cannot recycle the pulsar to a spin period of $2-3$ ms which requires accretion of $0.1\,M_\odot$. Recycling via a Case BB RLO, following CE evolution initiated while the progenitor of the donor star was on the RGB, is not possible either for PSR J1614−2230 since its orbital period is so large that its post-CE orbit must have been wide too and thus there could not have been released enough orbital energy during the in-spiral to eject the envelope.

Does such a high birth mass of $1.7\,M_\odot$ significantly exceed the neutron star birth masses in previously discovered radio pulsar systems? Below we test if other pulsars may have been born in a SN with a similarly high mass.

The interval of known radio pulsar masses ranges from $1.17\,M_\odot$ in the double neutron star binary PSR J1518+4909 ($3$-$\sigma$ upper limit, Janssen et al. 2008) to $1.97\,M_\odot$ in PSR J1614−2230, discussed in this paper. Prior to the measurement of PSR J1614−2230 the highest mass inferred for the birth mass of a radio pulsar was only $1.44\,M_\odot$ (PSR 1913+16, see Paper I). Whereas most of the pulsars in double neutron star systems have their mass determined accurately, only a few of the $\sim 140$ binary pulsars with WD companions have measured masses. These are listed in Table 10. At first sight, a handful of these are more massive than this previous limit of $1.44\,M_\odot$. However, in most of those cases the mass determinations have large uncertainties and in all cases also include mass accreted from the progenitor of their WD companion, i.e. these masses are pulsar masses *after* the recycling phase and not neutron star birth masses after the SN.

To take this effect into account we have estimated upper limits for the birth mass of the neutron star in all systems by subtracting the minimum mass needed to spin up the pulsar using equation (56). The results are shown in the fifth column of Table 10. Since we do not know the true age of the recycled pulsars, nor their braking index, we simply assumed in all cases that the presently observed pulsar spin period, $P$ is related to their original equilibrium spin period, $P_{eq}$ prior to the RLDP via: $P = \sqrt{2}P_{eq}$. For pulsars where the RLDP effect was small the initial spin period after recycling, $P_0 \simeq P_{eq}$ and hence $P \simeq \sqrt{2}P_0$, which corresponds to the case where the true age is half the characteristic age of a pulsar with braking index $n = 3$ (see Section 6.8 for further discussion on braking indices and true ages).

It is possible that some BMSPs have $P \simeq P_0$ (especially those BMSPs with small values of $B_0$). In this case the neutron star birth masses would be slightly larger than indicated here. However, recall that the limits in Table 10 are absolute upper limits obtained for *idealized* and efficient spin-up. Furthermore, once an accreting pulsar reaches its equilibrium spin period it can still accrete material but the net effect may not be further spin-up (depending on the mass-transfer rate, $\dot{M}$). Therefore, it should be emphasized that the actual SN birth mass could in many cases be substantially lower than estimated in Table 10. For example, we have demonstrated in Paper I that PSR J1614−2230 could have accreted $0.31\,M_\odot$, which is substantially more than shown in Table 10 where the pulsar was assumed to be born with a spin period of $3.15\,\mathrm{ms}/\sqrt{2} = 2.23\,\mathrm{ms}$, yielding $\Delta M_{eq} = 0.09\,M_\odot$.

Table 10: Measured masses of neutron stars in NS-WD systems in the Galactic disk, including PSR J1903+0327. The spin periods in the second column are in ms. The third, fourth and fifth column state the measured mass, the minimum accreted mass needed to spin-up the pulsar using equation (56) and the derived upper value for its birth mass, $M_{\mathrm{NS,max}}^{\mathrm{birth}} = M_{\mathrm{NS}} - \Delta M_{\mathrm{eq}}$, respectively. All masses are in quoted in $M_\odot$. All error bars are 1-$\sigma$ values.

| Pulsar | $P_{\mathrm{spin}}$ | $M_{\mathrm{NS}}$ | $\Delta M_{\mathrm{eq}}$ | $M_{\mathrm{NS,max}}^{\mathrm{birth}}$ |
|---|---|---|---|---|
| J0437−4715[1] | 5.76 | $1.76 \pm 0.20$ | 0.04 | 1.72 ± 0.20 |
| J0621+1002[2] | 28.9 | $1.70^{+0.10}_{-0.17}$ | 0.005 | 1.70 ± 0.14 |
| J0751+1807[2] | 3.48 | $1.26 \pm 0.14$ | 0.07 | 1.19 ± 0.13 |
| J1012+5307[3] | 5.26 | $1.64 \pm 0.22$ | 0.05 | 1.59 ± 0.21 |
| J1141−6545[4] | 394 | $1.27 \pm 0.01$ | — | 1.27 ± 0.01 |
| J1614−2230[5] | 3.15 | $1.97 \pm 0.04$ | 0.09 | 1.88 ± 0.04 |
| J1713+0747[6] | 4.57 | $1.53^{+0.08}_{-0.06}$ | 0.05 | 1.48 ± 0.07 |
| J1738+0333[7] | 5.85 | $1.46^{+0.06}_{-0.05}$ | 0.04 | 1.42 ± 0.05 |
| J1802−2124[8] | 12.6 | $1.24 \pm 0.11$ | 0.01 | 1.23 ± 0.11 |
| B1855+09[9] | 5.36 | $1.57^{+0.12}_{-0.11}$ | 0.04 | 1.53 ± 0.11 |
| J1903+0327[10] | 2.15 | $1.667 \pm 0.021$ | 0.15 | 1.52 ± 0.02 |
| J1909−3744[11] | 2.95 | $1.44 \pm 0.02$ | 0.09 | 1.35 ± 0.02 |
| B2303+46[12] | 1066 | $1.38^{+0.06}_{-0.10}$ | — | 1.38 ± 0.08 |

References: (1) Verbiest et al. (2008), (2) Nice, Stairs & Kasian (2008), (3) Callanan, Garnavich & Koester (1998), (4) Bhat, Bailes & Verbiest (2008), (5) Demorest et al. (2010), (6) Splaver et al. (2005), (7) Antoniadis et al. (2012), (8) Ferdman et al. (2010), (9) Nice, Splaver & Stairs (2003), (10) Freire et al. (2011b), (11) Jacoby et al. (2005), (12) Thorsett & Chakrabarty (1999).

### 6.10.1 PSR J0621+1002

Considering the derived SN birth masses in Table 10 we notice at first sight in particular PSR J0437−4715 and PSR J0621+1002 as further candidates for pulsars which were potentially born massive ($\sim 1.7\,M_\odot$), although the error bars are large. Given the uncertainty of the accretion efficiency (i.e. the possible difference between the actual amount of mass accreted and $\Delta M_{\mathrm{eq}}$) we see that in particular PSR J0621+1002 could have been born massive with $M_{\mathrm{NS,max}}^{\mathrm{birth}} \sim 1.7\,M_\odot$. PSR J0621+1002 (Camilo et al. 1996) has an orbital period of 8.3 days and a CO WD companion of mass $0.67\,M_\odot$ which is somewhat equivalent to PSR J1614−2230. However, its relatively slow spin period of 28.9 ms and large eccentricity of $2.5 \times 10^{-3}$ indicate that PSR J0621+1002 did not evolve through an X-ray phase with stable, and relatively long Case A RLO (unlike the situation for PSR J1614−2230). It seems much more evident that PSR J0621+1002 evolved through a rapid, less efficient, mass-transfer phase – such as early Case B RLO (Tauris, van den Heuvel & Savonije 2000). In this case the actual amount of accretion must have been at most a few $10^{-2}\,M_\odot$. Therefore, its measured mass is very close to its birth mass and *if* future mass measurements of PSR J0621+1002 yield smaller error bars, still centered on its current mass estimate, we conclude that this pulsar belongs to the same class of neutron stars as PSR J1614−2230, Vela X-1 and possibly the black widow pulsar which were all born massive (see Paper I). However, the error bar is quite large and PSR J0621+1002 could turn out to have an ordinary mass of about $1.4\,M_\odot$ (the 2-$\sigma$ lower mass limit).

### 6.10.2 PSR J0437−4715

PSR J0437−4715 is also a potential candidate for a pulsar which could have been born as a neutron star with a mass of $\sim 1.7\,M_\odot$. However, this pulsar has a fairly rapid spin period (like a few other relatively high-mass neutron star candidates in Table 10) which indicates a long spin-up phase. Hence, it may have accreted significantly more than suggested by $\Delta M_{\mathrm{eq}}$. In fact, as mentioned before, in the case of PSR J1614−2230 the pulsar might have accreted three times more than suggested by $\Delta M_{eq}$, see fig. 7 in Tauris, Langer & Kramer (2011). The problem is that as soon as the pulsar spin reaches $P_{eq}$ (which requires accumulation of mass $\Delta M_{eq}$) it may remain in near-spin equilibrium and keep accreting on and off while spinning up/down depending on the sign of the accretion torque. The situation is quite different for PSR J0621+1002 which formed through a short lived X-ray phase excluding evolution at near-spin equilibrium for a substantial amount of time. We therefore conclude that PSR J0437−4715 could originally have been born in a SN with a typical mass of $\sim 1.4\,M_\odot$, even if its present mass is indeed $1.72\,M_\odot$.

## 6.11  Summary

We have investigated in detail the recycling process of pulsars with respect to accretion, spin-up and the Roche-lobe decoupling phase, and discussed the implications for their spin periods, masses and true ages. In particular, we have discussed the concept of a spin-up line in the $P\dot{P}$–diagram and emphasize that such a line cannot be uniquely defined. Besides from the poorly known disk-magnetosphere physics, which introduces large uncertainties, for each individual pulsar the equilibrium spin period, $P_{eq}$ also depends on its magnetic inclination angle, $\alpha$ as well as its accretion history ($\dot{M}$) and its B-field strength. Furthermore, we have applied the Spitkovsky (2006) spin-down torque on radio pulsars which significantly changes the location of the spin-up lines compared to using the vacuum magnetic dipole model, especially for small values of $\alpha$.

We have derived a simple analytical expression (equation 56) to evaluate the amount of mass needed to be accreted to spin up a pulsar to any given equilibrium spin period, $P_{eq}$. Our result resembles that of Alpar et al. (1982) and approximately yields the same values (within a factor of two) as the expression derived by Lipunov & Postnov (1984). Using our formula we find, for example, that BMSPs with He WDs and $P_{eq} \simeq 2\,\mathrm{ms}$ must have accreted at least $0.10\,M_\odot$, whereas typical BSMPs with CO WDs and $P_{eq} \simeq 20\,\mathrm{ms}$ only needed to accrete $0.005\,M_\odot$.

Applying equation (56) enables us to explain the difference in spin distributions between BMSPs with He and CO WDs, respectively. The BMSPs with He WDs often evolved via an X-ray phase with stable RLO on a long timescale, allowing sufficient material to be accreted by the neutron star to spin it up efficiently to a short period, whereas the BMSPs with CO WDs evolved from IMXBs which often had a short phase of mass transfer via early Case B RLO or Case C RLO – the latter leading to a CE-evolution often followed by Case BB RLO. The only exception known so far is PSR J1614−2230 since this system produced a CO WD orbiting a fully recycled MSP and thus it must have evolved via Case A RLO of an IMXB.

It is not possible to recycle pulsars to become MSPs via wind accretion from $1.1 - 2.2\,M_\odot$ post-CE helium stars. However, we have demonstrated that Case BB RLO from such helium stars can spin up pulsars to at least $\sim 11$ ms. Further studies of these systems are needed.

There is an increasing number of recycled pulsars with WD companions which seem to fall outside the two main populations of BMSPs with He and CO WDs, respectively. These peculiar systems possibly have He WDs and always exhibit slow spin periods between 10 and 100 ms. We suggest that these systems with $P_{orb} \geq 1$ day may have formed via Case A RLO of IMXBs. We plan further studies on these binaries.

The Roche-lobe decoupling phase (RLDP), at the termination stage of the mass transfer (Tauris 2012), has been analysed and we have shown that while the RLDP effect is important in LMXBs – leading to significant loss of rotational energy of the recycled pulsars as well as characteristic ages at birth which may exceed the age of the Universe – it is not significant in IMXB systems where the duration of the RLDP is short.

In order to track the evolution of pulsars in the $P\dot{P}$–diagram we have introduced two types of true age isochrones – one which matches well with the banana shape of the observed distribution of known MSPs. We encourage further statistical population studies to better understand the formation and evolution of radio MSPs in the $P\dot{P}$–diagram (see e.g. Kiziltan & Thorsett 2010). The discrepancy between true ages and characteristic spin-down ages of recycled pulsars has been discussed and we confirm that the latter values are completely untrustworthy as true age indicators, leaving WD cooling ages as the only valid, although not accurate, measuring scale (Tauris 2012).

In the combined study presented here and in Paper I we have investigated the recycling of PSR J1614$-$2230 by detailed modelling of the mass exchanging X-ray phase of the progenitor system. Given the rapid spin of PSR J1614$-$2230 (3.15 ms) we argue that it is highly unlikely that it evolved through a CE, leaving Case A RLO in an IMXB as the only viable formation channel. We confirm the conclusion from Paper I that the neutron star in PSR J1614$-$2230 was born massive ($1.70 \pm 0.15 \, M_\odot$). We have demonstrated that PSR J1614$-$2230 could have been spun-up at $\dot{M} = \dot{M}_{\rm Edd}$ and subsequently evolve to its current position in the $P\dot{P}$–diagram within 2 Gyr (the estimated cooling age of its white dwarf companion).

Besides PSR J1614$-$2230, Vela X-1 and possibly the black widow pulsar, we have argued that also PSR J0621+1002 could belong to the same class of neutron stars born massive ($\geq 1.7 \, M_\odot$). The formation of such massive neutron stars in supernovae is in agreement with some supernova explosion models (e.g. Zhang, Woosley & Heger 2008; Ugliano et al. 2012).

## 6.12 Appendix: Identifying the nature of a pulsar companion star

For stellar evolutionary purposes, and for observers to judge the possibilities to detect a companion, it would be useful to have a tool to predict the most likely nature of the companion star of a given binary pulsar. As discussed in Section 6.3 radio pulsars are presently found with the following companions:

- Main sequence stars – $MS$

- Neutron stars – $NS$

- CO/ONeMg white dwarfs – $CO$

- He white dwarfs – $He$

- Ultra-light companions (or planets) – $UL$

Although still not detected, radio pulsars are also expected to exist in systems with either:

- Helium stars – $HS$

- (sub)Giant stars – $GS$

- Black Holes – $BH$

As explained in Section 6.3 pulsar systems found in globular clusters are not useful as probes of stellar evolution due to perturbations in their dense environment. We shall denote the companions of such pulsars by:

- Globular cluster companions – $GC$

The usual characteristic parameters measured of an observed binary pulsar are: the spin period ($P$), the period derivative ($\dot{P}$), the orbital period ($P_{\mathrm{orb}}$) and the projected semi-major axis of the pulsar ($a_1$). From the latter two parameters one can calculate the mass function ($f$) which is needed, but not sufficient, to estimate the mass of the companion star ($M_2$). Furthermore, in many cases it is also possible to estimate the eccentricity ($ecc$). Based on these parameters we state simple conditions in Table 11 which can be used to identify the most likely nature of pulsar companions at the 95 per cent level of confidence (based on statistics from current available data).

These conditions can be directly applied to the online, public available *ATNF Pulsar Catalogue*, http://www.atnf.csiro.au/research/pulsar/psrcat/ (Manchester et al. 2005) in order to select a subsample of pulsars with certain companion stars.

One example of the few pulsars where the identification fails is PSR J1614−2230 given its unusual combination of being fully recycled and having a CO WD companion, as discussed in this paper. The identification conditions can in principle be optimised further to increase the level of confidence even more. However, this would not only require these conditions to be updated frequently with new discoveries of somewhat peculiar binary pulsars but also lead to undesirable complexity in the conditions. The main uncertainty factor is the (most often) unknown orbital inclination angle of the binary system. The transition between systems with $UL$ and $He$ companions and, in particular, between $He$ and $CO$ companions is difficult to define satisfactory. Furthermore, WD+NS systems where the last formed compact object is the neutron star are very difficult to distinguish from NS+NS systems, *if* based solely on data from the last formed (non-recycled) NS. Finally, we note that the long-sought after radio pulsar in a binary with a black hole companion (NS+BH or BH+NS) will most likely be mildly recycled (i.e. NS+BH) due to its much longer lifetime as an observable pulsar after accretion. In this case it may be difficult to distinguish it from a system where the pulsar has a $NS$ companion orbiting with a small inclination angle. On the other hand, if the pulsar is not recycled then such a BH+NS system could resemble a pulsar with an (undetected) $MS$ companion.

Table 11: The most likely nature of a radio pulsar companion star based on observable characteristics. The selection conditions can be directly applied to the *ATNF Pulsar Catalogue* (Manchester et al. 2005). See also Section 6.3 for further discussions.

| Companion type | Conditions | | | | | | |
|---|---|---|---|---|---|---|---|
| *MS* | $M_2^* > 0.5\ M_\odot$ | and | $P_{\mathrm{orb}} > 50^{\mathrm{d}}$ | | | | |
| *NS* | | | $P_{\mathrm{orb}} < 50^{\mathrm{d}}$ | and | $ecc > 0.05$ | | |
| *CO* | $M_2^* > 0.335\ M_\odot$ | and | $P_{\mathrm{orb}} < 75^{\mathrm{d}}$ | and | $ecc < 0.05$ | and | $P > 8\ \mathrm{ms}$ |
| *He* | $M_2^* > 0.08\ M_\odot$ | and | $\{(P_{\mathrm{orb}} < 75^{\mathrm{d}}$ | and | $M_2^* < 0.335\ M_\odot)$ | or | $(P_{\mathrm{orb}} > 75^{\mathrm{d}}$ and $M_2^* < 0.46\ M_\odot)\}$ |
| *UL*** | $M_2^* < 0.08\ M_\odot$ | | | | | | |

\* The median companion mass $M_2$ is calculated for an orbital inclination angle $i = 60°$ and an assumed pulsar mass $M_{\mathrm{NS}} = 1.35\ M_\odot$.

\*\* Many pulsars with unmeasured values of $a_1$ are also expected to host an ultra-light companion if $P_{\mathrm{orb}} < 2$ days and $P < 8$ ms.

## 7. Timing of a Young Mildly Recycled Pulsar with a Massive White Dwarf Companion

**Lazarus, Tauris, Knispel, Freire, et al. (2014)**

### Abstract

We report on timing observations of the recently discovered binary pulsar PSR J1952+2630 using the Arecibo Observatory. The mildly recycled 20.7-ms pulsar is in a 9.4-hr orbit with a massive, $M_{WD} > 0.93\ M_\odot$, white dwarf (WD) companion. We present, for the first time, a phase-coherent timing solution, with precise spin, astrometric, and Keplerian orbital parameters. This shows that the characteristic age of PSR J1952+2630 is 77 Myr, younger by one order of magnitude than any other recycled pulsar–massive WD system. We derive an upper limit on the true age of the system of 150 Myr. We investigate the formation of PSR J1952+2630 using detailed modelling of the mass-transfer process from a naked helium star on to the neutron star following a common-envelope phase (Case BB Roche-lobe overflow). From our modelling of the progenitor system, we constrain the accretion efficiency of the neutron star, which suggests a value between 100 and 300% of the Eddington accretion limit. We present numerical models of the chemical structure of a possible oxygen-neon-magnesium WD companion. Furthermore, we calculate the past and the future spin evolution of PSR J1952+2630, until the system merges in about 3.4 Gyr due to gravitational wave emission. Although we detect no relativistic effects in our timing analysis we show that several such effects will become measurable with continued observations over the next 10 years; thus PSR J1952+2630 has potential as a testbed for gravitational theories.

### 7.1 Introduction

Since 2004, the Arecibo L-band Feed Array (ALFA), a 7-beam receiver at the focus of the 305-m William E. Gordon radio telescope at the Arecibo Observatory, is being used to carry out the Pulsar–ALFA (PALFA) survey, a deep pulsar survey of low Galactic latitudes (Cordes et al. 2006; Lazarus 2013; Nice et al. 2013). Given its short pointings, the PALFA survey is especially sensitive to binary pulsars in tight orbits, as demonstrated by the discovery of the relativistic binary pulsar PSR J1906+0746, which did not require any acceleration search techniques (Lorimer et al. 2006b). Another aspect of this and other modern Galactic plane surveys is the high time and frequency resolution, which allow the detection of millisecond pulsars (MSPs) at high dispersion measures (Champion et al. 2008; Deneva et al. 2012; Crawford et al. 2012) and therefore greatly expand the volume in which these can be discovered.

One of the innovative aspects of this survey is the use of distributed, volunteer computing. One of the main motivations is the detection of extremely tight (down to $P_b \sim 10$ minutes) binaries, for which acceleration and jerk searches become computationally challenging tasks. The analysis of survey data is distributed through the Einstein@Home (E@H) infrastructure (Knispel et al. 2010; Allen et al. 2013). Thus far, the E@H pipeline has discovered 24 new pulsars in the PALFA survey data alone, complementing the other data analysis pipelines the PALFA survey employs (Quicklook, and PRESTO; Stovall et. al, in prep., and Lazarus et. al, in prep., respectively).

PSR J1952+2630 was the first binary pulsar discovered with the E@H pipeline (Knispel et al. 2011). At that time the few observations available allowed only a rough estimate of the orbital parameters of this MSP based on Doppler measurements of the spin period. These already showed that PSR J1952+2630 has a massive WD companion ($M_{WD} > 0.945\ M_\odot$ assuming $M_p = 1.4\ M_\odot$), and may have evolved from an intermediate-mass X-ray binary (IMXB). Building on the analysis by Knispel et al. (2011), we present in this paper the phase-coherent timing solution of PSR J1952+2630 resulting from dedicated follow-up observations with the Arecibo telescope, which provides orbital parameters far more precise than those previously determined. Our timing solution also shows the system is relatively young ($\tau_c = 77$ Myr).

It is commonly accepted that MSPs are spun up to their high spin frequencies via accretion of mass and angular momentum from a companion star (Alpar et al. 1982; Radhakrishnan & Srinivasan 1982; Bhattacharya & van den Heuvel 1991). In this recycling phase the system is observable as an X-ray binary (e.g. Hayakawa 1985; Nagase 1989; Bildsten et al. 1997) and towards the end of this phase as an X-ray MSP (Wijnands & van der Klis 1998; Papitto et al. 2013).

The majority of MSPs have helium WD companions and their formation is mainly channeled through low-mass X-ray binaries (LMXBs) which have been well investigated in previous studies (e.g. Webbink, Rappaport & Savonije 1983; Pylyser & Savonije 1988; 1989; Rappaport et al. 1995; Ergma, Sarna & Antipova 1998; Tauris & Savonije 1999; Podsiadlowski, Rappaport & Pfahl 2002; Nelson, Dubeau & MacCannell 2004; van der Sluys, Verbunt & Pols 2005). In contrast, binary pulsars, such as PSR J1952+2630, with relatively heavy WDs (CO or ONeMg WDs) are less common in nature. Their formation and recycling process involves a more massive WD progenitor star in an IMXB (see Tauris, Langer & Kramer 2011; and references therein, for a discussion of their suggested formation channels). Here we distinguish IMXBs from other X-ray binaries as systems that leave behind a massive WD companion rather than a neutron star (NS). Some of these IMXB systems with donor stars of $6 - 7\ M_\odot$ could also be classified observationally as Be/X-ray binaries since these stars are of spectral class B3-4 with emission lines. Recently, Tauris, Langer & Kramer (2012) presented a detailed study of the recycling process of pulsars via both LMXBs and IMXBs and highlighted their similarities and differences. These authors also presented the first calculations of mild recycling in post common envelope systems where mass transfer proceeds via so-called Case BB Roche-lobe overflow (RLO; see Section 7.4 for details).

PSR J1952+2630's combination of a young, massive WD in a close orbit with a recycled pulsar poses interesting questions about its formation and future evolution. Binary MSPs represent the advanced stage of stellar evolution in close, interacting binaries. Their observed orbital and stellar properties are thus fossil records of their evolutionary history. Therefore by using the precise description of a pulsar binary system determined from phase-coherent timing, and binary evolution modelling, we use PSR J1952+2630 as a probe of stellar astrophysics.

We also demonstrate that PSR J1952+2630 is an interesting test case for Case BB RLO, enabling interesting constraints on the accretion physics from the combined modelling of binary stellar evolution and the spin kinematics of this young, mildly recycled pulsar.

The paper is presented as follows: Section 7.2 describes the observations of PSR J1952+2630, and details of the data reduction and timing analysis. Results from this analysis are presented in Section 7.3. The binary evolution of the system is detailed in Section 7.4. The implications of our results, and future prospects are described in Sections 7.5, and 7.6, respectively. Finally Section 7.7 summarizes the paper.

## 7.2   Observations and Data Analysis

Following its discovery in 2010 July PSR J1952+2630 was observed during PALFA survey observing sessions using the usual survey observing set-up: the 7-beam ALFA receiver with the Mock spectrometers[21]. In this set-up, $\sim$322 MHz ALFA observing band was split into two overlapping sub-bands centred at 1300.1680 MHz and 1450.1680 MHz, each with a bandwidth of 172.0625 MHz. All timing observations using this set-up were performed with ALFA's central beam, and were typically 5-10 minutes in duration.

A dedicated timing program at the Arecibo Observatory started in 2011 November. The dedicated timing observations took data using the "L-wide" receiver. These data were divided into four non-overlapping, contiguous sub-bands recorded by the Mock spectrometers in search-mode. Each sub-band has 172 MHz of bandwidth divided in 2048 channels, sampled every

---

[21]http://www.naic.edu/~astro/mock.shtml

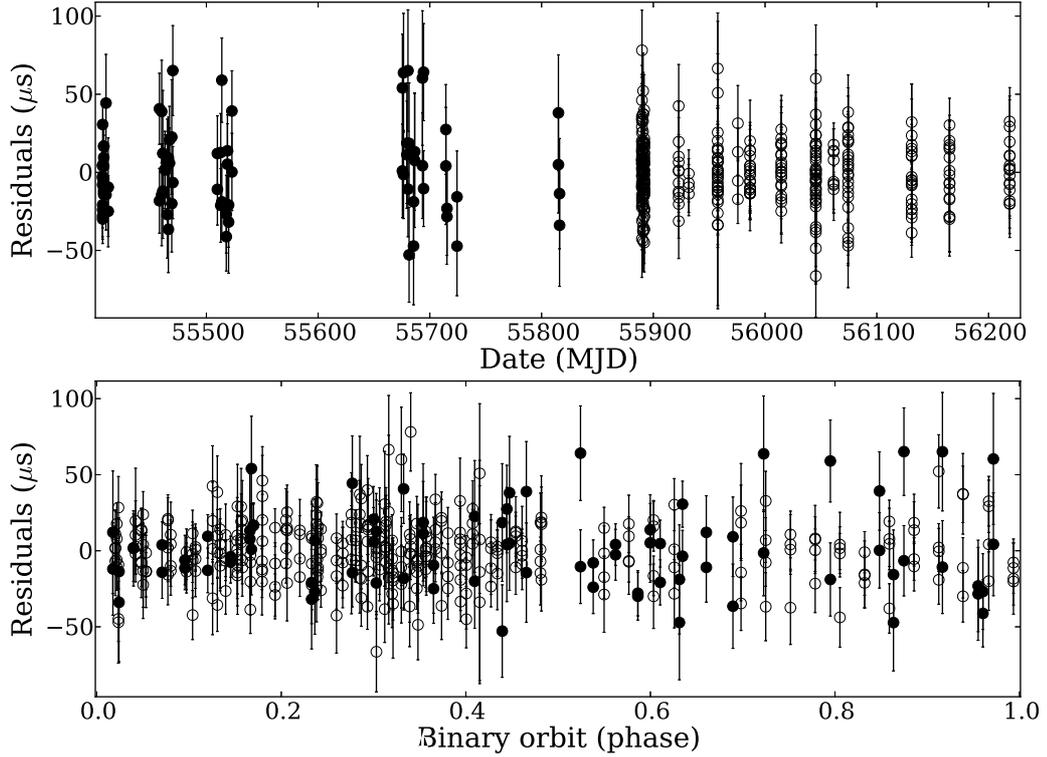

Figure 63: The difference between our pulsar TOAs and our timing solution. The filled circles are TOAs from data taken with the ALFA receiver, and the un-filled circles TOAs from data taken with the L-wide receiver. No systematic trends are visible as a function of epoch, or orbital phase.

∼83.3 $\mu$s. Together the four sub-bands cover slightly more than the maximum bandwidth of the receiver, 580 MHz, and are centred at 1444 MHz.

At the start of the dedicated timing campaign, three 3-hour observing sessions on consecutive days were conducted to obtain nearly complete orbital coverage, with the goal of detecting, or constraining a Shapiro delay signature caused by the pulsar's signal passing through its companions gravitational potential well.

Subsequently, monthly observations of 1 hour each were used for the next 11 months to monitor the pulsar, and refine our timing solution.

For analysis, each observation was divided into segments no longer than 15-minutes for each of the separate sub-bands. Each segment was folded offline using the appropriate topocentric spin period using `prepfold` program from `PRESTO`[22]. The resulting folded data files were converted to PSRFITS format (Hotan, van Straten & Manchester 2004) with `pam` from the `psrchive` suite of pulsar analysis tools[23]. Data files were fully time and frequency integrated, and times-of-arrival (TOAs) were computed using `pat` from `psrchive`. The result was a single TOA for each ∼ 15 minutes of observing for each sub-band.

TOAs for the different sub-bands were computed using a separate analytic template produced by fitting von Mises functions to the sum of profiles from the closely spaced three-day observing campaign. Profiles from different observations were aligned using the pulsar ephemeris.

Timing analysis was performed with the `TEMPO2` software (Hobbs, Edwards & Manchester 2006). The phase-coherent timing solution determined fits the 418 pulse times-of-arrival. Our timing solution accurately models the timing data, leaving no systematic trends as a function of epoch, or orbital phase (see Fig. 63).

---

[22]http://www.cv.nrao.edu/∼sransom/presto/
[23]http://psrchive.sourceforge.net/

## 7.3 Results

The timing analysis of PSR J1952+2630 has resulted in the determination of astrometric, spin, and Keplerian orbital parameters. The fitted timing parameters, as well as some derived parameters can be found in Table 12. The timing solution also includes a marginal detection of the proper motion of the pulsar.

### 7.3.1 The nature of the binary companion star

Given the combination of a relatively large observed mass function, $f = 0.153\ M_\odot$ (see Table 12) and a small orbital eccentricity, $e = 4.1 \times 10^{-5}$, it is clear that PSR J1952+2630 has a massive WD companion. The minimum companion mass is $M_{\rm WD}^{\rm min} \approx 0.93\ M_\odot$, obtained for an orbital inclination angle of $i = 90°$ and an assumed NS mass, $M_{\rm NS} = 1.35\ M_\odot$. The small eccentricity excludes a NS companion star since the release of the gravitational binding energy alone, during the core collapse, would make the post-SN eccentricity much larger ($e \gg 0.01$; Bhattacharya & van den Heuvel 1991), in contrast with the observed value. Furthermore, no known double-neutron star system has $e < 0.01$, according to the ATNF Pulsar Catalogue[24] (Manchester et al. 2005). Thus a double NS system is not possible and PSR J1952+2630 must have a massive WD companion, i.e. a carbon-oxygen (CO) or an oxygen-neon-magnesium (ONeMg) WD. The upper (Chandrasekhar) mass limit for a rigidly rotating WD is $\sim 1.48~M_\odot$ (e.g. Yoon & Langer 2005) and therefore we conclude that the WD companion star is in the mass interval $0.93 \lesssim M_{\rm WD}/M_\odot \lesssim 1.48$.

Table 12: Fitted and derived parameters for PSR J1952+2630.

| Parameter | Value[a] |
|---|---|
| *General Information* | |
| MJD Range | $55407 - 56219$ |
| Number of TOAs | 418 |
| Weighted RMS of Timing Residuals ($\mu$s) | 19 |
| Reduced-$\chi^2$ value[b] | 1.03 |
| MJD of Period Determination | 55813 |
| Binary Model Used | ELL1 |
| *Fitted Parameters* | |
| R.A., $\alpha$ (J2000) | 19:52:36.8401(1) |
| Dec., $\delta$ (J2000) | 26:30:28.073(2) |
| Proper motion in R.A., $\mu_\alpha$ (mas/yr) | -6(2) |
| Proper motion in Dec., $\mu_\delta$ (mas/yr) | 0(3) |
| Spin Frequency, $\nu$ (Hz) | 48.233774295845(7) |
| Spin Frequency derivative, $\dot\nu$ ($\times 10^{-15}$ Hz/s) | -9.9390(5) |
| Dispersion Measure, DM (pc cm$^{-3}$) | 315.338(2) |
| Projected Semi-Major Axis, $a \sin i$ (lt-s) | 2.798196(2) |
| Orbital Period, $P_b$ (days) | 0.39187863896(7) |
| Time of Ascending Node, $T_{asc}$ (MJD) | 55812.89716459(4) |
| $\epsilon_1$ | -0.000038(1) |
| $\epsilon_2$ | 0.000015(1) |
| *Derived Parameters* | |
| Spin Period, (ms) | 20.732360562672(3) |
| Spin Period Derivative ($\times 10^{-18}\,{\rm s\,s}^{-1}$) | 4.2721(2) |
| Galactic longitude, $l$ (°) | 63.254 |
| Galactic latitude, $b$ (°) | $-0.376$ |
| Distance (NE2001, kpc) | 9.6 |
| Orbital Eccentricity, $e$ ($\times 10^{-5}$) | 4.1(1) |
| Longitude of Periastron, $\omega$ (°) | 291(2) |
| Mass Function, $f$ ($M_\odot$) | 0.153184(1) |
| Characteristic Age, $\tau_c = P/(2\dot P)$ (Myr) | 77 |
| Inferred Surface Magnetic Field Strength, $B_S$ ($\times 10^9$ G) | 9.5 |
| Spin-down Luminosity, $\dot E$ ($\times 10^{35}$ ergs/s) | 0.19 |

[a]The numbers in parentheses are the 1-$\sigma$, `TEMPO2`-reported uncertainties on the last digit.

[b]The uncertainties of the ALFA and L-wide data sets were individually scaled such that the reduced $\chi^2$ of the data sets are 1.

The distance to PSR J1952+2630, estimated using the observed dispersion measure, and the NE2001 model of Galactic free electrons, is $d \simeq 9.6$ kpc (Cordes & Lazio 2002). The uncertainty in DM-derived distances using the NE2001 model can, in some cases, be up to a factor of $\sim 2$ off from the true distance. Unfortunately, this places the binary system too far away to hope to optically detect the WD companion with current telescopes. As expected, a search of optical and infrared catalogs yielded no counterpart.

### 7.3.2 The age of PSR J1952+2630

PSR J1952+2630 has a spin period of $P = 20.7$ ms and one of the highest values of the spin period derivative, $\dot{P} = 4.27 \times 10^{-18}\,{\rm s\,s^{-1}}$ for any known recycled pulsar[25] and by far the highest value for a mildly recycled pulsar with a massive WD companion. The observed value of $\dot{P}$ is contaminated by kinematic effects (see Section 7.6). However, the contamination is only $\sim$0.01%, assuming our current value of the proper motion. The combination of $P$ and $\dot{P}$ of PSR J1952+2630 yields a small characteristic age, $\tau \equiv P/2\dot{P} \simeq 77$ Myr. The characteristic age of a pulsar should only be considered a rough order-of-magnitude estimate of the true age of the pulsar (i.e. time since recycling terminated). Thus the *true* ages are quite uncertain for recycled pulsars with large $\tau$-values of several Gyr (Tauris 2012; Tauris, Langer & Kramer 2012), unless a cooling age of their WD companion can be determined. However, the true ages of recycled pulsars with small values of $\tau$ (less than a few 100 Myr) are relatively close to the characteristic age. Hence, we conclude that PSR J1952+2630 is young, for a recycled pulsar, and in Sections 7.5.1 and 7.5.1.1 we discuss its true age (i.e. its actual age since it switched-on as a recycled radio pulsar) and also constrain its spin evolution in the past and in the future.

### 7.3.3 Constraints on the binary system from timing

Given the current timing data, the binary motion of PSR J1952+2630 can be accurately modelled without requiring any relativistic, post-Keplerian parameters. Unfortunately, this means that the masses of the pulsar and its WD companion cannot be precisely determined by the current timing model.

Nevertheless, a $\chi^2$ analysis was performed to investigate what constraints the lack of a Shapiro delay detection imposes on the make-up and geometry of the binary system (i.e. the mass of the pulsar, $M_{\rm NS}$, and that of its companion, $M_{\rm WD}$, as well as the binary system inclination, $i$, by using the mass-function). A $\chi^2$-map was computed using the ChiSqCube plug-in[26] for TEMPO2 (see Fig. 64). The $\chi^2$-map was computed for $M_{\rm WD}$ between 0 and 1.5 $M_\odot$, in steps of $0.001~M_\odot$, and $\cos i$ from 0 to 1 in steps of 0.001.

For each point in the $\chi^2$-map the timing model was re-fit, holding the values of $M_{\rm WD}$ and $\cos i$ fixed. The resulting $\chi^2$ values were converted to probabilities by following Splaver et al. (2002), and then normalized. The 1-, 2-, and 3-$\sigma$ contours were chosen such that they contain $\sim$68.3, $\sim$95.5, and $\sim$99.7 % of the allowed binary system configurations. Based on this analysis we

---

[24]http://www.atnf.csiro.au/people/pulsar/psrcat/
[25]This comparison was made using the ATNF Pulsar Catalogue.
[26]The ChiSqCube plug-in populates a 3-dimensional $\chi^2$ space. For the purpose of the analysis presented here, the third dimension was not used.

*Thomas M. Tauris - Uni. Bonn*

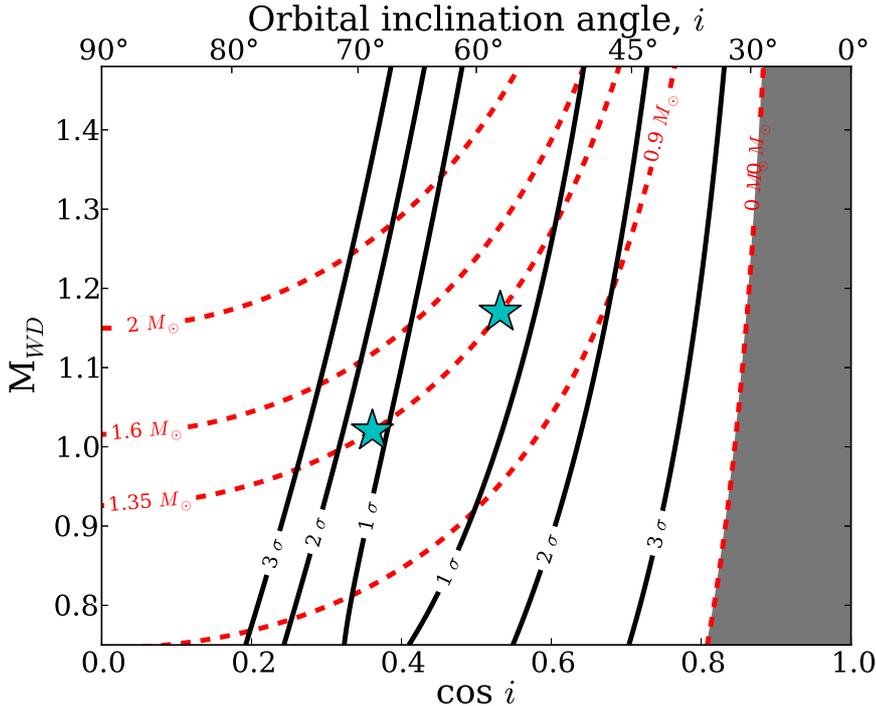

Figure 64: Map of companion-mass and inclination combinations allowed by the current timing data. The 1-,2-, and 3-σ contours, shown in black, enclose ∼68.3, ∼95.5, and ∼99.7 % of the allowed binary system configurations, respectively, given the current timing data, and requiring the companion's mass to lie in the range $0.75 \leq M_{WD} \leq 1.48 \, M_{\odot}$. The red dashed lines trace constant pulsar mass. The grey region in the right is excluded because the pulsar mass must be larger than 0. The two stars are the results from the two simulations of the binary system's evolution (see Section 7.4.1).

know the binary system cannot be edge-on ($i = 90°$). The inclination angle is constrained to be be $i \leq 75°$ for $M_{NS} \geq 1.35 \, M_{\odot}$, at the 3-σ level.

However, based on the massive companion, and small, but significantly non-zero eccentricity, we expect that with our current timing precision we will measure two (or possibly three) post-Keplerian parameters precisely enough to provide a stringent test of relativistic gravity, within the next 10 years (see Section 7.6).

## 7.4 Binary Evolution of the Progenitor

In general, binary pulsars with a CO WD companion[27] can form via different formation channels (see Tauris, Langer & Kramer 2012; and references therein).

Pulsars with CO WDs in orbits of $P_{orb} \leq 2$–3 days, like PSR J1952+2630 ($P_{orb} = 9.4$ hr), are believed to have formed via a common envelope (CE) scenario. Such systems originate from IMXBs which have very large values of $P_{orb}$ prior to the onset of the mass transfer. These systems are characterized by donor stars with masses between $2 < M_2/M_{\odot} < 7$, and very wide orbits up to $P_{orb} \simeq 10^3$ days. Donor stars near the tip of the red giant branch or on the asymptotic giant branch evolve via late Case B RLO or Case C RLO, respectively. As a result, these donor stars develop a deep convective envelope as they enter the giant phase, before filling their Roche lobe. These stars respond to mass loss by expanding, which causes them to overfill their Roche lobe even more. Binaries where mass transfer occurs from a more massive donor star to a less massive accreting NS shrink in size, causing further overfilling of the donor star's Roche lobe, which further enhances mass loss. This process leads to a dynamically unstable, runaway mass transfer and the formation of a CE (Paczyński 1976; Iben & Livio 1993; Ivanova et al. 2013). However, the wide orbit prior to the RLO is also the reason why these systems survive the CE and spiral-in phase, since the binding energy of donor star's envelope becomes weaker with advanced stellar age, and therefore the envelope is easier to eject, thereby avoiding a merger event.

---

[27]Here, and in the following, we simply write 'CO WD' for any massive WD whose exact chemical composition (CO or ONeMg) is unknown.

Given that the duration of the CE and spiral-in phase is quite short ($< 10^3$ yr; e.g. Podsiadlowski 2001; Passy et al. 2012; Ivanova et al. 2013) the NS can only accrete $\sim 10^{-5}$ $M_\odot$ during this phase, assuming that its accretion is limited by the Eddington accretion rate (a few $10^{-8}$ $M_\odot\,\mathrm{yr}^{-1}$). This small amount is not enough to even mildly recycle the pulsar. Instead, the NS is thought to be recycled during the subsequent so-called Case BB RLO (Tauris, Langer & Kramer 2012). This post-CE mass-transfer phase is a result of the naked helium star (the stripped core of the original IMXB donor star) filling its Roche lobe when it expands to become a giant during helium shell burning. Hence, for the purpose of understanding the recycling of PSR J1952+2630 we only have to consider this epoch of evolution in detail. A complete overview the full progenitor evolution of the system is illustrated in Fig. 65.

### 7.4.1 Calculations of Case BB RLO leading to recycling of PSR J1952+2630

Binary evolution for NS–massive WD systems have been studied using detailed calculations of the Case BB RLO we applied the Langer stellar evolution code (e.g. Tauris, Langer & Kramer 2011; 2012). However, none of the previously computed models produced systems sufficiently similar to that of PSR J1952+2630. We used the same code to study what progenitor systems can result in PSR J1952+2630-like binaries, the Case BB RLO of these systems, and the nature of the WD companion.

The masses of the WD and the NS ($M_{WD}$ and $M_{NS}$, respectively) are not known from timing at this stage (see Section 7.3.3). In order to limit the number of trial computations, we assumed $M_{NS} = 1.35$ $M_\odot$ and performed various calculations with different values of initial orbital period and initial mass of the helium donor star, $M_2$.

The young age of PSR J1952+2630 implies that $P_{orb}$ (now 9.4 hours) has not changed much by gravitational wave radiation since the termination of the mass transfer (it was at most ~9.6 hours; see Section 7.5). Therefore we only select progenitor solutions of our modelling that have similar orbital periods. We also impose the criterion $M_{WD}$ =0.93–1.48 $M_\odot$, to be consistent with the minimum companion mass derived from our timing solution.

Two solutions satisfying our selection criteria are shown in Fig. 66. The first solution (blue solid line) is for a 2.2 $M_\odot$ helium star leading to formation of a 1.17 $M_\odot$ ONeMg WD. The second solution (red dashed line) is for a 1.9 $M_\odot$ helium star leading to a 1.02 $M_\odot$ CO WD. In both cases we assumed a helium star metallicity of $Z = 0.02$ (solar metallicity), a reasonable assumption given that their $\sim 6 - 7$ $M_\odot$ progenitors had short lifetimes of ~100 Myr and thus belong to Galactic Population I stars. The second solution predicts a shorter pulsar spin period at the termination of accretion, and therefore imposes a less strict limit on the mass-accretion efficiency; see Section 7.5.1 for a discussion.

Unfortunately, given the current timing data it is not possible to place sufficiently stringent constraints on the binary system to be used to select either of the two simulated scenarios as the actual evolution of the binary (see Section 7.3.3, and Fig. 64).

For the remainder of Section 7.4 we will consider only the ONeMg WD solution to our modelling. In particular, we will highlight some of the more interesting characteristics of it.

The mass-transfer rate, $|\dot{M}_2|$, for the solution leading to the 1.17 $M_\odot$ ONeMg WD is shown in Fig. 67 as a function of time. The duration of the Case BB RLO is seen to last for about $\Delta t = 60$ kyr, which causes the NS to accrete an amount $\Delta M_{NS} \approx 0.7$–$6.4 \times 10^{-3}$ $M_\odot$, depending on the assumed accretion efficiency and the exact value of the Eddington accretion limit, $\dot{M}_{Edd}$. Here we assumed $\dot{M}_{Edd} = 3.9 \times 10^{-8}$ $M_\odot\,\mathrm{yr}^{-1}$ (a typical value for accretion of helium rich matter, Bhattacharya & van den Heuvel 1991) and allowed for the actual accretion rate to be somewhere in the interval 30–300% of this value. This is to account for the fact that the value of $\dot{M}_{Edd}$ is derived under idealized assumptions of spherical symmetry, steady-state accretion,

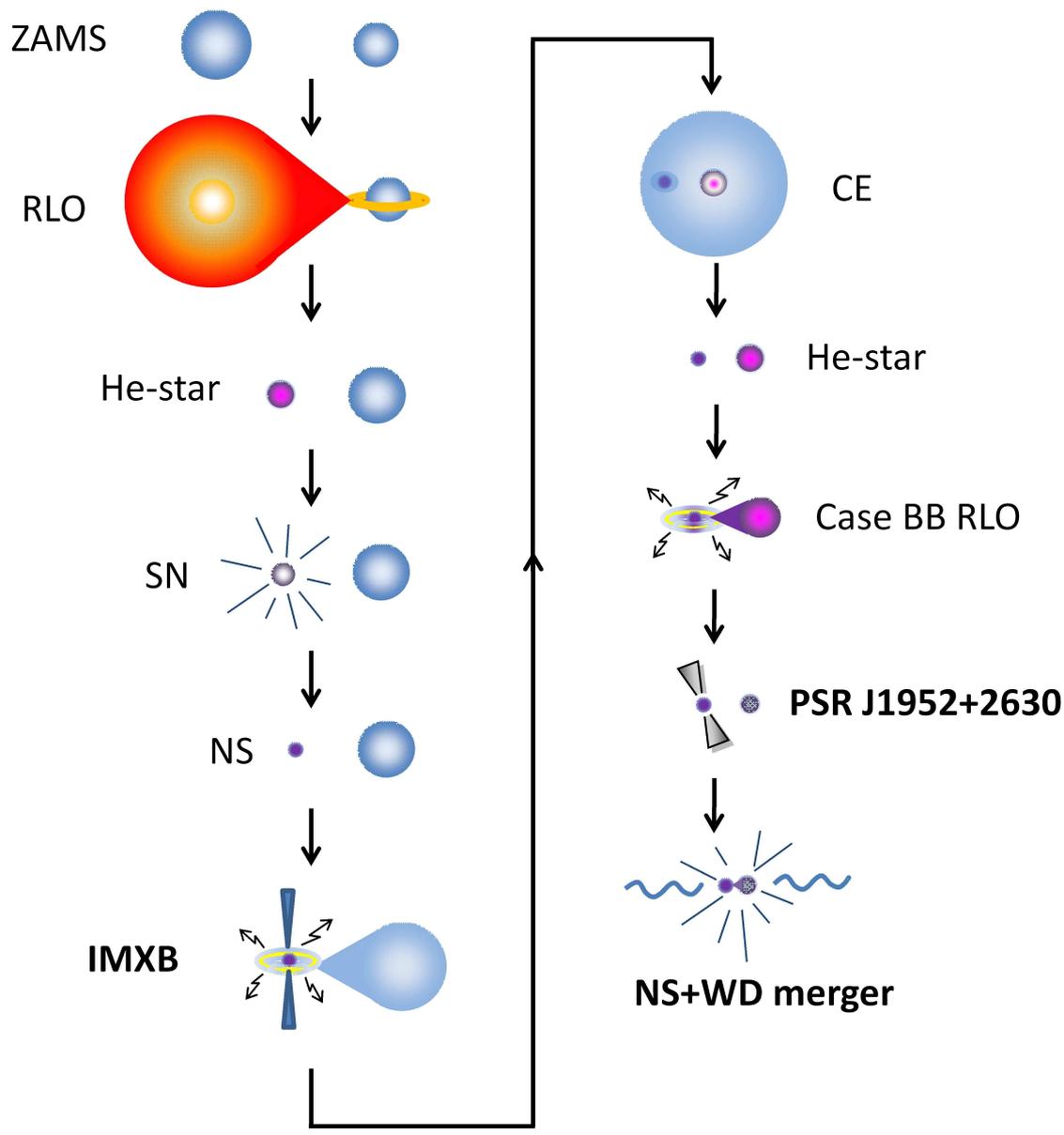

Figure 65: An illustration of the full binary stellar evolution from the zero-age main sequence (ZAMS) to the final merger stage. The initially more massive star evolves to initiate Roche-lobe overflow (RLO), leaving behind a naked helium core which collapses into a neutron star (NS) remnant, following a supernova explosion (SN). Thereafter, the system becomes a wide-orbit intermediate-mass X-ray binary (IMXB), leading to dynamically unstable mass transfer and the formation of a common envelope (CE), when the 6–7 $M_\odot$ donor star initiates RLO. The post-CE evolution, calculated in detail in this paper, is responsible for recycling the NS via Case BB RLO when the helium star companion expands to initiate a final mass-transfer episode. PSR J1952+2630 is currently observed as a mildly recycled radio pulsar orbiting a massive white dwarf (WD). The system will merge in $\sim 3.4$ Gyr, possibly leading to a $\gamma$-ray burst-like event and the formation of a single black hole.

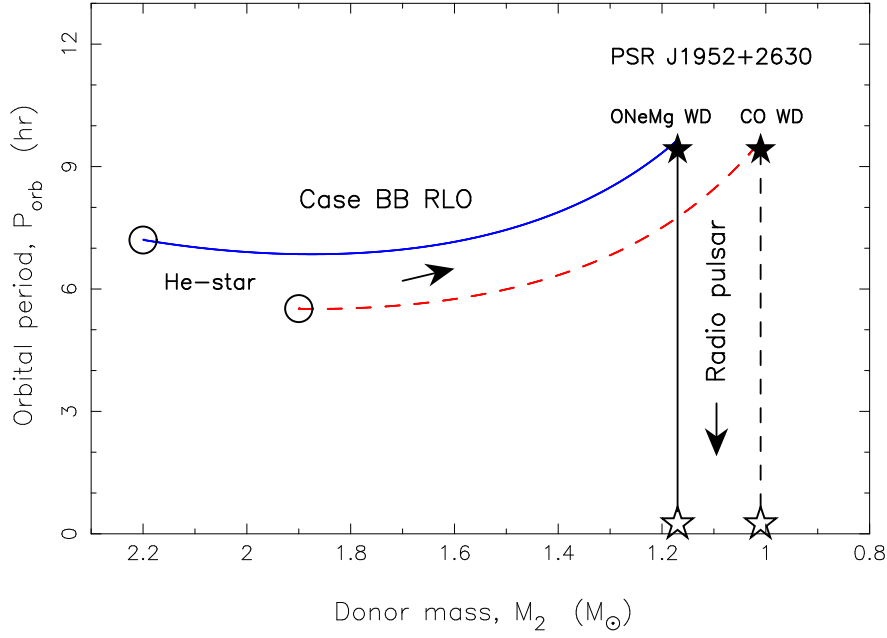

Figure 66: Progenitor evolution of the PSR J1952+2630 system in the $(M_2, P_{orb})$–plane during mass transfer via Case BB RLO. The blue solid line is the evolutionary track for a 2.2 $M_\odot$ helium star leaving a 1.17 $M_\odot$ ONeMg WD remnant. The red dashed line is for a 1.9 $M_\odot$ helium star leaving a 1.02 $M_\odot$ CO WD remnant. The open circles show the location of the helium star donors at the onset of the RLO. The solid stars indicate the termination of the RLO when the radio pulsar turns on. In about 3.4 Gyr the system will merge (open stars).

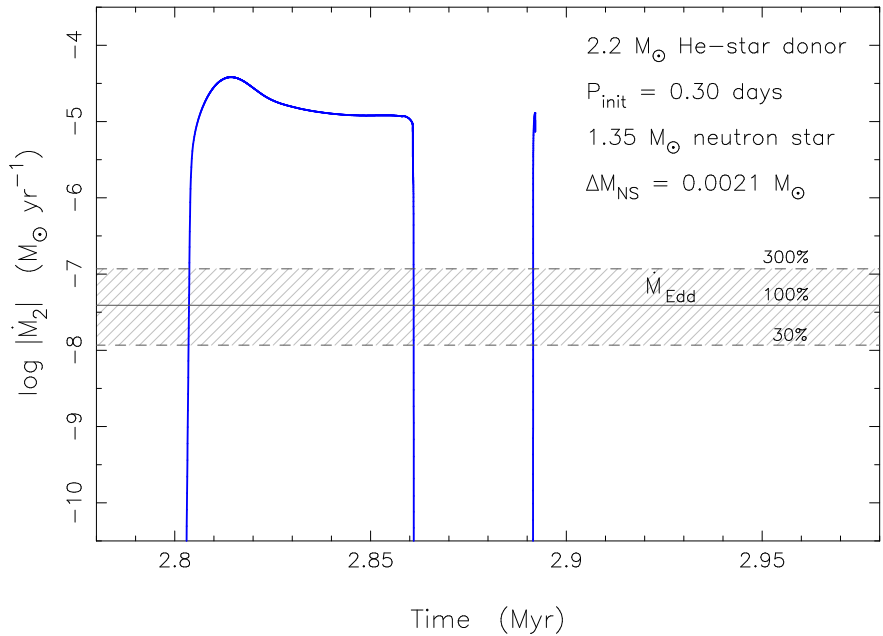

Figure 67: Mass-transfer rate as a function of stellar age for the progenitor evolution plotted in Fig. 66 leading to an ONeMg WD. The initial configuration is a 2.2 $M_\odot$ helium star orbiting a 1.35 $M_\odot$ NS with $P_{orb} = 0.30$ d. The Case BB RLO lasts for about 60 000 yr, then terminates for about 30 000 yr until a final vigorous helium shell flash is launched (the spike). The mass-transfer rate is seen to be highly super-Eddington ($\sim 10^3 \dot{M}_{Edd}$). The horizontal lines mark different values for the accretion efficiency in units of the Eddington accretion rate, $\dot{M}_{Edd}$. Depending on the exact value of $\dot{M}_{Edd}$, and the accretion efficiency, the NS accretes $0.7 - 6.4 \times 10^{-3} M_\odot$, which is sufficient to recycle PSR J1952+2630 – see text.

Thomson scattering opacity and Newtonian gravity. As we shall see, the accretion efficiency, and thus $\Delta M_{NS}$, is important for the spin period obtained by the NS during its spin-up phase. The mass-transfer rate from the helium star is highly super-Eddington ($|\dot{M}_2| \sim 10^3 \, \dot{M}_{Edd}$). The excess material (99.9%) is assumed to be ejected from the vicinity of the NS, in the form of a disc wind or a jet, with the specific orbital angular momentum of the NS following the so-called isotropic re-emission model (see Tauris, van den Heuvel & Savonije 2000; and references therein).

### 7.4.2 Detailed WD structure

The calculated interior structure and evolution of a likely progenitor of PSR J1952+2630, a 2.2 $M_\odot$ helium star which undergoes Case BB RLO and leaves behind an ONeMg WD, is illustrated in the "Kippenhahn diagram" (Kippenhahn & Weigert 1990) in Fig. 68. The plot shows the last 10 kyr of the mass-transfer phase ($t = 2.85$–$2.86$ Myr), followed by 32 kyr of evolution ($t = 2.860$–$2.892$ Myr) during which carbon is ignited in the detached donor star.

In our modelling of the companion star, there are four instances of off-centred carbon burning shells. These shells are the four blue regions underneath the green-hatched convection zones in Fig. 68. The ignition points are off-centre because these surrounding layers are hotter than the interior due to more efficient neutrino cooling in the higher-density inner core. The maximum temperature is near a mass coordinate of $m/M_\odot \simeq 0.4$.

The second carbon-burning shell penetrates to the centre of the proto-WD. However, at no point in the modelled evolution of the companion do the carbon burning shells, or the associated convection zones on top of these shells, reach the surface layers of the proto-WD. Therefore, the resulting WD structure is a hybrid, with a large ONeMg core engulfed by a thick CO mantle. The chemical abundance profile of the WD companion at the end of our modelling ($t = 2.892$ Myr) is demonstrated in Fig. 69. Notice the tiny layer ($2.7 \times 10^{-2} \, M_\odot$) of helium at the surface which gives rise to a vigorous helium shell flash at $t = 2.892$ Myr. This shell flash can also be seen in Figs. 67 and 68 and gives rise to numerical problems for our code. We therefore end our calculations without resolving this flash. However, since the NS is only expected to accrete of the order $\sim 10^{-5} \, M_\odot$ as a result of this flash (based on modelling of similar binaries where we managed to calculate through such a helium shell flash), its impact on the final binary and spin parameters will be completely negligible.

As far as we are aware, this is the first presentation in the literature of detailed calculations leading to an ONeMg WD orbiting a recycled pulsar.

### 7.5 Discussion

The future and past spin evolution of PSR J1952+2630 can be computed from the measured values of $P$ and $\dot{P}$ and assuming a (constant) braking index, $n$, which is defined by $\dot{\Omega} = -K\Omega^n$, where $\Omega = 2\pi/P$ and $K$ is a scaling factor (Manchester & Taylor 1977). The resulting future and the past spin evolution of PSR J1952+2630 are shown in Fig. 70, top and bottom panel, respectively.

Given our modelling, the true age of PSR J1952+2630 is $\lesssim 150$ Myr. If a cooling age of the WD companion of a pulsar in a similar system could be accurately determined (for which the WD mass is needed), it would be possible to constrain the braking index of a recycled pulsar. Although this is not likely possible in the case of PSR J1952+2630 due to its large (DM) distance of 9.6 kpc, it may be feasible for other systems in the future.

The orbital decay due to gravitational wave radiation has also been computed (see Fig. 70, top). It is evident that $P_{orb}$ has hardly decayed since the formation of PSR J1952+2630. Section 7.5.1 describes the implication for our binary stellar evolution modelling.

### 7.5.1 On the age and spin evolution of PSR J1952+2630

Our modelling of binary stellar evolution and accretion physics provides initial conditions for the spin of PSR J1952+2630, which must be consistent with current measurements. In particular, the spin period of PSR J1952+2630 predicted at the termination of accretion must be smaller than the observed spin period.

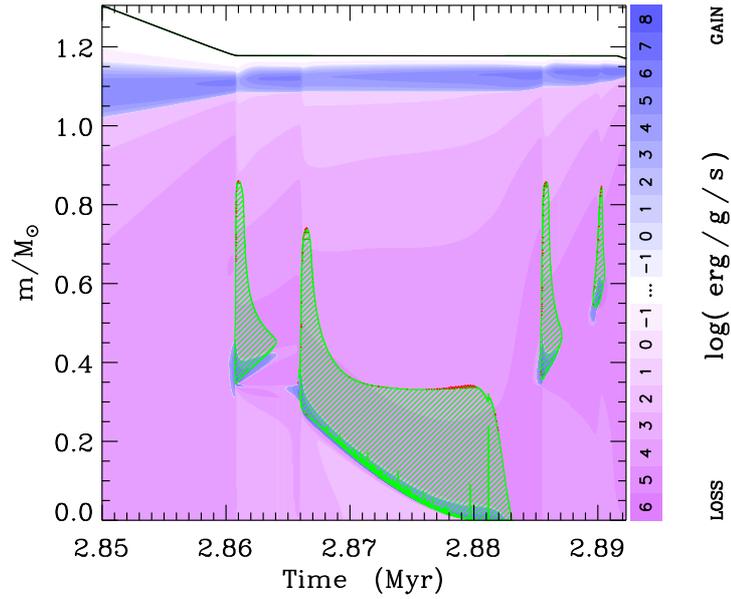

Figure 68: The Kippenhahn diagram showing the formation of an ONeMg WD companion to PSR J1952+2630. The plot shows cross-sections of the progenitor star in mass-coordinates from the centre to the surface, along the y-axis, as a function of stellar age (since the helium star ZAMS) on the x-axis. Only the last 42 kyr of our calculations are plotted. The Case BB RLO is terminated at time $t = 2.86$ Myr when the progenitor star has reduced its mass to $1.17\,M_\odot$. The green hatched areas denote zones where is convection. The intensity of the blue/purple colour indicates the net energy-production rate; the helium burning shell near the surface is clearly seen at $m/M_\odot \simeq 1.1$ as well the off-centred carbon ignition in shells, starting at $m/M_\odot \simeq 0.4$, defining the subsequent inner boundaries of the convection zones. The mixing of elements due to convection expands the ONeMg core out to a mass coordinate of about $m/M_\odot \simeq 0.85$ (see Fig. 69). Energy losses due to neutrino emission are quite dominant outside of the nuclear burning shells.

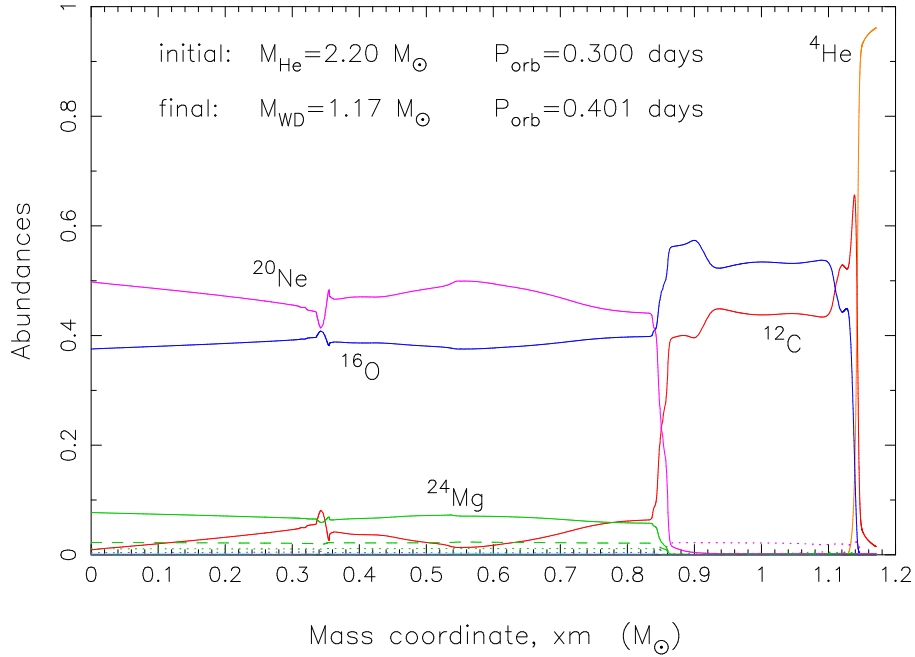

Figure 69: The chemical abundance structure of the ONeMg WD remnant (from our last calculated model at $t = 2.892$ Myr) of the Case BB RLO calculation shown in Figs. 66–68, one of the plausible solutions for PSR J1952+2630's companion found in our modelling. This $1.17\,M_\odot$ WD has a hybrid structure with an ONeMg core enclosed by a CO mantle and a tiny ($0.027\,M_\odot$) surface layer of helium.

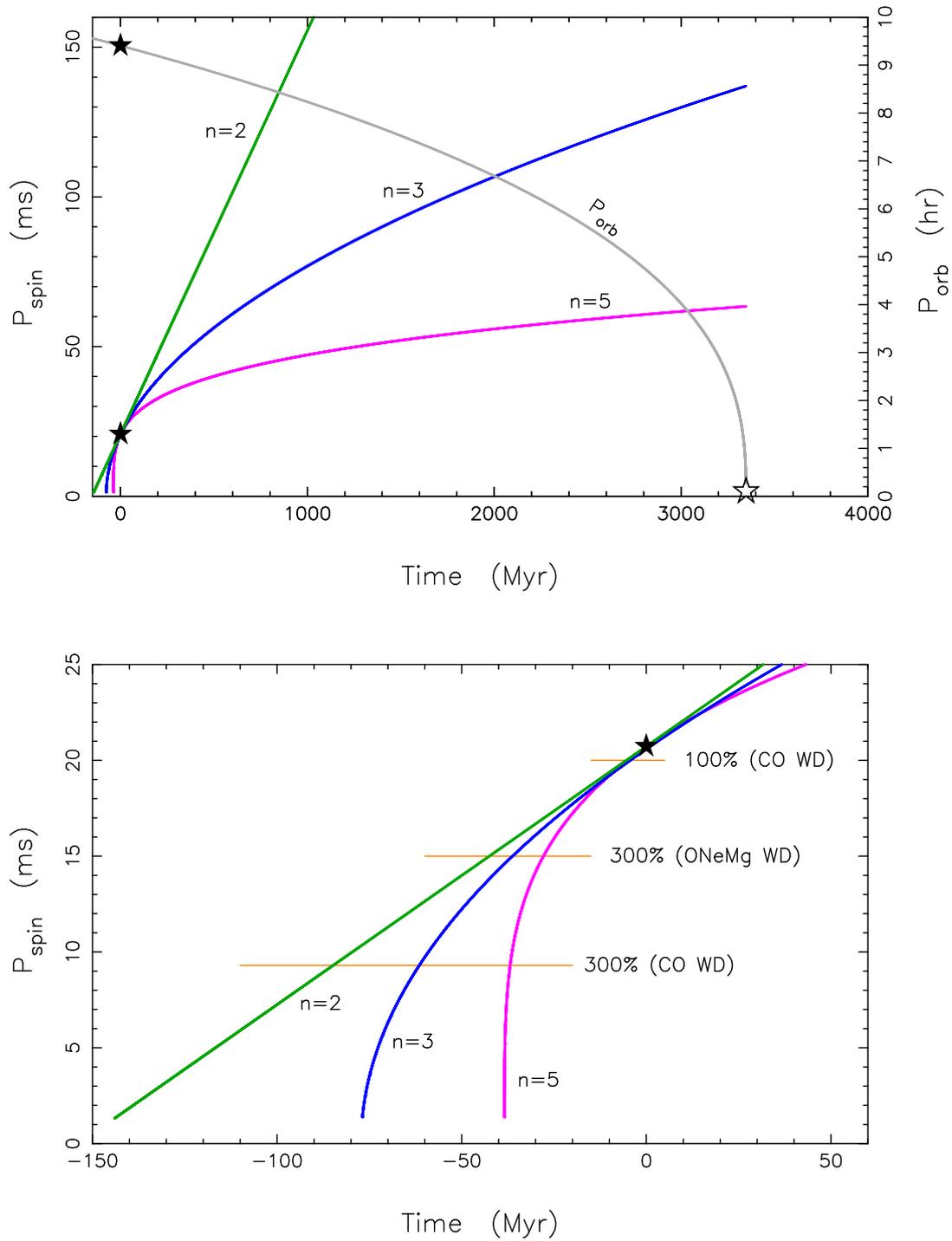

Figure 70: The future (top) and the past (bottom) spin evolution of PSR J1952+2630 for different values of the braking index, $n$. The location of PSR J1952+2630 at present is marked by a solid star. In the top panel, the grey curve shows the calculated orbital decay due to gravitational wave radiation until the system merges in about 3.4 Gyr (marked by an unfilled star). The lower panel is a zoom-in on the past spin evolution. Depending on $n$, the WD companion mass and the accretion efficiency of the NS during Case BB RLO, the pulsar could have been spun up to the initial spin periods indicated by the orange horizontal lines – see text for a discussion.

The minimum equilibrium spin period can be estimated given the amount of mass accreted by the NS, $\Delta M_{NS}$. This quantity depends on our binary evolution models and the assumed accretion efficiency (in units of $\dot{M}_{Edd}$, see Fig. 67). Following equation 14 of Tauris, Langer & Kramer (2012),

$$P_{ms} = \frac{(M_{NS}/M_\odot)^{1/4}}{(\Delta M_{NS}/0.22\ M_\odot)^{3/4}}, \tag{71}$$

where $P_{ms}$ is the equilibrium spin period in units of ms, we estimate the minimum equilibrium spin period of PSR J1952+2630. These results are tabulated in Table 13, and compared with PSR J1952+2630's current spin period in Fig. 70 (bottom). It is interesting to notice that we only obtain solutions for an accretion efficiency of 100% or 300% of $\dot{M}_{Edd}$[28]. If the accretion efficiency is smaller, the equilibrium spin period becomes larger that the present spin period of $P = 20.7$ ms, which is impossible.

The reason for the possibility of a lower initial spin period in case PSR J1952+2630 has a CO WD companion, is simply that the helium star progenitor of a CO WD has a lower mass and therefore evolves on a longer time-scale, thereby increasing $\Delta M_{NS}$. Hence, a cooling age estimate of the WD companion could, in principle, also help constrain the accretion physics, because it puts limitations on the possible values of the initial spin period.

This is the first time an accretion efficiency has been constrained for a recycled pulsar which evolved via Case BB RLO. In contrast, the accretion efficiency of millisecond pulsars formed in low-mass X-ray binaries (LMXBs) has been shown to be much lower, about 30% in some cases (Tauris & Savonije 1999; Jacoby et al. 2005; Antoniadis et al. 2012). The reason for this difference in accretion efficiencies may be related to the extremely high mass-transfer rates during Case BB RLO which could influence the accretion flow geometry and thus $\dot{M}_{Edd}$. Furthermore, accretion disc instabilities (Lasota 2001; Coriat, Fender & Dubus 2012), which act to decrease the accretion efficiency in LMXBs, do not operate in Case BB RLO binaries, due to the high value of $|\dot{M}_2|$.

Table 13: Equilibrium spin obtained via Case BB RLO.

| Acc. eff. | $\Delta t_{RLO}$ (kyr) | $\Delta M_{NS}$ $(M_\odot)$ | $P$ (ms) |
|---|---|---|---|
| 2.2 $M_\odot$ He star $\longrightarrow$ 1.17 $M_\odot$ ONeMg WD | | | |
| 30% | 57 | $0.7 \times 10^{-3}$ | 80 |
| 100% | 61 | $2.1 \times 10^{-3}$ | 35 |
| 300% | 61 | $6.4 \times 10^{-3}$ | 15 |
| 1.9 $M_\odot$ He star $\longrightarrow$ 1.02 $M_\odot$ CO WD | | | |
| 30% | 113 | $1.5 \times 10^{-3}$ | 45 |
| 100% | 119 | $4.4 \times 10^{-3}$ | 20 |
| 300% | 113 | $12.4 \times 10^{-3}$ | 9.3 |

#### 7.5.1.1 Evolution in the $P$-$\dot{P}$ diagram

By integrating the pulsar spin deceleration equation: $\dot{\Omega} = -K\Omega^n$, assuming a constant braking index, $n$ we obtain isochrones. The kinematic solution at time $t$ (positive in the future, negative in the past) is given by:

$$P = P_0 \left[ 1 + (n-1)\frac{\dot{P}_0}{P_0} t \right]^{1/(n-1)} \tag{72}$$

---

[28]Solutions requiring larger-than-$\dot{M}_{Edd}$ accretion efficiencies are in fact physically viable because assumptions made during the calculation of $\dot{M}_{Edd}$ mean it is only a rough measure of the true limiting accretion rate.

$$\dot{P} = \dot{P}_0 \left( \frac{P}{P_0} \right)^{2-n},  \tag{73}$$

where $P_0 = 20.7$ ms and $\dot{P}_0 = 4.27 \times 10^{-18}\,\mathrm{s\,s^{-1}}$ are approximately the present-day values of the spin period and its derivative. The past and future spin evolution of PSR J1952+2630 in the $P$-$\dot{P}$ diagram are plotted in Fig. 71. The isochrones are calculated using equations 72 and 73 where $n$ varies from 2 to 5, for different fixed values of $t$ in the future (rainbow colours) and past (brown). For each isochrone, the time is given by the well-known expression:

$$t = \frac{P}{(n-1)\dot{P}} \left[ 1 - \left( \frac{P_0}{P} \right)^{n-1} \right].  \tag{74}$$

These solutions, however, are purely based on rotational kinematics. As already discussed above, one must take the constraints obtained from binary evolution and accretion physics into account. Therefore, if the Case BB RLO is not able to spin up the pulsar to a value smaller than, for example, 15 ms, then the true age of PSR J1952+2630 cannot be much more than 40 Myr for all values of $2 \le n \le 5$. For nearby binary systems similar to PSR J1952+2630 where a cooling age determination of the WD is possible, such a measurement could be useful for constraining the birth period of the pulsar, as well as the system's accretion efficiency.

Also shown in Fig. 71 is the future spin evolution of PSR J1952+2630 until the system merges in about 3.4 Gyr. It is interesting to notice that a system like PSR J1952+2630 should be observable as a radio pulsar binary until it merges, suggesting the existence of similar NS–massive WD binaries with much shorter orbital periods, to which PALFA is sensitive (Allen et al. 2013; Lazarus 2013). The unique location of PSR J1952+2630 with respect to other known recycled pulsars with a massive WD companion (marked with blue diamonds) is also clear from this figure. It is seen that only three other systems may share a past location in the $P$-$\dot{P}$ diagram similar to that of PSR J1952+2630 (if $2 \le n \le 5$). This could suggest that such surviving post-CE systems are often formed with small values of $P_{\mathrm{orb}}$ which cause them to merge rapidly – either during the Case BB RLO or shortly thereafter due to gravitational wave radiation.

## 7.6  Future Prospects for PSR J1952+2630

We now investigate the future use of PSR J1952+2630 as a gravitational laboratory. Looking at Table 12, we can see that the eccentricity, $e$, and longitude of periastron, $\omega$, can be measured quite precisely, in the latter case to within $1.2°$, despite the small absolute value of the eccentricity, $4.1(1) \times 10^{-5}$. If we assume a mass of $1.35\,M_\odot$ for the pulsar and $1.1\,M_\odot$ for the WD, then general relativity predicts that $\omega$ should increase at a rate $\dot{\omega} = 1.72°\,\mathrm{yr^{-1}}$, which given the precision of $\omega$ implies that the effect should be detectable in the next few years. Measuring it will eventually give us an estimate of the total mass of the system (Weisberg & Taylor 1981). Furthermore, thanks to PSR J1952+2630's rather short orbital period, the shortest among recycled pulsar–massive WD systems, the rate of gravitational wave emission is much higher than for any other such system. This emission will cause the orbit to decay. For the same assumptions as above, the orbital period should change at a rate $\dot{P}_{b,\mathrm{pred}} = -1.14 \times 10^{-13}\,\mathrm{s\,s^{-1}}$. As previously mentioned, this will cause the system to merge within about 3.4 Gyr.

The orbital decay due to the emission of gravitational waves is not measurable at present but it should be detectable in the near future. In order to verify this, we made simulations of future timing of this pulsar that assume similar timing precision as at present (a single 18-$\mu$s TOA every 15 minutes for each of two 300-MHz bands centred around 1400 MHz). We assume two campaigns in 2015 and 2020, where the full orbit is sampled a total of 8 times, spread evenly for each of those years. The total observing time of 72 hours for each of those years is a realistic target. These simulations indicate that by the year 2020 $\dot{P}_b$ should be detectable with 12-$\sigma$ significance and $\dot{\omega}$ to about 9-$\sigma$ significance.

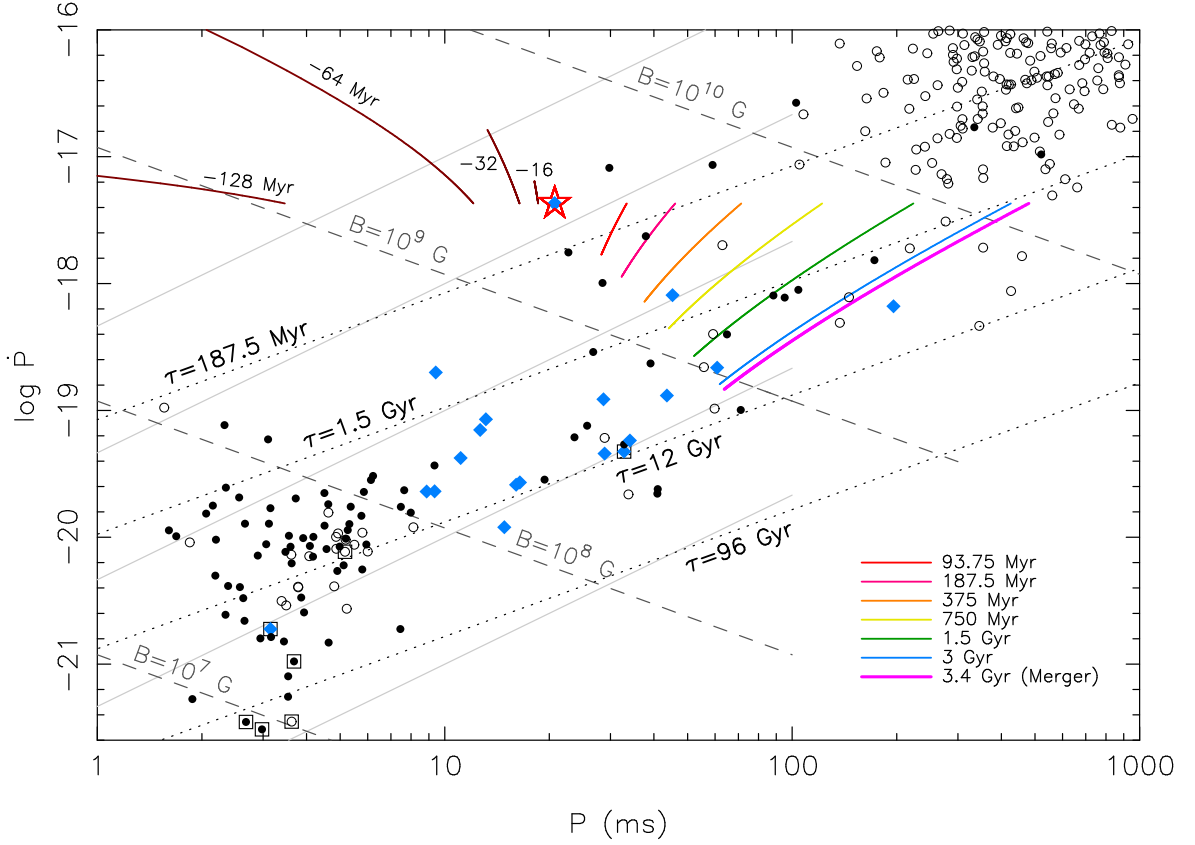

Figure 71: Isochrones of past (brown colour) and future evolution (rainbow colours) of PSR J1952+2630 in the $P$-$\dot{P}$ diagram. The present location is marked by the open red star. All isochrones were calculated for braking indices in the interval $2 \leq n \leq 5$. Also plotted are inferred constant values of $B$-fields (dashed lines) and characteristic ages, $\tau$ (dotted lines). The thin grey lines are spin-up lines with $\dot{M}/\dot{M}_{\mathrm{Edd}} = 1$, $10^{-1}$, $10^{-2}$, $10^{-3}$ and $10^{-4}$ (top to bottom), assuming a pulsar mass of 1.4 $M_\odot$. Observational data of the plotted Galactic field pulsars are taken from the ATNF Pulsar Catalogue in March 2013. Binary pulsars are marked with solid circles and isolated pulsars are marked with open circles. For further explanations of the calculations, and corrections to $\dot{P}$ values see Tauris, Langer & Kramer (2012). Binary pulsars with a massive (CO or ONeMg) WD companion are marked with a blue diamond. The past spin evolution of PSR J1952+2630 is particularly interesting as it constrains both the binary evolution and the recycling process leading to its formation – see text.

The resulting constraints on the masses of the components and system inclination are depicted graphically in Fig. 72. We plot the 1-$\sigma$ bands allowed by the 'measurement' of a particular parameter. The figure shows several interesting features. The first is that even with these measurements of $\dot{P}_b$ and $\dot{\omega}$, it is not possible to determine the two masses accurately from them, given the way their 1-$\sigma$ uncertainty bands intersect in the mass-mass diagram. The 1-$\sigma$ band of $h_3$ (the orthometric amplitude of the Shapiro delay, see Freire & Wex 2010) intersects the others at a rather sharp angle and can in principle be used to determine the masses more accurately. However, at the moment it is not clear whether $h_3$ is precisely measurable. A more massive companion in a less inclined orbit yields more pessimistic expectations, as depicted in Fig. 72.

Fortunately, very precise masses are not required to perform a test of general relativity using $\dot{\omega}$ and $\dot{P}_b$. As illustrated in Fig. 72, the bands allowed by $\dot{\omega}$ and $\dot{P}_b$ are nearly parallel. As the precision of these measurements improves general relativity is tested by the requirement that the two bands still overlap. Fig. 72 also shows that even if we could determine $h_3$ precisely, the precision of this $\dot{\omega}$-$h_3$-$\dot{P}_b$ test will be limited by the precision of the measurement of $\dot{\omega}$, because its uncertainty only decreases with $T^{-3/2}$ (where $T$ is the timing baseline), while for $\dot{P}_b$ the measurement uncertainty decreases with $T^{-5/2}$.

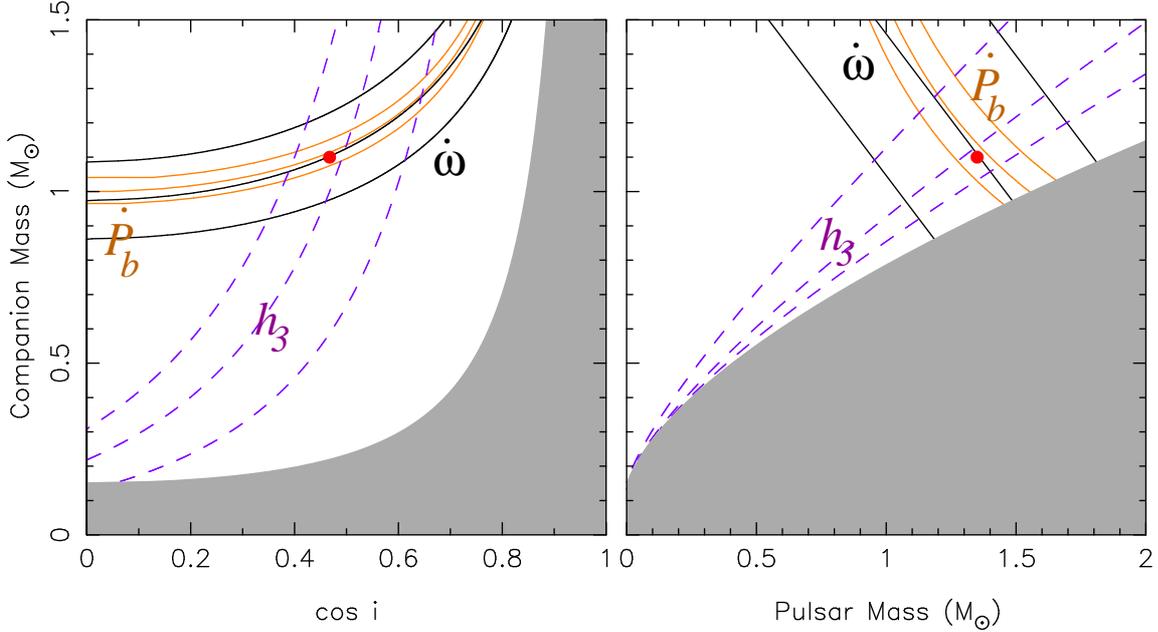

Figure 72: Constraints on system masses and orbital inclination from simulated radio timing of PSR J1952+2630. Each triplet of curves corresponds to the most likely value and standard deviations of the respective parameters; the region limited by it is the "1-$\sigma$" band for that parameter. The masses and inclination used in the simulation are indicated by the red points. Left-hand side: $\cos i$–$M_{WD}$ plane. The grey region is excluded by the condition $M_{NS} > 0$. Right-hand side: $M_{NS}$–$M_{WD}$ plane. The grey region is excluded by the condition $\sin i \leq 1$.

Eventually, though, the uncertainty of the measurement of $\dot{\omega}$ will become very small. At that stage the precision of this test will be limited by the precision of $\dot{P}_b$, which is limited by the lack of precise knowledge of the kinematic contributions (Damour & Taylor 1991):

$$\left(\frac{\dot{P}_b}{P_b}\right) = \left(\frac{\dot{P}_b}{P_b}\right)_{obs} - \frac{\mu^2 d}{c} - \frac{a(d)}{c}, \quad (75)$$

where the subscript "obs" indicated the observed quantity, $\mu$ is the total proper motion, $d$ is the distance to the pulsar and $a(d)$ is the (distance dependent) difference between the Galactic acceleration of the system and that of the Solar System Barycentre (SSB), projected along the direction from the SSB to the system. At present these cannot be estimated because the proper motion has not yet been measured precisely. However, they likely represent the ultimate constraint to the precision of this GR test, as in the case of PSR B1913+16 (Weisberg, Nice & Taylor 2010). For that reason, we now estimate the magnitude of these kinematic effects.

If we assume that the final proper motion is of the same order of magnitude as what is observed now ($\sim 6\ \text{mas yr}^{-1}$), then at the assumed DM-distance of 9.6 kpc the kinematic contribution to $\dot{P}_b$ will be about $3 \times 10^{-14}\ \text{s s}^{-1}$. This is four times smaller than $\dot{P}_{b,pred}$, as defined above, which means that if we can't determine the distance accurately, then the real value for the intrinsic orbital decay, $\dot{P}_{b,int}$, cannot be measured with a relative precision better than about 25 %. Things improve if the proper motion turns out to be smaller.

Is such a measurement useful? The surprising answer is that it is likely to be so. For pulsar-WD systems, alternative theories of gravity, like Scalar-Tensor theories (see Damour & Esposito-Farèse 1998; and references therein) predict the emission of dipolar gravitational waves, which would result in an increased rate of orbital decay. In the case of PSR J1738+0333, the orbital decay for that system was measured only with a significance of 8-$\sigma$. However, the absolute difference between the $\dot{P}_b$ predicted by GR and the observed value is so small that it introduces the most stringent constraints ever on these gravity theories (Freire et al. 2012). The implication is that for PSR J1952+2630, one should be able to derive similarly low limits. However, if the proper motion is significantly smaller, and/or if we are able to determine the distance independently, then this system can provide a much more stringent test of alternative theories of gravity. The main reason for this is that the limiting factor of the PSR J1738+0333 test is the limited precision in the measurement of the component masses (Antoniadis et al. 2012) which would not be an issue for PSR J1952+2630, given the tighter constraints on the total mass that will eventually be derived from $\dot{\omega}$.

## 7.7 Conclusions

We have presented phase-coherent timing of PSR J1952+2630. Our timing model includes precise determinations of parameters describing the pulsar spin-down, astrometry and binary motion. No post-Keplerian orbital parameters are required. However, detailed modelling suggests the current pulsar–massive WD binary system evolved via post-CE Case BB RLO with an accretion efficiency which exceeded the Eddington limit by a factor of 1–3. We presented, for the first time, a detailed chemical abundance structure of an ONeMg WD orbiting a pulsar. By projecting PSR J1952+2630's orbital evolution into the future we estimate it will merge with its WD companion in ∼3.4 Gyr due to the orbital decay from gravitational wave emission. Unfortunately, PSR J1952+2630 is too distant to make a detection of the cooling age of the pulsar's WD companion. In the case of the discovery of a less distant analog of PSR J1952+2630 such a measurement could make it possible to constrain the braking index of the recycled pulsar, and/or the accretion efficiency during the Case BB RLO-phase. Also, timing observations over the next 10 years will result in the detection of the advance of periastron, and the orbital decay, enabling a test of general relativity. Finally, additional timing may also further elucidate the nature of the companion, and will permit PSR J1952+2630 to be used to perform gravitational tests.

# FORMATION OF A TRIPLE MILLISECOND PULSAR

- **Chapter 8**
  Tauris & van den Heuvel (2014), ApJ Letters 781, L13
  *Formation of the Galactic Millisecond Pulsar Triple System PSR J0337+1715*
  *– A Neutron Star with Two Orbiting White Dwarfs*

## 8. Formation of the Galactic Millisecond Pulsar Triple System PSR J0337+1715 − A Neutron Star with Two Orbiting White Dwarfs

**Tauris & van den Heuvel (2014)**

**ApJ Letters 781, L13**

## Abstract

The millisecond pulsar in a triple system (J0337+1715, recently discovered by Ransom et al.) is an unusual neutron star with two orbiting white dwarfs. The existence of such a system in the Galactic field poses new challenges to stellar astrophysics for understanding evolution, interactions and mass-transfer in close multiple stellar systems. In addition, this system provides the first precise confirmation for a very wide-orbit system of the white dwarf mass−orbital period relation. Here we present a self-consistent, semi-analytical solution to the formation of PSR J0337+1715. Our model constrains the peculiar velocity of the system to be less than $160 \text{ km s}^{-1}$ and brings novel insight to, for example, common envelope evolution in a triple system, for which we find evidence for in-spiral of both outer stars. Finally, we briefly discuss our scenario in relation to alternative models.

### 8.1 Introduction

Stars are possibly always formed in multiple systems (e.g. Bonnell, Bate & Vine 2003) and observational estimates suggest that about 20%–30% of all binary stars are in fact members of triple systems (Tokovinin et al. 2006; Rappaport et al. 2013). Triple systems can remain bound with a long-term stability if they have a hierarchical structure (e.g. a close inner binary with a third star in relatively distant orbit). In addition, a number of peculiar binary pulsars have recently been discovered, such as PSR J1903+0327 (Champion et al. 2008), which require a triple system origin (e.g. Freire et al. 2011b; Portegies Zwart et al. 2011; Pijloo, Caputo & Portegies Zwart 2012).

The discoveries of binaries with a triple origin is not unexpected. Iben & Tutukov (1999) estimated that in ∼70% of the triple systems, the inner binary is close enough that the most massive star will evolve to fill its Roche lobe. Furthermore, in ∼15% of the triples, the outer third (tertiary) star may also fill its Roche lobe at some point, possibly leading to disintegration or production of rare configurations with three degenerate objects in the same system. Recently, Ransom et al. (2014) have reported the discovery of PSR J0337+1715, which is the first example of such an exotic system – a neutron star orbited by two white dwarfs.

PSR J0337+1715 is a triple system located at a distance of ∼1.3 kpc. It contains a 1.438 $M_\odot$ radio millisecond pulsar (MSP) with a spin period of $P = 2.73$ ms and two white dwarfs (WDs) with masses of $M_{\text{WD,2}} = 0.197 \ M_\odot$ and $M_{\text{WD,3}} = 0.410 \ M_\odot$, and orbital periods of $P_{\text{orb,12}} = 1.63$ days and $P_{\text{orb,3}} = 327$ days, respectively. Thus this triple system is highly hierarchical with a close inner binary and a distant tertiary star. In addition, the system is almost exactly coplanar ($\delta_i = 0.01°$), and the orbits are quite circular with eccentricities of $e_{12} = 6.9 \times 10^{-4}$ and $e_3 = 0.035$ (Ransom et al. 2014).

Here, we investigate the formation of such a triple compact object system and present a model which aims to explain and reconcile the observed data with current theories of stellar interactions.

### 8.2 Progenitor Evolution of PSR J0337+1715

To investigate the formation of PSR J0337+1715 we start with constraints obtained from the present-day triple system and trace the evolution backward. Before elaborating on the details, we briefly summarize the outline of our model which is illustrated in Figure 73. Numerical parameters are provided in Table 14.

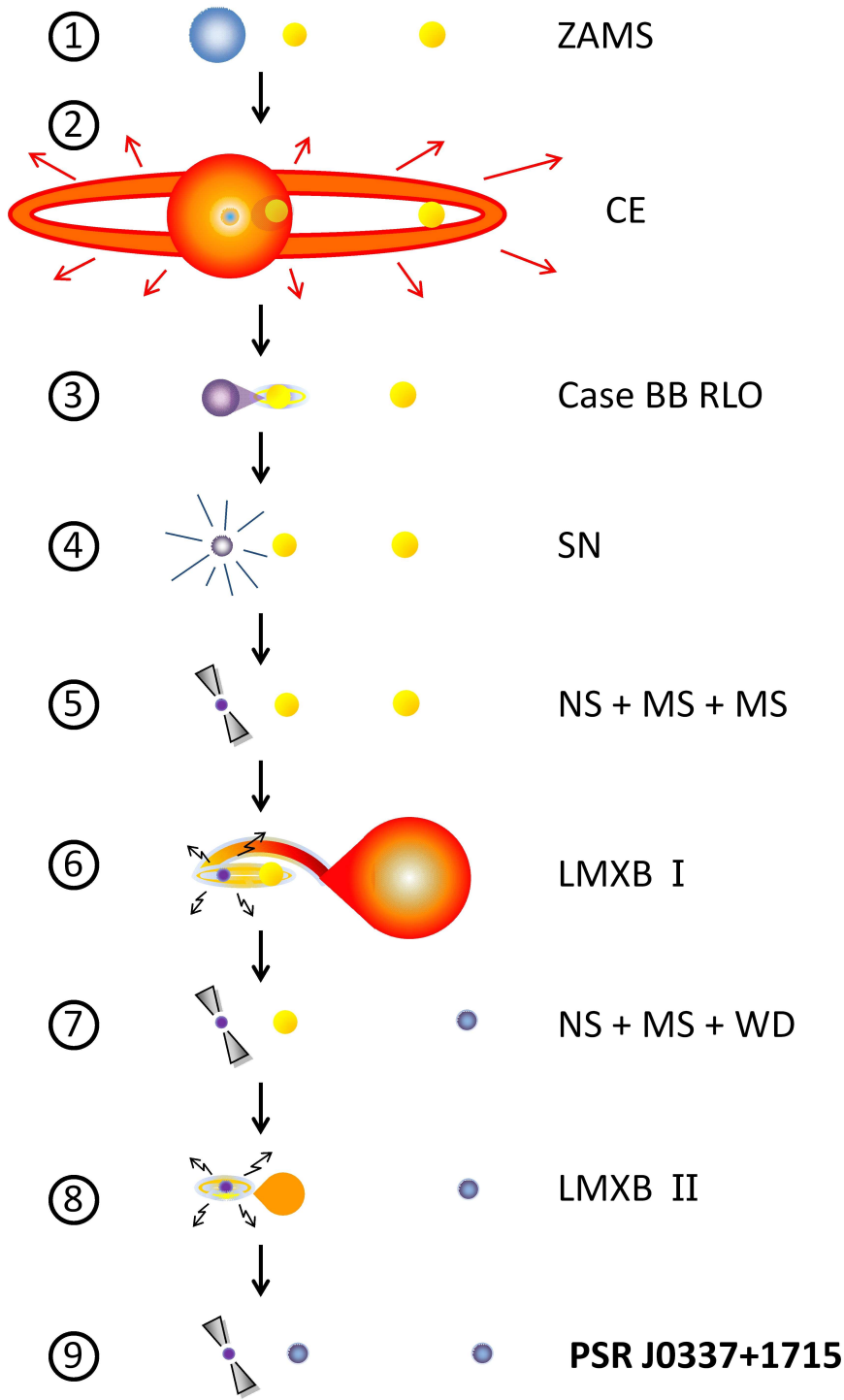

Figure 73: Illustration of our triple star evolution from the zero-age main sequence (ZAMS) to the present observed system PSR J0337+1715. Numerical parameters are given in Table 14. The initially massive B-star evolves to initiate Roche-lobe overflow (RLO) towards the inner G/F-star, leading to dynamical unstable mass transfer and the formation of a common envelope (CE), partially embedding the outer F-star. The resulting helium star (the naked core of the massive star) expands and initiates another phase of (Case BB) RLO, before it collapses into a neutron star (NS) in a supernova (SN) explosion. Thereafter, the system becomes visible as a young radio pulsar with two main-sequence (MS) stars. Given that the tertiary star is more massive than the secondary star, the outer LMXB phase (lasting $15 - 20$ Myr) occurs before the inner LMXB phase. The latter mass-transfer episode proceeds on a long timescale (2 Gyr), causing the NS to become a fully recycled MSP when finally orbited by two white dwarfs (WDs).

Table 14: Triple System Parameters at the Onset of each Stage in Our Scenario (Figure 73) for the Formation of PSR J0337+1715

| | | | Stage: | 1 | 2 | 3 | 4 | 5 | 6 | 7 | 8 | 9 |
|---|---|---|---|---|---|---|---|---|---|---|---|---|
| Age of the system | $t$ | [Myr] | | 0.0 | 23.3 | 23.3 | 25.1 | 25.2 | 5500 | 5517 | 8500 | 10500 |
| Mass of primary star | $M_1$ | [$M_\odot$] | | 10.0 | 9.90 | 2.90 | 1.70 | 1.28 | 1.28 | 1.30 | 1.30 | 1.438 |
| Mass of secondary star | $M_2$ | [$M_\odot$] | | 1.10 | 1.10 | 1.10 | 1.10 | 1.10 | 1.10 | 1.12 | 1.12 | 0.197 |
| Mass of tertiary star | $M_3$ | [$M_\odot$] | | 1.30 | 1.30 | 1.30 | 1.30 | 1.30 | 1.30 | 0.410 | 0.410 | 0.410 |
| Orbital period of inner binary | $P_{\mathrm{orb},12}$ | [days] | | 835 | 849 | 2.47 | 0.95 | 1.55 | 1.55 | 1.50 | 0.90 | 1.63 |
| Orbital period of tertiary star | $P_{\mathrm{orb},3}$ | [days] | | 4020 | 4080 | 17.1 | 15.7 | 15.3 | 14.2 | 250 | 250 | 327 |
| Eccentricity of inner binary | $e_{12}$ | | | 0.00 | 0.00 | 0.02 | 0.01 | 0.24 | 0.20 | 0.02 | 0.00 | 0.00 |
| Eccentricity of outer orbit | $e_3$ | | | 0.00 | 0.00 | 0.04 | 0.04 | 0.22 | 0.03 | 0.03 | 0.03 | 0.03 |
| Stability parameter | $(R_{\mathrm{peri}}/a_{\mathrm{in}})$ | | | 2.96 | 2.96 | 3.83 | 7.07 | 4.15 | 4.89 | 30.8 | 43.3 | 35.6 |
| Critical stability limit | $(R_{\mathrm{peri}}/a_{\mathrm{in}})_{\mathrm{crit}}$ | | | 2.93 | 2.93 | 3.21 | 3.44 | 3.93 | 3.80 | 3.04 | 3.04 | 3.13 |
| Temperature of outer WD | $T_{\mathrm{eff},3}$ | [K] | | | | | | | | 18000 | 5800 | 4300 |

### 8.2.1 Summary of our model

According to our model, the system started out on the zero-age main sequence (ZAMS) with a roughly 10 $M_\odot$ primary star and two companions with masses of about 1.10 $M_\odot$ and 1.30 $M_\odot$, for the secondary and the tertiary star, respectively (Table 14, stage 1). After a common envelope (CE) phase (stage 2) where the extended envelope of the primary engulfed the other two stars (initially only embedding the secondary star; later also partly the tertiary star), the orbital period of the inner system was $P_{\rm orb,12} = 2.47$ days and the orbital period of the outer star was $P_{\rm orb,3} = 17.1$ days. Following a second mass transfer (Case BB, stage 3) and a supernova (SN) explosion (stage 4) they became $P_{\rm orb,12} = 1.55$ days and $P_{\rm orb,3} = 15.3$ days, which after orbital circularization (before stage 6) became $P_{\rm orb,12} = 1.55$ days and $P_{\rm orb,3} = 14.2$ days. The last set of values were the orbital periods at the onset of the first (outer) low-mass X-ray binary (LMXB) phase, which ended with $P_{\rm orb,12} = 1.50$ days and $P_{\rm orb,3} = 250$ days, before the second (inner) LMXB phase left the system with its present observed properties. We now describe in more detail the physical properties of our model.

### 8.2.2 The $M_{\rm WD} - P_{\rm orb}$ relation

A close triple system like PSR J0337+1715 with one neutron star (NS) and two WDs requires two LMXB phases. Although PSR J0337+1715 was substantially less hierarchical earlier in its evolution, the system seems to have evolved through both of its LMXB phases, in effect, mainly via binary interactions, with only small dynamical perturbations from the second or third star. The important piece of evidence for this comes from the masses and orbital periods of the WDs which fall exactly as predicted by the $M_{\rm WD} - P_{\rm orb}$ relation for LMXB evolution (e.g. Savonije 1987; Rappaport et al. 1995; Tauris & Savonije 1999; van Kerkwijk et al. 2005). The match between this theoretical relation and the observational data for PSR J0337+1715 is excellent. This is demonstrated in Figure 74 where we plot all available data of helium WDs with masses measured to an accuracy $1\sigma < 0.1$ $M_{\rm WD}$. These helium WDs are companions to pulsars or found in binaries with A-type main sequence stars.

The eccentricities of both orbits are, although small, one to two orders of magnitude larger than expected theoretically for isolated binaries with similar components and orbital periods (Phinney & Kulkarni 1994). Although this may be a result of mutual triple interactions, a few binary pulsars with WD companions in the Galactic field have similar eccentricities (see Figure 4 in Tauris, Langer & Kramer 2012).

### 8.2.3 Evolution of the two LMXB phases

The two WDs orbiting PSR J0337+1715 are the remnants of two LMXB phases. Optical observations by Kaplan et al. (2014) show that the inner WD is quite hot (15 800 ± 100 K), whereas the outer WD is too cold to be detected. Therefore, we assume in the following that the inner WD formed last (see Section 8.4.2 for a discussion).

We can deduce that both LMXB phases evolved highly non-conservatively since the low MSP mass of 1.438 $M_\odot$ implies that it cannot have accreted much material (at most 0.1 – 0.2 $M_\odot$ in total; in our model we assume a NS birth mass of 1.28 $M_\odot$). Since the presently observed (post-LMXB) orbital period of the inner binary, $P_{\rm orb,12} = 1.63$ days is close to the so-called bifurcation period (between 1 and 2 days, Pylyser & Savonije 1989; Ma & Li 2009), below which magnetic braking is dominant in LMXBs (Rappaport, Verbunt & Joss 1983) and above which the widening of the inner binary is significant, we also conclude that the pre-LMXB (post-SN) orbital period of the inner binary must have been close to this value.

For the preceding outer LMXB phase there is further evidence for highly non-conservative evolution since the mass-transfer rate must have been super-Eddington for such a wide system (initially $P_{\rm orb,3} = 14.2$ days) where the donor star had a deep convective envelope at the onset of the Roche-lobe overflow (RLO; Tauris & Savonije 1999; Podsiadlowski, Rappaport & Pfahl 2002). During the rapid outer LMXB phase, we expect the pulsar only to be mildly recycled – possibly with a spin period between 25 ms and 1 sec. (a typical value for pulsars with a WD and $P_{orb} > 200$ days, Tauris, Langer & Kramer 2012). Effective recycling of the MSP was obtained in the subsequent long-lasting inner LMXB phase.

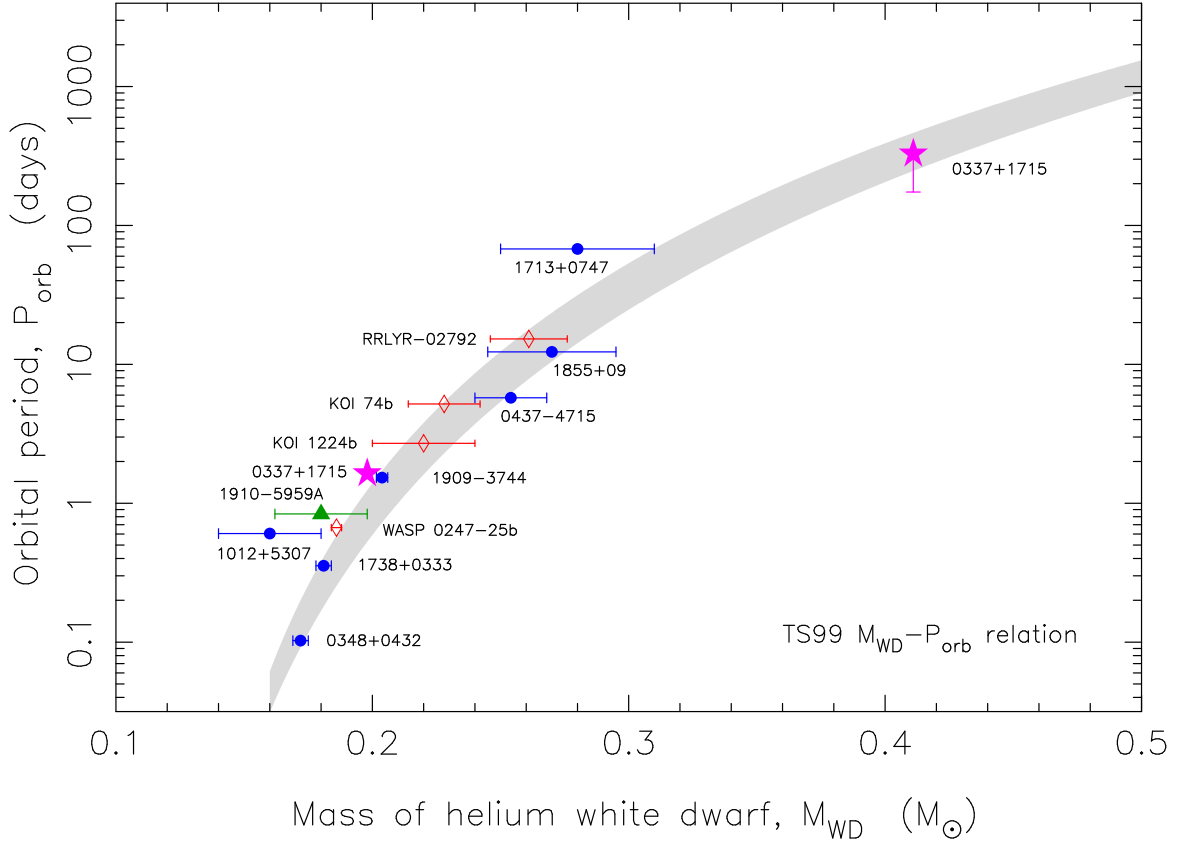

Figure 74: $M_{WD} - P_{orb}$ relation (TS99), as calculated by Tauris & Savonije (1999). The width of the relation is caused by using metallicities from $Z = 0.001 - 0.02$. Observational data is plotted for helium WD companions orbiting pulsars, for which the $1\sigma$ uncertainties are less than 10% of $M_{WD}$. Also included are four Galactic field proto-WDs orbiting an A-type main sequence star (WASP 0247-25b, KOI 1214b, KOI 74b and RRLYR−02792). The discovery of the triple MSP J0337+1715 adds two important high-precision data points to this graph and strengthens the validity of the relation. The error bar plotted for the outer WD is caused by an uncertainty in the widening of the outer orbit during the inner LMXB phase, see text. Theoretically, the relation becomes uncertain for $P_{orb} < 1$ day. (For references to data, with increasing $P_{orb}$, see: Antoniadis et al. 2013; 2012; van Kerkwijk et al. 2005; Maxted et al. 2013; Corongiu et al. 2012; Jacoby et al. 2005; Ransom et al. 2014; Breton et al. 2012; van Kerkwijk et al. 2010; Verbiest et al. 2008; Splaver 2004; Pietrzyński et al. 2012; Splaver et al. 2005; Ransom et al. 2014.)

The masses of the two WD progenitor stars are constrained, on the one hand, by requirements of the development of a degenerate helium core *and* dynamically stable mass transfer ($M_{2,3}^{ZAMS} \leq 1.6\ M_{\odot}$) and, on the other hand, by nuclear evolution and WD cooling within a Hubble time ($M_{2,3}^{ZAMS} > 1.0\ M_{\odot}$). The estimated component masses before/after the mass transfer then yield the amount of mass lost from the system.

The changes of the orbital separations as a result of LMXB mass transfer/loss can be found by solving the orbital angular momentum balance equation within the isotropic re-emission model (Soberman, Phinney & van den Heuvel 1997), but with some modification. For example, we cannot assume a pure fast (Jeans mode) wind mass loss from the inner binary with respect to the tertiary star during the inner LMXB phase (stage 8). Some of the inner binary material might be lost in a rather slow wind which only causes a moderate widening (if any) of $P_{orb,3}$ during this phase. Nevertheless, following the outer LMXB phase (stage 6), $P_{orb,3}$ could have been smaller than observed today (327 days). In the most extreme case (Jeans mode) it could be as short as 175 days, given that the semi-major axis of the outer orbit in this case changes according to $a_{3f} = a_{3i} (M_i/M_f)$, where $M$ is the total mass of the triple system and the indices $i$ and $f$ refer to initial and final values, respectively. Here we assume a more moderate value of 250 days.

The major uncertainties in our modelling are related to i) spin-orbit couplings mediated by tidal torques (e.g. magnetic braking) and ii) accretion onto the inner binary system during mass transfer from the tertiary star (stage 6), and the poorly known specific orbital angular momentum of the ejected mass. Presumably, an inner circumbinary disk (Dermine et al. 2013) will be formed which may subsequently influence the evolution of the inner orbit. However, investigating the dynamical effects of the SN explosion helps to constrain the properties of the pre-LMXB systems.

### 8.2.4 The dynamical effects of the SN explosion

A general discussion of dynamical effects of asymmetric SNe in hierarchical multiple star systems is found in Pijloo, Caputo & Portegies Zwart (2012), and references therein. Here we have simulated the dynamical effects of the SN explosion that created the NS in the PSR J0337+1715 system. In Figure 75 we have plotted the survival probability of our best model for the triple system as a function of the recoil velocity immediately imparted to the inner system (i.e. the "inner binary kick", $w_{12}$) as a consequence of the SN. All relevant pre-SN parameters are stated in the figure (see also Table 14, stage 4). Along the plotted curves are examples of average values of $w_{12}$ for kick velocities between 0 and 550 km s$^{-1}$ which were imparted to the newborn NS. For estimating the resulting values of $w_{12}$ we only considered systems which survived the SN in long-term stable orbits (Section 8.4), and for which the inner binary avoided merging[29].

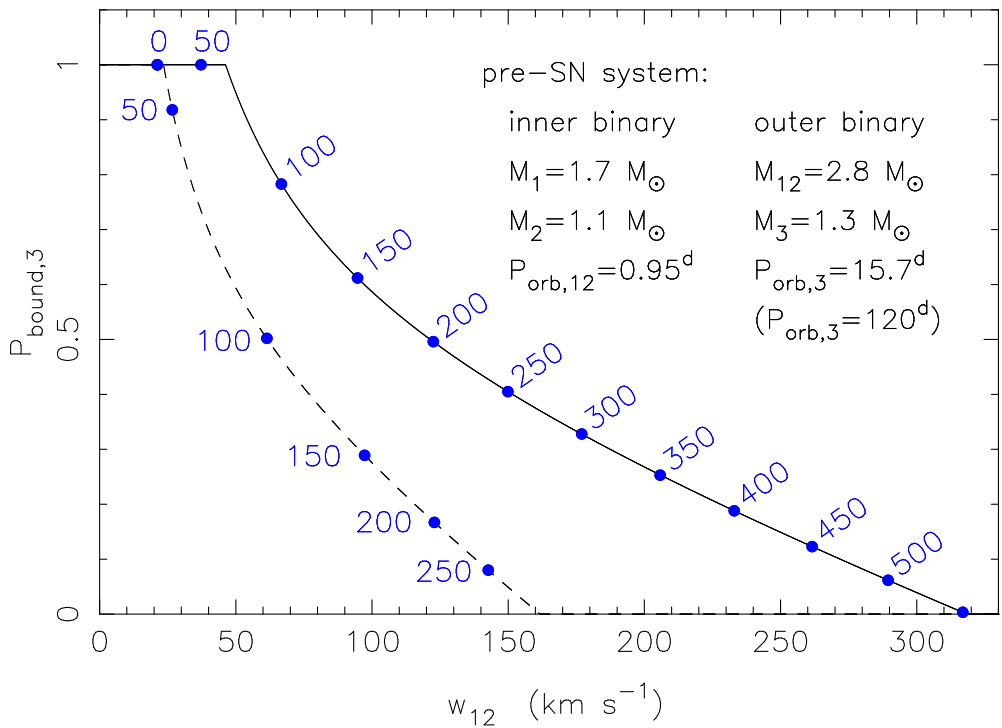

Figure 75: Probability for a triple system to survive a given recoil velocity, $w_{12}$ obtained by the inner binary due to a SN. The bullet points represent average values of $w_{12}$ for the stated kick magnitudes (in km s$^{-1}$) imparted on a newborn 1.28 $M_\odot$ NS in the inner binary. The kick directions were chosen from an isotropic distribution and simulations were done for random orbital phases between the mutual orbits. The solid (dashed) line is for a pre-SN $P_{orb,3} = 15.7$ days ($P_{orb,3} = 120$ days). When calculating $\langle w_{12} \rangle$ only triple systems surviving the SN in long-term dynamically stable orbits, and for which the inner binary avoided merging, were considered. The assumed pre-SN core mass is $M_1 = 1.7\ M_\odot$. All other relevant pre-SN triple parameters are given in the figure.

Given the constraints on the post-SN evolution to meet the requirements for PSR J0337+1715, we find that it may even have survived a NS kick up to 400 $\mathrm{km\,s^{-1}}$. The resulting peculiar velocities of the triple system range between 15 and 160 $\mathrm{km\,s^{-1}}$.

### 8.2.5 Pre-SN core mass and Case BB RLO

The triple system is much more likely to survive the explosion if the pre-collapsing core mass is low. A pre-SN core mass of only $\sim 1.7 M_\odot$ is indeed expected if the progenitor star (say, $M_1 = 10\ M_\odot$) lost its hydrogen-rich envelope on the red-giant branch (RGB), because the resulting naked helium core itself ($\sim 2.9\ M_\odot$) would expand and give rise to Case BB RLO (Habets 1986), leaving an even further stripped pre-SN core during stage 3. In case the NS progenitor did not lose its envelope until the asymptotic giant branch, the collapsing core mass could have been much larger. However, we find that a collapsing core mass of, for example, 3.2 $M_\odot$ decreases the survival probability considerably and leaves the triple system with $v \sim 150\ \mathrm{km\,s^{-1}}$.

### 8.2.6 The common envelope evolution – new lessons from a triple system

The outcome of the CE evolution is crucial for determining the pre-SN core mass and orbital periods (e.g. Tauris & Dewi 2001), and thus for the survival probability of the triple system. Unfortunately, CE evolution is the least understood of the important interactions in close binary systems (see Ivanova et al. 2013; for a recent review). For close triple systems, understanding the CE evolution is an even more complicated task. However, the existence of PSR J0337+1715 provides an important piece of information: namely that CE evolution (stage 2) not only leads to efficient orbital angular momentum loss of the inner binary orbit, also the tertiary star is subject to efficient in-spiral. The evidence for this conclusion is the following. On the one hand, the orbital period of the tertiary star could not be very large at the moment of the SN. There are two reasons for this: i) the post-SN $P_{\mathrm{orb},3}$ (after recircularization) must match the expected orbital period at the onset of the outer LMXB phase, and ii) to avoid a very small survival probability as a consequence of the SN. On the other hand, the system must have had $P_{\mathrm{orb},3} \gtrsim 4000$ days on the ZAMS. The evidence for this is that the ratio of the orbital periods ($P_{\mathrm{orb},3}/P_{\mathrm{orb},12}$) on the ZAMS must have been at least a factor of $\sim 5$, and even larger for non-circular orbits, in order for the triple system to remain dynamically stable on a long timescale (see Section 8.4). Furthermore, the onset of the CE could not have happened much earlier than near the tip of the RGB of the primary star, which corresponds to $P_{\mathrm{orb},12} \gtrsim 800$ days. The reason is that the binding energy of the hydrogen-rich envelope is simply too high to allow for ejection at earlier stages (Dewi & Tauris 2000). This constraint, in combination with the stability criteria of the triple system, sets the lower limit of $P_{\mathrm{orb},3} \sim 4000$ days on the ZAMS. Hence, we conclude that an efficient in-spiral of the tertiary star, and thus *both* of the outer stars, must have taken place.

Portegies Zwart et al. (2011) argued for a similar conclusion based on the tertiary F-dwarf orbiting the LMXB 4U 2129+47 (V1727 Cyg). As also pointed out by these authors, the SN explosion itself could also have decreased the orbital period of the tertiary star of the surviving triple system. In that case, the need for CE in-spiral is less extreme, but still highly demanded. From our simulations, we find that the pre-SN orbital period of the tertiary could have been up to about 120 days; still much shorter than the ZAMS $P_{\mathrm{orb},3}$ of about 4000 days.

---

[29]The plotted probabilities do not take into account the specific requirements on the value of the post-SN $P_{\mathrm{orb},3}$ necessary for forming PSR J0337+1715. If including this specific constraint, the probabilities shown would be much lower.

## 8.3 Discussion

## 8.4 Long-term Stability of a Triple System

Throughout our scenario, we have checked at each evolutionary stage that the mutual orbits of our solutions are expected to have a long-term dynamical stability. A number of stability criteria for triple systems have been proposed over the last four decades (see Mikkola 2008; for an overview). To be extra cautious and conservative, we only accepted our found solutions in case they fulfilled all criteria suggested by Harrington (1972); Bailyn (1987); Eggleton & Kiseleva (1995); Mardling & Aarseth (2001).

### 8.4.1 Comparison to PSR J1903+0327

The post-SN stages outlined here for the formation of PSR J0337+1715 differ somewhat from those proposed for PSR J1903+0327 (Champion et al. 2008; Liu & Li 2009; Freire et al. 2011b; Portegies Zwart et al. 2011; Pijloo, Caputo & Portegies Zwart 2012). The latter system most likely became dynamically unstable during a diverging LMXB evolution of the inner binary, as a result of the $(R_{peri}/a_{in})$–ratio decreasing below the critical limit (e.g. Mardling & Aarseth 2001) when the inner orbit expanded. This instability was possibly aided by cyclic perturbations of the inner binary by the unevolved tertiary star (Kozai 1962) while the critical orbital separation was approached. Hence, the J1903+0327 system may correspond to a disrupted case of an evolution which could otherwise have resulted in a triple MSP system.

### 8.4.2 Alternative models

Despite its cold temperature, it cannot be excluded entirely that the outer WD formed last. The reason for this is that some low-mass helium WDs take $1 - 2$ Gyr to reach the WD cooling track after detaching from their Roche lobe (A. Istrate et al., in preparation). Residual shell hydrogen burning cannot be ignored in these stars and keeps them hot on a long timescale (e.g. Alberts et al. 1996; Nelson, Dubeau & MacCannell 2004). Since the duration of the outer LMXB phase is much shorter than that of the inner LMXB phase ($15 - 20$ Myr versus $\sim 2$ Gyr, respectively, Tauris & Savonije 1999; Podsiadlowski, Rappaport & Pfahl 2002), and given that a $\sim 0.4\ M_\odot$ helium WD does not experience residual hydrogen burning and therefore cools faster, it seems conceivable to form the outer WD *after* the formation of the inner WD. Even a double LMXB phase is possible, depending on how close in mass the two WD progenitor stars were, or if the secondary star was forced into RLO during mass-transfer from the tertiary star.

de Vries, Portegies Zwart & Figueira (2014) recently presented a novel attempt to simulate the combined stellar evolution, gravitational dynamics, and hydrodynamical interactions of a triple system. Although it is too early to draw firm conclusion from such a study, it is interesting to notice that their finding of significant loss of orbital angular momentum in the inner binary during the RLO from the tertiary star would strengthen the possibility of a double LMXB phase. What are the alternative scenarios to the one presented here for producing PSR J0337+1715? A scenario with a massive star ($M_1$) orbited by a distant binary of low-mass stars ($M_2, M_3$) might be dynamically unstable during the CE stage. Instead, PSR J0337+1715 may possibly have formed in a quadruple system where the SN kick caused interactions with an outer binary and ejection of the fourth member.

The present triple system might also have evolved in a globular cluster and subsequently ejected into the Galactic field, for example, in a binary-binary encounter event. However, it is questionable if the triple system would survive such an ejection process. In addition, this scenario seems difficult to reconcile with the low eccentricities of PSR J0337+1715 and the fine match with the $M_{WD} - P_{orb}$ relation.

Finally, to obtain a small kick, one may advocate for the formation of the NS via accretion-induced collapse (AIC) of a WD (Nomoto et al. 1979). However, according to a recent study on AIC (Tauris et al. 2013b), the required donor star masses are considerably larger ($\gtrsim 2\ M_\odot$, depending on metallicity) than what is constrained here for $M_2$.

## 8.5 Future Prospects and Conclusions

We have presented a first self-consistent, semi-analytical solution to the formation of PSR J0337+1715 which constrains the peculiar velocity of the system to be less than $160\,\mathrm{km\,s^{-1}}$ and which requires double in-spiral during the common envelope evolution in a close triple system. We estimate that the uncertainties of our initial masses of all three stars are about $20\,\%$. We have briefly discussed a number of alternative models for which further calculations are needed. To probe the full parameter space with weighted probabilities for forming PSR J0337+1715 (depending on the pre-SN core mass, CE physics, orbital angular momentum losses, dynamical stability, geometry preferences for the mutual orbits, a possible quadruple origin, etc.) would require a full population synthesis investigation which is beyond the scope of this Letter.

PSR J0337+1715 is a unique example of a triple system which has survived *three* phases of RLO. The outcome of the two LMXB mass-transfer phases matches nicely with the theoretical expectations from the $M_{\mathrm{WD}}-P_{\mathrm{orb}}$ relation of WDs. The possibility of two RLO events in a triple system was first discussed by Eggleton & Kiseleva (1996) only two decades ago. These authors denoted such systems as 'doubly interesting' triples. PSR J0337+1715 has not only managed to experience *three* phases of RLO, it has also survived a SN explosion to evolve towards its present terminal stage containing three compact objects – a truly remarkable journey for a triple system. Our analysis of the formation of this system only allows for a plausible solution when current knowledge of stellar evolution and interactions is stretched to the limit. The existence of this system in the Galactic field has opened a new door to stellar astrophysics with resulting challenges to be met in the years to come.

# ACCRETION-INDUCED COLLAPSE OF MASSIVE WHITE DWARFS AND FORMATION OF MILLISECOND PULSARS

- **Chapter 9**
  Tauris, Sanyal, Yoon & Langer (2013), A&A 558, 39
  *Evolution towards and beyond accretion-induced collapse of massive white dwarfs and formation of millisecond pulsars*

- **Chapter 10**
  Freire & Tauris (2014), MNRAS Letters 438, 86
  *Direct formation of millisecond pulsars from rotationally delayed accretion-induced collapse of massive white dwarfs*

# 9. Evolution towards and beyond Accretion-Induced Collapse of Massive White Dwarfs and Formation of Millisecond Pulsars

**Tauris, Sanyal, Yoon & Langer (2013)**

**A&A 558, 39**

## Abstract

Millisecond pulsars (MSPs) are generally believed to be old neutron stars (NSs), formed via type Ib/c core-collapse supernovae (SNe), which have been spun up to high rotation rates via accretion from a companion star in a low-mass X-ray binary (LMXB). In an alternative formation channel, NSs are produced via the accretion-induced collapse (AIC) of a massive white dwarf (WD) in a close binary. Here we investigate binary evolution leading to AIC and examine if NSs formed in this way can subsequently be recycled to form MSPs and, if so, how they can observationally be distinguished from pulsars formed via the standard core-collapse SN channel in terms of their masses, spins, orbital periods and space velocities. Numerical calculations with a detailed stellar evolution code were used for the first time to study the combined pre- and post-AIC evolution of close binaries. We investigated the mass transfer onto a massive WD (treated as a point mass) in 240 systems with three different types of non-degenerate donor stars: main-sequence stars, red giants, and helium stars. When the WD is able to accrete sufficient mass (depending on the mass-transfer rate and the duration of the accretion phase) we assumed it collapses to form a NS and we studied the dynamical effects of this implosion on the binary orbit. Subsequently, we followed the mass-transfer epoch which resumes once the donor star refills its Roche lobe and calculated the continued LMXB evolution until the end. We show that recycled pulsars may form via AIC from all three types of progenitor systems investigated and find that the final properties of the resulting MSPs are, in general, remarkably similar to those of MSPs formed via the standard core-collapse SN channel. However, as a consequence of the finetuned mass-transfer rate necessary to make the WD grow in mass, the resultant MSPs created via the AIC channel preferentially form in certain orbital period intervals. In addition, their predicted small space velocities can also be used to identify them observationally. The production time of NSs formed via AIC can exceed 10 Gyr which can therefore explain the existence of relatively young NSs in globular clusters. Our calculations are also applicable to progenitor binaries of SNe Ia under certain conditions.

## 9.1 Introduction

The final outcome of close binary stellar evolution is a pair of compact objects if the system avoids disruption following a supernova (SN) explosion or a merger event in a common envelope (CE). The nature of the compact objects formed can be either black holes, neutron stars (NSs), or white dwarfs (WDs), depending primarily on the initial stellar masses and orbital period of the zero-age main-sequence (ZAMS) binary. Systems in which the primary star is not massive enough to end its life as a NS may leave behind an oxygen-neon-magnesium white dwarf (ONeMg WD) of mass $M_{WD} \simeq 1.1 - 1.3\ M_\odot$, which may stem from primaries with initial masses in the interval of $M_{ZAMS} \simeq 6 - 16\ M_\odot$, depending on the mass of the secondary star and, in particular, on the orbital period which affects both the important carbon/oxygen–ratio at the depletion of core helium burning and the occurrence of a second dredge-up phase at the beginning of the asymptotic giant branch (see e.g. Wellstein & Langer 1999; Podsiadlowski et al. 2004). In systems where an ONeMg WD forms, the WD may later accrete sufficient material, when the secondary star subsequently evolves and fills its Roche lobe, such that it reaches the Chandrasekhar-mass limit and implodes via an accretion-induced collapse (AIC) to form a NS (e.g. Nomoto et al. 1979; Taam & van den Heuvel 1986; Michel 1987; Canal, Isern & Labay 1990; Nomoto & Kondo 1991). The donor star in these systems can either be a main-sequence star, a low-mass giant, or a helium star. The first aim of this paper is to investigate in which binaries AIC can occur.

Millisecond pulsars (MSPs) are traditionally believed to be old NSs which have been spun up to high rotation rates via accretion of mass and angular momentum from a companion star in a low-mass X-ray binary, LMXB (e.g. Bhattacharya & van den Heuvel 1991; Tauris & van den Heuvel 2006; and references therein). It is important to investigate whether or not this standard recycling scenario is the sole formation channel of MSPs. Three key questions arise in the context of AIC:

1. Could the implosion of the ONeMg WD lead directly to the formation of an MSP?

2. When an additional accretion phase is necessary for NSs formed via AIC in order to explain the observed rapid spins and low B-fields of MSPs, is it possible at all to form MSPs via post-AIC accretion from the same donor star?

3. If so, can MSPs produced this way be distinguished observationally from MSPs that formed via the standard core-collapse SN channel?

In the context of direct and indirect formation of MSPs via AIC (i.e. prompt or after additional accretion, respectively), we point out that the origin of pulsar B-fields is not well understood. In direct MSP formation, B-fields could be created from conservation of magnetic flux of the collapsing core, as originally hypothesized for NSs by Woltjer (1964), or as suggested by theoretical work on thermomagnetic effects during or shortly after the NS is formed (Reisenegger 2003; Spruit 2008). In principle, a Chandrasekhar mass WD with an equatorial radius of about 3000 km and a B-field of $10^3$ G could, assuming flux conservation, undergo AIC and directly produce an MSP with $B \simeq 10^8$ G and a spin period of a few ms, equivalent to typical values of observed recycled radio MSPs (Tauris & van den Heuvel 2006) and accreting X-ray MSPs (Patruno & Watts 2012) near the spin-up line in the $P\dot{P}$–diagram. Producing the spin of the MSP via AIC is not a problem if just a small fraction of the spin angular momentum of the WD is conserved during the collapse (Dessart et al. 2006).

Until recently, it was thought that the distribution of WD B-fields is strongly bimodal, with a large majority of WDs being non-magnetic and a smaller fraction ($\sim$15%, primarily in binaries) having larger fields of typically $\sim 10^7$ G (Wickramasinghe & Ferrario 2000; Liebert, Bergeron & Holberg 2003). However, by using more sensitive instruments it has been demonstrated that some $15 - 20\%$ of WDs could have weak B-fields of the order of $10^3$ G (e.g. Jordan et al. 2007). Hence, AIC may lead to the formation of NSs with a potential large range of possible B-field values, if flux conservation is at work. One the other hand, regardless of its formation mechanism any newborn NS is extremely hot and liquid, and therefore differentially rotating, which may amplify any seed magnetic field such that the B-fields of NSs formed via AIC could be similar to those formed via an iron-core collapse. In this respect, it is important that none of the more than 40 known young NSs associated with SN remnants are observed with the characteristic properties of MSPs: low B-fields and fast spin. The B-fields of MSPs ($10^{7-9}$ $G$) are typically lower than that of young, normal pulsars ($10^{12-14}$ $G$) by five orders of magnitude. It is therefore questionable if an MSP would form directly from any type of collapse: iron-core collapse, electron capture SN, or AIC[30]. For the rest of this work, we therefore focus our attention exclusively on the possibility of indirect formation of MSPs via AIC, i.e. following the scenario where an AIC leads to the formation of a normal NS which then subsequently accretes matter, resulting in a weaker B-field and a faster spin.

The AIC route of forming MSPs has three main advantages: 1) it can explain the low space velocities of many recycled pulsars and the large fraction of NSs retained in globular clusters (Bailyn & Grindlay 1990) due to both the small amount of mass lost and the small momentum kick expected to be associated with the implosion (for details of simulations of the implosion, see e.g. Kitaura, Janka & Hillebrandt 2006; Dessart et al. 2006); 2) it can explain the existence of apparently *young* NSs in globular clusters (Lyne, Manchester & D'Amico 1996; Boyles et al. 2011); and finally 3) it may explain the presence of a number of peculiar high B-field and slowly spinning Galactic disk NSs in close binaries with (semi)degenerate companions (e.g. Yungelson, Nelemans & van den Heuvel 2002). In addition, direct formation of MSPs via AIC could possibly help explain the postulated birthrate problem (Kulkarni & Narayan 1988) between the small number of LMXB progenitor systems and the large observed number of MSPs.

---

[30]Although the argument for the AIC is less certain since it may not give rise to the formation of an observable remnant.

A recent, detailed population synthesis study by Hurley et al. (2010) concluded that one cannot ignore the AIC route to MSP formation and that some binary MSPs in wide orbits are best explained by an AIC scenario (see also Ivanova et al. 2008; for a specialized study on NS formation in globular clusters). There are, however, many uncertainties involved in even the best population synthesis studies and in particular in the applied physical conditions for making the ONeMg WD mass grow sufficiently.

The weaknesses of the AIC formation channel are that it lacks direct observational evidence of the AIC event itself and, as already mentioned, the difficulty in predicting the spin rate and the surface B-field associated with a newborn NS formed via AIC (e.g. Kitaura, Janka & Hillebrandt 2006; Dessart et al. 2006; 2007). It is not clear whether or not the AIC would be an observable transient event. According to Dessart et al. (2006), during the AIC only a few $0.001\ M_\odot$ of material is ejected (of which $\sim 25\%$ is $^{56}$Ni, decaying into $^{56}$Fe via $^{56}$CO) which quickly becomes optically thin. Hence, AIC events are most likely underluminous and very short-lived. The studies by Kitaura, Janka & Hillebrandt (2006); Metzger, Piro & Quataert (2009); Darbha et al. (2010) also yield somewhat small amounts ($< 0.015\ M_\odot$) of $^{56}$Ni ejected in the AIC process, possibly synthesized in a disk, which may result in a radioactively powered SN-like transient that peaks after $\leq 1$ day with a bolometric luminosity $\simeq 10^{41}$ erg s$^{-1}$. It is also possible that a transient radio source may appear, lasting for a few months, following the AIC event (Piro & Kulkarni 2013). In any case, these amounts are small enough to justify our assumption that the whole WD mass gets converted into the mass of the newborn NS (see Section 9.3).

Another issue is that any NS formed via AIC may shortly afterwards begin to accrete additional material from its companion star, once this donor star re-fills its Roche lobe when recovering from the dynamical effects of the implosion (partly caused by the released gravitational binding energy in the transition from a WD to a more compact NS). Therefore, regardless of its initial properties, any NS formed via AIC could in principle subsequently be spun up to become an MSP, as suggested by Helfand, Ruderman & Shaham (1983). The problem is, as pointed out by Hurley et al. (2010), that this post-AIC accretion phase should then resemble the conditions under which normal, old NSs are spun up to become MSPs via the conventional channel and, consequently, one cannot easily distinguish the outcome of this formation path from the standard scenario.

In this work, we therefore concentrate on answering the second and the third questions raised above, i.e. if MSPs can be produced via AIC events which are immediately followed by subsequent mass transfer and, if so, how they can be distinguished observationally from those MSPs formed via the standard SN channel. We aim at investigating which progenitor binaries lead to AIC in the first place and we present detailed modelling of both pre- and post-AIC evolution to predict the properties of MSPs formed via the indirect AIC channel. The structure of our paper is as follows: In Section 9.2 we review the suggested observational evidence for NS production via AIC. The computer code and our assumptions governing the pre- and post-AIC mass-transfer processes are given in Section 9.3. In Sections 9.4, 9.5, and 9.6 we present those of our calculated systems which successfully lead to AIC with main-sequence star, giant star, and helium star donors, respectively, and review the properties of the binary pulsars formed. We discuss our results in a broader context in Section 9.7 and summarize our conclusions in Section 9.8.

## 9.2 Observational evidence for AIC

The question of the origin of NSs is closely related to many of their observable parameters: spin, B-field, age, space velocity, and the nature of their companion star. As already pointed out, the observational evidence suggested in the literature for NSs formed via AIC can be categorized into three groups. We now review this evidence in more detail.

### 9.2.1 The role of NS kicks

It has been well established from observations of radio pulsar velocities that most NSs receive a momentum kick at birth (Lyne & Lorimer 1994; Hobbs et al. 2005). These kicks are possibly associated with SN explosion asymmetries and may arise from non-radial hydrodynamic instabilities (neutrino-driven convection and the standing accretion-shock instability) in the collapsing stellar core. These instabilities lead to large-scale anisotropies of the innermost SN ejecta, which interact gravitationally with the proto-NS and accelerate the nascent NS on a timescale of several seconds (e.g. Janka 2012; Wongwathanarat, Janka & Müller 2013). For the *entire* population of NSs, the range of required kick velocity magnitudes extends basically from a few $10 \, \mathrm{km \, s^{-1}}$ to more than $1000 \, \mathrm{km \, s^{-1}}$, in order to explain both the existence of NSs residing inside globular clusters (which have small escape velocities, $v_{\mathrm{esc}} < 50 \, \mathrm{km \, s^{-1}}$) as well as bow shocks and SN remnants associated with high-velocity pulsars. However, when considering only *young* NSs in the Galactic disk the study by Hobbs et al. (2005) is interesting; it revealed that the velocities of young ($< 3 \, \mathrm{Myr}$) radio pulsars are well described by a single Maxwellian distribution with a three-dimensional mean speed of $\sim 400 \, \mathrm{km \, s^{-1}}$. Furthermore, there are no detections of any low-velocity ($v_{\perp} < 60 \, \mathrm{km \, s^{-1}}$) single radio pulsars with a characteristic age, $\tau < 1 \, \mathrm{Myr}$. These facts indicate that NSs which formed recently in young stellar environments (the Galactic disk) received large kicks and that iron core-collapse SNe of type II and type Ib/c therefore, in general, result in these large kicks.

On the other hand, about half of the approximately 300 known MSPs are detected in globular clusters (Ransom et al. 2005; Freire et al. 2008). Obviously, pulsars retained in globular clusters (GCs) cannot have formed with large kicks since these clusters have small escape velocities, except in a few rare cases where an isolated low-velocity pulsar could form in a disrupted binary involving a large kick with a finetuned direction (cf. Fig. 5 in Tauris & Takens 1998). It is therefore tempting to believe that many of these MSPs in GCs were not formed by iron core-collapse SNe.

#### 9.2.1.1 Electron capture SNe

It seems clear that the lowest mass SN progenitors may not evolve all the way to form iron cores (see Langer 2012; for a recent review on pre-SN evolution of massive single and binary stars). The final fate of these stars with ONeMg cores is an electron-capture SN (EC SN), i.e. a collapse triggered by loss of pressure support owing to the sudden capture of electrons by neon and/or magnesium nuclei (e.g. Nomoto 1984; Wheeler, Cowan & Hillebrandt 1998). Work by Poelarends et al. (2008) shows that the initial mass range for EC SNe is quite narrow, only about $0.25 \, M_{\odot}$ wide, which would imply that some 4% of all single-star SNe would be of this type. However, it has been suggested by Podsiadlowski et al. (2004) that EC SNe could occur in close binaries for stars with masses between $8 - 11 \, M_{\odot}$ since these stars lose their envelopes via mass transfer before entering the AGB phase and thus avoid the dredge-up and the consequent erosion of the CO core by this process. Therefore, these stars undergo EC SNe rather than becoming ONeMg WDs, the likely outcome of most single stars of the same mass. Furthermore, these authors argue that EC SNe lead to prompt explosions (rather than slow, delayed neutrino-driven explosions) that naturally produce NSs with low-velocity kicks (see also van den Heuvel 2004; who proposed similar ideas). The idea of different NS kick magnitudes comes from the discovery of two classes of Be/X-ray binaries with significantly different orbital eccentricities (Pfahl et al. 2002). Furthermore, the low eccentricities and the low masses ($\sim 1.25 \, M_{\odot}$) of second-born NSs in double NS systems supports this picture (Schwab,

Podsiadlowski & Rappaport 2010; and references therein).

## 9.2.2 The role of young NSs in GCs

In Table 15 we list a number of apparently young NSs (characterized by slow spin and relatively high B-fields) that are found in GCs. The lifetime as an observable radio source is of the order of 100 Myr for a young (i.e. non-recycled) pulsar. Therefore, if these NSs had formed via iron core-collapse SNe their existence in GCs would not only be unlikely for kinematic reasons (as explained above), it would simply be impossible given that the stellar progenitor lifetimes of SNe II and SNe Ib/c are less than a few 10 Myr, much shorter than the age of the many Gyr old stellar populations in GCs. Similarly, the nuclear evolution timescales of stars undergoing EC SNe is of the order of $20-50$ Myr, which is still short compared to the age of GCs, and for this reason also EC SNe cannot explain the existence of young NSs in GCs today. It is therefore clear that these NSs in GCs, if they are truly young[31], are formed via a different channel.

### 9.2.2.1 A strong link to AIC – first piece of evidence

An AIC event is not very different from an EC SN and it is therefore expected that also NSs formed via AIC will receive a small kick (if any significant kick at all). For this reason formation via AIC could explain the many NSs in GCs and, more importantly, also the young ages of some of these NSs. As we shall demonstrate in this work, pre-AIC binaries may, in some cases, reach ages exceeding 10 Gyr before a low-mass giant companion star initiates Roche-lobe overflow (RLO) leading to the AIC event. Therefore, we would expect ongoing AIC events, and thus formation of newborn NSs, in GCs even today.

### 9.2.2.2 NS formation via the merger of two massive WDs?

For the sake of completeness, we also mention that the merger of two massive WDs may also produce a pulsar (Saio & Nomoto 1985; 2004). This could be an alternative way of producing young pulsars in an old stellar population like a GC. Even binary pulsars may be produced this way in a GC, since in a dense environment a single produced NS can capture a companion star later on (Ransom et al. 2005). It is more difficult for this scenario to produce MSPs in binaries in the Galactic disk. This scenario would not only require an initial triple system origin, which is somewhat rare in the Galactic disk (although recent work by Rappaport et al. (2013) suggests that at least 20% of all close binaries have tertiary companions), it would also require the remaining binary orbit to survive the dynamical effects of the merger event. However, it should be noted that the merging double CO WD event is also a key scenario (the so-called double-degenerate scenario) suggested as a progenitor of SNe Ia (cf. Webbink 1984; Iben & Tutukov 1984; Yoon, Podsiadlowski & Rosswog 2007; van Kerkwijk, Chang & Justham 2010; Pakmor et al. 2012).

---

[31]For an alternative view, see Verbunt & Freire (2014) who argue that these NSs that appear to be young, are not necessarily young.

Table 15: Neutron stars that are candidates for being formed via AIC in a globular cluster (a–d) or in the Galactic disk (e–h), respectively. See text for explanations and discussion.

| Object | $P$ ms | $B^*$ G | $P_{\rm orb}$ days | $M_{\rm comp}^{**}$ $M_\odot$ | Ref. |
|---|---|---|---|---|---|
| PSR B1718$-$19 | 1004 | $4.0 \times 10^{11}$ | 0.258 | $\sim 0.10$ | a |
| PSR J1745$-$20A | 289 | $1.1 \times 10^{11}$ | $-$ | $-$ | b |
| PSR J1820$-$30B | 379 | $3.4 \times 10^{10}$ | $-$ | $-$ | c |
| PSR J1823$-$3021C | 406 | $9.5 \times 10^{10}$ | $-$ | $-$ | d |
| GRO J1744$-$28 | 467 | $1.0 \times 10^{13}$ | 11.8 | $\sim 0.08$ | e |
| PSR J1744$-$3922 | 172 | $5.0 \times 10^{9}$ | 0.191 | $\sim 0.10$ | f |
| PSR B1831$-$00 | 521 | $2.0 \times 10^{10}$ | 1.81 | $\sim 0.08$ | g |
| 4U 1626$-$67 | 7680 | $3.0 \times 10^{12}$ | 0.028 | $\sim 0.02$ | h |

\* B-field values calculated from Eq.(5) in Tauris, Langer & Kramer (2012) which includes a spin-down torque due to a plasma-filled magnetosphere.
\*\* Median masses calculated for $i = 60°$ and $M_{\rm NS} = 1.35\,M_\odot$.
a) Lyne et al. (1993); b) Lyne, Manchester & D'Amico (1996); c) Biggs et al. (1994); d) Boyles et al. (2011). e) van Paradijs et al. (1997); f) Breton et al. (2007); g) Sutantyo & Li (2000); h) Yungelson, Nelemans & van den Heuvel (2002);

### 9.2.3  AIC candidates in the Galactic disk

The evidence for AIC is found not only in GCs. In Table 15 we list a number of Galactic disk binary NS systems which are postulated candidates for having formed via AIC. A common feature of these NSs is a slow spin and a relatively high B-field and an ultra-light ($\leq 0.10\ M_\odot$) companion star in a close orbit. The idea that the origin of some high B-field, slow spinning NSs (e.g. 4U 1626−67, Her X−1, and PSR B0820+08) is associated with AIC was originally suggested by Taam & van den Heuvel (1986). Although it was believed at that time that B-fields decay spontaneously on a timescale of only 50 Myr (and therefore these NSs could not have much larger ages), many of these sources remain good candidates for AIC today even though it has been demonstrated that pulsar B-fields can remain high on much longer timescales (Kulkarni 1986; Bhattacharya et al. 1992). One reason why these NS systems remain good AIC candidates is the very small masses of their companion stars which indicate that a significant amount of material ($0.5 − 1.0\ M_\odot$) was transfered towards the compact object[32]. The paradox is therefore that these NSs still have high B-fields and slow spins even though a significant mass transfer has occurred (see below).

#### 9.2.3.1  The role of accretion-induced B-field decay in NSs

There is solid observational evidence that the surface B-field strengths of NSs decrease with accretion (Taam & van den Heuvel 1986; Shibazaki et al. 1989; van den Heuvel & Bitzaraki 1994; 1995). The exact mechanism for this process is still unknown. It may be related to decay of crustal fields by ohmic dissipation and diffusion from heating via nuclear processing of accreted material (Romani 1990; Geppert & Urpin 1994; Konar & Bhattacharya 1997), burial (screening) of the field (Zhang 1998; Cumming, Zweibel & Bildsten 2001; Payne & Melatos 2007), or decay of core fields due to flux tube expulsion from the superfluid interior induced by rotational slow-down in the initial phases of mass accretion (Srinivasan et al. 1990); see also review by Bhattacharya (2002). Even a small amount of material accreted may lead to significant B-field decay, contradicting the observational evidence that these binary NSs have accreted large amounts of material.

#### 9.2.3.2  A strong link to AIC – second piece of evidence

A possible solution to the above-mentioned paradox of observing close binaries with ultra-light companions orbiting high B-field NSs, would be if these NSs were formed recently via AIC during the very final stages of the mass-transfer process in a binary.

The associated required finetuning of the AIC event to occur near the termination of the mass-transfer phase is important for preventing accretion of significant amounts of matter after the formation of the NS (resulting in low B-fields and fast spin). This finetuning problem will be investigated further in this paper. We note that the slow spins are expected from efficient loss of rotational energy due to the emission of magnetodipole waves from these high B-field NSs.

---

[32] A small companion-star mass suggests that the previous (or ongoing) mass-transfer episode was (is) dynamically stable (Tauris & Savonije 1999; Podsiadlowski, Rappaport & Pfahl 2002). However, even observed radio pulsars which are thought to have evolved via a CE phase have $B < 5 \times 10^9$ G (Tauris, Langer & Kramer 2012), in general much smaller that the B-fields of the NSs listed in Table 15.

In Section 9.7.5.1 we return to additional discussions of observational evidence for AIC in view of our theoretical calculations, and also comment on a handful of recycled radio pulsars in the Galactic disk with puzzling characteristics.

## 9.3 Numerical methods and physical assumptions of AIC

The numerical models presented in this work are divided into two parts: 1) evolution prior to AIC and 2) post-AIC LMXB evolution. Both parts are computed with a binary stellar evolution code originally developed by Braun (1997) on the basis of a single-star code (Langer 1998; and references therein). It is a one-dimensional implicit Lagrangian code which solves the hydrodynamic form of the stellar structure and evolution equations (Kippenhahn & Weigert 1990). The evolution of the donor star, the mass-transfer rate, and the orbital separation are computed simultaneously through an implicit coupling scheme (see also Wellstein & Langer 1999) using the Roche-approximation in the formulation of Eggleton (1983). To compute the mass-transfer rate, we use the prescription of Ritter (1988). In Section 9.5.3 we discuss the limitations of this description in wide-orbit LMXBs with giant donor stars. We employ the radiative opacities of Iglesias & Rogers (1996), which we interpolated in tables as function of density, temperature, and chemical element mass fractions, including carbon and oxygen. For the electron conduction opacity, we follow Hubbard & Lampe (1969) in the non-relativistic case, and Canuto (1970) in the relativistic case. The stellar models are computed using extended nuclear networks including the PP I, II, and III chains and the four CNO-cycles. In our default models we assumed a mixing-length parameter of $\alpha = l/H_{\rm p} = 1.5$ (Langer 1991) and a core convective overshooting parameter of $\delta_{\rm ov} = 0.10$. We tested several models using $\alpha = l/H_{\rm p} = 2.0$ which resulted in only slightly larger final WD masses ($\sim 1\%$) orbiting recycled pulsars in somewhat larger orbits (up to $\sim 3\%$ increase in $P_{\rm orb}$).

If the accreting ONeMg WD reached the limiting Chandrasekhar mass for a rigidly rotating WD (i.e. $M_{\rm Chan} = 1.48\,M_\odot$, e.g. Yoon & Langer 2005) we assumed it collapsed to form a NS. Differential rotation can persist if the timescale of angular momentum transport is smaller than the accretion timescale in accreting WDs, and leads to a critical mass that is significantly higher than the canonical Chandrasekhar mass (e.g. Yoon & Langer 2004). However, magnetic torques resulting from the Spruit-Tayler dynamo may enforce nearly rigid rotation in accreting WDs with the considered accretion rates in this study. Further research is needed to verify this. The WD collapse was modelled both with and without a momentum kick imparted to the newborn NS. Although AIC is generally believed to result in no kick, or possibly a small kick, we applied three different kick values ($w = 0, 50, 450\,{\rm km\,s^{-1}}$) to the newborn NS. Given that the physics behind the kick mechanism is still unclear, we have included a few extra models with a large kick of $450\,{\rm km\,s^{-1}}$ to probe the extreme boundary conditions of our calculations. We solved for the combined effects of sudden mass loss and imparted momentum kick on the orbital dynamics following Hills (1983). In all cases we assumed the mass equivalent of $0.20\,M_\odot$ was lost as released gravitational binding energy yielding a post-AIC NS gravitational mass of $1.28\,M_\odot$. The post-AIC LMXB evolution was followed using the same computer code (see also Tauris, Langer & Kramer 2011; 2012; for additional details). For a more general discussion of MSP formation from LMXBs we refer to e.g. Ergma, Sarna & Antipova (1998); Tauris & Savonije (1999); Podsiadlowski, Rappaport & Pfahl (2002); Deloye (2008).

### 9.3.1 Accretion onto a white dwarf

The mass-transfer process and the physics of accretion onto a WD has been described in a large number of papers, e.g. Whelan & Iben (1973); Nomoto & Sugimoto (1977); Nomoto (1982); Prialnik & Kovetz (1995); Iben & Tutukov (1996); Li & van den Heuvel (1997); Hachisu, Kato & Nomoto (1999); Langer et al. (2000); Livio (2000); Han & Podsiadlowski (2004); Yoon & Langer (2003; 2004); Nomoto et al. (2007), and more recently in Hachisu et al. (2012); Wheeler (2012); Idan, Shaviv & Shaviv (2012); Starrfield et al. (2012); Newsham, Starrfield & Timmes (2013); Denissenkov et al. (2013); Ma et al. (2013). Most of these papers aim at investigating progenitors of type Ia supernovae (SN Ia) which are important for cosmology studies of the accelerating Universe (e.g. Riess et al. 1998; Perlmutter et al. 1999). In these progenitor systems an accreting CO WD reaches the Chandrasekhar limit and explodes. Observationally, these systems manifest themselves as cataclysmic variables (Hellier 2001), symbiotic systems (Kenyon 1986), and supersoft X-ray sources (van den Heuvel et al. 1992; Kahabka & van den Heuvel 1997), depending on the companion star mass, its evolutionary status, and the mass-transfer rate. It should be noted that the classification scheme overlaps somewhat on certain aspects. The cataclysmic variables (CVs) can be further subdivided into several classes: (classical) novae, recurrent novae, dwarf novae, polars, AM CNn stars etc., again depending on the nature of the companion star, the orbital separation, as well as the magnetic field strength of the accreting WD and accretion disk morphology.

The classical novae have only been reported to erupt once whereas the recurrent novae usually erupt every one or two decades. The symbiotic variable star binaries have low-mass red giant donors which transfer mass at a fairly low rate via (beginning atmospheric) RLO. They undergo nova-like outbursts which last for a few decades before decaying back to their original luminosity. The luminous persistent supersoft X-ray sources, on the other hand, display luminosities between $10^{36} - 10^{39}$ erg s$^{-1}$ revealing high mass-transfer rates of the order of $10^{-7}$ $M_\odot$ yr$^{-1}$. These systems often have a more massive main-sequence donor star undergoing thermal timescale mass transfer.

It is generally believed that the accreting WDs in supersoft X-ray sources can grow in mass by accretion since the thermonuclear fusion of accreted hydrogen can be fairly stable with these high accretion rates, in contrast to the case of nova systems where nova explosions (caused by a violent ignition in a thin shell of degenerate hydrogen) may erode the WDs (Wiescher et al. 1986; Patterson et al. 2013). The supersoft X-ray sources are therefore believed to represent progenitors of SNe Ia, because many of these systems may produce a Chandrasekhar mass CO WD. However, existing multicycle computations of hydrogen accretion onto massive WDs at a high rate are still somewhat controversial. In a recent study Idan, Shaviv & Shaviv (2012) showed that the accumulated helium is completely lost in strong helium flashes, thereby making SN Ia and AIC impossible, whereas another study by Newsham, Starrfield & Timmes (2013) concluded that WDs continue to grow toward the Chandrasekhar limit.

A number of numerical simulations show that when a CO WD approaches the Chandrasekhar limit with accretion rates of the order of $\dot{M}_{WD} \simeq 10^{-7} - 10^{-6}$ $M_\odot$ yr$^{-1}$, a thermonuclear-runaway caused by carbon burning can occur at central densities of about $\rho_c = 2 - 5 \times 10^9$ g cm$^{-3}$ (e.g. Nomoto, Thielemann & Yokoi 1984; Yoon & Langer 2003; Lesaffre et al. 2006), which may result in a SN Ia explosion. In case an ONeMg WD mass grows to the Chandrasekhar limit, electron-captures onto $^{24}$Mg and $^{20}$Ne make the central density increase to about $10^{10}$ g cm$^{-3}$ before oxygen ignites at the center (Miyaji et al. 1980; Miyaji & Nomoto 1987; Nomoto 1987). The consequent oxygen deflagration with this high density cannot lead to a thermonuclear explosion because of very rapid electron-captures onto heavy elements produced by the oxygen burning (Miyaji et al. 1980; Timmes & Woosley 1992). Therefore, a NS is the most likely outcome in the case of the collapse of an ONeMg WD.

Although it is generally believed that accreting CO WDs reaching the Chandrasekhar mass limit lead to a SNe Ia, and that accreting ONeMg WDs reaching this limit undergo AIC and produce NSs, the outcome could in some cases be the opposite – see Section 9.7.2 for further discussion.

### 9.3.1.1 Dependence on mass-transfer rates

The response of the WD to mass transfer from the donor star depends on the mass-transfer rate, $|\dot{M}_2|$. In this work we assume that the WD receives (but does not necessarily accrete) mass at the same rate as it is lost from the donor star, $|\dot{M}_2|$. White dwarfs which accrete faster than the rate at which their cores can grow ($\dot{M}_{up}$) will puff up their envelope to giant star dimensions (see below). If the WD envelope does expand to such huge radii it will engulf the donor star and the system is likely to evolve through a common envelope (CE) and spiral-in phase (Paczyński 1976; Iben & Livio 1993). The relevant critical mass-transfer rates for our work are (see Fig. 76):

$\dot{M}_{CE}$         upper limit before the formation of a giant and a CE
$\dot{M}_{up}$          upper limit for steady hydrogen burning
$\dot{M}_{Edd,WD}$    Eddington limit for spherical accretion
$\dot{M}_{steady}$       lower limit for steady hydrogen burning
$\dot{M}_{accu}$        lower limit for WD mass accumulation

where we have adopted the following *ad hoc* condition:

$$\dot{M}_{CE} = 3\,\dot{M}_{Edd,WD}. \tag{76}$$

The Eddington accretion limit is found by equating the outward radiation pressure to the gravitational force per unit area acting on the nucleons of the accreted plasma. The radiation pressure (from photons that scatter on plasma electrons) is generated from both nuclear burning at the WD surface and from the release of gravitational binding energy of the accreted material. The total energy production is given by: $L = (\epsilon_{nuc} + \epsilon_{acc})\,\dot{M}_{WD}$, where $\dot{M}_{WD}$ is the accretion rate of the WD and $\epsilon_{nuc}$ and $\epsilon_{acc}$ denote the specific energy production from nuclear burning at the WD surface and release of gravitational binding energy, respectively. Hence, the Eddington accretion limit depends on the chemical composition of the accreted material and on the mass of the WD and is roughly given by:

$$\dot{M}_{Edd,WD} \approx (5.5 - 6.2) \times 10^{-7}\ M_{\odot}\,\text{yr}^{-1} \tag{77}$$

for the WDs studied in this work.

The value of $\dot{M}_{CE}$ is probably the largest uncertainty in our modelling. The rationale of even considering an accretion rate $|\dot{M}_2| > \dot{M}_{Edd,WD}$ is that the WD may drive a strong wind in a bipolar outflow which may prevent the WD envelope from otherwise expanding into giant star dimensions (Nomoto, Nariai & Sugimoto 1979). Many previous studies of accreting WDs have adapted the optically thick wind model of Kato & Iben (1992) and Kato & Hachisu (1994), without any restrictions on $|\dot{M_2}|$ (see Cassisi, Iben & Tornambe 1998; Langer et al. 2000; for a critique of this assumption). Given the uncertainties regarding the validity of this model, we adopt a maximum allowed mass-transfer rate limit of $|\dot{M_2}| = \dot{M}_{CE} = 3\,\dot{M}_{Edd,WD}$. For most of our models we stopped the calculations if $|\dot{M_2}| > \dot{M}_{CE}$ (assuming the system evolved into a CE and merged). To test the dependence of this limit we also computed some models by allowing $\dot{M}_{CE} = 10\,\dot{M}_{Edd,WD}$ for comparison. As we shall see, the role of adapting the optically thick wind model or not has important consequences for the progenitor parameter space leading to AIC. Constraints of the wind mass loss from an accreting WD can be determined directly from radio observations of SN Ia remnants (Chomiuk et al. 2012). Hence, there is some hope that future observations can clarify the situation.

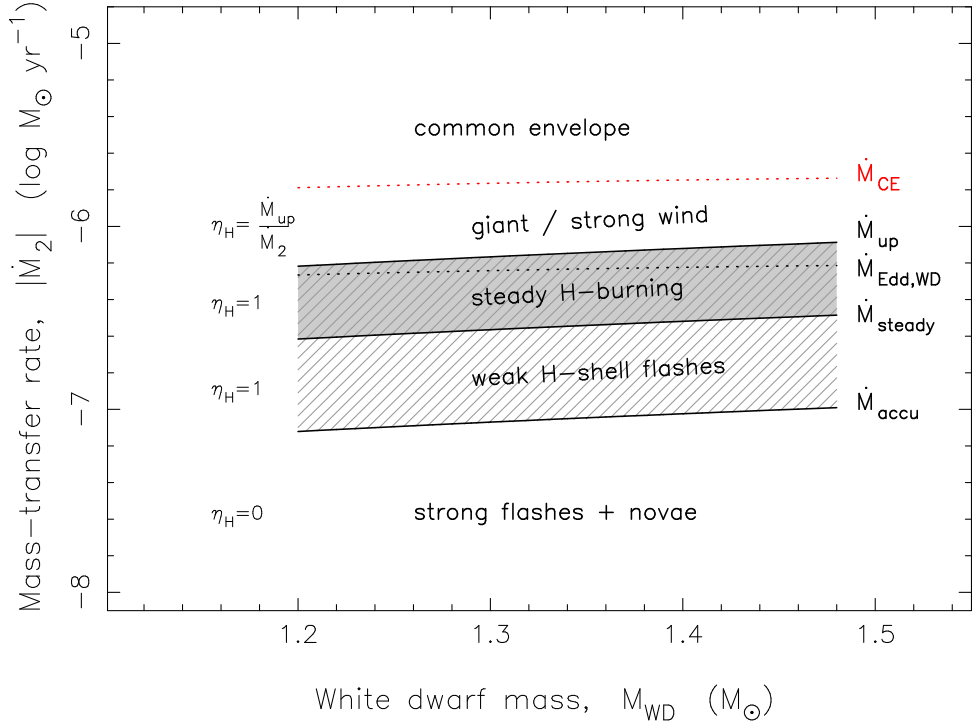

Figure 76: The accretion window of the WD, the corresponding critical mass-transfer rates, and our assumed mass-accumulation fractions, $\eta_H$, here calculated with a hydrogen abundance of $X = 0.70$ of the accreted matter (see text).

If a wind is driven from the WD (when $|\dot{M_2}| > \dot{M}_{up}$) we calculate its mass-accumulation fraction using: $\eta_H = \dot{M}_{up}/|\dot{M_2}|$ to restrict the accretion rate to a maximum of $\dot{M}_{up}$. This value represents the upper limit for *steady* shell hydrogen burning of a WD and can be estimated from the growth rate of the degenerate core in red giant stars undergoing hydrogen shell burning by applying the relation between core mass and luminosity (Iben & Tutukov 1989). As an example, Hachisu & Kato (2001) found:

$$\dot{M}_{up} = 5.3 \times 10^{-7}\ M_\odot\,\text{yr}^{-1}\ \cdot \left(\frac{M_{WD}}{M_\odot} - 0.40\right) \left(\frac{1.7 - X}{X}\right), \qquad (78)$$

which is valid for a hydrogen mass fraction $X \geq 0.10$. There is a variety of similar expressions for this critical limit in the literature (e.g. Nomoto 1982; Hachisu, Kato & Nomoto 1996; 1999; Nomoto et al. 2007). However, they do not differ by much and our results are stable against these minor variations. (We note that $\dot{M}_{up} \simeq \dot{M}_{Edd,WD}$).

The minimum value for steady hydrogen shell burning is given by Nomoto (1982):

$$\dot{M}_{steady} = 0.4\,\dot{M}_{up}. \qquad (79)$$

Although the hydrogen burning is not steady below this limit, the shell flashes are found to be weak for accretion rates just slightly below the limit. Following Hachisu, Kato & Nomoto (1999) we therefore assume $\eta_H = 1$ in the entire interval: $\dot{M}_{accu} < |\dot{M_2}| < \dot{M}_{up}$, where

$$\dot{M}_{accu} = 1/8\,\dot{M}_{up}. \qquad (80)$$

If the mass-accretion rate is below $\dot{M}_{accu}$ violent shell flashes and nova outbursts cannot be avoided and thus the WD is prevented from increasing its mass (i.e. $\eta_H = 0$), or may even erode.

Following hydrogen burning the helium is processed into carbon and oxygen. The mass accumulation efficiency in helium shell flashes was studied in detail by Kato & Hachisu (2004). We have adapted their mass accumulation efficiencies for helium burning, $\eta_{He}$ into our code. The long-term effective mass-accretion rate of the WD is therefore given by:

$$\dot{M}_{WD} = \eta_H \cdot \eta_{He} \cdot |\dot{M_2}|. \qquad (81)$$

For accretion of pure helium we used $\dot{M}_{WD} = \eta_{He} \cdot |\dot{M_2}|$ and

$$\dot{M}_{up,He} = 7.2 \times 10^{-6}\ M_\odot\,\text{yr}^{-1}\ \cdot \left(\frac{M_{CO}}{M_\odot} - 0.6\right), \qquad (82)$$

where $M_{CO}$ is the mass of the CO core of the helium donor star (Nomoto 1982). According to Jose, Hernanz & Isern (1993), $\eta_{He}$ might be somewhat smaller for direct accretion of helium, which leads to stronger shell flashes, compared to the case where helium is accumulated via multiple cycles of hydrogen burning (i.e. double shell burning).

For recent discussions on the WD growth rate and the dependence on the WD mass and the mixing of the accreted material, see e.g. Denissenkov et al. (2013) and Newsham, Starrfield & Timmes (2013).

### 9.3.2 Orbital dynamics

We consider close interacting binary systems which consist of a non-degenerate (evolved) donor star and a compact object, in our case initially a massive WD and later on, in the case of an AIC event, a NS. When the donor star fills its Roche lobe, any exchange and loss of mass from such an X-ray binary will also lead to alterations of the orbital dynamics, via modifications in the orbital angular momentum, and hence changes in the size of the critical Roche-lobe radius of the donor star. The stability of the mass-transfer process therefore depends on how these two radii evolve (i.e. the radius of the star and its Roche-lobe radius). The various possible modes of mass exchange and mass loss include, for example, fast wind mass loss (Jeans mode), Roche-lobe overflow (with or without isotropic re-emission), and common envelope evolution (e.g. van den Heuvel 1994a; Soberman, Phinney & van den Heuvel 1997; and references therein). The RLO mass transfer can be initiated while the donor star is still on the main sequence (Case A RLO), during hydrogen shell burning (Case B RLO), or during helium shell burning (Case C RLO). The corresponding evolutionary timescales for these different cases will in general proceed on a nuclear, thermal, or dynamical timescale, respectively, or a combination thereof. This timescale is important for the amount of mass that can be accreted and for the extent to which the NS produced in the AIC can be recycled after its formation.

The dynamical evolution of a binary system can be found by solving for the changes in the orbital separation, $a$. The orbital angular momentum of a circular binary system is given by: $J_{orb} = \mu \, \Omega \, a^2$, where $\mu$ is the reduced mass and the orbital angular velocity is: $\Omega = \sqrt{GM/a^3}$. A simple logarithmic differentiation of the orbital angular momentum equation yields the rate of change in orbital separation:

$$\frac{\dot{a}}{a} = 2\frac{\dot{J}_{orb}}{J_{orb}} - 2\frac{\dot{M}_1}{M_1} - 2\frac{\dot{M}_2}{M_2} + \frac{\dot{M}_1 + \dot{M}_2}{M}, \tag{83}$$

where the two stellar masses are given by $M_1$ and $M_2$, the total mass is $M = M_1 + M_2$, and the total change in orbital angular momentum per unit time is given by: $\dot{J}_{orb} = \dot{J}_{gwr} + \dot{J}_{mb} + \dot{J}_{ls} + \dot{J}_{ml}$. These four terms represent gravitational wave radiation, magnetic braking, other spin-orbit couplings, and mass loss, respectively (e.g. Tauris & van den Heuvel 2006; and references therein).

In this work we adopt the so-called isotropic re-emission mode[33] for modelling the mass-transfer and the mass loss from the binary (e.g. Bhattacharya & van den Heuvel 1991; Soberman, Phinney & van den Heuvel 1997). We let the mass ratio between the donor star ($M_2$) and the accretor ($M_1$) be denoted by $q = M_2/M_1$. Assuming that the direct wind mass loss of the donor star is negligible compared to the RLO mass-transfer rate from the donor star, $|\dot{M}_2|$, and ignoring mass stored in a circumbinary torus, one can show that a binary system always widens ($\dot{a} > 0$) as a result of mass transfer if $q < 1$. Similarly, a binary always decreases ($\dot{a} < 0$) if $q > (1 + \sqrt{17})/4 \simeq 1.28$, irrespective of the amount of mass ejected from the vicinity of the accretor (e.g. see Section 16.4.3 in Tauris & van den Heuvel 2006). This fact is worth remembering when we analyse our results.

In our calculations we have ignored any changes in $J_{orb}$ due to magnetic braking or any other tidal spin-orbit interactions. Magnetic braking is only known to operate efficiently in low-mass stars ($\lesssim 1.5 \, M_\odot$) which have convective envelopes. As we shall see, our calculations show that only main-sequence donor stars with masses in the range $1.4{-}2.6 \, M_\odot$ (depending on metallicity) transfer mass at the rates needed for AIC events to take place. Therefore, to be consistent with all our calculations, we have not included magnetic braking in the very few borderline cases. (These low-mass donors all have low-metallicities resulting in some stability against convective envelopes.) Our code includes gravitational wave radiation by calculating $\dot{J}_{gwr}$ and correcting for it. However, for these AIC progenitor systems the RLO mass-transfer timescales are always significantly shorter than the timescales on which gravitational wave radiation is important.

---

[33]Although at present there is not much observational evidence behind this model (not surprising given that systems with extremely high mass-transfer rates are short lived) there is some evidence of excess mass loss in the case of Cyg X-2, as demonstrated by King & Ritter (1999) and Podsiadlowski & Rappaport (2000). This mass loss could have been the cause in the past by a relativistic jet (as observed in SS433) or by coronal winds from the outer parts of the accretion disk (Blandford & Begelman 1999).

In all our calculations we assumed an initial (ONeMg) WD mass of $M_{\mathrm{WD}} = 1.20\ M_\odot$ prior to accretion. As mentioned earlier, for these calculations the WD is simply treated as a point mass.

### 9.3.2.1   The dynamical effect of the AIC on the orbit

If the WD mass reached $1.48\ M_\odot$ we assumed that the WD was subject to instantaneous AIC. The effect of sudden mass loss in a binary has been studied in detail by Hills (1983) for bound systems, and by Tauris & Takens (1998) for disrupted systems. Assuming a circular pre-AIC orbit we here follow Hills (1983) to find the changes of the binary orbital parameters. The change of the binary semi-major axis, as a result of an asymmetric AIC, is given by:

$$\frac{a}{a_0} = \left[\frac{1 - (\Delta M/M_0)}{1 - 2(\Delta M/M_0) - (w/v_c)^2 - 2\cos\theta\ (w/v_c)}\right], \qquad (84)$$

where $a_0$ is the pre-AIC semi-major axis (radius), $a$ the post-AIC semi-major axis, $\Delta M = 0.20\ M_\odot$ the effective mass loss during the AIC when the $1.48\ M_\odot$ WD is compressed to a NS with a gravitational mass of $1.28\ M_\odot$ (Zeldovich & Novikov 1971), $M_0 = M_{\mathrm{WD}} + M_2$ the pre-AIC total mass, $v_c = \sqrt{GM_0/a_0}$ the pre-AIC orbital velocity of the collapsing WD in a reference fixed on the companions star, $w$ the magnitude of the kick velocity, and $\theta$ the angle between the kick velocity vector, $\vec{w}$ and the pre-AIC orbital velocity vector, $\vec{v_c}$. The post-AIC eccentricity is given by:

$$e = \sqrt{1 + \frac{2E_{\mathrm{orb}}J_{\mathrm{orb}}^2}{\mu\ G^2 M_{\mathrm{NS}}^2 M_2^2}}, \qquad (85)$$

where the post-AIC orbital energy of the system is given by: $E_{\mathrm{orb}} = -GM_{\mathrm{NS}}M_2/2a$, and the orbital angular momentum is given by:

$$J_{\mathrm{orb}} = a_0\,\mu\sqrt{(v_c + w\cos\theta)^2 + (w\sin\theta\sin\phi)^2}, \qquad (86)$$

where $\mu$ is the post-AIC reduced mass and $\phi$ is the angle between the projection of the kick velocity vector onto a plane perpendicular to the pre-AIC velocity vector of the WD and the pre-AIC orbital plane. We neglected any shell impact effects on the companion star since the amount of material ejected in an AIC event is expected to be negligible. Even for SN Ib/c where a significant shell is ejected, the impact effect on the orbital dynamics is small if the pre-SN separation is larger than a few $R_\odot$ (Tauris & Takens 1998).

Following the AIC we checked if the post-AIC periastron separation, $a(1 - e)$ is smaller than the radius of the companion star, $R_2$. In that case we assumed that the system merges. The post-AIC orbit is expected to be circularized with time and we therefore assumed that the tidal interactions reduced the semi-major axis by a factor of $(1-e^2)$ in order to conserve $J_{\mathrm{orb}}$. When no momentum kick was added ($w = 0$) to the newborn NS, the relation between post-AIC orbital separation (including the subsequent effect of tidal circularization), $a_{\mathrm{circ}}$ and the pre-AIC orbital radius, $a_0$ is simply given by (Verbunt, Wijers & Burm 1990; Bhattacharya & van den Heuvel 1991):

$$a_{\mathrm{circ}} = a_0\frac{M_0}{M} = a_0\ \frac{1.48\ M_\odot + M_2}{1.28\ M_\odot + M_2}. \qquad (87)$$

Since post-AIC evolution calculations also require a significant amount of computational time, and each AIC event can lead to a large set of possible parameter outcomes for $(w, \theta, \phi)$, we have restricted ourselves to those cases which best probe the extreme cases for the post-AIC evolution with respect to orbital periods and systemic recoil velocities resulting from the AIC event.

### 9.3.3   Post-AIC LMXB evolution

The evolution of post-AIC binaries is, in principle, similar to normal LMXB evolution (i.e. a donor star which transfers matter and angular momentum to an accreting NS). For this recycling process we follow Tauris & Savonije (1999) and Tauris, Langer & Kramer (2012) for tracking the evolution of the LMXB (see also Section 9.3.2 above). The accretion rate onto the NS is assumed to be Eddington limited and is given by:

$$\dot{M}_{NS} = \left( |\dot{M}_2| - \max \left[ |\dot{M}_2| - \dot{M}_{Edd} , 0 \right] \right) \cdot e_{acc} \cdot k_{def}, \tag{88}$$

where $e_{acc}$ is the fraction of matter transfered to the NS which actually ends up being accreted and remains on the NS, and $k_{def}$ is a factor that expresses the ratio of gravitational mass to rest mass of the accreted matter (depending on the equation-of-state of supranuclear matter $k_{def} \simeq 0.85 - 0.90$; e.g. Lattimer & Prakash 2007). Here we assumed $e_{acc} \cdot k_{def} = 0.30$. Our motivation for this value is the increasing evidence of inefficient accretion in LMXBs, even in close systems where the mass-transfer rate is expected to be sub-Eddington ($|\dot{M}_2| < \dot{M}_{Edd}$) at all times (e.g. Jacoby et al. 2005; Antoniadis et al. 2012). Possible mechanisms for inefficient accretion include propeller effects, accretion disc instabilities, and direct irradiation of the donor's atmosphere from the pulsar (Illarionov & Sunyaev 1975; van Paradijs 1996; Dubus et al. 1999). For the post-AIC NS, we calculated the Eddington mass-accretion rate using:

$$\dot{M}_{Edd} = 2.3 \times 10^{-8} \ M_\odot \, \text{yr}^{-1} \ \cdot M_{NS}^{-1/3} \cdot \frac{2}{1 + X}. \tag{89}$$

In Section 9.4.3.3 we specify a relation between the amount of mass accreted and the final pulsar spin period.

### 9.3.4 Complete evolution in the ($P_{orb}, M_2$)-plane

To demonstrate the orbital evolution: 1) during pre-AIC mass transfer, followed by 2) post-AIC mass transfer in an LMXB, and to show the effect of possible kicks associated with the AIC, we have plotted complete evolutionary tracks in the ($P_{orb}, M_2$)-plane in Fig. 77. The two donor stars in these examples ($M_2 = 2.6 \ M_\odot$ and $M_2 = 2.2 \ M_\odot$) are both more massive than the accreting ONeMg WD (initially with mass ratio, $q \sim 2$.) This explains why both systems decrease in $P_{orb}$ prior to the AIC event. At the moment of the AIC the orbits widen instantaneously as a consequence of the sudden mass loss (c.f. Section 9.3.2.1). The larger the kick, $w$, the larger the post-AIC $P_{orb}$ becomes at which the donor star refills its Roche lobe and continues mass transfer to the newborn NS in the LMXB source. The final products are binary MSPs with He WDs. During the LMXB phase the orbit changes from a converging system to a diverging system when the mass ratio inverses (the exact value depends on the mass-transfer rate, c.f. Section 9.3.2).

It is interesting that for the original 2.2 $M_\odot$ donor star, the post-AIC mass transfer is not dynamically stable if the AIC was asymmetric (i.e. if $w = 50 \ \text{km s}^{-1}$ or $w = 450 \ \text{km s}^{-1}$). In this case the post-AIC binary becomes wide enough that the donor star develops a deep convective envelope before refilling its Roche lobe. The result is that the subsequent mass-transfer stage leads to excessive mass-transfer rates and thus to the formation of a CE. We did not follow the evolution of these systems further, although it is possible, in principle, that some of the donor star envelopes would be loosely enough bound to allow for ejection during spiral-in and thereby leave behind a mildly recycled pulsar orbiting a WD in a tight orbit. We return to this possibility in Section 9.7.5.

### 9.4 AIC in systems with main-sequence star donors

In the following three sections we present our results of AIC calculations in systems with main-sequence, giant, and helium star donors, respectively. We evolved a total of 240 binary systems. Key parameters from 36 examples of complete calculations (i.e. both pre-AIC and post-AIC LMXB computations leading to a recycled binary pulsar) are given in Table 16. Model names beginning with MS and MZ refer to main-sequence donor stars at solar metallicity ($Z = 0.02$) and $Z = 0.001$, respectively; giant donor star models are denoted by GS (or GSZ for $Z = 0.001$), and helium donor star models are denoted by He. The given parameters for each model are the following: $M_2^{ZAMS}$ and $P_{orb}^{ZAMS}$ refer to initial donor star mass and orbital period; $t_{RLO}$ is the age of the donor star when it initiates RLO; and $X_c$ is its central hydrogen content at that time; $\Delta t_{CV}$ is the duration of the mass-transfer phase until the AIC event; and $M_2^{AIC}$ and $P_{orb}^{AIC}$ are the donor star mass and the orbital period at the moment of the AIC. When a momentum kick ($w > 0$) is added to the newborn NS, its magnitude and direction are given. The resulting systemic recoil velocity of each post-AIC system is given by $v_{sys}$. As a result of the AIC the system temporarily detaches. The time it takes until the donor star refills its Roche lobe is denoted by $\Delta t_{detach}$, and $P_{orb}^{circ}$ is the orbital period at that time (after circularization). The duration of the subsequent post-AIC LMXB phase is given by $\Delta t_{LMXB}$. Finally, the parameters $M_{WD}$, $P_{orb}^{MSP}$, $M_{NS}$, $\Delta M_{NS}$, $P_{spin}$, and $t_{total}$ denote the WD mass, the orbital period, the NS mass, the amount of mass accreted by the NS during the post-AIC mass transfer, the final spin period of the recycled pulsar, and the total age of the binary system at this end point.

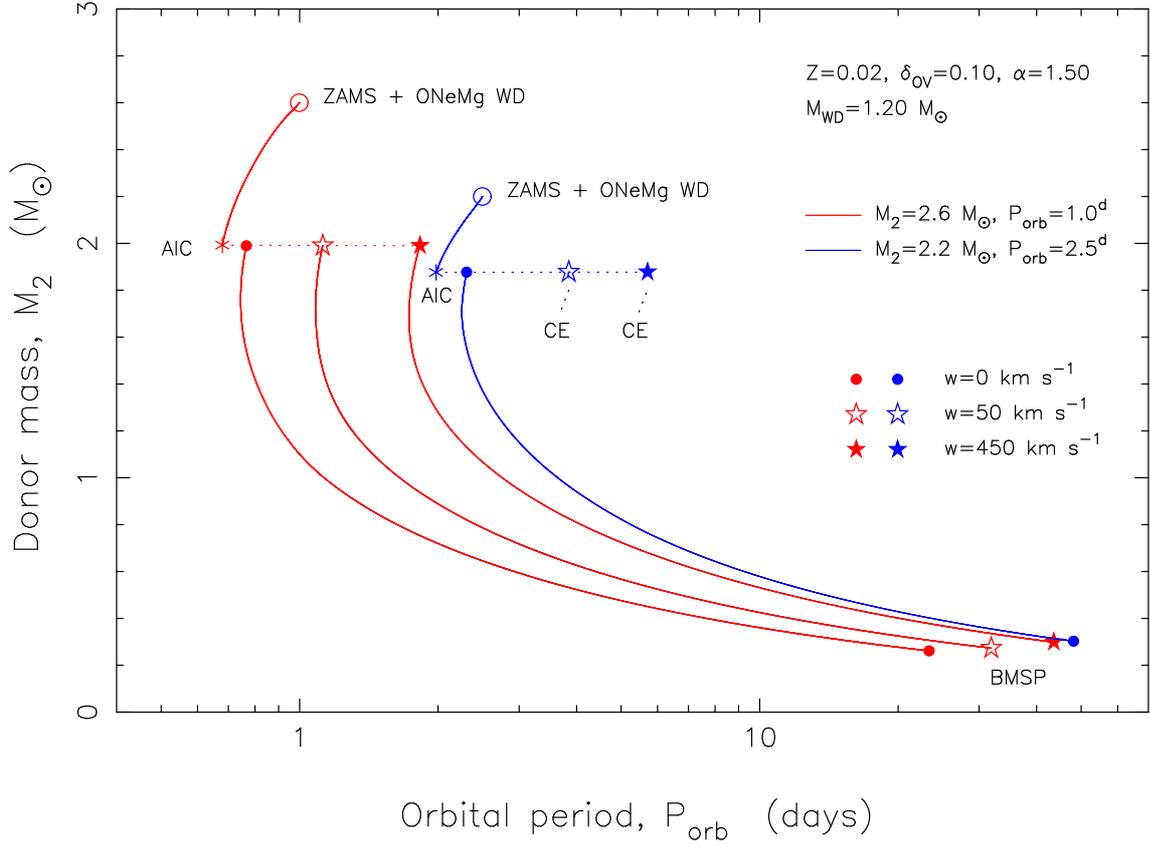

Figure 77: Evolutionary tracks in the $(P_{orb}, M_2)$-plane. The initial system configurations are: $M_2 = 2.6\,M_\odot$, $P_{orb} = 1.0$ days (red tracks) and $M_2 = 2.2\,M_\odot$, $P_{orb} = 2.5$ days (blue tracks). The ONeMg WD has an initial mass of $1.20\,M_\odot$ and grows to $1.48\,M_\odot$ when the AIC occurs. After the AIC three tracks were computed for each system depending on the kick velocity applied in the AIC. Here we applied $w = 0$, $w = 50\,\mathrm{kms}^{-1}$ ($\phi = \theta = 0°$), and $w = 450\,\mathrm{kms}^{-1}$ ($\phi = 107°$ and $\theta = 90°$) resulting in different values of $P_{orb}$ for the post-AIC system. In two post-AIC cases presented here, and in general if the donor star is quite evolved by the time it refills its Roche lobe (i.e. because the post-AIC $P_{orb}$ is large) *and* if the donor star is significantly more massive than the newborn NS ($1.28\,M_\odot$), the RLO becomes dynamically unstable and the system evolves through a common envelope (marked by CE).

For hydrogen-rich donors we have initiated our calculations assuming a ZAMS star orbiting an ONeMg WD. The error in placing the companion star on the ZAMS, and neglecting the evolution of this star while the ONeMg WD forms, is not very significant. As we shall see in a moment, our main-sequence companions have maximum masses of $M_2 = 2.6\,M_\odot$, and even these stars evolve on a much longer timescale (at least by a factor of $\sim 10$) compared to the typical $6 - 8\,M_\odot$ progenitor stars of ONeMg WDs.

Table 16: Summary of selected systems which evolved successfully to AIC and later produced recycled pulsars (see first paragraph of Section 9.4 for explanations of all the variables). [* = $Y_c$,   ** = (hybrid) CO WD,   *** = CO WD after He-flash on the RGB,   CE = onset of a common envelope].

| Model | $M_2^{\rm ZAMS}$ $M_\odot$ | $P_{\rm orb}^{\rm ZAMS}$ days | $t_{\rm RLO}$ Myr | $X_c$ | $\Delta t_{\rm CV}$ Myr | $M_2^{\rm AIC}$ $M_\odot$ | $P_{\rm orb}^{\rm AIC}$ days | kick* km s$^{-1}$ | angles deg. | $v_{\rm sys}$ km s$^{-1}$ | $\Delta t_{\rm detach}$ Myr | $P_{\rm orb}^{\rm circ}$ days | $\Delta t_{\rm LMXB}$ Myr | $M_{\rm WD}$ $M_\odot$ | $P_{\rm orb}^{\rm MSP}$ days | $M_{\rm NS}$ $M_\odot$ | $\Delta M_{\rm NS}$ $M_\odot$ | $P_{\rm spin}$ ms | $t_{\rm total}$ Myr |
|---|---|---|---|---|---|---|---|---|---|---|---|---|---|---|---|---|---|---|---|
| MS1W | 3.00 | 1.50 | 294 | 0.12 | 1.27 | 1.65 | 0.78 | 0 | | 12 | 0.054 | 0.88 | 335 | 0.310 | 17.8 | 1.46 | 0.18 | 1.2 | 630 |
| MS2W | 2.80 | 5.00 | 1250 | 0.00 | 0.51 | 1.56 | 3.19 | 0 | | 8 | 0.064 | 3.65 | 4 | 0.411** | 37.2 | 1.31 | 0.03 | 4.7 | 1250 |
| MS3 | 2.60 | 1.00 | 357 | 0.29 | 1.90 | 1.99 | 0.68 | 0 | | 13 | 0.003 | 0.76 | 1240 | 0.262 | 23.8 | 1.55 | 0.27 | 0.9 | 1600 |
| MS4 | | | | | | | | 50 | (107, 90) | 7 | 0.097 | 1.12 | 891 | 0.273 | 31.8 | 1.61 | 0.33 | 0.8 | 1250 |
| MS5 | | | | | | | | 450 | | 180 | 165 | 1.83 | 214 | 0.298 | 43.6 | 1.66 | 0.38 | 0.7 | 736 |
| MS6 | 2.50 | 1.00 | 403 | 0.28 | 1.98 | 1.99 | 0.72 | 0 | | 13 | 0.003 | 0.80 | 1180 | 0.264 | 24.8 | 1.56 | 0.28 | 0.9 | 1590 |
| MS7 | 2.50 | 1.50 | 504 | 0.08 | 1.68 | 1.97 | 1.07 | 0 | | 11 | 0.010 | 1.20 | 252 | 0.301 | 26.4 | 1.56 | 0.28 | 0.9 | 758 |
| MS8 | | | | | | | | 50 | (116, 90) | 9 | 0.19 | 1.87 | 159 | 0.307 | 41.0 | 1.64 | 0.36 | 0.7 | 665 |
| MS9 | | | | | | | | 450 | | 183 | 58 | 2.88 | 30 | 0.320 | 52.7 | 1.44 | 0.16 | 1.3 | 594 |
| MS10 | 2.50 | 2.50 | 539 | 0.00 | 0.67 | 1.88 | 1.76 | 0 | | 9 | 0.048 | 1.99 | 12 | 0.335** | 33.4 | 1.35 | 0.07 | 2.5 | 552 |
| MS11 | 2.50 | 3.00 | 545 | 0.00 | 0.63 | 1.92 | 2.12 | 0 | | 9 | 0.124 | 2.39 | 11 | 0.341** | 38.3 | 1.34 | 0.06 | 2.8 | 557 |
| MS12 | 2.20 | 1.00 | 584 | 0.26 | 2.08 | 1.85 | 0.79 | 0 | | 12 | 0.016 | 0.89 | 986 | 0.269 | 27.3 | 1.60 | 0.32 | 0.8 | 1570 |
| MS13 | | | | | | | | 50 | (112, 90) | 8 | 100 | 1.34 | 450 | 0.285 | 37.4 | 1.67 | 0.39 | 0.7 | 1140 |
| MS14 | | | | | | | | 450 | | 189 | 264 | 2.02 | 108 | 0.297 | 47.3 | 1.58 | 0.30 | 0.8 | 958 |
| MS15 | 2.20 | 2.00 | 765 | 0.00 | 1.05 | 1.84 | 1.58 | 0 | | 10 | 0.046 | 1.79 | 29 | 0.295 | 40.1 | 1.42 | 0.14 | 1.5 | 795 |
| MS16 | 2.20 | 2.50 | 770 | 0.00 | 0.89 | 1.88 | 1.98 | 0 | | 9 | 0.178 | 2.24 | 25 | 0.303 | 48.1 | 1.40 | 0.12 | 1.7 | 794 |
| MS17 | 2.00 | 1.00 | 788 | 0.24 | 4.92 | 1.59 | 0.84 | 0 | | 12 | 0.487 | 0.96 | 839 | 0.272 | 28.3 | 1.59 | 0.31 | 0.8 | 1630 |
| MS18 | 2.00 | 2.00 | 1010 | 0.00 | 2.02 | 1.63 | 1.68 | 0 | | 9 | 6.74 | 1.92 | 66 | 0.297 | 41.8 | 1.44 | 0.16 | 1.3 | 1090 |
| MZ1 | 2.00 | 1.00 | 594 | 0.00 | 1.00 | 1.30 | 0.99 | 0 | | 11 | 0.003 | 1.16 | 22.5 | 0.310 | 20.5 | 1.36 | 0.08 | 2.3 | 618 |
| MZ2 | 1.80 | 4.00 | 811 | 0.00 | 1.95 | 1.44 | 3.62 | 0 | | 7 | 1.62 | 4.17 | 56 | 0.345 | 58.6 | 1.44 | 0.16 | 1.3 | 871 |
| MZ3 | | | | | | | | 50 | (161, 90) | 16 | 15.0 | 8.19 | 47 | 0.367 | 95.3 | 1.44 | 0.16 | 1.3 | 875 |
| MZ4 | | | | | | | | 450 | | 218 | 21.4 | 10.57 | 41 | 0.377 | 114.7 | 1.44 | 0.16 | 1.3 | 875 |
| MZ5 | 1.40 | 1.00 | 1750 | 0.00 | 4.52 | 0.957 | 1.26 | 0 | | 9 | 0.228 | 1.51 | 165 | 0.285 | 21.7 | 1.45 | 0.17 | 1.3 | 1920 |
| GS1 | 1.10 | 80 | 8510 | 0.00 | 2.29 | 0.711 | 134 | 0 | | 2 | 3.83 | 163 | 5.01 | 0.408 | 552 | 1.31 | 0.02 | 6.4 | 8520 |
| GS2 | 1.10 | 180 | 8520 | 0.00 | 1.69 | 0.685 | 325 | 0 | | 1 | 1.91 | 394 | 2.40 | ~0.47*** | >919 | 1.29 | 0.011 | 14 | 8530 |
| GS3 | 1.00 | 400 | 13600 | 0.00 | 1.98 | 0.577 | 907 | 0 | | 1 | 2.51 | 1110 | ~0.01 | ~0.50*** | ~1400 | 1.280 | 0.000 | 500 | 13600 |
| GS4 | 0.90 | 160 | 19100 | 0.00 | 3.63 | 0.529 | 372 | 0 | | 1 | 0.80 | 459 | CE | ~0.46** | CE | 1.280 | 0.000 | – | – |
| GSZ1 | 1.00 | 250 | 6330 | 0.00 | 2.02 | 0.622 | 481 | 0 | | 1 | 0.30 | 587 | 2.32 | ~0.53*** | >819 | 1.32 | 0.010 | 11.0 | 6340 |
| He1 | 1.40 | 0.04 | 0.014 | 0.97* | 2.77 | 1.05 | 0.039 | 0 | | 30 | 3.59 | 0.046 | 7.96 | 0.670** | 0.072 | 1.32 | 0.04 | 3.8 | 14.3 |
| He2 | 1.40 | 0.05 | 2.49 | 0.51* | 5.87 | 1.03 | 0.044 | 0 | | 29 | 0.073 | 0.051 | 1.32 | 0.719** | 0.097 | 1.294 | 0.014 | 8.4 | 9.75 |
| He3 | 1.40 | 0.10 | 7.12 | 0.00* | 0.20 | 1.12 | 0.100 | 0 | | 23 | 0.043 | 0.118 | 0.392 | 0.821** | 0.199 | 1.284 | 0.004 | 21 | 7.75 |
| He4 | 1.20 | 0.05 | 4.26 | 0.00* | 5.60 | 0.797 | 0.054 | 0 | | 25 | 0.256 | 0.066 | 1.07 | 0.703** | 0.082 | 1.291 | 0.011 | 10 | 11.2 |
| He5 | 1.20 | 0.60 | 21.9 | 0.00* | 0.11 | 0.921 | 0.707 | 0 | | 11 | 0.005 | 0.84 | 0.121 | 0.833** | 1.03 | 1.281 | 0.001 | 60 | 22.1 |
| He6 | | | | | | | | 50 | (120, 90) | 18 | 0.017 | 1.30 | 0.113 | 0.836** | 1.53 | 1.281 | 0.001 | 60 | 22.1 |
| He7 | | | | | | | | 450 | | 267 | 0.023 | 1.66 | 0.108 | 0.838** | 1.95 | 1.281 | 0.001 | 60 | 22.1 |
| He8 | 1.10 | 1.20 | 32.4 | 0.00* | 0.16 | 0.829 | 1.52 | 0 | | 8 | 0.012 | 1.81 | 0.032 | 0.826** | 1.82 | 1.280 | 0.000 | 200 | 32.6 |

### 9.4.1 Pre-AIC evolution with main-sequence donors

In Fig. 78 we have plotted a grid of the initial orbital periods and donor star masses of our investigated systems with main-sequence donor stars (i.e. supersoft X-ray sources). The upper and lower panels are for different donor star metallicities. The type of symbol in each grid point represents the outcome of the computations, which we discuss in more detail below. We note that the term main-sequence donor star is slightly misleading here since many of these stars have passed the termination age of the main sequence (TAMS) by the time they fill their Roche lobes and become donors. Hence, many of these systems evolve via early Case B RLO from Hertzsprung-gap subgiant donors. The systems which successfully evolve to the AIC event stage have initial orbital periods between $0.5 - 4$ days and initial donor star masses between $2.0 - 2.6\,M_\odot$ for a metallicity of $Z = 0.02$ and between $1.4 - 2.2\,M_\odot$ for $Z = 0.001$.

The shift in parameters in Fig. 78 is interesting, in particular in allowed donor star masses, which lead to AIC depending on the chemical composition of the donor star. The shift to lower donor star masses for lower metallicity can be understood from the smaller radii of these stars (due to their lower opacities) compared to stars with higher metallicity. Therefore, these stars become more evolved when eventually initiating their RLO, leading to higher values of $|\dot{M}_2|$ (see also Langer et al. 2000).

At first it may seem peculiar that neighbouring grid points can lead to a mass-transfer rate that is too low/too high for the WD to grow sufficiently in mass (see e.g. $M_2 = 1.8\,M_\odot$, $P_{\rm orb} = 1.5^{\rm d} - 2.8^{\rm d}$, and $Z = 0.02$ along the horizontal part of the cross marked in the upper panel of Fig. 78). However, this behaviour can be understood from the required finetuning of the WD accretion rate. In Fig. 79 we see that while the donor star in an orbit with an initial $P_{\rm orb} = 1.5^{\rm d}$ delivers an insufficient mass-transfer rate, the same system with $P_{\rm orb} = 2.0^{\rm d}$ is seen to produce an excessive mass-transfer rate. The reason for this strong dependence on $P_{\rm orb}$ is due to the corresponding rapid increase in the depth of the convective envelope with increasing radius of these shell hydrogen burning donor stars (Paczyński & Sienkiewicz 1972). This is demonstrated in the Kippenhahn plot shown in Fig. 81. The negative mass-radius exponents ($\zeta = \partial \ln R/\partial \ln M$) of convective envelopes cause these stars to expand in response to mass loss and thereby result in excessive mass-transfer rates. Only in the first case for $P_{\rm orb} = 1.5^{\rm d}$ (upper-left panel of Fig. 81) will the mass transfer remain stable since here the mass ratio is inverted (causing the binary to widen) by the time the envelope has developed a deep convection zone. Similarly, in Fig. 80 we have shown the results of our mass-transfer calculations along four vertical, neighbouring points in the cross marked in the upper panel of Fig. 78. These points correspond to $1.6 \le M_2/M_\odot \le 2.2$, and in all cases $P_{\rm orb} = 2.0^{\rm d}$. Again the explanation for the different outcomes is the differences in the depth of the convective envelopes. The donor stars with $M_2 \lesssim 1.9\,M_\odot$ require a more advanced evolution to fill their Roche lobes and therefore they develop deep convective envelopes before, or during, the RLO which leads to $|\dot{M}_2|$ being too large. Hence, of those systems with $P_{\rm orb} = 2.0^{\rm d}$ and $Z = 0.02$, only those binaries with initial donor star masses of $2.0 \le M_2/M_\odot \le 2.5$ make it to the AIC event.

In general, for all initial $P_{\rm orb}$, donor stars with $M_2 \ge 2.6\,M_\odot$ result in $|\dot{M}_2|$ being too large (unless the criteria in Eq.(76) is relaxed to allow for $\dot{M}_{\rm CE} = 10\,\dot{M}_{\rm Edd,WD}$ in which case we get AIC solutions up to $M_2 \approx 3.0\,M_\odot$). The reason is that in these systems the mass ratio, $q > 2$, and therefore their orbits become significantly tighter with RLO, resulting in a large value of $|\dot{M}_2|$ (see Section 9.3.2).

### 9.4.2 Post-AIC LMXB evolution with main-sequence donors

The post-AIC binary mass transfer resembles that of normal LMXB evolution with an accreting NS. The only difference is that the donor star has already lost some of its mass during the pre-AIC evolution. To model these LMXB systems we proceeded as explained in Section 9.3.3. An example of the results of this modelling is shown in the right panel of Fig. 82 (the left panel shows the evolution of the pre-AIC binaries leading to these systems). An interesting feature becomes clear when comparing the four evolutionary tracks. Whereas the two donor stars in the widest orbits (initial $P_{\rm orb} = 2.5^{\rm d} - 3.0^{\rm d}$) are already undergoing shell hydrogen burning (Case AB RLO) at the time of the AIC, the two donors in the shortest period systems (initial $P_{\rm orb} = 1.0^{\rm d} - 1.5^{\rm d}$) are still undergoing Case A RLO (core hydrogen burning) at the moment of the AIC. Hence, these latter systems remain LMXB (post-AIC) sources on much longer timescales ($250\,{\rm Myr} - 1\,{\rm Gyr}$), initially via Case A RLO, and later via Case AB RLO once the hydrogen shell is ignited.

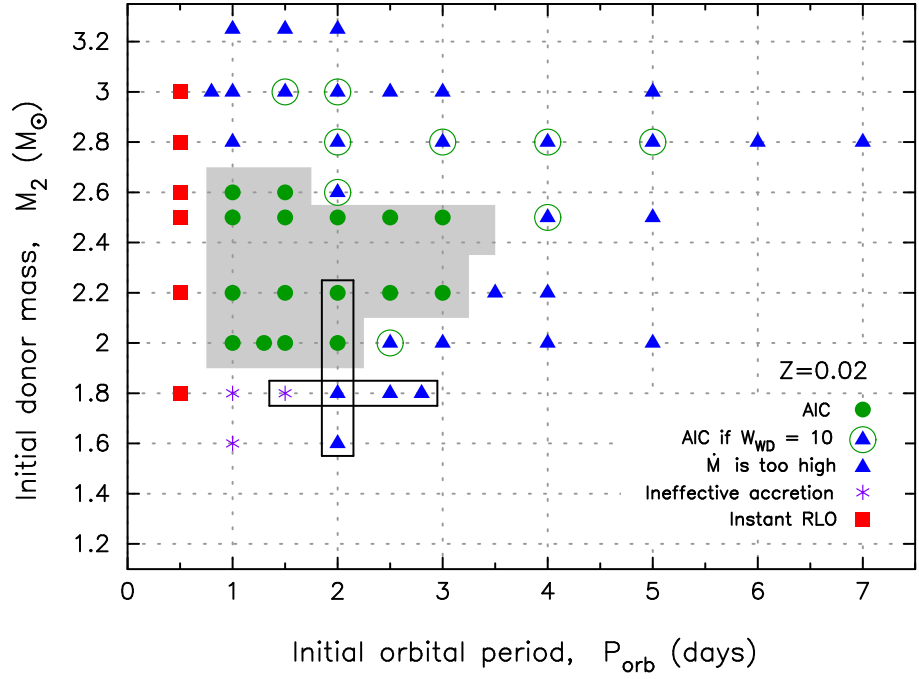

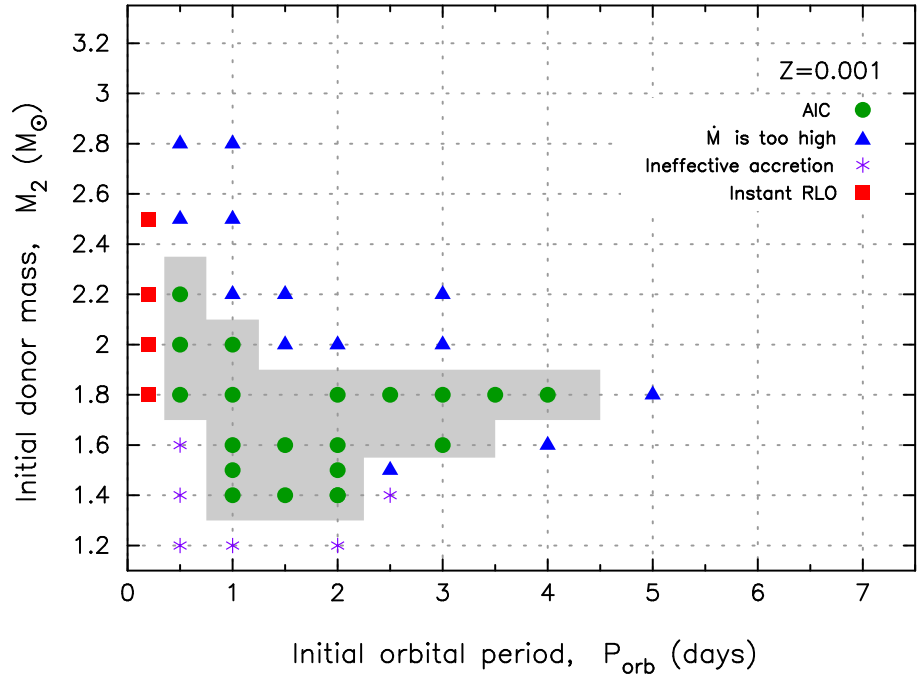

Figure 78: The grid of investigated initial orbital periods and masses for main-sequence donor stars with a metallicity of $Z = 0.02$ (upper panel) and $Z = 0.001$ (lower panel). The grey shaded region in each panel corresponds to systems which successfully evolve to the AIC stage (green circles). The blue triangles and purple asterisks correspond to cases where the mass-transfer rate was too high or too low, respectively, to allow for the WD to reach a critical mass of $1.48\ M_\odot$. The red squares indicate orbits which are too narrow to initially accommodate the ZAMS donor star. The blue triangles inside green circles in the upper panel are systems leading successfully to AIC assuming $\dot{M}_{CE} = 10\ \dot{M}_{Edd,WD}$. Details of the mass-transfer process of the binaries inside the marked cross in the upper panel are represented in Figs. 79 and 80.

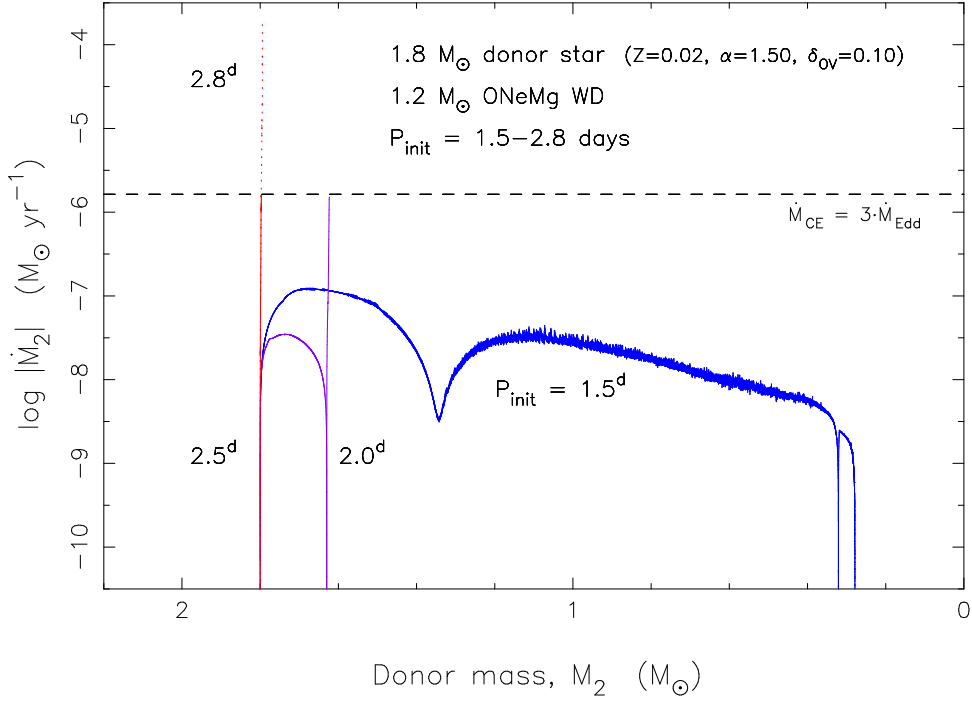

Figure 79: Mass-transfer rates of the four pre-AIC binaries shown in the horizontal part of the cross in the upper panel of Fig. 78. The mass-transfer-rate from the donor star in the system with $P_{\rm orb} = 1.5^{\rm d}$ is too low (see Fig. 76) to cause the accreting WD to grow sufficiently in mass and trigger an AIC event. For the donors in the other systems, on the other hand, the mass-transfer rate is too high to result in stable mass gain of the accretor. Hence, none of these models resulted in a successful AIC. The differences in these mass-transfer rates can be understood in terms of different thicknesses of convective envelopes (see text and Fig. 81).

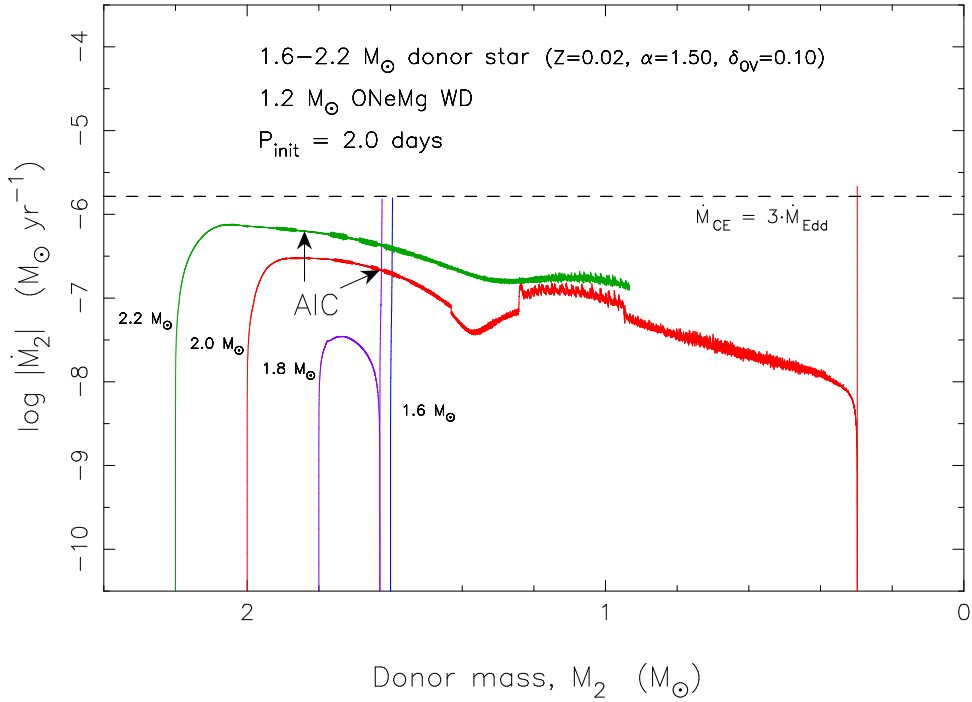

Figure 80: Mass-transfer rates of the pre-AIC binaries shown in the vertical part of the cross in the upper panel of Fig. 78. The four donor stars have masses of $1.6 - 2.2\ M_\odot$ and $P_{\rm orb} = 2.0^{\rm d}$. The arrows mark the collapse of the accreting ONeMg WD (AIC) for the two most massive stars. Donor stars with initial masses $M_2 \geq 2.6\ M_\odot$ result in excessive mass-transfer rates and thus do not produce AIC events (see text).

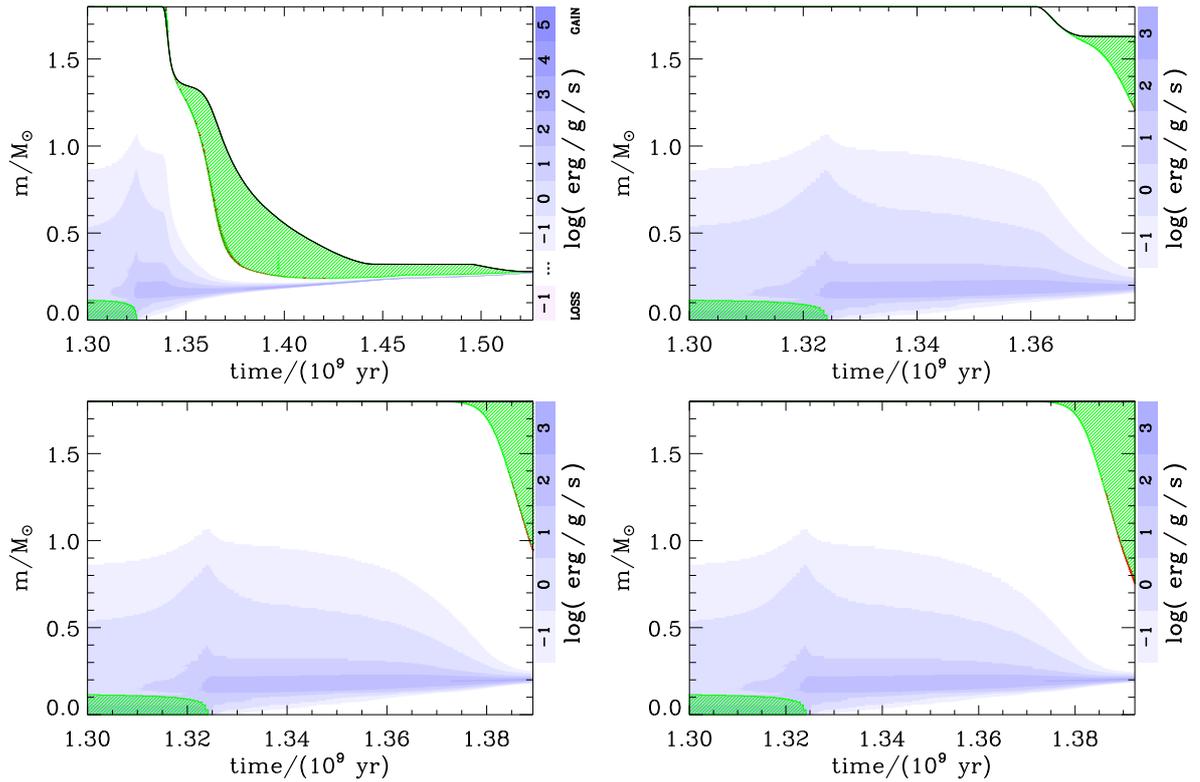

Figure 81: The Kippenhahn diagram of four $1.8\,M_\odot$ donor stars undergoing early Case B RLO in X-ray binaries with an accreting WD and orbital periods of 1.5, 2.0, 2.5, and 2.8 days, respectively. The plots show cross-sections of the stars in mass-coordinates from the centre to the surface of the star, along the y-axis, as a function of stellar age on the x-axis. The green hatched areas denote zones with convection (according to the Ledoux criterion) initially in the core and later in the envelope of the donor stars. The intensity of the blue/purple color indicates the net energy-production rate; the hydrogen burning shell is clearly seen in all panels at $m/M_\odot \simeq 0.2$. In the top-left panel, the donor star decreases its mass from $1.8\,M_\odot$ until it finally detaches from the Roche lobe and forms a $0.28\,M_\odot$ He WD. In the top-right panel, our calculation stops when the donor star reached $1.62\,M_\odot$ since the mass-transfer rate became too high when the donor star refilled its Roche lobe, following an initial phase of stable RLO with a lower value of $|\dot{M}_2|$. In the two lower panels the mass-transfer rate went immediately up to the critical value at the onset of the RLO. The response of a donor star to mass loss depends strongly on the depth of its convective envelope and thus, as seen here, it depends on $P_{\rm orb}$. See Fig. 79 and text for a discussion.

As a consequence of the AIC event the binaries detach for $3\,000 - 100\,000\,{\rm yr}$ before the donor stars refill their Roche lobes on a thermal timescale. (During pre-AIC mass loss the donor stars become smaller than their thermal equilibrium sizes. After the AIC they expand to recover thermal equilibrium). This is a very short (thermal) time interval compared to the typical lifetime of a young pulsar (about $50 - 100\,{\rm Myr}$) and therefore the possibility of detecting a system right at this epoch between the two long-lasting X-ray phases is quite small. Indeed, none of the 6 known radio pulsars with a main-sequence companion are candidates for being post-AIC systems: in all cases they have $P_{\rm orb} > 50^{\rm d}$ and their companions are largely underfilling their Roche lobes.

Next to each coloured graph in the right panel of Fig. 82 is listed the mass of the donor star at the moment of the AIC, $M_2^{\rm AIC}$, the final mass of the WD remnant orbiting the recycled pulsar, $M_{\rm WD}$, and the amount of mass accreted by the pulsar, $\Delta M_{\rm NS}$ (assuming an accretion efficiency of 30% at sub-Eddington mass-transfer rates, see Section 9.3.3). The differences between the two main mass-transfer histories mentioned above (post-AIC Case A RLO vs Case AB RLO) is reflected in both $\Delta M_{\rm NS}$ and $M_{\rm WD}$. The consequences for the final MSPs systems will be discussed below.

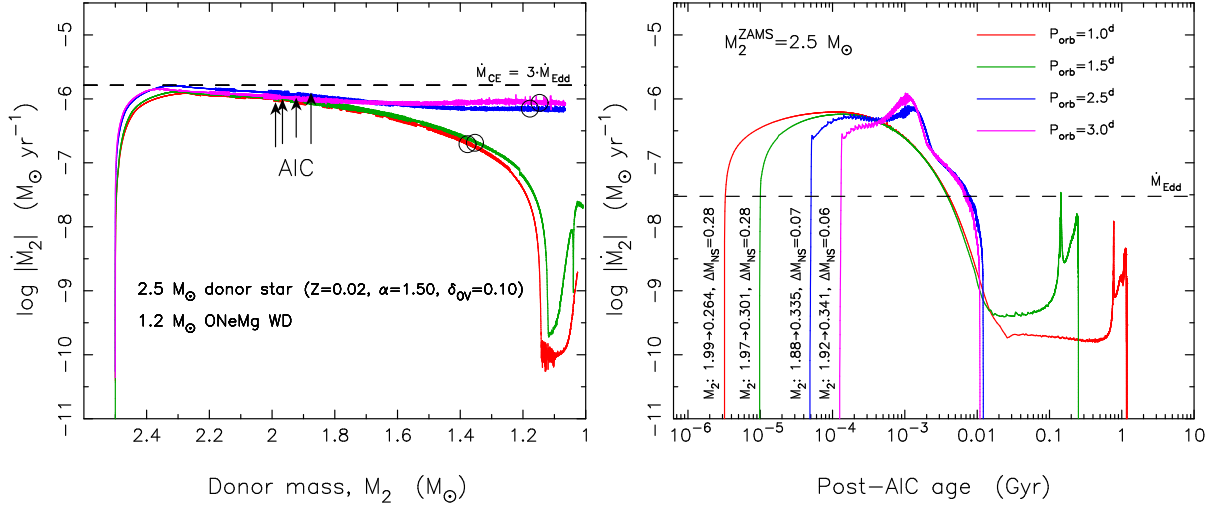

Figure 82: Left panel: Mass-transfer rates of four $2.5\,M_\odot$ donor stars in pre-AIC binaries with initial orbital periods between $1.0 - 3.0$ days leading to the post-AIC LMXB systems shown in the right panel. The arrows mark the collapse of the accreting WD (AIC) and the evolution from this point onwards can be ignored since it is now followed in the post-AIC LMXB systems shown in the right panel. The dashed line is the adopted upper limit for stable mass transfer (see Section 9.3.1.1). It is seen how the pre-AIC mass-transfer rates were quite close to our accepted upper limit. A slightly more massive donor star would lead to excessive mass-transfer rates and thus not result in an AIC event (see also the systematic effect of an increasing donor mass in Fig. 80). The open circles indicate hypothetical super-Chandrasekhar WD masses of $2.0\,M_\odot$. Right panel: Mass-transfer rates of the post-AIC LMXBs plotted as a function of time since the AIC event. The reason for the difference in time intervals between the AIC event and the donor stars refilling their Roche lobe (see initial interceptions of the four graphs with the x-axis) is mainly due to different expansions of the orbit due to the AIC event, caused by different $P_{\rm orb}$ at the moment of the WD collapse. The vertical text at the lower left side of each graph yields the mass of the donor star at the moment of the AIC, the mass of the final WD following the LMXB and the amount of mass accreted by the NS, $\Delta M_{\rm NS}$. The dashed line is the Eddington accretion limit for a NS. Only the systems with initial pre-AIC orbital periods of 1.0 days (red curve) and 1.5 days (green curve) experienced a long phase ($> 100\,{\rm Myr}$) of post-AIC mass-transfer driven by hydrogen shell burning. This phase leads to significant accretion onto the NS (and effective recycling) since $|\dot{M}_2| < \dot{M}_{\rm Edd}$. Hence, the NSs in these systems were able to accrete more mass which results in faster spinning MSPs (partly due to their smaller magnetospheres).

### 9.4.3 Resulting MSPs
#### 9.4.3.1 Final orbital periods
In Fig. 83 we have plotted our resulting binary MSPs in the final ($M_{\rm WD}$, $P_{\rm orb}$)-plane. The upper panel shows the resulting MSPs using a donor star metallicity of $Z = 0.02$. The lower panel is for $Z = 0.001$. For clarity we have not included all systems shown in Fig. 78 which successfully evolved to the AIC, but we have included most systems and made sure to display those that yield the more extreme values of $M_{\rm WD}$ and $P_{\rm orb}$. All the green filled circles were calculated assuming a symmetric AIC (i.e. $w = 0$). The pink open stars represent AIC with a small kick magnitude of $50\,{\rm km\,s^{-1}}$ and in all cases kick angles of $\theta = 0^\circ$ and $\phi = 0^\circ$. The red filled stars represent systems surviving AIC with a large kick $w = 450\,{\rm km\,s^{-1}}$. In these cases the kick angles were always chosen to yield the widest possible post-AIC orbits from a systematic trial procedure. See Table 16 for examples. The black lines connecting three symbols show the final results of three different kick values applied to the same AIC system. One should keep in mind that AICs are most likely not accompanied with a kick (see Section 9.1). However, we include the option here for the sake of completeness.

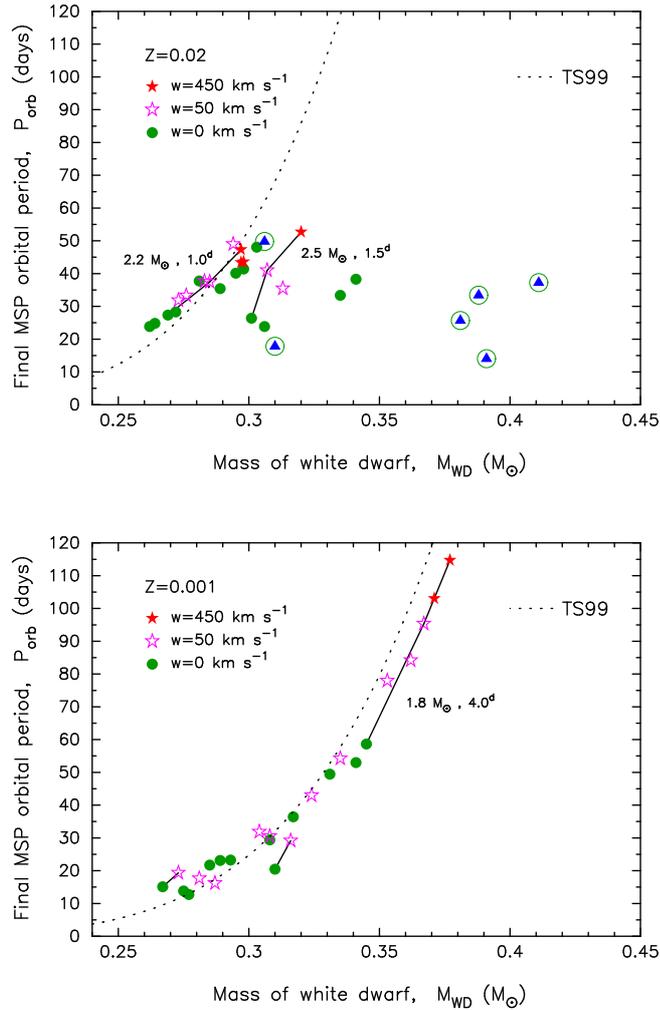

Figure 83: Final orbital period of the MSP binaries formed via AIC as a function of WD mass, evolving from main-sequence donor stars with a metallicity of $Z = 0.02$ (upper panel) and $Z = 0.001$ (lower panel). The different symbols refer to different kicks during the AIC. The filled green circles correspond to $w = 0$ (no kick), the open pink stars to $w = 50\,\mathrm{km\,s^{-1}}$, and the filled red stars are for $w = 450\,\mathrm{km\,s^{-1}}$. The symbols connected with a black line are for the same pre-AIC binaries but different values of the kick, $w$. The dotted line is the $(M_{\mathrm{WD}}, P_{\mathrm{orb}})$–relation taken from Tauris & Savonije (1999) and applies solely to low-mass donors with degenerate He cores. The blue triangles in green circles are explained in Fig. 78. See text for further details and discussions.

From the lower panel of Fig. 83 we note that for donor stars with low metallicity ($Z = 0.001$) all of the final MSP systems fall approximately on the well-known $(M_{\mathrm{WD}}, P_{\mathrm{orb}})$–relation (e.g. Savonije 1987; Tauris & Savonije 1999), shown as a dashed line. This relation follows from the relation between the radius of a giant star and the mass of its degenerate He-core (Refsdal & Weigert 1971). However, the more massive donors ($M_2 > 2.3\ M_\odot$) with non-degenerate cores do not obey this relation. This explains why many of the systems in the upper panel ($Z = 0.02$) deviate from the $(M_{\mathrm{WD}}, P_{\mathrm{orb}})$–relation. Most of these donor stars leave behind relatively massive (hybrid) CO WD remnants, see Section 9.4.3.2.

An interesting outcome of these calculations is that the binary MSPs only form within a limited interval of $P_{\mathrm{orb}}$. We conclude that the final orbital periods of MSPs, formed via AIC with main-sequence donor stars, are in the interval: $10^{\mathrm{d}} < P_{\mathrm{orb}} < 60^{\mathrm{d}}$. Only in the unlikely case where large AIC kicks were applied is it possible to form binary MSPs with $P_{\mathrm{orb}}$ up to 120 days.

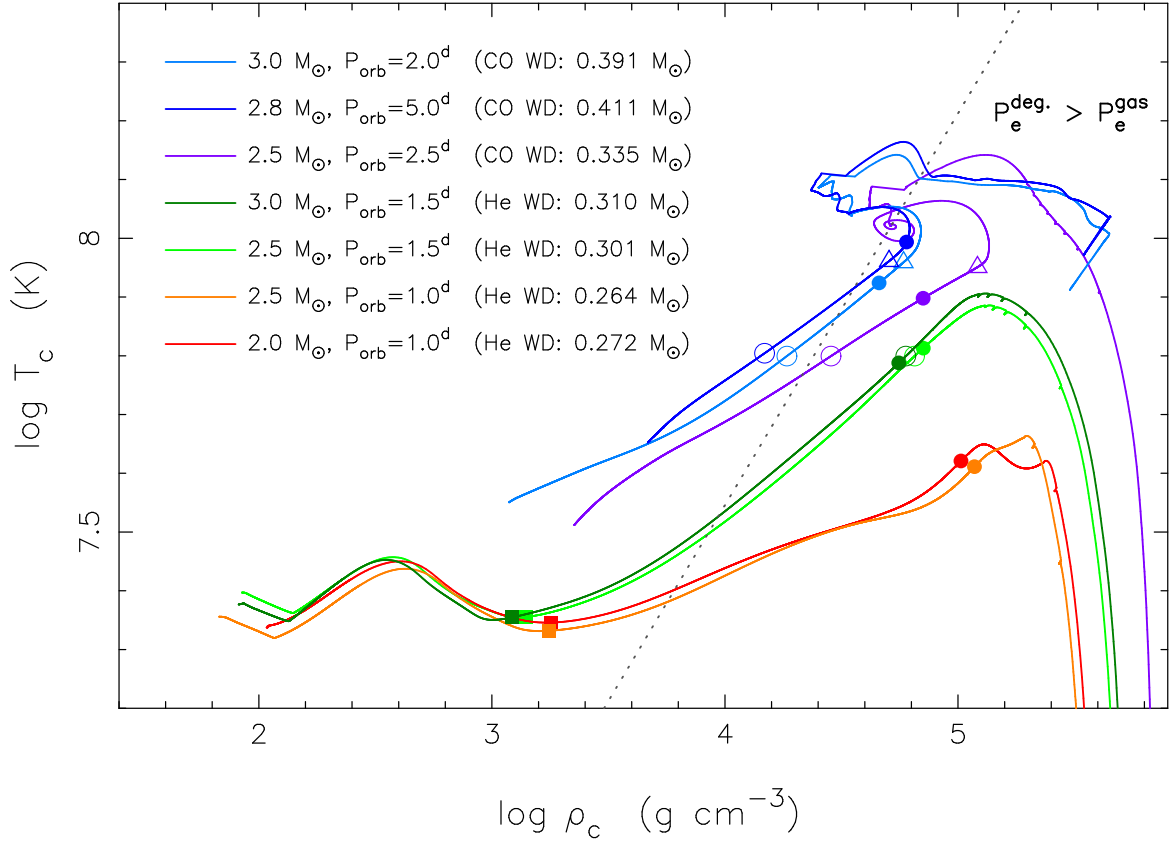

Figure 84: Evolutionary tracks in the core density–core temperature ($\rho_c$, $T_c$)-plane. For each graph the initial (ZAMS) values of the donor star mass, $M_2$, the orbital period, $P_{orb}$, and in parenthesis the main chemical composition and the mass of the final WD are listed. The dotted line separates regions where the stellar pressure is dominated by the gas pressure (left) and the degenerate electron pressure (right). On each track the symbols represent: the termination of core hydrogen burning (filled squares), the end of the RLO (filled circles), the onset of the $3\alpha$–process (open circles) and the onset of efficient helium burning (i.e. when the luminosity generated by the $3\alpha$–process exceeds the energy loss rate by neutrinos, open triangles). The calculations were followed to the WD cooling track, except for two cases causing numerical instabilities during hydrogen shell flashes.

### 9.4.3.2 Nature of the final WD orbiting the MSP

To illustrate which donors leave He WDs and which leave CO WDs, we have plotted in Fig. 84 evolutionary tracks in the ($\rho_c$, $T_c$)-plane of WD progenitor stars with different masses and different values of $P_{orb}$. It is seen that the donors with lower masses and/or shorter initial orbital periods are more exposed to a high pressure of degenerate electrons and a lower core temperature than the more massive donors. Hence, these lighter donors leave He WDs while the latter systems reach higher core temperatures and ignite helium to produce CO WDs. We find that the minimum threshold mass for efficient helium burning and production of a CO WD is about $0.33\ M_\odot$, in agreement with previous work by e.g. Kippenhahn & Weigert (1990); Tauris, van den Heuvel & Savonije (2000); Podsiadlowski, Rappaport & Pfahl (2002). A more correct description of these low-mass CO WDs in Fig. 84 of masses $0.335\ M_\odot$, $0.391\ M_\odot$, and $0.411\ M_\odot$ is hybrid WDs, since they have a composite composition of a CO core surrounded by a thick helium mantle (e.g. Iben & Tutukov 1985; Iben, Tutukov & Yungelson 1997). In these stars, the temperature was never high enough to cause helium burning throughout the outer layers and thus the growth of the CO core was stalled when it reached a mass fraction of $50-75\%$. Each of these WDs has a tiny hydrogen envelope of mass $1.2 - 3.7 \times 10^{-4}\ M_\odot$ and evolved via one or more final hydrogen shell flashes due to unstable CNO burning in a thin hydrogen layer near their surfaces before they settled on the WD cooling track (e.g. Althaus, Serenelli &

Benvenuto 2001; Nelson, Dubeau & MacCannell 2004). As an example, our 0.335 $M_\odot$ hybrid CO-He WD underwent one last, vigorous hydrogen shell flash lasting only 15 yr, which caused it to fill its Roche lobe and transfer $5 \times 10^{-5}$ $M_\odot$ towards the NS with a maximum rate of $1.1 \times 10^{-5}$ $M_\odot\,\mathrm{yr}^{-1}$ (almost $\sim 10^3$ $\dot{M}_{\mathrm{Edd}}$).

### 9.4.3.3   MSP spin periods

Given our calculated amounts of mass accreted by the NSs ($\Delta M_{\mathrm{NS}}$), we can constrain the possible pulsar spin periods after spin-up by using the formula (see Tauris, Langer & Kramer 2012; for details):

$$\Delta M_{\mathrm{NS}} \approx 0.22\,M_\odot\;\frac{(M_{\mathrm{NS}}/M_\odot)^{1/3}}{P_{\mathrm{ms}}^{4/3}} \tag{90}$$

or

$$P_{\mathrm{ms}} \approx 0.34\,(\Delta M_{\mathrm{NS}}/M_\odot)^{-3/4}, \tag{91}$$

where $P_{\mathrm{ms}}$ is the final equilibrium spin period in milliseconds, and $M_{\mathrm{NS}}$ is the initial NS mass (here, always 1.28 $M_\odot$ following the AIC). Obviously, it is not possible to spin up a NS to faster rotation than its break-up spin period at $\sim 0.6$ ms, and our values do not include the circumstance of being limited by either gravitational wave radiation or a relatively large magnetospheric radius of the pulsar. Similarly, we discard the potential possibility of initially preventing post-AIC accretion onto the young energetic pulsar as a consequence of the so-called radio ejection mechanism (i.e. ejection of material from the system caused by the outward magnetodipole radiation pressure exceeding the inward ram pressure of material at the first Lagrangian point; Burderi et al. 2001). In Table 16 we list $\Delta M_{\mathrm{NS}}$ and $P_{\mathrm{ms}}$ for our calculated post-AIC LMXB systems. It is seen that in almost all systems with main-sequence donor stars we obtain $\Delta M_{\mathrm{NS}} > 0.1\,M_\odot$ and therefore we find that these MSPs will be fully recycled.
*We conclude that in all our models where MSPs formed via AIC, the predicted equilibrium spin periods of a few ms are identical to those expected for MSPs formed via the conventional recycling channel where the NS was formed in a SN Ib/c.*

### 9.4.3.4   MSP systemic space velocities

We kept track of the post-AIC systemic velocities relative to the centre-of-mass rest frame of the pre-AIC binaries. The systems receive a recoil due to the sudden mass loss, possibly combined with a smaller kick, during the AIC. From conservation of momentum we obtain (e.g. following Tauris & Bailes 1996):

$$v_{\mathrm{cm}} = \sqrt{(\Delta P_x)^2 + (\Delta P_y)^2 + (\Delta P_z)^2}\,/\,M, \tag{92}$$

where the change in momentum is given by (cf. Section 9.3.2.1):

$$\begin{aligned}
\Delta P_x &= M_{\mathrm{NS}}\,w\cos\theta - \Delta M M_2 \sqrt{G/(r M_0)} \\
\Delta P_y &= M_{\mathrm{NS}}\,w\sin\theta\cos\phi \\
\Delta P_z &= M_{\mathrm{NS}}\,w\sin\theta\sin\phi.
\end{aligned} \tag{93}$$

In Table 16 we also list the calculated systemic velocities of post-AIC binaries. If large kicks (e.g. $w = 450\ \mathrm{km\,s^{-1}}$) were associated with AIC then the MSPs formed via this channel would reach velocities of up to 200 km s$^{-1}$, similar to the calculated systemic velocities of MSPs where the NS formed via a core-collapse SN (Tauris & Bailes 1996). However, for $w = 0 - 50\ \mathrm{km\,s^{-1}}$ we find that the expected velocities of the resultant MPS formed via AIC will be quite small and of the order of $10 - 30$ km s$^{-1}$ at maximum[34].

---

[34] In Table 16 we always chose $\theta = 0°$ for applied kicks of $w = 50\ \mathrm{km\,s^{-1}}$ which resulted in the largest possible post-SN values of $P_{\mathrm{orb}}$, but also in the smallest values of $v_{\mathrm{sys}}$. For $\theta = 180°$ we obtain typical values of $v_{\mathrm{sys}} \approx 30$ km s$^{-1}$, for $w = 50\ \mathrm{km\,s^{-1}}$.

## 9.5 AIC in systems with giant star donors

In Fig. 85 we have plotted a grid of the initial orbital periods and donor star masses of our investigated systems with giant star donors. These systems could potentially be detected as novae-like, symbiotic X-ray sources while the WD accretes material. The giant star donors which successfully lead to AIC have masses between $0.9 - 1.1\ M_\odot$. The upper limit is set by the mass-transfer rate (which is too high for larger values of $M_2$), and the lower limit is set by the age of our Universe ($0.9\ M_\odot$ donors only evolve into giant stars within a Hubble-timescale of $13.7$ Gyr if the metallicity is low enough). The initial parameter space dependence on metallicity is in general much weaker for these giant stars compared to the case of main-sequence donor stars shown in Fig. 78. The orbital periods, $P_{\rm orb}$ at the onset of the RLO must be at least $60 - 80$ days. If $P_{\rm orb}$ is smaller than this value the mass-transfer rate is too weak for the WD to grow efficiently in mass. The reason for different outcomes of $1.1\ M_\odot$ donor star models ($Z = 0.02$) with $P_{\rm orb} = 150^{\rm d} - 190^{\rm d}$ is that $\dot{M}_2$ is very close to, and fluctuating near, our hard threshold limit, $\dot{M}_{\rm CE}$.

### 9.5.1 Mass-transfer from giant star donors

The mass-transfer modelling with giant star donors is difficult to calculate for two reasons. First, giant stars have low surface gravities and thus extended atmospheres which make it non-trivial to estimate the turn-on of the RLO mass-transfer process. In some cases (see full discussion in Section 9.5.3) there is a significant amount of mass transfer through the inner Lagrangian point while the donor star is still underfilling its Roche lobe (the so-called optically thin mass transfer). Hence, for donor stars in the widest pre-AIC orbits ($P_{\rm orb} \geq 400^{\rm d}$) our code runs into problems, in particular for low-metallicity giant donors. Second, even low-mass giant stars have substantial wind mass loss (Reimers 1975) which causes the orbits to widen prior to the RLO. In some cases $P_{\rm orb}$ may increase by 20% prior to the RLO while the donor star loses up to 30% of its mass. In our modelling, we have neglected the wind mass loss of the giant prior to RLO in order to isolate and better investigate the above-mentioned effect of optically thin mass transfer from these donor stars with extended atmospheres.

As can be seen in Fig. 85 for giant star donors, the transition from low values of $|\dot{M}_2|$ to donors which result in excessive values of $|\dot{M}_2|$ is very narrow for donor star masses, $M_2 > 1.1\ M_\odot$. Again the explanation is the sudden set-in of rapid growth of the thickness of the convective envelope. Including the effect of stellar winds is expected to shift the green points (successful AIC) in Fig. 85 slightly to the left and upward.

### 9.5.2 Resulting MSPs

The binary pulsars formed via AIC from giant star donors all end up in wide systems with $P_{\rm orb} > 500^{\rm d}$. The WD companions are either $0.40 - 0.46\ M_\odot$ He WDs, or CO WDs more massive than $0.46\ M_\odot$. The pulsars are, in general, expected to be only mildly recycled with estimated spin periods of $10 - 500$ ms due to the short mass-transfer phase, lasting $\Delta t_{\rm LMXB} \simeq 10^4 - 10^6$ yr, following the AIC. Given this short phase of post-AIC mass transfer, combined with the effects of a young energetic pulsar (Section 9.4.3.3), the final spin periods are expected to be somewhat slower than indicated above and in Table 16. More notably, the systemic velocities of these binaries will be very small, $v_{\rm sys} \sim 1\ {\rm km\,s^{-1}}$, given the general assumption that AICs are not accompanied by a momentum kick. Hence, these will be at rest with respect to the local stellar population.

By the time the low-mass giant star donors initiate mass transfer towards the WD, they have already reached ages of the order of 10 Gyr (see Table 16). Therefore, we expect ongoing formation of NSs via the AIC channel in GCs today. In order for the giant star donors to deliver high enough mass-transfer rates to make the accreting WD grow sufficiently in mass, the initial $P_{\rm orb}$ must be large. Hence, the orbital period at the moment of the AIC and the subsequent post-AIC LMXB orbit are also quite large. Such wide binaries enhance the likelihood of disruption by stellar encounters in a GC (e.g., Verbunt & Freire 2014) and thus result in a number of isolated, young NSs in GCs as observed (or young NSs in close orbits formed after a three body exchange event), see Table 15.

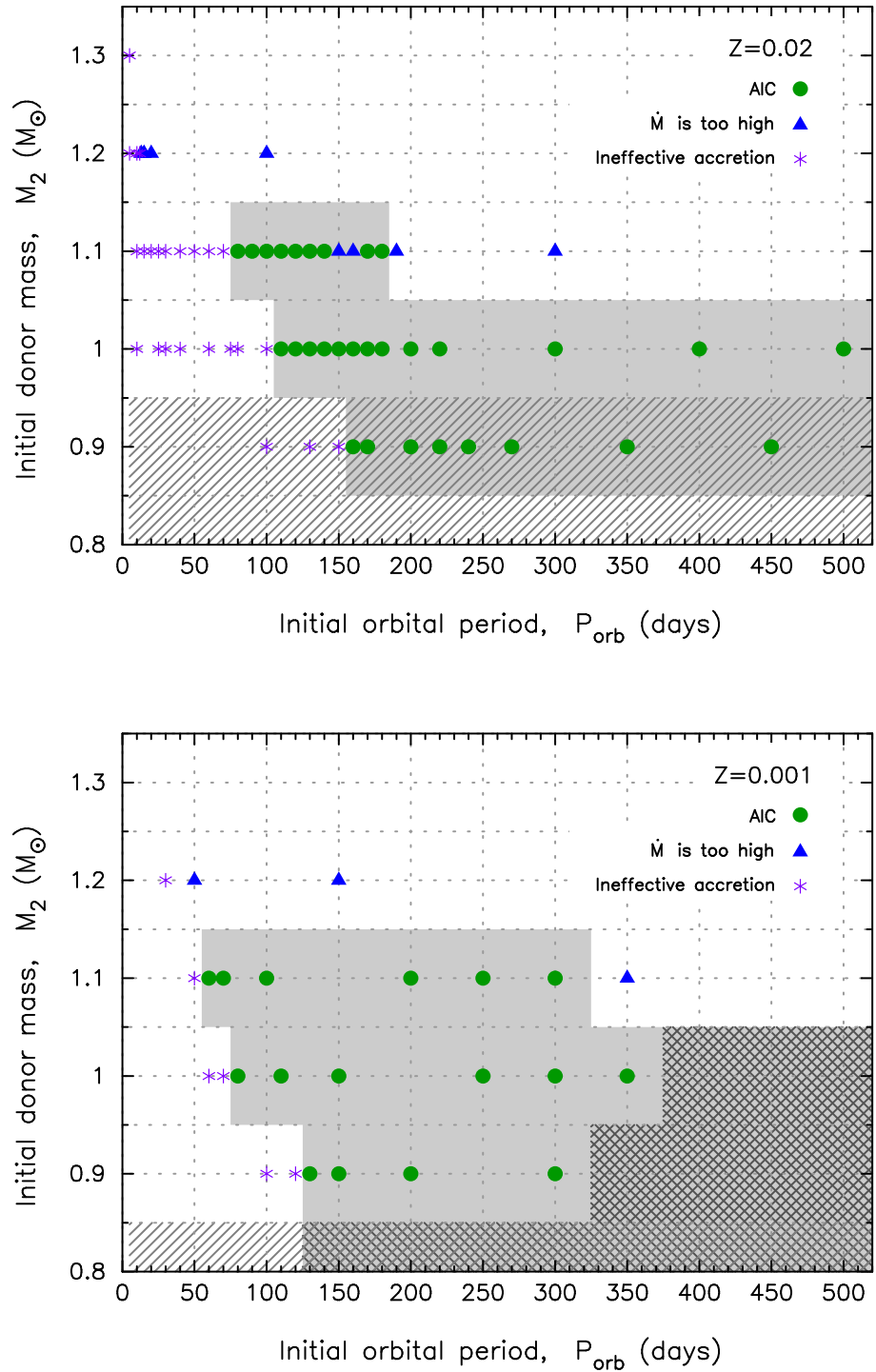

Figure 85: The grid of investigated initial orbital periods and masses for giant star donors with a metallicity of $Z = 0.02$ (top) and $Z = 0.001$ (bottom). The meaning of the various symbols is equivalent to those in Fig. 78. The grey shaded region in each panel corresponds to systems which successfully evolve to the AIC stage. Only giant donors with masses $0.9 \leq M_2/M_\odot \leq 1.1$ are able to produce MSPs via the AIC channel. The Hatched regions indicate donor stars which evolve on a timescale longer than the Hubble-timescale. The cross-hatched region corresponds to giant donors which have so large atmospheric, hydrogen pressure scale-heights that our adopted mass-transfer scheme breaks down – see text.

*We conclude that AIC is an attractive model to explain the small space velocities of some NSs and, in particular, the retention of NSs in GCs. Furthermore, the AIC channel can explain the existence of relatively young NSs in GCs.*

### 9.5.3 Break-down of the applied mass-transfer scheme for giant donors

According to the canonical criterion for mass-transfer (Kippenhahn & Weigert 1967) one simply has to ask whether the donor star radius, $R_2$ is larger or smaller than than its Roche-lobe radius, $R_{\mathrm{L}}$. Hence, one makes the implicit assumption that the edge of the star is infinitely sharp and therefore that the mass transfer starts/ends abruptly rather than following a gradual transition. Ritter (1988) improved this criterion by taking the finite scale height of the stellar atmosphere properly into account. The mass loss from the donor star was modelled as a stationary isothermal, subsonic flow of gas which reaches sound velocity near the nozzle at the first Lagrangian point, $L_1$. The accompanying mass-transfer rate is given by:

$$|\dot{M}_2| = \rho_{\mathrm{L1}}\, v_{\mathrm{s}}\, Q \; \simeq \; \frac{1}{\sqrt{e}}\rho_{\mathrm{ph}}\, v_{\mathrm{s}}\, Q \exp\left(-\frac{\Delta R}{H_{\mathrm{p}}}\right), \qquad (94)$$

where $\rho_{\mathrm{ph}}$ is the gas density at the donors' photosphere, $v_{\mathrm{s}} = \sqrt{kT/(\mu m_{\mathrm{H}})}$ the isothermal sound speed, $Q$ the effective cross section of the flow at $L_1$ (see Meyer & Meyer-Hofmeister 1983), and $\Delta R \equiv R_{\mathrm{L}} - R_2$. The last parameter,

$$H_{\mathrm{p}} = \frac{kT R_2^2}{\mu m_{\mathrm{H}} G M_2}, \qquad (95)$$

is the pressure scale-height of the stellar atmosphere ($\mu$ is the mean molecular weight). This scheme was developed to study the turn-on (turn-off) of mass transfer in nearly semi-detached systems, the so-called optically thin mass transfer for which $R_2 < R_{\mathrm{L}}$ (see also D'Antona, Mazzitelli & Ritter 1989; Kolb & Ritter 1990)[35]. However, this mass-transfer algorithm was derived for low-mass main-sequence donor stars in CV binaries for which $H_{\mathrm{p}} \ll R_2$, and therefore mass transfer was only assumed to occur for $\Delta R \ll R_2$.

For giant stars this picture has to change (Pastetter & Ritter 1989). These stars with low surface gravities often have $H_{\mathrm{p}}/R_2 \simeq 0.04$ and as a result we find that they can in some cases cause mass-transfer rates above $10^{-7}\,M_\odot\,\mathrm{yr}^{-1}$ even for $\Delta R = 0.3\,R_2$ (i.e. while the donor star is still underfilling its Roche lobe by 23% in radius). Hence, for giant star donors the assumptions behind the original Ritter scheme breaks down. As a consequence, we did not allow for mass transfer with $\Delta R > 0.3\,R_2$ and our calculations were abandoned if this limit was reached (see cross-hatched region in Fig. 85).

As mentioned previously, we did not include wind mass loss in our models. This effect would cause the binaries to widen further and thereby stabilize the systems against dynamical unstable mass transfer (Pastetter & Ritter 1989). Another uncertainty is the effect of irradiation feedback on the long-term evolution of a compact binary (e.g. Büning & Ritter 2004). However, the impact and the modelling of this effect, leading to cyclic accretion, is still unclear and is not included in our study. Recent work by Benvenuto, De Vito & Horvath (2012) on the evolution of ultra-compact X-ray binaries suggests that the inclusion of irradiation feedback is not very significant for the final properties of these systems. Furthermore, for the wide-orbit LMXBs with giant donors Ritter (2008) argued that irradiation-driven mass-transfer cycles cannot occur since these systems are transient because of disc instabilities.

---

[35]We note that $\Delta R$ and the mass ratio, $q$ are defined differently in Ritter (1988) and Kolb & Ritter (1990). The first paper has a typo in the last term in Eq. (A8), which should be $f_2^{-3}(q)$.

### 9.6    AIC in systems with helium star donors

#### 9.6.1    Mass transfer from helium star donors

Binary stars evolving from the ZAMS may lead to the formation of a tight binary with a massive WD, i.e. the remnant of the former primary star, and a helium star, i.e. the core of the secondary star that lost its envelope via CE evolution (e.g. Liu et al. 2010). These systems will thereafter proceed their post-CE evolution as described below, once the helium star fills its Roche-lobe (via so-called Case BB RLO).

We constructed helium star donors with $Z = 0.02$ (1.5% $^{14}N$, 0.2% $^{20}Ne$, and 0.3% isotopes of mainly Mg, Si, C, and O). As seen in Fig. 86, the WDs that successfully accreted mass up to the AIC limit (1.48 $M_\odot$) had helium star donors with masses between $1.1 - 1.5$ $M_\odot$ and initial $P_{orb}$ between 1 hr (0.04 days) and 1.2 days, see Table 16 for details of 8 models. The mass-transfer rates become higher than the critical limit ($\dot{M}_{CE}$) for helium stars in relatively wide orbits. However, for the lightest helium stars ($\leq 1.1$ $M_\odot$) this is not the case. Even in the widest orbits the mass-transfer rates become too low to yield significant WD accretion because the orbits expand further in response to mass-transfer, given that $q < 1$ at all times.

In the HR–diagram in Fig. 87 we compare the evolution of two 1.1 $M_\odot$ helium stars in binaries, with initial $P_{orb} = 0.3^d$ and $1.3^d$, respectively, with the evolution of an isolated helium star of the same mass. Whereas the single helium star evolves to a radius above 40 $R_\odot$ during shell helium burning, the binary helium stars which suffer from mass loss (and restriction in size due to their Roche lobes) only evolve up to a radius of maximum $\sim 2.6$ $R_\odot$ (for an initial $P_{orb} = 1.3^d$) before they evolve towards the WD cooling track as $0.80 - 0.82$ $M_\odot$ CO WDs.

We note that our models are computed with OPAL opacities which include carbon and oxygen abundances (cf. Section 9.3). Models computed with OPAL opacities (Iglesias, Rogers & Wilson 1992) that do not account for high C/O abundances (not shown here) resulted in final CO WD masses that were 4–9% larger than those in the models presented here.

#### 9.6.2    Resulting MSPs

It can be seen in Table 16 that the post-AIC LMXB phase, $\Delta t_{LMXB}$ lasts for less than about 1 Myr (typically only $10^5$ yr), except for one model (He1) where the RLO was initiated early on the helium star main sequence, resulting in $\Delta t_{LMXB} = 7.96$ Myr. Given the short lasting LMXB phase for these systems the final pulsar masses are all close to their original post-AIC mass of 1.28 $M_\odot$. For the same reason, these pulsars are only mildly recycled with minimum initial spin periods between $20 - 100$ ms and we expect their B-fields to be larger than those of the fully recycled MSPs. (We note that the spin periods could be slower than stated in Table 16 for the reasons given in Section 9.4.3.3.)

The pulsar companions are all CO WDs with masses between $0.6 - 0.9$ $M_\odot$ and $P_{orb} \simeq 2$ hr – 2 days, see Fig. 88. If a hypothetical large kick ($w > 450$ km s$^{-1}$) is applied the final $P_{orb}$ may reach 3 days. However, such large kicks are not expected for AIC (cf. Section 9.2).

The tight binaries with initial helium star companions cause the post-AIC systemic velocities to reach about $20 - 30$ km s$^{-1}$ (in models with no kicks) which is substantially faster than the recoil velocities imparted to the wider systems with main-sequence or giant star companions, although much smaller than the velocities calculated for systems evolving via the standard SN channel.

### 9.7    Discussion

Using a detailed stellar evolution code to probe the combined pre- and post-AIC evolution, we have demonstrated that it is possible to form MSPs indirectly via the AIC channel. In Fig. 89 we show the initial parameter space of all binaries which successfully evolve to AIC according to our calculations. In this section, we discuss our findings in more detail in relation to the predicted physical properties of the resultant NS binaries and compare them with observations.

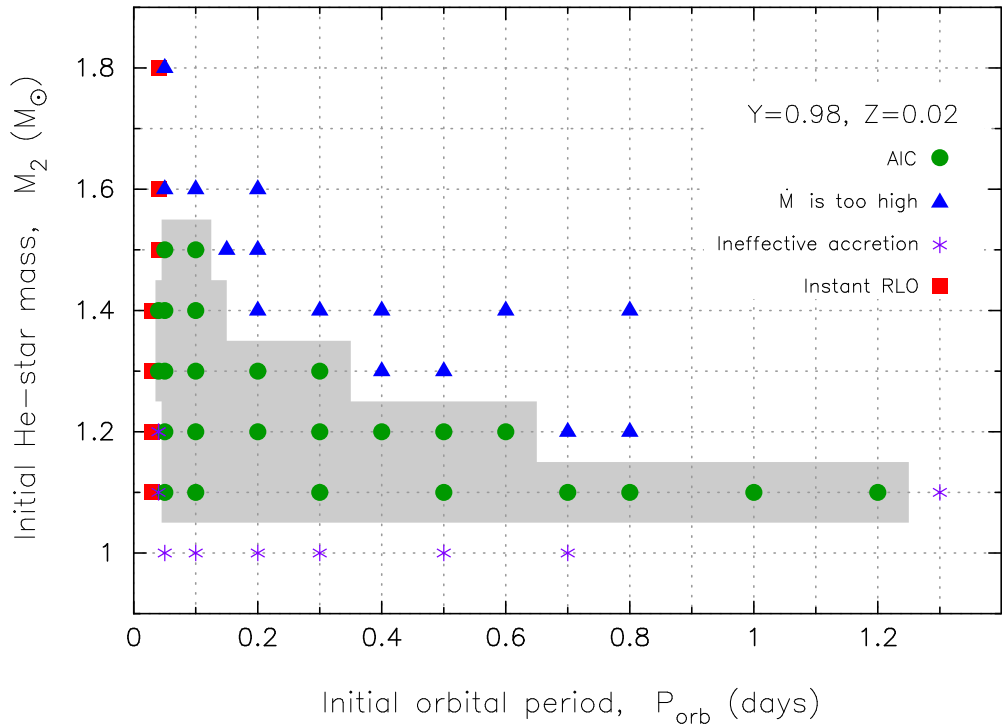

Figure 86: The grid of investigated initial orbital periods and masses for helium star donors with $Y = 0.98$ and a metallicity of $Z = 0.02$. The meaning of the various symbols is explained in Fig 78. The grey shaded region corresponds to systems that have successfully evolved to the AIC stage. See text for further discussions.

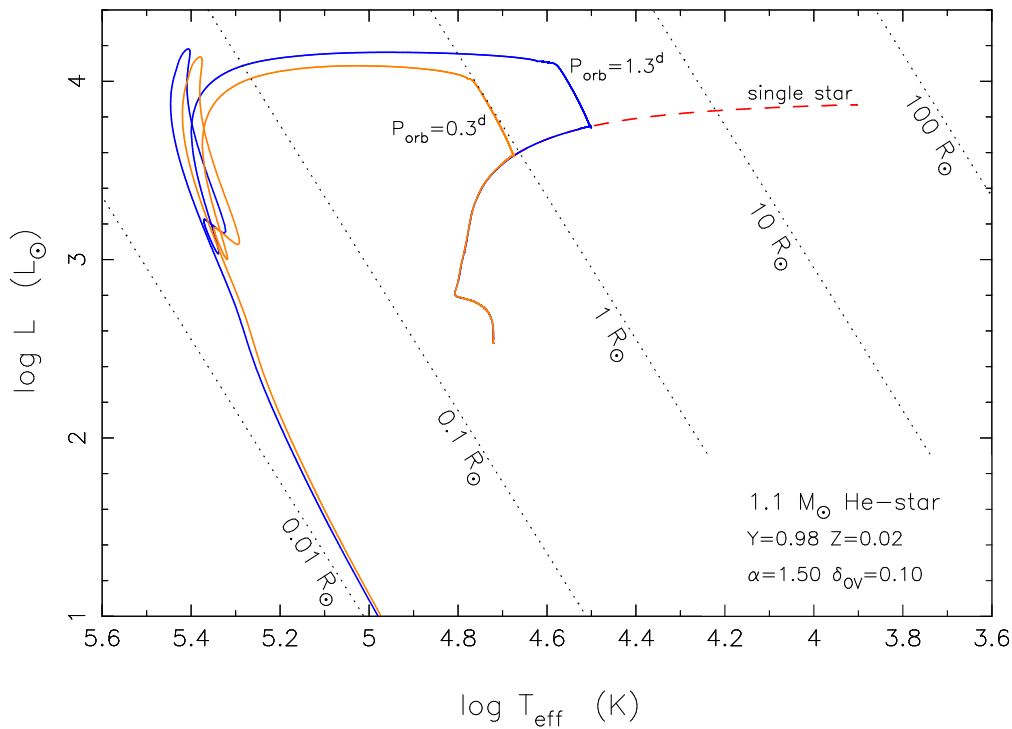

Figure 87: Evolutionary tracks in the HR–diagram for three 1.1 $M_\odot$ helium stars: two in binaries with a 1.2 $M_\odot$ WD and one evolving as a single star. The loops in the upper-left corner of the two binary helium star tracks are caused by the ignition of shell helium burning.

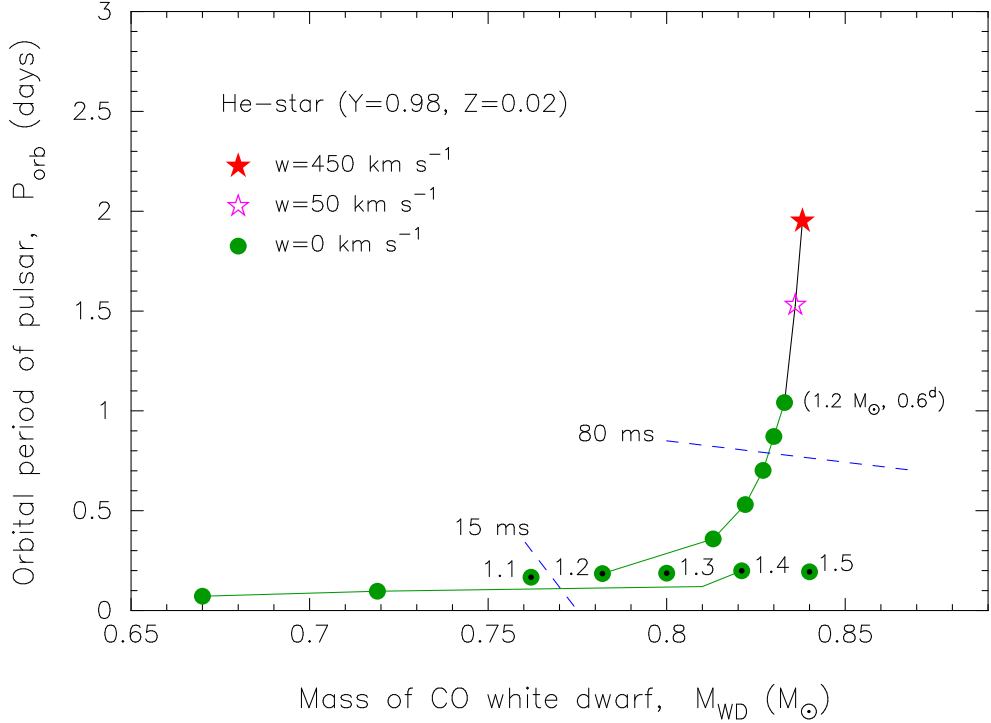

Figure 88: Final orbital periods of recycled radio pulsars as a function of their CO WD companion masses for systems evolving from helium star donors with a metallicity of $Z = 0.02$. Data points connected with a green line were calculated from the same initial helium star mass ($1.2\ M_\odot$ and $1.4\ M_\odot$, respectively). Data points with black dots were all calculated from a binary with an initial orbital period of 0.1 days. The solid red star and the open purple star, connected with a black line to the green data points, indicate similar calculations including an applied hypothetical kick velocity of $50\ \mathrm{km\,s^{-1}}$ and $450\ \mathrm{km\,s^{-1}}$, yielding the widest possible orbit. The less massive the initial helium star donor, and the more narrow the orbit, the more material the NS accretes and the faster spin it obtains. The dashed lines indicate roughly where the pulsar is spun up to 15 ms and 80 ms, respectively. See text for discussions.

We also discuss the connection between AIC and SNe Ia progenitors and compare our work to other recent studies.

### 9.7.1 Fully recycled MSPs or young NSs?

A general problem with postulating that a given observed high B-field NS was formed via AIC is that (as pointed out by Wijers 1997; and also demonstrated by Sutantyo & Li (2000)) it requires quite some finetuning to have the AIC occurring at the very final phases of the mass transfer in order to explain the high B-field of the NS. If the WD collapses earlier, the B-field of the newly formed NS (and its spin period, depending on $\dot{M}_2$) should decrease significantly when the donor star refills its Roche lobe following the AIC, thereby evolving through a relatively long lasting ($10^7 - 10^9$ yr) post-AIC LMXB phase. In that case, all traces of its origin will be erased and the final NS cannot be distinguished from those recycled pulsars formed via the standard SN channel. Even accretion of a few $0.01\ M_\odot$ is enough to decrease the B-field significantly according to some models (e.g. Wijers 1997; Zhang & Kojima 2006; and references therein). In all our model calculations the newborn NS undergoes post-AIC accretion, although in some extreme cases less than $10^{-3}\ M_\odot$ is accreted.

#### 9.7.1.1 The AIC Corbet diagram

In Fig. 90 we have plotted the Corbet diagram for a sample of our modelled NS systems which formed via the AIC channel (including the models listed in Table 16). For systems with main-sequence donor stars, the resultant NSs eventually become fully recycled MSPs (i.e. with spin periods of a few ms and most likely weak B-fields). Their orbital periods are confined to the interval: $10^d < P_{orb} < 60^d$ (up to 120 days when kicks are applied, cf. Section 9.4.3.1). In Section 9.7.5.1 we compare with the observed distribution of binary pulsars with He WD companions. For systems with either giant star or helium star donors, the spread in predicted final spin periods is much broader: from 4 ms to 0.5 sec. However, the orbital periods are constrained to values of $P_{orb} > 500^d$ and $P_{orb} \leq 2^d$, respectively.

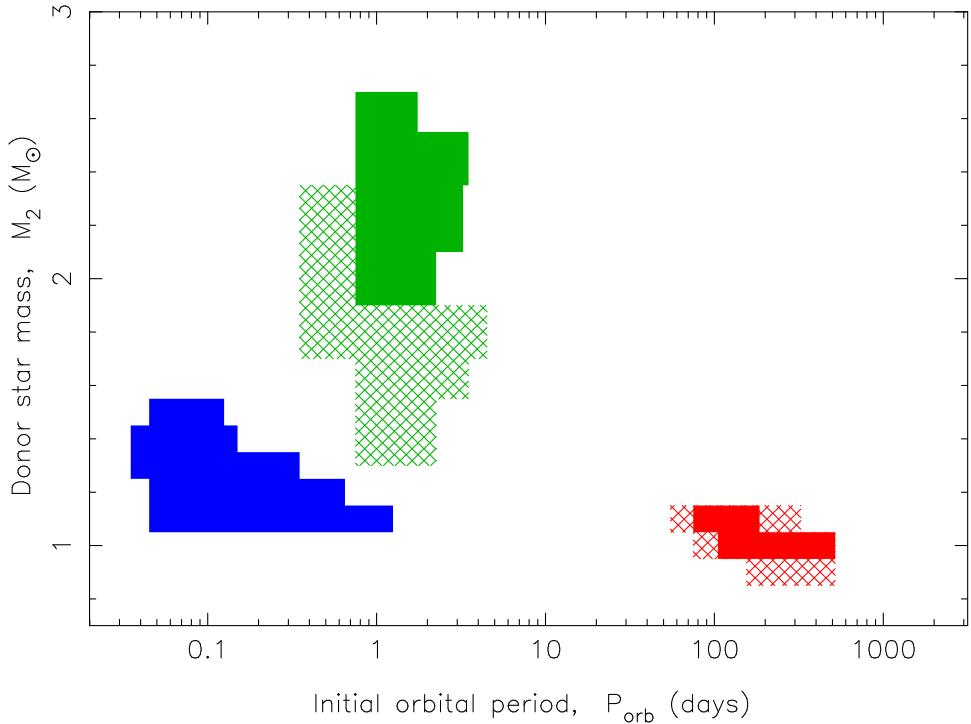

Figure 89: Grid showing all successful AIC systems, starting from an accreting 1.2 $M_\odot$ (ONeMg) WD orbiting a main-sequence star (green), a giant star (red), or a helium star (blue) with a metallicity of $Z = 0.02$. Hatched areas are calculated for $Z = 0.001$. The initial parameter space shown here is equivalent to that for progenitor binaries of SN Ia with an accreting 1.2 $M_\odot$ (CO) WD (see text).

The predictions with respect to the values of the final spin periods are uncertain for systems with either giant star or helium star donors, since in both of these cases the post-AIC LMXB phase is short-lived. On the one hand, if the NS in these systems only accretes $10^{-3} - 10^{-2}$ $M_\odot$ after its formation, we would expect it to form a mildly recycled pulsar with $P \simeq 4 - 500$ ms as shown in Fig. 90. On the other hand, in case the AIC produces (initially) very rapidly spinning and strong B-field pulsars, we cannot rule out that the radio ejection (Section 9.4.3.3) or the propeller mechanism will prevent these NSs from accreting matter during the short ($10^5 - 10^6$ yr) post-AIC LMXB phase. In that case, for these particular systems (red and blue symbols), the amount of material accreted by the post-AIC NS, $\Delta M_{NS}$ could be substantially smaller (and the resulting spin periods longer) than modelled here (see Fig. 91 for our modelled values for $\Delta M_{NS}$). Therefore, it might be possible to produce NSs via the AIC channel which retain relatively high B-fields and which therefore also appear young (similar to what is observed, cf. Table 15). To fully answer this question, one needs to model the accretion disk–magnetosphere interactions in greater detail combined with a decisive model for the decay of the NS B-field.

Nevertheless, if the AIC formation channel is significant in terms of numbers (Hurley et al. 2010; conclude that the AIC route is just as important as the standard SN route), then one may expect to observe a few cases where the AIC happened near the very end of the mass-transfer epoch, resulting in a young, high B-field pulsar.

The orbital period distribution in Fig. 90 is interesting. The resultant MSPs created via the AIC channel preferentially form in certain orbital period intervals. The fully recycled MSPs mainly form with $10^d < P_{orb} < 60^d$ (from systems with main-sequence donor stars), whereas mildly recycled pulsars form in either very wide orbit systems with $P_{orb} > 500^d$ (from systems with giant star donors) or end up in close binaries with $P_{orb} < 2.0^d$ (all with massive CO WD companions, left behind by helium stars donors). The tightest of these latter systems will merge within 1 Gyr. Hence, according to our modelling there is a gap in orbital periods roughly between $60^d < P_{orb} < 500^d$ where MSPs from AIC systems should not form (unless a kick is applied, in which case the lower limit increases to $100^d$). The reason why larger values of $P_{orb}$ are not possible for systems with main-sequence star donors is that the mass-transfer rate becomes too high for systems which could potentially widen to such large orbital periods. Only systems with low-mass giant star donors may produce MSPs in wider systems. However, the final values of $P_{orb}$ for these systems exceed 500 days. In Section 9.7.5 we compare this modelled $P_{orb}$ distribution with observed NS binaries.

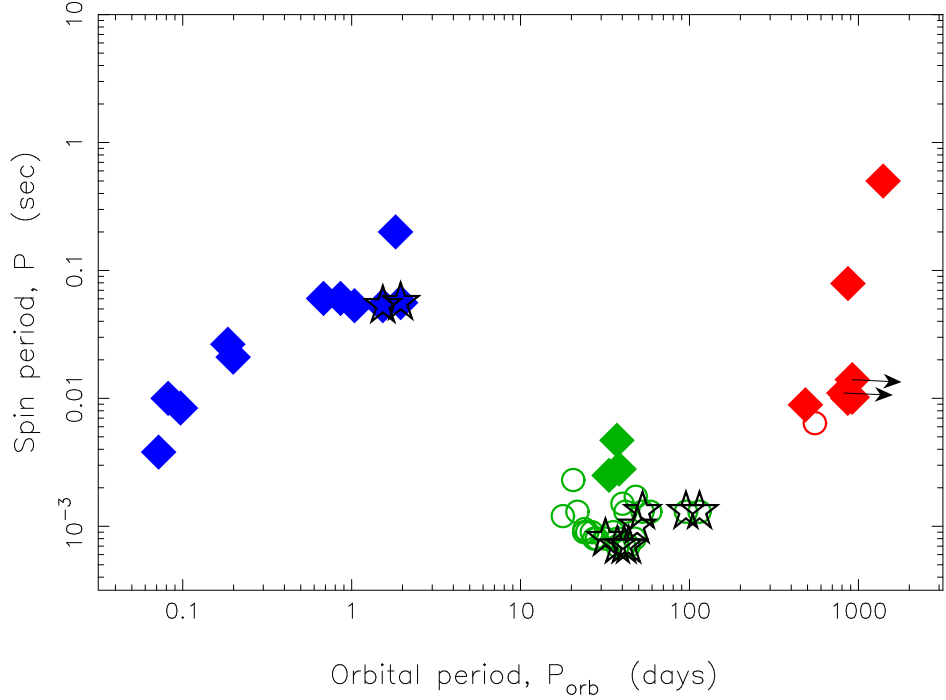

Figure 90: The Corbet diagram for 55 MSPs (or mildly recycled pulsars) produced via AIC formation and subsequent accretion, including those listed in Table 16. The three different donor star progenitor classes (green, red, and blue) are clearly distinguished, corresponding to main-sequence, giant star, and helium star donors, respectively. Open circles are MSPs with He WD companions, solid diamonds indicate MSPs with (hybrid) CO WD companions. Symbols with a superimposed open black star correspond to AIC models where a kick was applied. For a few systems with giant star donors only upper limits are given for $P_{orb}$ (and $P$) because of computational difficulties, cf. Section 9.5.3.

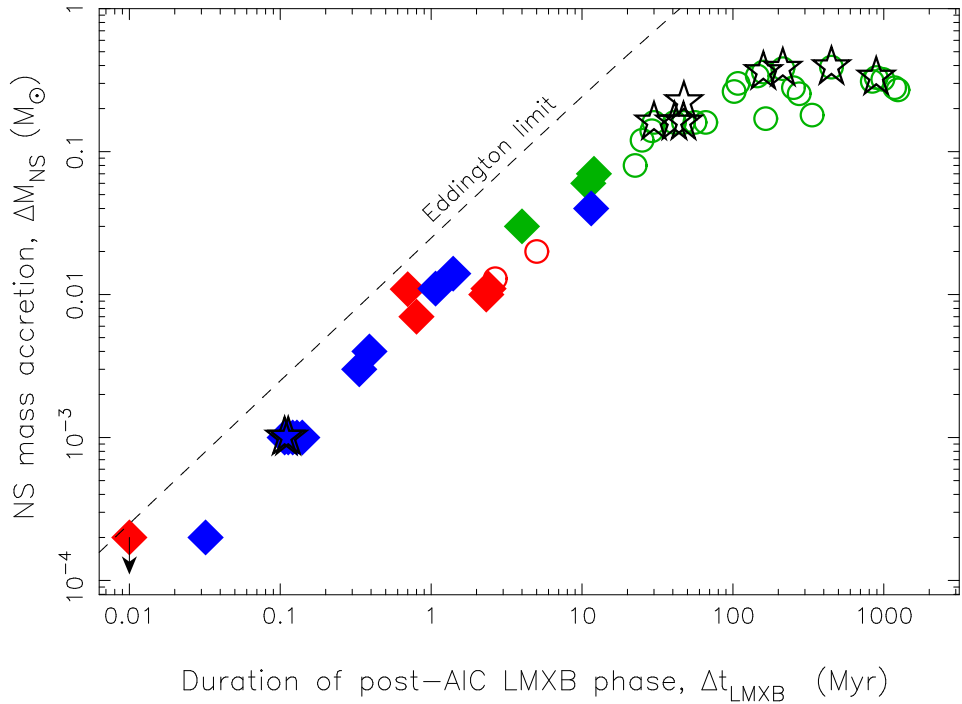

Figure 91: The amount of mass accreted by the NSs during the post-AIC LMXB phase, $\Delta M_{NS}$ as a function of the duration of the post-AIC LMXB phase, $\Delta t_{LMXB}$. The symbols correspond to those in Fig. 90. Our calculations assumed an accretion efficiency of 30%. The deviation from a straight line is caused by the lower mass-transfer rates $|\dot{M}_2| \ll \dot{M}_{Edd}$ in systems with long mass-transfer timescales. The more mass a NS accretes, the faster is the final spin rate of the resultant MSP (see text).

### 9.7.1.2 Production of young NSs in GCs

In GCs, the possibilities are more favourable for producing young NSs which (initially) avoid recycling. The reason is that some post-AIC binaries in GCs may become disrupted by an encounter event before the newborn NS experiences much accretion. We note that for these systems the duration of the post-AIC detachment phase is comparable to the duration of the subsequent LMXB phase (compare $\Delta t_{detach}$ and $\Delta t_{LMXB}$ in Table 16). Considering that GCs have an old stellar population, young NSs can only be produced from AIC with a giant star companion, since all possible main-sequence and helium star donors leading to AIC have nuclear evolution timescales much shorter than the age of the GCs (and thus these stars would already have evolved a long time ago). Therefore, AIC events in GCs only occur in wide binaries which also enhances the probability of disruption before the post-AIC RLO has terminated, thereby strengthening the case for observing a young NS in a GC.

On a much longer timescale it is, of course, quite likely that these isolated NSs may capture a companion star and become recycled. The LMXB IGR J17480−2446 in Terzan 5 might be an example of a NS formed via AIC that has captured a new companion star and is now undergoing recycling, as suggested by Patruno et al. (2012).

### 9.7.2 AIC vs SN Ia: implosion or explosion

All calculations in this work assumed an initial binary configuration with an accreting 1.2 $M_\odot$ WD, treated as a point mass. We implicitly assumed that when the (ONeMg) WD reaches the Chandrasekhar mass at 1.48 $M_\odot$ (the maximum mass of a rigidly rotating WD; e.g. Yoon & Langer 2005) it collapses and forms a NS. However, predicting the final fate of accreting massive WDs is not trivial (e.g. Nomoto & Kondo 1991). Accreting ONeMg WDs reaching the Chandrasekhar limit may not always lead to an implosion that leaves behind a NS. The outcome depends on whether or not the effects of electron captures dominate over nuclear burning (oxygen deflagration) and hence it is sensitive to the central density where explosive nuclear burning is ignited (Nomoto et al. 1979). If the central density is too low, the timescale of electron captures is too long compared to that of the nuclear energy release and the result may be explosive oxygen ignition and a SN Ia.

Similarly, it may not always be the case that accreting CO WDs explode in SNe Ia when reaching the Chandrasekhar limit. This requires that the interior temperature (affected by heat inflow from surface layers) is high enough to ignite carbon at a relatively low density. Therefore, the outcome depends on the duration of accretion phase and thus on the initial mass of the CO WD. If the initial CO WD mass is relatively high ($\geq 1.2\ M_\odot$), then the accretion phase leading to the Chandrasekhar limit is short and thus it may not result in a temperature high enough for such a (high-density) WD to explode in a SN Ia (Nomoto & Kondo 1991).

Nonetheless, using our stellar evolution code we find that WDs formed with an initial mass of 1.2 $M_\odot$ in close binaries are more likely to be ONeMg WDs (or at least hybrid WDs with ONeMg cores embedded in a thick CO mantle) compared to CO WDs.

#### 9.7.2.1 SN Ia progenitor calculations and shell impact

Assuming instead that our point mass WD is a CO WD leading to a SN Ia (despite the uncertainties mentioned above), we have probed the progenitor parameter space for the single degenerate scenario with the results presented in this work (see Fig. 89 and the discussion in Section 9.7.3). In addition, we have a library of many donor star structures, of different nature and at various evolutionary epochs, at the moment of the SN Ia events (cf. Table 16 for examples of stars for which we have detailed structures). These models can be used in future work to study the impact of the SN Ia explosion on the donor star and thus help predict the expected properties of former companions stars when searching for these in SN Ia remnants (see e.g. Marietta, Burrows & Fryxell 2000; Di Stefano & Kilic 2012; Liu et al. 2012; Pan, Ricker & Taam 2010; 2012; 2013). These authors demonstrated that whereas helium star donors may only lose about 5% of their mass, and main-sequence star donors lose some 10–20%, giant star donors may lose most of their envelope because of the SN impact. Since much less material is ejected in an AIC event, compared to a SN Ia, the impact effect will be less severe. However, even if there were somewhat less donor material to recycle the post-AIC pulsars, the main findings in this work would remain intact.

#### 9.7.2.2 Super-Chandrasekhar mass WDs

Recent observations of exceptionally luminous SNe Ia (e.g. Howell et al. 2006; Scalzo et al. 2010) suggest that their WD progenitors had a super-Chandrasekhar mass ($\sim 2.0 - 2.5\ M_\odot$). The possibility of super-Chandrasekhar mass WDs has been investigated theoretically for differentially rotating WDs (e.g. Yoon & Langer 2005), WDs with extreme magnetic fields (e.g. Das & Mukhopadhyay 2013) and merging WDs (e.g. Iben & Tutukov 1984).

Our binary calculations are able to produce such massive WDs, but only in systems with main-sequence star donors. In the left panel of Fig. 82, we show how WDs up to about 2.1 $M_\odot$ (if they exist in nature) can be produced in these systems (the open circles indicate when the WDs reach a mass of 2.0 $M_\odot$). The potential formation of these massive accreting WDs was also found in Langer et al. (2000). It is obvious that the initial parameter space for producing SNe Ia, or AIC events, is much more limited for such super-Chandrasekhar mass WDs.

### 9.7.3 Comparison to previous work

To explain the existence of high B-field NSs in old stellar populations, or in close binaries in the Galactic disk, it has been suggested in the literature that: 1) these NSs are formed via the AIC formation channel (cf. Section 9.2), and 2) post-AIC accretion, in general, does not affect these newborn NS significantly and therefore they remain to appear as young pulsars or, at most, mildly recycled pulsars (e.g. Ivanova et al. 2008). Our calculations partly disagree with the second hypothesis. We find that the majority of pulsars formed via AIC would have accreted significant amounts of matter following the AIC event, leading to the low B-fields of recycled pulsars (in particular, this is the most likely outcome if the donor is a main-sequence star). A similar conclusion was also found by Sutantyo & Li (2000) who argue that it is difficult to reproduce high B-field pulsars in close orbits (e.g. PSR B1831−00) via the AIC formation scenario. However, as pointed out in Section 9.7.1, in a few cases one would indeed expect to find pulsars that are not recycled, or are only very mildly recycled, as a consequence of late AIC towards the end of the mass-transfer process, disruption of the post-AIC binary by an encounter in a GC, or hampering of post-AIC accretion due to the propeller and/or the radio ejection mechanism for systems with giant star or helium star donors. A population synthesis investigation might help to study the probabilities of these events happening.

#### 9.7.3.1 The initial ($P_{orb}$, $M_2$)–parameter space

In Fig. 89 we summarize our results in terms of the initial parameter space of AIC progenitors in a ($P_{orb}$, $M_2$)–diagram. Langer et al. (2000) studied in detail the binary evolution of main-sequence star donors with $0.7 - 1.0\ M_\odot$ WD accretors, somewhat less massive than the initial WD mass of 1.2 $M_\odot$ assumed here. Three examples of studies which did have a 1.2 $M_\odot$ WD accretor include Li & van den Heuvel (1997) and Wang, Li & Han (2010), who studied the progenitors of SN Ia for main-sequence and giant star donors, and Han & Podsiadlowski (2004) who studied the SN Ia progenitors for main-sequence stars. One can compare Fig. 2 or Fig. 3 in these three papers to our Fig. 89. For giant stars, our results are generally in agreement, except for Wang et al. who did not reach the Chandrasekhar limit for giant stars in systems with $P_{orb} > 25^d$, possibly as a consequence of their inclusion of accretion disk instabilities. For the main-sequence donors, however, a key difference is that we are not able to produce systems leading to WD collapse for 3 $M_\odot$ donor stars. In our work, as discussed in Section 9.3.1.1, we did not allow for the optically thick wind model (Kato & Iben 1992; Kato & Hachisu 1994) to operate for high values of $|\dot{M}_2|$. With our assumptions, when the mass-transfer rate becomes very high for these massive donors, the fate of the system is a merger. Therefore, our maximum main-sequence donor star mass is $\sim 2.7 \pm 0.1$ $M_\odot$. Only if we let $\dot{M}_{CE} = 10$ $\dot{M}_{Edd,WD}$ are we able to produce AIC events with 3 $M_\odot$ donor stars.

Whereas Li & van den Heuvel (1997) and Han & Podsiadlowski (2004) found that SN Ia could result from systems having main-sequence donor stars with masses up to $\sim 3.5$ $M_\odot$, Hachisu, Kato & Nomoto (2008) propose solutions with main-sequence donor stars up to $\sim 7$ $M_\odot$ by introducing an efficient stripping of the donor star via impact of a strong wind from the accreting WD. By assuming a stripping rate up to a factor of 10 larger than the wind mass-loss rate of the WD ($\dot{M}_{strip} \lesssim 10\,\dot{M}_{wind}$) the systems can remain dynamically stable and avoid evolving into a CE. We find this scenario questionable, both with respect to its realization and its stability, although attractive if the main goal is to match the observed SN Ia rates with theoretical modelling based solely on the single-degenerate scenario. In a recent paper Bours, Toonen & Nelemans (2013) demonstrated the strong dependence on the predicted SN Ia rates on the wind-stripping effect and on the mass accumulation (retention) efficiency of the accreting WD. Wang & Han (2010) and Liu et al. (2010) studied the helium star donor channel for SNe Ia. From their Figs. 2 and 4, respectively, we can directly compare their results to ours shown in Fig. 86. The discrepancy is quite large. According to their work, systems with massive helium star donors (up to $\sim 3$ $M_\odot$) lead to a Chandrasekhar-mass WD, and Wang & Han (2010) even find that SNe Ia occur for low-mass helium star donors in very wide systems (up to 100 days). We find that only systems with initial helium star donor masses $M_2 \leq 1.5$ $M_\odot$ (and $P_{orb} \leq 1.2^d$) are able to produce WDs undergoing AIC (or SN Ia). Again, the main reason for this difference is their use of the optically thick wind model, as discussed above.

To investigate this question of wind dynamical instability for helium stars donors, we performed two test runs without restrictions from obtained values of $|\dot{M}_2|$. We considered a 2.0 $M_\odot$ and a 2.6 $M_\odot$ helium star in a binary with a 1.2 $M_\odot$ WD and $P_{orb} = 0.10^d$. Although the orbital size shrinks by up to 30 %, the systems remain dynamically stable in both cases. However, the mass-transfer rates become quite super-Eddington. For the 2.6 $M_\odot$ helium star, $|\dot{M}_2| \simeq 1.0 \times 10^{-4}$ $M_\odot\,yr^{-1}$ (corresponding to $|\dot{M}_2| > 100$ $\dot{M}_{Edd,WD}$). Whether or not the optically thick wind model remains valid in this regime is questionable and more research is needed to clarify this important question.

Another factor that determines the parameter space of systems evolving successfully to a critical WD mass (and therefore affects the Galactic SN Ia birthrate) is the assumed Chandrasekhar mass. Whereas we applied a threshold mass of 1.48 $M_\odot$ (to include the effect of rotation), a smaller limit of $\sim 1.38$ $M_\odot$ was applied by Han & Podsiadlowski (2004); Wang & Han (2010) and Wang, Li & Han (2010), which facilitates reaching the critical WD mass point.

### 9.7.3.2 Population synthesis studies of MSP formation via AIC

Hurley et al. (2010) performed a population synthesis study of binary MSPs formed via the AIC channel. A few interesting points become clear when comparing our results to their Figs. 1d, 2d, and 3d. They produce a large number of very wide systems with CO WDs and $P_{orb} = 10^3 - 10^6$ days. Although, it is expected that some systems widen significantly at the moment of the AIC when a small kick is applied, the numbers of these very wide systems is surprising to us. The donor stars in these systems are giants which later settle as CO WDs.

However, given that these giants have initial masses $0.9 \leq M_2/M_\odot \leq 1.1$, their maximum radii reach just a few $10^2 \, R_\odot$ and therefore they are unlikely to serve as donor stars in binaries expanding much beyond 1400 days. Hence, recycled pulsars are not expected in systems with $P_{orb} > 1400$ days (cf. our models GS1–4 and GSZ1). Similarly, we cannot reproduce the distribution of massive ($\sim 0.8 \, M_\odot$) CO WDs in a wide range of systems with $P_{orb} \simeq 0.1 - 100$ days. The progenitors of these systems are binaries with an accreting WD and a helium star donor. We find that for these systems the expected final orbital periods are $P_{orb} \simeq 0.05 - 2.0$ days. (Even applying large kicks will not change this picture much.) The reason for these apparent discrepancies is not clear to us and cannot simply be explained by using different limits for the WD accretion window, although we apply a different prescription for the mass accumulation of helium-rich material. It is also possible that the BSE code used by Hurley et al. (2010) fails to trace the mass-transfer process with sufficient accuracy.

There are, however, also similarities in our results. Most notably, Hurley et al. (2010) find that the minimum orbital period of MSPs with He WD companions, produced via AIC, is about 5 days (we find a minimum $P_{orb} \simeq 10$ days) and that there is a gap in the orbital period distribution between $\sim 50 - 200$ days (depending on their chosen model), where we find the gap is roughly between $60 - 500$ days (cf. Table 16 for models with no AIC kick, $w = 0$). Ideally, our work in this paper would be used as a basis for a new population synthesis study for comparison with the work of Hurley et al. (2010), for example.

### 9.7.4 Space velocities: distinguishing AIC from SNe Ib/c

In Fig. 92 (partly adapted from Tauris & Bailes 1996) we show examples of systemic recoil velocities expected for MSPs which formed indirectly via AIC. For systems with main-sequence donors, the pre-AIC orbital separations are typically $5 - 10 \, R_\odot$ and the resulting velocities are $v_{sys} \leq 50 \, km \, s^{-1}$, even when applying hypothetical kicks up to $w = 100 \, km \, s^{-1}$. For NSs formed via AIC with helium star donors the expected values of $v_{sys}$ are somewhat larger because of their narrower orbits. Neutron stars formed via AIC with giant star donors only survive in binaries if the associated kicks are very small ($w \ll 50 \, km \, s^{-1}$) as a result of their large pre-AIC orbital separations, $a_0 > 200 \, R_\odot$.

To summarize, we expect binary pulsars formed via AIC to have small systemic velocities relative to their local stellar environment; as opposed to what is expected for systems where the NS was formed via standard SN Ib/c SNe (Tauris & Bailes 1996).

In the histogram in Fig. 93 we have plotted the distribution of all 39 measured transverse velocities, $v_\perp$ (derived from proper motions) of binary radio pulsars with a WD companion. The average transverse velocity is $\langle v_\perp \rangle = 120 \, km \, s^{-1}$ and the median value is $89 \, km \, s^{-1}$. These values correspond to 3-D systemic space velocities which, in average for random orientation of their velocity vectors with respect to the line-of-sight, are larger by a factor of $4/\pi$, i.e. $\langle v_{sys} \rangle \approx 160 \, km \, s^{-1}$. (Here we assume that these space velocities correspond to typical peculiar velocities of the systems with respect to the local standard of rest at their birth locations in the Galactic plane.) It is seen that the vast majority of the observed systems have much larger $v_{sys}$ than the expected $10 - 30 \, km \, s^{-1}$ for NS binaries that were formed via the AIC channel. Based on this alone, one could be tempted to conclude that the fraction of binary pulsars formed via AIC must be rather small (at least less than 20 %), if AICs are symmetric or even accompanied by a small kick, $w < 50 \, km \, s^{-1}$. However, one must keep in mind the selection effects at work given that it is much easier to determine the proper motions of fast moving objects. In the future when the number of measured velocities has increased significantly it may be possible to identify structured peaks in the velocity distribution, reflecting NS origins from AIC (or EC SNe) and SN Ib/c.

### 9.7.5 Formation of high B-field NSs in close Galactic binaries

In this work, we have presented the suggested observational evidence for pulsars formed via AIC in GCs. Considering orbital parameters of binary pulsars, however, sources located in GCs are generally not suitable as tracers of their binary evolution history because of the frequent encounters and exchanges of companion stars in the dense environments (Sigurdsson & Phinney 1993; Heggie, Hut & McMillan 1996; Ransom et al. 2005). We therefore proceed to discuss only binary parameters of NSs in the Galactic disk.

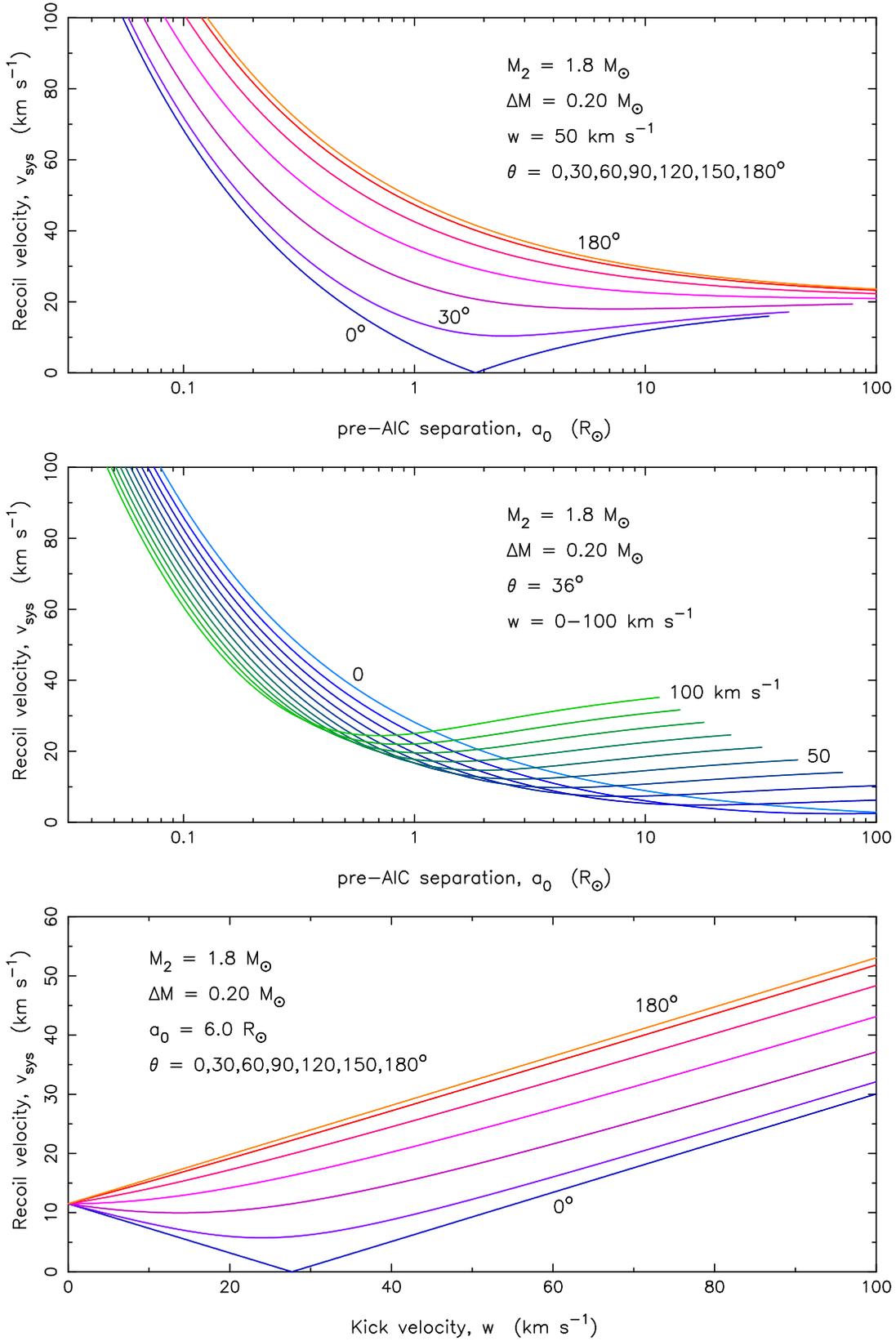

Figure 92: Systemic recoil velocities of MSPs formed indirectly via AIC. The two upper panels show sample calculations of the recoil velocities, $v_{\rm sys}$ obtained for a fixed kick of $w = 50\ {\rm km\,s^{-1}}$ and kick angles $\theta = 0 - 180°$ (upper panel), or a fixed value of $\theta = 36°$ and kicks $w = 0 - 100\ {\rm km\,s^{-1}}$ (central panel), in both cases as a function of the pre-AIC orbital separation, $a_0$. The lower panel shows $v_{\rm sys}$ as a function of kick velocities, $w$ for fixed values of $a_0$ and variable values of $\theta = 0 - 180°$. In all cases we assumed a companion star mass of $M_2^{\rm AIC} = 1.8\ M_\odot$ at the moment of the AIC. AIC events are generally believed to have $w \simeq 0$ (i.e. without a kick). See text for discussions.

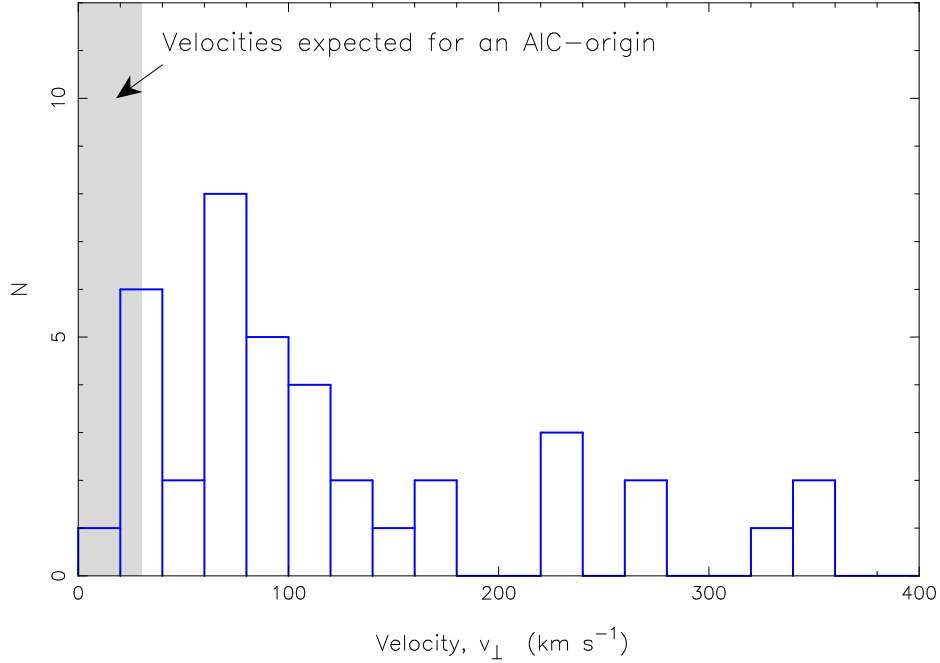

Figure 93: All measured transverse velocities of binary radio pulsars with a WD companion. Data taken from the *ATNF Pulsar Catalogue* in March 2013 (Manchester et al. 2005; http://www.atnf.csiro.au/research/pulsar/psrcat). AIC is expected to form binary pulsars with $v_\perp < 30\ \mathrm{km\,s^{-1}}$. However, the observed binary pulsars with such small velocities could also have formed by EC SNe.

In Table 15 we presented four NS binaries in the Galactic disk with unusual properties (i.e. relatively slow spins and high B-fields, despite being in close binaries with an ultra low-mass companion star), see Section 9.2.3. The interesting question now is if we can reproduce these systems with our AIC modelling. The answer is no. Whereas the slow spin periods and the large B-fields may be accounted for in some post-AIC binaries with similar orbital periods less than a few days (see Section 9.7.1.1), all our calculated close-orbit post-AIC systems have massive CO WD companions with masses $0.65 < M_{\mathrm{WD}}/M_\odot < 0.85$ This is in clear contrast to the four binaries in Table 15 which all have companion stars with masses $\leq 0.10\ M_\odot$. One way to reconcile this companion mass discrepancy is if the newborn NSs have evaporated their companion stars very efficiently after the AIC event. Although the available spin-down luminosity of these slow NSs is relatively small ($\dot{E} \propto P^{-3}$) at the present epoch, these NSs could have been much more energetic in the past. An evaporation scenario could possibly also apply to post-AIC systems that evolved into (and survived) a CE, in case of main-sequence star donors (cf. Fig. 77 and Section 9.3.4). The ultra-compact X-ray binary 4U 1626−67 exhibits UV and X-ray emission lines of C, O, and Ne which suggests an evolved helium star or even a WD donor (Nelemans et al. 2010). For further discussions on 4U 1626−67 and PSR J1744−3922, see Yungelson, Nelemans & van den Heuvel (2002) and Breton et al. (2007), respectively. The formation of these four abnormal NS binaries is one of the biggest challenges to our understanding of close binary evolution.

### 9.7.5.1 Puzzling radio pulsars in the Corbet diagram

Besides the four systems discussed above, one may ask if there are additional potential AIC candidates among the binary radio pulsars in the Galactic disk. Figure 94 shows the loca-tion of the known Galactic binary pulsars in the $P\dot{P}$–diagram. The mildly recycled MSPs ($10\,\text{ms} < P < 100\,\text{ms}$) are dominated by systems with CO/ONeMg WD and NS companions. As shown in Tauris, Langer & Kramer (2012), their relatively slow spin rates are expected from an evolutionary point of view as a consequence of the rapid mass-transfer phase from a relatively massive donor star.

As seen in Fig. 94 there is also a group of pulsars that have similarly slow spin periods, and relatively large B-fields, but with He WD companions. These pulsars (marked with a large open circle) may also have had a limited recycling phase which could hint at a possible alternative origin. Their puzzling nature is clearer when plotting all binary radio pulsars with He WDs in the Corbet diagram. As shown in Fig. 95, an interesting pattern is revealed and we now briefly discuss four regions in this diagram.

Region I shows that MSPs can be fully recycled over a spread of 3 orders of magnitude in final orbital period. If the $P_{\text{orb}} \gtrsim 200^{\text{d}}$ the pulsars are only partially recycled, as noticed from their slow spin periods (region III). This could be related to the relatively short mass-transfer phase in wide-orbit LMXBs where the donor star is highly evolved by the time it fills its Roche lobe (Tauris & Savonije 1999). In region II, one sees the small sub-population of puzzling systems (marked by circles and discussed above) with $1^{\text{d}} < P_{\text{orb}} < 200^{\text{d}}$, all of which are only mildly recycled MSPs with spin periods between 20 and 100 ms. Where do these systems come from? And why do they have much slower spin periods than the pulsars in region I with similar orbital periods? (In this group of marked pulsars in region II we exclude PSR J0348+0432 at $P_{\text{orb}} = 0.1024$ days and $P = 39$ ms, since this system can probably be explained by standard evolution of a converging LMXB system; Antoniadis et al. (2013)).

Given their slow spins these six pulsars might originate from progenitor systems where the amount of mass accreted was limited. Whereas the populations in regions I and III are thought to be produced from standard LMXBs with different initial orbital periods following a core-collapse SNe, the situation might be different for pulsars in region II. Could these systems then perhaps (as suggested in Tauris 2011) originate from AICs where the subsequent spin up of the newborn NS only resulted in a mild spin-up? The rationale is that if a limited amount of material remains in the donor star envelope by the time the ONeMg WD undergoes AIC, then only a limited amount of material is available to spin up the pulsar later on. Furthermore, the narrow range of orbital periods for these pulsars could then reflect the finetuned interval of allowed mass-transfer rates needed for the progenitor ONeMg WDs to accrete and grow in mass to the Chandrasekhar limit before their implosion. This is a tempting hypothesis but, as we argue below, we find that this cannot be the case.

In this paper we have demonstrated that MSPs with He WD companions and $P_{\text{orb}} < 10^{\text{d}}$ do not form via AIC for the following reason: Main sequence donor stars must have masses above at least 1.6 $M_\odot$ to ensure sufficiently high mass-transfer rates for the WD mass to grow to the AIC limit, and therefore magnetic braking is not expected to operate in these systems since such massive main-sequence stars do not have convective envelopes. As a consequence of this, the systems leading to AIC avoid significant loss of orbital angular momentum and their final orbital periods will always exceed 10 days. Given that three out of the six puzzling pulsars have $P_{\text{orb}} < 10$ days we conclude that the AIC channel cannot explain the origin of their characteristics. Last but not least, two of these six binary pulsars have measured transverse velocities and their values are extraordinarily large: $v_\perp = 269$ km s$^{-1}$ for PSR J1745$-$0952 and $v_\perp = 326$ km s$^{-1}$ for PSR J1810$-$2005, according to the *ATNF Pulsar Catalogue*, based on measurements by Janssen et al. (2010). Such large transverse velocities are clearly in contradiction with the expectations of an AIC origin.

### 9.7.5.2 Observed pulsars with CO WDs in the Corbet diagram

For the observed distribution of recycled binary pulsars with CO WDs in the Corbet diagram, we refer to Fig. 3 in Tauris, Langer & Kramer (2012). We note that some of our modelled systems with an AIC origin, plotted in Fig. 90, share properties of a subpopulation of the observed systems. These are systems with helium star donors which leave behind recycled pulsars with

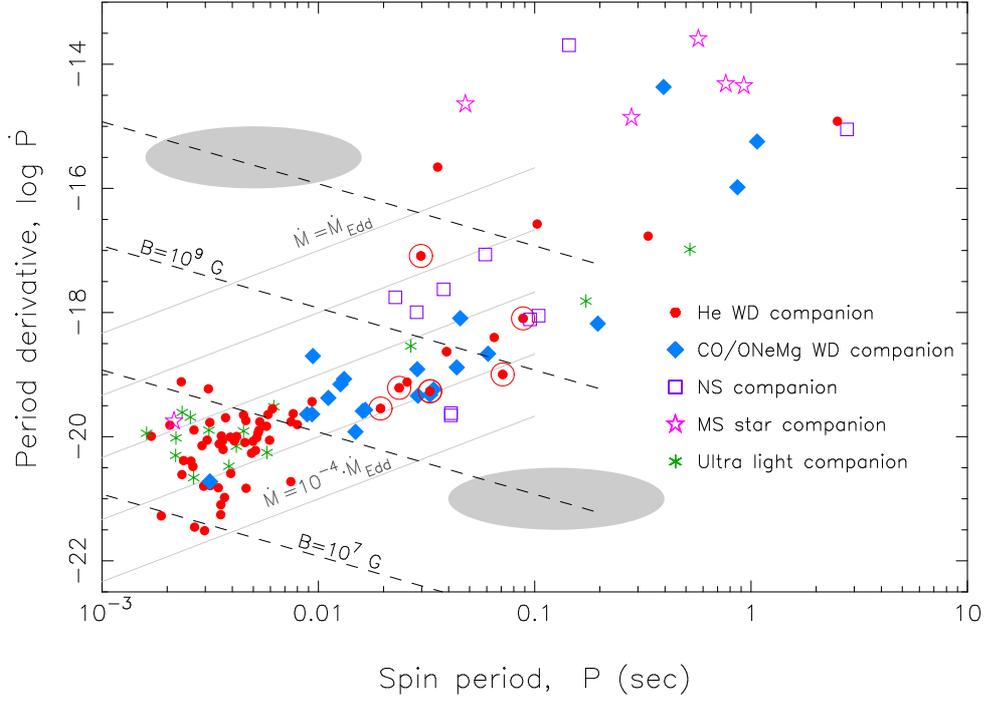

Figure 94: The distribution of 115 Galactic binary radio pulsars in the $P\dot{P}$–diagram. The error bars are much smaller than the size of each symbol. The caption to the right explains the nature of the companion stars. Pulsars with a He WD companion marked by a circle are discussed in Section 9.7.5.1. The B-fields (dashed lines) and the spin-up lines (grey lines) were calculated following Tauris, Langer & Kramer (2012) and assuming $M_{NS} = 1.4M_{\odot}$ and $\sin\alpha = \phi = \omega_c = 1$. Data taken from the *ATNF Pulsar Catalogue* in March 2013. All observed $\dot{P}$ values were corrected for kinematic effects. The grey-shaded areas mark empty regions that cannot be populated according to current recycling scenarios.

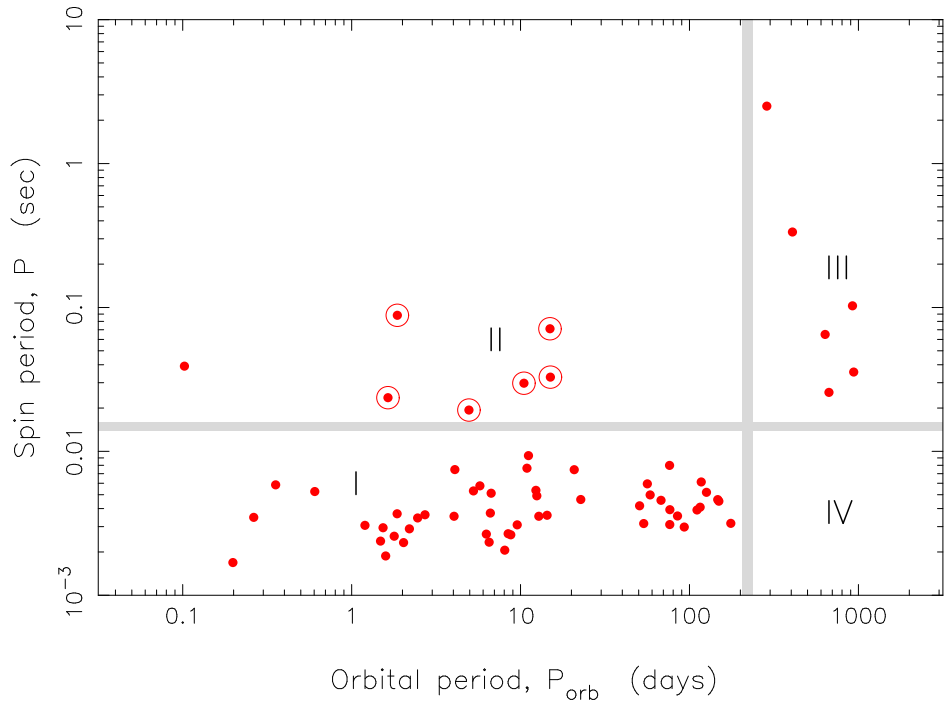

Figure 95: The Corbet diagram for the observed distribution of the 64 binary radio pulsars with a He WD companion. The plot reveals four regions, labelled by I, II, III, and IV, that may be understood from an evolutionary point of view (see text). The pulsars in region II are also marked by a circle in the $P\dot{P}$–diagram in Fig. 94 and their potential link to an AIC origin is discussed in the text (figure updated after Tauris 2011).

4 ms $< P <$ 200 ms and $P_{orb} < 2^d$. The very wide orbit radio pulsars with CO WDs, which result from our AIC channel modelling in systems with giant star donors (e.g. model GS3), also resemble an observed system: PSR B0820+02. This pulsar is in a 1232 day orbit, has a spin period of 0.86 sec, and a CO WD companion of mass $> 0.52\ M_\odot$ (Koester & Reimers 2000). Hence, it is indeed possible that some of the observed radio pulsars with CO WDs could have an AIC origin. If this is the case, then these pulsars should all have a mass close to 1.28 $M_\odot$ since they do not accrete much in the post-AIC LMXB phase. This is in contrast to the binary pulsars with He WDs, which may accrete $\sim 0.4\ M_\odot$ following the AIC event. One observed pulsar with a CO WD and a low-mass NS is PSR J1802$-$2124 which has $M_{NS} = 1.24 \pm 0.11\ M_\odot$, $P = 12.6$ ms, $P_{orb} = 0.70^d$, and a $0.78 \pm 0.04\ M_\odot$ CO WD (Ferdman et al. 2010). This pulsar shares some properties of our AIC models He3 and He4 (cf. Table 16). However, binary pulsars with CO WDs formed via the standard SN Ib/c channel are not expected to have accreted much either (since here too the mass-transfer phase is short lived, either in intermediate-mass X-ray binaries in relatively close orbits or LMXBs with very wide orbits and giant star donors). However, these NSs could have formed in SN Ib/c with birth masses substantially larger than 1.28 $M_\odot$ and hence their measured masses could be much larger too.

### 9.7.6 Optimal evidence for NSs formed via AIC?

One may ask what observational evidence would be needed to firmly prove the AIC formation channel of NSs? This is a difficult question to answer since we do not know the exact NS properties expected from AIC. However, we could point to a few hypothetical cases of NSs which are not expected to exist from current knowledge of pulsar recycling. For example, it would be interesting if future observational surveys discover either a very slowly spinning ($\sim 100$ ms) radio pulsar associated with a very low B-field ($\sim 10^8$ G), or a close binary MSP with a very rapid spin ($< 5$ ms) and a high B-field ($> 10^{10}$ G), orbiting a companion star which has experienced mass loss (see grey-shaded areas in Fig. 94). None of these pulsars is expected to form according to present recycling and spin-up theory (e.g. Tauris, Langer & Kramer 2012). However, even if the second kind of pulsar existed it would be very unlikely to be detected given that its strong magnetic dipole radiation would slow down its rapid spin rate within a few 100 kyr. In our view, the best evidence for AIC is the apparently young NSs detected in GCs.

### 9.8 Conclusions

I) We have demonstrated, using a detailed stellar evolution code modelling both the pre- and the post-AIC binary evolution (to our knowledge, for the first time), that MSPs can be formed indirectly via the AIC formation channel. In this scenario, a normal NS is formed in an AIC event and which subsequently undergoes recycling by accretion (from the same donor star, now in a post-AIC LMXB) leading to the formation of an MSP. This scenario is possible for systems with donor stars that are either main-sequence stars, giants stars, or helium stars. The first type of donor stars lead to fully recycled MSPs with He WD companions, whereas the other two types of donors lead to more mildly recycled pulsars with mainly CO WD companions. The parameter space of the successful progenitor systems is restricted to three limited areas in the initial ($P_{orb}, M_2$)-plane.

II) Millisecond pulsars formed via AIC are difficult to distinguish from MSPs formed via the standard SN Ib/c channel with respect to their masses, B-fields, spin periods, and WD companions. The reason for this is that the donor stars in most cases transfer more than 0.1 $M_\odot$ of material to the NS after the AIC event, in a process which mimics the standard recycling scenario.

III) Nevertheless, we identify two parameters which can, at least in some cases, be used to differentiate the two formation channels. First, MSPs formed via AIC and which have He WD companions have $P_{orb}$ between 10 and 60 days (or $P_{orb} > 500^d$ for giant star donors). Second, the velocities of pulsars formed via AIC are predicted to be less than 30 km s$^{-1}$, which agrees well with the hypothesis of AIC being the origin of some pulsars retained in GCs. In contrast, MSPs formed via the standard SN Ib/c channel and which have He WD companions may obtain $P_{\rm orb}$ in a continuous wide range between a few hours and up to $\sim 1000$ days, and their space velocities typically exceed 100 km s$^{-1}$.

IV) In a few cases where the post-AIC LMXB phase is short-lived, and particularly in GCs where the probability of disruption of a wide binary by an encounter event is large, pulsars formed via AIC will be largely unaffected by, or will avoid, subsequent accretion from the donor star. This could, for example, be the explanation for the existence of some GC pulsars which appear to be young.

V) More research is needed on optically thick wind modelling, and direct observations of SN Ia remnants to constrain pre-SN Ia wind mass loss, as well as modelling of the prescription for calculating the mass-transfer rates from giant star donors with small surface gravities and extended atmospheres. As demonstrated in this paper, the application of the optically thick wind model is very important for the estimated event rate of both AIC and SNe Ia.

# 10. Direct Formation of Millisecond Pulsars from Rotationally Delayed Accretion-Induced Collapse of Massive White Dwarfs

**Freire & Tauris (2014)**

**MNRAS Letters 438, L86**

## Abstract

Millisecond pulsars (MSPs) are believed to be old neutron stars, formed via type Ib/c core-collapse supernovae, which have subsequently been spun up to high rotation rates via accretion from a companion star in a highly circularised low-mass X-ray binary. The recent discoveries of Galactic field binary MSPs in eccentric orbits, and mass functions compatible with that expected for helium white dwarf companions, PSR J2234+06 and PSR J1946+3417, therefore challenge this picture. Here we present a hypothesis for producing this new class of systems, where the MSPs are formed directly from a rotationally-delayed accretion-induced collapse of a super-Chandrasekhar mass white dwarf. We compute the orbital properties of the MSPs formed in such events and demonstrate that our hypothesis can reproduce the observed eccentricities, masses and orbital periods of the white dwarfs, as well as forecasting the pulsar masses and velocities. Finally, we compare this hypothesis to a triple star scenario.

## 10.1 Introduction

Almost since the discovery of PSR B1937+21, the first millisecond pulsar (MSP, Backer et al. 1982), it has been suggested that these objects are old neutron stars spun up to high spin frequencies of several hundred Hz via accretion of mass and angular momentum from a companion star (Alpar et al. 1982; Bhattacharya & van den Heuvel 1991). In this so-called recycling phase, the system is first observable as a low-mass X-ray binary (LMXB, e.g. Bildsten et al. 1997), later as an accreting X-ray MSP (Wijnands & van der Klis 1998), or even as an MSP in the transition phase between an accretion powered MSP and a rotation powered radio MSP (Archibald et al. 2009; Papitto et al. 2013; Tauris 2012).

An inevitable consequence of a long phase ($10^8 - 10^9$ yr) of recycling in an LMXB, where tidal forces operate, is that it should leave a fossil record of a highly circular system (Phinney & Kulkarni 1994). And indeed, until recently, *all* of the more than 100 observed, fully recycled MSPs (here defined as pulsars with spin periods less than 20 ms), in binaries with helium white dwarf (He WD) companions and located outside of globular clusters, have very small eccentricities in the range $e = 10^{-7} - 10^{-3}$ (*ATNF Pulsar Catalogue*, Manchester et al. 2005). Pulsar systems in globular clusters, on the other hand, often have their orbits perturbed after the recycling phase terminates because of their location in a dense environment (Rasio & Heggie 1995; Heggie & Rasio 1996). Until the start of 2013, the only known fully recycled MSP with a high eccentricity, and located in the Galactic field, was PSR J1903+0327 ($e = 0.44$, Champion et al. 2008). This MSP has a G-type main-sequence companion star and is thought to have originated from a hierarchical triple system that ejected one of its members (Freire et al. 2011b; Portegies Zwart et al. 2011; Pijloo, Caputo & Portegies Zwart 2012).

### 10.1.1 Discovery of MSPs in eccentric orbits

Recently, Deneva et al. (2013) presented the discovery of PSR J2234+06 and soon afterwards Barr et al. (2013) announced the discovery of PSR J1946+3417. PSR J2234+06 and PSR J1946+3417 are of special interest because they resemble each other and share very unusual properties. Both of these Galactic field pulsars (see Table 17) have a spin period, $P \simeq 3$ ms, an orbital period, $P_{\rm orb} \simeq 30$ days and a companion mass, $M_2 \simeq 0.24\ M_\odot$. All these values are within typical ranges expected for MSPs with He WD companions. However, both of these MSP binaries are also eccentric, $e \simeq 0.13$, which is unusual and unexpected from current formation theories of MSPs, as explained above. Therefore, it is clear that these system must have a formation history which is different from the "normal" MSP-WD systems in the Galactic field. We notice that the two median expectations for the companion masses of PSR J2234+06 and PSR J1946+3417 are very similar to each other and close to the values expected from the correlation between WD mass and orbital period for post-LMXB systems (e.g. Tauris & Savonije 1999); in particular if the slight widening of the orbit from the event that imparted the eccentricity is accounted for, see Section 10.3. This provides confidence that the current companions are indeed He WDs which have lost their hydrogen envelopes via stable Roche-lobe overflow. Optical detections would confirm this.

Table 17: Physical parameters of two newly discovered binary MSPs (data taken from Deneva et al. 2013; Barr et al. 2013).

| Pulsar | $P$ | $P_{\mathrm{orb}}$ | $M_{2,\mathrm{med}}$ | $e$ |
|--------|-----|--------------------|-----------------------|-----|
| J2234+06 | 3.6 ms | 32 days | 0.23 $M_\odot$ | 0.13 |
| J1946+3417 | 3.1 ms | 27 days | 0.24 $M_\odot$ | 0.14 |

### 10.1.2 A triple system formation scenario?

By analogy with PSR J1903+0327, one could advance the hypothesis that both PSRs J2234+06 and J1946+3417 originated as hierarchical triple systems, which evolved to produce a neutron star orbited by two F/G-type dwarfs. Because of the widening of the inner orbit during the subsequent neutron star accretion in the LMXB phase, the systems later became dynamically unstable (e.g. Mikkola 2008) and one of the components was eventually ejected. The only difference being that it was the donor star (the WD progenitor) in the inner binary which was ejected in the case of PSR J1903+0327 (Freire et al. 2011b; Portegies Zwart et al. 2011; Pijloo, Caputo & Portegies Zwart 2012), whereas it would have been the outer tertiary star in the cases of PSR J2234+06 and PSR J1946+3417. In a triple system, the Kozai process (Kozai 1962) may lead to large cyclic variations in the inner orbital eccentricity prior to ejection of the tertiary star (e.g., Mardling & Aarseth 2001). Hence, one may expect a wide range of eccentricities of the surviving MSP-WD binaries. Whether or not triple star evolution, or formation and ejection of a binary system from a dense cluster, is plausible for the relatively small eccentricities ($e \simeq 0.13$) observed in PSRs J2234+06 and J1946+3417 (compared to $e = 0.44$ for PSR J1903+0327) requires detailed modelling beyond the scope of this Letter.

Here we advocate for another solution. In Section 10.2 we present a hypothesis of a new direct formation channel of MSPs which can exactly explain both the unusual properties and the similarities of the recently discovered MSPs. In Section 10.3 we present simulations and make further falsifiable predictions about these systems which can be tested in the near future. In Section 10.4 we discuss future perspectives and we summarise our conclusions in Section 10.5.

## 10.2 Direct formation of millisecond pulsars via delayed AIC

Besides from formation via core-collapse supernovae, it has been suggested for many years that neutron stars may also be produced from accretion-induced collapse (AIC) of a massive ONeMg WD in a close binary (Nomoto et al. 1979; Taam & van den Heuvel 1986). The properties of such neutron stars are unknown. It has been suggested that AIC events cannot produce MSPs directly since r-mode instabilities would spin-down any young, hot MSP on a very short timescale (Andersson, Kokkotas & Schutz 1999). However, if the scenario described here is confirmed by further observations, then the role of r-mode instabilities has to be revised. In the following, we rely on the results of the recent modelling by Tauris et al. (2013b). They investigated a scenario where MSPs are produced *indirectly* via AIC, i.e. the AIC leaves behind a normal neutron star which is subsequently recycled to become an MSP, once the mass-transfer resumes after the donor star refills its Roche lobe and continues LMXB evolution until the end.

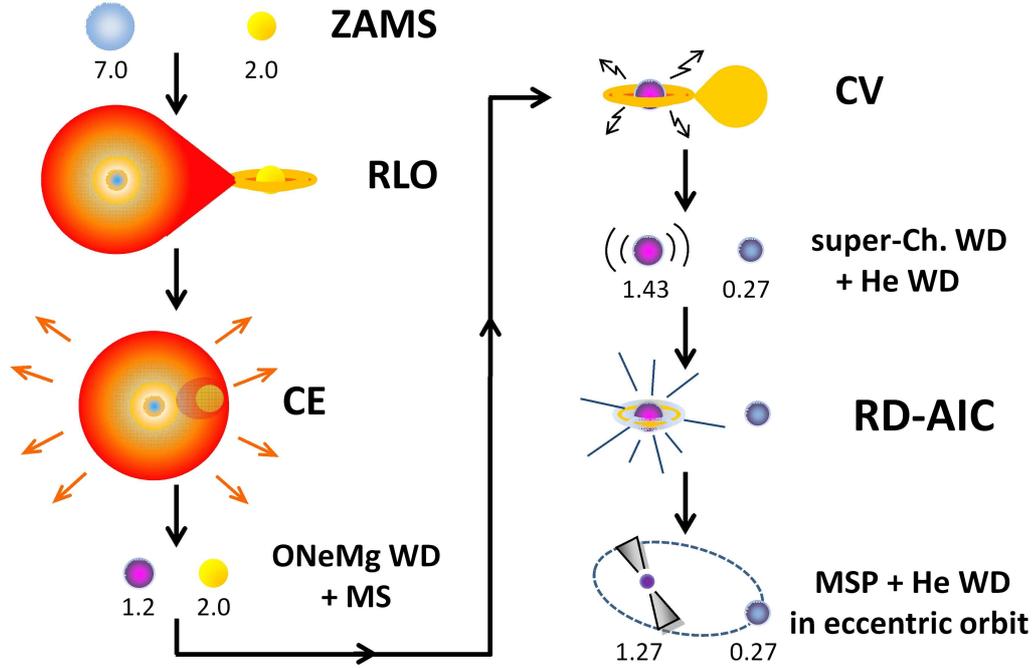

Figure 96: Illustration of the binary stellar evolution from the zero-age main sequence (ZAMS) to the final millisecond pulsar (MSP) stage. A primary $6-8\ M_\odot$ star evolves to initiate Roche-lobe overflow (RLO) towards the $\sim 2\ M_\odot$ secondary star, leading to dynamically unstable mass transfer and the formation of a common envelope (CE; Ivanova et al. 2013). The envelope ejection leads to formation of an oxygen-neon-magnesium white dwarf (ONeMg WD) from the naked core of the primary star (possibly after a stage of Case BB RLO from the naked core – not shown). When the secondary star evolves, it initiates RLO leading to a cataclysmic variable (CV) X-ray binary system. As a result of accretion the WD becomes a rapidly spinning super-Chandrasekhar mass WD. After accretion has terminated it loses spin angular momentum and eventually undergoes a rotationally-delayed accretion-induced collapse (RD-AIC) to directly form an MSP with a helium white dwarf (He WD) companion in an eccentric orbit. Stellar masses given in units of $M_\odot$.

Their main result is that as a consequence of the finetuned mass-transfer rate necessary to make the WD grow in mass, the resultant MSPs created via the AIC channel preferentially form with $10 < P_{\rm orb} < 60$ days, clustering more at $P_{\rm orb} \simeq 20-40$ days. Furthermore, the modelling of these systems produced He WD companions with masses, $M_{\rm WD} \simeq 0.24-0.31\ M_\odot$. These values are interesting since they match exactly the observed values of $P_{\rm orb}$ and $M_{\rm WD}$ for the newly discovered MSPs in eccentric orbits (Table 17). However, in the Tauris et al. (2013b) scenario of indirect formation of MSPs, continued post-AIC mass transfer leads to highly circularised systems. Therefore, that scenario cannot explain the newly discovered MSPs with $e \sim 0.1$.

### 10.2.1 Rotationally-delayed accretion-induced collapse (RD-AIC)

In case a mass-gaining WD is spun up to rapid rotation via near-Keplerian disk accretion (Langer et al. 2000), it can avoid AIC (Yoon & Langer 2004; 2005) and evolve further to super-Chandrasekhar mass values via continuous accretion (cf. fig. 7 in Tauris et al. 2013b; for the possible growth up to $\gtrsim 2\ M_\odot$).

Here we propose a scenario, where accretion leads to the formation of a super-Chandrasekhar mass ONeMg WD which initially avoids AIC as a result of rapid rotation. Only after the accretion has terminated, and the WD loses sufficient spin angular momentum (see below), does it undergo AIC to *directly* produce an MSP. We shall refer to this event as rotationally-delayed accretion-induced collapse (RD-AIC), see Fig. 96.

It is important to notice that under the new hypothesis presented here, accretion ceases completely *before* AIC occurs. At that stage the detached system consists of two WDs: a low-mass He WD (the remnant of the former donor star) and an ONeMg WD with a mass above the Chandrasekhar limit, and which later undergoes RD-AIC. Hence, in this case there will be no re-circularisation after the AIC event.

### 10.2.1.1 Observational and theoretical support for RD-AIC

Observations of binaries confirm that accreting WDs rotate much faster than isolated ones (Sion 1999); in one case, HD 49798/RX J0648, there is even evidence for a WD rotating with a spin period of only $P_{WD} = 13.2$ s (Mereghetti et al. 2011), corresponding to $\sim 50$ per cent of its critical (break-up) rotation frequency. The observational evidence for such fast rotation supports the increase of the mass stability limit above the standard value for non-rotating WDs (1.37 $M_\odot$), as required by our scenario. An analogous idea of rotationally-delayed SNe Ia explosions has been proposed by Justham (2011) and Di Stefano, Voss & Claeys (2011) for massive CO WDs.

For WDs with rigid body rotation, the resulting super-Chandrasekhar masses are in the range $1.37 - 1.48$ $M_\odot$ (Yoon & Langer 2004; and references therein). For differentially rotating WDs, the stability limit may in principle reach $\sim 4.0$ $M_\odot$, although it is quite possible that efficient transport of angular momentum by magnetic torques and/or baroclinic instabilities acts to ensure rigid rotation (Piro 2008). On the other hand, recent observations of exceptionally luminous SNe Ia (e.g. Howell et al. 2006; Scalzo et al. 2010) suggest that their WD progenitors had a mass of $\sim 2.0 - 2.5$ $M_\odot$. If the critical rotation frequency is obtained during accretion then further mass accumulation is prohibited, unless angular momentum is transported from the WD back to the disk by viscous effects (Popham & Narayan 1991; Saio & Nomoto 2004).

The final fate of super-Chandrasekhar ONeMg WDs depends on whether or not the effects of electron captures dominate over nuclear burning (Nomoto et al. 1979; Nomoto & Kondo 1991). The onset of electron captures on $^{24}$Mg and $^{20}$Ne occurs at a density of $\rho \sim 4 \times 10^9$ g cm$^{-3}$, whereas the density for the ignition of explosive nuclear burning (oxygen deflagration) depends on the central temperature. Therefore, after accretion has terminated, the final fate of a super-Chandrasekhar WD depends on the competition between its cooling rate and its loss of angular momentum, as demonstrated in detail by Yoon & Langer (2004; 2005). If the WD interior has crystallised by the time its spin angular momentum decreases below the critical level (corresponding to $J_{crit}^{AIC} \simeq 0.4 \times 10^{50}$ erg s, for a 1.48 $M_\odot$ WD) it undergoes RD-AIC.

Yoon & Langer (2004; 2005) discussed the loss of WD spin angular momentum due to gravitational wave emission caused by so-called CFS instabilities to non-axisymmetric perturbations. In their second paper, these authors investigated 2-dimensional models and found that only r-mode instabilities (Andersson 1998) are relevant for accreting WDs, whereas bar-mode instabilities (Chandrasekhar 1970; Friedman & Schutz 1978) are irrelevant because the ratio of rotational to potential energy cannot reach the critical limit of $T/W = 0.14$ (corresponding to $J = 4 \times 10^{50}$ erg s). The estimated timescale of removing (or redistributing) angular momentum has been estimated to be in the range $10^5 - 10^9$ yr, depending on $T/W$ and the degree of differential rotation of the WD (Lindblom 1999; Yoon & Langer 2004; 2005). However, recent work by Ilkov & Soker (2012) questions the efficiency of r-mode instabilities and hence they advocate for a very long delay timescale $> 1$ Gyr. This would give the super-Chandrasekhar mass WD plenty of time to cool down, crystallise and undergo RD-AIC, thus favouring our scenario.

To summarize, we postulate that MSPs can be formed directly (without any need for further spin up from a companion star) in an RD-AIC event that happens up to $\sim 1$ Gyr after termination of the mass-transfer phase.

In Fig. 97 we show an evolutionary track of a rapidly spinning WD undergoing RD-AIC (see fig. 11 in Yoon & Langer 2005; for more detailed tracks). The WD is assumed to be non-spinning initially and have a mass of 1.2 $M_\odot$ prior to accretion from its companion star. We assumed rigid rotation and efficient angular momentum accretion at the Keplerian disk value. The r-mode instabilities (giving rise to loss of rotational energy via gravitational waves) were calculated during accretion following Lindblom (1999). If the timescale of loss of spin angular momentum, from the termination of the accretion phase until the WD has a spin angular momentum, $J < J_{\mathrm{crit}}^{\mathrm{AIC}}$, is sufficiently long ($\sim 10^9 \mathrm{yr}$) then the result is an AIC event (Yoon & Langer 2005).

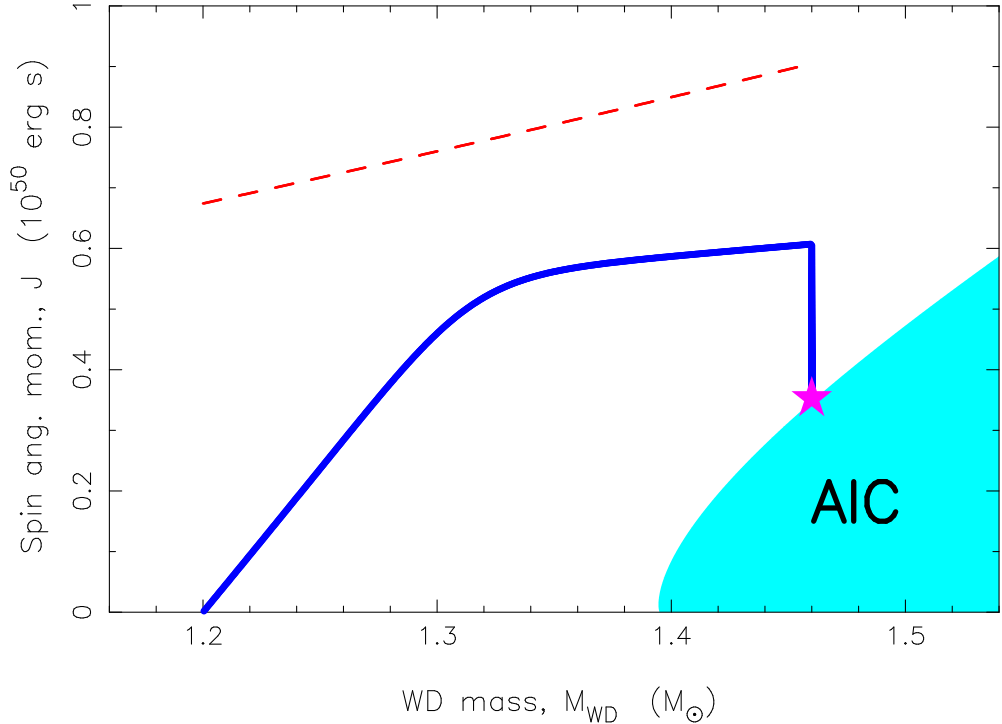

Figure 97: Schematic evolutionary track (blue solid line) in the $(M_{\mathrm{WD}}, J)$–plane calculated for an accreting ONeMg WD in a close binary system. After termination of the mass-transfer process, the super-Chandrasekhar mass WD is rapidly spinning, which prevents its collapse. From this point, the spin evolution is solely determined by loss of angular momentum (e.g. as caused by r-mode instabilities and magnetodipole radiation). The light-blue hatched area marks the critical region, $J \leq J_{\mathrm{crit}}^{\mathrm{AIC}}$ (Yoon & Langer 2005) at which boundary the WD undergoes a RD-AIC event and produces a MSP. The red dashed line indicates critical break-up rotation.

## 10.3 Properties of the RD-AIC events and resultant MSP-WD systems

The implosion of a WD with a radius of about 3000 km and an assumed surface magnetic flux density, $B \sim 10^3$ G (e.g. Jordan et al. 2007) into a neutron star with a radius of $\sim 10$ km should produce, by conservation of magnetic flux, an MSP surface B-field of $10^3$ G $\times (3000/10)^2 \sim 10^8$ G. The resultant neutron star must have a spin rate below the break-up limit and for a typical MSP spin period of a few ms, it is expected that it must lose spin angular momentum during the AIC, possibly by ejection of a few 0.01 $M_\odot$ of baryonic matter in a circumstellar disk. According to modelling by Dessart et al. (2006); Kitaura, Janka & Hillebrandt (2006); Metzger, Piro & Quataert (2009); Darbha et al. (2010), up to a few 0.01 $M_\odot$ of material is ejected in the AIC event, possibly leading to synthesis of $^{56}$Ni in the disk which may result in a radioactively powered, short-lived SN-like transient (that peaks within $\leq 1$ day and with a bolometric luminosity $\simeq 10^{41} \mathrm{erg\,s^{-1}}$).

The RD-AIC hypothesis makes several very precise, easily falsifiable predictions:

- As already mentioned, the He WD companions in our RD-AIC scenario are expected to have masses in the range $M_2 \simeq 0.24 - 0.31$ $M_\odot$ (up to 0.35 $M_\odot$ for low-metallicity WD progenitors) and orbital periods of $10 - 60$ days. In rare cases, we expect WD masses up to $\sim 0.41$ $M_\odot$, if the donor star had a ZAMS mass $> 2.3$ $M_\odot$ (Tauris et al. 2013b).

- The binding energy of a neutron star can be expressed as: $E_b \simeq 0.084 \, (M_{NS}/M_\odot)^2 \, M_\odot \, c^2$ (Lattimer & Yahil 1989), where $M_{NS}$ is its gravitational mass. The collapse of super-Chandrasekhar mass WDs of $1.37 - 1.48 \, M_\odot$ (for rigid rotation) therefore leads to MSPs with gravitational masses of $1.22 - 1.31 \, M_\odot$, if we assume that $0.02 \, M_\odot$ of baryonic material is lost during the AIC.

- The sudden release of gravitational binding energy (and mass ejection into a disk) increases the orbital period and imposes an eccentricity to the system given by (Bhattacharya & van den Heuvel 1991): $e = \Delta M/(M_{NS} + M_2)$, if the AIC is symmetric and no kick is imparted to the newborn MSP (see below). Here we assume that the pre-AIC binary orbit is circular, which is a good assumption for X-ray binaries were tidal torques circularise the system on a short timescale. For the ranges of $M_{NS}$ and $M_2$ given above, this leads to a remarkable narrow range of post-AIC eccentricities: $0.09 - 0.12$. (The exact values depend on the still unknown equation-of-state of neutron stars.) This result is in excellent agreement with the systems presented in Table 17, cf. Section 10.3.1 for a discussion.

- The momentum kick imparted to a newborn neutron star via an AIC event is expected to be small. This follows from detailed simulations of AIC events which imply explosion energies significantly smaller than those inferred for standard iron-core collapse supernovae (Kitaura, Janka & Hillebrandt 2006; Dessart et al. 2006), and also because of the small ejecta mass and the short timescale of the event (compared to the timescales of the non-radial hydrodynamic instabilities producing large kicks), e.g. Podsiadlowski et al. (2004); Janka (2012). Our hypothesis therefore predicts that eccentric binary MSPs with He WDs will have small peculiar space velocities.

### 10.3.1 Simulations of the $(P_{orb}, e)$–plane

The spread of eccentricities and orbital periods of the resultant systems formed via RD-AIC is extremely sensitive to any kick given to the MSP during the AIC event. In Fig. 98 we demonstrate this by showing a Monte Carlo simulation of the expected eccentricities and orbital periods using the range of pre-AIC parameters given above and adding small kick velocities of $w \leq 10 \text{ km s}^{-1}$. The dynamical effects were calculated following the formulae of Hills (1983). The properties of systems undergoing RD-AIC events are seen to be surprisingly similar to the characteristics of the recently discovered MSPs in eccentric orbits (Table 17).

## 10.4 Future perspectives and tests

If the WD companions happen to be bright, then a study of their spectral lines will yield the mass ratio of the binary components, $q$. Furthermore, given the eccentric orbits of these MSPs, we will certainly be able to measure the rate of advance of periastron ($\dot{\omega}$) for these systems. If the radius of the companion is small compared to the size of the orbit (which is the case for a WD), then $\dot{\omega}$ is solely due to the effects of general relativity and can be used to estimate the total mass of the system (Weisberg & Taylor 1981). The combination of $\dot{\omega}$ and $q$ would be enough to determine the masses of the components. Another possible solution is the measurement of the Shapiro delay for these systems. Even a relatively low-precision measurement of $h_3$ (Freire & Wex 2010) can, when combined with the measurement of $\dot{\omega}$, yield very precise component masses, as in the cases of PSR J1903+0327 (Freire et al. 2011b) and PSR J1807−2500B (Lynch et al. 2012). These mass measurements are very important for testing the RD-AIC hypothesis, which predicts MSP masses between $1.22 - 1.31 \, M_\odot$. Measuring a higher MSP mass would, if not falsifying our hypothesis, require differential rotation of the progenitor WD, which may be a problem with respect to the need of a long delay timescale (Ilkov & Soker 2012).

The unusual MSPs discussed in this Letter were discovered in recent pulsar surveys (e.g. Cordes et al. 2006; Deneva et al. 2013; Barr et al. 2013) with high time and frequency resolution that have greatly increased the number of known MSPs, revealing new rare pulsar populations. If on-going and future surveys detect many eccentric MSPs with WD companions with $e \sim 0.1$ (and $P_{orb} = 10 - 60$ days), this would not only support our RD-AIC hypothesis; it would also imply that AIC events do not produce kicks (or at least $w \leq 5\ \mathrm{km\,s^{-1}}$, cf. Fig. 98) and that WDs rotate rigidly. Furthermore, it would imply that r-mode instabilities do not necessarily slow down young, hot MSPs, as previously suggested (Andersson, Kokkotas & Schutz 1999).

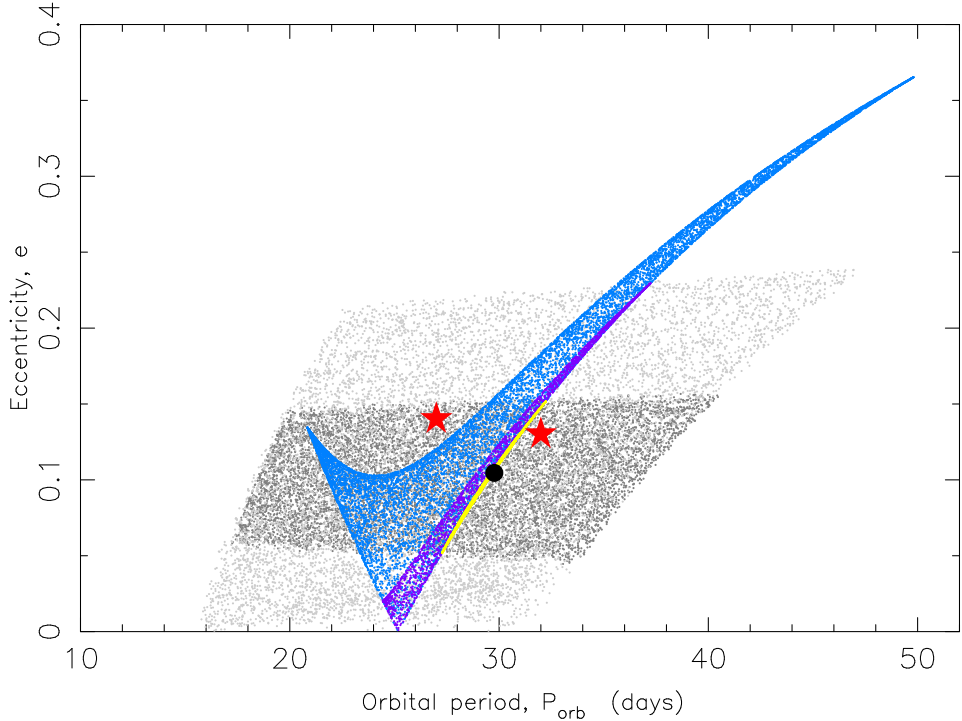

Figure 98: Distribution in the $(P_{orb}, e)$–plane of systems formed via the RD-AIC scenario. The two red stars are the recently discovered eccentric MSPs (Table 17). The black solid circle indicates a symmetric ($w = 0$) AIC from a 1.43 $M_\odot$ WD with a pre-AIC orbital period, $P_{orb,0} = 24$ days and a He WD companion star of mass, $M_2 = 0.27\ M_\odot$. The V-shaped light-blue distribution is for the same system but applying a small kick of $w = 10\ \mathrm{km\,s^{-1}}$ in a random (isotropic) direction. The indigo-violet and the yellow distributions superimposed are also for the same system but with $w = 5$ and $2\ \mathrm{km\,s^{-1}}$, respectively. The wide light grey distribution is for $w = 5\ \mathrm{km\,s^{-1}}$ and a random selection of pre-AIC systems (assuming equal probabilities), with $1.37 - 1.48\ M_\odot$ WDs and $P_{orb,0} = 15 - 30$ days (corresponding to $M_2 = 0.26 - 0.28\ M_\odot$). The dark grey distribution is similar but restricted to $w = 2\ \mathrm{km\,s^{-1}}$.

Note, there may also be eccentric MSPs with WDs formed via the triple scenario outlined in Section 10.1.2, which will have a much wider distribution in the $(P_{orb}, e)$–plane and possibly more massive companions. Detection of an MSP with a main-sequence companion and $e \sim 0.1$ would support a triple star scenario for the formation of MSPs with WDs and $e \sim 0.1$, and thus significantly weaken the need for our RD-AIC hypothesis.

Population synthesis investigations of MSP formation via AIC have been performed by Hurley et al. (2010) and Chen et al. (2011). The former study concluded that, in general, the AIC channel to MSP formation is important. The latter study investigated direct MSP formation via AIC and concluded that the probability of forming eccentric MSPs can be ruled out (Even using high kicks they could not produce eccentric MSPs with $P_{orb} \geq 20$ days), in contradiction with the new discoveries, cf. Table 17. We recommend new population synthesis modelling using our RD-AIC scenario in order to probe more carefully the expected number of such eccentric MSP systems to be detected, and for the statistics of their resulting parameter space. Ideally, the triple system scenario should be modelled for comparison as well.

Finally, it should be investigated under which circumstances a binary evolves via RD-AIC or follows the Tauris et al. (2013b) path. The latter was calculated using a point mass accreting WD which did not allow for detailed spin angular momentum modelling. For the resulting

MSPs with He WD companions, the values of $P_{orb}$ and $M_2$ are expected to be roughly similar. The RD-AIC scenario, however, produces eccentric systems.

## 10.5 Summary

The common scenario for the formation of MSPs via recycling in LMXBs is well established with plenty of observational evidence, as discussed in Section 10.1. The RD-AIC hypothesis presented in this Letter provides an additional formation channel of MSPs that makes very specific predictions about future discoveries and the existence of a separate population of eccentric MSPs. If this hypothesis is confirmed by future observations, it would also have interesting consequences for better understanding the direct AIC channel to produce MSPs, i.e. with respect to WD progenitor masses, (absence of) momentum kicks in AIC, and possibly even constraining neutron star equations-of-state given that the post-AIC eccentricities depend on the released gravitational binding energy.

## 11.1   Summary of thesis work

This Habilitationsschrift is composed of 9 papers which cover a wide range of topics related to the formation, evolution and final fate of millisecond pulsars (MSPs) in close binaries. We have investigated the formation channels of neutron stars (NSs) which eventually become observable as recycled pulsars. Within the standard scenario, MSPs are produced via old NSs (formed in a close binary via a Type Ib/c supernova (SN)) which undergo a long phase of accretion in a low-mass X-ray binary (LMXB). Although this scenario has been confirmed by convincing observational evidence and is now widely accepted, it cannot be the only formation channel. For example, it cannot explain the recent discoveries of MSPs with white dwarf (WD) companions in eccentric binary systems, nor the apparent presence of young NSs in some globular clusters. In this thesis, we have investigated two alternative formation channels.

The first one of these is *indirect* formation of MSPs via the accretion-induced collapse (AIC) of a massive WD (Tauris et al. 2013b). In this model, the implosion of a massive ONeMg WD leads to the production of a regular pulsar which is then, shortly thereafter, recycled to become an MSP via additional mass transfer from the very same donor star which delivered mass to the imploding WD. This model has been known for several decades but is still somewhat controversial due to the lack of direct observational evidence. We performed a systematic study of detailed mass-transfer modelling, using three different types of donor stars, and confirmed that this model can work – i.e. within a certain parameter range, the WD is able to grow in mass to the critical Chandrasekhar limit and implode. We found, however, that the MSPs formed via this channel cannot easily be differentiated from MSPs formed via the standard channel. The reason is that they share many characteristics in terms of their modelled spin periods, orbital periods, masses and WD companions. On the other hand, from these AIC calculations we can indeed produce young NSs in old globular clusters (as observed) if their wide-orbit progenitor binaries are disrupted shortly after the AIC, thus preventing their former giant donor stars from refilling their Roche lobes (which is not an impossible condition given the many encounter events in a dense stellar environment like a globular cluster).

In the second alternative model that we investigated (a model that we developed ourselves), the MSP is produced *directly* in a rotationally-delayed AIC event (Freire & Tauris 2014). The rationale behind this model is that the WD can gain a mass beyond the usual Chandrasekhar limit without collapsing, if it obtains a very high rotation rate via accretion. Hence, such a WD will only collapse after the mass transfer has terminated (and after sufficient loss of spin angular momentum from magnetodipole radiation or r-mode instabilities). At this point, its companion star (the former donor) has become a helium WD and therefore the post-AIC orbit with the newborn MSP will remain eccentric after the implosion event – unlike the case in the standard scenario where the orbit circularizes efficiently during the recycling process while the NS accretes material from its companion star. The advantage of this new model is therefore that it can explain why some MSPs are found in eccentric binaries with WD companions. In addition, the observed eccentricities and orbital periods match nicely with our theoretical expectations for this model. Finally, it is attractive is that our hypothesis is readily falsifiable. We should soon have measurements of the masses of these MSPs and their WDs, which can then be confronted with our theoretical predictions to confirm or reject our alternative scenario.

Another topic related to NS formation is extreme mass stripping of pre-SN stars via so-called Case BB mass transfer to a compact companion star in a tight binary. This stripping affects the final mass of the exploding star, and thus the SN ejecta mass and the observed light curve of the SN. We have demonstrated for the first time (Tauris et al. 2013a), that it is possible to produce pre-SN stars in close binaries which are basically naked metal cores barely above the Chandrasekhar mass limit of $\sim 1.4\ M_\odot$. We predict that these ultra-stripped SNe occur once for every 100–1000

SNe. Given that SN light curves are seen out to cosmological distances these ultra-stripped SNe should therefore be observable at a reasonable rate. We have produced synthetic light curves of such ultra-stripped SNe. As a result of their often tiny ejecta mass ($\sim 0.1 \ M_\odot$), and small amounts of Ni produced, the expected light curves should be rapidly decaying and relatively weak, but nevertheless observable. Among the recent detections of unusual optical transients is the reported light curve of SN 2005ek which matches very well with our predictions for an ultra-stripped SN. Drout et al. (2013) reported that SN 2005ek has the smallest ratio of ejecta mass to remnant mass of any SN observed to date. Our model predicts indeed such an extreme mass ratio.

A key question now is what kick magnitude is expected for these ultra-stripped SNe? The ultra-stripped SNe may produce low-mass NSs ($1.1 - 1.3 \ M_\odot$), depending on the mass cut during the explosion. Hence, under these circumstances, it is plausible that also iron core-collapse SNe produce low-mass NSs with small kicks – similar to the expectations for electron capture SNe – in accordance with the often small eccentricities observed among the double NS systems. Further investigations are currently initiated to answer this question.

At the moment we are finalizing a manuscript on the progenitors of electron capture SNe versus iron core-collapse SNe in close binaries. This theoretical work will be followed by an investigation of the expected properties of double NS systems. The long term goal is to implement these results into our new population synthesis code (see Section 11.2.2) and compare with observational data.

Whereas ultra-stripped SNe produce low-mass NSs, one may ask what is the upper limit of the NS birth mass from normal Type Ib/c or Type II SNe? This question is interesting from the point of view of both stellar evolution, explosion physics and the equation-of-state (EoS). Hence, we have investigated the formation of the two most massive (precisely measured) NSs known: PSR J1614−2230 and J0348+0432, and asked how much mass (if any significant) they must have accreted in order to end up with their current masses of $2 \ M_\odot$. The result of this exercise then constrains the birth masses of these NSs.

We found that in the case of PSR J1614−2230, it must have been born with a large mass of about 1.7 $M_\odot$ (Tauris, Langer & Kramer 2011). In addition, there is observational evidence from the accreting NS in Vela X-1 that it was also born with a similar value. This is interesting since the upper limit on birth mass inferred from all double NS systems is only 1.39 $M_\odot$. This is a puzzling result because, from a stellar evolution point of view, nothing prevents the birth mass of NSs in double NS systems from being larger. Keep in mind that whereas ultra-stripping effects in close binaries can produce very low-mass exploding cores, not all cores are stripped efficiently enough to be close to the Chandrasekhar mass limit. More massive helium star progenitors lead to larger metal cores prior to the explosion. In addition, massive helium stars expand less and can therefore avoid ultra stripping all together. A final note on PSR J1614−2230 is that we have argued that this system is the first case known which formed via Case A RLO in an intermediate-mass X-ray binary (IMXB), i.e. the Roche-lobe overflow (RLO) was initiated already when the $4 - 5 \ M_\odot$ donor star was still on the main sequence.

In the case of PSR J0348+0432 (Antoniadis et al. 2013), on the other hand, there does not seem to be a strong argument for the NS being born massive. Depending on its accretion efficiency, the NS could possibly have been born within a wide range of masses prior to the LMXB phase (see also Istrate et al., in prep.). What is really peculiar about PSR J0348+0432, however, is its relatively slow spin period of 39 ms and unusual strong B-field of $\sim 10^9$ G. This is very different from that of other binary pulsars with a He WD in a close orbit. These characteristics could suggest a formation via a common envelope (CE). However, as we have argued, this scenario is very difficult to reconcile with the small mass ($\sim 0.17 \ M_\odot$) of its helium WD companion. Instead, we postulate that the slow spin period might be explained by a strong braking of the NS rotation during the Roche-lobe decoupling phase (RLDP) in an LMXB system.

The effects of the RLDP on the spin evolution of a NS were calculated for the first time in Tauris (2012). I used a combination of detailed stellar evolution modelling and calculations of the accretion torque acting on the spinning NS. I demonstrated that some MSPs may lose up to 50% of their rotational energy during the RLDP – possibly even more for relatively high B-field MSPs such as PSR J0348+0432. The RLDP calculations were repeated for both an LMXB and an IMXB system and we found that the RLDP effect is negligible for the IMXB systems. The reason is that these relatively massive donors terminate their RLO on a short timescale such that the braking torque (due to the expanding magnetosphere) does not have enough time to decelerate the MSP during the RLDP.

Many detailed aspects of the recycling process of MSPs were investigated and reassessed in Tauris, Langer & Kramer (2012). In many ways, I consider this work to be the principal paper of this thesis. We demonstrated the clear differences in the evolution of LMXBs and IMXBs, leading to fully recycled MSPs with He WDs and mildly recycled MSPs with CO/ONeMg WDs, respectively. The relationships between donor masses and resulting WD masses, RLO lifetimes and MSP spin periods, were explored and clearly outlined – and shown to be nicely supported by observational data. (Yet another confirmation of why the properties of binary MSPs are fossil records of their evolutionary history.) We also discussed the concept of a spin-up line in the $P\dot{P}$– diagram and found that such a line cannot be uniquely defined. Instead, one should talk about a *spin-up valley* since the location of such a line depends on uncertain disk-magnetosphere interactions, as well as the individual mass and magnetic inclination angle of each accreting MSP. Finally, in that same publication we presented the first detailed spin-up calculations of post-CE IMXBs where a naked helium star transfers matter to an accreting NS. (It was these Case BB RLO calculations that eventually led us to perform further investigations, resulting in the discovery of the ultra-stripped SNe discussed above.)

The post-CE IMXB calculations were brought one step further with the analysis of the formation of the young, mildly recycled pulsar PSR J1952+2630 (Lazarus et al. 2014). Given the age constraint that we obtained for this system, we could place limits on its initial spin period after recycling. Therefore, we could also constrain the accretion efficiency parameter of the Case BB RLO by modelling the mass transfer in detail and combining it with our previously obtained relation between amount of mass accreted and final spin period. Interestingly enough, we found the accretion efficiency parameter for this Case BB RLO system to result in a slightly larger accretion rate than the classical Eddington accretion limit of a NS – in contrast to the efficiency of the NSs in LMXBs which often seem to accrete less than 30% of the transfered material even at sub-Eddington levels.

Last, but not least, in Tauris & van den Heuvel (2014) we had great fun trying to explain the formation of the newly discovered PSR J0337+1715 – a triple system MSP with two white dwarf companions and located in the Galactic disk (Ransom et al. 2014). This is the first time a system is observed with *three* compact objects! Therefore, this system must have experienced (at least) three major phases of mass transfer *and* survived one SN explosion as well. And yet this system remained dynamically stable throughout its entire evolution – a truly remarkable journey for a multiple stellar system. We argued that this system was not ejected from a cluster and we managed to develop a plausible formation scenario for this system based on semi-analytical calculations. Of course, there are still many uncertainties involved in this model. For example, it requires that the system underwent a CE phase where *both* the secondary and the tertiary star were captured and suffered significant spiral-in. A nice bonus of the discovery of this triple system is that it confirms the correlation between mass and orbital period of He WDs (Tauris & Savonije 1999), and it even extents its observational validation out to an orbital period of 327 days (the orbital period of the outer WD).

Many spin-off projects resulted from this thesis work and below (Section 11.2) I discuss some of these ongoing projects.

## 11.2 Ongoing projects and outlook

- *LMXB Evolution near the Bifurcation Period: Millisecond Pulsars in Tight Orbits and Formation of very Low-mass Helium White Dwarfs*
  Collaborators: A. Istrate, N. Langer, J. Antoniadis
  (PhD project of Alina Istrate)

- *Merging Neutron Stars and Black Holes*
  Collaborators: M. Kruckow, N. Langer, M. Kramer
  (PhD project of Matthias Kruckow, based on my DFG Grant: TA 964/1-1)

- *Formation of Double Neutron Star Systems via Ultra-stripped SNe*
  Collaborators: M. Kramer, N. Langer, T. Janka, Ph. Podsiadlowski, P. Freire

- *Searching for a Mildly Recycled Pulsar Orbiting a Black Hole*
  Collaborators: D. J. Champion, M. Kruckow, M. Kramer, N. Wex, N. Langer

- *Prospects of Detecting Isolated Black Holes with eROSITA*
  Collaborators: L. Grygosch
  (M.Sc. project of Lars Grygosch)

- *Hypervelocity Stars Originating from Disrupted Binaries [not discussed here]*
  Collaborators: in prep.

- *Shell Impact on Companion Stars from Type Ib/c Supernovae [not discussed here]*
  Collaborators: Z.-W. Liu, et al.

- *The Square Kilometre Array [SKA] and the Neutron Star Population*
  Collaborators: SKA Pulsar Science Working Group

### 11.2.1  LMXB evolution near the bifurcation period: millisecond pulsars in tight orbits and formation of very low-mass helium white dwarfs

MSPs are generally believed to be old NSs which have been spun up to high rotation rates via accretion from a companion star in an LMXB. This scenario has been strongly supported by various pieces of observational evidence. However, many details of this recycling scenario remain to be understood. In this project, we investigate binary evolution in converging LMXBs to study the formation of radio MSPs in tight binaries ($P_{orb} \simeq 2-15$ hr) with low-mass He WD companions. In particular, we examine:

  i) if such observed systems can be reproduced from standard prescriptions of orbital angular momentum losses (i.e. the strength and the switch-off of magnetic braking),

 ii) if our modelling of the Roche-lobe detachments can match the observed orbital periods,

iii) if the ($M_{WD}$, $P_{orb}$)–relation is valid for systems with $P_{orb} \leq 1$ d,

 iv) the thermal evolution and radial contraction of the detached donor star until it reaches the WD cooling track and investigate how the further evolution depends on the WD mass and the amount of hydrogen left in the WD envelope,

  v) the impact on early WD evolution by hydrogen flashes due to unstable CNO burning,

 vi) the formation of PSR J0348+0432 (cf. Chapter 4) and the question of its birth mass.

Numerical calculations with a detailed stellar evolution code (BEC) were used to trace the mass-transfer phase in more than 300 close LMXB systems with different initial values of donor star mass, NS mass, orbital period and so-called $\gamma$-value of magnetic braking. Subsequently, we followed the evolution of the low-mass (proto) He WDs, including stages with residual shell hydrogen burning and the associated vigorous shell flashes.

Three papers are shortly to be submitted with the results from this project. The next step is to compare and extend our results with those obtained by the new stellar evolution code MESA.

This project is undertaken by my PhD student Alina Istrate.

### 11.2.2  Merging neutron stars and black holes [DFG project grant: TA 964/1-1]

The expected detection of gravitational waves will open a new window to the Universe and compliment our knowledge of astrophysical sources obtained from photons and neutrinos. The advanced LIGO (Laser Interferometer Gravitational-wave Observatory) detector, together with its European sister VIRGO and the German GEO 600, will be operational in 2015 and reach full sensitivity in 2016. The most promising candidate sources for transient burst detections of high-frequency gravitational waves (10 Hz – 10 kHz) are merging neutron stars (NSs) and black holes (BHs). These compact objects are formed in tight binaries and undergo spiral-in due to continuous emission of gravitational waves until they finally merge in a violent event.

The main aim of this theoretical project is to calculate the rate of such merging NS–BH binaries in the local Universe using our best up-to-date knowledge of stellar and binary evolution. These results will be compared to the upcoming LIGO detection rates which will enable us to further constrain the input physics behind key binary stellar interactions. A second aim of this project is to better understand – and to quantify – the formation, location, properties and lifetimes of Galactic recycled radio pulsars in binaries with a NS or a BH. Whereas we know of 13 radio pulsars orbiting another NS we still have no detections of a radio pulsar orbiting a BH. With this project, we hope to gain theoretical knowledge about such systems which will enhance the chance of success in future radio surveys. A third outcome of this study is therefore a better knowledge of the evolution of their progenitors, the high-mass X-ray binaries (HMXBs).

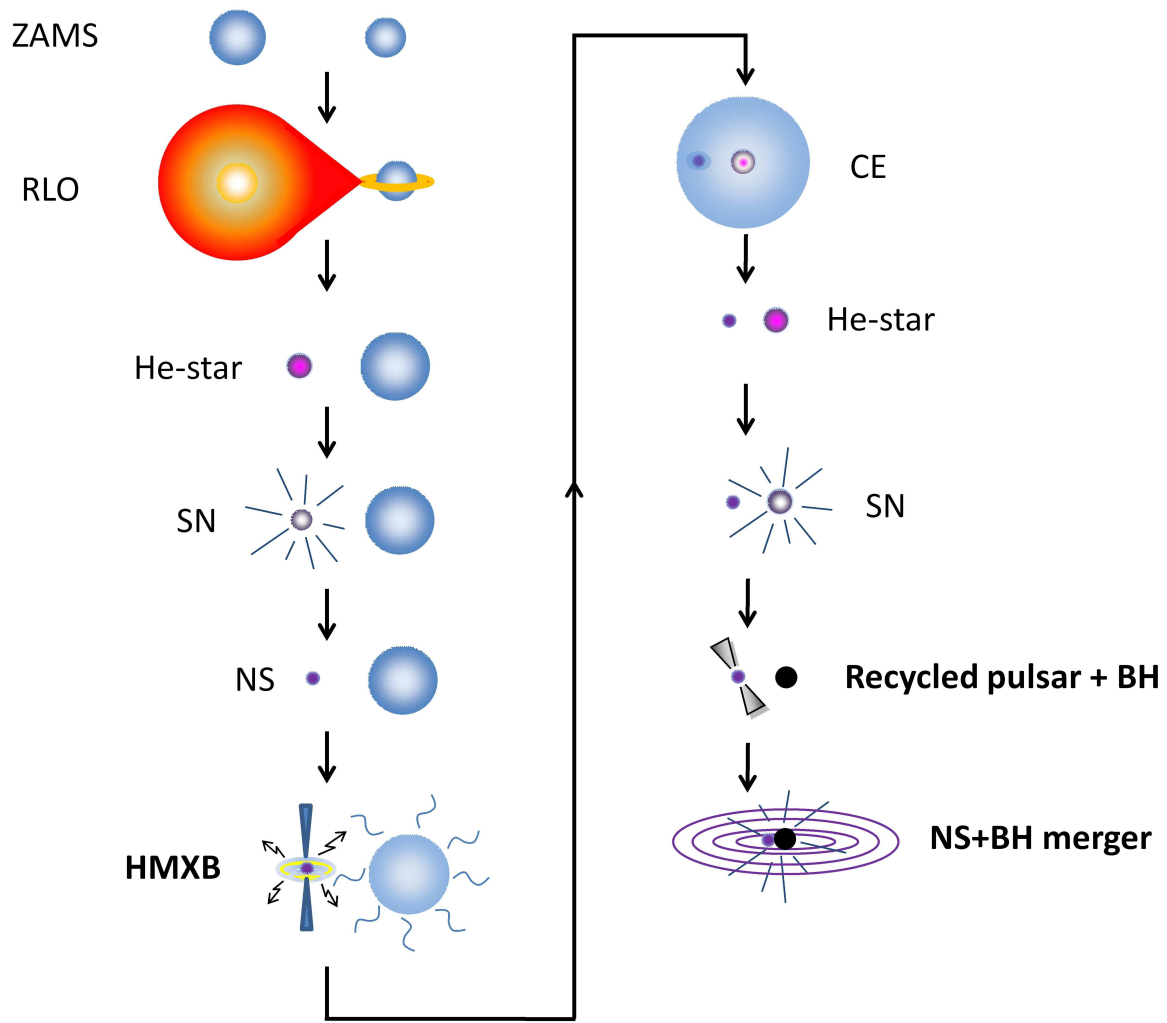

Figure 99: An illustration of the main steps during the binary evolution from the zero-age main sequence (ZAMS) to the final NS–BH merger event. The main uncertainty of the evolution from the ZAMS to the merging compact objects is the common envelope (CE) phase.

To model the population of compact binaries, the evolution of a large sample of binary stellar systems is needed in order to predict the number of observable radio pulsars orbiting NSs and BHs, as well as the LIGO detection rate of mergers. To achieve these goals we make use of advanced population synthesis techniques with Monte Carlo simulations and explore new aspects of massive binary stellar evolution for the first time. In Fig. 99 we have depicted some of the interceding steps during the binary evolution from the zero-age main sequence (ZAMS) to the final NS–BH merger event. The main uncertainty of the evolution is the common envelope (CE) phase (see e.g. Ivanova et al. 2013; for a review).

The population synthesis code used for this project was originally developed by myself and my former collaborators and students. The code has previously been applied to a variety of projects, e.g. related to: the formation of MSPs and their velocities (Tauris & Bailes 1996), dynamical effects of asymmetric SNe (Tauris & Takens 1998), and the formation of binaries where a WD forms prior to its NS companion (Tauris & Sennels 2000). The code was then updated and used for a novel study of calculating the expected merger rates of NS–BH systems in different host galaxies (Voss & Tauris 2003). I was recently awarded a DFG grant under which my PhD student Matthias Kruckow is now rewriting the code for a major revision.

### 11.2.3 Formation of double neutron star systems via ultra-stripped SNe

The magnitude of the momentum kick added to newborn NSs and BHs has been investigated in many studies (see, Janka 2012; and references therein), also in relation to the predicted merger rate of NS–BH binaries. Knowledge of kicks is important since it affects the fraction of binaries surviving the SN explosion and the post-SN orbital separations. Whereas the distribution of kicks imparted to newborn NSs is fairly well-constrained from observations (e.g. Lyne & Lorimer 1994; Hobbs et al. 2005), it is still uncertain which fraction of pulsars originate from the different types of SNe: Type II, Type Ib/c, electron capture SNe or the accretion-induced collapse of a massive ONeMg WD. The current consensus is that the latter two events result in small kicks ($< 50\ \mathrm{km\,s^{-1}}$), whereas the iron core-collapse SNe lead to large kicks of the order $400 - 500\ \mathrm{km\,s^{-1}}$. However, this picture might be different in close orbit binaries where iron core-collapse SNe of ultra-stripped cores can occur.

Podsiadlowski et al. (2004) first investigated the formation of NSs via electron capture SNe in binaries. We have recently performed a systematic investigation of this formation channel in double NS systems and constrained the initial parameter space leading to such events (Tauris, Langer & Podsiadlowski, in prep.). The next step is to explode some of these ultra-stripped stars to learn about which kicks to expect from such SNe, and then include this knowledge in simulations to learn about the expected properties of the double NS population in terms of their orbital separations, eccentricities, NS masses and systemic recoil velocities. These predictions will be compared to observational data.

Ultimately, we want to integrate our updated population synthesis code with the outcomes of the detailed binary stellar evolution calculations to revise the predictions about the double NS population and the theoretically estimated merger rates of NS binaries.

Collaborators: M. Kramer, N. Langer, T. Janka, Ph. Podsiadlowski, P. Freire

### 11.2.4 Searching for a mildly recycled pulsar orbiting a black hole

Binary pulsars are currently the only gravity laboratories that allow for high-precision tests of possible strong field deviations from general relativity (Kramer et al. 2006b; Kramer & Wex 2009; Freire, Kramer & Wex 2012; Antoniadis et al. 2013; Wex 2014). The outcome of these tests is a vital part of our quest to understand the fundamental nature of gravity and space-time. A particular interesting system would be a recycled radio pulsar orbiting a BH. Such a system would provide a clean test of a BH space-time (i.e. without circumstellar material or tidal interactions). Moreover, even for gravity theories that predict the same properties for BHs as general relativity, like scalar-tensor theories, a NS–BH binary would constitute an outstanding test system, due to the large difference in the strong field properties of these two components. Despite many efforts, until this date, no pulsar binary with a BH companion has been discovered. It is therefore important to model the expected properties (masses, orbital periods, spin periods, eccentricities) and the most likely environments (birth places and systemic velocities) of such systems in order to improve targeted survey strategies in the coming years, for example, in selected star forming regions of our Galaxy or the Magellanic Clouds. A priori knowledge (predictions) of the properties of NS–BH binaries will also help realizing a detection since the search for the (otherwise unknown) Doppler accelerated radio signals in a close binary is limited by computational power. Hence, modelling NS–BH binary properties is important – not only from a binary evolutionary point of view but also for optimizing radio surveys and for comparison with the rate of NS–BH mergers detected by LIGO/VIRGO.

Collaborators: D. J. Champion, M. Kruckow, M. Kramer, N. Wex, N. Langer

### 11.2.5 Prospects of detecting isolated black holes with eROSITA

The German/Russian satellite eROSITA will be launched ultimo 2015 (Merloni et al. 2012). Given that as many as 35 000 isolated BHs may be present within only 250 pc from our Sun (Fender, Maccarone & Heywood 2013) it is interesting to investigate if some of these isolated BHs accrete sufficient gas from the interstellar medium (molecular gas clouds) to make them detectable with the eROSITA all-sky survey. The aims of this project are to:

i) reassess the estimate for the local number density of isolated BHs,

ii) perform Monte Carlo simulations of their X-ray emission, and

iii) highlight the prospects for detection using eROSITA.

This project is undertaken by my Master's student Lars Grygosch.

### 11.2.6 The Square Kilometre Array (SKA) and the neutron star population

*This section was partly adapted from the introduction of Tauris, Kaspi, Breton, Deller, et al. (2014), SKA Science Book Chapter "Understanding the Neutron Star Population".*

The surveys that will be performed with SKA-low and SKA-mid will produce a tenfold increase in the number of Galactic NSs known. Moreover, the SKA's broad spectral coverage, sub-arraying and multi-beaming capabilities will allow us to characterise these sources with unprecedented accuracy, in turn enabling a giant leap in the understanding of their properties. It is important to outline our strategies for studying each of the growing number of diverse classes that are populating the 'NS zoo'. Some of the main scientific questions (to name just a few) that will be addressed by the much larger statistical samples and vastly improved timing accuracy provided by SKA Phase I include: (i) the mass distribution of NSs; (ii) questions as to the origin of high eccentricity MSPs; (iii) the formation channels for recently identified triple systems whose creation was thought to be impossible; (iv) how isolated MSPs are formed; (v) the limits on pulsar spin rates – the smallest possible period for a recycled pulsar reflects allowable magnetosphere-accretion disc interactions, the longest possible pulsar period reveals the conditions where pulsars die; (vi) the period and spin-down rate distributions at NS birth, and the associated information about the SNe wherein they are formed; and (vii) proper motions which reveal the distribution of SN kick magnitudes. As well as this lengthy (but not exhaustive) scientific shopping list, we can expect that SKA1, and in particular the full SKA Phase II, will break new ground unveiling exotic and heretofore unknown systems that will challenge our current knowledge and theories, thus fostering the development of new research areas. While predicting the future is difficult, the following discoveries would be possible with the SKA (should they exist), and each in their own right would represent significant milestones in the astrophysics of compact objects: (i) NS–BH binaries; (ii) NS–NS–WD triple systems; (iii) sub-ms MSPs; and (iv) NSs with a mass below 1.1 or above 2.5 $M_\odot$.

The observations from the SKA will provide new insight to better understand the formation, evolution and interaction of pulsars in a broad range of areas: intrinsic birth properties (including SN physics), torque evolution, emission mechanisms, binary and multiple system interactions and stellar physics, internal structure of NSs, and various fundamental gravity tests.

What can be expected from a tenfold increase in the number of radio pulsars? Assume that the SKA (under the best circumstances) will roughly tenfold the number of known pulsars. It can be difficult to imagine what is to be expected other than simply 10 times more of the same kind of pulsars. But numbers matter! Moreover, and more importantly, an interesting comparison is look back in history. In the late 1970's there were about 1/10 the number of known pulsars compared to today. The following NSs were *not* known at that time (see Chapter 1 for further descriptions of these classes):

- MSPs (radio MSPs, X-ray and γ-ray MSPs)

- Binary pulsars (besides from one system; today we know more than 200 pulsar binaries)

- Triple pulsars

- Pulsars in globular clusters (today we know 143 pulsars in globular clusters)

- Eclipsing MSPs (i.e. the *spiders*) and transitional MSPs

- Intermittent pulsars and RRATs

- Magnetars

- NSs more massive than 1.4 $M_\odot$

It would have been an impossible task in the late 1970's (even for the wildest speculator) to foresee this diversity of upcoming NS discoveries. Similarly, we can expect that the SKA will break new ground and reveal exotic NSs that will challenge our current knowledge and theories, and thus foster continued progress and the development of new research topics.

Collaborators: SKA Pulsar Science Working Group

## 11.3  CV

**Short CV for Thomas M. Tauris (born 1968, Viborg, Denmark)**

### EDUCATION

| | |
|---|---|
| 2002−2003 | Diploma in Higher Education Teaching Practice, University of Copenhagen, A course for university assistant/associate professors. |
| 1993−1997 | Ph.D., Aarhus University (Denmark) / ATNF-CSIRO Sydney (Australia), Astrophysics: "Formation and Evolution of Binary Millisecond Pulsars". |
| 1988−1993 | M.Sc., Aarhus University. Astrophysics (minor degree in science history). |
| 1984−1988 | Baccalaureate, Silkeborg Gymnasium / High School Exchange Student, USA. |

### CAREER

| | |
|---|---|
| 2010−present | Visiting Research Professor, AIfA, University of Bonn / MPIfR. |
| 2007−2010 | External Associate Professor, Niels Bohr Institute, University of Copenhagen. |
| 2003−2010 | International Baccalaureate (IB) teacher in physics/math at Herlufsholm Skole. |
| 2002−2003 | Assistant Professor, Niels Bohr Institute, University of Copenhagen. |
| 2000−2002 | Research Fellow, NORDITA (Nordic Institute for Theoretical Physics). |
| 1997−2000 | Marie Curie Fellow, University of Amsterdam (individual EU research grant). |

# References

Aasi J. et al., 2013, LIGO-P1200087, VIR-0288A-12, ArXiv astro-ph:1304.0670

Abadie J. et al., 2010, Classical and Quantum Gravity, 27, 173001

Alberts F., Savonije G. J., van den Heuvel E. P. J., Pols O. R., 1996, Nature, 380, 676

Allen B., Knispel B., Cordes J. M., et al., 2013, ApJ, 773, 91

Alpar M. A., Cheng A. F., Ruderman M. A., Shaham J., 1982, Nature, 300, 728

Alsing J., Berti E., Will C. M., Zaglauer H., 2012, Phys. Rev. D, 85, 064041

Althaus L. G., Serenelli A. M., Benvenuto O. G., 2001, MNRAS, 324, 617

Andersson N., 1998, ApJ, 502, 708

Andersson N., Kokkotas K., Schutz B. F., 1999, ApJ, 510, 846

Antoniadis J. et al., 2013, Science, 340, 448

Antoniadis J., van Kerkwijk M. H., Koester D., Freire P. C. C., Wex N., Tauris T. M., Kramer M., Bassa C. G., 2012, MNRAS, 423, 3316

Appenzeller I. et al., 1998, The Messenger, 94, 1

Applegate J. H., 1992, ApJ, 385, 621

Archibald A. M. et al., 2009, Science, 324, 1411

Arnett W. D., 1982, ApJ, 253, 785

Baade W., Zwicky F., 1934, Physical Review, 46, 76

Backer D. C., Kulkarni S. R., Heiles C., Davis M. M., Goss W. M., 1982, Nature, 300, 615

Bagchi M., 2011, MNRAS, 413, L47

Bagnulo S., Jehin E., Ledoux C., Cabanac R., Melo C., Gilmozzi R., ESO Paranal Science Operations Team, 2003, The Messenger, 114, 10

Bailes M. et al., 2011, Science, 333, 1717

Bailes M. et al., 1994, ApJ, 425, L41

Bailyn C. D., 1987, PhD thesis, Harvard University, Cambridge, MA.

Bailyn C. D., Grindlay J. E., 1990, ApJ, 353, 159

Barnes J., Kasen D., 2013, ApJ, 775, 18

Barr E. D. et al., 2013, MNRAS, 435, 2234

Barziv O., Kaper L., Van Kerkwijk M. H., Telting J. H., Van Paradijs J., 2001, A&A, 377, 925

Bassa C. G., van Kerkwijk M. H., Koester D., Verbunt F., 2006, A&A, 456, 295

Bassa C. G., van Kerkwijk M. H., Kulkarni S. R., 2003, A&A, 403, 1067

Bates S. D. et al., 2011, MNRAS, 416, 2455

Bejger M., Zdunik J. L., Haensel P., Fortin M., 2011, A&A, 536, A92

Benvenuto O. G., De Vito M. A., Horvath J. E., 2012, ApJ, 753, L33

Benvenuto O. G., De Vito M. A., Horvath J. E., 2014, ApJ, 786, L7

Berger E., 2010, ApJ, 722, 1946

Berger E., Fong W., Chornock R., 2013, ApJ, 774, L23

Bertotti B., Iess L., Tortora P., 2003, Nature, 425, 374

Beskin V. S., Gurevich A. V., Istomin I. N., 1988, Ap&SS, 146, 205

Bhalerao V. B., Kulkarni S. R., 2011, ApJ, 737, L1

Bhat N. D. R., Bailes M., Verbiest J. P. W., 2008, Phys. Rev. D, 77, 124017

Bhattacharya D., 2002, Journal of Astrophysics and Astronomy, 23, 67

Bhattacharya D., van den Heuvel E. P. J., 1991, Physics Reports, 203, 1

Bhattacharya D., Wijers R. A. M. J., Hartman J. W., Verbunt F., 1992, A&A, 254, 198

Biggs J. D., Bailes M., Lyne A. G., Goss W. M., Fruchter A. S., 1994, MNRAS, 267, 125

Bildsten L., 1998, ApJ, 501, L89

Bildsten L. et al., 1997, ApJS, 113, 367

Bildsten L., Shen K. J., Weinberg N. N., Nelemans G., 2007, ApJ, 662, L95

Bisnovatyi-Kogan G. S., Komberg B. V., 1974, AZh, 51, 373

Blanchet L., 2006, Living Reviews in Relativity, 9, 4

Blandford R. D., Begelman M. C., 1999, MNRAS, 303, L1

Blandford R. D., Payne D. G., 1982, MNRAS, 199, 883

Blinnikov S. I., Novikov I. D., Perevodchikova T. V., Polnarev A. G., 1984, Soviet Astronomy Letters, 10, 177

Blinnikov S. I., Röpke F. K., Sorokina E. I., Gieseler M., Reinecke M., Travaglio C., Hillebrandt W., Stritzinger M., 2006, A&A, 453, 229

Bonnell I. A., Bate M. R., Vine S. G., 2003, MNRAS, 343, 413

Bours M. C. P., Toonen S., Nelemans G., 2013, A&A, 552, A24

Bowers R. L., Deeming T., 1984, Astrophysics. Volume 1 - Stars, Jones and Bartlett Publishers

Boyles J., Lorimer D. R., Turk P. J., Mnatsakanov R., Lynch R. S., Ransom S. M., Freire P. C., Belczynski K., 2011, ApJ, 742, 51

Boyles J. et al., 2013, ApJ, 763, 80

Bozzo E., Stella L., Vietri M., Ghosh P., 2009, A&A, 493, 809

Braun H., 1997, PhD thesis, , Ludwig-Maximilians-Univ. München, (1997)

Breton R. P., Rappaport S. A., van Kerkwijk M. H., Carter J. A., 2012, ApJ, 748, 115

Breton R. P., Roberts M. S. E., Ransom S. M., Kaspi V. M., Durant M., Bergeron P., Faulkner A. J., 2007, ApJ, 661, 1073

Breton R. P. et al., 2013, ApJ, 769, 108

Brown G. E., 1995, ApJ, 440, 270

Brown G. E., Heger A., Langer N., Lee C., Wellstein S., Bethe H. A., 2001, New Astronomy, 6, 457

Büning A., Ritter H., 2004, A&A, 423, 281

Burderi L., D'Antona F., Burgay M., 2002, ApJ, 574, 325

Burderi L. et al., 2001, ApJ, 560, L71

Callanan P. J., Garnavich P. M., Koester D., 1998, MNRAS, 298, 207

Camenzind M., 2007, Compact objects in astrophysics : white dwarfs, neutron stars, and black holes

Camero-Arranz A., Pottschmidt K., Finger M. H., Ikhsanov N. R., Wilson-Hodge C. A., Marcu D. M., 2012, A&A, 546, A40

Camilo F., 1996, in Astronomical Society of the Pacific Conference Series, Vol. 105, IAU Colloq. 160: Pulsars: Problems and Progress, S. Johnston, M. A. Walker, & M. Bailes, ed., pp. 539–+

Camilo F. et al., 2001, ApJ, 548, L187

Camilo F., Nice D. J., Shrauner J. A., Taylor J. H., 1996, ApJ, 469, 819

Camilo F., Ransom S. M., Chatterjee S., Johnston S., Demorest P., 2012, ApJ, 746, 63

Camilo F., Thorsett S. E., Kulkarni S. R., 1994, ApJ, 421, L15

Campana S., Colpi M., Mereghetti S., Stella L., Tavani M., 1998, A&A Rev., 8, 279

Canal R., Isern J., Labay J., 1990, ARA&A, 28, 183

Canuto V., 1970, ApJ, 159, 641

Casares J., Charles P. A., Kuulkers E., 1998, ApJ, 493, L39+

Cassisi S., Iben, Jr. I., Tornambe A., 1998, ApJ, 496, 376

Chakrabarty D., 2008, in American Institute of Physics Conference Series, Vol. 1068, American Institute of Physics Conference Series, Wijnands R., Altamirano D., Soleri P., Degenaar N., Rea N., Casella P., Patruno A., Linares M., eds., pp. 67–74

Chakrabarty D., Morgan E. H., Muno M. P., Galloway D. K., Wijnands R., van der Klis M., Markwardt C. B., 2003, Nature, 424, 42

Champion D. J., Ransom S. M., Lazarus P., et al., 2008, Science, 320, 1309

Chandrasekhar S., 1970, Physical Review Letters, 24, 611

Chaty S., 2013, Advances in Space Research, 52, 2132

Chen H.-L., Chen X., Tauris T. M., Han Z., 2013, ApJ, 775, 27

Chen K., Ruderman M., 1993, ApJ, 402, 264

Chen W.-C., Liu X.-W., Xu R.-X., Li X.-D., 2011, MNRAS, 410, 1441

Chevalier R. A., 1993, ApJ, 411, L33

Chomiuk L. et al., 2012, ApJ, 750, 164

Claret A., 2007, A&A, 475, 1019

Contopoulos I., Spitkovsky A., 2006, ApJ, 643, 1139

Corbet R. H. D., 1984, A&A, 141, 91

Cordes J. M., 1979, Space Sci. Rev., 24, 567

Cordes J. M. et al., 2006, ApJ, 637, 446

Cordes J. M., Lazio T. J. W., 2002, ArXiv: astro-ph/0207156

Coriat M., Fender R. P., Dubus G., 2012, MNRAS, 424, 1991

Corongiu A. et al., 2012, ApJ, 760, 100

Córsico A. H., Romero A. D., Althaus L. G., Hermes J. J., 2012, A&A, 547, A96

Crawford F., Stovall K., Lyne A. G., et al., 2012, ApJ, 757, 90

Cumming A., Zweibel E., Bildsten L., 2001, ApJ, 557, 958

Dai H.-L., Li X.-D., 2006, A&A, 451, 581

D'Amico N., 2000, in Astronomical Society of the Pacific Conference Series, Vol. 202, IAU
    Colloq. 177: Pulsar Astronomy - 2000 and Beyond, M. Kramer, N. Wex, & R. Wielebinski,
    ed., p. 27

Damour T., 2009, in Astrophysics and Space Science Library, Vol. 359, Astrophysics and Space
    Science Library, Colpi M., Casella P., Gorini V., Moschella U., Possenti A., eds., pp. 1–4020

Damour T., Deruelle N., 1985, Ann. Inst. Henri Poincaré Phys. Théor., Vol. 43, No. 1, p. 107 -
    132, 43, 107

Damour T., Deruelle N., 1986, Ann. Inst. Henri Poincaré Phys. Théor., Vol. 44, No. 3, p. 263 -
    292, 44, 263

Damour T., Esposito-Farese G., 1992, Classical and Quantum Gravity, 9, 2093

Damour T., Esposito-Farese G., 1993, Physical Review Letters, 70, 2220

Damour T., Esposito-Farèse G., 1996, Phys. Rev. D, 54, 1474

Damour T., Esposito-Farèse G., 1998, Phys. Rev. D, 58, 042001

Damour T., Gibbons G. W., Taylor J. H., 1988, Physical Review Letters, 61, 1151

Damour T., Taylor J. H., 1991, ApJ, 366, 501

Damour T., Taylor J. H., 1992, Phys. Rev. D, 45, 1840

D'Angelo C. R., Spruit H. C., 2010, MNRAS, 406, 1208

D'Angelo C. R., Spruit H. C., 2011, MNRAS, 416, 893

D'Angelo C. R., Spruit H. C., 2012, MNRAS, 420, 416

D'Antona F., Mazzitelli I., Ritter H., 1989, A&A, 225, 391

Darbha S., Metzger B. D., Quataert E., Kasen D., Nugent P., Thomas R., 2010, MNRAS, 409,
    846

Das U., Mukhopadhyay B., 2013, Physical Review Letters, 110, 071102

Davidson K., Ostriker J. P., 1973, ApJ, 179, 585

de Kool M., 1990, ApJ, 358, 189

de Marco O., Passy J., Moe M., Herwig F., Mac Low M., Paxton B., 2011, MNRAS, 28

de Vries N., Portegies Zwart S., Figueira J., 2014, MNRAS, 438, 1909

Deloye C. J., 2008, in American Institute of Physics Conference Series, Vol. 983, 40 Years of Pulsars: Millisecond Pulsars, Magnetars and More, C. Bassa, Z. Wang, A. Cumming, & V. M. Kaspi, ed., pp. 501–509

Deloye C. J., Bildsten L., 2003, ApJ, 598, 1217

Demorest P. B., Pennucci T., Ransom S. M., Roberts M. S. E., Hessels J. W. T., 2010, Nature, 467, 1081

Deneva J. S., Freire P. C. C., Cordes J. M., et al., 2012, ApJ, 757, 89

Deneva J. S., Stovall K., McLaughlin M. A., Bates S. D., Freire P. C. C., Martinez J. G., Jenet F., Bagchi M., 2013, ApJ, 775, 51

Denissenkov P. A., Herwig F., Bildsten L., Paxton B., 2013, ApJ, 762, 8

Dermer C. D., Atoyan A., 2006, ApJ, 643, L13

Dermine T., Izzard R. G., Jorissen A., Van Winckel H., 2013, A&A, 551, A50

Dessart L., Burrows A., Livne E., Ott C. D., 2007, ApJ, 669, 585

Dessart L., Burrows A., Ott C. D., Livne E., Yoon S.-C., Langer N., 2006, ApJ, 644, 1063

Dessart L., Hillier D. J., Livne E., Yoon S.-C., Woosley S., Waldman R., Langer N., 2011, MNRAS, 414, 2985

Dewi J. D. M., Podsiadlowski P., Sena A., 2006, MNRAS, 368, 1742

Dewi J. D. M., Pols O. R., 2003, MNRAS, 344, 629

Dewi J. D. M., Pols O. R., Savonije G. J., van den Heuvel E. P. J., 2002, MNRAS, 331, 1027

Dewi J. D. M., Tauris T. M., 2000, A&A, 360, 1043

Dewi J. D. M., Tauris T. M., 2001, in Astronomical Society of the Pacific Conference Series, Vol. 229, Evolution of Binary and Multiple Star Systems, P. Podsiadlowski, S. Rappaport, A. R. King, F. D'Antona, & L. Burderi , ed., p. 255

Dhillon V. S. et al., 2007, MNRAS, 378, 825

Di Stefano R., Kilic M., 2012, ApJ, 759, 56

Di Stefano R., Voss R., Claeys J. S. W., 2011, ApJ, 738, L1

Dowd A., Sisk W., Hagen J., 2000, in Astronomical Society of the Pacific Conference Series, Vol. 202, IAU Colloq. 177: Pulsar Astronomy - 2000 and Beyond, Kramer M., Wex N., Wielebinski R., eds., p. 275

Driebe T., Schoenberner D., Bloecker T., Herwig F., 1998, A&A, 339, 123

Drout M. R., Soderberg A. M., Mazzali P. A., et al., 2013, ApJ, 774, 58

Dubus G., Lasota J.-P., Hameury J.-M., Charles P., 1999, MNRAS, 303, 139

Eddington A. S., 1926, The Internal Constitution of the Stars, Cambridge University Press

Eggleton P., Kiseleva L., 1995, ApJ, 455, 640

Eggleton P. P., 1983, ApJ, 268, 368

Eggleton P. P., Kiseleva L. G., 1996, in NATO ASIC Proc. 477: Evolutionary Processes in Binary Stars, Wijers R. A. M. J., Davies M. B., Tout C. A., eds., p. 345

Eichler D., Livio M., Piran T., Schramm D. N., 1989, Nature, 340, 126

Eldridge J. J., Izzard R. G., Tout C. A., 2008, MNRAS, 384, 1109

Elsner R. F., Ghosh P., Lamb F. K., 1980, ApJ, 241, L155

Ergma E., Sarna M. J., Antipova J., 1998, MNRAS, 300, 352

Fender R. P., Maccarone T. J., Heywood I., 2013, MNRAS, 430, 1538

Ferdman R. D. et al., 2010, ApJ, 711, 764

Folkner W. M., Williams J. G., Boggs D. H., 2009, Interplanetary Network Progress Report, 178, C1

Fox D. B., Frail D. A., Price P. A., et al., 2005, Nature, 437, 845

Frank J., King A., Raine D. J., 2002, Accretion Power in Astrophysics: Third Edition. Cambridge University Press

Freire P. C. C., Abdo A. A., Ajello M., et al., 2011a, Science, 334, 1107

Freire P. C. C. et al., 2011b, MNRAS, 412, 2763

Freire P. C. C., Kramer M., Wex N., 2012, Classical and Quantum Gravity, 29, 184007

Freire P. C. C., Ransom S. M., Bégin S., Stairs I. H., Hessels J. W. T., Frey L. H., Camilo F., 2008, ApJ, 675, 670

Freire P. C. C., Tauris T. M., 2014, MNRAS, 438, L86

Freire P. C. C., Wex N., 2010, MNRAS, 409, 199

Freire P. C. C. et al., 2012, MNRAS, 423, 3328

Friedman J. L., Schutz B. F., 1978, ApJ, 222, 281

Fruchter A. S., Gunn J. E., Lauer T. R., Dressler A., 1988, Nature, 334, 686

Fruchter A. S., Stinebring D. R., Taylor J. H., 1988, Nature, 333, 237

Fryer C. L., 1999, ApJ, 522, 413

Fryer C. L., 2006, New A Rev., 50, 492

Fryer C. L., Benz W., Herant M., 1996, ApJ, 460, 801

Fujii Y., Maeda K.-I., 2003, The Scalar-Tensor Theory of Gravitation

Geppert U., Urpin V., 1994, MNRAS, 271, 490

Ghosh P., Lamb F. K., 1979a, ApJ, 232, 259

Ghosh P., Lamb F. K., 1979b, ApJ, 234, 296

Ghosh P., Lamb F. K., 1992, in X-Ray Binaries and Recycled Pulsars, E.P.J. van den Heuvel, S.A. Rappaport, ed., Springer, pp. 487–510

Glendenning N. K., ed., 2000, Compact stars : nuclear physics, particle physics, and general relativity

Goenner H., 2012, General Relativity and Gravitation, 44, 2077

Goldreich P., Julian W. H., 1969, ApJ, 157, 869

Gray D. F., 2005, The Observation and Analysis of Stellar Photospheres

Guillemot L., Tauris T. M., 2014, MNRAS, 439, 2033

Habets G. M. H. J., 1986, A&A, 165, 95

Hachinger S., Mazzali P. A., Taubenberger S., Hillebrandt W., Nomoto K., Sauer D. N., 2012, MNRAS, 422, 70

Hachisu I., Kato M., 2001, ApJ, 558, 323

Hachisu I., Kato M., Nomoto K., 1996, ApJ, 470, L97+

Hachisu I., Kato M., Nomoto K., 1999, ApJ, 522, 487

Hachisu I., Kato M., Nomoto K., 2008, ApJ, 679, 1390

Hachisu I., Kato M., Saio H., Nomoto K., 2012, ApJ, 744, 69

Haensel P., Proszynski M., Kutschera M., 1981, A&A, 102, 299

Haensel P., Zdunik J. L., 1989, Nature, 340, 617

Han Z., Podsiadlowski P., 2004, MNRAS, 350, 1301

Han Z., Podsiadlowski P., Eggleton P. P., 1994, MNRAS, 270, 121

Han Z., Podsiadlowski P., Eggleton P. P., 1995, MNRAS, 272, 800

Harrington R. S., 1972, Celestial Mechanics, 6, 322

Hayakawa S., 1985, Phys. Rep., 121, 317

Heger A., Fryer C. L., Woosley S. E., Langer N., Hartmann D. H., 2003, ApJ, 591, 288

Heger A., Langer N., Woosley S. E., 2000, ApJ, 528, 368

Heggie D. C., Hut P., McMillan S. L. W., 1996, ApJ, 467, 359

Heggie D. C., Rasio F. A., 1996, MNRAS, 282, 1064

Helfand D. J., Ruderman M. A., Shaham J., 1983, Nature, 304, 423

Hellier C., 2001, Cataclysmic Variable Stars. Springer

Henyey L., Vardya M. S., Bodenheimer P., 1965, ApJ, 142, 841

Hermes J. J., Kilic M., Brown W. R., Montgomery M. H., Winget D. E., 2012a, ApJ, 749, 42

Hermes J. J. et al., 2013, ApJ, 765, 102

Hermes J. J., Montgomery M. H., Winget D. E., Brown W. R., Kilic M., Kenyon S. J., 2012b, ApJ, 750, L28

Hessels J., Ransom S., Roberts M., Kaspi V., Livingstone M., Tam C., Crawford F., 2005, in Astronomical Society of the Pacific Conference Series, Vol. 328, Binary Radio Pulsars, F. A. Rasio & I. H. Stairs, ed., p. 395

Hessels J. W. T., 2008, in American Institute of Physics Conference Series, Vol. 1068, American Institute of Physics Conference Series, Wijnands R., Altamirano D., Soleri P., Degenaar N., Rea N., Casella P., Patruno A., Linares M., eds., pp. 130–134

Hessels J. W. T., Ransom S. M., Stairs I. H., Freire P. C. C., Kaspi V. M., Camilo F., 2006, Science, 311, 1901

Hewish A., Bell S. J., Pilkington J. D. H., Scott P. F., Collins R. A., 1968, Nature, 217, 709

Hills J. G., 1983, ApJ, 267, 322

Hjellming M. S., Webbink R. F., 1987, ApJ, 318, 794

Ho W. C. G., Heinke C. O., 2009, Nature, 462, 71

Ho W. C. G., Maccarone T. J., Andersson N., 2011, ApJ, 730, L36

Hobbs D., Holl B., Lindegren L., Raison F., Klioner S., Butkevich A., 2010a, in IAU Symposium, Vol. 261, IAU Symposium, Klioner S. A., Seidelmann P. K., Soffel M. H., eds., pp. 315–319

Hobbs G., Archibald A., Arzoumanian Z., et al., 2010b, Classical and Quantum Gravity, 27, 084013

Hobbs G. et al., 2012, MNRAS, 427, 2780

Hobbs G. et al., 2004, MNRAS, 352, 1439

Hobbs G., Lorimer D. R., Lyne A. G., Kramer M., 2005, MNRAS, 360, 974

Hobbs G. B., Edwards R. T., Manchester R. N., 2006, MNRAS, 369, 655

Hofmann F., Müller J., Biskupek L., 2010, A&A, 522, L5

Hotan A. W., van Straten W., Manchester R. N., 2004, PASA, 21, 302

Howell D. A. et al., 2006, Nature, 443, 308

Hubbard W. B., Lampe M., 1969, ApJS, 18, 297

Hurley J. R., Tout C. A., Wickramasinghe D. T., Ferrario L., Kiel P. D., 2010, MNRAS, 402, 1437

Iben, Jr. I., Livio M., 1993, PASP, 105, 1373

Iben, Jr. I., MacDonald J., 1985, ApJ, 296, 540

Iben, Jr. I., Tutukov A. V., 1984, ApJS, 54, 335

Iben, Jr. I., Tutukov A. V., 1985, ApJS, 58, 661

Iben, Jr. I., Tutukov A. V., 1989, ApJ, 342, 430

Iben, Jr. I., Tutukov A. V., 1996, ApJS, 105, 145

Iben, Jr. I., Tutukov A. V., 1999, ApJ, 511, 324

Iben, Jr. I., Tutukov A. V., Yungelson L. R., 1997, ApJ, 475, 291

Idan I., Shaviv N. J., Shaviv G., 2012, Journal of Physics Conference Series, 337, 012051

Iglesias C. A., Rogers F. J., 1996, ApJ, 464, 943

Iglesias C. A., Rogers F. J., Wilson B. G., 1992, ApJ, 397, 717

Ikhsanov N. R., Beskrovnaya N. G., 2012, Astronomy Reports, 56, 589

Ilkov M., Soker N., 2012, MNRAS, 419, 1695

Illarionov A. F., Sunyaev R. A., 1975, A&A, 39, 185

Ivanova N., 2011, ApJ, 730, 76

Ivanova N., Belczynski K., Kalogera V., Rasio F. A., Taam R. E., 2003, ApJ, 592, 475

Ivanova N., Heinke C. O., Rasio F. A., Belczynski K., Fregeau J. M., 2008, MNRAS, 386, 553

Ivanova N. et al., 2013, A&A Rev., 21, 59

Jacoby B. A., Hotan A., Bailes M., Ord S., Kulkarni S. R., 2005, ApJ, 629, L113

Janka H.-T., 2012, Annual Review of Nuclear and Particle Science, 62, 407

Janssen G. H., Stappers B. W., Bassa C. G., Cognard I., Kramer M., Theureau G., 2010, A&A, 514, A74

Janssen G. H., Stappers B. W., Kramer M., Nice D. J., Jessner A., Cognard I., Purver M. B., 2008, A&A, 490, 753

Jeffery C. S., Hamann W.-R., 2010, MNRAS, 404, 1698

Jones S. et al., 2013, ApJ, 772, 150

Jordan S., Aznar Cuadrado R., Napiwotzki R., Schmid H. M., Solanki S. K., 2007, A&A, 462, 1097

Jose J., Hernanz M., Isern J., 1993, A&A, 269, 291

Justham S., 2011, ApJ, 730, L34

Kaaret P. et al., 2007, ApJ, 657, L97

Kahabka P., van den Heuvel E. P. J., 1997, ARA&A, 35, 69

Kalogera V., 1998, ApJ, 493, 368

Kalogera V., Webbink R. F., 1996, ApJ, 458, 301

Kaplan D. L., van Kerkwijk M. H., Koester D., Stairs I. H., Ransom S. M., Archibald A. M., Hessels J. W. T., Boyles J., 2014, ApJ, 783, L23

Kasliwal M. M., Kulkarni S. R., Gal-Yam A., et al., 2010, ApJ, 723, L98

Kasliwal M. M., Nissanke S., 2013, ArXiv astro-ph:1309.1554

Kaspi V. M., 2010, Proceedings of the National Academy of Science, 107, 7147

Kaspi V. M., Johnston S., Bell J. F., Manchester R. N., Bailes M., Bessell M., Lyne A. G., D'Amico N., 1994, ApJ, 423, L43

Kato M., Hachisu I., 1994, ApJ, 437, 802

Kato M., Hachisu I., 2004, ApJ, 613, L129

Kato M., Iben, Jr. I., 1992, ApJ, 394, 305

Keith M. J. et al., 2010, MNRAS, 409, 619

Kenyon S. J., 1986, The symbiotic stars. Cambridge University Press

Kilic M., Brown W. R., Allende Prieto C., Kenyon S. J., Panei J. A., 2010, ApJ, 716, 122

Kim C., Kalogera V., Lorimer D. R., White T., 2004, ApJ, 616, 1109

King A. R., Begelman M. C., 1999, ApJ, 519, L169

King A. R., Ritter H., 1999, MNRAS, 309, 253

King A. R., Schenker K., Kolb U., Davies M. B., 2001, MNRAS, 321, 327

Kippenhahn R., Weigert A., 1967, ZAp, 65, 251

Kippenhahn R., Weigert A., 1990, Stellar Structure and Evolution. Springer, Berlin

Kitaura F. S., Janka H.-T., Hillebrandt W., 2006, A&A, 450, 345

Kiziltan B., Kottas A., Thorsett S. E., 2011, ArXiv e-prints, 1011.4291

Kiziltan B., Thorsett S. E., 2010, ApJ, 715, 335

Kleiser I. K. W., Kasen D., 2014, MNRAS, 438, 318

Kluzniak W., Ruderman M., Shaham J., Tavani M., 1988, Nature, 334, 225

Knigge C., Coe M. J., Podsiadlowski P., 2011, Nature, 479, 372

Knispel B., Allen B., Cordes J. M., et al., 2010, Science, 329, 1305

Knispel B. et al., 2011, ApJ, 732, L1

Koester D., 2010, Mem. Soc. Astron. Italiana, 81, 921

Koester D., Reimers D., 2000, A&A, 364, L66

Kolb U., Ritter H., 1990, A&A, 236, 385

Konar S., Bhattacharya D., 1997, MNRAS, 284, 311

Kozai Y., 1962, AJ, 67, 591

Kramer M., Lyne A. G., O'Brien J. T., Jordan C. A., Lorimer D. R., 2006a, Science, 312, 549

Kramer M. et al., 2006b, Science, 314, 97

Kramer M., Wex N., 2009, Classical and Quantum Gravity, 26, 073001

Kramer M., Xilouris K. M., Lorimer D. R., Doroshenko O., Jessner A., Wielebinski R., Wolszczan A., Camilo F., 1998, ApJ, 501, 270

Kuijken K., Gilmore G., 1989, MNRAS, 239, 605

Kulkarni S. R., 1986, ApJ, 306, L85

Kulkarni S. R., Narayan R., 1988, ApJ, 335, 755

Lamb F., Yu W., 2005, in Astronomical Society of the Pacific Conference Series, Vol. 328, Binary Radio Pulsars, F. A. Rasio & I. H. Stairs, ed., p. 299

Lamb F. K., Pethick C. J., Pines D., 1973, ApJ, 184, 271

Lange C., Camilo F., Wex N., Kramer M., Backer D. C., Lyne A. G., Doroshenko O., 2001, MNRAS, 326, 274

Langer N., 1989, A&A, 210, 93

Langer N., 1991, A&A, 252, 669

Langer N., 1998, A&A, 329, 551

Langer N., 2012, ARA&A, 50, 107

Langer N., Deutschmann A., Wellstein S., Höflich P., 2000, A&A, 362, 1046

Lasota J.-P., 2001, New Astron. Rev., 45, 449

Lattimer J. M., Prakash M., 2004, Science, 304, 536

Lattimer J. M., Prakash M., 2007, Phys. Rep., 442, 109

Lattimer J. M., Prakash M., 2010, ArXiv e-prints, 1012.3208

Lattimer J. M., Yahil A., 1989, ApJ, 340, 426

Lazaridis K. et al., 2011, MNRAS, 584

Lazaridis K. et al., 2009, MNRAS, 400, 805

Lazarus P., 2013, in IAU Symposium, Vol. 291, IAU Symposium, van Leeuwen J., ed., pp. 35–40

Lazarus P. et al., 2014, MNRAS, 437, 1485

Lesaffre P., Han Z., Tout C. A., Podsiadlowski P., Martin R. G., 2006, MNRAS, 368, 187

Lewin W. H. G., van der Klis M., 2006, Compact stellar X-ray sources. Cambridge University Press

Li J., Wickramasinghe D. T., 1998, MNRAS, 300, 1015

Li L.-X., Paczyński B., 1998, ApJ, 507, L59

Li X.-D., 2002, ApJ, 564, 930

Li X.-D., van den Heuvel E. P. J., 1997, A&A, 322, L9

Liebert J., Bergeron P., Holberg J. B., 2003, AJ, 125, 348

Lin J., Rappaport S., Podsiadlowski P., Nelson L., Paxton B., Todorov P., 2011, ApJ, 732, 70

Lindblom L., 1999, Phys. Rev. D, 60, 064007

Linden T., Kalogera V., Sepinsky J. F., Prestwich A., Zezas A., Gallagher J. S., 2010, ApJ, 725, 1984

Lipunov V. M., Postnov K. A., 1984, Ap&SS, 106, 103

Liu Q. Z., van Paradijs J., van den Heuvel E. P. J., 2006, A&A, 455, 1165

Liu Q. Z., van Paradijs J., van den Heuvel E. P. J., 2007, A&A, 469, 807

Liu W.-M., Chen W.-C., Wang B., Han Z. W., 2010, A&A, 523, A3

Liu X.-W., Li X.-D., 2009, ApJ, 692, 723

Liu Z. W., Pakmor R., Röpke F. K., Edelmann P., Wang B., Kromer M., Hillebrandt W., Han Z. W., 2012, A&A, 548, A2

Livio M., 2000, in Type Ia Supernovae, Theory and Cosmology, J. C. Niemeyer & J. W. Truran, ed., pp. 33–+

Locsei J. T., Melatos A., 2004, MNRAS, 354, 591

Lorimer D. R., 1994, PhD thesis, The University of Manchester

Lorimer D. R. et al., 2006a, MNRAS, 372, 777

Lorimer D. R., Kramer M., 2004, Handbook of Pulsar Astronomy, Cambridge University Press

Lorimer D. R., Lyne A. G., Festin L., Nicastro L., 1995, Nature, 376, 393

Lorimer D. R., Lyne A. G., McLaughlin M. A., Kramer M., Pavlov G. G., Chang C., 2012, ApJ, 758, 141

Lorimer D. R., Stairs I. H., Freire P. C., et al., 2006b, ApJ, 640, 428

Loveridge A. J., van der Sluys M. V., Kalogera V., 2011, ApJ, 743, 49

Lynch R. S., Boyles J., Ransom S. M., er al., 2013, ApJ, 763, 81

Lynch R. S., Freire P. C. C., Ransom S. M., Jacoby B. A., 2012, ApJ, 745, 109

Lyne A. G., Biggs J. D., Harrison P. A., Bailes M., 1993, Nature, 361, 47

Lyne A. G., Lorimer D. R., 1994, Nature, 369, 127

Lyne A. G., Manchester R. N., 1988, MNRAS, 234, 477

Lyne A. G., Manchester R. N., D'Amico N., 1996, ApJ, 460, L41

Ma B., Li X.-D., 2009, ApJ, 691, 1611

Ma X., Chen X., Chen H.-l., Denissenkov P. A., Han Z., 2013, ApJ, 778, L32

Maggiore M., 2008, Gravitational Waves: Volume 1: Theory and Experiments. Cambridge University Press

Manchester R. N., 1995, Journal of Astrophysics and Astronomy, 16, 107

Manchester R. N., Hobbs G. B., Teoh A., Hobbs M., 2005, AJ, 129, 1993

Manchester R. N., Taylor J. H., 1977, Pulsars. W. H. Freeman, San Francisco

Mardling R. A., Aarseth S. J., 2001, MNRAS, 321, 398

Marietta E., Burrows A., Fryxell B., 2000, ApJS, 128, 615

Maxted P. F. L. et al., 2013, Nature, 498, 463

Mereghetti S., La Palombara N., Tiengo A., Pizzolato F., Esposito P., Woudt P. A., Israel G. L., Stella L., 2011, ApJ, 737, 51

Merloni A. et al., 2012, ArXiv e-prints

Metzger B. D. et al., 2010, MNRAS, 406, 2650

Metzger B. D., Piro A. L., Quataert E., 2009, MNRAS, 396, 1659

Meyer F., Meyer-Hofmeister E., 1983, A&A, 121, 29

Michel F. C., 1987, Nature, 329, 310

Michel F. C., 1991, Theory of neutron star magnetospheres. The University of Chigaco Press

Mikkola S., 2008, in Multiple Stars Across the H-R Diagram, Hubrig S., Petr-Gotzens M., Tokovinin A., eds., p. 11

Mirabel I. F., Dijkstra M., Laurent P., Loeb A., Pritchard J. R., 2011, A&A, 528, A149+

Miyaji S., Nomoto K., 1987, ApJ, 318, 307

Miyaji S., Nomoto K., Yokoi K., Sugimoto D., 1980, PASJ, 32, 303

Modjaz M., Li W., Butler N., et al., 2009, ApJ, 702, 226

Moriya T., Tominaga N., Tanaka M., Nomoto K., Sauer D. N., Mazzali P. A., Maeda K., Suzuki T., 2010, ApJ, 719, 1445

Motz L., 1952, ApJ, 115, 562

Nagase F., 1989, PASJ, 41, 1

Navarro J., de Bruyn A. G., Frail D. A., Kulkarni S. R., Lyne A. G., 1995, ApJ, 455, L55+

Nelemans G., Yungelson L. R., van der Sluys M. V., Tout C. A., 2010, MNRAS, 401, 1347

Nelson L. A., Dubeau E., MacCannell K. A., 2004, ApJ, 616, 1124

Nelson R. W. et al., 1997, ApJ, 488, L117

Newsham G., Starrfield S., Timmes F., 2013, ArXiv astro-ph: 1303.3642

Nice D. J., Altiere E., Bogdanov S., et al., 2013, ApJ, 772, 50

Nice D. J., Splaver E. M., Stairs I. H., 2003, in Astronomical Society of the Pacific Conference Series, Vol. 302, Radio Pulsars, M. Bailes, D. J. Nice, & S. E. Thorsett, ed., pp. 75–+

Nice D. J., Stairs I. H., Kasian L. E., 2008, in American Institute of Physics Conference Series, Vol. 983, 40 Years of Pulsars: Millisecond Pulsars, Magnetars and More, C. Bassa, Z. Wang, A. Cumming, & V. M. Kaspi, ed., pp. 453–458

Nomoto K., 1982, ApJ, 253, 798

Nomoto K., 1984, ApJ, 277, 791

Nomoto K., 1987, ApJ, 322, 206

Nomoto K., Kondo Y., 1991, ApJ, 367, L19

Nomoto K., Miyaji S., Sugimoto D., Yokoi K., 1979, in IAU Colloq. 53: White Dwarfs and Variable Degenerate Stars, van Horn H. M., Weidemann V., eds., pp. 56–60

Nomoto K., Nariai K., Sugimoto D., 1979, PASJ, 31, 287

Nomoto K., Saio H., Kato M., Hachisu I., 2007, ApJ, 663, 1269

Nomoto K., Sugimoto D., 1977, PASJ, 29, 765

Nomoto K., Thielemann F.-K., Yokoi K., 1984, ApJ, 286, 644

Nomoto K., Yamaoka H., Pols O. R., van den Heuvel E. P. J., Iwamoto K., Kumagai S., Shigeyama T., 1994, Nature, 371, 227

Nordtvedt K., 1990, Physical Review Letters, 65, 953

Olausen S. A., Kaspi V. M., 2014, ApJS, 212, 6

Ouyed R., Dey J., Dey M., 2002, A&A, 390, L39

Özel F., Psaltis D., Ransom S., Demorest P., Alford M., 2010, ApJ, 724, L199

Pacini F., 1967, Nature, 216, 567

Paczyński B., 1976, in IAU Symposium, Vol. 73, Structure and Evolution of Close Binary Systems, P. Eggleton, S. Mitton, & J. Whelan, ed., Dordrecht, Holland, pp. 75–+

Paczynski B., 1986, ApJ, 308, L43

Paczyński B., Sienkiewicz R., 1972, Acta Astronomica, 22, 73

Pakmor R., Kromer M., Taubenberger S., Sim S. A., Röpke F. K., Hillebrandt W., 2012, ApJ, 747, L10

Pan K.-C., Ricker P. M., Taam R. E., 2010, ApJ, 715, 78

Pan K.-C., Ricker P. M., Taam R. E., 2012, ApJ, 750, 151

Pan K.-C., Ricker P. M., Taam R. E., 2013, ApJ, 773, 49

Panei J. A., Althaus L. G., Chen X., Han Z., 2007, MNRAS, 382, 779

Papitto A. et al., 2011, A&A, 535, L4

Papitto A. et al., 2013, Nature, 501, 517

Papitto A., Torres D. F., Rea N., Tauris T., 2014, ArXiv astro-ph: 1403.6775

Passy J.-C. et al., 2012, ApJ, 744, 52

Pastetter L., Ritter H., 1989, A&A, 214, 186

Patruno A., 2010a, ArXiv astro.ph/1007.1108

Patruno A., 2010b, ApJ, 722, 909

Patruno A., Alpar M. A., van der Klis M., van den Heuvel E. P. J., 2012, ApJ, 752, 33

Patruno A., Watts A. L., 2012, in Timing Neutron Stars: Pulsations, Oscillations and Explosions, T. Belloni, M. Mendez, ed., Astrophysics and Space Science, Springer

Patterson J. et al., 2013, ArXiv astro-ph: 1303.0736

Paxton B., Bildsten L., Dotter A., Herwig F., Lesaffre P., Timmes F., 2011, ApJS, 192, 3

Payne D. J. B., Melatos A., 2007, MNRAS, 376, 609

Perets H. B., Gal-Yam A., Mazzali P. A., et al., 2010, Nature, 465, 322

Perlmutter S. et al., 1999, ApJ, 517, 565

Pfahl E., Rappaport S., Podsiadlowski P., 2003, ApJ, 597, 1036

Pfahl E., Rappaport S., Podsiadlowski P., Spruit H., 2002, ApJ, 574, 364

Phinney E. S., 1992, Royal Society of London Philosophical Transactions Series A, 341, 39

Phinney E. S., Kulkarni S. R., 1994, ARA&A, 32, 591

Pietrzyński G. et al., 2012, Nature, 484, 75

Pijloo J. T., Caputo D. P., Portegies Zwart S. F., 2012, MNRAS, 424, 2914

Piro A. L., 2008, ApJ, 679, 616

Piro A. L., Kulkarni S. R., 2013, ApJ, 762, L17

Podsiadlowski P., 1991, Nature, 350, 136

Podsiadlowski P., 2001, in Astronomical Society of the Pacific Conference Series, Vol. 229, Evolution of Binary and Multiple Star Systems, P. Podsiadlowski, S. Rappaport, A. R. King, F. D'Antona, & L. Burderi , ed., pp. 239–+

Podsiadlowski P., Dewi J. D. M., Lesaffre P., Miller J. C., Newton W. G., Stone J. R., 2005, MNRAS, 361, 1243

Podsiadlowski P., Langer N., Poelarends A. J. T., Rappaport S., Heger A., Pfahl E., 2004, ApJ, 612, 1044

Podsiadlowski P., Rappaport S., 2000, ApJ, 529, 946

Podsiadlowski P., Rappaport S., Pfahl E. D., 2002, ApJ, 565, 1107

Poelarends A. J. T., Herwig F., Langer N., Heger A., 2008, ApJ, 675, 614

Popham R., Narayan R., 1991, ApJ, 370, 604

Portegies Zwart S., van den Heuvel E. P. J., van Leeuwen J., Nelemans G., 2011, ApJ, 734, 55

Possenti A., 2013, in IAU Symposium, Vol. 291, IAU Symposium, van Leeuwen J., ed., pp. 121–126

Prialnik D., Kovetz A., 1995, ApJ, 445, 789

Pringle J. E., 1981, ARA&A, 19, 137

Pringle J. E., Rees M. J., 1972, A&A, 21, 1

Pylyser E., Savonije G. J., 1988, A&A, 191, 57

Pylyser E. H. P., Savonije G. J., 1989, A&A, 208, 52

Radhakrishnan V., Srinivasan G., 1982, Current Science, 51, 1096

Rankin J. M., 1983, ApJ, 274, 333

Rankin J. M., 1990, ApJ, 352, 247

Ransom S. M., Hessels J. W. T., Stairs I. H., Freire P. C. C., Camilo F., Kaspi V. M., Kaplan D. L., 2005, Science, 307, 892

Ransom S. M., Stairs I. H., Archibald A. M., Hessels J., Kaplan D. L., van Kerkwijk M. H., et al., 2014, Nature, 505, 520

Rappaport S., Deck K., Levine A., Borkovits T., Carter J., El Mellah I., Sanchis-Ojeda R., Kalomeni B., 2013, ApJ, 768, 33

Rappaport S., Podsiadlowski P., Joss P. C., Di Stefano R., Han Z., 1995, MNRAS, 273, 731

Rappaport S., Verbunt F., Joss P. C., 1983, ApJ, 275, 713

Rappaport S. A., Fregeau J. M., Spruit H., 2004, ApJ, 606, 436

Rasio F. A., Heggie D. C., 1995, ApJ, 445, L133

Rawls M. L., Orosz J. A., McClintock J. E., Torres M. A. P., Bailyn C. D., Buxton M. M., 2011, ApJ, 730, 25

Refsdal S., Weigert A., 1971, A&A, 13, 367

Reimers D., 1975, Circumstellar envelopes and mass loss of red giant stars, Springer-Verlag, New York, pp. 229–256

Reisenegger A., 2003, ArXiv astro-ph: 0307133

Riess A. G. et al., 1998, AJ, 116, 1009

Ritter H., 1988, A&A, 202, 93

Ritter H., 2008, New A Rev., 51, 869

Roberts M. S. E., 2013, in IAU Symposium, Vol. 291, IAU Symposium, van Leeuwen J., ed., pp. 127–132

Romani R. W., 1990, Nature, 347, 741

Romani R. W., Filippenko A. V., Silverman J. M., Cenko S. B., Greiner J., Rau A., Elliott J., Pletsch H. J., 2012, ApJ, 760, L36

Rosenfeld L., 1974, in Proceedings of the Solvay Conference on Physics, Vol. 16, Astrophysics and Gravitation. Proceedings of the Sixtenth Solvay Conference on Physics, Edoardo Amaldi, ed., p. 174

Ruderman M., Shaham J., Tavani M., 1989, ApJ, 336, 507

Ruderman M., Shaham J., Tavani M., Eichler D., 1989, ApJ, 343, 292

Rutledge R. E., Bildsten L., Brown E. F., Pavlov G. G., Zavlin V. E., 1999, ApJ, 514, 945

Saio H., Nomoto K., 1985, A&A, 150, L21

Saio H., Nomoto K., 2004, ApJ, 615, 444

Sapir N., Katz B., Waxman E., 2013, ApJ, 774, 79

Sathyaprakash B. S., Schutz B. F., 2009, Living Reviews in Relativity, 12, 2

Savonije G. J., 1987, Nature, 325, 416

Scalzo R. A. et al., 2010, ApJ, 713, 1073

Schlegel D. J., Finkbeiner D. P., Davis M., 1998, ApJ, 500, 525

Schwab J., Podsiadlowski P., Rappaport S., 2010, ApJ, 719, 722

Serenelli A. M., Althaus L. G., Rohrmann R. D., Benvenuto O. G., 2001, MNRAS, 325, 607

Serenelli A. M., Althaus L. G., Rohrmann R. D., Benvenuto O. G., 2002, MNRAS, 337, 1091

Shapiro S. L., Teukolsky S. A., 1983, Black holes, white dwarfs, and neutron stars: The physics of compact objects. Wiley-Interscience, New York

Shibazaki N., Murakami T., Shaham J., Nomoto K., 1989, Nature, 342, 656

Shklovskii I. S., 1970, Soviet Ast., 13, 562

Shklovsky I. S., 1967, ApJ, 148, L1

Sigurdsson S., Phinney E. S., 1993, ApJ, 415, 631

Sion E. M., 1999, PASP, 111, 532

Smarr L. L., Blandford R., 1976, ApJ, 207, 574

Soberman G. E., Phinney E. S., van den Heuvel E. P. J., 1997, A&A, 327, 620

Soderberg A. M., Berger E., Page K. L., et al., 2008, Nature, 453, 469

Spitkovsky A., 2006, ApJ, 648, L51

Spitkovsky A., 2008, in American Institute of Physics Conference Series, Vol. 983, 40 Years of Pulsars: Millisecond Pulsars, Magnetars and More, C. Bassa, Z. Wang, A. Cumming, & V. M. Kaspi, ed., pp. 20–28

Splaver E. M., 2004, PhD thesis, Princeton University

Splaver E. M., Nice D. J., Arzoumanian Z., Camilo F., Lyne A. G., Stairs I. H., 2002, ApJ, 581, 509

Splaver E. M., Nice D. J., Stairs I. H., Lommen A. N., Backer D. C., 2005, ApJ, 620, 405

Spruit H. C., 2008, in American Institute of Physics Conference Series, Vol. 983, 40 Years of Pulsars: Millisecond Pulsars, Magnetars and More, Bassa C., Wang Z., Cumming A., Kaspi V. M., eds., pp. 391–398

Spruit H. C., Taam R. E., 1993, ApJ, 402, 593

Srinivasan G., Bhattacharya D., Muslimov A. G., Tsygan A. J., 1990, Current Science, 59, 31

Stairs I. H., 2003, Living Reviews in Relativity, 6, 5

Stairs I. H., Arzoumanian Z., Camilo F., Lyne A. G., Nice D. J., Taylor J. H., Thorsett S. E., Wolszczan A., 1998, ApJ, 505, 352

Stairs I. H. et al., 2005, ApJ, 632, 1060

Stairs I. H., Thorsett S. E., Taylor J. H., Wolszczan A., 2002, ApJ, 581, 501

Stappers B. W. et al., 1996, ApJ, 465, L119

Starrfield S., Timmes F. X., Iliadis C., Hix W. R., Arnett W. D., Meakin C., Sparks W. M., 2012, Baltic Astronomy, 21, 76

Steiner A. W., Lattimer J. M., Brown E. F., 2013, ApJ, 765, L5

Steinfadt J. D. R., Bildsten L., Arras P., 2010, ApJ, 718, 441

Sutantyo W., 1974, A&A, 35, 251

Sutantyo W., Li X.-D., 2000, A&A, 360, 633

Taam R. E., Sandquist E. L., 2000, ARA&A, 38, 113

Taam R. E., van den Heuvel E. P. J., 1986, ApJ, 305, 235

Tanvir N. R., Levan A. J., Fruchter A. S., Hjorth J., Hounsell R. A., Wiersema K., Tunnicliffe R. L., 2013, Nature, 500, 547

Tauris T. M., 1996, A&A, 315, 453

Tauris T. M., 1997, PhD thesis, Aarhus University

Tauris T. M., 2011, in Astronomical Society of the Pacific Conference Series, Vol. 447, Evolution of Compact Binaries, Schmidtobreick L., Schreiber M. R., Tappert C., eds., p. 285

Tauris T. M., 2012, Science, 335, 561

Tauris T. M., Bailes M., 1996, A&A, 315, 432

Tauris T. M., Dewi J. D. M., 2001, A&A, 369, 170

Tauris T. M., Konar S., 2001, A&A, 376, 543

Tauris T. M., Langer N., Kramer M., 2011, MNRAS, 416, 2130

Tauris T. M., Langer N., Kramer M., 2012, MNRAS, 425, 1601

Tauris T. M., Langer N., Moriya T. J., Podsiadlowski P., Yoon S.-C., Blinnikov S. I., 2013a, ApJ, 778, L23

Tauris T. M., Manchester R. N., 1998, MNRAS, 298, 625

Tauris T. M., Sanyal D., Yoon S.-C., Langer N., 2013b, A&A, 558, A39

Tauris T. M., Savonije G. J., 1999, A&A, 350, 928

Tauris T. M., Savonije G. J., 2001, in The Neutron Star - Black Hole Connection, C. Kouveliotou, J. Ventura, & E. van den Heuvel, ed., pp. 337–+

Tauris T. M., Sennels T., 2000, A&A, 355, 236

Tauris T. M., Takens R. J., 1998, A&A, 330, 1047

Tauris T. M., van den Heuvel E. P. J., 2006, Formation and evolution of compact stellar X-ray sources, Cambridge University Press, pp. 623–665

Tauris T. M., van den Heuvel E. P. J., 2014, ApJ, 781, L13

Tauris T. M., van den Heuvel E. P. J., Savonije G. J., 2000, ApJ, 530, L93

Taylor J. H., 1992, Royal Society of London Philosophical Transactions Series A, 341, 117

Taylor J. H., Weisberg J. M., 1989, ApJ, 345, 434

Thorne K. S., Żytkow A. N., 1977, ApJ, 212, 832

Thorsett S. E., Chakrabarty D., 1999, ApJ, 512, 288

Thoul A. A., Bahcall J. N., Loeb A., 1994, ApJ, 421, 828

Timmes F. X., Woosley S. E., 1992, ApJ, 396, 649

Timmes F. X., Woosley S. E., Weaver T. A., 1996, ApJ, 457, 834

Tokovinin A., Thomas S., Sterzik M., Udry S., 2006, A&A, 450, 681

Tolstov A. G., Blinnikov S. I., Nadyozhin D. K., 2013, MNRAS, 429, 3181

Tremblay P.-E., Bergeron P., 2009, ApJ, 696, 1755

Tremblay P.-E., Ludwig H.-G., Steffen M., Bergeron P., Freytag B., 2011, A&A, 531, L19

Tremblay P.-E., Ludwig H.-G., Steffen M., Freytag B., 2013, A&A, 552, A13

Trümper J. E., 2005, in NATO ASIB Proc. 210: The Electromagnetic Spectrum of Neutron Stars, Baykal A., Yerli S. K., Inam S. C., Grebenev S., eds., p. 117

Ugliano M., Janka H.-T., Marek A., Arcones A., 2012, ApJ, 757, 69

Umeda H., Yoshida T., Takahashi K., 2012, Prog. Theor. Exp. Phys., DOI: 10.1093/ptep/pts017

van den Heuvel E. P. J., 1994a, in Saas-Fee Advanced Course 22: Interacting Binaries, Shore S. N., Livio M., van den Heuvel E. P. J., Nussbaumer H., Orr A., eds., pp. 263–474

van den Heuvel E. P. J., 1994b, A&A, 291, L39

van den Heuvel E. P. J., 2004, in ESA Special Publication, Vol. 552, 5th INTEGRAL Workshop on the INTEGRAL Universe, V. Schoenfelder, G. Lichti, & C. Winkler, ed., p. 185

van den Heuvel E. P. J., Bhattacharya D., Nomoto K., Rappaport S. A., 1992, A&A, 262, 97

van den Heuvel E. P. J., Bitzaraki O., 1994, Mem. Soc. Astron. Italiana, 65, 237

van den Heuvel E. P. J., Bitzaraki O., 1995, A&A, 297, L41

van der Klis M., 2006, Rapid X-ray Variability, Cambridge University Press, pp. 39–112

van der Sluys M. V., Verbunt F., Pols O. R., 2005, A&A, 431, 647

van Haaften L. M., Nelemans G., Voss R., Jonker P. G., 2012a, A&A, 541, A22

van Haaften L. M., Nelemans G., Voss R., Wood M. A., Kuijpers J., 2012b, A&A, 537, A104

van Kerkwijk M. H., Bassa C. G., Jacoby B. A., Jonker P. G., 2005, in Astronomical Society of the Pacific Conference Series, Vol. 328, Binary Radio Pulsars, F. A. Rasio & I. H. Stairs, ed., pp. 357–+

van Kerkwijk M. H., Breton R. P., Kulkarni S. R., 2011, ApJ, 728, 95

van Kerkwijk M. H., Chakrabarty D., Pringle J. E., Wijers R. A. M. J., 1998, ApJ, 499, L27

van Kerkwijk M. H., Chang P., Justham S., 2010, ApJ, 722, L157

van Kerkwijk M. H., Rappaport S. A., Breton R. P., Justham S., Podsiadlowski P., Han Z., 2010, ApJ, 715, 51

van Paradijs J., 1996, ApJ, 464, L139+

van Paradijs J., van den Heuvel E. P. J., Kouveliotou C., Fishman G. J., Finger M. H., Lewin W. H. G., 1997, A&A, 317, L9

Vasyliunas V. M., 1979, Space Sci. Rev., 24, 609

Verbiest J. P. W. et al., 2008, ApJ, 679, 675

Verbunt F., Freire P. C. C., 2014, A&A, 561, A11

Verbunt F., Wijers R. A. M. J., Burm H. M. G., 1990, A&A, 234, 195

Vink J. S., de Koter A., Lamers H. J. G. L. M., 2001, A&A, 369, 574

Voss R., Tauris T. M., 2003, MNRAS, 342, 1169

Wade L., Siemens X., Kaplan D. L., Knispel B., Allen B., 2012, Phys. Rev. D, 86, 124011

Wang B., Han Z., 2010, A&A, 515, A88

Wang B., Li X.-D., Han Z.-W., 2010, MNRAS, 401, 2729

Wang J., Zhang C. M., Zhao Y. H., Kojima Y., Yin H. X., Song L. M., 2011, A&A, 526, A88

Wang Y.-M., 1997, ApJ, 475, L135+

Webbink R. F., 1984, ApJ, 277, 355

Webbink R. F., Rappaport S., Savonije G. J., 1983, ApJ, 270, 678

Weber F., 2005, Progress in Particle and Nuclear Physics, 54, 193

Weisberg J. M., Nice D. J., Taylor J. H., 2010, ApJ, 722, 1030

Weisberg J. M., Taylor J. H., 1981, General Relativity and Gravitation, 13, 1

Wellstein S., Langer N., 1999, A&A, 350, 148

Wex N., 2014, ArXiv astro-ph: 1402.5594

Wheeler J. C., 2012, ApJ, 758, 123

Wheeler J. C., Cowan J. J., Hillebrandt W., 1998, ApJ, 493, L101

Whelan J., Iben, Jr. I., 1973, ApJ, 186, 1007

Wickramasinghe D. T., Ferrario L., 2000, PASP, 112, 873

Wiescher M., Gorres J., Thielemann F.-K., Ritter H., 1986, A&A, 160, 56

Wijers R. A. M. J., 1997, MNRAS, 287, 607

Wijnands R., van der Klis M., 1998, Nature, 394, 344

Will C. M., 1993, Theory and Experiment in Gravitational Physics

Will C. M., 1994, Phys. Rev. D, 50, 6058

Will C. M., 2006, Living Reviews in Relativity, 9, 3

Will C. M., 2009, Space Sci. Rev., 148, 3

Will C. M., 2011, Proceedings of the National Academy of Science, 108, 5938

Wolszczan A. et al., 2000, ApJ, 528, 907

Woltjer L., 1964, ApJ, 140, 1309

Wongwathanarat A., Janka H.-T., Müller E., 2013, A&A, 552, A126

Woosley S. E., Heger A., Weaver T. A., 2002, Reviews of Modern Physics, 74, 1015

Woosley S. E., Langer N., Weaver T. A., 1995, ApJ, 448, 315

Woosley S. E., Weaver T. A., 1995, ApJS, 101, 181

Worley A., Krastev P. G., Li B., 2008, ApJ, 685, 390

Xu X., Li X., 2010a, ApJ, 722, 1985

Xu X., Li X., 2010b, ApJ, 716, 114

Yakovlev D. G., Haensel P., Baym G., Pethick C., 2013, Physics Uspekhi, 56, 289

Yi I., Wheeler J. C., Vishniac E. T., 1997, ApJ, 481, L51

Yoon S., Langer N., Norman C., 2006, A&A, 460, 199

Yoon S., Woosley S. E., Langer N., 2010, ApJ, 725, 940

Yoon S.-C., Gräfener G., Vink J. S., Kozyreva A., Izzard R. G., 2012, A&A, 544, L11

Yoon S.-C., Langer N., 2003, A&A, 412, L53

Yoon S.-C., Langer N., 2004, A&A, 419, 623

Yoon S.-C., Langer N., 2005, A&A, 435, 967

Yoon S.-C., Podsiadlowski P., Rosswog S., 2007, MNRAS, 380, 933

Yungelson L. R., Nelemans G., van den Heuvel E. P. J., 2002, A&A, 388, 546

Zeldovich Y. B., Novikov I. D., 1971, Relativistic astrophysics. Vol.1: Stars and relativity. University of Chicago Press, Chicago

Zhang C. M., 1998, A&A, 330, 195

Zhang C. M., Kojima Y., 2006, MNRAS, 366, 137

Zhang W., Woosley S. E., Heger A., 2008, ApJ, 679, 639

Zorotovic M., Schreiber M. R., Gänsicke B. T., Nebot Gómez-Morán A., 2010, A&A, 520, A86

*Thomas M. Tauris - Uni. Bonn*